\documentclass[krantz2]{krantz}

\usepackage{amssymb, amsmath, bbm, bm, mathtools, enumitem, url, float}
\usepackage{graphicx, color, subfig}
\usepackage{makeidx}
\usepackage{multicol}
\usepackage[all,import]{xy}
\usepackage{natbib}

\definecolor{mygray}{gray}{0.6}
\usepackage{listings}
 
\lstnewenvironment{rc}[1][]{\lstset{language=R}}{}
\newcommand{\ri}[1]{\lstinline{#1}}

\newenvironment{myproof}[2] {\noindent {\bfseries Proof of {#1} {#2}:}}{\hfill$\square$}

\makeindex

\usepackage{hyperref}
\makeatletter
\providecommand*{\Hy@tocdestname}{}
\makeatother

\newtheorem{lemma}{Lemma}
\newtheorem{proposition}{Proposition}
\newtheorem{assumption}{Assumption}

\newtheorem{condition}{Condition}
\newtheorem{corollary}{Corollary}

\newtheorem{theorem}{Theorem}

\newtheorem{example}{Example}
\newtheorem{definition}{Definition}

\def\ind{\begin{picture}(9,8)
         \put(0,0){\line(1,0){9}}
         \put(3,0){\line(0,1){8}}
         \put(6,0){\line(0,1){8}}
         \end{picture}
        }

\def\var{\textup{var}}
\def\cov{\textup{cov}}
\def\sumn{\sum_{i=1}^n}

\def\iidsim{\stackrel{\textsc{iid}}{\sim}}

\def\pr{\textup{pr}}

\def\d{\textnormal{d}}
\def\N{\textsc{N}}

\def\asim{\stackrel{\textup{a}}{\sim}}
\def\diff{\textup{d}}

\def\T{\tiny\textsc{t}}
\def\rss{\textsc{rss}}
\def\ehw{\textsc{ehw}}

\begin{document}

\frontmatter

\title{Linear Model and Extensions}
\author{Peng Ding\\
Department of Statistics\\
University of California, Berkeley}

\maketitle

\cleardoublepage
\thispagestyle{empty}
\vspace*{\stretch{1}}
\begin{center}
\Large\itshape

To students and readers\\
who are interested in linear models\\
and their applications to real-world problems

\end{center}
\vspace{\stretch{2}}


\tableofcontents

  \chapter*{Acronyms}

I try hard to avoid using acronyms to reduce the unnecessary burden for reading. However, the following acronyms are standard and will be used repeatedly. 

\begin{tabular}{ll}
ANOVA & (Fisher's) analysis of variance \\ 
BLUE & best linear unbiased estimator (in Gauss--Markov Theorem) \\ 
CATE & conditional average treatment effect \\
CDF& cumulative distribution function\\ 
CLT & central limit theorem\\
CV& cross-validation\\ 
EHW& Eicker--Huber--White (robust covariance matrix or standard error)\\
FDA &  U.S. Food and Drug Administration \\  
FWL & Frisch--Waugh--Lovell (theorem)\\ 
GEE & generalized estimating equation\\ 
GLM& generalized linear model\\
HC& heteroskedasticity-consistent (covariance matrix or standard error)\\ 
IID &  independent and identically distributed\\
LAD & least absolute deviations\\ 
lasso & least absolute shrinkage and selection operator \\
MLE& maximum likelihood estimate\\ 
OLS& ordinary least squares\\
PCA& principal component analysis\\ 
RCT& randomized controlled trial\\ 
RSS& residual sum of squares\\ 
SVD& singular value decomposition\\ 
WLS& weighted least squares
\end{tabular}

  \chapter*{Symbols}

All vectors are column vectors as in  \ri{R} unless stated otherwise. Let the superscript ``$^{\T}$''
denote the transpose of a vector or matrix.

\begin{tabular}{ll}
$\mathbb{R}$ & the set of all real numbers \\ 
$\mathbb{R}^p$ & the set of $p$-dimensional vectors with real components \\ 
$\ind$& independence and conditional independence\\ 
$\stackrel{\textup{IID}}{\sim}$ & independent and identically distributed (IID)\\
$\asim$ & approximation in distribution \\ 
$I_n$ & identity matrix of dimension $n\times n$\\ 
$x_i$ & covariate vector for unit $i$ \\
$y_i$ & outcome for unit $i$ \\
$X$ & covariate matrix \\
$Y$ & outcome vector \\
$H$ & hat matrix $H = X(X^{\T}X)^{-1}X^{\T}$\\ 
$h_{ii}$ & leverage score: the $(i, i)$the element of the hat matrix $H$ \\
$\beta$ & regression coefficient\\
$\varepsilon$ & error term 
\end{tabular}
  \chapter*{Useful R packages}

This book uses the following \ri{R} packages and functions. 

\begin{tabular}{lll}
package & function or data & use \\
\hline 
\texttt{car} & \texttt{hccm} & Eicker--Huber--White robust standard error\\
& \texttt{linearHypothesis} & testing linear hypotheses in linear models\\
\texttt{foreign} & \texttt{read.dta} & read stata data\\
\texttt{gee} & \texttt{gee} & Generalized estimating equation\\
\texttt{HistData} & \texttt{GaltonFamilies} & Galton's data on parents' and children's heights\\ 
\texttt{MASS} & \texttt{lm.ridge} & ridge regression\\
& \texttt{glm.nb} & Negative-Binomial regression \\
\texttt{glmnet}  & \texttt{cv.glmnet}  & Lasso with cross-validation \\ 
\texttt{mlbench}  & \texttt{BostonHousing}  & Boston housing data \\ 
& \texttt{polr} & proportional odds logistic regression \\ 
\texttt{Matching} & \texttt{lalonde} & LaLonde data\\
\texttt{nnet} & \texttt{multinom} & Multinomial logistic regression\\
\texttt{quantreg} & \texttt{rq} & quantile regression\\
\texttt{survival} & \texttt{coxph} & Cox proportional hazards regression\\ 
& \texttt{survdiff} & log rank test\\
& \texttt{survfit} & Kaplan--Meier curve \\
\texttt{ElemStatLearn} & \texttt{prostate} & data for \citet{hastie2009elements} \\
\texttt{wooldridge} &  \texttt{mroz} & data for \citet{wooldridge2016introductory}
\end{tabular}

\chapter*{Preface}

%
%

\subsection*{The importance of studying the linear model}

A central task in statistics is to use data to build models to make inferences about the underlying data-generating processes or make predictions of future observations. Although real problems are very complex, the linear model can often serve as a good approximation to the true data-generating process. Sometimes, although the true data-generating process is nonlinear, the linear model can be a useful approximation if we properly transform the data based on domain knowledge. Even in highly nonlinear problems, the linear model can still be a useful first attempt in the data analysis process.

Moreover, the linear model has many elegant algebraic and geometric properties. Under the linear model, we can derive many explicit formulas to gain insights about various aspects of statistical modeling. In more complicated models, deriving explicit formulas may be impossible. Nevertheless, we can use the linear model to build intuition and make conjectures about more complicated models.

Pedagogically, the linear model serves as a building block in the whole statistical training. This book builds on my lecture notes for a master's level ``Linear Model'' course at UC Berkeley, taught over the past ten years. Most students are master's students in statistics. Some are undergraduate students with strong technical preparations. Some are Ph.D. students in statistics. Some are master's or Ph.D. students in other departments. This book requires the readers to have basic training in linear algebra, probability theory, and statistical inference.

\subsection*{Recommendations for instructors}

This book has twenty-seven chapters in the main text and four appendices. As I mentioned before, this book grows out of my teaching of ``Linear Model'' at UC Berkeley. In different years, I taught the course in different ways, and this book is a union of my lecture notes over the past ten years. Below I make some recommendations for instructors based on my own teaching experience. Since UC Berkeley is on the semester system, instructors on the quarter system should make some adjustments to my recommendations below.

\paragraph*{Version 1: a basic linear model course assuming minimal technical preparations}
If you want to teach a basic linear model course without assuming strong technical preparations from the students, you can start from the appendices by reviewing basic linear algebra, probability theory, and statistical inference. Then you can cover Chapters \ref{chapter::ols-1d}--\ref{chapter::interaction}. If time permits, you can consider covering Chapter \ref{chapter::binary-logit} due to the importance of the logistic model for binary data.

\paragraph*{Version 2: an advanced linear model course assuming strong technical preparations}
If you want to teach an advanced linear model course assuming strong technical preparations from the students, you can start with the main text directly. When I did this, I asked my teaching assistants to review the appendices in the first two lab sessions and assigned homework problems from the appendices to remind the students to review the background materials. Then you can cover Chapters \ref{chapter::ols-1d}--\ref{chapter::sandwich}. You can omit Chapter \ref{chapter::rols} and some sections in other chapters due to their technical complications. If time permits, you can consider covering Chapter \ref{chapter::gee} due to the importance of the generalized estimating equation as well as its byproduct called the ``cluster-robust standard error,'' which is important for many social science applications. 
Furthermore, you can consider covering Chapter \ref{chapter::survival-analysis} due to the importance of the Cox proportional hazards model.

\paragraph*{Version 3: an advanced generalized linear models course} 
If you want to teach a course on generalized linear models, you can use Chapters \ref{chapter::binary-logit}--\ref{chapter::survival-analysis}.

\subsection*{Additional recommendations for readers and students}

Readers and students can first read my recommendations for instructors above.  In addition, I have three other recommendations.

\paragraph*{More simulation studies}
This book contains some basic simulation studies. I encourage the readers to conduct more intensive simulation studies to deepen their understanding of the theory and methods.

\paragraph*{Practical data analysis}
Box wrote wisely that ``all models are wrong but some are useful.'' The usefulness of models strongly depends on the applications. When teaching ``Linear Model,'' I sometimes replaced the final exam with the final project to encourage students to practice data analysis and make connections between the theory and applications.

\paragraph*{Homework problems}
This book contains many homework problems. It is important to try some homework problems. Moreover, some homework problems contain useful theoretical results, and some are even stated as theorems. Even if you do not have time to figure out the details for those problems,  it is helpful to at least read the statements of the problems.

\subsection*{Omitted topics}

Although ``Linear Model'' is a standard course offered by most statistics departments, it is not entirely clear what we should teach as the field of statistics is evolving. When I was teaching ``Linear Model,'' I asked myself the following question many times: 
\begin{quote}
What are the most widely used statistics methods? 
\end{quote}
Arguably, in \texttt{R}, the following five functions are the top choices by empirical researchers:
\begin{itemize}
\item
\texttt{lm()},
\item
\texttt{glm()},
\item
\texttt{coxph()} in the package \texttt{survival},
\item
\texttt{gee()} in the package \texttt{gee},
\item
\texttt{rq()} in the package \texttt{quantreg},
\end{itemize}
which are for
\begin{itemize}
\item
linear model,
\item
generalized linear model including logistic regression and Poisson regression,
\item
Cox proportional hazards model,
\item
generalized estimating equations,
\item
quantile regression,
\end{itemize}
respectively. Readers can view this book as a tutorial for these five functions. However, my selection of the topics was biased by my own experience with applied statistics. Some readers may feel that this book has omitted some important topics related to the linear model. I make some brief comments below.

\paragraph*{Bootstrap}

\citet{efron1979bootstrap} proposed the bootstrap as a powerful method to estimate the variance of general estimators. Of course, it is also useful for estimating the variances of all estimators discussed in this book. For instance, the bootstrap works well for generalized linear models. If we sample with replacement from the data $\{ x_i, y_i \}_{i=1}^n$, we can recalculate the maximum likelihood estimators to obtain the bootstrap variance estimators. It turns out that the bootstrap variance estimators approximate the sandwich variance estimators discussed in Chapter \ref{chapter::sandwich}. Therefore, bootstrap variance estimators are robust to misspecified models \citep{buja2019models, buja2019models2}. The discussion extends to clustered data. In particular, we can  the cluster bootstrap by resampling the clusters to approximate the cluster-robust standard error in Chapter \ref{sec::crse-econometrics}. Overall, the bootstrap seems like a magic!

I largely ignore the discussion of the bootstrap for at least the following four reasons. First, the statistical models in this book are all ``nice'' models that enjoy explicit analytic approximations of the variances. The bootstrap is not crucial for them. Second, this book aims to provide insights into specific models, including deriving explicit formulas of the variance estimators. Third, this book can be viewed as a tutorial for the \ri{R} functions mentioned above. Those \ri{R} functions use the explicit formulas. Fourth, although the bootstrap is intuitive, its rigorous theory requires advanced proving techniques. These technical issues are beyond the scope of this book; see \citet{wu1986jackknife, mammen1992does, shao1995jackknife} for further discussion.

Nevertheless, it is impossible to completely ignore the bootstrap in this book. In Chapter \ref{section::numerical-examples-qr}, the small simulation study on quantile regression demonstrates the advantage of the bootstrap: it performs better than other existing variance estimators based on analytic formulas because those estimators are sensitive to the choices of tuning parameters. If your computational resource permits, the bootstrap is attractive for variance estimation.

\paragraph*{Bayesian methods}
Bayesian methods are powerful in applied data analysis. 
However, Bayesian methods are more demanding for computation and require specifications of prior distributions by the users. Therefore, I feel that they are more advanced statistical methods and thus are beyond the scope of this book. \citet{hoff2009first} and \citet{gelman2013bayesian} are two excellent books on Bayesian statistics.

\paragraph*{Advanced econometric models}
After the linear model, many econometric textbooks cover the instrumental variable models and panel data models. For these more specialized topics, \citet{wooldridge2010econometric} is a canonical textbook.

\paragraph*{Advanced biostatistics models}
This book covers the generalized estimating equation in Chapter \ref{chapter::gee}. For analyzing longitudinal data, linear and generalized linear mixed effects models are powerful tools. \citet{fitzmaurice2012applied} is a canonical textbook on applied longitudinal data analysis.  
\citet{peter2007correlated} is a textbook on general correlated data analysis.

This book also covers the Cox proportional hazards model in Chapter \ref{chapter::survival-analysis}. For more advanced methods for survival analysis, \citet{kalbfleisch2011statistical} is a canonical textbook.

\paragraph*{Causal inference}
I do not cover causal inference in this book intentionally. To minimize the overlap of the materials, I wrote another textbook on causal inference \citep{ding2023first}. However,  at UC Berkeley, I did teach a version of ``Linear Model'' with a causal inference unit after introducing the basics of linear model and logistic model. Students seemed to like it because of the connections between statistical models and causal inference.

\subsection*{Features of the book}

The linear model is an old topic in statistics. There are already many excellent textbooks on the linear model. This book has the following features.

\begin{itemize}
\item
This book provides an intermediate-level introduction to the linear model. It balances rigorous proofs and intuitive explanations.

\item
This books introduces the theory of misspecified models and emphasizes its implications for practical data analysis. The mathematical theory of misspecified models dated back at least to \citet{huber::1967} and attracted research interest from academic statisticians ever since. However, it is not common to see the theory introduced in standard textbooks on statistical modeling and data analysis.

\item
This book provides not only theory but also simulation studies and case studies.

\item
This book provides the \texttt{R} code and data to replicate all simulation studies and case studies at Harvard Dataverse:
\begin{center}
\href{https://doi.org/10.7910/DVN/DBDYVJ}{https://doi.org/10.7910/DVN/DBDYVJ}
\end{center}

\item
This book covers the theory of the linear model related to social sciences and biomedical studies. 

\item
This book provides homework problems with different technical difficulties. 

\item
For instructors who teach related courses using this book,  the solutions to the problems are available upon request. 
\end{itemize}

Other textbooks may also have one or two of the above features. This book has the above features simultaneously. I hope that instructors and readers find these features attractive.

\subsection*{Acknowledgments}

Many students at UC Berkeley made critical and constructive comments on early versions of my lecture notes.  As teaching assistants for the ``Linear Model'' course,  
Sizhu Lu,
Yaxuan Huang, 
Andy Shen, 
Chaoran Yu, 
and
Jason Wu read early versions of my book carefully and helped me to improve the book a lot.

Professors 
Hongyuan Cao, 
Zhichao Jiang, 
and Richard Guo taught related courses based on an early version of the book. They made very valuable suggestions.

My research collaborations with 
Anqi Zhao, 
Lihua Lei, 
Zhichao Jiang, 
Dennis Shen, 
Wen Zhou, 
Zifeng Zhang,
and Mingrui Zhang used a lot of results related to this book. They helped me sharpen many statements in this book.

I am also very grateful for the suggestions from Professors 
Nianqiao Ju, 
Alan Agresti, 
and Peter XK Song. 
My colleague, Professor Bin Yu,  and more broadly, the intellectual environment at UC Berkeley encouraged me to teach and write more about the theory of misspecified statistical models and its implications for practical data analysis.

When I was a student, I took a linear model course based on \citet{weisberg2005applied}. 
In my early years of teaching, I used \citet{christensen2002plane} and \citet{agresti2015foundations} as reference books. 
When I was a junior faculty at UC Berkeley, I sat in Professor Jim Powell's econometrics courses and got access to his wonderful lecture notes. They all heavily impacted my understanding and formulation of the linear model.

The U.S. National Science Foundation partially supported my research over the years (grant numbers \# 1712714, \# 1745640, and \# 1945136).

\subsection*{Contacting me}

Please feel free to email me at 
\begin{center}
\href{mailto:pengdingpku@berkeley.edu}{pengdingpku@berkeley.edu}
\end{center}
if you identify any errors in the book, or if you use the book for teaching and want the solutions to the homework problems.


\mainmatter

\makeatletter
\@addtoreset{example}{chapter}
\@addtoreset{assumption}{chapter}
\@addtoreset{example}{chapter}
\@addtoreset{lemma}{chapter}
\@addtoreset{remark}{chapter}
\@addtoreset{theorem}{chapter}
\@addtoreset{proposition}{chapter}
\@addtoreset{corollary}{chapter}
\@addtoreset{condition}{chapter}
\makeatother

\renewcommand {\theexample} {\thechapter.\arabic{example}}
\renewcommand {\theassumption} {\thechapter.\arabic{assumption}}
\renewcommand {\thelemma} {\thechapter.\arabic{lemma}}
\renewcommand {\thetheorem} {\thechapter.\arabic{theorem}}
\renewcommand {\theremark} {\thechapter.\arabic{remark}}
\renewcommand {\theproposition} {\thechapter.\arabic{proposition}}
\renewcommand {\thecorollary} {\thechapter.\arabic{corollary}}
 \renewcommand {\thecondition} {\thechapter.\arabic{condition}}
 \renewcommand {\thedefinition} {\thechapter.\arabic{definition}}
\renewcommand{\theparagraph}{\arabic{chapter}.\arabic{paragraph}}


\part{Introduction}

\chapter{Motivations for Statistical Models}\label{chapter::motivation}

This book is about the linear model and its extensions. Before delving into the mathematical details of specific models, I will briefly provide some motivations for studying statistical models.

\section{Data and statistical models}
A wide range of problems in statistics and machine learning have the following data structure:

\begin{tabular}{|c|c|p{1cm}|p{1cm}|p{1cm}|p{1cm}|}
Unit & outcome/response & \multicolumn{4}{c|}{covariates/features/predictors}\tabularnewline
\hline 
\hline 
$i$ & $Y$ & $X_{1}$ & $X_{2}$ & $\cdots$ & $X_{p}$\tabularnewline
\hline 
$1$ & $y_{1}$ & $x_{11}$ & $x_{12}$ & $\cdots$ & $x_{1p}$\tabularnewline
\hline 
$2$ & $y_{2}$ & $x_{21}$ & $x_{22}$ & $\cdots$ & $x_{2p}$\tabularnewline
\hline 
$\vdots$ & $\vdots$ & $\vdots$ & $\vdots$ &  & $\vdots$\tabularnewline
\hline 
$n$ & $y_{n}$ & $x_{n1}$ & $x_{n2}$ & $\cdots$ & $x_{np}$\tabularnewline
\end{tabular}

For each unit $i$, we observe the outcome of interest (also called the response), $y_i$, as well as $p$  covariates (also called features or predictors), $x_{i1}, \ldots, x_{ip}$. 
We often use 
\[
Y=\left(\begin{array}{c}
y_{1}\\
y_{2}\\
\vdots\\
y_{n}
\end{array}\right)
\]
to denote the $n$-dimensional outcome vector, and 
\[
X=\left(\begin{array}{cccc}
x_{11} & x_{12} & \cdots & x_{1p}\\
x_{21} & x_{22} & \cdots & x_{2p}\\
\vdots & \vdots &  & \vdots\\
x_{n1} & x_{n2} & \cdots & x_{np}
\end{array}\right)
\]
to denote the $n\times p$ covariate matrix, also called the {\it design matrix}. In
most cases, the first column of $X$ contains constants $1$s. 

Based on the data $(X,Y)$, we can ask the following questions:
\begin{enumerate}[label=(Q\arabic*), ref=Q\arabic*]
\item Describe the relationship between $X$ and $Y$, i.e., their association
or correlation. For example, how is the patients' average height related
to the children's average height? How is one's height related to one's
weight? How are one's education and working experience related to one's income?

\item Predict $Y^{*}$ with new data $X^{*}$, based on the old data $(X, Y)$. In particular,
we want to use the current data $(X,Y)$ to train a predictor, and
then use it to predict future $Y^*$ based on future $X^*$. This is called
{\it supervised learning} in the field of machine learning. For example,
how do we predict whether an email is spam or not based on the frequencies of the most commonly occurring words and punctuation marks in the email? How do we predict cancer patients' survival time based on their clinical measures?

\item Estimate the causal effect of some components in $X$ on $Y$. What if we change
some components of $X$? How do we measure the impact of the hypothetical intervention of some components of $X$ on $Y$? This is a much harder question because most statistical tools are designed to infer association, not causation. For example, the U.S. Food and Drug Administration (FDA) approves drugs based on randomized controlled trials (RCT) because RCTs are most credible to infer causal effects of drugs on health outcomes. Economists are interested in evaluating the effect of a job training program on employment and wages. However, this is a notoriously difficult problem because participation in the job training program is not randomized in observational data. 
\end{enumerate}

The above descriptions are about generic $X$ and $Y$, which can have many different types. We often use different statistical models to capture the features of different types of data. Below I give a brief overview
of models that will appear in later parts of this book.
\begin{enumerate}[label=(T\arabic*), ref=T\arabic*]
\item $X$ and $Y$ are univariate and continuous. In Francis Galton's\footnote{Who was Francis Galton? He was Charles Darwin's nephew and was famous
for his pioneer work in statistics and for devising a method for classifying
fingerprints that proved useful in forensic science. He also invented the term {\it eugenics}, a field that causes a lot of controversies nowadays.} classic example, $X$ is the parents' average height and $Y$ is the children's
average height \citep{galton1886regression}. 
Let $\hat{y}_i$ denote the ``fitted value'' of the outcome for unit $i$ with covariate value $x_i$. 
Galton derived the following formula:
\[
\hat{y}_i = \bar{y} + \hat\rho \frac{\hat{\sigma}_{y}}{\hat{\sigma}_{x}} (x_i -\bar{x})
\]
which is equivalent to
\begin{equation}
\label{eq::galtonian-formula-1}
\frac{\hat{y}-\bar{y}}{\hat{\sigma}_{y}}
= \hat\rho \frac{x_i-\bar{x}}{\hat{\sigma}_{x}},
\end{equation}
where 
$$
\bar{x} = n^{-1} \sumn x_i,\qquad \bar{y} = n^{-1} \sumn y_i
$$ 
are the sample means, 
$$
\hat{\sigma}_{x}^2 = (n-1)^{-1} \sumn (x_i - \bar{x})^2,\qquad \hat{\sigma}_{y}^2 =  (n-1)^{-1} \sumn (y_i - \bar{y})^2
$$ 
are the sample variances, and $ \hat\rho =  \hat{\sigma}_{xy} / ( \hat{\sigma}_{x} \hat{\sigma}_{y} ) $ is the
sample Pearson correlation coefficient with the sample covariance 
$$
\hat{\sigma}_{xy}  =  (n-1)^{-1} \sumn(x_{i}-\bar{x})(y_{i}-\bar{y}).
$$
The identity \eqref{eq::galtonian-formula-1} is the famous formula of ``regression towards mediocrity'' or ``regression towards the mean''. Galton first introduced the terminology ``regression.''\footnote{The name ``regression'' is widely used in statistics now. For instance, we sometimes use ``linear regression'' interchangeably with ``linear model.'' We also extend the name to ``logistic regression'' or ``Cox regression,'' which will be discussed in later chapters of this book.} 
Galton called regression because the relative deviation of the children's average height is smaller than that of the parents' average height if $ | \hat\rho  | < 1$. We will derive \eqref{eq::galtonian-formula-1} in Chapter \ref{chapter::ols-1d}.

\item $Y$ is univariate and continuous, and $X$ is multivariate of mixed types.
In the \ri{R} package \ri{ElemStatLearn}, the dataset \ri{prostate}
has an outcome of interest as the log of the prostate-specific antigen \ri{lpsa} and some potential predictors
including the log cancer volume \ri{lcavol}, the log prostate weight \ri{lweight}, age \ri{age}, etc. 
The main chapters of this book, 
Chapters \ref{chapter::ols-vector}--\ref{chapter::WLS}, will discuss linear regression with multidimensional covariates.

\item $Y$ is binary or indicator of two classes, and $X$ is multivariate of
mixed types. For example, in the \ri{R} package \ri{wooldridge}, the dataset \ri{mroz} contains
an outcome of interest being the binary indicator for whether a woman
was in the labor force in the year 1975, and some useful covariates are in the table below: 
\begin{center}
\begin{tabular}{|c|c|}
\hline 
covariate name & covariate meaning\tabularnewline
\hline 
\hline 
kidslt6 & number of kids younger than six years old\tabularnewline
\hline 
kidsge6 & number of kids between six and eighteen years old\tabularnewline
\hline 
age & age\tabularnewline
\hline 
educ & years of education\tabularnewline
\hline 
husage & husband's age\tabularnewline
\hline 
huseduc & husband's years of education\tabularnewline
\hline 
\end{tabular} 
\end{center}
Chapter \ref{chapter::binary-logit} will discuss {\it logistic regression} for binary outcomes.

\item $Y$ is categorical without ordering. For example, the choice of housing type, single-family house, townhouse, or condominium, is a categorical variable. 
Chapter \ref{chapter::logit-categorical} will discuss {\it multinomial logistic regression} for categorical outcomes without ordering.

\item $Y$ is categorical and ordered. For example, the final course evaluation at UC Berkeley can take value in $\{ 1,2,3,4,5,6,7 \}$. These numbers have clear ordering but they are not the usual real numbers. 
Chapter \ref{chapter::logit-categorical} will discuss {\it 
proportional odds regression} for ordered categorical outcomes.

\item $Y$ represents counts. For example, the number of times one went to the gym last week is a non-negative integer representing counts. 
Chapter \ref{chapter::count} will discuss regression models for count outcomes.

\item $Y$ is  multivariate and correlated. In medical trials, the data are often longitudinal, meaning that the patient's outcomes are measured repeatedly over time. So each patient has a multivariate outcome. In field experiments of public health and development economics, the randomized interventions are often at the village level but the outcome data are collected at the household level. So within villages, the outcomes are correlated. 
Chapter \ref{chapter::gee} will discuss the {\it generalized estimating equation} for correlated data.

\item $Y$ represent time-to-event outcomes. For example, in medical trials, a major outcome of interest is the survival time; in labor economics, a major outcome of interest is the time to find the next job. The former is called {\it survival analysis} in biostatistics and the latter is called {\it duration analysis} in econometrics.
Chapter \ref{chapter::survival-analysis} will discuss {\it Cox proportional hazards regression}.

\end{enumerate}

\section{Why linear models?}

Why do we study linear models if many real problems may have nonlinear structures? There are important reasons.

\begin{enumerate}[label=(R\arabic*), ref=R\arabic*]
\item Linear models are simple but non-trivial starting points for learning.

\item Linear models can provide insights because we can derive explicit
formulas based on elegant algebra and geometry.

\item Linear models can handle nonlinearity by incorporating nonlinear terms of covariates, for example, $X$ can contain the polynomials or nonlinear transformations of the original covariates. In statistics, ``linear'' means {\it linear in parameters}, not {\it linear in covariates}.

\item Linear models can be good approximations of nonlinear data-generating
processes.
\item Linear models are simpler than nonlinear models, but they do not necessarily
perform worse than more complicated nonlinear models. We have finite data so we cannot fit arbitrarily complicated models.

\end{enumerate}

If you are interested in nonlinear models, you can read another book on machine learning.

\chapter{Ordinary Least Squares with a Univariate Covariate}
 \label{chapter::ols-1d}

This chapter discusses ordinary least squares (OLS) with a single covariate. It can provide insights into later chapters because it is the building block of OLS with multiple covariates.

\section{Ordinary least squares with a univariate covariate}
\begin{figure}[ht]
\centering
\includegraphics[width= 0.7\textwidth]{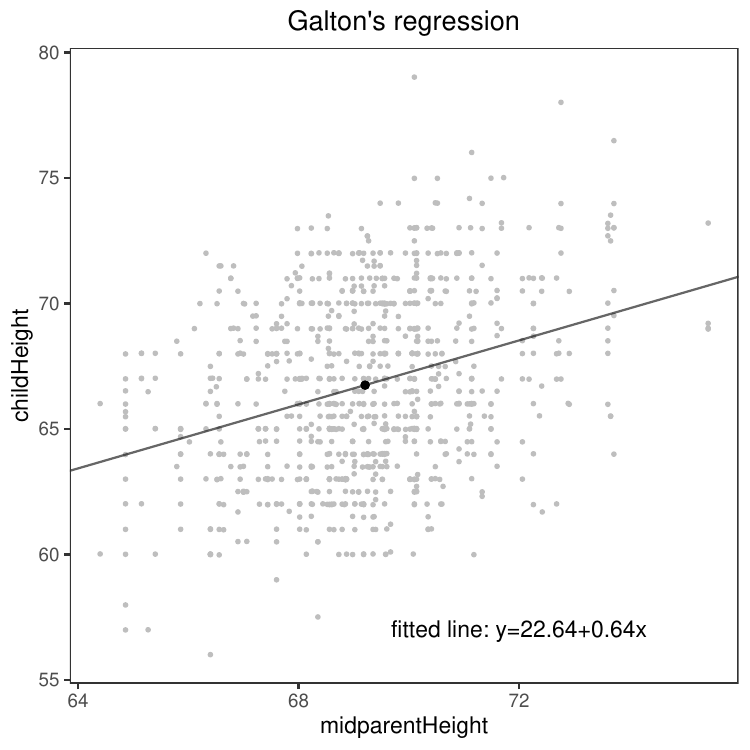}
\caption{Galton's dataset}\label{fig::galton_data_scatterplot}
\end{figure}

Figure \ref{fig::galton_data_scatterplot} shows the scatterplot of Galton's dataset which can be found in the \ri{R} package \ri{HistData} as \ri{GaltonFamilies}. In this dataset, \ri{father} denotes the height of the father and \ri{mother} denotes the height of the mother. 
 The x-axis denotes the mid-parent height, calculated as (\ri{father} + 1.08*\ri{mother})/2, and the y-axis denotes the height of a child.

With $n$ data points $(x_{i,}y_{i})_{i=1}^{n}$, our goal is to find the best linear fit of the data
\[
(x_{i,}\hat{y}_{i}=\hat{\alpha}+\hat{\beta}x_{i})_{i=1}^{n},
\]
where the coefficients $\hat{\alpha}$ and $\hat{\beta}$ are determined from the data. 
What do we mean by the ``best'' fit? Gauss proposed to use the following OLS criterion:\footnote{The idea of OLS is often attributed to Gauss and Legendre. Gauss used it in the process of discovering Ceres, and his work was published in 1809. Legendre's work appeared in 1805 but Gauss claimed that he had been using it since 1794 or 1795. \citet{stigler1981gauss} reviews the history of OLS.}
\begin{equation}
\label{eq::ols-univariate-objective}
(\hat{\alpha},\hat{\beta})=\arg\min_{a,b}n^{-1}\sumn(y_{i}-a-b x_{i})^{2}.
\end{equation}

The OLS criterion is based on the squared ``misfits'' $y_{i}- a - b  x_{i}$. Another intuitive criterion is based on the absolute values of those misfits, which is called the least absolute deviation (LAD). However, OLS is simpler because the objective function is smooth in $(a, b)$, and we can obtain close-form solutions. I will discuss LAD in Chapter \ref{chapter::quantile-regression} as a special case of {\it quantile regression}.

How do we solve the OLS minimization problem in \eqref{eq::ols-univariate-objective}? The objective function
is quadratic, and as $a$ and $b$ diverge, it diverges to
infinity. So it must have a minimizer $(\hat{\alpha},\hat{\beta})$, which satisfies the first-order condition:
\begin{eqnarray}
-\frac{2}{n}\sumn(y_{i}-\hat{\alpha}-\hat{\beta}x_{i}) & = &0,\label{eq::univariate-foc-1} \\
-\frac{2}{n}\sumn x_{i}(y_{i}-\hat{\alpha}-\hat{\beta}x_{i}) & =&0.\label{eq::univariate-foc-2}
\end{eqnarray}
The equations \eqref{eq::univariate-foc-1} and \eqref{eq::univariate-foc-2} are called the Normal Equations of OLS. The first
equation \eqref{eq::univariate-foc-1} implies 
\begin{equation}
\bar{y}=\hat{\alpha}+\hat{\beta}\bar{x},\label{eq:normal1}
\end{equation}
that is, the OLS line must go through the sample mean of the data $(\bar{x},\bar{y}).$ The second equation \eqref{eq::univariate-foc-2} implies
\begin{equation}
\overline{xy}=\hat{\alpha}\bar{x}+\hat{\beta}\overline{x^{2},}\label{eq:normal2}
\end{equation}
where $\overline{xy}$ is the sample mean of the $x_{i}y_{i}$'s, and $\overline{x^{2}}$
is the sample mean of the $x_{i}^{2}$'s. Subtracting (\ref{eq:normal1})$\times\bar{x}$
from (\ref{eq:normal2}), we have
$$
\overline{xy} - \bar{x} \bar{y} = \hat{\beta} ( \overline{x^{2}} - \bar{x}^2 ) ,
$$
which is
$$
\hat{\sigma}_{xy} =\hat{\beta} \hat{\sigma}_{x}^2 ,
$$
and implies
\begin{equation}
\label{eq::ols-slope}
\hat{\beta} =   \frac{\hat{\sigma}_{xy}}{\hat{\sigma}_{x}^2}. 
\end{equation}
So the OLS coefficient of $x$ equals the sample covariance between
$x$ and $y$ divided by the sample variance of $x$. From (\ref{eq:normal1}),
we obtain that 
\begin{equation}
\label{eq::ols-intercept}
\hat{\alpha}=\bar{y}-\hat{\beta}\bar{x}.
\end{equation}
By \eqref{eq::ols-intercept}, the fitted line $\hat{y}_i  =\hat{\alpha}+\hat{\beta}x_i$ simplifies to $\hat{y}_i  = \bar{y}-\hat{\beta}\bar{x}+\hat{\beta}x_i$, and more symmetrically, $\hat{y}_i -\bar{y}=\hat{\beta}(x_i-\bar{x})$. With \eqref{eq::ols-slope}, we can further simplify the fitted line as
\begin{eqnarray*}
\hat{y}_i -\bar{y} &=& \frac{   \hat{\sigma}_{xy}  }{  \hat{\sigma}_{x}^2  }(x_i-\bar{x}) \\
 &=&\frac{  \hat{\rho}_{xy} \hat{\sigma}_{x}\hat{\sigma}_{y}}{\hat{\sigma}_{x}^{2}}(x_i-\bar{x}),
\end{eqnarray*}
which implies
$$
\frac{\hat{y}_i -\bar{y}}{\hat{\sigma}_{y}}= \hat{\rho}_{xy}  \frac{x_i-\bar{x}}{\hat{\sigma}_{x}},
$$
the Galtonian formula mentioned in Chapter \ref{chapter::motivation}.

%
%
%

We can obtain the fitted line based on Galton's data using the \ri{R} code below. 

\begin{rc}
> library("HistData")
> xx = GaltonFamilies$midparentHeight
> yy = GaltonFamilies$childHeight
> 
> center_x = mean(xx)
> center_y = mean(yy)
> sd_x     = sd(xx)
> sd_y     = sd(yy)
> rho_xy   = cor(xx, yy)
> 
> beta_fit  = rho_xy*sd_y/sd_x
> alpha_fit = center_y - beta_fit*center_x
> alpha_fit
[1] 22.63624
> beta_fit
[1] 0.6373609
\end{rc}

I then generate Figure \ref{fig::galton_data_scatterplot} based on the original data and the OLS coefficients.

\section{Final comments}\label{section::1d-comments}

I make two final comments on OLS. 
\begin{enumerate}[label=(C\arabic*), ref=C\arabic*]
\item
We can write the sample mean as the solution to the OLS with only the intercept:
\begin{equation}\label{eq::ols-mean}
\bar{y} = \arg\min_{\mu}n^{-1}\sumn(y_{i} - \mu)^{2}.
\end{equation}

\item
We can fit OLS of $y_i$ on $x_i$ without the intercept: 
\[
 \hat{\beta}=\arg\min_{b}n^{-1}\sumn(y_{i} -b x_{i})^{2} 
\]
which equals 
\begin{equation}
\label{eq::ols-without-intercept}
 \hat{\beta} 
 = \frac{ \sumn x_iy_i }{  \sumn x_i^2 } = \frac{  \langle x, y \rangle   }{ \langle x, x \rangle } , 
\end{equation}
 where $x = (x_1, \ldots, x_n)^{\T}$ and $y = (y_1, \ldots, y_n)^{\T}$ are the $n$-dimensional vectors containing all observations, and $\langle x, y \rangle = \sumn x_iy_i$ denotes the inner product between $x$ and $y$. 
 \end{enumerate}

Although it is rare to fit the above OLS in practical data analysis, the formulas in \eqref{eq::ols-mean} and \eqref{eq::ols-without-intercept} will be the building blocks for many discussions in later parts of the book. I leave the proof of \eqref{eq::ols-mean} and \eqref{eq::ols-without-intercept} as Problem \ref{hw2::univariate-ols-2}.

\section{Homework problems}

\paragraph{Univariate OLS}\label{hw2::univariate-ols-2}

Prove   \eqref{eq::ols-mean} and \eqref{eq::ols-without-intercept}. 

\paragraph{Pairwise slopes}\label{hw2::pairwise-slope}

Prove Theorem \ref{thm::ols-pairwise-formula} below. 

\begin{theorem}
\label{thm::ols-pairwise-formula}
Given $(x_i, y_i)_{i=1}^n$ with univariate $x_i$ and $y_i$, show that $\hat\beta$ in \eqref{eq::ols-slope} equals
$$
\hat{\beta} = \sum_{(i,j)} w_{ij} b_{ij},
$$
where the summation is over all pairs of observations $(i,j)$, 
$$
b_{ij} = \frac{ y_i - y_j }{  x_i - x_j } 
$$
is the slope determined by two points $(x_i,y_i)$ and $(x_j, y_j)$, and 
$$
w_{ij} = \frac{  x_i - x_j)^2 }{   \sum_{(i',j')} (x_{i'} - x_{j'})^2 } 
$$
is the weight proportional to the squared distance between $x_i$ and $x_j$. In the above formulas, if $x_i=x_j$, then we define $b_{ij} = 0$, and the corresponding weight $w_{ij}$ equals $0$. 
\end{theorem}

Remark: \citet{wu1986jackknife} and \citet{gelman2009splitting} used Theorem \ref{thm::ols-pairwise-formula}. 
Problem \ref{hw03::jacobi} gives a more general result.

\paragraph{No regression}\label{hw2::no-regression}

\begin{figure}
\centering
\includegraphics[width =\textwidth]{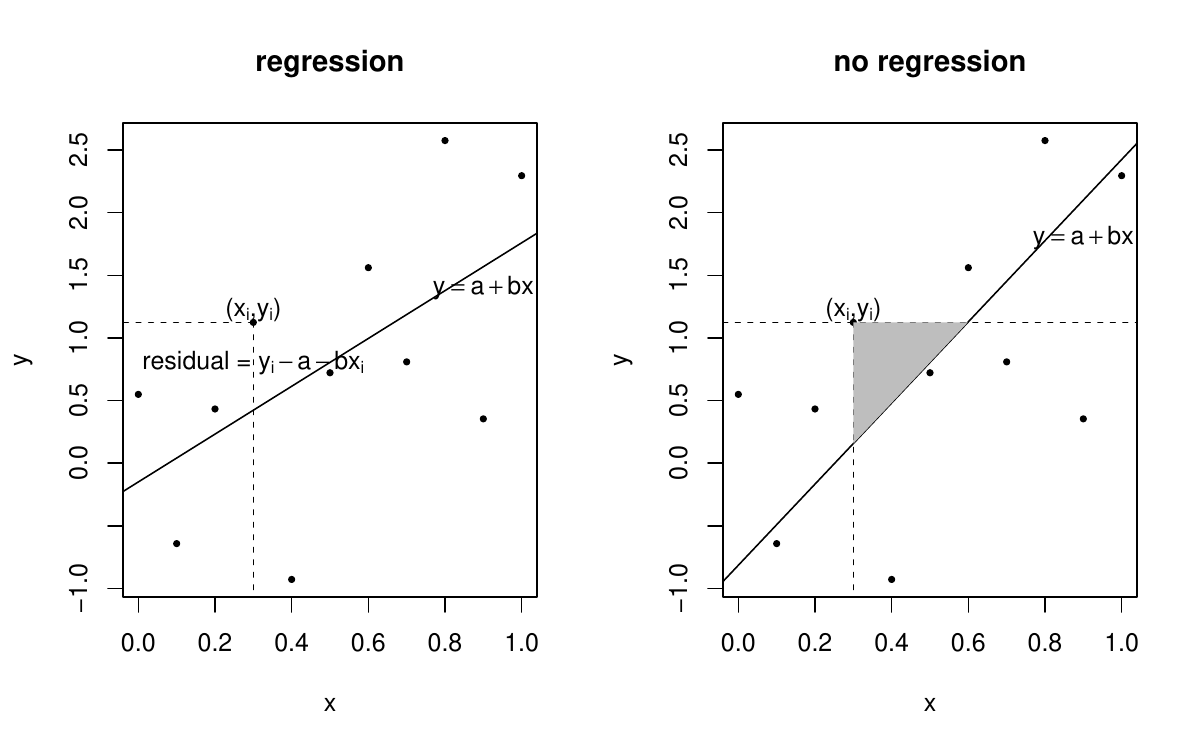}
\caption{Regression (left) and no regression (right)}
\label{fig::no-regression}
\end{figure}

\citet{woolley1941method} proposed a method to minimize the sum of the areas formed between the data points and fitted line. The right panel of Figure \ref{fig::no-regression} illustrates the area formed between data point $(x_i, y_i)$ and the line $y=a+bx$. 

Prove that if $\hat{\rho}_{xy}>0$, then the minimizer $(\hat\alpha', \hat\beta')$ satisfies
$$
\hat\beta '  = \frac{\hat\sigma_y}{\hat\sigma_x}
$$
and
$$
\hat\alpha '  = \bar{y} - \hat\beta  \bar{x}.
$$

Remark: The fitted line is
\begin{eqnarray*}
\hat{y}_i 
&=& \hat\alpha '  + \hat\beta '  x_i  \\
&=& \bar{y} - \hat\beta  \bar{x} + \hat\beta '  x_i , 
\end{eqnarray*}
which is equivalent to
$$
\frac{\hat{y}_i-\bar{y}}{\hat{\sigma}_{y}}=   \frac{x_i-\bar{x}}{\hat{\sigma}_{x}} .
$$
It does not have the regression factor $\hat{\rho}_{xy}$, compared with the Galtonian formula, which is derived by minimizing the residual sum of squares as illustrated by the left panel of Figure \ref{fig::no-regression}.

\part{Ordinary Least Squares and Statistical Inference}

\chapter{Ordinary Least Squares with Multiple Covariates}
 \label{chapter::ols-vector}

This chapter provides algebraic results about ordinary least squares (OLS). The results in this chapter do not rely on any stochastic assumptions.

\section{The OLS formula}

Recall that we have the outcome vector and covariate matrix: 
\[
Y=\left(\begin{array}{c}
y_{1}\\
y_{2}\\
\vdots\\
y_{n}
\end{array}\right)
,\qquad 
 X=\left(\begin{array}{cccc}
x_{11} & x_{12} & \cdots & x_{1p}\\
x_{21} & x_{22} & \cdots & x_{2p}\\
\vdots & \vdots &  & \vdots\\
x_{n1} & x_{n2} & \cdots & x_{np}
\end{array}\right) .
\]
Depending on the purpose, it is convenient to view $X$ as a collection of row or column vectors: 
\[
X =\left(\begin{array}{c}
x_{1}^{\T}\\
x_{2}^{\T}\\
\vdots\\
x_{n}^{\T}
\end{array}\right)
=(X_1,\ldots, X_p)
\]
where $x_{i}^{\T}=(x_{i1},\ldots,x_{ip})$ is the row vector consisting
of the covariates of unit $i$, and $X_j = (x_{1j}, \ldots, x_{nj})^{\T} $ is the column vector of the $j$-th covariate for all units.

We want to find the ``best'' linear fit of the data
$
(x_{i}, \hat{y}_{i} )_{i=1}^{n}
$
with
$$
\hat{y}_{i}=x_{i}^{\T}\hat{\beta} = \hat{\beta}_1 x_{i1} + \cdots + \hat{\beta}_p x_{ip} 
$$
 in the sense that 
\begin{eqnarray} 
\hat{\beta} 
&=&\arg\min_{ b \in\mathbb{R}^{p} }n^{-1}\sumn(y_{i}-x_{i}^{\T} b )^{2} \label{eq::ols-beta-def1} \\
&=&\arg\min_{ b \in\mathbb{R}^{p} }n^{-1}\|Y-X b\|^{2},\label{eq::ols-beta-def2}
\end{eqnarray} 
 where $\hat{\beta} = (\hat\beta_1, \ldots, \hat\beta_p)^{\T}$  is called the OLS coefficient, the $\hat{y}_{i}$'s are called the fitted values, and the $\hat\varepsilon_i = y_i-\hat{y}_{i}$'s are called the residuals. 

The objective function in \eqref{eq::ols-beta-def1} is quadratic in $b$, which diverges to infinity when $b$ diverges to infinity. So it must have a minimizer
$\hat{\beta}$ satisfying the first-order condition
\[
-\frac{2}{n}\sumn x_{i}(y_{i}-x_{i}^{\T}\hat{\beta})=0,
\]
which simplifies to 
\begin{equation}
\sumn x_{i}(y_{i}-x_{i}^{\T}\hat{\beta})=0 ,  \label{eq:normalequation-1}
\end{equation}
or, equivalently, in matrix form:
\begin{equation}
X^{\T}(Y-X\hat{\beta})=0 .  \label{eq:normalequation-2}
\end{equation}
The above equations \eqref{eq:normalequation-1} and \eqref{eq:normalequation-2} are called the {\it Normal equation} of the OLS, which implies the main theorem:
\begin{theorem}
\label{thm:The-OLS-coefficient}
The OLS coefficient in \eqref{eq::ols-beta-def1} and \eqref{eq::ols-beta-def2} equals
\begin{eqnarray*}
\hat{\beta}
&=&\left(\sumn x_{i}x_{i}^{\T}\right)^{-1}\left(\sumn x_{i}y_{i}\right) \\
&=& (X^{\T}X)^{-1}X^{\T}Y
\end{eqnarray*}
if $X^{\T}X = \sumn x_{i}x_{i}^{\T}$ is non-degenerate. 
\end{theorem}

\emph{Comment on the two equivalent forms in Theorem \ref{thm:The-OLS-coefficient}.}
The equivalence of the two forms of the OLS coefficient follows from
$$
X^{\T}X = (x_1,\ldots, x_n)  \left(\begin{array}{c}
x_{1}^{\T}\\
x_{2}^{\T}\\
\vdots\\
x_{n}^{\T}
\end{array}\right)
= \sumn x_{i}x_{i}^{\T},
$$
and
$$
X^{\T}Y = (x_1,\ldots, x_n)\left(\begin{array}{c}
y_{1}\\
y_{2}\\
\vdots\\
y_{n}
\end{array}\right)
= \sumn x_{i}y_{i}. 
$$
Depending on the purpose, both forms can be useful in later discussions.

\emph{Comment on the condition in Theorem \ref{thm:The-OLS-coefficient}.}
The non-degeneracy of $X^{\T}X$ in Theorem \ref{thm:The-OLS-coefficient}  requires that for any non-zero vector $\alpha = (\alpha_1, \ldots, \alpha_p)^{\T} \in\mathbb{R}^{p},$ we must have
\[
\alpha^{\T}X^{\T}X\alpha=\|X\alpha\|^{2}\neq 0 
\]
which is equivalent to
$$
X\alpha = \alpha_1 X_1 + \cdots + \alpha_p X_p \neq 0,
$$
i.e., the columns of $X$ are {\it linearly independent}.\footnote{This book uses different notions of ``independence,'' which can be confusing sometimes. In linear algebra, a set of vectors is linearly \emph{independent} if any nonzero linear combination of them is not zero; see Appendix \ref{chapter::linear-algebra}. In probability theory, two random variables are \emph{independent} if their joint density factorizes into the product of the marginal distributions; see Appendix \ref{chapter:appendix-rvs}.} This effectively
rules out redundant columns in the design matrix $X$. If $X_1$ can be represented by other columns $X_1 = c_2 X_2 + \cdots + c_p X_p$ for some $(c_2,\ldots, c_p)$, then $X^{\T} X$ is degenerate. 

Throughout the book, we invoke the following condition unless stated otherwise.
\begin{condition}\label{condition::ols-non-degenerate}
The column vectors of $X$ are linearly independent. 
\end{condition}

\section{The geometry of OLS}\label{sec::geometry-ols}

\begin{figure}
\centering
\includegraphics[width = 0.9 \textwidth]{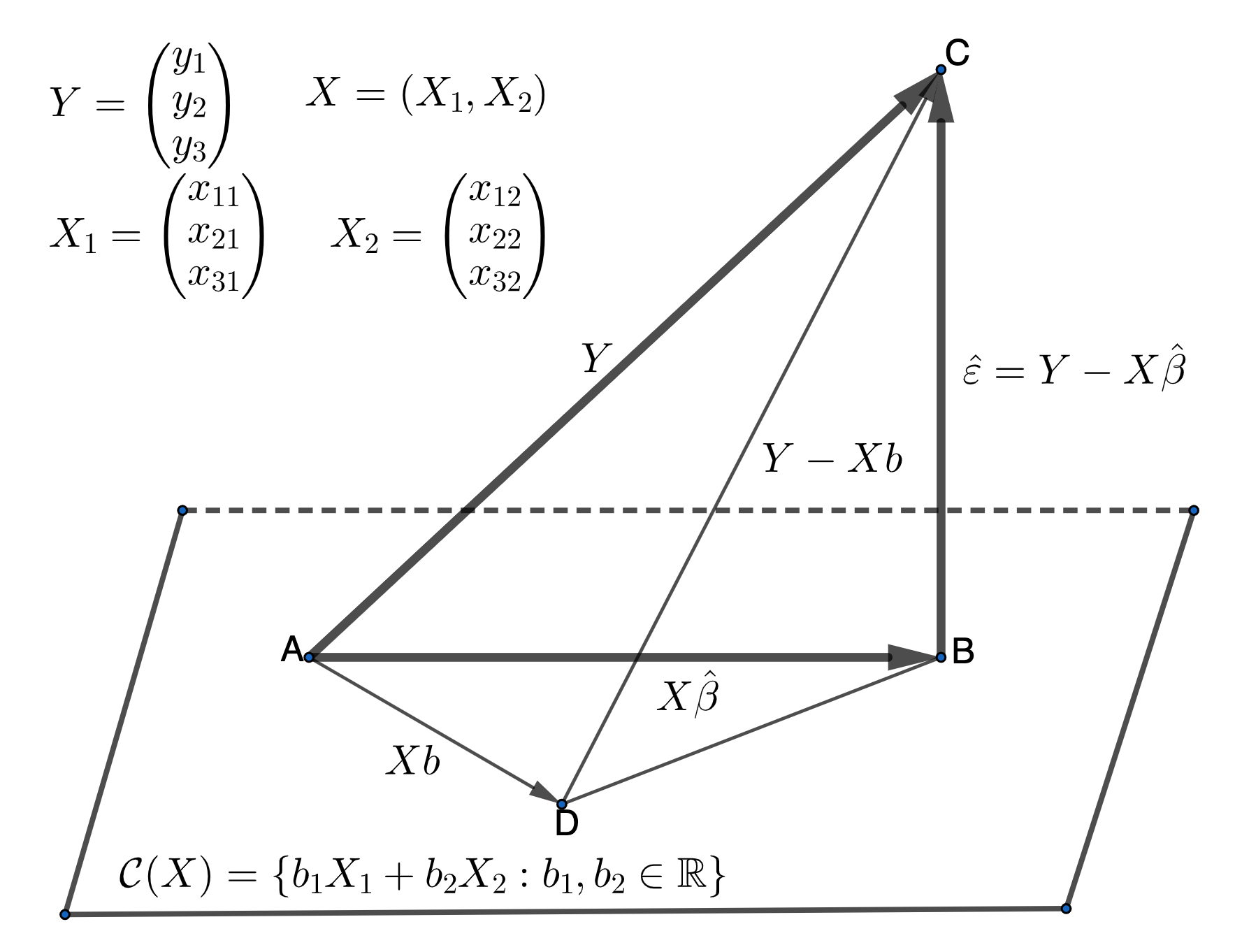}
\caption{The geometry of OLS}\label{fig::olsgeometry}
\end{figure}

The OLS has clear geometric interpretations. Figure \ref{fig::olsgeometry} illustrate its geometry with $n=3$ and $p=2$. 
For any $b=(b_{1},\ldots,b_{p})^{\T}\in\mathbb{R}^{p}$ and $X=(X_{1},\ldots,X_{p})\in\mathbb{R}^{n\times p}$,
$$
Xb
=(X_{1},\ldots,X_{p}) \begin{pmatrix}
b_1\\
\vdots \\
b_p
\end{pmatrix}
=b_{1}X_{1}+\cdots+b_{p}X_{p}
$$ 
represents a linear combination
of the column vectors of the design matrix $X$. So the OLS problem
is to find the best linear combination of the column vectors of $X$
to approximate the response vector $Y$. Recall that all linear combinations
of the column vectors of $X$ constitute the column space of $X$,
denoted by\footnote{Please review Appendix \ref{chapter::linear-algebra} for some basic linear algebra background.} 
$$
\mathcal{C}(X) = \{  b_{1}X_{1}+\cdots+b_{p}X_{p} : b_1, \ldots, b_p \in  \mathbb{R}   \}.
$$
So the OLS problem is to find the vector
in $\mathcal{C}(X)$ that is the closest to $Y$. Geometrically, the vector
must be the projection of $Y$ onto $\mathcal{C}(X).$ By projection,
the residual vector $\hat{\varepsilon}=Y-X\hat{\beta}$ must be orthogonal
to $\mathcal{C}(X)$, or, equivalently, the residual
vector is orthogonal to $X_{1},\ldots,X_{p}$. This geometric intuition implies that
$$
X_{1}^{\T} \hat{\varepsilon} =0,\ldots,X_{p}^{\T}  \hat{\varepsilon} =0 .
$$
In matrix form, we have 
$$
 X^{\T} \hat{\varepsilon}
= \begin{pmatrix}
X_1^{\T} \hat{\varepsilon} \\
\vdots \\
X_p^{\T}\hat{\varepsilon}
\end{pmatrix}
=0,
$$
which is equivalent to
$$
 X^{\T} (Y-X\hat{\beta}) = 0 ,
$$
the Normal equation in (\ref{eq:normalequation-2}). 
The above argument gives a geometric derivation of the OLS formula
in Theorem \ref{thm:The-OLS-coefficient}.

In Figure \ref{fig::olsgeometry}, since the triangle ABC is rectangular, the fitted vector $\hat{Y} = X\hat{\beta}$ is orthogonal to the residual vector $\hat{\varepsilon}$, and moreover, the Pythagorean Theorem implies that
$$
\| Y\|^2 = \| X\hat{\beta} \|^2 + \| \hat{\varepsilon} \|^2.
$$

The following theorem states an algebraic fact that gives an alternative proof of the OLS formula. It is essentially the Pythagorean Theorem for the rectangular
triangle BCD in Figure \ref{fig::olsgeometry}. 

\begin{theorem}\label{thm::geometryofols}
For any $b\in\mathbb{R}^{p},$ we have the following decomposition
\[
\|Y-Xb\|^{2}=\|Y-X\hat{\beta}\|^{2}+\|X(\hat{\beta}-b)\|^{2},
\]
where implies that $\|Y-Xb\|^{2}\geq\|Y-X\hat{\beta}\|^{2}$ with
equality holding if and only if $b=\hat{\beta}.$
\end{theorem}
\begin{myproof}{Theorem}{\ref{thm::geometryofols}}
We have the following decomposition: 
\begin{eqnarray*}
\|Y-Xb\|^{2} 
& =& (Y-Xb)^{\T}(Y-Xb)\\
 & =& (Y-X\hat{\beta}+X\hat{\beta}-Xb)^{\T}(Y-X\hat{\beta}+X\hat{\beta}-Xb)\\
 & =& (Y-X\hat{\beta})^{\T}(Y-X\hat{\beta})+(X\hat{\beta}-Xb)^{\T}(X\hat{\beta}-Xb)\\
 & & +(Y-X\hat{\beta})^{\T}(X\hat{\beta}-Xb)+(X\hat{\beta}-Xb)^{\T}(Y-X\hat{\beta}).
\end{eqnarray*}
The first term equals $\|Y-X\hat{\beta}\|^{2}$ and the second term equals $\|X(\hat{\beta}-b)\|^{2}$. 
We need to show the last two terms are zero. By symmetry of these
two terms, we only need to show that the last term is zero. This is
true by the Normal equation (\ref{eq:normalequation-2}) of the OLS:
\begin{align*}
(X\hat{\beta}-Xb)^{\T}(Y-X\hat{\beta}) & =(\hat{\beta}-b)^{\T}X^{\T}(Y-X\hat{\beta})=0.\\
\end{align*}
\end{myproof}

I end this section by commenting on the role of the intercept in OLS. 

\emph{The role of the intercept in OLS.}
In most applications,  $X$ contains a column of $1_{n}=(1,\ldots,1)^{\T}$. In those cases, we have
\[
1_{n}^{\T}\hat{\varepsilon}=0,
\]
and therefore,
\[ 
n^{-1}\sumn\hat{\varepsilon}_{i}=0.
\]
That is, with the intercept in OLS,  the residuals are automatically centered to have mean 0.

\section{The projection matrix from OLS}

The geometry  in  Section \ref{sec::geometry-ols} also shows that $\hat{Y}=X\hat{\beta}$ is the  solution
to the following problem
\[
\hat{Y}=\arg\min_{v\in\mathcal{C}(X)}\|Y-v\|^{2}.
\]
Using Theorem \ref{thm:The-OLS-coefficient}, we have $\hat{Y}=X\hat{\beta}=HY$, 
where 
\[
H=X(X^{\T}X)^{-1}X^{\T}
\]
is an $n\times n$ matrix. It is called the {\it hat matrix} because it
puts a hat on $Y$ when multiplying $Y$. Algebraically, we can show that
$H$ is a projection matrix\footnote{
Review the definition and properties of projection matrices in Appendix \ref{chapter::linear-algebra}.
} because
\begin{eqnarray*}
H^{2} & = & X(X^{\T}X)^{-1}X^{\T}X(X^{\T}X)^{-1}X^{\T} \\
&=& X(X^{\T}X)^{-1}X^{\T} \\
&=& H,
\end{eqnarray*}
and
\begin{eqnarray*}
H^{\T} & = &\{ X(X^{\T}X)^{-1}X^{\T}\} ^{\T} \\
&=& X(X^{\T}X)^{-1}X^{\T} \\
&=& H.
\end{eqnarray*}
Its rank equals its trace, so 
\begin{eqnarray*}
\text{rank}(H) = 
\text{trace}(H)&=&\text{trace}\{ X(X^{\T}X)^{-1}X^{\T}\} \\
&=&\text{trace}\{ (X^{\T}X)^{-1}X^{\T}X\} \\
&=&\text{trace}(I_{p})\\
&=&p.
\end{eqnarray*}
The projection matrix $H$ has the following geometric interpretations. 

\begin{proposition}\label{prop::projectionmatrix-geometry}
The projection matrix $H=X(X^{\T}X)^{-1}X^{\T}$ satisfies
\begin{enumerate}[label=(G\arabic*), ref=G\arabic*]
\item \label{enu:proj1}$Hv=v$ if and only if $ v\in\mathcal{C}(X);$
\item \label{enu:proj2}$Hw=0$ if and only if $ w\perp\mathcal{C}(X).$
\end{enumerate}
\end{proposition}

Recall that $\mathcal{C}(X)$ is the column space of $X$. 
\eqref{enu:proj1} states that projecting any vector in $\mathcal{C}(X)$ onto $\mathcal{C}(X)$ does not change the vector. 
\eqref{enu:proj2} states that projecting any vector orthogonal to $\mathcal{C}(X)$ onto $\mathcal{C}(X)$ results in a zero vector. 

\begin{myproof}{Proposition}{\ref{prop::projectionmatrix-geometry}}
I first prove \eqref{enu:proj1}.
If $v\in\mathcal{C}(X),$ then $v=Xb$ for some $b$, which implies
 $$
 Hv=X(X^{\T}X)^{-1}X^{\T}Xb=Xb=v .
 $$ 
 Conversely, if $v=Hv$, then $v=X(X^{\T}X)^{-1} X^{\T}v = Xb$ with $b=(X^{\T}X)^{-1} X^{\T}v$, which ensures $v\in \mathcal{C}(X)$. 

I then prove \eqref{enu:proj2}. 
If $w\perp\mathcal{C}(X)$, then $w$ is orthogonal to all column
vectors of $X$, that is, $X_{j}^{\T}w=0\quad(j=1,\ldots,p)$. In matrix form, we have $X^{\T}w=0$, which implies 
$$ 
Hw=X(X^{\T}X)^{-1}X^{\T}w=0.
$$ 
Conversely, if $Hw=  X(X^{\T}X)^{-1}X^{\T}w =0$, then $w^{\T} X(X^{\T}X)^{-1}X^{\T}w=0$. Because $(X^{\T}X)^{-1}$ is positive definite under Condition \ref{condition::ols-non-degenerate}, we have $X^{\T}w=0$, which implies  $w\perp\mathcal{C}(X)$. 
\end{myproof}

Writing $H=(h_{ij})_{1\leq i,j\leq n}$ and $\hat{y}=(\hat{y}_{1},\ldots,\hat{y}_{n})^{\T}$,
we have another basic identity 
\begin{eqnarray*}
\hat{y}_{i} &=&\sum_{j =1}^{n}h_{ij}y_{j} \\
 &=& h_{ii}y_{i}+\sum_{j\neq i}h_{ij}y_{j}.
\end{eqnarray*}
 It shows that the predicted value $\hat{y}_{i}$ is a linear combination
of  the outcomes of all units and the coefficients depend on $H$. Moreover, if $X$ contains a column of intercepts
$1_{n}=(1,\ldots,1)^{\T},$ then 
\begin{equation}\label{eq::projection-constant1}
H1_{n}=1_{n},
\end{equation}
which implies 
\begin{equation}\label{eq::projection-constant2}
\sum_{j=1}^{n}h_{ij}=1\quad(i=1,\ldots,n)
\end{equation}
and therefore, $\hat{y}_{i}$ is a weighted average of the
outcomes of all units. Although the sum of the weights is 1, some of them can be negative. Readers, make sure the claims of \eqref{eq::projection-constant1} and \eqref{eq::projection-constant2} make sense to you. See Problem \ref{hw03::hat-matrix-intercept}.

In general, the hat matrix has complex forms, but when the covariates are dummy variables for group indicators, it has more explicit forms. I give two examples below. 

\begin{example}
\label{eg::treatment-control-H}
In a treatment-control experiment with $n_1$ treated and $n_0$ control units, the matrix $X$ contains $1$ and a dummy variable for the treatment:
$$
X = \begin{pmatrix}
1_{n_1}& 1_{n_1}\\
1_{n_0}& 0_{n_0} 
\end{pmatrix}.
$$
We can show that 
$$
H = \textup{diag}\{ n_1^{-1} 1_{n_1} 1_{n_1}^{\T} , n_0^{-1} 1_{n_0} 1_{n_0}^{\T} \}.
$$
\end{example}

\begin{example}
\label{eg::anova-H}
In an experiment with $n_j$ units receiving treatment level $j$ $(j=1,\ldots, J)$, the covariate matrix $X$ contains $J$ dummy variables for the treatment levels:
$$
X =  \textup{diag}\{ 1_{n_1}, \ldots, 1_{n_J} \} .
$$
We can show that 
$$
H = \textup{diag}\{ n_1^{-1} 1_{n_1} 1_{n_1}^{\T} , \ldots, n_J^{-1} 1_{n_J} 1_{n_J}^{\T} \}.
$$
\end{example}

I leave the proofs of Examples \ref{eg::treatment-control-H} and \ref{eg::anova-H} as Problem \ref{hw03::special-H}.

\section{Homework problems}

\paragraph{Univariate and multivariate OLS}

Derive the univariate OLS based on the multivariate OLS formula with
$$
X = \begin{pmatrix}
1& x_1 \\
\vdots & \vdots \\
1 & x_n
\end{pmatrix}, 
$$
where the $x_i$'s are scalars.

\paragraph{OLS via vector and matrix calculus}

Use vector and matrix calculus to prove that the OLS coefficient $\hat\beta$ minimizes
$(Y-Xb)^{\T}(Y-Xb)$.

\paragraph{OLS based on pseudo inverse}\label{hw3::ols-pseudoinverse}

Prove that $\hat\beta = X^{+} Y$.

Remark: Recall the definition of the pseudo inverse in Appendix \ref{chapter::linear-algebra}. Under Condition \ref{condition::ols-non-degenerate}, we have $X^{+} = (X^{\T} X)^{-1} X^{\T} .$

\paragraph{Invariance of OLS}\label{hw3::invariance-ols}

Theorem \ref{thm::ols-invariance} below states the invariance properties of OLS. Prove Theorem \ref{thm::ols-invariance}.

\begin{theorem}
\label{thm::ols-invariance}
Assume that $X^{\T} X$ is non-degenerate and $\Gamma$ is a $p\times p$ non-degenerate matrix. Define $\tilde{X}=X\Gamma$. 
From the OLS fit of $Y$ on $X$, we obtain the coefficient $\hat{\beta}$, the fitted value $\hat{Y}$, and the residual $\hat{\varepsilon}$; from the OLS fit of $Y$ on $\tilde{X}$, we obtain the coefficient $\tilde{\beta}$, the fitted value $\tilde{Y}$, and the residual $\tilde{\varepsilon}$. 

We have 
$$
\hat{\beta} = \Gamma \tilde{\beta},\quad
\hat{Y} = \tilde{Y},\quad 
\hat{\varepsilon} = \tilde{\varepsilon}. 
$$
\end{theorem}

Remark: From a linear algebra perspective, $X$ and $X\Gamma$ have the same column space if $\Gamma$ is a non-degenerate matrix:
$$ 
\{   Xb : b\in\mathbb{R}^p \} = \{   X\Gamma c : c\in\mathbb{R}^p \} .
$$
Consequently, there must be a unique projection of $Y$ onto the common column space.

\paragraph{Invariance of the hat matrix}\label{hw03::invariance-of-H}

This problem extends Theorem \ref{thm::ols-invariance} in Problem \ref{hw3::invariance-ols}. 

Prove that $H$ does not change if we change $X$ to $X\Gamma$ where $\Gamma \in \mathbb{R}^{p\times p}$ is a non-degenerate matrix.

\paragraph{Hat matrix with the intercept}\label{hw03::hat-matrix-intercept}

Prove \eqref{eq::projection-constant1} and \eqref{eq::projection-constant2}.

\paragraph{Special hat matrices}\label{hw03::special-H}

Verify the formulas of the hat matrices in Examples \ref{eg::treatment-control-H} and \ref{eg::anova-H}.

\paragraph{OLS with multiple responses}
\label{hw03::multiple-responses}
 
For each unit $i=1,\ldots, n$, we have multiple responses
$y_i=(y_{i1},\ldots, y_{iq})^{\T}   \in \mathbb{R}^q  $ and multiple covariates $x_i=(x_{i1}, \ldots,x_{ip})^{\T}   \in \mathbb{R}^p $. Define
$$
Y =\begin{pmatrix}
y_{11} &\cdots & y_{1q} \\
\vdots & & \vdots \\
y_{n1} & \cdots & y_{nq}
\end{pmatrix} = \begin{pmatrix}
y_{1}^{\T} \\
\vdots \\
y_{n}^{\T}
\end{pmatrix} 
= (Y_1,\ldots,Y_q) 
\in \mathbb{R}^{n\times q}
$$
and
$$
X = \begin{pmatrix}
x_{11} &\cdots & x_{1p} \\
\vdots & & \vdots \\
x_{n1} & \cdots & x_{np}
\end{pmatrix} = \begin{pmatrix}
x_{1}^{\T} \\
\vdots \\
x_{n}^{\T}
\end{pmatrix} 
= (X_1,\ldots,X_p)
\in \mathbb{R}^{n\times p}
$$
as the response and covariate matrices, respectively. Define the multiple OLS coefficient matrix as
\[ \hat{B} = \arg\min_{B \in \mathbb{R}^{p \times q}}\sum_{i=1}^n \| y_i-B^{\T}x_i \|^2 \]
Show that
$
\hat{B} =(
\hat{B}_1, 
\ldots, 
\hat{B}_q 
)
$
has column vectors 
\begin{eqnarray*}
\hat{B}_1 &=& (X^{\T}  X)^{-1}X^{\T}  Y_1 ,\\
 &\vdots & \\ 
\hat{B}_q &=&(X^{\T}  X)^{-1}X^{\T} Y_q.
\end{eqnarray*}

Remark: 
This result tells us that the OLS fit with a vector outcome reduces to multiple separate OLS fits, or, the OLS fit of a matrix $Y$ on a matrix $X$ reduces to the column-wise OLS fits of $Y$ on $X$.

\paragraph{Full sample and subsample OLS coefficients}\label{hw3::full-subsample-ols}

Partition the full sample into $K$ subsamples:
$$
X=\left(\begin{array}{c}
X_{(1)}\\
\vdots\\
X_{(K)}
\end{array}\right),\quad 
Y=\left(\begin{array}{c}
Y_{(1)}\\
\vdots\\
Y_{(K)}
\end{array}\right),
$$
where the $k$th sample consists of $(X_{(k)}, Y_{(k)})$ with $X_{(k)} \in \mathbb{R}^{n_k\times p}$ and $ Y_{(k)} \in \mathbb{R}^{n_k} $ being the covariate matrix and outcome vector. The sample sizes satisfy $n = \sum_{k=1}^K n_k$. Let $\hat{\beta}$ be the OLS coefficient based on the full sample, and $\hat{\beta}_{(k)}$ be the OLS coefficient based on the $k$th sample. 

Prove that 
$$
\hat{\beta} = \sum_{k=1}^{K}W_{(k)}\hat{\beta}_{(k)},
$$
where the weight matrix equals
$$
W_{(k)} =(X^{\T} X)^{-1} X_{(k)}^{\T}X_{(k)} .
$$

Remark: In the special case of a univariate $y_i$ and $x_i$, the OLS of $y_i$ on $x_i$ without the intercept gives the coefficient
$$
\hat\beta = \frac{ \sumn  x_i y_i }{  \sumn x_i^2 } . 
$$
Partition the units into $K$ disjoint parts: $\{ 1, \ldots, n \} = I_1 \cup \cdots \cup I_K$. Run OLS of $y_i$ on $x_i$ without the intercept using units in $I_k$ to obtain the coefficient $\hat{\beta}_{(k)}$. The above formula implies that
$$
\hat\beta =  \sum_{k=1}^K  W_{(k)} \hat{\beta}_{(k)}
$$
where
$$
W_{(k)}  = \frac{   \sum_{i \in I_k} x_i^2 }{ \sumn x_i^2 } 
$$
is proportional to the sum of squares of the regressor $x_i$'s in $I_k$.

\paragraph{Jacobi's theorem}\label{hw03::jacobi}

Prove Theorem \ref{thm::jacobi-theorem-ols} below. 

\begin{theorem}[Jacobi's Theorem]
\label{thm::jacobi-theorem-ols}
The set $\{1 ,\ldots, n\}$ has $\binom{n}{p}$ size-$p$ subsets. Each subset $S$ defines a linear equation for $b\in \mathbb{R}^p$: 
$$
Y_S = X_S b   
$$
where $Y_S \in \mathbb{R}^{p}$ is the subvector of $Y$ and $X_S \in \mathbb{R}^{p\times p}$ is the submatrix of $X$, corresponding to the units in $S$. Define the subset coefficient 
$$
\hat{\beta}_S = X_S^{-1} Y_S
$$
if $X_S$ is invertible and  $\hat{\beta}_S =  0 $ otherwise. 

The OLS coefficient equals a weighted average of these subset coefficients: 
$$
\hat{\beta} = \sum_S   w_S  \hat{\beta}_S
$$
where the summation is over all subsets, and the weights are 
$$
w_S =  \frac{  |  \det(X_S) | ^2  }{  \sum_{S'}  |  \det(X_{S'}) | ^2 } . 
$$
\end{theorem}

Remark: 
Theorem \ref{thm::jacobi-theorem-ols} extends Problem \ref{hw2::pairwise-slope}. \citet{subrahmanyam1972property} reported Theorem \ref{thm::jacobi-theorem-ols} although
\citet{berman1988theorem} attributed it to Jacobi. 
\citet{wu1986jackknife} used it in analyzing the statistical properties of OLS. 

To prove Theorem \ref{thm::jacobi-theorem-ols}, you can use Cramer's rule to express the OLS coefficient and use the Cauchy--Binet formula to expand the determinant of $X^{\T} X$.

\chapter{Gauss--Markov Model and Gauss--Markov Theorem}\label{chapter::gauss-markov}

 Without any stochastic assumptions, the OLS in Chapter \ref{chapter::ols-vector}  is purely algebraic. If we want to discuss the statistical properties of OLS, we must invoke some statistical modeling assumptions. This chapter focuses on the Gauss--Markov model.

\section{Gauss--Markov model}

  A simple starting point is the
following Gauss--Markov model with a fixed design matrix $X$ and unknown
parameters $(\beta,\sigma^{2})$. 

\begin{assumption}[Gauss--Markov model]\label{assume::gm-model}
We have
\[
Y=X\beta+\varepsilon
\]
where the design matrix $X$ is fixed with linearly independent column vectors, and the random error term $\varepsilon$ has the first two moments: 
\begin{eqnarray*}
E(\varepsilon)  &=&  0,\\ 
\cov(\varepsilon) &=& \sigma^{2}I_{n} . 
\end{eqnarray*}
The unknown parameters are $ (\beta,   \sigma^{2} )$.
\end{assumption}

The Gauss--Markov model assumes that $Y$ has mean $X\beta$ and covariance
matrix $\sigma^{2}I_{n}.$ At the individual level, we can also write
it as
\[
y_{i}=x_{i}^{\T}\beta+\varepsilon_{i},\qquad(i=1,\ldots,n)
\]
where the error terms are uncorrelated with mean $0$ and variance
$\sigma^{2}$.

The assumption that $X$ is fixed is not essential, because we can condition on $X$ even if we think $X$ is random. The mean of each $y_i$ is linear in $x_i$ with the same $\beta$ coefficient, which can be a strong assumption. So is the {\it homoskedasticity}\footnote{In this book, I do not spell it as {\it homoscedasticity} since ``k'' better indicates the meaning of variance. \citet{mcculloch1985miscellanea} gave a convincing argument. See also \citet{paloyo2014did}.
} assumption that the error terms have the same variance $\sigma^2$. The critiques on the assumptions aside, I will  derive the properties of $\hat{\beta}$
under the Gauss--Markov model. 

\section{Properties of the OLS estimator}

I first derive the mean and covariance of $\hat{\beta}=(X^{\T}X)^{-1}X^{\T}Y$.
\begin{theorem}
\label{thm:varOLS}Under Assumption  \ref{assume::gm-model}, we have 
\begin{eqnarray*}
E(\hat{\beta}) &=& \beta,\\ 
\cov(\hat{\beta}) &=&  \sigma^{2}(X^{\T}X)^{-1}.
\end{eqnarray*}
\end{theorem}

\begin{myproof}{Theorem}{\ref{thm:varOLS}}
Because $E(Y)=X\beta,$ we have
\begin{eqnarray*}
E(\hat{\beta}) &=& E\{ (X^{\T}X)^{-1}X^{\T}Y\} \\
 &=& (X^{\T}X)^{-1}X^{\T}E(Y) \\
 &=&(X^{\T}X)^{-1}X^{\T}X\beta \\
 &=&\beta.
\end{eqnarray*}
 Because $\cov(Y)=\sigma^{2}I_{n}$, we have 
\begin{align*}
\cov(\hat{\beta}) & =\cov\{ (X^{\T}X)^{-1}X^{\T}Y\}  \\
&=(X^{\T}X)^{-1}X^{\T}\cov(Y)X(X^{\T}X)^{-1}\\
 & =\sigma^{2}(X^{\T}X)^{-1}X^{\T}X(X^{\T}X)^{-1} \\
 & =\sigma^{2}(X^{\T}X)^{-1}.
\end{align*}
\end{myproof}

We can decompose the response vector as 
\[
Y=\hat{Y}+\hat{\varepsilon},
\]
 where the fitted vector is 
 $$
 \hat{Y}=X\hat{\beta}=HY
 $$ and the
residual vector is 
$$
\hat{\varepsilon}=Y-\hat{Y}=(I_{n}-H)Y.
$$ The two
matrices $H$ and $I_{n}-H$ are the keys, which have the following
properties.

\begin{lemma}
\label{lem:projectionms}
Both $H$ and $I_{n}-H$ are projection matrices. They satisfy 
\begin{eqnarray*}
HX &=&X,\\ 
(I_{n}-H)X &=& 0 .  
\end{eqnarray*}
They are orthogonal: 
$$
H(I_{n}-H)=(I_{n}-H)H=0.
$$
\end{lemma}

Lemma \ref{lem:projectionms} follows from simple linear algebra, and I leave its proof as Problem \ref{hw::gauss-markov-projectionm}. It states that $H$ and $I_{n}-H$ are projection matrices onto the column space of $X$ and its complement. 
 Algebraically, $\hat{Y}$ and $\hat{\varepsilon}$ are orthogonal
by the OLS projection because Lemma \ref{lem:projectionms} implies
\begin{eqnarray*}
\hat{Y}^{\T}\hat{\varepsilon} &=& Y^{\T}H^{\T}(I_{n}-H)Y  \\
&=&  Y^{\T}H(I_{n}-H)Y \\
&=& 0.
\end{eqnarray*}
This is also coherent with the geometry in Figure \ref{fig::olsgeometry}.

Moreover, we can derive the mean and covariance matrix of $\hat{Y}$
and $\hat{\varepsilon}$.
\begin{theorem}
\label{thm:GMcov}Under Assumption \ref{assume::gm-model}, we have 
\[
E\left(\begin{array}{c}
\hat{Y}\\
\hat{\varepsilon}
\end{array}\right)=\left(\begin{array}{c}
X\beta\\
0
\end{array}\right)
\]
and 
\[
\cov\left(\begin{array}{c}
\hat{Y}\\
\hat{\varepsilon}
\end{array}\right)=\sigma^{2}\left(\begin{array}{cc}
H & 0 \\
0 & I_{n}-H
\end{array}\right).
\]
So $\hat{Y}$ and $\hat{\varepsilon}$
are uncorrelated. 
\end{theorem}

Please do not be confused with the two statements about $\hat{Y}$ and $\hat{\varepsilon}$:
\begin{enumerate}[label=(S\arabic*), ref=S\arabic*]
\item\label{hats-statement1}
$\hat{Y}$ and $\hat{\varepsilon}$ are orthogonal. 
\item\label{hats-statement2}
$\hat{Y}$ and $\hat{\varepsilon}$ are uncorrelated.
\end{enumerate}
They have different meanings. 
The first statement \eqref{hats-statement1} is an algebraic fact of the OLS procedure. It is about a relationship between two vectors $\hat{Y}$ and $\hat{\varepsilon}$ which holds without assuming the Gauss--Markov model. 
The second statement \eqref{hats-statement2} is stochastic. It is about a relationship between two random vectors $\hat{Y}$ and $\hat{\varepsilon}$, which requires the Gauss--Markov model assumption.

\begin{myproof}{Theorem}{\ref{thm:GMcov}}
The conclusion follows from the simple fact that 
\begin{eqnarray*}
\left(\begin{array}{c}
\hat{Y}\\
\hat{\varepsilon}
\end{array}\right) 
&=& 
\left(\begin{array}{c}
HY\\
(I_{n}-H)Y
\end{array}\right) \\
&=&
\left(\begin{array}{c}
H\\
I_{n}-H
\end{array}\right)Y
\end{eqnarray*}
 is a linear transformation of $Y$.

 It has mean
\begin{eqnarray*}
E\left(\begin{array}{c}
\hat{Y}\\
\hat{\varepsilon}
\end{array}\right)
&=& \left(\begin{array}{c}
H\\
I_{n}-H
\end{array}\right)E(Y) \\
&=&\left(\begin{array}{c}
H\\
I_{n}-H
\end{array}\right)X\beta \\
&=&\left(\begin{array}{c}
HX\beta\\
\left(I_{n}-H\right)X\beta
\end{array}\right) \\
&=&\left(\begin{array}{c}
X\beta\\
0
\end{array}\right),
\end{eqnarray*}
because $HX=X$ and $(I_n-H)X=0$ by Lemma \ref{lem:projectionms}.

It has covariance matrix
\begin{align*}
\cov\left(\begin{array}{c}
\hat{Y}\\
\hat{\varepsilon}
\end{array}\right) & =\left(\begin{array}{c}
H\\
I_{n}-H
\end{array}\right)\cov(Y)\left(\begin{array}{cc}
H^{\T} & (I_{n}-H)^{\T}\end{array}\right)\\
 & =\sigma^{2}\left(\begin{array}{c}
H\\
I_{n}-H
\end{array}\right)\left(\begin{array}{cc}
H & I_{n}-H\end{array}\right)\\
 & =\sigma^{2}\left(\begin{array}{cc}
H^{2} & H(I_{n}-H)\\
(I_{n}-H)H & (I_{n}-H)^{2}
\end{array}\right)\\
 & =\sigma^{2}\left(\begin{array}{cc}
H & 0\\
0 & I_{n}-H
\end{array}\right),
\end{align*}
because $H^2=2, (I_{n}-H)^{2} = I_{n}-H$, and $H(I_{n}-H) = 0$ by Lemma \ref{lem:projectionms}. 
\end{myproof}

Assume the Gauss--Markov model. 
Although the original responses and error terms are uncorrelated between
units with 
$$
\cov(y_i, y_j)=0,\quad 
\cov(\varepsilon_i, \varepsilon_j)=0    \text{ for } i\neq j,
$$ 
the fitted values and the residuals are correlated with 
$$
\cov(\hat{y}_i, \hat{y}_j) 
= \sigma^2 h_{ij},\quad 
\cov(\hat{\varepsilon}_i, \hat{\varepsilon}_j) 
=   - \sigma^2 h_{ij}   \text{ for } i\neq j,
$$ 
based on Theorem \ref{thm:GMcov}.

\section{Variance estimation}

Theorem \ref{thm:varOLS} quantifies the uncertainty of $\hat{\beta}$
by its covariance matrix. However, it is not directly useful because
$\sigma^{2}$ is still unknown. Our next task is to estimate $\sigma^{2}$ based on
the observed data. It is the variance of each $\varepsilon_{i}$,
but the $\varepsilon_{i}$'s are not observable either. Their empirical
analogues are the residuals $\hat{\varepsilon}_{i}=y_{i}-x_{i}^{\T}\hat{\beta}$.
It seems intuitive to estimate $\sigma^{2}$ by 
\[
\tilde{\sigma}^{2}=\textsc{rss}/n , 
\]
where
\[
\textsc{rss}=\sumn\hat{\varepsilon}_{i}^{2}
\]
is the residual sum of squares. However, Theorem \ref{thm:GMcov}
shows that $\hat{\varepsilon}_{i}$ has mean zero and variance $\sigma^{2}(1-h_{ii}),$
which is not the same as the variance of the original $\varepsilon_{i}.$ Consequently,  $\textsc{rss}$ has mean 
\begin{eqnarray*}
E(\textsc{rss}) 
&=& 
\sumn\sigma^{2}(1-h_{ii}) \\
&=& \sigma^{2} \{ n-\text{trace}(H) \} \\
&=& \sigma^{2}(n-p),
\end{eqnarray*}
which implies Theorem \ref{thm:varianceestOLS} below.

\begin{theorem}
\label{thm:varianceestOLS}Define 
$$
\hat{\sigma}^{2}=\textsc{rss}/(n-p)=\sumn\hat{\varepsilon}_{i}^{2}/(n-p).
$$
Then $E(\hat{\sigma}^{2})=\sigma^{2}$ under Assumption \ref{assume::gm-model}. 
\end{theorem}

Theorem \ref{thm:varianceestOLS} implies that $\tilde{\sigma}^{2}$
is a biased estimator for $\sigma^{2}$ because $E(\tilde{\sigma}^{2})=\sigma^{2}(n-p)/n.$
It underestimates $\sigma^{2}$ but with a large sample size $n$, the bias is
small. 

\section{Gauss--Markov Theorem}

So far, we have focused on the OLS estimator. It is intuitive, but
we have not answered the fundamental question yet. Why should we focus
on it? Are there any other better estimators? Under the Gauss--Markov
model, the answer is definite: we focus on the OLS estimator because
it is optimal in the sense of having the smallest covariance matrix
among all linear unbiased estimators. The following famous Gauss--Markov
theorem quantifies this claim, which was named after Carl Friedrich Gauss and Andrey Markov.\footnote{\citet{david1938extensions} used the name ``Markoff Theorem''. \citet{lehmann1951general} appeared to first use the name ``Gauss--Markov Theorem.''} It is for this reason that I call the
corresponding model the Gauss--Markov model. The textbook by \citet{monahan2008primer} also uses this name.

\begin{theorem}
[Gauss--Markov Theorem]
\label{thm:GMtheorem}Under Assumption \ref{assume::gm-model}, the OLS estimator
$\hat{\beta}$ for $\beta$ is the best linear unbiased estimator
(BLUE) in the sense that\footnote{We write $M_1 \succeq   M_2$  if $M_1 - M_2$ is positive semi-definite. See Appendix \ref{chapter::linear-algebra} for a review.} 
$$
\cov(\tilde{\beta})  \succeq 
\cov(\hat{\beta})
$$
for any estimator $\tilde{\beta}$ satisfying 
\begin{enumerate}[label=(\textup{C}\arabic*), ref=\textup{C}\arabic*]
\item\label{item::gm-linear} $\tilde{\beta}=AY$ for some $A\in\mathbb{R}^{p\times n}$ not depending
on $Y$;
\item\label{item::gm-unbiased} $ E (\tilde{\beta} ) = \beta$   for any
$\beta$.
\end{enumerate}
\end{theorem}

Before proving Theorem \ref{thm:GMtheorem}, we need to understand its meaning and immediate implications. We do not compare the OLS estimator with any arbitrary estimators. In fact, we restrict to the estimators that are linear and unbiased.  Condition \eqref{item::gm-linear} requires that  $\tilde{\beta}$ is a linear estimator. More precisely, it is a linear transformation of the
response vector $Y$, where $A$ can be any complex and possibly nonlinear function of $X$. Condition \eqref{item::gm-unbiased} requires that $\tilde{\beta}$  is an unbiased estimator for $\beta$, no matter what true value $\beta$ takes.

Why do we restrict the estimator to be linear? The class of linear estimator is actually quite large because $A$ can be any nonlinear function of $X$, and the only requirement is that the estimator is linear in $Y$. The unbiasedness is a natural requirement for many problems. However, in modern applications with many covariates, some biased estimators can perform better than unbiased estimators if they have smaller variances. We will discuss these estimators in Part \ref{part::modelselection} of this book.

We compare the estimators based on their covariances, which are natural extensions of variances for scalar random variables. The conclusion $\cov(\tilde{\beta}) \succeq  \cov(\hat{\beta})$ implies that for any vector $c  \in \mathbb{R}^p$, we have
$$
c ^{\T}\cov(\tilde{\beta}) c 
\geq 
c ^{\T}  \cov(\hat{\beta})  c  
$$
which is equivalent to
\begin{equation}\label{eq::gauss-markov-1dim}
\var(  c ^{\T} \tilde{\beta} ) \geq   \var(c ^{\T}  \hat{\beta}) . 
\end{equation}
So any linear transformation of the OLS estimator has a variance smaller than or equal to the same linear transformation of any other estimator. In particular, if $c  = (0,\ldots, 1,\ldots, 0)^{\T}$ with only the $j$th coordinate being $1$, then the inequality \eqref{eq::gauss-markov-1dim} implies that 
$$
\var(\tilde{\beta}_j) \geq 
\var(\hat{\beta}_j)   ,\quad (j=1,\ldots, p).
$$
So the OLS estimator has a smaller variance than other estimators for all coordinates. 

Now we prove Theorem \ref{thm:GMtheorem}. 

\begin{myproof}{Theorem}{\ref{thm:GMtheorem}}
We must verify that the OLS estimator itself satisfies \eqref{item::gm-linear} and \eqref{item::gm-unbiased}. We have $\hat{\beta}=\hat{A}Y$ with  $\hat{A}=(X^{\T}X)^{-1}X^{\T}$, and it is unbiased by Theorem \ref{thm:varOLS}. 

First, the unbiasedness requires that $E(\tilde{\beta}) =\beta$ for any value of $\beta$. Under the Gauss--Markov Model, it requires that 
$$
E(AY)=AE(Y)=AX\beta=\beta
$$
for any value of $\beta$. The requirement $AX\beta=\beta$ for any value of $\beta$ implies that 
\begin{eqnarray}\label{eq::gaussmarkov-unbiasedcondition}
 AX=I_{p}
\end{eqnarray}
 must hold. In particular,
the OLS estimator satisfies $\hat{A}X = (X^{\T}X)^{-1}X^{\T} X =I_{p}.$

Second, we can decompose the covariance of $\tilde{\beta}$ as
\begin{eqnarray*}
\cov(\tilde{\beta}) 
&=& \cov(\hat{\beta} + \tilde{\beta} - \hat{\beta} ) \\
&=&  \cov(\hat{\beta} )  + \cov ( \tilde{\beta} - \hat{\beta} ) 
+ \cov(\hat{\beta} ,  \tilde{\beta} - \hat{\beta})   +  \cov(   \tilde{\beta} - \hat{\beta}, \hat{\beta}) .
\end{eqnarray*}
The last two terms are in fact 0. By symmetry, we only need to
show that the third term is 0:
\begin{eqnarray*}
  \cov(\hat{\beta} ,  \tilde{\beta} - \hat{\beta})   
& = &\cov\{ \hat{A}Y,(A-\hat{A})Y \} \\
 & =&\hat{A}\cov(Y)(A-\hat{A})^{\T}\\
 & =&\sigma^{2}\hat{A}(A-\hat{A})^{\T}\\
 & =&\sigma^{2}(\hat{A}A^{\T}-\hat{A}\hat{A}^{\T})\\
 & =&\sigma^{2}\left\{ (X^{\T}X)^{-1}X^{\T}A^{\T}-(X^{\T}X)^{-1}X^{\T}X(X^{\T}X)^{-1}\right\} \\
 & =&\sigma^{2}\left\{ (X^{\T}X)^{-1}I_{p}-(X^{\T}X)^{-1}\right\}  \qquad (\text{by } \eqref{eq::gaussmarkov-unbiasedcondition}) \\
 & =&0.
\end{eqnarray*}
The above covariance decomposition simplifies to
\[
\cov(\tilde{\beta})= \cov(\hat{\beta} )  + \cov ( \tilde{\beta} - \hat{\beta} )  , 
\]
which implies 
\[
\cov(\tilde{\beta})-\cov(\hat{\beta}) = \cov(\tilde{\beta} - \hat{\beta}) \succeq 0. 
\]
\end{myproof}

In the process of the proof, we have shown two stronger results
$$
\cov(\tilde{\beta} - \hat{\beta}, \hat{\beta}) = 0
$$
and
$$
\cov(\tilde{\beta} - \hat{\beta})  =  \cov(\tilde{\beta})-\cov(\hat{\beta}).
$$
They hold only when $\hat{\beta}$ is BLUE. They do not hold when comparing two general estimators.

Theorem \ref{thm:GMtheorem} is elegant but abstract. It says that
in some sense, we can focus on the OLS estimator because it is
the best one in terms of the covariance among all linear unbiased estimators. Then
we do not need to consider other estimators. However, we have not
mentioned any other estimators for $\beta$ yet, which makes Theorem
\ref{thm:GMtheorem} not concrete enough. From the proof above, a
linear unbiased estimator $\tilde{\beta}=AY$ only needs to satisfy
$AX=I_{p}$, which imposes $p^{2}$ constraints on the $p\times n$
matrix $A$. Therefore, we have $p(n-p)$ free parameters to choose
from and have infinitely many linear unbiased estimators in general. A
class of linear unbiased estimators that will be discussed more thoroughly in Chapter \ref{chapter::WLS} are the weighted least squares estimators
\[
\tilde{\beta}=(X^{\T}\Sigma^{-1}X)^{-1}X^{\T}\Sigma^{-1}Y,
\]
 where $\Sigma$ is a positive definite matrix not depending on $Y$
such that $\Sigma$ and $X^{\T}\Sigma^{-1}X$ are invertible. It is linear in $Y$, and we
can show that it is unbiased for $\beta$:
\begin{eqnarray*}
E(\tilde{\beta}) 
&=& E\{ (X^{\T}\Sigma^{-1}X)^{-1}X^{\T}\Sigma^{-1}Y\} \\
&=& (X^{\T}\Sigma^{-1}X)^{-1}X^{\T}\Sigma^{-1}X\beta \\
&=&\beta.
\end{eqnarray*}
Different choices of $\Sigma$ give different $\tilde{\beta},$ but
Theorem \ref{thm:GMtheorem} states that the OLS estimator with $\Sigma=I_{n}$
has the smallest covariance matrix under the Gauss--Markov model. 

I will give an extension and some applications of the Gauss--Markov Theorem in Problems \ref{hw04::blue-mean}--\ref{hw4::gaussmarkov-prediction}.

\section{Homework problems}

\paragraph{Projection matrices}\label{hw::gauss-markov-projectionm}

Prove Lemma \ref{lem:projectionms}.

\paragraph{Univariate OLS and the optimal design}
\label{hw::gaussmarkov-1d}

Assume the Gauss--Markov model $y_i = \alpha + \beta x_i + \varepsilon_i$ $(i=1,\ldots, n)$ with a scalar $x_i$. Show that the variance of the OLS coefficient for $x_i$ equals
$$
\var(  \hat{\beta} ) = \sigma^2 \Big / \sumn (x_i - \bar{x})^2 .
$$

Assume $x_i$ is in the interval $[0, 1]$. We want to choose their values to minimize $\var(  \hat{\beta} )$. Assume that $n$ is an even number. Find the $x_i$'s that minimize $\var(  \hat{\beta} ) $. 

Remark: You may find the following probability result useful. For a random variable $\xi$ in the interval $[0, 1]$, we have the following inequality
\begin{eqnarray*}
\var(\xi) &=& E(\xi^2) - \{ E(\xi) \}^2 \\
 &\leq & E(\xi) - \{ E(\xi) \}^2  \\
 &=& E(\xi)\{ 1 - E(\xi) \} \\
 &\leq & 1/4. 
\end{eqnarray*}
The first inequality becomes an equality if and only if $\xi=0$ or $1$; the second inequality becomes an equality if and only if $E(\xi) = 1/2.$

\paragraph{BLUE estimator for the mean}\label{hw04::blue-mean}

Assume that $y_i$ has mean $\mu$ and variance $\sigma^2$, and the $y_i$'s are uncorrelated $(i=1, \ldots, n)$. A linear estimator of the mean $\mu$ has the form $\hat{\mu} = \sumn a_i y_i$, which is unbiased as long as $ \sumn a_i = 1$. So there are infinitely many linear unbiased estimators for $\mu$. 

Find the BLUE for $\mu$ and prove why it is BLUE.

\paragraph{More variance estimators under the Gauss-Markov model}\label{hw04::variance-estimation-A}

Assume the Gauss-Markov model in Assumption \ref{assume::gm-model}.

With the OLS estimator $\hat\beta = (X^{\T}X)^{-1} X^{\T} Y$, we can obtain the residual vector $\hat\varepsilon = Y - X\hat{\beta}$. For any symmetric matrix $A$, prove that
$$
\hat{\sigma}_A^2 =  \frac{ \hat\varepsilon^{\T} A \hat\varepsilon  }{  \text{trace}(A(I_n -  H))  } 
$$
is unbiased for $\sigma^2$ as long as $\text{trace}(A(I_n -  H)) \neq 0.$

Remark: This chapter focuses on  $\hat{\sigma}_A^2$ with $A = I_n$. In general, $\hat{\sigma}_A^2$ gives infinitely many unbiased estimators for $\sigma^2$. A natural question is: what is the optimal choice of $A$ with minimum variance? This question is more complicated because the variance of $\hat{\sigma}_A^2$ depends on not only the mean and covariance of $\varepsilon$ as specified in Assumption \ref{assume::gm-model} but also higher-order moments of $\varepsilon$. We will revisit this problem in Problem \ref{hw5:optimal-A-variance-normal}.

\paragraph{Consequence of useless regressors}\label{hw4::useless-regressor}

Partition the covariate matrix and parameter into 
$$
X=(X_1, X_2),\quad \beta = \begin{pmatrix}
\beta_1\\
\beta_2
\end{pmatrix},
$$
where $X_1 \in \mathbb{R}^{n\times k}, X_2 \in \mathbb{R}^{n\times l}, \beta_1 \in \mathbb{R}^k$ and $\beta_2 \in \mathbb{R}^l$ with $k+l=p$. Assume the Gauss--Markov model with $\beta_2 = 0$. Let $\hat{\beta}_1$ be the first $k$ coordinates of $\hat{\beta} = (X^{\T}X)^{-1} X^{\T} Y$ and $\tilde{\beta}_1 = (X_1^{\T}X_1)^{-1} X_1^{\T} Y$ be the coefficient based on the OLS fit of $Y$ on $X_1$ only. 

Prove that 
$$
\cov(\hat{\beta}_1) \succeq \cov(\tilde{\beta}_1).
$$

 \paragraph{Simple average of subsample OLS coefficients}\label{hw4::average-subsample-ols}
 
 Inherit the setting of Problem \ref{hw3::full-subsample-ols}. Define the simple average of the subsample OLS coefficients as 
 $$
 \bar{\beta} = K^{-1}\sum_{k=1}^{K} \hat{\beta}_{(k)}.
 $$ 
 
 Assume the Gauss--Markov model. 
  Prove that 
 $$
\cov(\bar{\beta}) \succeq \cov(\hat{\beta}).
$$

 \paragraph{Gauss--Markov Theorem for prediction}\label{hw4::gaussmarkov-prediction}

Theorem \ref{thm::GM-prediction} below extends Theorem \ref{thm:GMtheorem}. Prove Theorem \ref{thm::GM-prediction}.

\begin{theorem}[Gauss--Markov Theorem for Prediction]
\label{thm::GM-prediction}
Under Assumption \ref{assume::gm-model}, the OLS predictor 
$\hat{Y}=X\hat{\beta}$ for the mean $X\beta$ is the best linear unbiased predictor in the sense that $\cov(\tilde{Y}) \succeq \cov( \hat{Y} )$
for any predictor $\tilde{Y}$ satisfying 
\begin{enumerate}[label=(\textup{C}\arabic*), ref=\textup{C}\arabic*]
\item\label{item::gm-linear-pred} $\tilde{Y}=\tilde{H}Y$ for some $ \tilde{H} \in\mathbb{R}^{n \times n}$ not depending
on $Y$;
\item\label{item::gm-unbiased-pred} $E ( \tilde{Y} ) = X \beta$   for any $\beta$. 
\end{enumerate}
\end{theorem}

\paragraph{Nonlinear unbiased estimator under the Gauss--Markov model}
\label{hw4::nonlinear-unbiased-underGM}

Under Assumption \ref{assume::gm-model}, prove that if the matrices $Q_j$'s satisfy  
\begin{equation}
\label{eg::quadratic-GM}
X^{\T} Q_j X = 0
,\quad 
\text{trace}(Q_j) = 0  \text{ for all } j=1,\ldots, p
\end{equation}
then
$$
\tilde\beta = \hat\beta + \begin{pmatrix}
Y^{\T} Q_1 Y \\
\vdots \\
Y^{\T} Q_p Y 
\end{pmatrix}
$$
is unbiased for $\beta$.

Remark: The above estimator $\tilde\beta $ is a quadratic function of $Y$. It is a nonlinear unbiased estimator for $\beta$. It is not difficult to show the unbiasedness. More remarkably, \citet[][Theorem 4.3]{koopmann1982parameterschatzung} showed that under Assumption \ref{assume::gm-model}, any unbiased estimator for $\beta$ must have the form of $\tilde\beta $.

The condition in \eqref{eg::quadratic-GM} is not a trivial condition. Can you find $Q_j$'s to satisfy \eqref{eg::quadratic-GM}?

\chapter{Normal Linear Model: Inference and Prediction}
 \label{chapter::normal-linear-model}

Under the Gauss--Markov model in Assumption \ref{assume::gm-model} in Chapter \ref{chapter::gauss-markov}, we have calculated the first two moments of the OLS estimator $\hat{\beta} = (X^{\T} X)^{-1} X^{\T} Y$:
\begin{eqnarray*}
E(\hat{\beta}) &=& \beta,\\
\cov(\hat{\beta}) &=& \sigma^{2}(X^{\T}X)^{-1},
\end{eqnarray*}
and have shown that $\hat{\sigma}^{2} =   \hat{\varepsilon}^{\T}  \hat{\varepsilon} / (n-p)$ is unbiased for $\sigma^{2}$, where $\hat{\varepsilon} = Y - X \hat{\beta}$ is the residual vector. The Gauss--Markov theorem further ensures that the OLS estimator is the best linear unbiased estimator (BLUE). 
Although these results characterize the nice properties of the OLS estimator,
they do not fully determine its distribution and are thus inadequate
for statistical inference.

 This chapter will derive the joint distribution
of $(\hat{\beta},\hat{\sigma}^{2})$ under the Normal linear model with stronger distribution assumptions.

\begin{assumption}
[Normal linear model] 
\label{assume::nlm}
We have
$$
Y\sim\N(X\beta,\sigma^{2}I_{n}) , 
$$
or, equivalently,
$$
y_{i} \stackrel{ \textsc{ind} }{ \sim}\N(x_{i}^{\T}\beta,\sigma^{2}),\qquad(i=1,\ldots,n),
$$ 
where the design matrix $X$ is fixed with linearly independent column vectors. The unknown parameters are $(\beta,\sigma^{2})$. 
\end{assumption}

We can also write the Normal linear model as a linear function of covariates with error terms:
$$
Y = X\beta +\varepsilon \text{ with } \varepsilon \sim \N(0,\sigma^2 I_n) 
$$
or, equivalently, 
$$
y_i = x_{i}^{\T}\beta  +\varepsilon_i \text{ with } \varepsilon_i \iidsim \N(0,\sigma^2) \qquad(i=1,\ldots,n).
$$

Assumption \ref{assume::nlm} implies Assumption \ref{assume::gm-model}. Beyond the Gauss--Markov model, it further requires independent and identically distributed (IID) Normal error terms. 
Assumption \ref{assume::nlm} is strong, but it is canonical in statistics. It allows us to derive elegant formulas and also justifies the outputs of the linear regression functions in many statistical packages.

More interestingly, even though Assumption \ref{assume::nlm} is strong, the statistical procedures derived in this chapter are robust to various violations. 
\footnote{
As another example, \citet{zhang2025random} discussed the robustness of the $t$ test in this chapter even when the error terms are correlated in an unknown way.
}  
Chapter \ref{chapter::EHW} will discuss the robustness of the $t$ test in this chapter even when the error terms are not Normally distributed. Moreover,  Chapter \ref{chapter::EHW} will relax Assumption \ref{assume::nlm} and propose a modified procedure that is particularly robust to heteroskedasticity.

\section{Joint distribution of the OLS coefficient and variance estimator}

We first state the main theorem on the joint distribution of $(\hat{\beta},\hat{\sigma}^{2})$
via the joint distribution of $(\hat{\beta},\hat{\varepsilon}).$
\begin{theorem}
\label{thm:normalexactdistribution}Under Assumption \ref{assume::nlm}, we have 
\[
\left(\begin{array}{c}
\hat{\beta}\\
\hat{\varepsilon}
\end{array}\right)\sim\N\left\{ \left(\begin{array}{c}
\beta\\
0
\end{array}\right),\sigma^{2}\left(\begin{array}{cc}
(X^{\T}X)^{-1} & 0\\
0 & I_{n}-H
\end{array}\right)\right\} ,
\]
and
$$
\hat{\sigma}^{2}/\sigma^{2}\sim\chi_{n-p}^{2}/(n-p).
$$
So 
$$
\hat{\beta}\ind\hat{\varepsilon} ,\quad 
 \hat{\beta}\ind\hat{\sigma}^{2}  .
$$
\end{theorem}
 
\begin{myproof}{Theorem}{\ref{thm:normalexactdistribution}}
First, 
\begin{eqnarray*}
\left(\begin{array}{c}
\hat{\beta}\\
\hat{\varepsilon}
\end{array}\right)
&=&
\left(\begin{array}{c}
(X^{\T}X)^{-1}X^{\T}Y\\
(I_{n}-H)Y
\end{array}\right) \\
&=&\left(\begin{array}{c}
(X^{\T}X)^{-1}X^{\T}\\
I_{n}-H
\end{array}\right)Y
\end{eqnarray*}
is a linear transformation of $Y$, so they are jointly Normal. We have verified their means and variances in Chapter \ref{chapter::gauss-markov}, so we only need to show that their covariance is zero:
\begin{eqnarray*}
\cov(\hat{\beta},\hat{\varepsilon})
&=&(X^{\T}X)^{-1}X^{\T}\cov(Y)(I_{n}-H)^{\T}\\
&=&\sigma^{2}(X^{\T}X)^{-1}X^{\T}(I_{n}-H^{\T}) \\
&=&0
\end{eqnarray*}
which holds because $(I_{n}-H)X=0$ by Lemma \ref{lem:projectionms}.
The joint Normality with 0 covariance implies $\hat{\beta}\ind\hat{\varepsilon}$.

Second, because $\hat{\sigma}^{2}=\textsc{rss}/(n-p)=\hat{\varepsilon}^{\T}\hat{\varepsilon}/(n-p)$
is a quadratic function of $\hat{\varepsilon}$, it is independent
of $\hat{\beta}$. We only need to show that it is a scaled chi-squared
distribution. This follows from Theorem \ref{thm::normal-chisq} in Appendix \ref{chapter:appendix-rvs}  due to the Normality of $\hat{\varepsilon}/\sigma$
with the projection matrix $I_{n}-H$ as its covariance matrix. 
\end{myproof}

The second theorem is on the joint distribution of $(\hat{Y},\hat{\varepsilon}).$
We have shown their means and covariance matrix in Chapter \ref{chapter::gauss-markov}.
Because they are linear transformations of $Y$, they are jointly Normal
and independent. 
\begin{theorem}\label{thm::gaussianlm-independence-ye}
Under Assumption \ref{assume::nlm}, we have 
\[
\left(\begin{array}{c}
\hat{Y}\\
\hat{\varepsilon}
\end{array}\right)\sim\N\left\{ \left(\begin{array}{c}
X\beta\\
0
\end{array}\right),\sigma^{2}\left(\begin{array}{cc}
H & 0\\
0 & I_{n}-H
\end{array}\right)\right\} ,
\]
so 
$$
\hat{Y}\ind\hat{\varepsilon}.
$$
\end{theorem}

Now we have seen several properties of $\hat{Y} $ and $  \hat{\varepsilon}$:
\begin{enumerate}[label=(P\arabic*), ref=P\arabic*]
\item\label{fitted-residual-chapter5-p1}
In Chapter \ref{chapter::ols-vector}, we have shown that $Y = \hat{Y}  + \hat{\varepsilon}$ with $\hat{Y}  \perp  \hat{\varepsilon}$ by the OLS properties, which is a pure linear algebra fact without assumptions.
\item\label{fitted-residual-chapter5-p2}
In Chapter \ref{chapter::gauss-markov}, Theorem \ref{thm:GMcov} ensures that $\hat{Y} $ and $  \hat{\varepsilon}$ are uncorrelated under Assumption \ref{assume::gm-model}.
\item\label{fitted-residual-chapter5-p3}
Now Theorem \ref{thm::gaussianlm-independence-ye} further ensures that $\hat{Y}  \ind  \hat{\varepsilon}$ under Assumption \ref{assume::nlm}.
\end{enumerate}

The first result \eqref{fitted-residual-chapter5-p1} states that $\hat{Y} $ and $  \hat{\varepsilon}$
are orthogonal. 
The second result \eqref{fitted-residual-chapter5-p2} states that $\hat{Y} $ and $  \hat{\varepsilon}$ are uncorrelated. 
The third result \eqref{fitted-residual-chapter5-p3} states $\hat{Y} $ and $  \hat{\varepsilon}$ are independent. 
They hold under different assumptions.

\section{Pivotal quantities and statistical inference}

\subsection{Scalar parameters}
We first consider statistical inference for $c ^{\T}\beta$,
a one-dimensional linear function of $\beta$ where $c \in\mathbb{R}^{p}.$
For example, if $c = e_j \equiv   (0,\ldots,1,\ldots,0)^{\T}$ with only the
$j$th element being one, then $c ^{\T}\beta=\beta_{j}$ is the
$j$th element of $\beta$, which measures the impact of $x_{ij}$
on $y_{i}$ on average. Standard software packages report statistical
inference for each element of $\beta$. Sometimes we may also be interested in $\beta_j - \beta_{j'}$, the difference between the coefficients of two covariates, which corresponds to $c=(0, \ldots,0,1,0,\ldots,0,-1,0,\ldots,0)^{\T} = e_j - e_{j'}$. 

Theorem \ref{thm:normalexactdistribution} implies that 
\[
c ^{\T}\hat{\beta}\sim\N\{ c ^{\T}\beta,\sigma^{2}c ^{\T}(X^{\T}X)^{-1}c \} .
\]
However, this is not directly useful because $\sigma^{2}$ is unknown.
With $\sigma^{2}$ replaced by $\hat{\sigma}^{2}$, the standardized
distribution
\[
T_c  = 
\frac{c ^{\T}\hat{\beta}-c ^{\T}\beta}{\sqrt{\hat{\sigma}^{2}c ^{\T}(X^{\T}X)^{-1}c }}
\]
does not follow $\N(0,1)$ anymore. In fact, it is a $t$ distribution as shown in Theorem \ref{thm:1dimpivotal} below. 

\begin{theorem}
\label{thm:1dimpivotal}Under Assumption \ref{assume::nlm}, for a fixed vector $c  \in \mathbb{R}^p$, we have 
$$
T_c  \sim t_{n-p}.
$$
\end{theorem}

\begin{myproof}{Theorem}{\ref{thm:1dimpivotal}}
From Theorem \ref{thm:normalexactdistribution}, the standardized
distribution with the true $\sigma^{2}$ follows
\[
\frac{c ^{\T}\hat{\beta}-c ^{\T}\beta}{\sqrt{\sigma^{2}c ^{\T}(X^{\T}X)^{-1}c }}\sim\N(0,1),
\]
 $\hat{\sigma}^{2}/\sigma^{2}\sim\chi_{n-p}^{2}/(n-p),$ and they
are independent. These facts imply that
\begin{align*}
T_c &= 
\frac{c ^{\T}\hat{\beta}-c ^{\T}\beta}{\sqrt{\hat{\sigma}^{2}c ^{\T}(X^{\T}X)^{-1}c }} \\
& =\frac{c ^{\T}\hat{\beta}-c ^{\T}\beta}{\sqrt{\sigma^{2}c ^{\T}(X^{\T}X)^{-1}c }}\Big/\sqrt{\frac{\hat{\sigma}^{2}}{\sigma^{2}}}\\
 & \sim \frac{\N(0,1)}{\sqrt{\chi_{n-p}^{2}/(n-p)}} ,
\end{align*}
where $\N(0,1)$ and $\chi_{n-p}^{2}$ denote independent standard Normal and $\chi_{n-p}^{2}$ random variables, respectively. Therefore, 
$T_c \sim t_{n-p}$ by the definition of the $t$ distribution in Appendix \ref{chapter:appendix-rvs}. 
\end{myproof}

In Theorem \ref{thm:1dimpivotal}, the $T_c $ on the left-hand side
depends on the observed data and the unknown true parameters, but the $t_{n-p}$ on the right-hand side is a random variable depending on only the dimension
$(n,p)$ of $X$, but neither the data nor the true parameters. Because of this, we call the quantity on the left-hand side a {\it pivotal quantity}. Based
on the quantiles of the $t_{n-p}$ random variable, we can tie the
data and the true parameter via the following probability statement
\[
\pr\left\{ \left|\frac{c ^{\T}\hat{\beta}-c ^{\T}\beta}{\sqrt{\hat{\sigma}^{2}c ^{\T}(X^{\T}X)^{-1}c }}\right|\leq  
t_{1-\alpha/2,n-p}
\right\} =1-\alpha
\]
 for any $0<\alpha<1$, where $t_{1-\alpha/2,n-p}  $ is the $1-\alpha/2$
quantile of $t_{n-p}$. 
When $n-p$ is large (e.g. larger than $30$), the $1-\alpha/2$ quantile of $t_{n-p}$ is close to that of $\N(0,1)$. 
In particular, when $n-p$ is large, $t_{97.5\%,n-p}\approx 1.96$, the $97.5\%$ quantile of $\N(0,1)$, which is the critical value for the $95\%$ confidence interval.

Define 
$$
\hat{\text{se}}_{c } = 
\sqrt{\hat{\sigma}^{2}c ^{\T}(X^{\T}X)^{-1}c }
$$
which is often called the (estimated) standard error of $c ^{\T}\hat{\beta}$. Using the definition of $\hat{\text{se}}_{c } $,
we can equivalently write the above probability statement as
\[
\pr\left\{ c ^{\T}\hat{\beta}- t_{1-\alpha/2,n-p} \hat{\text{se}}_{c }\leq c ^{\T}\beta\leq c ^{\T}\hat{\beta}+ t_{1-\alpha/2,n-p} \hat{\text{se}}_{c }\right\} =1-\alpha.
\]
We use 
\[
c ^{\T}\hat{\beta} \pm t_{1-\alpha/2,n-p} \hat{\text{se}}_{c } 
\]
as a $1-\alpha$ level confidence interval for $c ^{\T}\beta$. By duality of confidence
interval and hypothesis testing, we can also construct a level $\alpha$
test for $c ^{\T}\beta$. More precisely, we reject the null hypothesis $c ^{\T}\beta = d$ if the above confidence interval does not cover $d$, for a fixed number $d$.

As an important case, $c=e_j$ so $c^{\T} \beta = \beta_j$. Standard software packages, for example, \ri{R}, report the point estimator $\hat{\beta}_j$, the standard error $\hat{\text{se}}_j = \sqrt{ \hat{\sigma}^2 [(X^{\T}X)^{-1}]_{jj} }$, the $t$ statistic $T_j = \hat{\beta}_j / \hat{\text{se}}_j $, and the two-sided $p$-value $\pr( | t_{n-p} | \geq  |T_j|  )$ for testing whether $\beta_j$ equals zero or not. Section \ref{sec::normallinearmodel-r} below gives some examples.

\subsection{Vector parameters}
We then consider statistical inference for $C\beta$, a multi-dimensional
linear function of $\beta$ where $C\in\mathbb{R}^{l\times p}.$ If
$l=1$, then it reduces to the one-dimensional case. If 
$$
C = \begin{pmatrix}
c_1^{\T} \\
\vdots \\
c_l^{\T}
\end{pmatrix} 
$$
with $l>1$, then 
$$
C\beta = \begin{pmatrix}
c_1^{\T} \beta \\
\vdots \\
c_l^{\T} \beta
\end{pmatrix}
$$
corresponds to the joint values of $l$ parameters $ c_1^{\T} \beta, \ldots,c_l^{\T} \beta$.

\begin{example}\label{eg::testing-1}
If
\[
C=\left(\begin{array}{ccccc}
0 & 1 & 0 & \cdots & 0\\
0 & 0 & 1 & \cdots & 0\\
\vdots & \vdots & \vdots & \cdots & \vdots\\
0 & 0 & 0 & \cdots & 1
\end{array}\right) , 
\]
then 
\[
C\beta=\left(\begin{array}{c}
\beta_{2}\\
\vdots\\
\beta_{p}
\end{array}\right) 
\]
contains all the coefficients except for the first one (the
intercept in most cases). Most software packages report the test of
the joint significance of $(\beta_{2}, \ldots, \beta_{p})$. Section \ref{sec::normallinearmodel-r} below gives some examples. 
\end{example}

\begin{example}\label{eg::partition-reg-F}
Another
leading application is to test whether $\beta_{2}=0$ in the following
regression partitioned by $X=(X_{1},X_{2})$, where $X_{1}$ and $X_{2}$
are $n\times k$ and $n\times l$ matrices:
\[
Y=X_{1}\beta_{1}+X_{2}\beta_{2}+\varepsilon,
\]
 with
\[
C=\left(\begin{array}{cc}
0_{l\times k} & I_{l}\end{array}\right)\left(\begin{array}{c}
\beta_{1}\\
\beta_{2}
\end{array}\right)
\]
and therefore, $ C\beta=\beta_{2}.$
 We will discuss this partitioned regression in more detail in Chapters \ref{chapter::FWL-theorem} and \ref{chapter::FWL-application}.
 \end{example}
 
Now we will focus on the generic problem of inferring $C\beta$.
To avoid degeneracy, we assume that $C$ does not have redundant rows, quantified below. 

\begin{assumption}\label{assume::c-non-degenerate}
 $C$ has linearly independent
rows. 
\end{assumption}

Theorem \ref{thm:normalexactdistribution} implies that 
\[
C\hat{\beta}-C\beta\sim\N\{ 0,\sigma^{2}C(X^{\T}X)^{-1}C^{\T}\} ,
\]
and therefore, by Theorem \ref{thm::normal-chisq} in Appendix \ref{chapter:appendix-rvs},  the standardized quadratic form has a chi-squared distribution\footnote{
Technically, we need to prove that $\sigma^{2}C(X^{\T}X)^{-1}C^{\T}$ is a positive definite matrix.

Because $X$ has linearly independent columns, $X^{\T}X$ is a non-degenerate and thus positive definite
matrix. Since $ u^{\T} C(X^{\T}X)^{-1}C^{\T}u \geq 0$, to show that $C(X^{\T}X)^{-1}C^{\T}$ is non-degenerate, we only need to show that $u^{\T} C(X^{\T}X)^{-1}C^{\T}u  = 0$ must imply  $u = 0$. 
From $u^{\T} C(X^{\T}X)^{-1}C^{\T}u  = 0$, we know  $C^{\T}u = u_1 c_1 + \cdots u_l c_l = 0$. Since the rows of $C$ are linearly independent, we must have $u = 0$. The proof is complete.}
\[
(C\hat{\beta}-C\beta)^{\T}\left\{ \sigma^{2}C(X^{\T}X)^{-1}C^{\T}\right\} ^{-1}(C\hat{\beta}-C\beta)\sim\chi_{l}^{2}.
\]
Again this is not directly useful with unknown $\sigma^{2}.$ Replacing
$\sigma^{2}$ with the unbiased estimator $\hat{\sigma}^{2}$ and
using a scaling factor $l$, we can obtain a pivotal quantity that
has an $F$ distribution as summarized in  Theorem \ref{thm:ldimpivotal} below.

\begin{theorem}
\label{thm:ldimpivotal}Under Assumptions \ref{assume::nlm} and \ref{assume::c-non-degenerate}, the $F$ statistic
\[
F_C  = 
\frac{(C\hat{\beta}-C\beta)^{\T}\left\{ C(X^{\T}X)^{-1}C^{\T}\right\} ^{-1}(C\hat{\beta}-C\beta)}{l\hat{\sigma}^{2}}
\]
follows the $F$ distribution with degrees of freedom $l$ and $n-p$:
$$
F_C  \sim F_{l,n-p} . 
$$
\end{theorem}

\begin{myproof}{Theorem}{\ref{thm:ldimpivotal}}
Similar to the proof of Theorem \ref{thm:1dimpivotal}, we apply Theorem
\ref{thm:normalexactdistribution} to derive that
\[
\begin{aligned} 
F_C
= & \frac{(C\hat{\beta}-C\beta)^{\T}\left\{ \sigma^{2}C(X^{\T}X)^{-1}C^{\T}\right\} ^{-1}(C\hat{\beta}-C\beta)/l}{\hat{\sigma}^{2}/\sigma^{2}}\\
\sim & \frac{\chi_{l}^{2}/l}{\chi_{n-p}^{2}/(n-p)} ,
\end{aligned}
\]
where $\chi_{l}^{2} $ and $\chi_{n-p}^{2} $ denote independent $\chi_{l}^{2} $ and $\chi_{n-p}^{2} $ random variables, respectively. Therefore, $F_C  \sim  F_{l,n-p}$ by the definition of the $F$ distribution in Appendix \ref{chapter:appendix-rvs}. 
\end{myproof}

Theorem \ref{thm:ldimpivotal} motivates the following confidence
region for $C\beta$:
\[
\left\{ r  :(C\hat{\beta}- r )^{\T}\left\{ C(X^{\T}X)^{-1}C^{\T}\right\} ^{-1}(C\hat{\beta}- r)\leq l\hat{\sigma}^{2}
f_{1-\alpha, l, n-p}
\right\} ,
\]
 where $f_{1-\alpha, l, n-p}$ is the upper $\alpha$ quantile of the
$F_{l,n-p}$ distribution. By duality of the confidence region and hypothesis
testing, we can also construct a level $\alpha$ test for $C\beta$. Most statistical packages automatically report the $p$-value based on the $F$ statistic in Example \ref{eg::testing-1}.

As a final remark, the statistics in Theorems \ref{thm:1dimpivotal}
and \ref{thm:ldimpivotal} are called the Wald-type statistics. 

\section{Prediction based on pivotal quantities}\label{sec::prection-normallinear}

Practitioners use OLS not only to infer $\beta$ but also to predict
future outcomes. For the pair of future data $(x_{n+1},y_{n+1})$, we observe
only $x_{n+1}$ and want to predict $y_{n+1}$ based on $(X,Y)$ and $x_{n+1}$.
Assume a stable relationship between $y_{n+1}$ and $x_{n+1}$, that is,
\[
y_{n+1}\sim\N(x_{n+1}^{\T}\beta,\sigma^{2})
\]
with the same $(\beta,\sigma^{2}).$ 

First, we can predict the mean of $y_{n+1}$ which is $x_{n+1}^{\T}\beta$.
It is just a one-dimensional linear function of $\beta$, so the theory
in Theorem \ref{thm:1dimpivotal} is directly applicable. A natural
unbiased predictor is $x_{n+1}^{\T}\hat{\beta}$ with $1-\alpha$ level
prediction interval
\[
 x_{n+1}^{\T}\hat{\beta} \pm t_{1-\alpha/2,  n-p} \hat{\text{se}}_{x_{n+1}} .
\]

Second, we can predict $y_{n+1}$ itself, which is a random variable.
We can still use $x_{n+1}^{\T}\hat{\beta}$ as a natural unbiased predictor
but need to modify the prediction interval. Because $y_{n+1}\ind\hat{\beta}$,
we have
\[
y_{n+1}-x_{n+1}^{\T}\hat{\beta}\sim\N\{ 0,\sigma^{2}+\sigma^{2}x_{n+1}^{\T}(X^{\T}X)^{-1}x_{n+1}\} ,
\]
and therefore
\begin{align*}
\frac{y_{n+1}-x_{n+1}^{\T}\hat{\beta}}{\sqrt{\hat{\sigma}^{2}+\hat{\sigma}^{2}x_{n+1}^{\T}(X^{\T}X)^{-1}x_{n+1}}} & =\frac{y_{n+1}-x_{n+1}^{\T}\hat{\beta}}{\sqrt{\sigma^{2}+\sigma^{2}x_{n+1}^{\T}(X^{\T}X)^{-1}x_{n+1}}}\Big/\sqrt{\frac{\hat{\sigma}^{2}}{\sigma^{2}}}\\
 & \sim\frac{\N(0,1)}{\sqrt{\chi_{n-p}^{2}/(n-p)}} , 
\end{align*}
where $\N(0,1)$ and $\chi_{n-p}^{2}$ denote independent standard Normal and $\chi_{n-p}^{2}$ random variables, respectively. Therefore,  
$$
\frac{y_{n+1}-x_{n+1}^{\T}\hat{\beta}}{\sqrt{\hat{\sigma}^{2}+\hat{\sigma}^{2}x_{n+1}^{\T}(X^{\T}X)^{-1}x_{n+1}}}
  \sim t_{n-p}
$$
is a pivotal quantity. Define  the squared prediction error  as
\begin{align*}
\hat{\text{pe}}_{x_{n+1}}^{2} & =\hat{\sigma}^{2}+\hat{\sigma}^{2}x_{n+1}^{\T}(X^{\T}X)^{-1}x_{n+1}\\
 & =\hat{\sigma}^{2}\left\{ 1+n^{-1}x_{n+1}^{\T}\left(n^{-1}\sumn x_{i}x_{i}^{\T}\right)^{-1}x_{n+1}\right\}  ,
\end{align*}
which has two components. 
The first one has magnitude close to $\sigma^{2}$, which is of constant order. The second one has a magnitude that is decreasing in $n$, if $n^{-1}\sumn x_{i}x_{i}^{\T}$ converges to a finite limit
with large $n$. Therefore, the first component dominates the second
one with large $n$, which results in the main difference between
predicting the mean of $y_{n+1}$ and predicting $y_{n+1}$ itself. Using
the notation $\hat{\text{pe}}_{x_{n+1}}$, we can construct the following
$1-\alpha$ level prediction interval: 
\[
 x_{n+1}^{\T}\hat{\beta} \pm  t_{1-\alpha/2,  n-p} \hat{\text{pe}}_{x_{n+1}} .
\]

\section{Examples}
\label{sec::normallinearmodel-r}

Below I illustrate the theory in this chapter with two classic datasets.

\subsection{Univariate regression}

Revisiting Galton's data, we have the following result: 

\begin{lstlisting}
> GaltonFamilies = read.table("GaltonFamilies.txt", header = TRUE)
> 
> ## OLS fit by the "lm" function
> galton_fit = lm(childHeight ~ midparentHeight,
+                 data = GaltonFamilies)
> 
> ## OLS coefficients and inference
> summary(galton_fit)$coef
                  Estimate Std. Error   t value     Pr(>|t|)
(Intercept)     22.6362405  4.2651074  5.307308 1.390930e-07
midparentHeight  0.6373609  0.0616076 10.345491 8.053865e-24
\end{lstlisting}

With the fitted line, we can predict \ri{childHeight} at different values of \ri{midparentHeight}. In the \ri{predict} function, if we specify \ri{interval = "confidence"}, it gives the {\it confidence} intervals for the means of the new outcomes; if we specify \ri{interval = "prediction"}, it gives the {\it prediction} intervals for the new outcomes themselves. 

\begin{lstlisting}
> ## predictions: confidence and prediction intervals
> new_mph  = seq(60, 80, by = 0.5)
> new_data = data.frame(midparentHeight = new_mph)
> new_ci   = predict(galton_fit, new_data, 
+                         interval = "confidence")
> new_pi   = predict(galton_fit, new_data, 
+                         interval = "prediction")
> head(round(new_ci, 2))
    fit   lwr   upr
1 60.88 59.74 62.01
2 61.20 60.12 62.27
3 61.52 60.50 62.53
4 61.83 60.88 62.79
5 62.15 61.25 63.05
6 62.47 61.63 63.31
> head(round(new_pi, 2))
    fit   lwr   upr
1 60.88 54.13 67.63
2 61.20 54.45 67.94
3 61.52 54.78 68.25
4 61.83 55.11 68.56
5 62.15 55.44 68.87
6 62.47 55.76 69.18
\end{lstlisting}

\begin{figure}[ht]
\centering
\includegraphics[width = \textwidth]{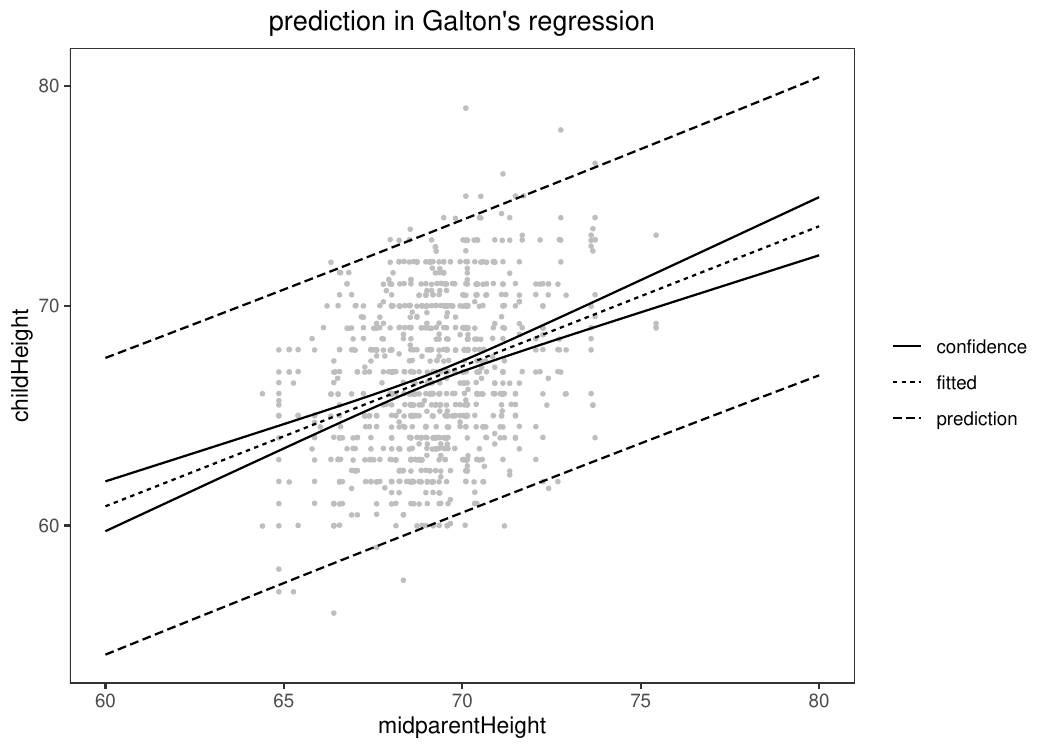}
\caption{Prediction in Galton's regression}\label{fig::prediction-galton}
\end{figure}

Figure \ref{fig::prediction-galton} plots the fitted line as well as the confidence intervals and prediction intervals at level $95\%$.  

\subsection{Anscombe's Quartet: the importance of graphical diagnostics}
\label{sec::anscombe-quartet}

\citet{anscombe1973graphs} used four simple datasets to illustrate the importance of graphical diagnostics in linear regression. 
His datasets are in \texttt{anscombe} in the \texttt{R} package \texttt{datasets}: \texttt{x1} and \texttt{y1} constitute the first dataset, and so on.

\begin{lstlisting}
> library(datasets)
> ## Anscombe's Quartet
> anscombe
   x1 x2 x3 x4    y1   y2    y3    y4
1  10 10 10  8  8.04 9.14  7.46  6.58
2   8  8  8  8  6.95 8.14  6.77  5.76
3  13 13 13  8  7.58 8.74 12.74  7.71
4   9  9  9  8  8.81 8.77  7.11  8.84
5  11 11 11  8  8.33 9.26  7.81  8.47
6  14 14 14  8  9.96 8.10  8.84  7.04
7   6  6  6  8  7.24 6.13  6.08  5.25
8   4  4  4 19  4.26 3.10  5.39 12.50
9  12 12 12  8 10.84 9.13  8.15  5.56
10  7  7  7  8  4.82 7.26  6.42  7.91
11  5  5  5  8  5.68 4.74  5.73  6.89
\end{lstlisting}

The four datasets have similar sample moments. 

\begin{lstlisting}
> ## mean of x
> c(mean(anscombe$x1),
+   mean(anscombe$x2),
+   mean(anscombe$x3),
+   mean(anscombe$x4))
[1] 9 9 9 9
> ## variance of x
> c(var(anscombe$x1),
+   var(anscombe$x2),
+   var(anscombe$x3),
+   var(anscombe$x4))
[1] 11 11 11 11
> ## mean of y
> c(mean(anscombe$y1),
+   mean(anscombe$y2),
+   mean(anscombe$y3),
+   mean(anscombe$y4))
[1] 7.500909 7.500909 7.500000 7.500909
> ## variance of y
> c(var(anscombe$y1),
+   var(anscombe$y2),
+   var(anscombe$y3),
+   var(anscombe$y4))
[1] 4.127269 4.127629 4.122620 4.123249
\end{lstlisting}

The results based on linear regression are almost identical. 

\begin{lstlisting}
> ols1 = lm(y1 ~ x1, data = anscombe)
> summary(ols1)

Call:
lm(formula = y1 ~ x1, data = anscombe)

Residuals:
     Min       1Q   Median       3Q      Max 
-1.92127 -0.45577 -0.04136  0.70941  1.83882 

Coefficients:
            Estimate Std. Error t value Pr(>|t|)   
(Intercept)   3.0001     1.1247   2.667  0.02573 * 
x1            0.5001     0.1179   4.241  0.00217 **

Residual standard error: 1.237 on 9 degrees of freedom
Multiple R-squared:  0.6665,	Adjusted R-squared:  0.6295 
F-statistic: 17.99 on 1 and 9 DF,  p-value: 0.00217

> ols2 = lm(y2 ~ x2, data = anscombe)
> summary(ols2)

Call:
lm(formula = y2 ~ x2, data = anscombe)

Residuals:
    Min      1Q  Median      3Q     Max 
-1.9009 -0.7609  0.1291  0.9491  1.2691 

Coefficients:
            Estimate Std. Error t value Pr(>|t|)   
(Intercept)    3.001      1.125   2.667  0.02576 * 
x2             0.500      0.118   4.239  0.00218 **

Residual standard error: 1.237 on 9 degrees of freedom
Multiple R-squared:  0.6662,	Adjusted R-squared:  0.6292 
F-statistic: 17.97 on 1 and 9 DF,  p-value: 0.002179

> ols3 = lm(y3 ~ x3, data = anscombe)
> summary(ols3)

Call:
lm(formula = y3 ~ x3, data = anscombe)

Residuals:
    Min      1Q  Median      3Q     Max 
-1.1586 -0.6146 -0.2303  0.1540  3.2411 

Coefficients:
            Estimate Std. Error t value Pr(>|t|)   
(Intercept)   3.0025     1.1245   2.670  0.02562 * 
x3            0.4997     0.1179   4.239  0.00218 **

Residual standard error: 1.236 on 9 degrees of freedom
Multiple R-squared:  0.6663,	Adjusted R-squared:  0.6292 
F-statistic: 17.97 on 1 and 9 DF,  p-value: 0.002176

> ols4 = lm(y4 ~ x4, data = anscombe)
> summary(ols4)

Call:
lm(formula = y4 ~ x4, data = anscombe)

Residuals:
   Min     1Q Median     3Q    Max 
-1.751 -0.831  0.000  0.809  1.839 

Coefficients:
            Estimate Std. Error t value Pr(>|t|)   
(Intercept)   3.0017     1.1239   2.671  0.02559 * 
x4            0.4999     0.1178   4.243  0.00216 **

Residual standard error: 1.236 on 9 degrees of freedom
Multiple R-squared:  0.6667,	Adjusted R-squared:  0.6297 
F-statistic:    18 on 1 and 9 DF,  p-value: 0.002165
\end{lstlisting}

\begin{figure}[th]
\centering
\includegraphics[width = \textwidth]{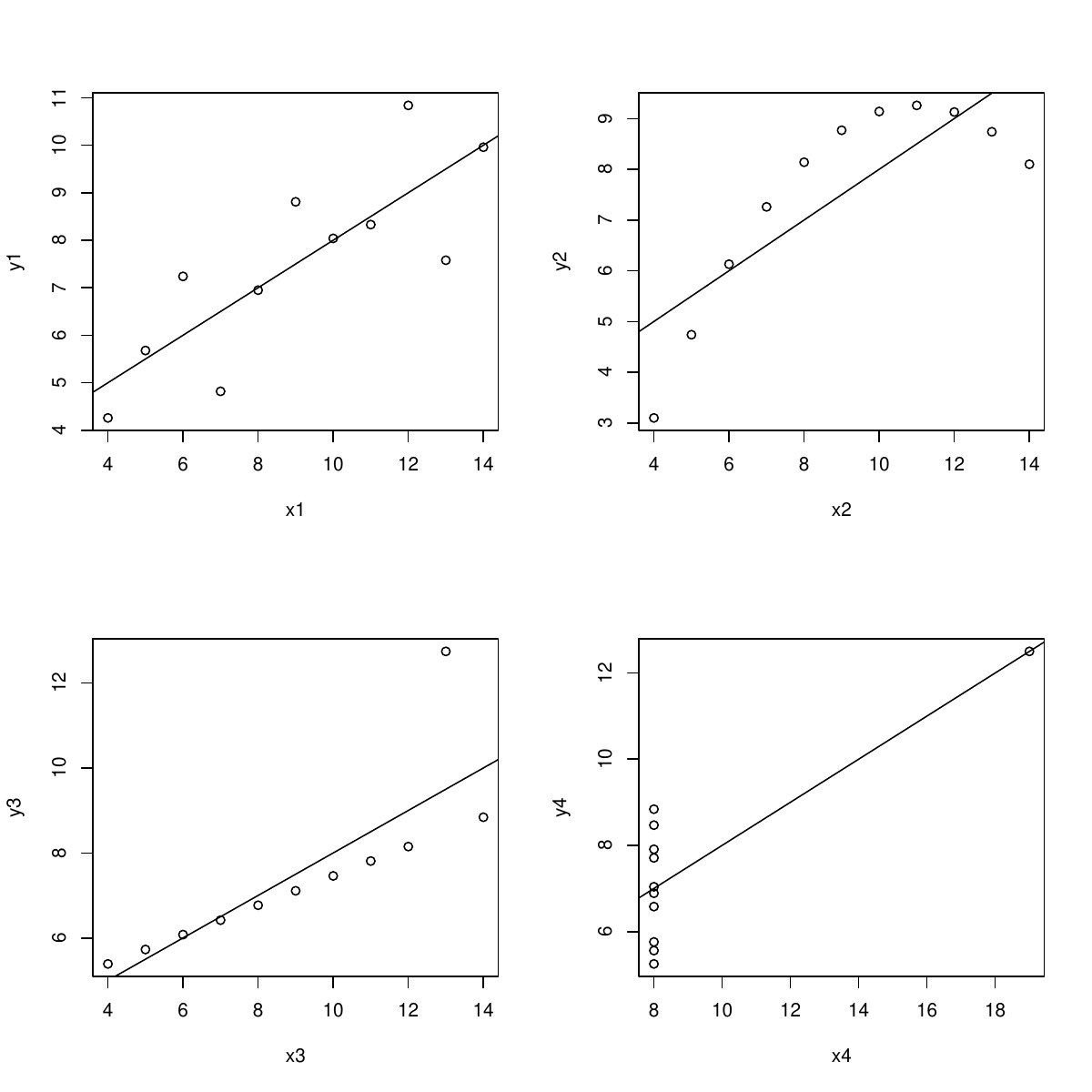}
\caption{Anscombe's Quartet: scatter plots}\label{fig::AnscombeQuartetplots}
\end{figure}

However, the scatter plots of the datasets in Figure \ref{fig::AnscombeQuartetplots} reveal fundamental differences between the datasets. The first dataset seems ideal for linear regression. The second dataset shows a quadratic form of $y$ versus $x$, and therefore, the linear model is misspecified. The third dataset shows a linear trend of $y$ versus $x$, but an outlier has severely distorted the slope of the linear line. The fourth dataset is supported on only two values of $x$ and thus may suffer from severe extrapolation.

\subsection{Multivariate regression}\label{section::normal-lm-lalonde}

The \ri{R} package \ri{Matching} contains an experimental dataset \ri{lalonde} from \citet{lalonde1986evaluating}. Units were randomly assigned to a job training program, with \ri{treat} being the treatment indicator. The outcome \ri{re78} is the real earnings in the year 1978, and other variables are pretreatment covariates. From the simple OLS, the treatment has a significant positive effect, whereas none of the covariates are predictive of the outcome.

\begin{lstlisting}
> library("Matching")
> data(lalonde)
> lalonde_fit = lm(re78 ~ ., data = lalonde)
> summary(lalonde_fit)

Call:
lm(formula = re78 ~ ., data = lalonde)

Residuals:
   Min     1Q Median     3Q    Max 
 -9612  -4355  -1572   3054  53119 

Coefficients:
              Estimate Std. Error t value Pr(>|t|)   
(Intercept)  2.567e+02  3.522e+03   0.073  0.94193   
age          5.357e+01  4.581e+01   1.170  0.24284   
educ         4.008e+02  2.288e+02   1.751  0.08058 . 
black       -2.037e+03  1.174e+03  -1.736  0.08331 . 
hisp         4.258e+02  1.565e+03   0.272  0.78562   
married     -1.463e+02  8.823e+02  -0.166  0.86835   
nodegr      -1.518e+01  1.006e+03  -0.015  0.98797   
re74         1.234e-01  8.784e-02   1.405  0.16079   
re75         1.974e-02  1.503e-01   0.131  0.89554   
u74          1.380e+03  1.188e+03   1.162  0.24590   
u75         -1.071e+03  1.025e+03  -1.045  0.29651   
treat        1.671e+03  6.411e+02   2.606  0.00948 **

Residual standard error: 6517 on 433 degrees of freedom
Multiple R-squared:  0.05822,	Adjusted R-squared:  0.0343 
F-statistic: 2.433 on 11 and 433 DF,  p-value: 0.005974
\end{lstlisting}

The above result shows that none of the pretreatment covariates is significant {\it  marginally}. It is also of interest to test whether they are {\it jointly} significant. The result below shows that they are only weakly significant at the level $0.05$ based on a joint test (the $p$-value is almost $0.05$!). 

\begin{lstlisting}
> library("car")
> linearHypothesis(lalonde_fit,
+                  c("age=0", "educ=0", "black=0",
+                    "hisp=0", "married=0", "nodegr=0",
+                    "re74=0", "re75=0", "u74=0",
+                    "u75=0"))
Linear hypothesis test

Hypothesis:
age = 0
educ = 0
black = 0
hisp = 0
married = 0
nodegr = 0
re74 = 0
re75 = 0
u74 = 0
u75 = 0

Model 1: restricted model
Model 2: re78 ~ age + educ + black + hisp + married + nodegr + re74 + 
    re75 + u74 + u75 + treat

  Res.Df        RSS Df Sum of Sq      F  Pr(>F)  
1    443 1.9178e+10                              
2    433 1.8389e+10 10 788799023 1.8574 0.04929 *
\end{lstlisting}

Below I create two pseudo datasets: one with all units assigned to the treatment, and the other with all units assigned to the control, fixing all the pretreatment covariates. The predicted outcomes are the {\it counterfactual outcomes} under the treatment and control. I further calculate their means and verify that their difference equals the OLS coefficient of \ri{treat}.

\begin{lstlisting}
> new_treat          = lalonde
> new_treat$treat    = 1
> predict_lalonde1   = predict(lalonde_fit, new_treat, 
+                            interval = "none")
> new_control        = lalonde
> new_control$treat  = 0
> predict_lalonde0   = predict(lalonde_fit, new_control, 
+                              interval = "none")
> mean(predict_lalonde1)
[1] 6276.91
> mean(predict_lalonde0)
[1] 4606.201
> 
> mean(predict_lalonde1) - mean(predict_lalonde0)
[1] 1670.709
\end{lstlisting}

\section{Homework problems}

\paragraph{Maximum likelihood estimator and OLS}
\label{hw5:normal-mle}

Under the Normal linear model, prove that the maximum likelihood
estimator for $\beta$ is the OLS estimator, but the maximum likelihood
estimator for
$\sigma^{2}$ is $\tilde{\sigma}^{2}=\textsc{rss}/n$. Compare the
mean squared errors of $\hat{\sigma}^{2}$ and $\tilde{\sigma}^{2}$ for estimating $\sigma^2$.

Remark: The definitions of the mean squared errors are $E\{ (\hat{\sigma}^{2} - \sigma^2)^2 \} $ and $E\{ (\tilde{\sigma}^{2} - \sigma^2)^2 \} $.

\paragraph{Optimal estimation of the variance}
\label{hw5:optimal-A-variance-normal}

This problem extends Problem \ref{hw04::variance-estimation-A}.

Under the Normal linear model, calculate $\var(\hat{\sigma}_A^2)$ and prove that $A = I_n$ minimizes $\var(\hat{\sigma}_A^2)$.

Remark: The minimizer of $\var(\hat{\sigma}_A^2)$ is not unique. For instance, $A = (I_n - H)^{+}$ also minimizes $\var(\hat{\sigma}_A^2)$, where $+$ denotes the pseudoinverse. There are other minimizers.

\paragraph{Maximum likelihood estimator with Laplace errors}\label{hw5:laplace-mle}

Assume that $y_{i}=x_{i}^{\T}\beta+\sigma\varepsilon_{i}$, where the
$\varepsilon_{i}$'s are IID Laplace distribution with density
$f(\varepsilon)=2^{-1}e^{-|\varepsilon|}\ (i=1,\ldots,n)$. Find the
Maximum likelihood estimators of $(\beta,\sigma^{2}).$

Remark: We will revisit this problem in Chapter \ref{chapter::quantile-regression}.

\paragraph{Joint prediction}\label{hw5:joint-preduction}

With multiple future data points $(X_{n+1},Y_{n+1})$ where $X_{n+1}\in\mathbb{R}^{l\times p}$
and $Y_{n+1}\in\mathbb{R}^{l},$ construct the joint predictors and
prediction region for $Y_{n+1}$ based on $(X,Y)$ and $X_{n+1}$.

As a starting point, you can assume that $l\leq p$ and the rows of $X_{n+1}$ are linearly independent. You can then consider the case in which the rows of $X_{n+1}$ are not linearly independent.

Remark: Assume the Normal linear model for all observations and apply Theorem \ref{thm::normal-chisq} in Appendix \ref{chapter:appendix-rvs}.

\paragraph{Two-sample problem}\label{hw5::two-sample}

\begin{enumerate}
\item Assume that $z_{1},\ldots,z_{m}\iidsim\N(\mu_{1},\sigma^{2})$ and
$w_{1},\ldots,w_{n}\iidsim\N(\mu_{2},\sigma^{2})$, and test
$H_0: \mu_{1}=\mu_{2}$. Under $H_0$, the $t$ statistic with pooled variance
estimator equals 
\[
t_{\text{equal}}=\frac{\bar{z}-\bar{w}}{  \sqrt{ \hat{\sigma}^2  (m^{-1} + n^{-1})  } } 
\] 
where 
$$
\hat{\sigma}^2 = \left\{ (m-1)S_{z}^{2}+(n-1)S_{w}^{2}\right\} /(m+n-2) 
$$
with the sample means  
\[
\bar{z}=m^{-1}\sum_{i=1}^{m}z_{i},\qquad\bar{w}=n^{-1}\sumn w_{i},
\]
and the sample variances
\[
S_{z}^{2}=(m-1)^{-1}\sum_{i=1}^{m}(z_{i}-\bar{z})^{2},\qquad S_{w}^{2}=(n-1)^{-1}\sumn(w_{i}-\bar{w})^{2}.
\]

Prove that under $H_0$, the $t$ statistic has the following distribution:
\[
t_{\text{equal}} \sim t_{m+n-2},
\] 

Remark: 
The name ``\ri{equal}'' is motivated by the ``\ri{var.equal}''
parameter of the \ri{R} function \ri{t.test}.

\item We can write the above problem as testing hypothesis $H_{0}:\beta_{1}=0$ in the linear regression $Y=X\beta+\varepsilon$
with 
\[
Y=\left(\begin{array}{c}
z_{1}\\
\vdots\\
z_{m}\\
w_{1}\\
\vdots\\
w_{n}
\end{array}\right),\quad X=\left(\begin{array}{cc}
1 & 1\\
\vdots & \vdots\\
1 & 1\\
1 & 0\\
\vdots & \vdots\\
1 & 0
\end{array}\right),\quad\beta=\left(\begin{array}{c}
\beta_{0}\\
\beta_{1}
\end{array}\right),\quad\varepsilon=\left(\begin{array}{c}
\varepsilon_{1}\\
\vdots\\
\varepsilon_{m}\\
\varepsilon_{m+1}\\
\vdots\\
\varepsilon_{m+n}
\end{array}\right).
\]
Based on the Normal linear model, we can compute the $t$ statistic for the coefficient of $\beta_1$. 

Prove that the $t$ statistic from OLS is numerically identical to $t_{\text{equal}}$. 
\end{enumerate}

\paragraph{Analysis of Variance (ANOVA) with a multi-level treatment}\label{hw5:anova-f}

Let $x_{i}$ be the indicator vector for $J$ treatment levels in
a completely randomized experiment, for example, $x_{i} = e_j=(0,\ldots,1,\ldots,0)^{\T} $
with the $j$th element being one if unit $i$ receives treatment
level $j$ $(j=1,\ldots,J)$. Let $y_{i}$ be the outcome of unit
$i$ $(i=1,\ldots,n)$. Let $\mathcal T_{j}$ be the indices of units receiving
treatment $j$, and let $n_{j}=|\mathcal T_{j}|$ be the sample size and $\bar{y}_{j} = n_j^{-1} \sum_{i\in \mathcal T_j} y_i$ be the sample
mean of the outcomes under treatment $j$. Define $\bar{y} = n^{-1} \sumn y_i$ as the grand mean. We can test whether the treatment
has any effect on the outcome by testing the null hypothesis 
\[
H_{0}:\beta_{1}=\cdots=\beta_{J}
\] 
in the Normal linear model $Y=X\beta+\varepsilon$.
This is a special case of testing $C\beta=0$. 

Find $C$ and prove that the corresponding $F$ statistic is identical to 
\[
F=\frac{\sum_{j=1}^{J}n_{j}(\bar{y}_{j}-\bar{y})^{2}/(J-1)}{\sum_{j=1}^{J}\sum_{i\in \mathcal T_{j}}(y_{i}-\bar{y}_{j})^{2}/(n-J)}\sim F_{J-1,n-J}.
\]

Remarks: (1)
This is Fisher's $F$ statistic. (2) In this linear model formulation, $X$ does not contain a column of 1's. (3) The choice of $C$ is not
unique, but the final formula for $F$ is. (4) You may use the Sherman--Morrison formula in Problem \ref{hwmath1::inverse-block-matrix} in Appendix \ref{chapter::linear-algebra}.

\paragraph{Confidence interval for $\sigma^2$}\label{hw5::confidence-interval-sigma2}

Based on Theorem \ref{thm:normalexactdistribution}, construct a $1-\alpha$ level confidence interval for $\sigma^2$.

\paragraph{Relationship between $t$ and $F$}\label{hw5::T-F-1dim}

Prove that when $C$ containing only one row $c^{\T}$, then $T_c^2 = F_C$, where $T_c$ is defined in Theorem \ref{thm:1dimpivotal} and $F_C$ is defined in Theorem \ref{thm:ldimpivotal}.

\paragraph{\textsc{rss} and $t$-statistic in univariate OLS}\label{hw5::t-ratio-univariateOLS}

Focus on univariate OLS discussed in Chapter \ref{chapter::ols-1d}: $y_i = \hat{\alpha}  +  \hat{\beta} x_i + \hat{\varepsilon}_i$ $(i=1,\ldots, n)$. 

Prove that \textsc{rss}  equals
$$
\sumn \hat{\varepsilon}_i^2 = \sumn (y_i -  \bar{y}  )^2 (1- \hat{\rho}_{xy} ^2)
$$ 
and under the homoskedasticity assumption, the $t$-statistic associated with $ \hat{\beta}$ equals
$$
\frac{  \hat{\rho}_{xy}   }{  \sqrt{    (1-\hat{\rho}_{xy} ^2) /(n-2)  }   }.
$$

\paragraph{Equivalence of the $t$-statistics}\label{hw5::t-stat-equivalent}

Consider data $(x_i, y_i)_{i=1}^n$ with scalar $x_i$ and $y_i$. Run OLS fit of $y_i$ on $(1,x_i)$ to obtain $t_{y\mid x}$, the $t$-statistic of the coefficient of $x_i$, under the homoskedasticity assumption. Run OLS fit of $x_i$ on $(1,y_i)$ to obtain $t_{x\mid y}$, the $t$-statistic of the coefficient of $y_i$, under the homoskedasticity assumption. 

Prove $t_{y\mid x} = t_{x\mid y}$.

Remark: This is a numerical result that holds without any stochastic assumptions. I give an example below.

\begin{lstlisting}
> library(MASS)
> #simulate bivariate normal distribution
> xy = mvrnorm(n=100, mu=c(0, 0), 
+              Sigma=matrix(c(1, 0.5, 0.5, 1), ncol=2))
> xy = as.data.frame(xy)
> colnames(xy) = c("x", "y")
> ## OLS
> reg.y.x = lm(y ~ x, data = xy)
> reg.x.y = lm(x ~ y, data = xy)
> ## compare t statistics based on homoskedastic errors
> summary(reg.y.x)$coef[2, 3]
[1] 4.470331
> summary(reg.x.y)$coef[2, 3]
[1] 4.470331
\end{lstlisting}
The equivalence of the $t$-statistics from the OLS fit of $y$ on x and that of $x$ on $y$ demonstrates that based on OLS, the data do not contain any information about the direction of the relationship between $x$ and $y$.

\paragraph{An application}

The \ri{R} package \ri{sampleSelection} \citep{toomet2008sample} describes the dataset \ri{RandHIE} as follows: ``The RAND Health Insurance Experiment was a comprehensive study of health care cost, utilization and outcome in the United States. It is the only randomized study of health insurance, and the only study which can give definitive evidence as to the causal effects of different health insurance plans.'' You can find more detailed information about other variables in this package. The main outcome of interest \ri{lnmeddol} means the log of medical expenses. 

Use OLS to investigate the relationship between the outcome and various important covariates. 

Remark: The solution to this problem is not unique, but you need to justify your choice of covariates and model, and need to interpret the results.

\chapter{Asymptotic Inference in OLS: Eicker--Huber--White (EHW) robust standard error}\label{chapter::EHW}
  \chaptermark{EHW standard error in OLS}

The results in Chapter \ref{chapter::normal-linear-model} rely on the assumption of the Normal linear model, which imposes strong distributional assumptions. If we think the Normal linear model is unlikely to hold, how much should we trust the results in Chapter \ref{chapter::normal-linear-model}?

This chapter will show that the Normality assumption is not that crucial but the homoskedasticity assumption of the errors is. The theory will show that if the homoskedasticity assumption fails, we must modify the covariance estimator of the OLS to be
\begin{equation}\label{eq::ehw-first-appear}
\hat{V}_{\textsc{ehw}} =(X^{\T}X)^{-1}(X^{\T}\hat{\Omega}X)(X^{\T}X)^{-1}
\end{equation}
with $\hat{\Omega}=\text{diag}\left\{ \hat{\varepsilon}_{1}^{2},\ldots,\hat{\varepsilon}_{n}^{2}\right\} $, where the $\hat{\varepsilon}_i$'s are the residuals.  The matrix $\hat{V}_{\textsc{ehw}}$ is called the Eicker--Huber--White (EHW) robust covariance matrix estimator.

Before diving into the theory, I will first present some numerical examples.

\section{Motivation}

\subsection{Numerical examples}

The  first one is the ideal Normal linear model:

\begin{lstlisting}
> library(car)
> n     = 200
> x     = runif(n, -2, 2)
> beta  = 1
> xbeta = x*beta 
> Simu1  = replicate(5000,
+                   {y = xbeta + rnorm(n)
+                   ols.fit = lm(y ~ x)
+                   c(summary(ols.fit)$coef[2, 1:2],
+                     sqrt(hccm(ols.fit)[2, 2]))
+                   })
\end{lstlisting}

\begin{figure} 
\centering
\includegraphics[width=\textwidth]{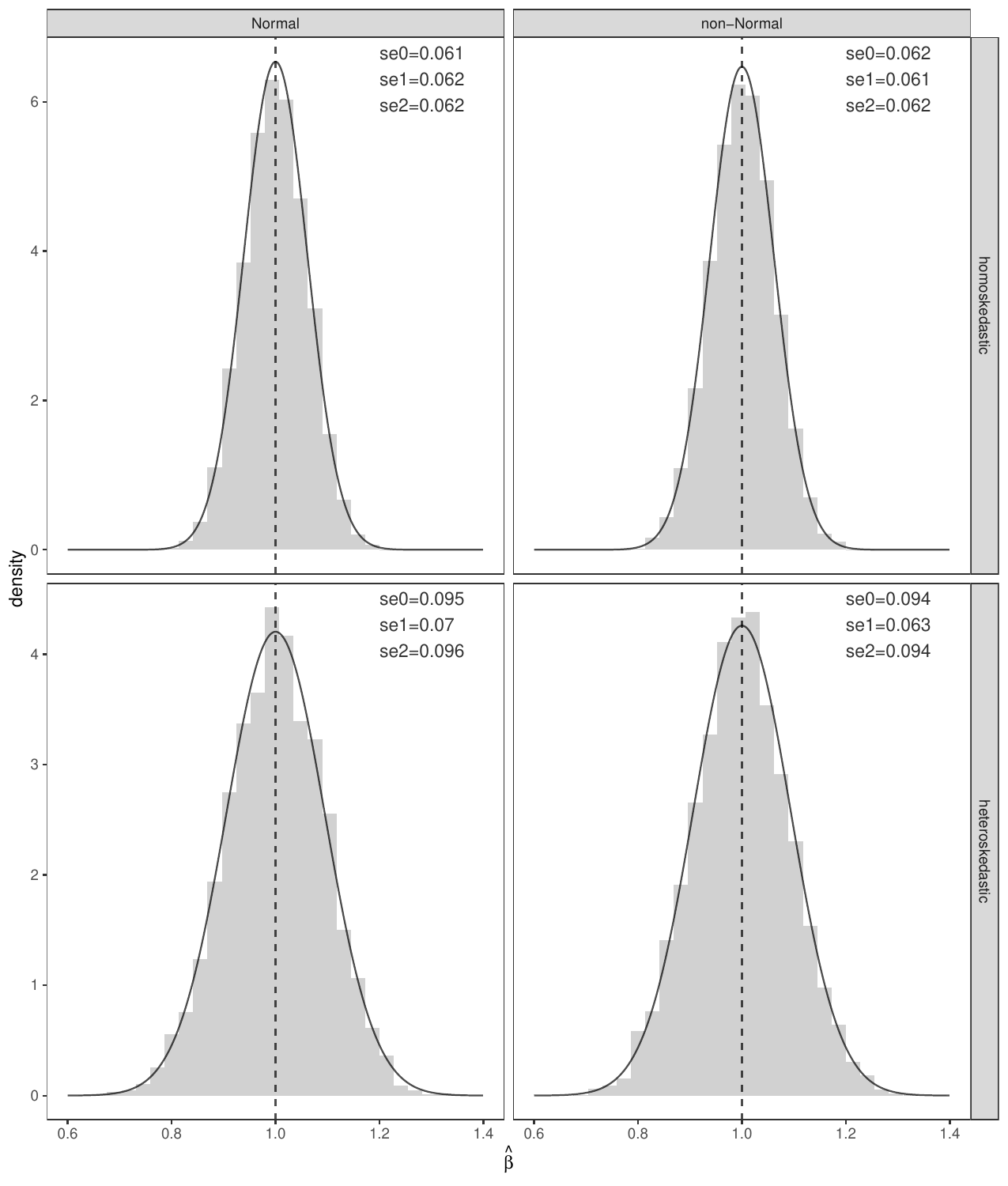}
\caption{Simulation with 5000 replications: ``se0'' denotes the true standard error of $\hat{\beta}$, ``se1'' denotes the estimated standard error based on the homoskedasticity assumption, and ``se2'' denotes the Eicker--Huber--White standard error allowing for heteroskedasticity. The density curves are Normal with mean 1 and standard deviation se0.}
\label{fig::simulation-nonnormal-heteroskedasticity}
\end{figure}

In the above, I generate outcomes from a simple linear model $y_i = x_i + \varepsilon_i$ with $\varepsilon_i\iidsim \N(0,\sigma^2=1)$. Over 5000 replications of the data, we computed the OLS coefficient $\hat{\beta}$ of $x_i$ and reported two standard errors. One is the standard error discussed in Chapter \ref{chapter::normal-linear-model} under the Normal linear model, which is also the default choice of the \ri{lm} function of \ri{R}. The other one is the square root of $(2,2)$th element of $\hat{V}_{\textsc{ehw}}$ in \eqref{eq::ehw-first-appear}, which can be computed by the \ri{hccm} function in the \ri{R} package \ri{car} The $(1,1)$ the panel of
Figure \ref{fig::simulation-nonnormal-heteroskedasticity} shows the histogram of the estimator and reports the standard error (se0), as well as two estimated standard errors (se1 and se2). The distribution of $\hat{\beta}$ is symmetric and bell-shaped around the true parameter 1, and both of the two estimated standard errors are close to the true one. 

To investigate the impact of Normality, I change the error terms to be IID exponential with mean 1 and variance 1. 

\begin{lstlisting}
> Simu2  = replicate(5000,
+                   {y = xbeta + rexp(n)
+                   ols.fit = lm(y ~ x)
+                   c(summary(ols.fit)$coef[2, 1:2],
+                     sqrt(hccm(ols.fit)[2, 2]))
+                   })
\end{lstlisting}

The $(1,2)$ panel of Figure \ref{fig::simulation-nonnormal-heteroskedasticity} corresponds to this setting. With non-Normal errors, $\hat{\beta}$ is still symmetric and bell-shaped around the true parameter 1, and the estimated standard errors are close to the true one. So the Normality of the error terms does not seem to be a crucial assumption for the validity of the inference procedure under the Normal linear model.

I then generate errors from Normal with variance depending on $x$:
\begin{lstlisting}
> Simu3  = replicate(5000,
+                   {y = xbeta + rnorm(n, 0, abs(x))
+                   ols.fit = lm(y ~ x)
+                   c(summary(ols.fit)$coef[2, 1:2],
+                     sqrt(hccm(ols.fit)[2, 2]))
+                   })
\end{lstlisting}

The $(2,1)$ panel of Figure \ref{fig::simulation-nonnormal-heteroskedasticity} corresponds to this setting. With heteroskedastic Normal errors, $\hat{\beta}$ is symmetric and bell-shaped around the true parameter 1, se2 is close to se0, but se1 underestimates se0. So the heteroskedasticity of the error terms does not change the Normality of the OLS estimator dramatically, although the statistical inference discussed in Chapter \ref{chapter::normal-linear-model} is invalid.

Finally, I generate heteroskedastic non-Normal errors:
\begin{lstlisting}
> Simu4  = replicate(5000,
+                   {y = xbeta + runif(n, -x^2, x^2)
+                   ols.fit = lm(y ~ x)
+                   c(summary(ols.fit)$coef[2, 1:2],
+                     sqrt(hccm(ols.fit)[2, 2]))
+                   })
\end{lstlisting} 
The  $(2,2)$ panel of Figure \ref{fig::simulation-nonnormal-heteroskedasticity} corresponds to this setting, which has a similar pattern as the $(2,1)$ panel. So the Normality of the error terms is not crucial, but the homoskedasticity is.

\subsection{Goal of this chapter}

This chapter will still impose the linearity assumption, but
relax the distributional assumption on the error terms. Assume the following heteroskedastic linear model.

\begin{assumption}
[Heteroskedastic linear model]
\label{assume::heteroskedasticity-lm}
We have
\[
y_{i}=x_{i}^{\T}\beta+\varepsilon_{i},
\]
where the $\varepsilon_{i}$'s are independent
with mean zero and variance $\sigma_{i}^{2}$ $(i=1,\ldots,n)$.
The design matrix $X = (x_1^{\T}, \ldots, x_n^{\T})^{\T}$ is fixed with linearly independent column vectors, and $(\beta,\sigma^{2}_1, \ldots, \sigma^2_n)$
are unknown parameters. 
\end{assumption}

 Because
the error terms can have different variances, they are not IID
in general under the heteroskedastic linear model. Their variances can be functions of the $x_{i}$'s, and
the variances $\sigma_{i}^{2}$ are $n$ free unknown numbers. Again, treating the
$x_{i}$'s as fixed is not essential, because we can condition on
them if they are random.

Without imposing Normality on the error terms,
we cannot determine the finite sample exact distribution of the OLS
estimator. This chapter will use the asymptotic analysis,
assuming that the sample size $n$ is large so that certain laws of large numbers and central limit theorems (CLTs) hold. 

The asymptotic analysis later will show that if the error terms are
homoskedastic, i.e., $\sigma_{i}^{2}=\sigma^{2}$ for all $i=1,\ldots,n$,
we can still trust the statistical inference discussed in Chapter \ref{chapter::normal-linear-model} based on the Normal
linear model as long as the CLT for the OLS estimator
holds as $n\rightarrow\infty$. If the error terms are heteroskedastic,
i.e., their variances are different, we must modify the standard
error as the so-called Eicker--Huber--White (EHW) heteroskedasticity robust standard error introduced in \eqref{eq::ehw-first-appear}. 
I will give the technical details below. 
If you are unfamiliar with the asymptotic analysis, please first review the basics in Appendix \ref{chapter::limiting-theorems}.

\section{Consistency of OLS}

Under the heteroskedastic linear model,  the OLS estimator $\hat{\beta}$ is still unbiased for $\beta$ because the error terms have mean zero. Moreover, we can show that it is consistent
for $\beta$ with large $n$ and some regularity conditions. We start
with a lemma.
\begin{lemma}
\label{lem:The-OLS-estimator}
Under Assumption \ref{assume::heteroskedasticity-lm}, the OLS estimator has the
representation
$
\hat{\beta}-\beta=\ B_{n}^{-1} \xi_n,
$
where
\begin{eqnarray*}
B_{n} &=& n^{-1}\sumn x_{i}x_{i}^{\T},\\ 
\xi_n &=& n^{-1}\sumn x_{i}\varepsilon_{i} . 
\end{eqnarray*}
\end{lemma}

\begin{myproof}{Lemma}{\ref{lem:The-OLS-estimator}}
Since $y_{i}=x_{i}^{\T}\beta+\varepsilon_{i}$, we have
\begin{align*}
\hat{\beta} & =B_{n}^{-1}n^{-1}\sumn x_{i}y_{i}\\
 & =B_{n}^{-1}n^{-1}\sumn x_{i}(x_{i}^{\T}\beta+\varepsilon_{i})\\
 & =B_{n}^{-1}B_{n}\beta+B_{n}^{-1}n^{-1}\sumn x_{i}\varepsilon_{i}\\
 & =\beta+ B_{n}^{-1} \xi_n. 
\end{align*}
Therefore, $\hat{\beta}-\beta=\ B_{n}^{-1} \xi_n$. 
\end{myproof}

In the representation of Lemma \ref{lem:The-OLS-estimator}, $B_n$ is fixed and $\xi_n$ is random. Since $E(\xi_n) = 0$, we know that $E(\hat{\beta})=\beta$, so the OLS estimator is unbiased. Moreover, 
\begin{eqnarray*}
\cov(\xi_n)  
&=&  \cov\left( n^{-1}\sumn x_{i}\varepsilon_{i} \right)   \\
&=& n^{-2} \sumn \sigma_{i}^{2}x_{i}x_{i}^{\T} \\
&=&  M_n/n,
\end{eqnarray*}
where
$$
M_{n}=n^{-1}\sumn\sigma_{i}^{2}x_{i}x_{i}^{\T}.
$$
So the covariance of the OLS estimator is 
$$
\cov(\hat{\beta})=n^{-1}B_{n}^{-1}M_{n}B_{n}^{-1}.
$$
It has a sandwich form, justifying
the choice of notation $B_n$ for the ``bread matrix'' and $M_n$ for the ``meat matrix.'' Vegetarians can read $M_n$ as the ``middle matrix.''

Intuitively, if $B_{n}$ and $M_{n}$ have finite limits, then the
covariance of $\hat{\beta}$ shrinks to zero with large $n$, implying
that $\hat{\beta}$ will concentrate near its mean $\beta$. This
is the idea of consistency, formally stated below. 
\begin{assumption}
\label{assume::ols-basic-asymptotic-conditions}
$B_{n}\rightarrow B$ and $M_{n}\rightarrow M$
where $B$ and $M$ are finite with $B$ invertible.
\end{assumption}

\begin{theorem}
\label{thm:consistencyofOLS} Under Assumptions \ref{assume::heteroskedasticity-lm} and \ref{assume::ols-basic-asymptotic-conditions}, we have $\hat{\beta}\rightarrow\beta$
in probability.
\end{theorem}

\begin{myproof}{Theorem}{\ref{thm:consistencyofOLS}}
We only need to show that $ \xi_n \rightarrow0$
in probability. It has mean zero and covariance matrix $M_{n}/n$,
so it converges to zero in probability using Proposition \ref{prop::markov-lln} in Appendix \ref{chapter::limiting-theorems}. 
\end{myproof}

\section{Asymptotic Normality of the OLS estimator}

Intuitively, $ \xi_n$ is the sample average
of some independent terms, and therefore, the classic Lindberg--Feller
theorem (see Proposition \ref{prop::lf-clt} in Appendix \ref{chapter::limiting-theorems}) guarantees that it enjoys a CLT under some regularity conditions.
Consequently, $\hat{\beta}$ also enjoys a CLT with mean $\beta$
and covariance matrix $n^{-1}B_{n}^{-1}M_{n}B_{n}^{-1}.$ The asymptotic results in this chapter require rather tedious regularity conditions. I give them for generality, and they hold automatically if we are willing to assume that the covariates and error terms are all bounded by a constant not depending on $n$. These general conditions are basically moment conditions required by the law of large numbers and CLT.  You do not have to pay too much attention to the conditions when you first read this chapter or you focus on applied statistics.

The CLT relies on an additional condition on a higher-order moment 
\[
d_{2+\delta,n}=n^{-1}\sumn\|x_{i}\|^{2+\delta}E( | \varepsilon_{i} | ^{2+\delta}).
\]

\begin{theorem}
\label{thm:clt-ols} Under Assumptions \ref{assume::heteroskedasticity-lm} and \ref{assume::ols-basic-asymptotic-conditions}, if there exist
a $\delta>0$ and $C>0$ not depending on $n$ such that $d_{2+\delta,n} \leq C$,
then 
$$
\sqrt{n}(\hat{\beta}-\beta)\rightarrow\N(0,B^{-1}MB^{-1})
$$
in distribution. 
\end{theorem}

\begin{myproof}{Theorem}{\ref{thm:clt-ols}}
The key is to show the CLT for $ \xi_n$, and the CLT for $\hat{\beta}$ holds due to the Slutsky's Theorem; see Appendix \ref{chapter::limiting-theorems} for a review. Define 
\[
z_{n,i}=n^{-1/2}x_{i}\varepsilon_{i},\qquad(i=1,\ldots,n)
\]
with mean zero and finite covariance, and we need to verify the two conditions required by the Lindeberg--Feller CLT stated as Proposition \ref{prop::lf-clt} in Appendix \ref{chapter::limiting-theorems}. 

First, the Lyapunov condition holds because
\begin{align*}
\sumn E\left(  \|z_{n,i}\|^{2+\delta} \right) 
& =\sumn E\left(  n^{-(2+\delta)/2}\|x_{i}\|^{2+\delta} | \varepsilon_{i} | ^{2+\delta}  \right)\\
 & =n^{-\delta/2} \times n^{-1}\sumn\|x_{i}\|^{2+\delta}E( | \varepsilon_{i} | ^{2+\delta}) \\
 & =n^{-\delta/2} \times  d_{2+\delta,n} \\
  &\rightarrow 0,
\end{align*}
by the assumption that $d_{2+\delta,n}$ is bounded by a constant $C$.

Second, 
\begin{eqnarray*}
\sumn\cov(z_{n,i})  &=&  n^{-1}\sumn\sigma_{i}^{2}x_{i}x_{i}^{\T} \\
&=& M_{n} \\
&\rightarrow & M.
\end{eqnarray*}
So the Lindberg--Feller CLT implies that $n^{-1/2}\sumn x_{i}\varepsilon_{i}=\sumn z_{n,i}\rightarrow\N(0,M)$
in distribution. 
\end{myproof}

\section{Eicker--Huber--White standard error}

\subsection{Sandwich variance estimator}
The CLT in Theorem \ref{thm:clt-ols} shows that 
\[
\hat{\beta}\asim\N(\beta,n^{-1}B^{-1}MB^{-1}),
\]
where $\asim$ denotes ``approximation in distribution.'' However,
the asymptotic covariance is unknown, and we need to use the data
to construct a reasonable estimator for it to conduct statistical inference. It is relatively easy to replace $B$ with its sample analog $B_{n}$,
but
\[
\tilde{M}_{n}=n^{-1}\sumn\varepsilon_{i}^{2}x_{i}x_{i}^{\T}
\]
as the sample analog for $M$ is not directly useful because the error
terms are unknown either. It is natural to use $\hat{\varepsilon}_{i}^{2}$
to replace $\varepsilon_{i}^{2}$ to obtain the following estimator
for $M$: 
\[
\hat{M}_{n}=n^{-1}\sumn\hat{\varepsilon}_{i}^{2}x_{i}x_{i}^{\T}.
\]

Although each $\hat{\varepsilon}_{i}^{2}$ is a poor estimator for
$\sigma_{i}^{2}$, the sample average $\hat{M}_{n}$ turns out to
be well-behaved with large $n$ and the regularity conditions below.

\begin{theorem}\label{thm::ehw-fixeddesign-consistency}
Under Assumptions \ref{assume::heteroskedasticity-lm} and \ref{assume::ols-basic-asymptotic-conditions}, 
we have $\hat{M}_{n}\rightarrow M$ in probability
if
\begin{equation}
\label{thm::ehw-consistency}
n^{-1}\sumn\var(\varepsilon_{i}^{2})x_{ij_1}^2 x_{ij_2}^2,\quad
n^{-1}\sumn |  x_{ij_{1}}x_{ij_{2}}x_{ij_{3}}x_{ij_{4}} |  ,\quad
n^{-1}\sumn\sigma_{i}^{2}x_{ij_{1}}^{2}x_{ij_{2}}^{2}x_{ij_{3}}^{2}
\end{equation}
are bounded from above by a constant $C$ not depending on $n$ for any $j_{1},j_{2},j_{3},j_{4}=1,\ldots,p$. 
\end{theorem}
\begin{myproof}{Theorem}{\ref{thm::ehw-fixeddesign-consistency}}
Assumption \ref{assume::ols-basic-asymptotic-conditions} ensures that $\hat{\beta}\rightarrow\beta$
in probability by Theorem \ref{thm:consistencyofOLS}. Markov's inequality and the boundedness of the first term in \eqref{thm::ehw-consistency}
ensure that $\tilde{M}_{n}-M_{n}\rightarrow0$ in probability. So
we only need to show that $\hat{M}_{n}-\tilde{M}_{n}\rightarrow0$
in probability. The $(j_{1},j_{2})$th element of their difference
is
\begin{align*}
(\hat{M}_{n}-\tilde{M}_{n})_{j_{1},j_{2}} & =n^{-1}\sumn\hat{\varepsilon}_{i}^{2}x_{i,j_{1}}x_{i,j_{2}}-n^{-1}\sumn\varepsilon_{i}^{2}x_{i,j_{1}}x_{i,j_{2}}\\
 & =n^{-1}\sumn\left[\left(\varepsilon_{i}+x_{i}^{\T}\beta-x_{i}^{\T}\hat{\beta}\right)^{2}-\varepsilon_{i}^{2}\right]x_{i,j_{1}}x_{i,j_{2}}\\
 & =n^{-1}\sumn\left[\left(x_{i}^{\T}\beta-x_{i}^{\T}\hat{\beta}\right)^{2}+2\varepsilon_{i}\left(x_{i}^{\T}\beta-x_{i}^{\T}\hat{\beta}\right)\right]x_{i,j_{1}}x_{i,j_{2}}\\
 & = \textup{I}  + 2 \cdot \textup{II} , 
\end{align*}
where
\begin{eqnarray}
\textup{I} &=& (\beta-\hat{\beta})^{\T}n^{-1}\sumn x_{i}x_{i}^{\T}x_{i,j_{1}}x_{i,j_{2}}(\beta-\hat{\beta}) ,  \\
\textup{II} &=& (\beta-\hat{\beta})^{\T}n^{-1}\sumn x_{i}x_{i,j_{1}}x_{i,j_{2}}\varepsilon_{i}.
\end{eqnarray}
The $(\hat{M}_{n}-\tilde{M}_{n})_{j_{1},j_{2}}$ converges to 0 in probability because both term I and term II converge to 0 in probability. I will show these two facts below. 

First,  $\textup{I} \rightarrow 0$ in probability because $\beta-\hat{\beta} \rightarrow 0$ in probability and $ | n^{-1}\sumn x_{i}x_{i}^{\T}x_{i,j_{1}}x_{i,j_{2}} | $ is bounded by the assumption in \eqref{thm::ehw-consistency}.

Second, $\textup{II} \rightarrow 0$ in probability because $\beta-\hat{\beta} \rightarrow 0$ in probability and $n^{-1}\sumn x_{i}x_{i,j_{1}}x_{i,j_{2}}\varepsilon_{i} \rightarrow 0$ in probability due to\footnote{Technically, we only need this term to be bounded in probability to ensure that $\textup{II} \rightarrow 0$ in probability. A weaker condition is that $n^{-2}\sumn\sigma_{i}^{2}x_{ij_{1}}^{2}x_{ij_{2}}^{2}x_{ij_{3}}^{2}$ is bounded by a constant. Nevertheless, I invoke a stronger condition because the sample mean is easier to interpret.} 
\begin{eqnarray*}
E(n^{-1}\sumn x_{i}x_{i,j_{1}}x_{i,j_{2}}\varepsilon_{i}) &=& 0,\\
\cov(n^{-1}\sumn x_{i}x_{i,j_{1}}x_{i,j_{2}}\varepsilon_{i}) &=& n^{-2} \sumn \sigma_i^2 x_{i} x_{i}^{\T}  x_{i,j_{1}}^2 x_{i,j_{2}}^2 \rightarrow 0, 
\end{eqnarray*}
and Markov's inequality (see Proposition \ref{prop::markov-lln}). 
\end{myproof}

The final variance estimator for $\hat{\beta}$ is 
\[
\hat{V}_{\textsc{ehw}}=n^{-1}\left(n^{-1}\sumn x_{i}x_{i}^{\T}\right)^{-1}\left(n^{-1}\sumn\hat{\varepsilon}_{i}^{2}x_{i}x_{i}^{\T}\right)\left(n^{-1}\sumn x_{i}x_{i}^{\T}\right)^{-1},
\]
 which is called the Eicker--Huber--White (EHW) heteroskedasticity robust covariance matrix.
In matrix form, it equals 
\[
\hat{V}_{\textsc{ehw}} =(X^{\T}X)^{-1}(X^{\T}\hat{\Omega}X)(X^{\T}X)^{-1},
\]
with $\hat{\Omega}=\text{diag}\left\{ \hat{\varepsilon}_{1}^{2},\ldots,\hat{\varepsilon}_{n}^{2}\right\} $, which was introduced in \eqref{eq::ehw-first-appear} at the beginning of this chapter.  
\citet{eicker1967limit} first proposed to use $\hat{V}_{\textsc{ehw}} $.  \citet{white::1980} popularized it in economics, which has been influential in empirical research.  Related estimators appeared in many other contexts of statistics.
\citet{cox1961tests} and \citet{huber::1967} discussed the sandwich variance in the context of misspecified parametric models; see Appendix \ref{sec::mle}. 
\citet{fuller1975regression} proposed a more general form of $\hat{V}_{\textsc{ehw}} $ in the context of survey sampling. 
The square root of the diagonal terms of $\hat{V}_{\textsc{ehw}} $, denoted by $\hat{\text{se}}_{\textsc{ehw},j}\ (j=1,\ldots,p)$, are called the heteroskedasticity-consistent standard errors, heteroskedasticity-robust standard errors, White standard errors, Huber--White standard errors, or Eicker--Huber--White standard errors, among many other names. 

We can conduct statistical inference based on Normal approximations. For example, we can test linear hypotheses based on
$$
\hat{\beta} \asim \N(\beta,  \hat{V}_{\textsc{ehw}}  ),
$$
and in particular, we can infer each element of the coefficient based on 
$$
\hat{\beta}_j \asim \N(\beta_j, \hat{\text{se}}_{\textsc{ehw}, j}^2 ).
$$

\subsection{Other heteroskedasticity-consistent (HC) standard errors}
\label{sec::HCs}

Statistical inference based on the EHW standard error relaxes the parametric assumptions of the Normal linear model. However, its validity relies strongly on the asymptotic argument. In finite samples, it can have poor behavior. 
Since \citet{white::1980} published his paper, several modifications of $\hat{V}_{\textsc{ehw}}$ appeared aiming for better finite-sample properties. I summarize some of them below. They all rely on the $h_{ii}$'s, which are the diagonal elements of the projection matrix $H$ and called the {\it leverage scores}. Define
\[
\hat{V}_{\textsc{ehw},k} = n^{-1}\left(n^{-1}\sumn x_{i}x_{i}^{\T}\right)^{-1}\left(n^{-1}\sumn\hat{\varepsilon}_{i,k}^{2}x_{i}x_{i}^{\T}\right)\left(n^{-1}\sumn x_{i}x_{i}^{\T}\right)^{-1},
\]
where  
\[
\hat{\varepsilon}_{i,k}=\begin{cases}
\hat{\varepsilon}_{i}, & (k=0, \text{ HC}0); \\
\hat{\varepsilon}_{i}\sqrt{\frac{n}{n-p},} & (k=1, \text{ HC}1) ; \\
\hat{\varepsilon}_{i}/\sqrt{1-h_{ii}}, & (k=2, \text{ HC}2) ; \\
\hat{\varepsilon}_{i}/(1-h_{ii}), & (k=3, \text{ HC}3); \\
\hat{\varepsilon}_{i}/(1-h_{ii})^{\min\left\{ 2,nh_{ii}/(2p)\right\} }, & (k=4, \text{ HC}4).
\end{cases}
\]
The HC1 correction is similar to the degrees of freedom correction in the OLS covariance estimator. The HC2 correction was motivated by the unbiasedness of covariance when the error terms have the same variance; see Problem \ref{hw::ehw-unbiased-hc2} for more details. The HC3 correction was motivated by a method called {\it jackknife}, which will be discussed in Chapter \ref{chapter::leave-one-out}. This version appeared even earlier than  \citet{white::1980}; see \citet{miller1974unbalanced}, \citet{hinkley1977jackknifing}, and \citet{reeds1978jackknifing}. 
See \citet{mackinnon1985some}, \citet{long2000using} and \citet{cribari2004asymptotic} for reviews. Based on simulation studies, \citet{long2000using} recommended HC3.

\subsection{Special case with homoskedasticity}
As an important special case with $\sigma_{i}^{2}=\sigma^{2}$ for
all $i=1,\ldots,n$, we have 
\[
M_{n}=\sigma^{2}n^{-1}\sumn x_{i}x_{i}^{\T}=\sigma^{2}B_{n},
\]
which simplifies the covariance of $\hat{\beta}$ to $\cov(\hat{\beta})=\sigma^{2}B_{n}^{-1} / n,$
and the asymptotic Normality to $\sqrt{n}(\hat{\beta}-\beta)\rightarrow\N(0,\sigma^{2}B^{-1})$
in distribution. We have shown that under the Gauss--Markov model,
$\hat{\sigma}^{2}=(n-p)^{-1}\sumn\hat{\varepsilon}_{i}^{2}$ is unbiased
for $\sigma^{2}$. Moreover, $\hat{\sigma}^{2}$ is consistent for
$\sigma^{2}$ under the same condition as Theorem \ref{thm:consistencyofOLS},
justifying the use of 
\[
\hat{V} = 
\hat{\sigma}^{2}\left(\sumn x_{i}x_{i}^{\T}\right)=\hat{\sigma}^{2}\left(X^{\T}X\right)^{-1}
\]
as the covariance estimator. So under homoskedasticity, we can conduct statistical inference based on the following approximate Normality:
\[
\hat{\beta}\asim\N(\beta,\hat{\sigma}^{2}\left(X^{\T}X\right)^{-1}).
\]
It is slightly different from the inference based on $t$ and $F$ distributions. But with large $n$, the difference is very small.

I will end this section with a formal result on the consistency of
$\hat{\sigma}^{2}.$

\begin{theorem}\label{thm::homoskeda-asymptotic}
Under Assumptions \ref{assume::heteroskedasticity-lm} and \ref{assume::ols-basic-asymptotic-conditions},  we have $\hat{\sigma}^{2}\rightarrow\sigma^{2}$ in probability if 
 $\sigma_{i}^{2}=\sigma^{2}<\infty$ for all $i=1,\ldots,n$, and $n^{-1} \sumn \var(\varepsilon_i^2  ) $ is bounded above by a constant not depending on $n$. 
\end{theorem}

\begin{myproof}{Theorem}{\ref{thm::homoskeda-asymptotic}}
Using Markov's inequality, we can show that $n^{-1}\sumn\varepsilon_{i}^{2}\rightarrow\sigma^{2}$ 
in probability. In addition, $n^{-1}\sumn\hat{\varepsilon}_{i}^{2}$ has the
same probability limit as $\hat{\sigma}^{2}$. So we only need to
show that $n^{-1}\sumn\hat{\varepsilon}_{i}^{2}-n^{-1}\sumn\varepsilon_{i}^{2}\rightarrow0$
in probability. Their difference is
\begin{align*}
&n^{-1}\sumn\hat{\varepsilon}_{i}^{2}-n^{-1}\sumn\varepsilon_{i}^{2} \\
& =n^{-1}\sumn\left\{ \left(\varepsilon_{i}+x_{i}^{\T}\beta-x_{i}^{\T}\hat{\beta}\right)^{2}-\varepsilon_{i}^{2}\right\} \\
 & =n^{-1}\sumn\left\{ \left(x_{i}^{\T}\beta-x_{i}^{\T}\hat{\beta}\right)^{2}+2\left(x_{i}^{\T}\beta-x_{i}^{\T}\hat{\beta}\right)\varepsilon_{i}\right\} \\
 & =(\beta-\hat{\beta})^{\T}n^{-1}\sumn x_{i}x_{i}^{\T}(\beta-\hat{\beta})+2(\beta-\hat{\beta})^{\T}n^{-1}\sumn x_{i}\varepsilon_{i} 
 \\
 &= -(\beta-\hat{\beta})^{\T}n^{-1}\sumn x_{i}x_{i}^{\T}(\beta-\hat{\beta}),
\end{align*}
where the last step follows from Lemma \ref{lem:The-OLS-estimator}. So the difference converges to zero in probability because $\hat{\beta}-\beta\rightarrow0$
in probability by Theorem \ref{thm:consistencyofOLS} and $B_{n}\rightarrow B$ by Assumption \ref{assume::ols-basic-asymptotic-conditions}. 
\end{myproof}

\section{Examples}

I use three examples to compare various standard errors for the regression coefficients. The \ri{car} package contains the \ri{hccm} function that implements the EHW standard errors.  

\begin{lstlisting}
> library("car")
\end{lstlisting}

\subsection{LaLonde experimental data}

First, I revisit the \ri{lalonde} data, which were analyzed in Chapter \ref{section::normal-lm-lalonde}. In the following analysis,  different standard errors give similar $t$ statistics. Only \ri{treat} is significant, but none of the other pretreatment covariates are significant.

\begin{lstlisting}
> library("Matching")
> data(lalonde)
> ols.fit = lm(re78 ~ ., data = lalonde)
> ols.fit.hc0 = sqrt(diag(hccm(ols.fit, type = "hc0")))
> ols.fit.hc1 = sqrt(diag(hccm(ols.fit, type = "hc1")))
> ols.fit.hc2 = sqrt(diag(hccm(ols.fit, type = "hc2")))
> ols.fit.hc3 = sqrt(diag(hccm(ols.fit, type = "hc3")))
> ols.fit.hc4 = sqrt(diag(hccm(ols.fit, type = "hc4")))
> ols.fit.coef =summary(ols.fit)$coef
> tvalues = ols.fit.coef[,1]/
+   cbind(ols.fit.coef[,2], ols.fit.hc0, ols.fit.hc1, 
+         ols.fit.hc2, ols.fit.hc3, ols.fit.hc4)
> colnames(tvalues) = c("ols", "hc0", "hc1", "hc2", "hc3", "hc4")
> round(tvalues, 2)
              ols   hc0   hc1   hc2   hc3   hc4
(Intercept)  0.07  0.07  0.07  0.07  0.07  0.07
age          1.17  1.29  1.28  1.27  1.25  1.25
educ         1.75  2.03  2.00  1.99  1.94  1.92
black       -1.74 -2.00 -1.97 -1.95 -1.91 -1.91
hisp         0.27  0.30  0.30  0.30  0.29  0.29
married     -0.17 -0.17 -0.17 -0.17 -0.16 -0.16
nodegr      -0.02 -0.01 -0.01 -0.01 -0.01 -0.01
re74         1.40  0.98  0.96  0.92  0.87  0.77
re75         0.13  0.14  0.14  0.13  0.13  0.12
u74          1.16  0.89  0.88  0.87  0.85  0.83
u75         -1.05 -0.76 -0.75 -0.75 -0.74 -0.74
treat        2.61  2.49  2.46  2.45  2.41  2.40
\end{lstlisting}

\subsection{Data from King and Roberts (2015)}
\label{sec::example-ehw-kingdata1}

The following example uses the data from \citet{king2015robust}. The outcome variable is the multilateral aid flows, and the covariates include log population, log population squared, gross domestic product, former colony status, distance from the Western world, political freedom, military expenditures, arms imports, and the indicators for the years. Different standard errors give very different $t$ statistics for some coefficients.

\begin{lstlisting}
> library(foreign)
> dat = read.dta("isq.dta")
> dat = na.omit(dat[,c("multish", "lnpop", "lnpopsq", 
+                      "lngdp", "lncolony", "lndist", 
+                      "freedom", "militexp", "arms", 
+                      "year83", "year86", "year89", "year92")])
> ols.fit = lm(multish ~ lnpop + lnpopsq + lngdp +  lncolony 
+              + lndist + freedom + militexp + arms 
+              + year83 + year86 + year89 + year92, data=dat)
> ols.fit.hc0 = sqrt(diag(hccm(ols.fit, type = "hc0")))
> ols.fit.hc1 = sqrt(diag(hccm(ols.fit, type = "hc1")))
> ols.fit.hc2 = sqrt(diag(hccm(ols.fit, type = "hc2")))
> ols.fit.hc3 = sqrt(diag(hccm(ols.fit, type = "hc3")))
> ols.fit.hc4 = sqrt(diag(hccm(ols.fit, type = "hc4")))
> ols.fit.coef =summary(ols.fit)$coef
> tvalues = ols.fit.coef[,1]/
+   cbind(ols.fit.coef[,2], ols.fit.hc0, ols.fit.hc1, 
+         ols.fit.hc2, ols.fit.hc3, ols.fit.hc4)
> colnames(tvalues) = c("ols", "hc0", "hc1", "hc2", "hc3", "hc4")
> round(tvalues, 2)
              ols   hc0   hc1   hc2   hc3   hc4
(Intercept)  7.40  4.60  4.54  4.43  4.27  4.14
lnpop       -8.25 -4.46 -4.40 -4.30 -4.14 -4.01
lnpopsq      9.56  4.79  4.72  4.61  4.44  4.31
lngdp       -6.39 -6.14 -6.06 -6.01 -5.88 -5.86
lncolony     4.70  4.75  4.69  4.64  4.53  4.47
lndist      -0.14 -0.16 -0.16 -0.16 -0.15 -0.16
freedom      2.25  1.80  1.78  1.75  1.69  1.65
militexp     0.51  0.59  0.59  0.57  0.55  0.52
arms         1.34  1.17  1.15  1.10  1.03  0.91
year83       1.05  0.85  0.84  0.83  0.80  0.79
year86       0.35  0.40  0.39  0.39  0.38  0.38
year89       0.70  0.81  0.80  0.80  0.78  0.79
year92       0.31  0.40  0.40  0.40  0.39  0.40
\end{lstlisting}

However, if we apply the log transformation on the outcome, then all standard errors give similar $t$ statistics. 

\begin{lstlisting}
> ols.fit = lm(log(multish + 1) ~ lnpop + lnpopsq + lngdp +  lncolony 
+              + lndist + freedom + militexp + arms 
+              + year83 + year86 + year89 + year92, data=dat)
> ols.fit.hc0 = sqrt(diag(hccm(ols.fit, type = "hc0")))
> ols.fit.hc1 = sqrt(diag(hccm(ols.fit, type = "hc1")))
> ols.fit.hc2 = sqrt(diag(hccm(ols.fit, type = "hc2")))
> ols.fit.hc3 = sqrt(diag(hccm(ols.fit, type = "hc3")))
> ols.fit.hc4 = sqrt(diag(hccm(ols.fit, type = "hc4")))
> ols.fit.coef =summary(ols.fit)$coef
> tvalues = ols.fit.coef[,1]/
+   cbind(ols.fit.coef[,2], ols.fit.hc0, ols.fit.hc1, 
+         ols.fit.hc2, ols.fit.hc3, ols.fit.hc4)
> colnames(tvalues) = c("ols", "hc0", "hc1", "hc2", "hc3", "hc4")
> round(tvalues, 2)
              ols   hc0   hc1   hc2   hc3   hc4
(Intercept)  2.96  2.81  2.77  2.72  2.63  2.53
lnpop       -2.87 -2.63 -2.60 -2.54 -2.45 -2.35
lnpopsq      4.21  3.72  3.67  3.59  3.46  3.32
lngdp       -8.02 -7.49 -7.38 -7.38 -7.27 -7.33
lncolony     6.31  6.19  6.11  6.08  5.97  5.95
lndist      -0.16 -0.14 -0.14 -0.14 -0.14 -0.14
freedom      1.47  1.53  1.51  1.50  1.47  1.46
militexp    -0.32 -0.32 -0.31 -0.31 -0.30 -0.29
arms         1.27  1.12  1.10  1.05  0.98  0.86
year83       0.10  0.10  0.10  0.10  0.10  0.10
year86      -0.14 -0.14 -0.14 -0.14 -0.14 -0.14
year89       0.46  0.45  0.44  0.44  0.44  0.44
year92       0.03  0.03  0.03  0.03  0.03  0.03
\end{lstlisting}

  In general, the difference between the OLS and EHW standard errors may be due to the heteroskedasticity or the poor approximation of the linear model. The above two analyses based on the original and transformed outcomes suggest that the linear approximation works better for the log-transformed outcome. 
  We will discuss the issues of model misspecification and transformation in Chapters \ref{chapter::populationOLS} and \ref{chapter::transformation}, respectively.

\subsection{Boston housing data}

I also re-analyze the classic Boston housing data \citep{harrison1978hedonic}. The outcome variable is the median value of owner-occupied homes in US dollars 1000, and the covariates include per capita crime rate by town, the proportion of residential land zoned for lots over 25,000 square feet, the proportion of non-retail business acres per town, etc. You can find more details in the \ri{R} package \ri{mlbench}. In this example, different standard errors give very different $t$ statistics.

\begin{lstlisting}
> library("mlbench")
> data(BostonHousing)
> ols.fit = lm(medv ~ ., data = BostonHousing)
> summary(ols.fit)

Call:
lm(formula = medv ~ ., data = BostonHousing)

Residuals:
    Min      1Q  Median      3Q     Max 
-15.595  -2.730  -0.518   1.777  26.199 

Coefficients:
              Estimate Std. Error t value Pr(>|t|)    
(Intercept)  3.646e+01  5.103e+00   7.144 3.28e-12 ***
crim        -1.080e-01  3.286e-02  -3.287 0.001087 ** 
zn           4.642e-02  1.373e-02   3.382 0.000778 ***
indus        2.056e-02  6.150e-02   0.334 0.738288    
chas1        2.687e+00  8.616e-01   3.118 0.001925 ** 
nox         -1.777e+01  3.820e+00  -4.651 4.25e-06 ***
rm           3.810e+00  4.179e-01   9.116  < 2e-16 ***
age          6.922e-04  1.321e-02   0.052 0.958229    
dis         -1.476e+00  1.995e-01  -7.398 6.01e-13 ***
rad          3.060e-01  6.635e-02   4.613 5.07e-06 ***
tax         -1.233e-02  3.760e-03  -3.280 0.001112 ** 
ptratio     -9.527e-01  1.308e-01  -7.283 1.31e-12 ***
b            9.312e-03  2.686e-03   3.467 0.000573 ***
lstat       -5.248e-01  5.072e-02 -10.347  < 2e-16 ***

Residual standard error: 4.745 on 492 degrees of freedom
Multiple R-squared:  0.7406,	Adjusted R-squared:  0.7338 
F-statistic: 108.1 on 13 and 492 DF,  p-value: < 2.2e-16

> 
> ols.fit.hc0 = sqrt(diag(hccm(ols.fit, type = "hc0")))
> ols.fit.hc1 = sqrt(diag(hccm(ols.fit, type = "hc1")))
> ols.fit.hc2 = sqrt(diag(hccm(ols.fit, type = "hc2")))
> ols.fit.hc3 = sqrt(diag(hccm(ols.fit, type = "hc3")))
> ols.fit.hc4 = sqrt(diag(hccm(ols.fit, type = "hc4")))
> ols.fit.coef =summary(ols.fit)$coef
> tvalues = ols.fit.coef[,1]/
+   cbind(ols.fit.coef[,2], ols.fit.hc0, ols.fit.hc1, 
+         ols.fit.hc2, ols.fit.hc3, ols.fit.hc4)
> colnames(tvalues) = c("ols", "hc0", "hc1", "hc2", "hc3", "hc4")
> round(tvalues, 2)
               ols   hc0   hc1   hc2   hc3   hc4
(Intercept)   7.14  4.62  4.56  4.48  4.33  4.25
crim         -3.29 -3.78 -3.73 -3.48 -3.17 -2.58
zn            3.38  3.42  3.37  3.35  3.27  3.28
indus         0.33  0.41  0.41  0.41  0.40  0.40
chas1         3.12  2.11  2.08  2.05  2.00  2.00
nox          -4.65 -4.76 -4.69 -4.64 -4.53 -4.52
rm            9.12  4.57  4.51  4.43  4.28  4.18
age           0.05  0.04  0.04  0.04  0.04  0.04
dis          -7.40 -6.97 -6.87 -6.81 -6.66 -6.66
rad           4.61  5.05  4.98  4.91  4.76  4.65
tax          -3.28 -4.65 -4.58 -4.54 -4.43 -4.42
ptratio      -7.28 -8.23 -8.11 -8.06 -7.89 -7.93
b             3.47  3.53  3.48  3.44  3.34  3.30
lstat       -10.35 -5.34 -5.27 -5.18 -5.01 -4.93
\end{lstlisting}

The log transformation of the outcome does not remove the discrepancy among the standard errors. In this example, heteroskedasticity seems an important problem.

\begin{lstlisting}
> ols.fit = lm(log(medv) ~ ., data = BostonHousing)
> summary(ols.fit)

Call:
lm(formula = log(medv) ~ ., data = BostonHousing)

Residuals:
     Min       1Q   Median       3Q      Max 
-0.73361 -0.09747 -0.01657  0.09629  0.86435 

Coefficients:
              Estimate Std. Error t value Pr(>|t|)    
(Intercept)  4.1020423  0.2042726  20.081  < 2e-16 ***
crim        -0.0102715  0.0013155  -7.808 3.52e-14 ***
zn           0.0011725  0.0005495   2.134 0.033349 *  
indus        0.0024668  0.0024614   1.002 0.316755    
chas1        0.1008876  0.0344859   2.925 0.003598 ** 
nox         -0.7783993  0.1528902  -5.091 5.07e-07 ***
rm           0.0908331  0.0167280   5.430 8.87e-08 ***
age          0.0002106  0.0005287   0.398 0.690567    
dis         -0.0490873  0.0079834  -6.149 1.62e-09 ***
rad          0.0142673  0.0026556   5.373 1.20e-07 ***
tax         -0.0006258  0.0001505  -4.157 3.80e-05 ***
ptratio     -0.0382715  0.0052365  -7.309 1.10e-12 ***
b            0.0004136  0.0001075   3.847 0.000135 ***
lstat       -0.0290355  0.0020299 -14.304  < 2e-16 ***

Residual standard error: 0.1899 on 492 degrees of freedom
Multiple R-squared:  0.7896,	Adjusted R-squared:  0.7841 
F-statistic: 142.1 on 13 and 492 DF,  p-value: < 2.2e-16

> 
> ols.fit.hc0 = sqrt(diag(hccm(ols.fit, type = "hc0")))
> ols.fit.hc1 = sqrt(diag(hccm(ols.fit, type = "hc1")))
> ols.fit.hc2 = sqrt(diag(hccm(ols.fit, type = "hc2")))
> ols.fit.hc3 = sqrt(diag(hccm(ols.fit, type = "hc3")))
> ols.fit.hc4 = sqrt(diag(hccm(ols.fit, type = "hc4")))
> ols.fit.coef =summary(ols.fit)$coef
> tvalues = ols.fit.coef[,1]/
+   cbind(ols.fit.coef[,2], ols.fit.hc0, ols.fit.hc1, 
+         ols.fit.hc2, ols.fit.hc3, ols.fit.hc4)
> colnames(tvalues) = c("ols", "hc0", "hc1", "hc2", "hc3", "hc4")
> round(tvalues, 2)
               ols   hc0   hc1   hc2   hc3   hc4
(Intercept)  20.08 14.29 14.09 13.86 13.43 13.13
crim         -7.81 -5.31 -5.24 -4.85 -4.39 -3.56
zn            2.13  2.68  2.64  2.62  2.56  2.56
indus         1.00  1.46  1.44  1.43  1.40  1.41
chas1         2.93  2.69  2.66  2.62  2.56  2.56
nox          -5.09 -4.79 -4.72 -4.67 -4.56 -4.54
rm            5.43  3.31  3.26  3.20  3.10  3.02
age           0.40  0.33  0.32  0.32  0.31  0.31
dis          -6.15 -6.12 -6.03 -5.98 -5.84 -5.82
rad           5.37  5.23  5.16  5.05  4.87  4.67
tax          -4.16 -5.05 -4.98 -4.90 -4.76 -4.69
ptratio      -7.31 -8.84 -8.72 -8.67 -8.51 -8.55
b             3.85  2.80  2.76  2.72  2.65  2.59
lstat       -14.30 -7.86 -7.75 -7.63 -7.40 -7.28
\end{lstlisting}

\section{Final remarks}

The beauty of the asymptotic analysis and the EHW standard error is that they hold under weak parametric assumptions on the error terms. We do not need to modify the OLS estimator but only need to modify the covariance estimator. 
However, this framework has limitations. 
\begin{enumerate}[label=(L\arabic*), ref=L\arabic*]
\item
The proofs are based on limiting theorems that require the sample size to be infinity. We are often unsure whether the sample size is large enough for a particular application we have.
\item
The EHW standard errors can be severely biased and have large variability in finite samples. Problem \ref{hw::ehw-unbiased-hc2} shows that the HC2 correction is unbiased for the true covariance matrix of $\hat\beta$ under the Gauss--Markov model. However, no such result exists for the heteroskedastic linear model.
\item
Under the heteroskedastic linear model, the Gauss--Markov theorem does not hold, so the OLS can be inefficient. We will discuss possible improvements in Chapter \ref{chapter::WLS}. 
\item
Unlike Section \ref{sec::prection-normallinear}, we cannot create any reasonable prediction intervals for a future observation $y_{n+1}$ based on $(X,Y, x_{n+1})$ since its variance $\sigma_{n+1}^2$ is fundamentally unknown without further assumptions.  Chapter \ref{sec::conformal-prediction} will discuss the problem of prediction under a modified statistical framework. 
\end{enumerate}

\section{Homework problems}

\paragraph{Testing linear hypotheses under heteroskedasticity} Under the heteroskedastic linear model, how do we test the hypotheses
$$
H_0: c^{\T} \beta = 0,  
$$
for $c\in \mathbb{R}^p$,
and
$$
H_0:  C\beta = 0
$$
for $ C\in \mathbb{R}^{l\times p}$ with $l$ linearly independent rows?

\paragraph{Two-sample problem continued}\label{hw08::twosample-ehw}

This problem extends Problem \ref{hw5::two-sample}. 

\begin{enumerate}
\item Assume that $z_{1},\ldots,z_{m}$ are IID with mean $\mu_{1}$ and
variance $\sigma_{1}^{2}$, and $w_{1},\ldots,w_{n}$ are IID with
mean $\mu_{2}$ and variance $\sigma_{2}^{2}$, and test
$H_0: \mu_{1}=\mu_{2}$. 

Prove that under $H_0$, the following
$t$ statistic has an asymptotically Normal distribution:
\[
t_{\text{unequal}}=\frac{\bar{z}-\bar{w}}{\sqrt{S_{z}^{2}/m+S_{w}^{2}/n}}\rightarrow \N(0,1)
\]
in distribution. 

Remark: 
The name  ``\ri{unequal}'' is motivated by the ``\ri{var.equal}''
parameter of the \ri{R} function \ri{t.test}.

\item We can write the above problem as testing hypothesis $H_{0}:\beta_{1}=0$ in the heteroskedastic linear regression. Based on the EHW standard error, we can compute the $t$ statistic.

Prove that $t_{\text{unequal}}$ is numerically identical to the $t$ statistic based on the EHW robust standard error with the HC2 correction.
\end{enumerate}

\paragraph{ANOVA with heteroskedasticity}\label{hw8::anova-ols-hc02}

This problem extends Problem \ref{hw5:anova-f}.

Assume $y_i \mid i \in \mathcal{T}_j $ has mean $\beta_j$ and variance $\sigma_j^2$, which can be rewritten as a linear model without the Normality and homoskedasticity. In the process of solving Problem \ref{hw5:anova-f}, you have derived the estimator of the covariance matrix of the OLS estimator under homoskedasticity. Find the  HC0 and HC2 versions of the EHW covariance matrix. Which covariance matrices do you recommend, and why?


 \paragraph{Invariance of the EHW covariance estimator}
\label{hw08::invariance-ehw01234}

Theorem \ref{thm::invariance-ehw} below extends Theorem \ref{thm::ols-invariance} in Problem \ref{hw3::invariance-ols}. 
Prove Theorem \ref{thm::invariance-ehw}. 

\begin{theorem}
\label{thm::invariance-ehw}
If we transform $X$ to $\tilde{X} = X\Gamma$ where $\Gamma$ is a $p\times p$ non-degenerate matrix, the OLS fit changes from
$$
Y = X\hat{\beta} + \hat{\varepsilon}
$$
to
$$
Y = \tilde{X}\tilde{\beta} + \tilde{\varepsilon},
$$
and the associated EHW covariance estimator changes from $\hat{V}_\textsc{ehw}$ to $\tilde{V}_\textsc{ehw}$. 

Then
$$
\hat{V} = \Gamma \tilde{V}  \Gamma^{\T},
$$ 
and the above result holds for HC$j$ $(j=0,1,2,3,4)$. The relationship also holds for the covariance estimator assuming homoskedasticity. 
\end{theorem}

Remark: You can use the results in Problems \ref{hw3::invariance-ols} and \ref{hw03::invariance-of-H}.

\paragraph{Breakdown of the equivalence of the $t$-statistics based on the EHW standard error}\label{hw6::t-stat-equivalent-breakdown}

This problem parallels Problem \ref{hw5::t-stat-equivalent}.

Consider data $(x_i, y_i)_{i=1}^n$, where both $x_i$ and $y_i$ are scalars. Run OLS fit of $y_i$ on $(1,x_i)$ to obtain $t_{y\mid x}$, the $t$-statistic of the coefficient of $x_i$, based on the EHW standard error. Run OLS fit of $x_i$ on $(1,y_i)$ to obtain $t_{x\mid y}$, the $t$-statistic of the coefficient of $y_i$, based on the EHW standard error. 

Give a counterexample with $t_{y\mid x} \neq  t_{x\mid y}$.

\paragraph{Empirical comparison of the standard errors}
\citet{long2000using} reviewed and compared several commonly-used standard errors in OLS. Redo their simulation and replicate their Figures 1--4. They specified more details of their covariate generating process in a technical report \citep{long1998correcting}.

\paragraph{Robust standard error in practice}
\citet{king2015robust} gave three examples where the EHW standard errors differ from the OLS standard error. I have replicated one example in Section \ref{sec::example-ehw-kingdata1}. Replicate another one using linear regression although the original analysis used Poisson regression. You can find the datasets used by \citet{king2015robust} at Harvard Dataverse (\texttt{https://dataverse.harvard.edu/}).

Remark: You may encounter the issue of $h_{ii} = 1$ for some observations. How would you deal with it? See Chapter \ref{chapter::leave-one-out} for discussions of  $h_{ii}$.

\paragraph{Unbiased sandwich variance estimator under the Gauss--Markov model}\label{hw::ehw-unbiased-hc2}

Under the Gauss--Markov model with $\sigma_i^2 = \sigma^2$ for $i=1,\ldots, n$, show that the HC0 version of $\hat{V}_{\textsc{ehw}} $ is biased but the HC2 version of $\tilde{V}_{\textsc{ehw}} $ is unbiased  for $\cov(\hat{\beta})$.

  \part{Interpretation of Ordinary Least Squares Based on Partial Regressions}\label{part::FWL-Cochran}

\chapter{Frisch--Waugh--Lovell Theorem}
 \label{chapter::FWL-theorem}

The Frisch--Waugh--Lovell (FWL) Theorem is a powerful theorem about OLS. It allows us to reduce complicated, multi-dimensional OLS to simpler, often one-dimensional OLS. It helps to interpret the OLS coefficients. Moreover, it is a theoretical tool to derive many other results about OLS.
This chapter introduces the FWL Theorem, and Chapter \ref{chapter::FWL-application} will discuss its applications.

\section{Long and short regressions}

If we partition $X$ and $\beta$ as
\[
X=\left(\begin{array}{cc}
X_{1} & X_{2}\end{array}\right),\qquad\beta=\left(\begin{array}{c}
\beta_{1}\\
\beta_{2}
\end{array}\right),
\]
where $X_{1}\in\mathbb{R}^{n\times k},X_{2}\in\mathbb{R}^{n\times l},\beta_{1}\in\mathbb{R}^{k}$
and $\beta_2 \in  \mathbb{R}^{l},$ then we can consider the {\it long regression} 
\begin{eqnarray*}
Y &=& X\hat{\beta}+\hat{\varepsilon}  \\
&=&\left(\begin{array}{cc}
X_{1} & X_{2}\end{array}\right)\left(\begin{array}{c}
\hat{\beta}_{1}\\
\hat{\beta}_{2}
\end{array}\right)+\hat{\varepsilon} \\
&=&X_{1}\hat{\beta}_{1}+X_{2}\hat{\beta}_{2}+\hat{\varepsilon},
\end{eqnarray*}
 and the {\it short regression} 
\[
Y=X_{2}\tilde{\beta}_{2}+\tilde{\varepsilon},
\]
where $\hat{\beta}=\left(\begin{array}{c}
\hat{\beta}_{1}\\
\hat{\beta}_{2}
\end{array}\right)$ and $\tilde{\beta}_{2}$ are the OLS coefficients, and $\hat{\varepsilon}  = Y - X\hat{\beta}$
and $\tilde{\varepsilon} = Y - X_{2}\tilde{\beta}_{2} $ are the residual vectors from the long
and short regressions, respectively. These two regressions are of great interest
in practice. For example, we can ask the following questions: 
\begin{enumerate}[label=(Q\arabic*), ref=Q\arabic*]
\item\label{question::fwl1} if the true $\beta_{1}$ is zero, then what is the consequence of including $X_1$ in the long regression?
\item\label{question::fwl2} if the true $\beta_{1}$ is not zero, then what is the consequence
of omitting $X_1$ in the short regression?
\item\label{question::fwl3} what is the difference between $\hat{\beta}_{2}$ and $\tilde{\beta}_{2}$?
Both of them are measures of the ``impact'' of $X_{2}$ on $Y$. Then
why are they different? Does their difference give us any information
about $\beta_{1}$?
\end{enumerate}

Many problems in statistics are related to the long and short regressions. We will discuss some applications in Chapter \ref{chapter::FWL-application} and give a related result in Chapter \ref{chapter::cochran-ovb}.

\section{FWL Theorem for the regression coefficients}

Theorem \ref{thm:fwl} below helps to answer the questions in \eqref{question::fwl1}--\eqref{question::fwl3}.
 
\begin{theorem}[FWL Theorem]
\label{thm:fwl}The OLS estimator for $\beta_2$ in the short regression is $\tilde{\beta}_{2}=(X_{2}^{\T}X_{2})^{-1}X_{2}^{\T}Y$, and the OLS estimator for $\beta_{2}$ in the long regression has
the following equivalent forms
\begin{align}
\hat{\beta}_{2} & =\left[(X^{\T}X)^{-1}X^{\T}Y\right]_{\textup{last }l\textup{ elements}}\label{eq:fwl1}\\
 & =\left\{ X_{2}^{\T}(I_{n}-H_{1})X_{2}\right\} ^{-1}X_{2}^{\T}(I_{n}-H_{1})Y\quad\text{ where }H_{1}=X_{1}(X_{1}^{\T}X_{1})^{-1}X_{1}^{\T}  \label{eq:fwl2}\\
 & =(\tilde{X}_{2}^{\T}\tilde{X}_{2})^{-1}\tilde{X}_{2}^{\T}Y\quad\text{ where }\tilde{X}_{2}=(I_{n}-H_{1})X_{2}\label{eq:fwl3}\\
 & =(\tilde{X}_{2}^{\T}\tilde{X}_{2})^{-1}\tilde{X}_{2}^{\T}\tilde{Y}\quad\text{ where }\tilde{Y}=(I_{n}-H_{1})Y.\label{eq:fwl4}
\end{align}
\end{theorem}

Theorem \ref{thm:fwl} is often called the  Frisch--Waugh--Lovell (FWL) Theorem in econometrics \citep{frisch1933partial, lovell1963seasonal}, although its equivalent forms were also known in classic statistics.\footnote{Professor Alan Agresti gave me the reference of \citet{yule1907theory}.}

Before proving Theorem \ref{thm:fwl}, I will first discuss its meanings
and interpretations. Equation (\ref{eq:fwl1}) follows from the definition
of the OLS coefficient. The matrix $I_{n}-H_{1}$ in equation (\ref{eq:fwl2})
is the projection matrix onto the space orthogonal to the column
space of $X_{1}$. Equation (\ref{eq:fwl3}) states that $\hat{\beta}_{2}$
equals the OLS coefficient of $Y$ on $\tilde{X}_{2}=(I_{n}-H_{1})X_{2}$,
which is the residual matrix from the column-wise OLS fit of $X_{2}$
on $X_{1}$.\footnote{See Problem \ref{hw03::multiple-responses} for more details. } So $\hat{\beta}_{2}$ measures the ``impact'' of $X_{2}$
on $Y$ after ``adjusting'' for the impact of $X_{1}$, that is, it
measures the partial or pure ``impact'' of $X_{2}$ on $Y$. Equation
(\ref{eq:fwl4}) is a slight modification of Equation (\ref{eq:fwl3}),
stating that $\hat{\beta}_{2}$ equals the OLS coefficient of $\tilde{Y}$
on $\tilde{X}_{2}$, where $\tilde{Y}=(I_{n}-H_{1})Y$ is the residual
vector from the OLS fit of $Y$ on $X_{1}$. From (\ref{eq:fwl3}) and
(\ref{eq:fwl4}), it is not crucial to residualize $Y$, but
it is crucial to residualize $X_{2}$. 

The forms (\ref{eq:fwl3}) and (\ref{eq:fwl4}) suggest the interpretation of
$\hat{\beta}_{2}$ as the ``impact'' of $X_{2}$ on $Y$ holding $X_{1}$
constant, or in an econometric term, the ``impact'' of $X_{2}$ on
$Y$ \emph{ceteris paribus}. \citet{marshall1890principles} used the Latin phrase \emph{ceteris paribus}. Its English meaning is ``with other conditions remaining the same.'' 
However, the algebraic meaning of the FWL Theorem is that the OLS coefficient of a variable equals the {\it partial regression} coefficient based on the residuals. 
Therefore, taking the Latin phase too
seriously may be problematic because Theorem \ref{thm:fwl} is a pure algebraic
result without any distributional assumptions. We cannot hold $X_{1}$
constant using pure linear algebra. Sometimes, we can manipulate the value
of $X_{1}$ in an experimental setting, but this relies on the assumption of the data-collecting process. 

There are many ways to prove Theorem \ref{thm:fwl}. Below I first take
a detour to give an unnecessarily complicated proof because some intermediate
steps will be useful for later parts of the book. I will then give a
simpler proof, which requires a deep understanding of OLS as a linear projection.

The first proof relies on the following lemma. 

\begin{lemma}
\label{lem:blockinverse}The inverse of $X^{\T}X$ is
\[
(X^{\T}X)^{-1}=\left(\begin{array}{cc}
 S_{11}  &  S_{12} \\
 S_{21}  &  S_{22} 
\end{array}\right),
\]
 where 
\begin{align*}
 S_{11}  & =(X_{1}^{\T}X_{1})^{-1}+(X_{1}^{\T}X_{1})^{-1}X_{1}^{\T}X_{2}(\tilde{X}_{2}^{\T}\tilde{X}_{2})^{-1}X_{2}^{\T}X_{1}(X_{1}^{\T}X_{1})^{-1},\\
 S_{12}  & =-(X_{1}^{\T}X_{1})^{-1}X_{1}^{\T}X_{2}(\tilde{X}_{2}^{\T}\tilde{X}_{2})^{-1},\\
 S_{21}  & = S_{12}^{\T},\\
 S_{22}  & =(\tilde{X}_{2}^{\T}\tilde{X}_{2})^{-1},
\end{align*}
with $\tilde{X}_{2}=(I_{n}-H_{1})X_{2}$ and $H_{1}=X_{1}(X_{1}^{\T}X_{1})^{-1}X_{1}^{\T}$ defined in Theorem \ref{thm:fwl}. 
\end{lemma}

I leave the proof of Lemma \ref{lem:blockinverse} as Problem \ref{hw6::inverse-block-gram}. With Lemma \ref{lem:blockinverse}, we can easily prove Theorem \ref{thm:fwl}.

\begin{myproof}{Theorem}{\ref{thm:fwl}} (Version 1) 
The OLS coefficient is
\[
\left(\begin{array}{c}
\hat{\beta}_{1}\\
\hat{\beta}_{2}
\end{array}\right)=(X^{\T}X)^{-1}X^{\T}Y=\left(\begin{array}{cc}
 S_{11}  &  S_{12} \\
 S_{21}  &  S_{22} 
\end{array}\right)\left(\begin{array}{c}
X_{1}^{\T}Y\\
X_{2}^{\T}Y
\end{array}\right).
\]
Then using Lemma \ref{lem:blockinverse}, we can simplify $\hat{\beta}_{2}$
as 
\begin{align}
\hat{\beta}_{2} & = S_{21} X_{1}^{\T}Y+ S_{22} X_{2}^{\T}Y\nonumber \\
 & =-(\tilde{X}_{2}^{\T}\tilde{X}_{2})^{-1}X_{2}^{\T}X_{1}(X_{1}^{\T}X_{1})^{-1}X_{1}^{\T}Y+(\tilde{X}_{2}^{\T}\tilde{X}_{2})^{-1}X_{2}^{\T}Y\nonumber \\
 & =-(\tilde{X}_{2}^{\T}\tilde{X}_{2})^{-1}X_{2}^{\T}H_{1}Y+(\tilde{X}_{2}^{\T}\tilde{X}_{2})^{-1}X_{2}^{\T}Y\nonumber \\
 & =(\tilde{X}_{2}^{\T}\tilde{X}_{2})^{-1}X_{2}^{\T}(I_{n}-H_{1})Y\label{eq:form2}\\
 & =(\tilde{X}_{2}^{\T}\tilde{X}_{2})^{-1}\tilde{X}_{2}^{\T}Y.\label{eq:form3}
\end{align}
Equation (\ref{eq:form2}) is the form (\ref{eq:fwl2}), and Equation
(\ref{eq:form3}) is the form (\ref{eq:fwl3}). Because we also have
$X_{2}^{\T}(I_{n}-H_{1})Y=X_{2}^{\T}(I_{n}-H_{1})^{2}Y=\tilde{X}_{2}^{\T}\tilde{Y}$,
we can write $\hat{\beta}_{2}$ as $\hat{\beta}_{2}=(\tilde{X}_{2}^{\T}\tilde{X}_{2})^{-1}\tilde{X}_{2}^{\T}\tilde{Y}$,
giving the form (\ref{eq:fwl4}). 
\end{myproof}

The second proof does not invert the block matrix of $X^{\T} X$ directly.

\begin{myproof}{Theorem}{\ref{thm:fwl}} (Version 2)
First, the OLS decomposition $Y =X_{1}\hat{\beta}_{1}+X_{2}\hat{\beta}_{2}+\hat{\varepsilon}$
satisfies $X^{\T}\hat{\varepsilon}= (X_1, X_2)^{\T}\hat{\varepsilon} =  0$, which implies 
$$ 
X_{1}^{\T}\hat{\varepsilon}=0,
\quad 
X_{2}^{\T}\hat{\varepsilon}=0.
$$

Second, multiplying $I_{n}-H_{1}$ on both sides of the OLS decomposition $Y =X_{1}\hat{\beta}_{1}+X_{2}\hat{\beta}_{2}+\hat{\varepsilon}$,
we have
\[
(I_{n}-H_{1}) Y =(I_{n}-H_{1})X_{1}  \hat{\beta}_1 +(I_{n}-H_{1})X_{2} \hat{\beta}_2 +(I_{n}-H_{1})\hat{\varepsilon},
\]
which reduces to 
\begin{equation}\label{eq::fwl-proof-intermediate}
(I_{n}-H_{1})Y=(I_{n}-H_{1})X_{2}\hat{\beta}_{2}+\hat{\varepsilon}
\end{equation}
because $(I_{n}-H_{1})X_{1}=0$ and 
$
(I_{n}-H_{1})\hat{\varepsilon}=\hat{\varepsilon}-H_{1}\hat{\varepsilon}=\hat{\varepsilon}-X_{1}(X_{1}^{\T}X_{1})^{-1}X_{1}^{\T}\hat{\varepsilon}=\hat{\varepsilon}.
$

Third, multiplying $X_2^{\T}$ on both sides of \eqref{eq::fwl-proof-intermediate}, we have
$$
X_2^{\T}(I_{n}-H_{1})Y = X_2^{\T}(I_{n}-H_{1})X_{2} \hat{\beta}_{2}
$$
because $X_{2}^{\T}\hat{\varepsilon}=0$. 
The FWL Theorem follows immediately with $\hat{\beta}_{2} = (X_2^{\T}(I_{n}-H_{1})X_{2})^{-1} X_2^{\T}(I_{n}-H_{1})Y$. 

A subtle issue in this proof is to verify that $X_{2}^{\T}(I_{n}-H_{1})X_{2}$ is invertible. It is relatively easy to show that matrix $X_{2}^{\T}(I_{n}-H_{1})X_{2}$ is positive semi-definite. To show it has rank $l$, we only need to show that 
$$
u_{2}^{\T}X_{2}^{\T}(I_{n}-H_{1})X_{2}u_{2}  =0 \text{ implies } u_2 = 0.
$$
We have $u_{2}^{\T}X_{2}^{\T}(I_{n}-H_{1})X_{2}u_{2}   = \|(I_{n}-H_{1})X_{2}u_{2} \|^2=0$, so $(I_{n}-H_{1})X_{2}u_{2} = 0$, which further implies $X_{2}u_{2}\in\mathcal{C}(X_{1})$ by Proposition \ref{prop::projectionmatrix-geometry}. That is, $ X_{2}u_{2} = X_1 u_1$ for some $u_1$. So $X_1 u_1 - X_2 u_2 = 0$. Since the columns of $X$ are linearly independent, we must have $u_1 = 0$ and $u_2 = 0$. 
\end{myproof}

I will end this section with two byproducts of the FWL Theorem. First, $\tilde{X}_{2}$ is the residual matrix from the OLS fit of $X_2$ on $X_1$. It is an $n\times l$ matrix with linearly independent columns as shown in the proof of Theorem \ref{thm:fwl} (Version 2)  and induces a projection matrix
$$
\tilde{H}_2 = \tilde{X}_{2} (\tilde{X}_{2}^{\T} \tilde{X}_{2})^{-1} \tilde{X}_{2}^{\T}. 
$$
This projection matrix is closely related to the projection matrices induced by $X$ and $X_1$ as shown in the following lemma. 

\begin{lemma}
\label{lemma::decompose-projection-matrices}
We have
\begin{equation}\label{eq::decompose-projection-matrices1}
H_1 \tilde{H}_2  = \tilde{H}_2  H_1 = 0
\end{equation}
and
\begin{equation}\label{eq::decompose-projection-matrices2}
H = H_1 +  \tilde{H}_2. 
\end{equation}
\end{lemma}

Lemma \ref{lemma::decompose-projection-matrices} is purely algebraic. I leave the proof as Problem \ref{hw08::decompose-projection-m}. 
The  identities in \eqref{eq::decompose-projection-matrices1} imply that the column space of $\tilde{X}_{2}$ is orthogonal to the column space of $X_1$.  
The identity in \eqref{eq::decompose-projection-matrices2} has a clear geometric interpretation. For any vector $v \in \mathbb{R}^n$, we have $Hv = H_1v +  \tilde{H}_2v$, so the projection of $v$ onto the column space of $X$ equals the summation of the projection of $v$ onto the column space of $X_1$ and the projection of $v$ onto the column space of $\tilde{X}_{2}$. Importantly, $H \neq H_1 + H_2$ in general.

Second, we can obtain $\hat{\beta}_2$ from \eqref{eq:fwl3} or \eqref{eq:fwl4}, which corresponds to the partial regression of $Y$ on $\tilde{X}_{2}$ or the partial regression of $\tilde{Y}$ on $\tilde{X}_{2}$. However, subtle issues arise with the residuals. 
Corollary \ref{corollary::fwl-residuals} below states that the residual vector from the second partial regression equals the residual vector from the full regression. The conclusion does not hold if we only residualize $X_2$. See Problem \ref{hw06-fwl-residual}. Therefore, to ensure the residual vectors are the same, it is important to residualize both $Y$ and $X_2$. 

\begin{corollary}
\label{corollary::fwl-residuals}
We have $\hat{\varepsilon} = \hat{e}$, where $\hat{\varepsilon}$ is the residual vector from the OLS fit of $Y$ on $X$ and $\hat{e}$ is the residual vector from the OLS fit of $\tilde{Y}$ on $\tilde{X}_{2}$, respectively. 
\end{corollary}

\begin{myproof}{Corollary}{\ref{corollary::fwl-residuals}}
We have $\hat{ \varepsilon} = (I-H)Y$ and 
$$
\hat{e}= (I-\tilde{H}_2) \tilde{Y} = (I- \tilde{H}_2)(I-H_1) Y .
$$ 
It suffices to show that $I-H = (I- \tilde{H}_2)(I-H_1)$, or, equivalently, $ I-H =  I - H_1 - \tilde{H}_2 + \tilde{H}_2 H_1$. This holds due to Lemma \ref{lemma::decompose-projection-matrices}. 
\end{myproof}

%
%
%

\section{FWL Theorem for standard errors}

Based on the OLS fit of $Y$ on $X$, we have two estimated covariances for the second component $ \hat{\beta}_2$: $\hat{V}$ assuming homoskedasticity and $\hat{V}_{\textsc{ehw}}$
allowing for heteroskedasticity.

The FWL Theorem demonstrates that we can obtain $ \hat{\beta}_2$ from the OLS fit of $\tilde{Y}$ on $\tilde{X}_2$. Then based on this partial regression, we have two estimated covariances for $ \hat{\beta}_2$: 
$
\tilde{V}  
$ 
assuming homoskedasticity and 
$
\tilde{V}_{\textsc{ehw}}  
$
 allowing for heteroskedasticity. 
 
Theorem \ref{thm::fwl-se} below establishes the equivalence between the estimated covariances from the long and partial regressions.

 \begin{theorem}
 \label{thm::fwl-se}
 $(n-k-l)\hat{V} = (n-l)  \tilde{V} $ and $\hat{V}_{\textsc{ehw}} = \tilde{V}_{\textsc{ehw}}.$
 \end{theorem}

Theorem \ref{thm:fwl} is well known for a long time but Theorem \ref{thm::fwl-se} is less well known.
Theorem \ref{thm::fwl-se} is a numeric result that does not depend on the statistical assumptions. 
\citet{lovell1963seasonal} hinted at the first identity in Theorem \ref{thm::fwl-se}, and 
\citet{ding2021frisch} proved Theorem \ref{thm::fwl-se}. 
In Theorem \ref{thm::fwl-se}, the equivalence of the EHW covariances only holds for the original version, and it breaks down for other modified versions discussed in Chapter \ref{sec::HCs}.

\begin{myproof}{Theorem}{\ref{thm::fwl-se}}
By Corollary \ref{corollary::fwl-residuals}, the full regression and partial regression have the same residual vector, denoted by $\hat{ \varepsilon}$. Therefore, 
$\hat{\Omega}_\textsc{ehw} = \tilde{\Omega}_\textsc{ehw} = \text{diag}\{ \hat{\varepsilon}^2\}$ in the EHW covariance estimators. 

Based on the full regression, define $\hat{\sigma}^2 = \| \hat{ \varepsilon} \|_2^2/(n-k-l)$.  Then $\hat{V}$ equals the $(2,2)$th block of
$
  \hat{\sigma}^2 (X^{\T} X)^{-1} ,
$
and $\hat{V}_{\textsc{ehw}}$ equals the $(2,2)$th block of
$
  (X^{\T} X)^{-1} X^{\T} \hat{\Omega}_\textsc{ehw} X (X^{\T}X)^{-1}.
$ 

Based on the partial regression, define $\tilde{\sigma}^2 = \| \hat{\varepsilon}\|_2^2/(n-l)$.  Then $
\tilde{V} =   \tilde{\sigma}^2 ( \tilde{X}_2 ^{\T} \tilde{X}_2)^{-1} 
$  and $
\tilde{V}_{\textsc{ehw}} = ( \tilde{X}_2 ^{\T} \tilde{X}_2)^{-1}  \tilde{X}_2^{\T}  \tilde{\Omega}_\textsc{ehw} \tilde{X}_2  ( \tilde{X}_2 ^{\T} \tilde{X}_2)^{-1}. 
$

Let $\hat{\sigma}^2 = \| \hat{ \varepsilon} \|^2/(n-k-l)$ and $\tilde{\sigma}^2 = \|  \hat{ \varepsilon} \|^2/(n-l)$ be the common variance estimators. They are identical up to the degrees of freedom correction. 
Under homoskedasticity, the covariance estimator for $\hat{\beta}_2$ is the $(2,2)$th block of $\hat{\sigma}^2 (X ^{\T} X)^{-1}$, that is, $ \hat{\sigma}^2   S_{22} = \hat{\sigma}^2  ( \tilde{X}_2 ^{\T} \tilde{X}_2 )^{-1}$ by Lemma \ref{lem:blockinverse}, which is identical to the covariance estimator for $\tilde{\beta}_2$ up to the degrees of freedom correction.

The EHW covariance estimator from the full regression is the $(2,2)$ block of
$
 \hat{A}\hat{\Omega}_\textsc{ehw} \hat{A}^{\T},
$
where  
\begin{eqnarray*}
\hat{A} &=& (X ^{\T} X)^{-1} X ^{\T} \\
&=& \begin{pmatrix}
* & * \\
- (\tilde{X}_{2}^{\T}\tilde{X}_{2})^{-1}X_{2}^{\T} X_1(X_{1}^{\T}X_{1})^{-1} & (\tilde{X}_{2}^{\T}\tilde{X}_{2})^{-1}
\end{pmatrix}
\begin{pmatrix}
X_{1}^{\T}\\
X_{2}^{\T}
\end{pmatrix} \\
&=& \begin{pmatrix}
*\\
 -  ( \tilde{X}_2 ^{\T} \tilde{X}_2 )^{-1} X_2 ^{\T} H_1 + ( \tilde{X}_2 ^{\T}  \tilde{X}_2 )^{-1}X_2 ^{\T}
\end{pmatrix} \\
&=&\begin{pmatrix}
*\\
 ( \tilde{X}_2 ^{\T} \tilde{X}_2 )^{-1}\tilde{X}_2 ^{\T}
\end{pmatrix},
\end{eqnarray*}
by Lemma \ref{lem:blockinverse}. I omit the  $*$ terms because they do not affect the final calculation.
Define $\tilde{A}_2 =   ( \tilde{X}_2 ^{\T} \tilde{X}_2 )^{-1}\tilde{X}_2^{\T}$, and then
$$
\hat{V}_{\textsc{ehw}} 
= \tilde{A}_2 \hat{\Omega}_\textsc{ehw} \tilde{A}_2^{\T}
=  \tilde{A}_2 \tilde{\Omega}_\textsc{ehw}  \tilde{A}_2 ^{\T},
$$
which equals the EHW covariance estimator $\tilde{V}_{\textsc{ehw}}$ from the partial regression. 
\end{myproof}

\section{Gram--Schmidt orthogonalization, QR decomposition, and  computation of OLS}

 When the regressors are orthogonal, the coefficients from the long and short regressions are identical, which simplifies the calculation and theoretical discussion. 

\begin{corollary}
\label{cor:orthogonal}If $X_{1}^{\T}X_{2}=0$, i.e., the columns
of $X_{1}$ and $X_{2}$ are orthogonal, then $\tilde{X}_{2}=X_{2}$
and $\hat{\beta}_{2}=\tilde{\beta}_{2}.$ 
\end{corollary}

\begin{myproof}{Corollary}{\ref{cor:orthogonal}}
We can directly prove Corollary \ref{cor:orthogonal} by verifying that $X^{\T} X$ is block diagonal. 

Alternatively, 
Corollary \ref{cor:orthogonal} follows from
\[
\tilde{X}_{2}=(I_{n}-H_{1})X_{2}=X_{2}-X_{1}(X_{1}^{\T}X_{1})^{-1}X_{1}^{\T}X_{2}=X_{2},
\]
 and Theorem \ref{thm:fwl}.
\end{myproof}

The simple fact of Corollary \ref{cor:orthogonal} motivates us to orthogonalize the columns of the covariate matrix $X$, which in turn gives the famous QR decomposition in linear algebra. Interestingly, the \ri{lm} function in \ri{R} uses the QR decomposition to compute the OLS estimator. To facilitate the discussion, I will use the notation
$$
 \hat{\beta}_{V_2|V_1} V_1
$$
as the linear projection of the vector $V_2 \in \mathbb{R}^n$ onto the vector $V_1 \in \mathbb{R}^n$, where $\hat{\beta}_{V_2|V_1}  =  V_2^{\T} V_1  /   V_1^{\T} V_1 $. This is from the univariate OLS of $V_2$ on $ V_1$ (recall Chapter \ref{section::1d-comments}).

With a slight abuse of notation, partition the covariate matrix into column vectors $X = (X_1, \ldots, X_p)$. The goal is to find orthogonal vectors $(U_1, \ldots, U_p)$ that generate the same column space as $X$. Start with 
$$
X_1 = U_1.
$$ 
Regress $X_2$ on $U_1$ to obtain the fitted and residual vector  
$$
X_2 = \hat{\beta}_{X_2|U_1} U_1  + U_2; 
$$
by OLS, $U_1$ and $U_2$ must be orthogonal. 
Regress $X_3$ on $(U_1, U_2)$ to obtain the fitted and residual vector  
 $$
X_3 = \hat{\beta}_{X_3|U_1} U_1 +   \hat{\beta}_{X_3|U_2} U_2 + U_3; 
$$
by Corollary \ref{cor:orthogonal}, the OLS reduces to two univariate OLS by $U_1 \perp U_2$ and ensures that $U_3$ is orthogonal to both $U_1$ and $U_2$.  This justifies the notation $\hat{\beta}_{X_3|U_1} $ and $\hat{\beta}_{X_3|U_2}$. Continue this procedure to the last column vector:
$$
X_p = \sum_{j=1}^{p-1} \hat{\beta}_{X_p|U_j} U_j + U_p;
$$
by OLS, $U_p$ is orthogonal to all $U_j$ $(j=1,\ldots, p-1)$. We further normalize the $U$ vectors to have unit length:
$$
Q_j = U_j / \|  U_j \|,\quad (j= 1, \ldots, p).
$$
The whole process is called the Gram--Schmidt orthogonalization, which is essentially the sequential OLS fits. This process generates an $n\times p$ matrix with orthonormal column vectors 
$$
Q = (Q_1, \ldots, Q_p). 
$$
More interestingly, the column vectors of $X$ and $Q$ can linearly represent each other because 
\begin{eqnarray*}
X &=&  (X_1, \ldots, X_p) \\
&=& (U_1, \ldots, U_p) \begin{pmatrix}
1 & \hat{\beta}_{X_2|U_1} & \hat{\beta}_{X_3|U_1} &\cdots & \hat{\beta}_{X_p|U_1} \\
0& 1& \hat{\beta}_{X_3|U_2} &\cdots & \hat{\beta}_{X_p|U_2} \\
\vdots & \vdots & \vdots & \cdots & \vdots \\
0&0&0& \cdots & 1
\end{pmatrix} \\
&=& Q
\text{diag}\{ \|  U_j \| \}_{j=1}^p
\begin{pmatrix}
1 & \hat{\beta}_{X_2|U_1} & \hat{\beta}_{X_3|U_1} &\cdots & \hat{\beta}_{X_p|U_1} \\
0& 1& \hat{\beta}_{X_3|U_2} &\cdots & \hat{\beta}_{X_p|U_2} \\
\vdots & \vdots & \vdots & \cdots & \vdots \\
0&0&0& \cdots & 1
\end{pmatrix}  . 
\end{eqnarray*}
We can verify that the product of the second and the third matrix is an upper triangular matrix, denoted by $R$. By definition, the $j$th diagonal element of $R$ equals $\|  U_j \|$, and the $(j, j')$th element of $R$ equals $\|  U_j \|  \hat{\beta}_{X_{j'}|U_j} $ for $j' > j.$
Therefore, we can decompose $X$ as
$$
X = QR
$$
where $Q$ is an $n\times p$ matrix with orthonormal columns and $R$ is a $p\times p$ upper triangular matrix. This is called the QR decomposition of $X$.

Most software packages, for example, \texttt{R}, do not calculate the inverse of $X^{\T} X$ directly. Instead, they first find the QR decomposition of $X = QR$. Since the Normal equation simplifies to 
\begin{align*}
X^{\T}X\hat{\beta} & =X^{\T}Y,\\
R^{\T}Q^{\T}QR\hat{\beta} & =R^{\T}Q^{\T}Y,\\
R\hat{\beta} & =Q^{\T}Y,
\end{align*}
they then backsolve the last linear equation since $R$ is upper triangular.

In \ri{R}, the \ri{qr} function returns the QR decomposition of a matrix. 

\begin{lstlisting}
> X   = matrix(rnorm(7*3), 7, 3)
> X
            [,1]       [,2]       [,3]
[1,] -0.57231223  0.1196325  0.8087505
[2,] -1.76090225  1.0627631  1.8170361
[3,] -0.04144281 -0.2904749 -1.8372247
[4,] -0.37627821  0.4476932 -0.9629320
[5,] -1.40848027  0.2735408 -0.8047917
[6,]  1.84878518  0.7290005  1.2688929
[7,]  0.06432856  0.2256284  0.3972229
> qrX = qr(X)
> qr.Q(qrX)
            [,1]        [,2]        [,3]
[1,] -0.19100878 -0.03460617  0.30340481
[2,] -0.58769981 -0.60442928  0.23753900
[3,] -0.01383151  0.21191991 -0.55839928
[4,] -0.12558257 -0.28728403 -0.62864750
[5,] -0.47007924 -0.07020076 -0.36640938
[6,]  0.61703067 -0.68778411 -0.09999859
[7,]  0.02146961 -0.16748246  0.01605493
> qr.R(qrX)
         [,1]       [,2]       [,3]
[1,] 2.996261 -0.3735673  0.0937788
[2,] 0.000000 -1.3950642 -2.1217223
[3,] 0.000000  0.0000000  2.4826186
\end{lstlisting}

If we specify \ri{qr = TRUE} in the \ri{lm} function, it will also return the QR decomposition of the covariate matrix.\footnote{The ``\texttt{0 +}'' in the code below forces the OLS to exclude the constant term.} 

\begin{lstlisting}
> Y   = rnorm(7)
> lmfit = lm(Y ~ 0 + X, qr = TRUE)
> qr.Q(lmfit$qr)
            [,1]        [,2]        [,3]
[1,] -0.43535054 -0.25679823 -0.65480400
[2,] -0.47091275 -0.13639459  0.14746444
[3,]  0.66494532  0.07725435 -0.39436265
[4,]  0.21136347 -0.78814737  0.34611820
[5,] -0.04493356 -0.40413829 -0.01156273
[6,] -0.28046504  0.10655755 -0.16193251
[7,]  0.14561808 -0.33708219 -0.49780561
> qr.R(lmfit$qr)
        X1         X2         X3
1 3.190035 -0.6964269  1.8693260
2 0.000000  2.0719787  1.9210212
3 0.000000  0.0000000 -0.9261921
\end{lstlisting}

\section{Homework problems}

\paragraph{Inverse of a block matrix}\label{hw6::inverse-block-gram}

Prove Lemma \ref{lem:blockinverse} and the following alternative form: 
$$
(X^{\T} X)^{-1} = \begin{pmatrix}
Q_{11} & Q_{12} \\
Q_{21} & Q_{22}
\end{pmatrix},
$$
where $H_2 = X_2(X_2^{\T} X_2)^{-1} X_2 ^{\T} $, $\tilde{X}_1 = (I_n - H_2) X_1$, and
\begin{eqnarray*}
Q_{11} &=& (\tilde{X}_1^{\T} \tilde{X}_1)^{-1} ,\\
Q_{12} &=& - (\tilde{X}_1^{\T} \tilde{X}_1)^{-1} X_1^{\T} X_2 (X_2^{\T} X_2)^{-1} ,\\
Q_{21} &=& Q_{12}^{\T},\\
Q_{22} &=& (X_2^{\T} X_2)^{-1}  + (X_2^{\T} X_2)^{-1} X_2^{\T} X_1 (\tilde{X}_1^{\T} \tilde{X}_1)^{-1} X_1^{\T} X_2 (X_2^{\T} X_2)^{-1}.
\end{eqnarray*}

Remark: Use the formula in Problem \ref{hwmath1::inverse-block-matrix}.

\paragraph{Residuals in the FWL Theorem}\label{hw06-fwl-residual}

Give an example in which the residual vector from the partial regression of $Y$ on $\tilde{X}_{2}$ does not equal to the residual vector from the full regression.

%

%

\paragraph{Projection matrices}\label{hw08::decompose-projection-m}

 Prove Lemma \ref{lemma::decompose-projection-matrices}.

 Remark: Use Lemma \ref{lem:blockinverse}.

\paragraph{FWL Theorem and leverage scores}\label{hw7::fwl-hat-matrix}

Consider the partitioned regression $Y = X_1 \hat{\beta}_1 + X_2 \hat{\beta}_2 + \hat{\varepsilon}$.
To obtain the coefficient $\hat{\beta}_2$, we can run two OLS fits: 
\begin{enumerate}[label=(R\arabic*), ref=R\arabic*]
\item\label{reg1-leverage}
regress $X_2$ on $X_1$ to obtain the residual $\tilde{X}_2$;
\item\label{reg2-leverage}
regress $Y$ on $\tilde{X}_2$ to obtain the coefficient, which equals $\hat{\beta}_2$ by the FWL Theorem. 
\end{enumerate}

Although partial regression \eqref{reg2-leverage} can recover the OLS coefficient, the leverage scores from  \eqref{reg2-leverage} are not the same as those from the long regression. Prove that the summation of the corresponding leverage scores from \eqref{reg1-leverage} and \eqref{reg2-leverage} equals the leverage scores from the long regression.

Remark: The leverage scores are the diagonal elements of the hat matrix from OLS fits. Chapter \ref{chapter::EHW} before mentioned them and Chapter \ref{chapter::leave-one-out} later will discuss them in more detail.

\paragraph{Another invariance property of the OLS coefficient}\label{hw7::invariance-ols-fwl}

Partition the covariate matrix as $X = (X_1, X_2)$ where $X_1 \in \mathbb{R}^{n\times k}$ and $X_2 \in \mathbb{R}^{n\times l}$. Given any $A\in \mathbb{R}^{k\times l}$, define $\tilde{X}_2 = X_2 - X_1A$. Fit two OLS:
$$
Y =   X_1 \hat{\beta}_1 +   X_2 \hat{\beta}_2  + \hat\varepsilon
$$
and
$$
Y =   X_1 \tilde{\beta}_1 +   \tilde{X}_2 \tilde{\beta}_2 + \tilde\varepsilon .
$$

Prove that
$$
\hat{\beta}_2 = \tilde{\beta}_2, \quad \hat\varepsilon = \tilde\varepsilon.
$$

Remark: You can use the result in Problem \ref{hw3::invariance-ols} to prove the result in this problem. As a special case, if we choose $A = (X_1^{\T}X_1)^{-1} X_1^{\T} X_2$ to be the coefficient matrix of the OLS fit of $X_2$ on $X_1$, then the above result ensures that $\hat{\beta}_2$ from the OLS fit of $Y$ on $X_1$ and $X_2$ equals $\tilde{\beta}_2$ from the OLS fit of $Y$ on  $X_1$ and $(I_n-H_1)X_2$, which is coherent with the FWL Theorem since $X_1^{\T} (I_n-H_1)X_2 = 0$.

\paragraph{Alternative formula for the EHW standard error}\label{hw7::alternative-ehw-se}

Consider the partition regression $Y = X_1 \hat\beta_1 + X_2 \hat\beta_2  + \hat{\varepsilon}$ with $X_1$ is an $n\times (p-1)$ matrix and  $X_2 $ is an $n$ dimensional vector. So $ \hat\beta_2$ is a scalar, and the $(p,p)$th element of $\hat{V}_\textsc{ehw}$ equals $\hat{\text{se}}_{\textsc{ehw}, 2}^2$, the squared EHW standard error for $ \hat\beta_2$. 

Define 
$$
\tilde{X}_2 = (I_n - H_1) X_2 = \begin{pmatrix}
\tilde{x}_{12}\\
\vdots \\
\tilde{x}_{n2}
\end{pmatrix} . 
$$
Prove that under Assumption \ref{assume::heteroskedasticity-lm},  we have 
\begin{eqnarray*}
\var( \hat\beta_2  ) &=& \sumn  w_i \sigma_i^2,  \\
\hat{\text{se}}_{\textsc{ehw}, 2}^2  &=&  \sumn  w_i   \hat\varepsilon_i^2
\end{eqnarray*}
where 
$$
w_i = \frac{  \tilde{x}_{i2}^2  }{   ( \sumn    \tilde{x}_{i2}^2 )^2  } .
$$

Remark: You can use Theorems \ref{thm:fwl} and \ref{thm::fwl-se} to prove the result. 
The original formula of the EHW covariance matrix has a complex form. However, using the FWL theorems, we can simplify each of the squared EHW standard errors as a weighted average of the squared residuals, or, equivalently, a simple quadratic form of the residual vector.

\paragraph{A counterexample to the Gauss--Markov Theorem}\label{hw3::counterexample-gauss-markov-heteroskedasticity}

The Gauss--Markov Theorem does not hold under the heteroskedastic linear model. This problem gives a counterexample in a simple linear model.

Assume $y_i = \beta x_i + \varepsilon_i$ without the intercept and with potentially different $\var(\varepsilon_i) = \sigma_i^2$ across $i=1,\ldots, n$. 
Consider two OLS estimators: the first OLS estimator does not contain the intercept $\hat{\beta} = \sumn x_i y_i / \sumn x_i^2$; the second OLS estimator contains the intercept $\tilde{\beta} =\sumn (x_i - \bar{x}) y_i / \sumn (x_i - \bar{x})^2 $ even though the true linear model does not contain the intercept.

The Gauss-Markov Theorem ensures that if $\sigma_i^2=\sigma^2$ for all $i$'s, then the variance of $\hat{\beta}$ is smaller than or equal to the variance of $\tilde{\beta} $. However, it does not hold when $\sigma_i^2$'s vary. 

Give a counterexample in which the variance of $\hat{\beta}$ is larger than the variance of $\tilde{\beta} $.

 \paragraph{QR decomposition of $X$ and the computation of OLS}\label{hw3::ols-qr-decomposition}
 
Verify that the $R$ matrix equals 
 $$
R = 
\begin{pmatrix}
Q_1^{\T} X_1 & Q_1^{\T} X_2 & \cdots &  Q_1^{\T} X_p  \\
0 & Q_2^{\T} X_2& \cdots &  Q_2^{\T} X_p  \\
\vdots & \vdots &  \cdots &  \vdots \\
0&0& \cdots & Q_p^{\T} X_p
\end{pmatrix}.
$$

Based on the QR decomposition of $X$, prove that 
$$
H = QQ^{\T},
$$ 
and the its $(i,i)$the diagonal element $h_{ii}$ equals the squared length of the $i$-th row of $Q$.

 \paragraph{Uniqueness of the QR decomposition}\label{hw6::uniqueQR}
 
 Prove that if $X$ has linearly independent column vectors, the QR decomposition must be unique. That is, if $X = QR = Q_1 R_1$ where $Q$ and $Q_1$ have orthonormal columns and $R$ and $R_1$ are upper triangular, then we must have
 $$
 Q = Q_1,\quad  R = R_1.
 $$

\chapter{Applications of the Frisch--Waugh--Lovell Theorem}\label{chapter::FWL-application}
 
The Frisch--Waugh--Lovell (FWL) theorem has many applications. I will highlight some of them in this chapter.

\section{Centering regressors}

\subsection{Intercept and centering regressors}

As a special case, partition the covariate matrix into $X=(X_1,X_2)$ with  $X_{1}=1_{n}$. This is the usual case including the constant as the first regressor.   
The projection matrix 
\[
H_{1}=1_{n}(1_{n}^{\T}1_{n})^{-1}1_{n}^{\T} 
=n^{-1}1_{n}1_{n}^{\T} = \begin{pmatrix}
n^{-1} & \cdots & n^{-1} \\
\vdots && \vdots \\
n^{-1} &\cdots & n^{-1} 
\end{pmatrix} \equiv A_n
\]
 contains $n^{-1}$'s as its elements, and 
 $$
 C_{n}=I_{n}-n^{-1}1_{n}1_{n}^{\T}
 $$
is the projection matrix orthogonal to $1_{n}$. 
The matrices $A_n$ and $C_n$ have convenient properties:
\begin{enumerate}[label=(P\arabic*), ref=P\arabic*]
\item
Multiplying any vector by $A_n$ is equivalent to obtaining the average of its components.
\item
Multiplying any vector 
by $C_{n}$ is equivalent to centering that vector.
\end{enumerate}
For example, 
\[
A_n Y = \left(\begin{array}{c}
 \bar{y}\\
\vdots\\
 \bar{y}
\end{array}\right) =  \bar{y} 1_n ,
\]
and
\[
C_{n}Y=\left(\begin{array}{c}
y_{1}-\bar{y}\\
\vdots\\
y_{n}-\bar{y}
\end{array}\right).
\]
More generally, multiplying any matrix by $A_n$ is equivalent to averaging each column, and multiplying any matrix by $C_{n}$ is equivalent to centering each
column of that matrix. For example,
\[
A_n X_2 = \left(\begin{array}{c}
 \bar{x}_{2}^{\T}\\
\vdots\\
 \bar{x}_{2}^{\T}
\end{array}\right) = 1_n  \bar{x}_{2}^{\T} , 
\]
and
\[
C_{n}X_{2}=\left(\begin{array}{c}
x_{12}^{\T}-\bar{x}_{2}^{\T}\\
\vdots\\
x_{n2}^{\T}-\bar{x}_{2}^{\T}
\end{array}\right),
\]
where $X_{2}$ contains row vectors $x_{12}^{\T},\ldots,x_{n2}^{\T}$ with average $\bar{x}_2 = n^{-1} \sumn x_{i2}$. The FWL Theorem implies that the coefficient of $X_2$ in the OLS fit of $Y$ on $(1_{n},X_{2})$ equals the coefficient of $C_{n}X_{2}$ 
 in the OLS fit of $C_{n}Y$ on $C_{n}X_{2}$,
that is, the OLS fit of the centered response vector on the column-wise
centered $X_{2}.$ An immediate consequence is that if each column is centered in the design matrix, then to obtain the OLS coefficients, it does not matter whether to
include the column $1_{n}$ or not. 

The centering matrix $C_{n}$ has another property: 
\begin{enumerate}[label=(P\arabic*), ref=P\arabic*, start = 3]
\item
The quadratic
form of $C_{n}$ equals the sample variance multiplied by $n-1$. 
\end{enumerate}
For example,
\begin{eqnarray*}
Y^{\T}C_{n}Y&=&Y^{\T} C_{n}^{\T} C_{n}Y \\
&=& (y_{1}-\bar{y},\ldots, y_{n}-\bar{y})  \left(\begin{array}{c}
y_{1}-\bar{y}\\
\vdots\\
y_{n}-\bar{y}
\end{array}\right) \\
&=& \sumn(y_{i}-\bar{y})^{2} \\
&=& (n-1) \hat{\sigma}^2_y,
\end{eqnarray*}
where $ \hat{\sigma}^2_y $ is the sample variance of the outcomes. 
For an $n\times p$ matrix $X$, 
\begin{align*}
X^{\T}C_{n}X 
 & =\left(\begin{array}{c}
X_{1}^{\T}\\
\vdots\\
X_{p}^{\T}
\end{array}\right) C_{n}  \left(\begin{array}{ccc}
X_{1} & \cdots & X_{p}\end{array}\right)\\
&= \left(\begin{array}{ccc}
X_1^{\T} C_n X_1 & \cdots & X_1^{\T} C_n X_p \\
\vdots &  & \vdots\\
X_p^{\T} C_n X_1  & \cdots & X_p^{\T} C_n X_p 
\end{array}\right) \\ 
 & =(n-1)\left(\begin{array}{ccc}
\hat{\sigma}_{11} & \cdots & \hat{\sigma}_{1p}\\
\vdots &  & \vdots\\
\hat{\sigma}_{p1} & \cdots & \hat{\sigma}_{pp}
\end{array}\right),
\end{align*}
where 
$$
\hat{\sigma}_{j_{1}j_{2}}=(n-1)^{-1}\sumn(x_{ij_{1}}-\bar{x}_{\cdot j_{1}})(x_{ij_{2}}-\bar{x}_{\cdot j_{2}})
$$
is the sample covariance between $X_{j_{1}}$ and $X_{j_{2}}$. 
So $(n-1)^{-1} X^{\T} C_n X$ equals the sample covariance matrix of $X$. 
For these reason, I choose the notation $C_n$ with ``$C$'' for both ``centering'' and ``covariance.''

\subsection{Dummy variables and centering regressors within groups}

As another important special case, $X_{1}$ contains the dummies for
a discrete variable, for example, the indicators for different treatment
levels or groups. 
See Example \ref{eg::anova-H} for the background. 
With $k$ groups, $X_{1}$ can take the following
two forms:
\begin{eqnarray}\label{eq::discrete-regressor-dummy}
X_{1}=\left(\begin{array}{cccc}
1 & 1 & \cdots & 0\\
\vdots & \vdots &  & \vdots\\
1 & 1 & \cdots & 0\\
\vdots & \vdots &  & \vdots\\
1 & 0 & \cdots & 1\\
\vdots & \vdots &  & \vdots\\
1 & 0 &  & 1\\
1 & 0 &  & 0\\
\vdots & \vdots &  & \vdots\\
1 & 0 & \cdots & 0
\end{array}\right)_{n\times k}\qquad\text{or}\qquad X_{1}=\left(\begin{array}{ccc}
1 & \cdots & 0\\
\vdots &  & \vdots\\
1 & \cdots & 0\\
\vdots &  & \vdots\\
0 & \cdots & 1\\
\text{\ensuremath{\vdots}} &  & \vdots\\
0 & \cdots & 1
\end{array}\right)_{n\times k},
\end{eqnarray}
where the first form of $X_1$ contains $1_{n}$ and $k-1$ dummy variables,
and the second form of $X_1$ contains $k$ dummy variables. In both forms
of $X_{1}$, the observations are sorted according to the group indicators.
If we regress $Y$ on $X_{1}$, the residual vector is 
\begin{eqnarray}\label{eq::centeringbyK}
Y - 
\left(\begin{array}{c}
 \bar{y}_{[1]}\\
\vdots\\
 \bar{y}_{[1]}\\
\vdots\\
 \bar{y}_{[k]}\\
\vdots\\
 \bar{y}_{[k]}
\end{array}\right),
\end{eqnarray}
where $\bar{y}_{[1]},\ldots,\bar{y}_{[k]}$ are the averages of the
outcomes within groups $1,\ldots,k$. Effectively, we center $Y$ by
group-specific means. Similarly, if we regress $X_{2}$ on $X_{1}$,
we center each column of $X_{2}$ by the group-specific means. Let
$Y^{\textsc{c}}$ and $X_{2}^{\textsc{c}}$ be the centered response
vector and design matrix. The FWL Theorem implies that the OLS coefficient
of $X_{2}$ in the long regression of $Y$ on $(X_1, X_2)$ equals the OLS coefficient of $X_{2}^{\textsc{c}}$
in the partial regression of $Y^{\textsc{c}}$ on $X_{2}^{\textsc{c}}$.
When $k$ is large, running the OLS with centered variables can reduce
the computational cost. 
See Problem \ref{hw7::lfe-panel} for an application in econometrics.

\section{Partial correlation coefficient and Simpson's paradox\label{sec:Partial-correlation-coefficient}}

The sample Pearson correlation coefficient between $n$ observations of two
scalars $(x_{i},y_{i})_{i=1}^n $,
\[
 \hat{\rho}_ {yx}=\frac{\sumn(x_{i}-\bar{x})(y_{i}-\bar{y})}{\sqrt{\sumn(x_{i}-\bar{x})^{2}}\sqrt{\sumn(y_{i}-\bar{y})^{2}}},
\]
measures the linear relationship between $x$ and $y$. How do we
measure the linear relationship between $x$ and $y$ after controlling
for some other variables $w\in\mathbb{R}^{k-1}$? Intuitively, we
can measure it with the sample Pearson correlation coefficient based on the
residuals from the following two OLS fits:
\begin{enumerate}[label=(R\arabic*), ref=R\arabic*]
\item \label{enu:partialols1}run OLS of $Y$ on $(1,W)$ and obtain residual
vector $\hat{\varepsilon}_{y}$ and residual sum of squares $\textsc{rss}_{y}$;
\item \label{enu:partialols2}run OLS of $X$ on $(1,W)$ and obtain residual
vector $\hat{\varepsilon}_{x}$ and residual sum of squares $\textsc{rss}_{x}$.
\end{enumerate}
With $\hat{\varepsilon}_{y}$ and $\hat{\varepsilon}_{x}$, we can
define the sample partial correlation coefficient between $x$ and $y$ given
$w$ as 
\begin{equation}
\label{eq::sample-partial-CC}
 \hat{\rho}_ {yx\mid w}=\frac{\sumn\hat{\varepsilon}_{x,i}\hat{\varepsilon}_{y,i}}{\sqrt{\sumn\hat{\varepsilon}_{x,i}^{2}}\sqrt{\sumn\hat{\varepsilon}_{y,i}^{2}}}.
\end{equation}
 In \eqref{eq::sample-partial-CC}, we do not center the residuals because they
have zero sample means due to the inclusions of the intercepts in the
OLS fits (R\ref{enu:partialols1}) and (R\ref{enu:partialols2}). The sample partial correlation coefficient determines the coefficient of $\hat{\varepsilon}_{x}$ in the OLS fit  of $\hat{\varepsilon}_{y}$ on $\hat{\varepsilon}_{x}$:
\begin{eqnarray}
\label{eq::galtonianformula-multiple}
\hat{\beta}_{yx\mid w}=\frac{\sumn\hat{\varepsilon}_{x,i}\hat{\varepsilon}_{y,i}}{\sumn\hat{\varepsilon}_{x,i}^{2}}= \hat{\rho}_ {yx\mid w}\sqrt{\frac{\sumn\hat{\varepsilon}_{y,i}^{2}}{\sumn\hat{\varepsilon}_{x,i}^{2}}}= \hat{\rho}_ {yx\mid w}\frac{\hat{\sigma}_{y \mid  w}}{\hat{\sigma}_{x \mid  w}},
\end{eqnarray}
where $\hat{\sigma}_{y \mid  w}^{2}=\textsc{rss}_{y}/(n-k)$ and
$\hat{\sigma}_{x \mid  w}^{2}=\textsc{rss}_{x}/(n-k)$ are the
variance estimators based on regressions (R\ref{enu:partialols1}) and
(R\ref{enu:partialols2}) motivated by the Gauss--Markov model.  Based on the FWL Theorem, $\hat{\beta}_{yx\mid w}$
equals the OLS coefficient of $X$ in the long regression of $Y$
on $(1,X,W).$ Therefore,  \eqref{eq::galtonianformula-multiple} is the Galtonian formula for multiple regression, which is analogous to that for univariate regression \eqref{eq::galtonian-formula-1}.

To investigate the relationship between $y$ and $x$, different researchers
may run different regressions. One may run OLS of $Y$ on $(1, X ,W)$, and the other may run OLS of $Y$ on $(1, X ,W')$,
where $W'$ is a subset of $W$. Let $\hat{\beta}_{yx\mid w}$ be
the coefficient of $X$ in the first regression, and let $\hat{\beta}_{yx\mid w'}$
be the coefficient of $X$ in the second regression. Mathematically,
it is possible that these two coefficients have different signs, which
is called {\it Simpson's paradox}.\footnote{The usual form of Simpson's paradox is in terms of a $2\times 2\times 2$ table with all binary variables. Here we focus on its continuous version.} It is a paradox because we expect both
coefficients to measure the ``impact'' of $X$ on $Y$. Because these two coefficients have the same signs as the partial
correlation coefficients $ \hat{\rho}_ {yx\mid w}$ and $ \hat{\rho}_ {yx\mid w'}$, Simpson's
paradox is equivalent to 
\[
 \hat{\rho}_ {yx\mid w} \hat{\rho}_ {yx\mid w'}<0.
\]
To simplify the presentation, we discuss the special case with $w$ being a scalar and $w'$
being an empty set. Simpson's
paradox is then equivalent to 
\[
 \hat{\rho}_ {yx\mid w} \hat{\rho}_ {yx}<0.
\]
Theorem \ref{thm::sample-partialcorr-corr} below gives an expression that links $ \hat{\rho}_ {yx\mid w}$ and $ \hat{\rho}_ {yx}$.

\begin{theorem}\label{thm::sample-partialcorr-corr}
For $Y,X,W\in\mathbb{R}^{n}$, we have 
\[
 \hat{\rho}_ {yx\mid w}=\frac{ \hat{\rho}_ {yx}- \hat{\rho}_ {yw} \hat{\rho}_ {xw}}{\sqrt{1- \hat{\rho}_ {yw}^{2}}\sqrt{1- \hat{\rho}_ {xw}^{2}}}.
\]
\end{theorem}

\begin{figure}
\centering
\includegraphics[width = 0.9\textwidth]{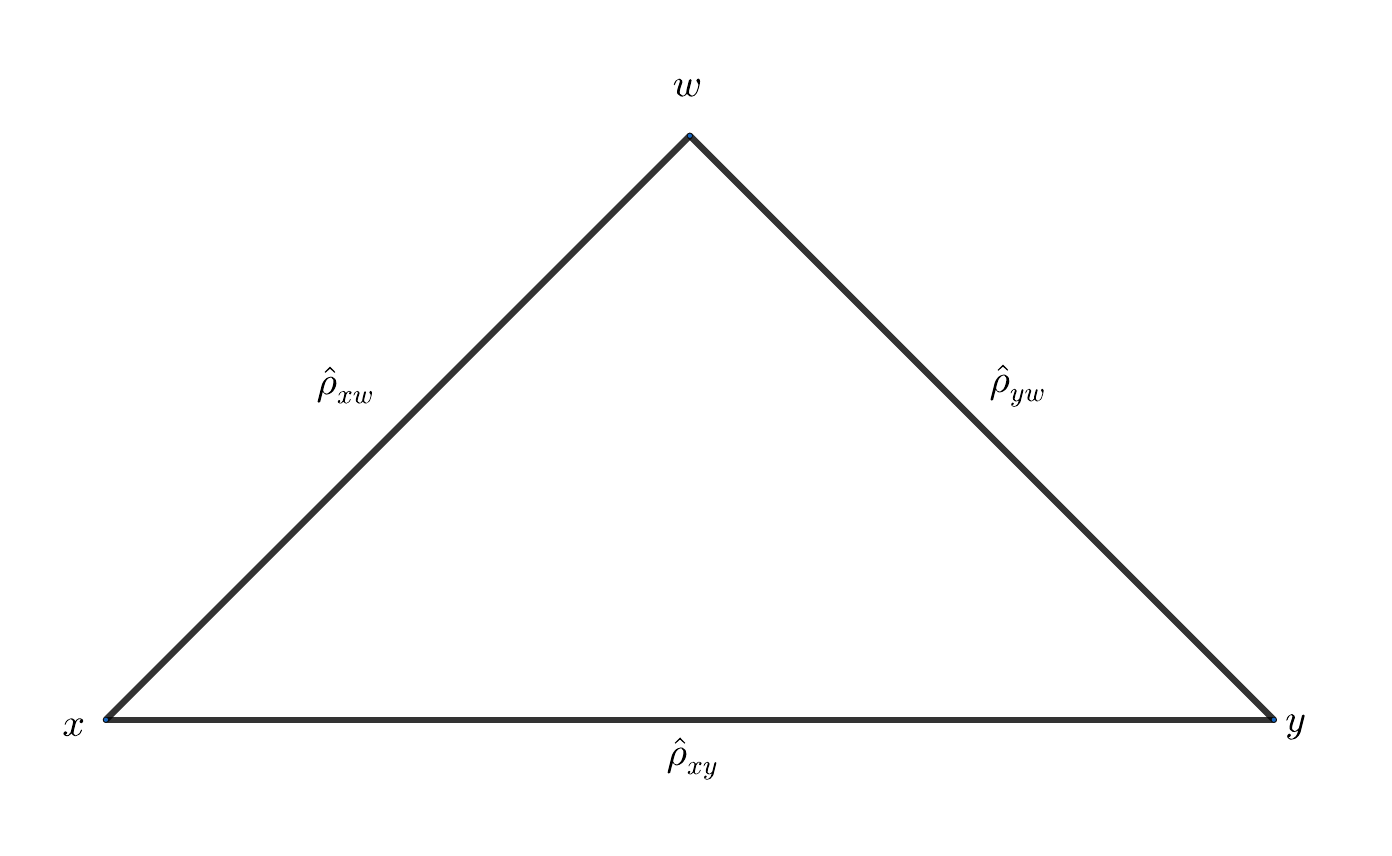}
\caption{Correlations among three variables}\label{fig::correlations3}
\end{figure} 
 
The proof of Theorem \ref{thm::sample-partialcorr-corr} is purely algebraic, so I leave it as Problem \ref{hw7::sample-partial-correlation}. Theorem \ref{thm::sample-partialcorr-corr} states that we can obtain the sample partial correlation coefficient based on the three pairwise correlation coefficients. Figure \ref{fig::correlations3} illustrates the interplay among three variables. In particular, the correlation between $x$ and $y$ is due to two ``pathways'': the one acting through $w$ and the one acting independent of $w$. The first path way is related to the product term $ \hat{\rho}_ {yw} \hat{\rho}_ {xw}$, and the second pathway is related to $ \hat{\rho}_ {yx\mid w}$. This gives some intuition for Theorem \ref{thm::sample-partialcorr-corr}.

Based on data $(y_{i},x_{i},w_{i})_{i=1}^{n}$, we can compute the
sample correlation matrix
\[
 \left(\begin{array}{ccc}
1 &  \hat{\rho}_ {yx} &  \hat{\rho}_ {yw}\\
 \hat{\rho}_ {xy} & 1 &  \hat{\rho}_ {xw}\\
 \hat{\rho}_ {wy} &  \hat{\rho}_ {wx} & 1
\end{array}\right),
\]
which is symmetric and positive semi-definite. 
Simpson's paradox happens if and only if
\[
 \hat{\rho}_ {yx}( \hat{\rho}_ {yx}- \hat{\rho}_ {yw} \hat{\rho}_ {xw})<0 ,
 \]
 which is equivalent to
 \[  \hat{\rho}_ {yx}^{2}< \hat{\rho}_ {yx} \hat{\rho}_ {yw} \hat{\rho}_ {xw}.
\]

We can observe Simpson's Paradox in the following simulation. 

\begin{lstlisting}
> n  = 1000
> w  = rbinom(n, 1, 0.5)
> x1 = rnorm(n, -1, 1)
> x0 = rnorm(n, 2, 1)
> x  = ifelse(w, x1, x0)
> y  = x + 6*w + rnorm(n)
> fit.xw = lm(y ~ x + w)$coef
> fit.x  = lm(y ~ x)$coef
> fit.xw 
(Intercept)           x           w 
 0.05655442  0.97969907  5.92517072 
> fit.x 
(Intercept)           x 
  3.6422978  -0.3743368 
\end{lstlisting}

\begin{figure}[ht]
\centering
\includegraphics[width = 0.8\textwidth]{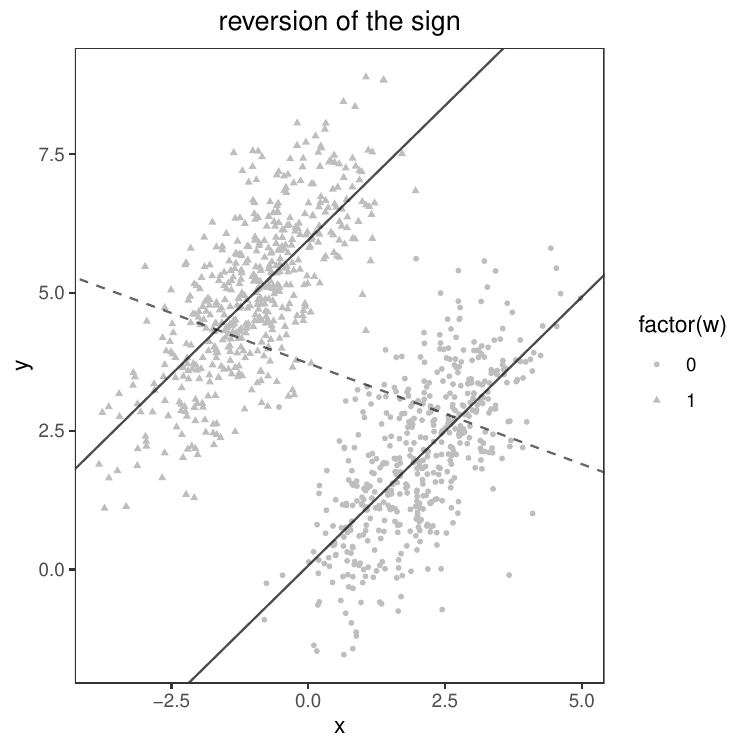}
\caption{An Example of Simpson's Paradox. The two solid regression lines are fitted separately using the data from two groups, and the dash regression line is fitted using the pooled data. }\label{fig::simpsonparadoxexample}
\end{figure}

Because $w$ is binary, we can plot $(x,y)$ in each group of $w=1$ and $w=0$ in Figure \ref{fig::simpsonparadoxexample}. In both groups, $y$ and $x$ are positively associated with positive regression coefficients; but in the pooled data, $y$ and $x$ are negatively associated with a negative regression coefficient. 

\section{Hypothesis testing and analysis of variance}\label{sec::fwl-anova}

Partition $X$ and $\beta$ as
\[
X=\left(\begin{array}{cc}
X_{1} & X_{2}\end{array}\right),\qquad\beta=\left(\begin{array}{c}
\beta_{1}\\
\beta_{2}
\end{array}\right),
\]
where $X_{1}\in\mathbb{R}^{n\times k},X_{2}\in\mathbb{R}^{n\times l},\beta_{1}\in\mathbb{R}^{k}$
and $ \beta_2 \in  \mathbb{R}^{l}.$ We are often interested in testing
\[
H_{0}:\beta_{2}=0
\]
in the long regression
\begin{equation}
Y=X\beta+\varepsilon=X_{1}\beta_{1}+X_{2}\beta_{2}+\varepsilon,\label{eq:longreg}
\end{equation}
where $\varepsilon\sim\N(0,\sigma^{2}I_{n}).$ If $H_{0}$ holds,
then $X_{2}$ is redundant and a short regression suffices:
\begin{equation}
Y=X_{1}\beta+\varepsilon. \label{eq:shortreg}
\end{equation}
This is a special case of testing $C\beta=0$
with 
\[
C=\left(\begin{array}{cc}
0_{l\times k} & I_{l\times l}\end{array}\right).
\]
As discussed in Chapter \ref{chapter::normal-linear-model}, we can use 
\[
\hat{\beta}_{2}\sim\N(0,\sigma^{2}S_{22})
\]
with $S_{22} = (\tilde{X}_{2}^{\T}\tilde{X}_{2})^{-1}$ being the $(2,2)$th block of $(X^{\T}X)^{-1}$ by Lemma \ref{lem:blockinverse}, to construct the Wald-type statistic for hypothesis testing:
\begin{eqnarray*}
F_{\text{Wald}}  
&=& \frac{\hat{\beta}_{2}^{\T}(S_{22})^{-1}\hat{\beta}_{2}}{l\hat{\sigma}^{2}}
\\
&=& \frac{\hat{\beta}_{2}^{\T}  \tilde{X}_{2}^{\T}\tilde{X}_{2}  \hat{\beta}_{2}}{l\hat{\sigma}^{2}}
\\
&\sim & F_{l,n-p}.
\end{eqnarray*}

Now I will discuss testing  $H_{0}$ from an alternative perspective based on comparing the residual sum of squares in the long regression (\ref{eq:longreg}) and the short regression (\ref{eq:shortreg}). This technique is called the analysis of variance (ANOVA), pioneered by R. A. Fisher in the design and analysis of experiments. Intuitively, if $\beta_{2}=0$, then the residual vectors from the long regression (\ref{eq:longreg})
and the short regression (\ref{eq:shortreg}) should not be ``too
different.'' However, with the error term $\varepsilon$, these residuals are
random, then the key is to quantify the magnitude of the difference.
Define 
\[
\textsc{rss}_{\text{long}}=Y^{\T}(I_{n}-H)Y
\]
and 
\[
\textsc{rss}_{\text{short}}=Y^{\T}(I_{n}-H_{1})Y
\]
as the residual sums of squares from the long and short regressions,
respectively. By the definition of OLS, it must be true that
$$
\textsc{rss}_{\text{long}}\leq\textsc{rss}_{\text{short}} 
$$
and therefore, 
\begin{equation}
\textsc{rss}_{\text{short}}-\textsc{rss}_{\text{long}}
=Y^{\T}(H-H_{1})Y
\geq 0.\label{eq:diffvar}
\end{equation}
To understand the magnitude of the change in the residual sum of squares,
we can standardize the above difference and define
\[
F_\textsc{anova} = \frac{(\textsc{rss}_{\text{short}}-\textsc{rss}_{\text{long}})/l}{\textsc{rss}_{\text{long}}/(n-p)},
\]
In the definition of the above statistic, $l$ and $n-p$ are the
degrees of freedom to make the mathematics more elegant, but
they do not change the discussion fundamentally. The denominator of $F_\textsc{anova} $
is $\hat{\sigma}^{2}$, so we can also write it as 
\begin{equation}
F_\textsc{anova} =\frac{\textsc{rss}_{\text{short}}-\textsc{rss}_{\text{long}}}{l\hat{\sigma}^{2}}.\label{eq:Fformula}
\end{equation}

Theorem \ref{thm::f-two-proofs} below states that these two perspectives yield an identical test statistic.

\begin{theorem}\label{thm::f-two-proofs}
Under Assumption \ref{assume::nlm}, if $\beta_{2}=0$, then $F_\textsc{anova} \sim F_{l,n-p}.$
In fact, $F_\textsc{anova} =F_\textup{Wald}  $, which is a numerical result without Assumption \ref{assume::nlm}. 
\end{theorem}

I divide the proof into two parts. The first part derives the exact distribution of $F_\textsc{anova}$ under the Normal linear model. It relies on the following lemma on the basic properties of the projection matrices. I relegate its proof to Problem \ref{problem07::projection-matrices}. 

\begin{lemma}\label{lemma::projection-m-F}
We have
$$
HX_{1}=X_{1}, \quad HX_2 = X_2, \quad  HH_{1} =H_{1},\quad H_{1}H=H_{1}
$$
Moreover, 
$H-H_{1}$ is a projection matrix of rank $p-k=l$, 
$I_{n}-H$ is a projection matrix of rank $n-p$, and they are orthogonal:
\begin{equation}
(H-H_{1})(I_{n}-H)=0.\label{eq:basic4}
\end{equation}
\end{lemma}

\begin{myproof}{Theorem}{\ref{thm::f-two-proofs} (Part I)}

The residual vector from the long regression equals 
$$
\hat{\varepsilon}=(I_{n}-H)Y=(I_{n}-H)(X\beta+\varepsilon)=(I_{n}-H)\varepsilon ,
$$
so the residual sum of squares equals 
\[
\textsc{rss}_{\text{long}}=\hat{\varepsilon}^{\T}\hat{\varepsilon}=\varepsilon^{\T}(I_{n}-H)\varepsilon . 
\]
Since $\beta_{2}=0$, the residual vector from the short regression
equals
$$
\tilde{\varepsilon}=(I_{n}-H_{1})Y=(I_{n}-H_{1})(X_{1}\beta_{1}+\varepsilon)=(I_{n}-H_{1})\varepsilon , 
$$
so the residual sum of squares equals
\[
\textsc{rss}_{\text{short}}=\tilde{\varepsilon}^{\T}\tilde{\varepsilon}=\varepsilon^{\T}(I_{n}-H_{1})\varepsilon.
\]
Let $\varepsilon_{0}=\varepsilon/\sigma\sim\N(0,I_{n})$ be a standard
Normal random vector, then we can write $F_\textsc{anova} $ as
\begin{eqnarray}
F_\textsc{anova}  &=& \frac{\varepsilon^{\T}(H-H_{1})\varepsilon/l}{\varepsilon^{\T}(I_{n}-H)\varepsilon/(n-p)} \nonumber  \\
&=&\frac{\varepsilon_{0}^{\T}(H-H_{1})\varepsilon_{0}/l}{\varepsilon_{0}^{\T}(I_{n}-H)\varepsilon_{0}/(n-p)} \nonumber\\
&=&\frac{\|(H-H_{1})\varepsilon_{0}\|^{2}/l}{\|(I_{n}-H)\varepsilon_{0}\|^{2}/(n-p)}.\label{eq:FratioofQ}
\end{eqnarray}

We have the following joint Normality using the basic fact
(\ref{eq:basic4}): 
\begin{align*}
\left(\begin{array}{c}
(H-H_{1})\varepsilon_{0}\\
(I_{n}-H)\varepsilon_{0}
\end{array}\right) & =\left(\begin{array}{c}
H-H_{1}\\
I_{n}-H
\end{array}\right)\varepsilon_{0}\\
 & \sim\N\left\{ \left(\begin{array}{c}
0\\
0
\end{array}\right),\left(\begin{array}{cc}
H-H_{1} & 0\\
0 & I_{n}-H
\end{array}\right)\right\} .
\end{align*}
So $(H-H_{1})\varepsilon_{0}$ and $(I_{n}-H)\varepsilon_{0}$ are
Normal with mean zero and two projection matrices $H-H_{1}$ and $I_{n}-H$
as covariances, respectively, and moreover, they are independent.
These imply that their squared lengths are chi-squared (by Theorem \ref{thm::normal-chisq} in Appendix \ref{chapter:appendix-rvs}):
\begin{eqnarray*}
\|(H-H_{1})\varepsilon_{0}\|^{2} &\sim & \chi_{l}^{2},\\
\|(I_{n}-H)\varepsilon_{0}\|^{2} &\sim & \chi_{n-p}^{2},
\end{eqnarray*} 
and they are independent. These facts, coupled with (\ref{eq:FratioofQ}),
imply that $F_\textsc{anova}  \sim F_{l,n-p}.$ 
\end{myproof}

The second part demonstrates that $F_\textsc{anova}  = F_\text{Wald}$ without assuming the Normal linear model, which gives an indirect proof for the exact distribution of $F_\textsc{anova}$ under the Normal linear model. 

\begin{myproof}{Theorem}{\ref{thm::f-two-proofs} (Part II)}
Using the FWL Theorem that $\hat{\beta}_{2}=(\tilde{X}_{2}^{\T}\tilde{X}_{2})^{-1}\tilde{X}_{2}^{\T}Y$,
we can rewrite $F_\text{Wald} $ as 
\begin{align}
F_\text{Wald} & =\frac{Y^{\T}\tilde{X}_{2}(\tilde{X}_{2}^{\T}\tilde{X}_{2})^{-1}\tilde{X}_{2}^{\T}\tilde{X}_{2}(\tilde{X}_{2}^{\T}\tilde{X}_{2})^{-1}\tilde{X}_{2}^{\T}Y}{l\hat{\sigma}^{2}}\nonumber \\
 & =\frac{Y^{\T}\tilde{X}_{2}(\tilde{X}_{2}^{\T}\tilde{X}_{2})^{-1}\tilde{X}_{2}^{\T}Y}{l\hat{\sigma}^{2}}\nonumber \\
 & =\frac{Y^{\T}\tilde{H}_{2}Y}{l\hat{\sigma}^{2}},\label{eq:Wformula}
\end{align}
recalling that  
$
\tilde{H}_{2}=\tilde{X}_{2}(\tilde{X}_{2}^{\T}\tilde{X}_{2})^{-1}\tilde{X}_{2}^{\T}
$
is the projection matrix onto the column space of $\tilde{X}_{2}.$ Therefore, $F_\textsc{anova}  = F_\text{Wald}$ follows from the basic identity $H-H_{1}=\tilde{H}_{2}$ ensured by Lemma \ref{lemma::decompose-projection-matrices}. 
\end{myproof}

We can use the \ri{anova} function in \ri{R} to compute the $F$ statistic and the $p$-value. Below I revisit the \ri{lalonde} data, first analyzed in Chapter \ref{section::normal-lm-lalonde} of this book.  The result is identical as in Section \ref{section::normal-lm-lalonde}.

 \begin{lstlisting}
> library("Matching")
> data(lalonde)
> lalonde_full  = lm(re78 ~ ., data = lalonde)
> lalonde_treat = lm(re78 ~ treat, data = lalonde)
> anova(lalonde_treat, lalonde_full)
Analysis of Variance Table

Model 1: re78 ~ treat
Model 2: re78 ~ age + educ + black + hisp + married + nodegr + re74 + 
    re75 + u74 + u75 + treat
  Res.Df        RSS Df Sum of Sq      F  Pr(>F)  
1    443 1.9178e+10                              
2    433 1.8389e+10 10 788799023 1.8574 0.04929 *
\end{lstlisting}

In fact, we can conduct an analysis of variance in a sequence of models. For example, we can supplement the above analysis with a model containing only the intercept. The function \ri{anova} works for a sequence of nested models with increasing complexities. 
\begin{lstlisting}
> lalonde1   = lm(re78 ~ 1, data = lalonde)
> anova(lalonde1, lalonde_treat, lalonde_full)
Analysis of Variance Table

Model 1: re78 ~ 1
Model 2: re78 ~ treat
Model 3: re78 ~ age + educ + black + hisp + married + nodegr + re74 + 
    re75 + u74 + u75 + treat
  Res.Df        RSS Df Sum of Sq      F   Pr(>F)   
1    444 1.9526e+10                                
2    443 1.9178e+10  1 348013456 8.1946 0.004405 **
3    433 1.8389e+10 10 788799023 1.8574 0.049286 * 
\end{lstlisting}

Overall, the treatment variable is significantly related to the outcome,  but none of the pretreatment covariates are.

\section{Homework problems}

\paragraph{FWL Theorem with an intercept}\label{hw7::fwl-intercept}

Theorem \ref{thm:fwl2} below extends Theorem \ref{thm:fwl} and highlights the inclusion of the intercept in OLS. 
Prove Theorem \ref{thm:fwl2}.

With the intercept in OLS, partition $X$ and $\beta$ as 
\[
X=\left(\begin{array}{ccc}
1_n & X_{1} & X_{2}\end{array}\right),\qquad
\beta=\left(\begin{array}{c}
\beta_0 \\
\beta_{1}\\
\beta_{2}
\end{array}\right),
\]
where $X_{1}\in\mathbb{R}^{n\times k},X_{2}\in\mathbb{R}^{n\times l},  \beta_0 \in \mathbb{R},  \beta_{1}\in\mathbb{R}^{k}$ and $\beta_2 \in  \mathbb{R}^{l}$.  Consider the {\it long regression} 
\begin{eqnarray*}
Y &=& X\hat{\beta}+\hat{\varepsilon} \\
&=& \left(\begin{array}{ccc}
1_n & X_{1} & X_{2}\end{array}\right)\left(\begin{array}{c}
\hat{\beta}_0 \\
\hat{\beta}_{1}\\
\hat{\beta}_{2}
\end{array}\right)+\hat{\varepsilon} \\
&=& 1_n \hat{\beta}_0 +  X_{1}\hat{\beta}_{1}+X_{2}\hat{\beta}_{2}+\hat{\varepsilon},
\end{eqnarray*}
 and the {\it short regression} 
\[
Y= 1_n \tilde{\beta}_0  +   X_{2}\tilde{\beta}_{2}+\tilde{\varepsilon},
\]
where $\hat{\beta}=\left(\begin{array}{c}
\hat{\beta}_0 \\ 
\hat{\beta}_{1}\\
\hat{\beta}_{2}
\end{array}\right)$ and
$
\left(\begin{array}{c}
\tilde{\beta}_{0}\\
\tilde{\beta}_{2}
\end{array}\right)
$ are the OLS coefficients, and $\hat{\varepsilon}$
and $\tilde{\varepsilon}$ are the residual vectors from the long
and short regressions, respectively.

\begin{theorem}
\label{thm:fwl2}The OLS estimator for $\beta_{2}$ in the long regression equals the coefficient of $\tilde{X}_{2}$ in the OLS fit of $Y$ on $(1_n,\tilde{X}_{2} )$, where $\tilde{X}_{2}$ is the residual matrix of the column-wise OLS fit of $X_2$ on $(1_n, X_1)$, and also equals the coefficient of $\tilde{X}_{2}$ in the OLS fit of $\tilde{Y}$ on $(1_n,\tilde{X}_{2} )$, where $\tilde{Y}$ is the residual vector of the OLS fit of $Y$ on $(1_n, X_1)$.
\end{theorem}

\paragraph{General centering}\label{hw7::general-centering}

Verify \eqref{eq::centeringbyK}.

\paragraph{Two-way centering of a matrix}\label{hw7::2way-centering}

Given $X \in \mathbb{R}^{n\times p}$, show that all rows and columns of $ C_n X C_p $ have mean $0$, where $C_n = I_n - n^{-1} 1_n 1_n^{\T}$ and  $C_p = I_p - p^{-1} 1_p 1_p^{\T}$.

\paragraph{Linear fixed-effects regression for panel data}
\label{hw7::lfe-panel}

Assume that we have data over $n$ units and $T$ time points: $\{ y_{it} \in \mathbb{R}, x_{it} \in \mathbb{R}^p : i=1, \ldots, n; t = 1, \ldots, T \}$. Prove that the solution of the problem
$$
(\hat{\alpha}_1, \ldots, \hat{\alpha}_n, \hat{\beta})
= \arg\min_{a_1, \ldots, a_n, b\in \mathbb{R}^p} \sum_{i=1}^n \sum_{t=1}^T
( y_{it} - a_i  - x_{it}^{\T} b )^2
$$
is $\hat{\alpha}_i = \bar{y}_{i\cdot} - \bar{x}_{i\cdot}^{\T} \hat{\beta}$, with 
$$
\hat{\beta} = \arg\min_{  b\in \mathbb{R}^p } \sum_{i=1}^n \sum_{t=1}^T
( \dot{y}_{it} -  \dot{x}_{it}^{\T} b )^2
$$
and 
\begin{eqnarray*}
\bar{y}_{i \cdot}  = T^{-1} \sum_{t=1}^T y_{it}, \quad
\dot{y}_{it} = y_{it} - \bar{y}_{i\cdot}, \\ 
\bar{x}_{i\cdot} =  T^{-1} \sum_{t=1}^T x_{it}, \quad
\dot{x}_{it} = x_{it} - \bar{x}_{i\cdot}.
\end{eqnarray*}

Remark: In econometrics, cross-section data contain observations over units at one time point, whereas panel data contain observations over units and time. The OLS problem is motivated by the following linear fixed-effects  panel data model:
$$
y_{it} = \alpha_i + x_{it}^{\T} \beta + \varepsilon_{it},
$$
where the $\alpha_i$'s and $\beta$ are the unknown parameters. The total number of parameters grows with $n$, so solving the original OLS problem directly can be computationally inefficient.

\paragraph{Linear two-way fixed-effects regression for panel data}
\label{hw7::lfe-panel-TWFE}

This problem extends Problem \ref{hw7::lfe-panel}.

Assume that we have data over $n$ units and $T$ time points: $\{ y_{it} \in \mathbb{R}, x_{it} \in \mathbb{R}^p : i=1, \ldots, n; t = 1, \ldots, T \}$. Prove that the solution of the problem
\begin{eqnarray*}
(\hat{\mu}, \hat{\alpha}_1, \ldots, \hat{\alpha}_n, \hat{\gamma}_1, \ldots, \hat{\gamma}_T,  \hat{\beta} \in \mathbb{R}^p)
&=& \arg\min_{\mu, a_1, \ldots, a_n, r_1, \ldots, r_T,  b} \sum_{i=1}^n \sum_{t=1}^T
( y_{it} - \mu -  a_i - r_t  - x_{it}^{\T} b )^2 \\
&&\text{such that } \sum_{i=1}^n a_i = \sum_{t=1}^T r_t = 0
\end{eqnarray*}
is 
$\hat{\mu} = \bar{y}_{\cdot\cdot} -\bar{x}_{\cdot\cdot}^{\T} \hat{\beta} $, 
$\hat{\alpha}_i = (\bar{y}_{i\cdot} - \bar{y}_{\cdot\cdot}) - (  \bar{x}_{i\cdot}  - \bar{x}_{\cdot \cdot} )^{\T} \hat{\beta}$, 
and $\hat{\lambda}_t = (\bar{y}_{\cdot t} - \bar{y}_{\cdot\cdot}) - (  \bar{x}_{\cdot t}  - \bar{x}_{\cdot \cdot} )^{\T} \hat{\beta}$, with 
$$
\hat{\beta} = \arg\min_{  b \in \mathbb{R}^p } \sum_{i=1}^n \sum_{t=1}^T
( \ddot{y}_{it} -  \ddot{x}_{it}^{\T} b )^2
$$
and 
\begin{eqnarray*}
\bar{y}_{i\cdot}  = T^{-1} \sum_{t=1}^T y_{it}, \quad
\bar{y}_{\cdot t}  = n^{-1} \sum_{i=1}^n y_{it}, \quad
\bar{y}_{\cdot\cdot} = (nT)^{-1} \sum_{i=1}^n\sum_{t=1}^T y_{it},\quad 
\ddot{y}_{it} = y_{it} - \bar{y}_{i\cdot}  - \bar{y}_{\cdot t}  + \bar{y}_{\cdot\cdot}, \\ 
\bar{x}_{i\cdot}  = T^{-1} \sum_{t=1}^T x_{it}, \quad
\bar{x}_{\cdot t}  = n^{-1} \sum_{i=1}^n x_{it}, \quad
\bar{x}_{\cdot\cdot} = (nT)^{-1} \sum_{i=1}^n\sum_{t=1}^T x_{it},\quad 
\ddot{x}_{it} = x_{it} - \bar{x}_{i\cdot}  - \bar{x}_{\cdot t}  + \bar{x}_{\cdot\cdot} . 
\end{eqnarray*}

Remark: The OLS problem is motivated by the following linear two-way fixed-effects  panel data model:
$$
y_{it} = \mu +  \alpha_i + \gamma_t +  x_{it}^{\T} \beta + \varepsilon_{it},
$$
where the $\mu$, $\alpha_i$'s, $\gamma_t$'s and $\beta$ are the unknown parameters. The total number of parameters grows with $n$ and $T$, so solving the original OLS problem directly can be computationally inefficient.

Before solving this problem, you can first consider a simpler problem without covariates:
\begin{eqnarray*}
(\hat{\mu}, \hat{\alpha}_1, \ldots, \hat{\alpha}_n, \hat{\gamma}_1, \ldots, \hat{\gamma}_T)
&=& \arg\min_{\mu, a_1, \ldots, a_n, r_1, \ldots, r_T} \sum_{i=1}^n \sum_{t=1}^T
( y_{it} - a_i - r_t  )^2\\
&&\text{such that } \sum_{i=1}^n a_i = \sum_{t=1}^T r_t = 0
\end{eqnarray*}
which is called the balanced two-way analysis of variance (ANOVA) model in statistics.

\paragraph{$t$-statistic in multivariate OLS}
\label{hw7::tstat-partialcorr-OLS}

This problem extends Problem \ref{hw5::t-ratio-univariateOLS}.

Focus on multivariate OLS discussed in Chapter \ref{chapter::normal-linear-model}: $y_i = \hat{\alpha}  +  \hat{\beta}_1 x_{i1} + \hat{\beta}_2^{\T} x_{i2} +  \hat{\varepsilon}_i$ $(i=1,\ldots, n)$, where $x_{i1}$ is a scalar and $x_{i2}$ can be a vector.  
Show that under homoskedasticity, the $t$-statistic associated with $ \hat{\beta}_1$ equals
$$
\frac{  \hat{\rho}_{yx_1|x_2}   }{  \sqrt{    (1  -  \hat{\rho}_{yx_1|x_2} ^2) /(n-p)  }   },
$$
where $p$ is the total number of regressors and $\hat{\rho}_{yx_1|x_2}$ is the sample partial correlation coefficient between $y$ and $x_1$ given $x_2$.

Remark: \citet{frank2000impact} applied this formula to study causal inference.

\paragraph{Equivalence of the $t$-statistics in multivariate OLS}\label{hw7::t-stat-equivalent-multivariateOLS}

This problem extends Problems \ref{hw5::t-stat-equivalent} and \ref{hw6::t-stat-equivalent-breakdown}.

Consider data $(x_{i1}, x_{i2}, y_i)_{i=1}^n$, where both $x_{i1}$ and $y_i$ are scalars and $x_{i2}$ can be a vector. Run OLS fit of $y_i$ on $(1,x_{i1}, x_{i2})$ to obtain $t_{y\mid x_1, x_2}$, the $t$-statistic of the coefficient of $x_{i1}$, under the homoskedasticity assumption. Run OLS fit of $x_{i1}$ on $(1, y_i, x_{i2})$ to obtain $t_{x_1\mid y, x_2}$, the $t$-statistic of the coefficient of $y_i$, under the homoskedasticity assumption. 

Show $t_{y\mid x_1, x_2} = t_{x_1\mid y, x_2} $. Give a counterexample in which the numerical equivalence of the $t$-statistics breaks down based on the EHW standard error.

\paragraph{Formula of the partial correlation coefficient}\label{hw7::sample-partial-correlation}

Prove Theorem \ref{thm::sample-partialcorr-corr} based on the definition in \eqref{eq::sample-partial-CC}.

\paragraph{Examples of Simpson's Paradox}

Give three numerical examples of $(Y,X,W)$ that cause Simpson's Paradox. Report the mean and covariance matrix for each example.

\paragraph{Simpson's Paradox in reality}

Find a real-life dataset with Simpson's Paradox.

\paragraph{Basic properties of projection matrices}\label{problem07::projection-matrices}
 
Prove Lemma \ref{lemma::projection-m-F}.

\paragraph{Correlation of the regression coefficients}\label{hw::7:corr-reg-coef}

\begin{enumerate}
\item
Regress $Y$ on $(1_n,X_1, X_2)$ where $X_1$ and $X_2$ are two $n$-vectors with positive sample Pearson correlation $\hat\rho_{x_1x_2} > 0$. 

Prove that the corresponding OLS coefficients $\hat{\beta}_1$ and $\hat{\beta}_2$ are negatively correlated under the Gauss--Markov model of $Y$ on $(1_n,X_1, X_2)$. 

\item
Regress $Y$ on $(1_n,X_1, X_2, X_3)$ where $X_1$ and $X_2$ are two $n$-vectors and $X_3$ is an $n\times L$ dimensional matrix. Assume the partial correlation coefficient between $X_1$ and $X_2$ given $X_3$ is positive.

Prove that the corresponding OLS coefficients $\hat{\beta}_1$ and $\hat{\beta}_2$ are negatively correlated under the Gauss--Markov model $Y$ on $(1_n,X_1, X_2, X_3)$. 
\end{enumerate}

\paragraph{Inverse of sample covariance matrix and partial correlation coefficient}\label{hw::7:inverse-samplecov-pcc}

This is the sample version of Problem \ref{hwmath2::inverse-cov-conind-normal} in Appendix \ref{chapter:appendix-rvs}.

Based on $X \in \mathbb{R}^{n\times p}$, we can compute the sample covariance matrix $\hat\Sigma$. Denote its inverse by $\hat\Sigma^{-1} = (\hat{\sigma}^{jk})_{1\leq j,k\leq p}$. 
Prove Theorem \ref{thm::sample-graphical-model} below.

\begin{theorem}
\label{thm::sample-graphical-model}
For any pair $j\neq k$, we have 
$$
\hat{\sigma}^{jk} = 0 \text{ if and only if } \hat{\rho}_{x_jx_k\mid x_{\backslash (j,k)}} = 0
$$
where $\hat{\rho}_{x_jx_k\mid x_{\backslash (j,k)}} $ is the partial correlation coefficient of $X_j$ and $X_k$ given all other variables. 
\end{theorem}

\chapter{Cochran's Formula and Omitted-Variable Bias}
 \label{chapter::cochran-ovb}

Frisch--Waugh--Lovell (FWL) Theorem in Chapter \ref{chapter::FWL-theorem} and Cochran's formula in this chapter are sister results about OLS. The FWL Theorem states the equivalence between the coefficients in the long regression and partial regressions. Cochran's formula compares the coefficients in the  long and short regressions. They are both useful for interpreting the OLS coefficients.

\section{Cochran's formula} 
 
 Consider an $n\times 1$ vector $Y$, an $n\times k$ matrix $X_1$, and an $n\times l$ matrix $X_2$. Similar to the FWL Theorem, we do not impose any statistical models. 
We can fit the following OLS:
\begin{eqnarray}
Y &=& X_1 \hat{\beta}_1 + X_2 \hat{\beta}_2+ \hat{\varepsilon},  \label{eq::cochran-1} \\
Y &=& X_2 \tilde{\beta}_2 + \tilde{\varepsilon} ,\label{eq::cochran-2} \\
X_1 &=& X_2 \hat{\delta} + \hat{U}, \label{eq::cochran-3} 
\end{eqnarray}
where $\hat{\varepsilon}, \tilde{\varepsilon}$ are the residual vectors, and $\hat{U} $ is the residual matrix from the column-wise OLS of $X_1$ on $X_2$. Therefore, $\hat{U}$ is an $n\times k$ matrix.

 \begin{theorem}
 \label{thm::cochran-formula}
 Under the OLS fits \eqref{eq::cochran-1}--\eqref{eq::cochran-3}, we have 
 $$
\tilde{\beta}_2 = \hat{\beta}_2 +  \hat{\delta} \hat{\beta}_1.
$$
 \end{theorem}
 
This is a pure linear algebra fact similar to the FWL Theorem. It is called {\it Cochran's formula} in statistics. Sir David Cox \citep{cox2007generalization} attributed the formula to \citet{cochran1938omission} although Cochran himself attributed the formula to \citet{fisher1925statistical}.

Cochran's formula may seem familiar to readers who know the chain rule in calculus. In a deterministic world with scalar $y, x_1, x_2$, if
$$
y(x_1, x_2) = x_1  \beta_1 + x_2  \beta_2
$$
is a function of $x_1$ and $x_2$, 
and
$$
x_1(x_2) = x_2  \delta
$$
is a function of $x_2$, 
then the derivative of $y$ with respect to $x_2$ equals 
\begin{eqnarray*}
\frac{  \text{d} y }{  \text{d}x_2 } 
&=&   \frac{  \partial y }{ \partial x_1 }   \frac{  \partial x_1 }{ \partial x_2 }  + \frac{  \partial y }{ \partial x_2 }  \\
&=&  \delta   \beta_1 + \beta_2. 
\end{eqnarray*}
But the OLS decompositions in \eqref{eq::cochran-1}--\eqref{eq::cochran-3} do not establish any deterministic relationships among $Y$ and $(X_1, X_2)$.

 In some sense, the formula in Theorem \ref{thm::cochran-formula} is obvious. From the OLS fits \eqref{eq::cochran-1} and \eqref{eq::cochran-3}, we have
\begin{align} 
Y & =X_{1}\hat{\beta}_{1}+X_{2}\hat{\beta}_{2}+\hat{\varepsilon}  \nonumber  \\
 & =(X_{2}\hat{\delta}+\hat{U})\hat{\beta}_{1}+X_{2}\hat{\beta}_{2}+\hat{\varepsilon}  \nonumber\\
 & =X_{2}\hat{\delta}\hat{\beta}_{1}+\hat{U}\hat{\beta}_{1}+X_{2}\hat{\beta}_{2}+\hat{\varepsilon}  \nonumber\\
 & =X_{2}(\hat{\delta}\hat{\beta}_{1}+\hat{\beta}_{2})+(\hat{U}\hat{\beta}_{1}+\hat{\varepsilon}).
 \label{eq::cochran-proof1}
\end{align} 
This suggests that $\tilde{\beta}_{2}=\hat{\beta}_{2}+\hat{\delta}\hat{\beta}_{1}$. The above derivation follows from simple algebraic manipulations and does not use any properties of the OLS. To prove Theorem \ref{thm::cochran-formula}, we need to verify that the last line is indeed the OLS fit of $Y$ on $X_2.$ The proof is indeed very simple.

 \begin{myproof}{Theorem}{\ref{thm::cochran-formula}}
Based on the above discussion, 
we only need to show that \eqref{eq::cochran-proof1} is the OLS fit of $Y$
on $X_{2}$, which is equivalent to show that $\hat{U}\hat{\beta}_{1}+\hat{\varepsilon}$
is orthogonal to all columns of $X_{2}.$ This follows from
\[
X_{2}^{\T}(\hat{U}\hat{\beta}_{1}+\hat{\varepsilon})=X_{2}^{\T}\hat{U}\hat{\beta}_{1}+X_{2}^{\T}\hat{\varepsilon}=0,
\]
because $X_{2}^{\T}\hat{U}=0$ based on the OLS fit in \eqref{eq::cochran-3} and $X_{2}^{\T}\hat{\varepsilon}=0$
based on the OLS fit in \eqref{eq::cochran-1}. 
\end{myproof}

\begin{figure}
\centering
\includegraphics[width = 0.7\textwidth]{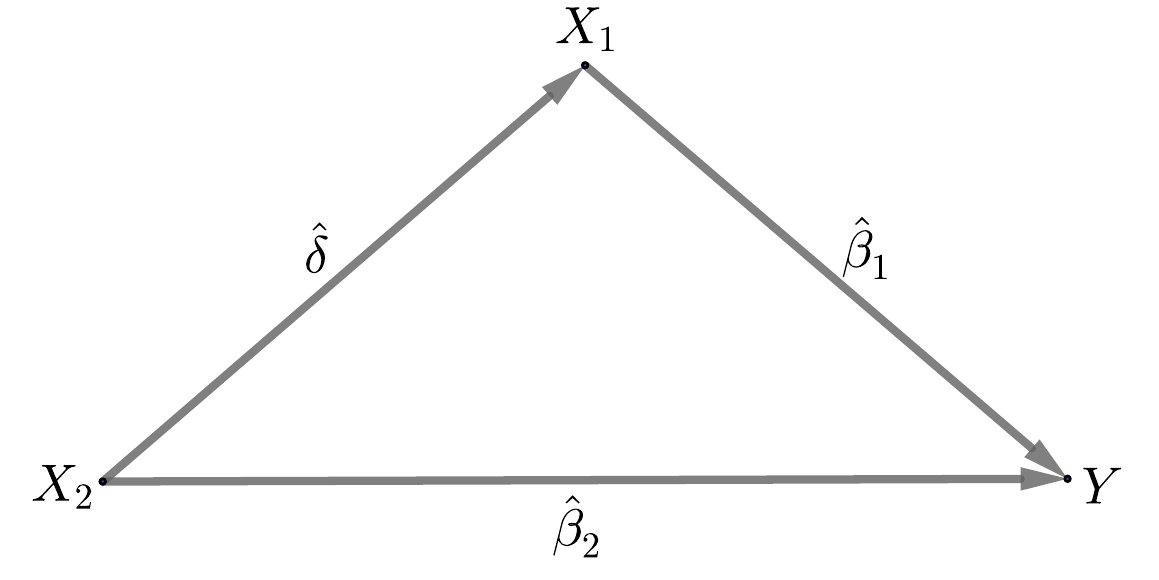}
\caption{A diagram for Cochran's formula}\label{fig::cochranformula}
\end{figure}

Figure \ref{fig::cochranformula} illustrates Theorem \ref{thm::cochran-formula}. Intuitively, $\tilde{\beta}_{2}$  measures the total impact of $X_2$ on $Y$, which has two channels: 
\begin{enumerate}[label=(C\arabic*), ref=C\arabic*]
\item
$\hat{\beta}_{2}$ measures the impact acting directly;
\item
$\hat{\delta}\hat{\beta}_{1}$ measures the impact acting indirectly through $X_1$. 
\end{enumerate}
This interpretation is closely related to {\it mediation analysis} in causal inference. See Problem \ref{hw07::baron-kenney} for more details. 
If you are interested in more discussions on mediation analysis, you can read Chapter 27 of \citet{ding2023first} or the monograph of \citet{Vanderweele::2015}.

Figure \ref{fig::cochranformula} shows the interplay among three variables. Theoretically, we can discuss a system of more than three variables which is called the {\it path model}. This more advanced topic is beyond the scope of this book. \citet{sewall1921correlation, wright1934method}'s initial discussion of this approach was motivated by genetic studies.
Problems \ref{hw07::baron-kenney} and \ref{hw07cochran::path} are two examples. See \citet{freedman2009statistical} for a critical textbook introduction.

\section{Omitted-variable bias} 
\label{sec::ovb}

The proof of Theorem \ref{thm::cochran-formula} is very simple. However, it is one of the most insightful formulas in statistics. Econometricians often call it the {\it omitted-variable bias} formula because it quantifies the bias of the OLS coefficient of $X_2$ in the short regression omitting possibly important variables in $X_1$.  If the OLS coefficient from the long regression is unbiased then the OLS coefficient from the short regression has a biased term
$
 \hat{\delta} \hat{\beta}_1,
$
which equals the product of the coefficient of $X_2$ in the OLS fit of $X_1$ on $X_2$ and the coefficient of $X_1$ in the long regression.

Below I will discuss a canonical example of using OLS to estimate the treatment effect in observational studies. For unit $i\  (i = 1, \ldots, n)$, let $y_i$ be the outcome, $z_i$ be the binary treatment indicator ($1$ for the treatment group and $0$ for the control group) and $x_i$ be the observed covariate vector. Practitioners often fit the following OLS:
$$
y_i = \tilde{\beta}_0 + \tilde{\beta}_1 z_i + \tilde{\beta}^{\T}_2 x_i + \tilde{\varepsilon}_i
$$
and interpret $ \tilde{\beta}_1$ as the treatment effect estimate. However, observational studies may suffer from unmeasured confounding, that is, the treatment and control units differ in unobserved but important ways. In the simplest case, the above OLS may have omitted a variable $u_i$ for each unit $i$, which is called a {\it confounder}. The oracle OLS is
$$
y_i = \hat{\beta}_0 + \hat{\beta}_1 z_i + \hat{\beta}^{\T}_2 x_i  + \hat{\beta}_3 u_i + \hat{\varepsilon}_i
$$
and the coefficient $ \hat{\beta}_1$ is an unbiased estimator if the model with $u_i$ is correct. 
With $X_1$ containing the values of the $u_i$'s and $X_2$ containing the values of the $(1, z_i, x_i^{\T})$'s, Cochran's formula implies that
$$
\begin{pmatrix}
\tilde{\beta}_0 \\
 \tilde{\beta}_1 \\
  \tilde{\beta}_2
\end{pmatrix}
=
\begin{pmatrix}
\hat{\beta}_0 \\
 \hat{\beta}_1 \\
  \hat{\beta}_2
\end{pmatrix}
+ 
\hat{\beta}_3 \begin{pmatrix}
 \hat{\delta}_0 \\
 \hat{\delta}_1 \\
  \hat{\delta}_2
\end{pmatrix}
$$
where $(  \hat{\delta}_0,  \hat{\delta}_1, \hat{\delta}_2^{\T})^{\T}$ is the coefficient vector in the OLS fit of $u_i$ on $(1, z_i, x_i)$. Therefore, we can
quantify the difference between the observed estimate $ \tilde{\beta}_1$ and oracle estimate $ \hat{\beta}_1$:
$$
\tilde{\beta}_1  - \hat{\beta}_1 =  \hat{\beta}_3 \hat{\delta}_1,
$$
which is sometimes called the {\it confounding bias}. 

Using the basic properties of OLS, we can show that $\hat{\delta}_1$ equals the difference in means of $e_i =  u_i -   \hat{\delta}_2^{\T} x_i$ across the treatment and control groups:
$$
\hat{\delta}_1 = \bar{e}_1 - \bar{e}_0,
$$
where the bar and subscript jointly denote the sample mean of a particular variable within a treatment group.
So
\begin{eqnarray} \label{eq::sensitivity-1}
\tilde{\beta} - \hat{\beta}_1 =  \hat{\beta}_3  ( \bar{e}_1 - \bar{e}_0).
\end{eqnarray} 
Moreover, we can obtain a more explicit formula for $\hat{\delta}_1$: 
$$
\hat{\delta}_1 = \bar{u}_1 - \bar{u}_0 -  \hat{\delta}_2^{\T}  ( \bar{x}_1 - \bar{x}_0 ).
$$
So
\begin{eqnarray} \label{eq::sensitivity-2}
\tilde{\beta}_1 - \hat{\beta}_1 =  \hat{\beta}_3(\bar{u}_1 - \bar{u}_0) -  \hat{\beta}_3\hat{\delta}_2^{\T}  ( \bar{x}_1 - \bar{x}_0 ).
\end{eqnarray} 

Both \eqref{eq::sensitivity-1} and \eqref{eq::sensitivity-2} give some insights into the bias due to omitting an important covariate $u$. It is clear that the bias depends on $ \hat{\beta}_3$, which quantifies the relationship between $u$ and $y$. The formula \eqref{eq::sensitivity-1} shows that the bias also depends on the imbalance in means of $u$ across the treatment and control groups, after adjusting for the observed covariates $x$, that is, the imbalance in means of the residual confounding. The formula \eqref{eq::sensitivity-2} shows a more explicit formula of the bias. The above discussion is often called {\it bias analysis} in epidemiology or {\it sensitivity analysis} in statistics and econometrics.

\section{Homework problems}

\paragraph{Baron--Kenny method for mediation analysis}
\label{hw07::baron-kenney}

The Baron--Kenny method, popularized by \citet{baron1986moderator}, is one of the most cited methods in social science. 
It concerns the interplay among three variables $z, m, y$, after controlling for some other variables $x$. Let $Z, M, Y$ be  $n\times 1$ vectors representing the observed values of $z, m, y$, and let $X$ be the $n\times p$ matrix representing the observations of $x$. The question of interest is to assess the ``direct'' and ``indirect'' effects of $z$ on $y$, acting independently and through $m$, respectively. We do not need to define these notions precisely since we are only interested in the algebraic property below.

\begin{figure}
\centering
\includegraphics[width = 0.5\textwidth]{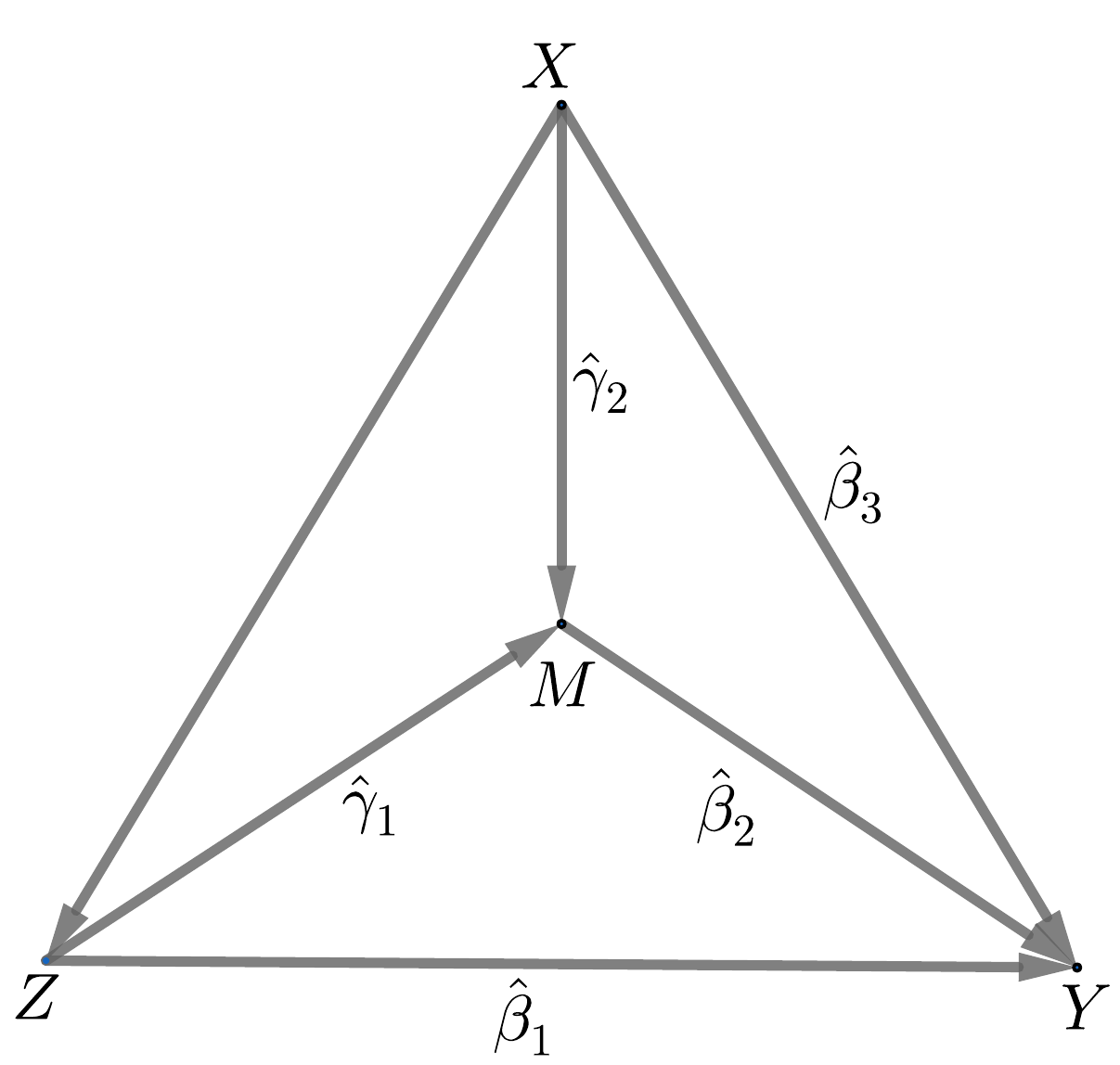}
\caption{The graph for the Baron--Kenny method}\label{fig::bkmethod}
\end{figure}

The Baron--Kenny method runs the OLS
$$
Y = \hat{\beta}_0 1_n   +  \hat{\beta}_1 Z  + \hat{\beta}_2 M  + X \hat{\beta}_3 + \hat{\varepsilon}_Y
$$
and interprets $\hat{\beta}_1$ as the estimator of the ``direct effect'' of $z$ on $y$. 
The ``indirect effect'' of $z$ on $y$ through $m$ has two estimators. First, based on the OLS
$$
Y = \tilde{\beta}_0 1_n   +  \tilde{\beta}_1 Z     + X \tilde{\beta}_3 + \tilde{\varepsilon}_Y,
$$
define the {\it difference estimator} as 
$
 \tilde{\beta}_1 -   \hat{\beta}_1 .
$
Second, based on the OLS
$$
M = \hat{\gamma}_0 1_n   +  \hat{\gamma}_1 Z    + X \hat{\gamma}_2 + \hat{\varepsilon}_M , 
$$
define the {\it product estimator} as
$
  \hat{\gamma}_1   \hat{\beta}_2 .
$
Figure \ref{fig::bkmethod} illustrates the OLS fits used in defining the estimators. 

Prove that 
$$ 
\tilde{\beta}_1 -   \hat{\beta}_1 =    \hat{\gamma}_1   \hat{\beta}_2
$$ 
that is, the difference estimator and product estimator are numerically identical.

\paragraph{A special case of path analysis}\label{hw07cochran::path}

\begin{figure}
\centering
\includegraphics[width = 0.8\textwidth]{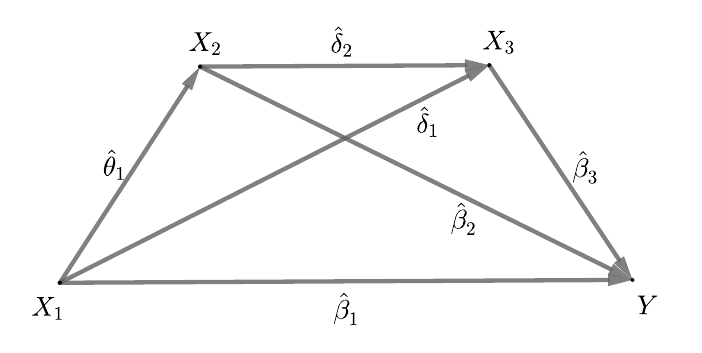}
\caption{A path model}\label{fig::path}
\end{figure}

Figure \ref{fig::path} represents the order of the variables $X_1, X_2, X_3, Y \in \mathbb{R}^n$. Run the following OLS:
\begin{eqnarray*}
Y &=& \hat{\beta}_0 1_n  + \hat{\beta}_1 X_1 + \hat{\beta}_2 X_2 + \hat{\beta}_3 X_3 + \hat{\varepsilon}_Y, \\
X_3 &=&   \hat{\delta}_0 1_n+ \hat{\delta}_1 X_1 + \hat{\delta}_2 X_2 +  \hat{\varepsilon}_3, \\
X_2 &=&  \hat{\theta}_0 1_n+ \hat{\theta}_1 X_1 +  \hat{\varepsilon}_2,
\end{eqnarray*}
and
$$
Y = \tilde{\beta}_0 1_n+ \tilde{\beta}_1 X_1 + \tilde{\varepsilon}_Y.
$$

Prove that 
$$
 \tilde{\beta}_1 =  \hat{\beta}_1  +  \hat{\beta}_2\hat{\theta}_1  +  \hat{\beta}_3 \hat{\delta}_1 +  \hat{\beta}_3 \hat{\delta}_2  \hat{\theta}_1 .
$$

Remark: The OLS coefficient of $X_1$ in the short regression of $Y$ on $(1_n, X_1)$ equals the summation of all the path coefficients from $X_1$ to $Y$ as illustrated by Figure \ref{fig::path}:
\begin{eqnarray*}
X_1 \longrightarrow Y ,\\
X_1 \longrightarrow X_2  \longrightarrow Y ,\\
X_1 \longrightarrow X_3  \longrightarrow Y ,\\
X_1 \longrightarrow X_2  \longrightarrow X_3  \longrightarrow Y .
\end{eqnarray*} 

This problem is a special case of the path model, but the conclusion holds in general. 

  \paragraph{EHW in long and short regressions}\label{hw08::ehw-long-short-ols}
Theorem \ref{thm::cochran-formula} gives Cochran's formula related to the coefficients from three OLS fits. 
This problem concerns the covariance estimation. 
There are at least two ways to estimate the covariance of $\tilde \beta_2$ in the short regression \eqref{eq::cochran-2}. First, from the short regression \eqref{eq::cochran-2}, the EHW covariance estimator is
$$
\tilde V_2 = (X_2^{\T} X_2)^{-1}   X_2^{\T}  \text{diag}(\tilde \varepsilon^2)  X_2 (X_2^{\T} X_2)^{-1}.
$$
Second, Cochran's formula implies that 
$$
\tilde \beta_2 = (\hat \delta, I_l)  \begin{pmatrix}
\hat{\beta}_1 \\
\hat{\beta}_2
\end{pmatrix}
$$ is a linear transformation of the coefficient from the long regression, which further justifies the EHW covariance estimator 
$$
\tilde V_2 ' = (\hat \delta, I_l) (X^{\T} X)^{-1}   X^{\T}  \text{diag}(\hat \varepsilon^2)  X (X^{\T} X)^{-1}  \begin{pmatrix}
\hat \delta^{\T} \\
I_l
\end{pmatrix} .
$$

Prove that 
$$
\tilde V_2 ' = (X_2^{\T} X_2)^{-1}   X_2^{\T}  \text{diag}(\hat \varepsilon^2)  X_2 (X_2^{\T} X_2)^{-1}.
$$

Remark: 
This problem states that $\tilde V_2  \neq \tilde V_2 ' $ in general. To prove the result, you can use the result in Problem \ref{hw6::inverse-block-gram}. 

Moreover, based on Theorem \ref{thm::fwl-se}, the EHW covariance estimator for $\hat{\beta}_2$ is
$$
\hat{V}_2 = (\tilde X_2^{\T} \tilde X_2)^{-1}   \tilde X_2^{\T}  \text{diag}(\hat \varepsilon^2)  \tilde X_2 (\tilde X_2^{\T} \tilde X_2)^{-1},
$$
where $ \tilde X_2 = (I_n - H_1) X_2$. It differs from $\tilde V_2 $ and $ \tilde V_2 ' $ in general.

 \paragraph{Statistical properties of under-fitted OLS}
\label{hw::underfit-ols}

Assume that $Y = X \beta + \varepsilon = X_1 \beta_1 + X_2 \beta_2 + \varepsilon$ follows the Gauss--Markov model, where $X_1 \in \mathbb{R}^{n\times k}$, $X_2 \in \mathbb{R}^{n\times l}$, and $\cov(\varepsilon)=\sigma^2 I_n$. However, we only fit the OLS of $Y$ on $X_2$ with coefficient $\tilde{\beta}_2$ and estimated variance $\tilde{\sigma}^2$. 

Prove that
\begin{eqnarray*}
E(\tilde{\beta}_2) &=&  (X_2^{\T} X_2)^{-1} X_2^{\T} X_1 \beta_1 +   \beta_2 ,\\ 
\var( \tilde{\beta}_2) &=& \sigma^2 (X_2^{\T} X_2)^{-1} , \\
E(\tilde{\sigma}^2) &=& \sigma^2 + \beta_1^{\T} X_1^{\T} (I_n - H_2) X_1\beta_1/(n-l) \geq \sigma^2 . 
\end{eqnarray*}

\part{Model Fitting, Checking, and Misspecification}
   
\chapter{Multiple Correlation Coefficient}
 
This chapter will introduce the $R^2$,  the {\it multiple correlation coefficient}, also called the {\it coefficient of determination} \citep{sewall1921correlation}. It can achieve two goals: 
\begin{enumerate}[label=(G\arabic*), ref=G\arabic*]
\item
it extends the sample Pearson correlation
coefficient between two scalars to a measure of correlation between a
scalar outcome and a vector covariate;

\item 
it measures how well multiple covariates
can linearly represent an outcome.
\end{enumerate}

\section{Equivalent definitions of the multiple correlation coefficient}

I start with the standard definition of $R^{2}$ between $Y$ and $X$.  
Slightly different from other chapters, $X$ in this chapter excludes the column of $1$'s, so now $X$ is an $n\times (p-1)$ matrix. Based on the OLS of $Y$ on $(1_n, X)$, we define 
\begin{equation}
\label{eq::definition-r^2-1}
R^{2}=\frac{\sumn(\hat{y}_{i}-\bar{y})^{2}}{\sumn(y_{i}-\bar{y})^{2}}.
\end{equation}
You may wonder why the numerator of \eqref{eq::definition-r^2-1} does not use $\bar{\hat{y}} = n^{-1}\sumn\hat{y}_{i}$. This is because the average of the fitted values equals the average of the original observed
outcomes, i.e., $\bar{\hat{y}}  = \bar{y}$, if we include $1_{n}$ in the OLS; see Problem \ref{hw09::averagey-averagefitted}. 
With scaling factor $(n-1)^{-1}$, the denominator
of $R^{2}$ is the sample variance of the outcomes, and the numerator
of $R^{2}$ is the sample variance of the fitted values. We can verify
the following decomposition:
\begin{lemma} 
\label{lemma:The-total-sum}
Based on the OLS of $Y$ on $(1_n, X)$, we have the following variance decomposition: 
\[
\sumn(y_{i}-\bar{y})^{2}=\sumn(\hat{y}_{i}-\bar{y})^{2}+\sumn(y_{i}-\hat{y}_{i})^{2}.
\]
\end{lemma}

I leave the proof of Lemma \ref{lemma:The-total-sum} as Problem \ref{hw09::var-decompose}. 
Lemma \ref{lemma:The-total-sum} states that the total sum of squares $\sumn(y_{i}-\bar{y})^{2}$
equals  the regression sum of squares $\sumn(\hat{y}_{i}-\bar{y})^{2}$
plus the residual sum of squares (\textsc{rss}) $\sumn(y_{i}-\hat{y}_{i})^{2}$. 
From Lemma
\ref{lemma:The-total-sum}, $R^{2}$ must be lie within the interval
$[0,1]$ which measures the proportion of the regression sum of squares
in the total sum of squares. An immediate consequence of Lemma \ref{lemma:The-total-sum} is that
$$
\textsc{rss} = (1-R^2)\times  \sumn(y_{i}-\bar{y})^{2}.
$$

We can also verify that $R^{2}$ is the squared sample Pearson correlation
coefficient between $Y$ and $\hat{Y}$. 
\begin{theorem}\label{theorem::r2=rho2}
We have $R^{2}= \hat{\rho}_{y\hat{y}}^2$, where
\begin{equation}
\hat{\rho}_{y\hat{y}}
= \frac{\sumn(y_{i}-\bar{y})(\hat{y}_{i}-\bar{y})}{\sqrt{\sumn(y_{i}-\bar{y})^{2}}\sqrt{\sumn(\hat{y}_{i}-\bar{y})^{2}}} .
\label{eq:r2PCC}
\end{equation}
\end{theorem}

I leave the proof of Theorem \ref{theorem::r2=rho2} to Problem \ref{hw9::r2-pcc}.  Theorem \ref{theorem::r2=rho2} states that the multiple correlation coefficient equals the squared Pearson correlation coefficient between the observed outcome $y_i$ and its fitted value $\hat{y}_i$. 
Although the sample Pearson correlation coefficient can be positive or negative, $R^2$ is always non-negative.  
Geometrically, $R^2$ equals the squared cosine of the angle between the centered vectors $Y-\bar{y} 1_n$ and $\hat{Y} - \bar{y} 1_n$; see Appendix \ref{sec::basics-vectors-matrices} for the geometric interpretation of the Pearson correlation coefficient.

In terms of long and short regressions, we can partition the design matrix into   $ 1_{n}$ and $X$, then the OLS fit of the long regression is
\begin{equation}
Y=1_{n}\hat{\beta}_0+X \hat{\beta} +\hat{\varepsilon},\label{eq:longregR2}
\end{equation}
 and the OLS fit of the short regression is
\begin{equation}
Y=1_{n}\tilde{\beta}_0 +\tilde{\varepsilon},\label{eq:shortregR2}
\end{equation}
with $\tilde{\beta}_0 =\bar{y}$.
The total sum of squares is the residual sum of squares from the short
regression so by Lemma \ref{lemma:The-total-sum}, $R^{2}$ also equals 
\begin{equation}
R^{2}  =  \frac{\textsc{rss}_{\text{short}}-\textsc{rss}_{\text{long}}}{\textsc{rss}_{\text{short}}} . \label{eq::residuals-R2} 
\end{equation}

\section{The multiple correlation coefficient and F statistic}

Under the Normal linear model 
\begin{equation}
Y=1_{n}\beta_0+X \beta +\varepsilon,\qquad\varepsilon\sim\N(0,\sigma^{2}I_{n}),\label{eq:gaussianolsR2}
\end{equation}
we can use the $F$ statistic to test whether $\beta =0$:
$$
F=\frac{(\textsc{rss}_{\text{short}}-\textsc{rss}_{\text{long}})/(p-1)}{\textsc{rss}_{\text{long}}/(n-p)} . 
$$ 
This
$F$ statistic is a monotone function of $R^{2}$ in \eqref{eq::residuals-R2} as shown in Theorem \ref{theorem:F-R2} below. Most standard software
packages report both $F$ and $R^2$.  

\begin{theorem}
\label{theorem:F-R2}We have 
\[
F=\frac{n-p}{p-1}\times\frac{R^{2}}{1-R^{2}}.
\]
\end{theorem}

I leave the proof of Theorem \ref{theorem:F-R2} as Problem \ref{hw09::F-R2}.Theorem \ref{theorem:F-R2} is a numeric result without assuming that model \eqref{eq:gaussianolsR2} is correct. 
Under the Normal linear model, we can derive the exact distribution of $R^2$. 

\begin{corollary}\label{coro::dist-r2}
Under the Normal linear model (\ref{eq:gaussianolsR2}), if $\beta =0$,
then 
\[
R^{2}\sim\textup{Beta}\left(\frac{p-1}{2},\frac{n-p}{2}\right).
\]
\end{corollary}
\begin{myproof}{Corollary}{\ref{coro::dist-r2}}
By definition, the $F$ statistic can be represented as 
\[
F=\frac{\chi_{p-1}^{2}/(p-1)}{\chi_{n-p}^{2}/(n-p)} , 
\]
where $\chi_{p-1}^{2}$ and $\chi_{n-p}^{2}$ denote independent $\chi_{p-1}^{2}$ and $\chi_{n-p}^{2}$ random variables, respectively. Using Theorem \ref{theorem:F-R2},
we have
\[
\frac{R^{2}}{1-R^{2}}=F\times\frac{p-1}{n-p}=\frac{\chi_{p-1}^{2}}{\chi_{n-p}^{2}} , 
\]
which implies
\[ 
R^{2}=\frac{\chi_{p-1}^{2}}{\chi_{p-1}^{2}+\chi_{n-p}^{2}}\text{.}
\]
Because $\chi_{p-1}^{2}\sim\textup{Gamma}\left(\frac{p-1}{2},\frac{1}{2}\right)$
and $\chi_{n-p}^{2}\sim\textup{Gamma}\left(\frac{n-p}{2},\frac{1}{2}\right)$ by Proposition \ref{prop::chi2-gamma} in Appendix \ref{chapter:appendix-rvs},
we have
$$
R^{2} 
=\frac{\textup{Gamma}\left(\frac{p-1}{2},\frac{1}{2}\right)}{\textup{Gamma}\left(\frac{p-1}{2},\frac{1}{2}\right)+\textup{Gamma}\left(\frac{n-p}{2},\frac{1}{2}\right)} 
$$
where $\textup{Gamma}\left(\frac{p-1}{2},\frac{1}{2}\right)$ and $\textup{Gamma}\left(\frac{n-p}{2},\frac{1}{2}\right)$ denote independent Gamma random variables. 
The $R^2$ follows the Beta distribution by the Beta--Gamma duality in Theorem \ref{thm:beta-gamma-duality} in Appendix \ref{chapter:appendix-rvs}. 
\end{myproof}

\section{Numerical examples}

I first revisit the LaLonde data to verify Theorems \ref{theorem::r2=rho2} and \ref{theorem:F-R2} numerically. 
\begin{lstlisting}
> library("Matching")
> data(lalonde)
> ols.fit = lm(re78 ~ ., y = TRUE, data = lalonde)
> ols.summary = summary(ols.fit)
> r2 = ols.summary$r.squared
> all.equal(r2, (cor(ols.fit$y, ols.fit$fitted.values))^2,
+           check.names = FALSE)
[1] TRUE
> 
> fstat = ols.summary$fstatistic
> all.equal(fstat[1], fstat[3]/fstat[2]*r2/(1-r2),
+           check.names = FALSE)
[1] TRUE
\end{lstlisting}

I then revisit the data from \citet{king2015robust} to verify Theorems \ref{theorem::r2=rho2} and \ref{theorem:F-R2} numerically. 
\begin{lstlisting}
> library(foreign)
> dat = read.dta("isq.dta")
> dat = na.omit(dat[,c("multish", "lnpop", "lnpopsq", 
+                      "lngdp", "lncolony", "lndist", 
+                      "freedom", "militexp", "arms", 
+                      "year83", "year86", "year89", "year92")])
> 
> ols.fit = lm(log(multish + 1) ~ lnpop + lnpopsq + lngdp +  lncolony 
+              + lndist + freedom + militexp + arms 
+              + year83 + year86 + year89 + year92, 
+              y = TRUE, data=dat)
> ols.summary = summary(ols.fit)
> r2 = ols.summary$r.squared
> all.equal(r2, (cor(ols.fit$y, ols.fit$fitted.values))^2,
+           check.names = FALSE)
[1] TRUE
> 
> fstat = ols.summary$fstatistic
> all.equal(fstat[1], fstat[3]/fstat[2]*r2/(1-r2),
+           check.names = FALSE)
[1] TRUE
\end{lstlisting}

\section{Homework problems}

\paragraph{Average outcome equals average fitted values}\label{hw09::averagey-averagefitted}

Prove that from the OLS of $Y$ on $(1_n, X)$, we have $\bar{\hat{y}}  = \bar{y}$.

\paragraph{Variance decomposition}\label{hw09::var-decompose}
Prove Lemma \ref{lemma:The-total-sum}.

\paragraph{$R^2$ and the sample Pearson correlation coefficient}\label{hw9::r2-pcc}
Prove Theorem \ref{theorem::r2=rho2}.

\paragraph{$F$ and $R^2$}\label{hw09::F-R2}

Prove Theorem \ref{theorem:F-R2}.

\paragraph{Exact distribution of $\hat{\rho}$}\label{hw9::exact-d-rho}

Assume the Normal linear model $y_i = \alpha + \beta x_i + \varepsilon_i$ with
 a univariate $x_i$ with $\beta = 0$ and $\varepsilon_i$'s IID $\N(0,  \sigma^2 )$. 
 Find the exact distribution of $\hat{\rho}_{xy}$.

\paragraph{Partial $R^2$}\label{hw09::partial-R2}

The form \eqref{eq::residuals-R2} of $R^2$ is well defined in more general long and short regressions:
$$
 Y  = 1_n \hat{\beta}_0 + X \hat{\beta} + W \hat{\gamma} + \hat{\varepsilon}_Y
$$
and
$$
 Y  = 1_n \tilde{\beta}_0   + W \tilde{\gamma} + \tilde{\varepsilon}_Y
$$
where $X$ is an $n\times k$ matrix and $W$ is an $n\times l$ matrix. Define the {\it partial $R^2$} between $Y$ and $X$ given $W$ as
$$
R^2_{Y.X|W} = \frac{ \textsc{rss}(Y\sim 1_n+W) -   \textsc{rss}(Y\sim 1_n+X+W) }{ \textsc{rss}(Y\sim 1_n+W) }
$$
which spells out the formulas of the long and short regressions. This is an intuitive measure of the multiple correlation between $Y$ and $X$ after controlling for $W$. The following properties make this intuition more explicit.

\begin{enumerate}
\item
The partial $R^2$ equals 
$$
R^2_{Y.X|W} = \frac{  R^2_{Y.XW}  - R^2_{Y.W}  }{ 1- R^2_{Y.W}} , 
$$
where $R^2_{Y.XW} $ is the multiple correlation between $Y$ and $(X,W)$, and $R^2_{Y.W}$ is the multiple correlation between $Y$ and $W$.

\item
The partial $R^2$ equals the $R^2$ between $ \tilde{\varepsilon}_Y $ and $\tilde{\epsilon}_X$:
$$
R^2_{Y.X|W} =  R^2_{  \tilde{\varepsilon}_Y . \tilde{\varepsilon}_X }, 
$$
where $\tilde{\varepsilon}_X$ is the residual matrix from the OLS fit of $X$ on $(1_n, W)$. 
\end{enumerate}

Prove the above two results. 

Do the following two results hold? 
\begin{eqnarray*}
R^2_{Y.XW} &=& R^2_{Y.W} +  R^2_{Y.X|W} ,\\
R^2_{Y.XW} &=& R^2_{Y.W|X} +  R^2_{Y.X|W}.
\end{eqnarray*}
For each result, give a proof if it is correct, or give a counterexample if it is incorrect in general.

\paragraph{Omitted-variable bias in terms of the partial $R^2$}\label{hw09::cochran-partial-R2}

Revisit Section \ref{sec::ovb} on the following three OLS fits. The first one involves only the observed variables:
$$
y_i  =  \tilde{\beta}_0 + \tilde{\beta}_1 z_i + \tilde{\beta}^{\T}_2 x_i + \tilde{\varepsilon}_i,
$$
and the second and third ones involve the unobserved $u$:
\begin{eqnarray*}
y_i &=& \hat{\beta}_0 + \hat{\beta}_1 z_i + \hat{\beta}^{\T}_2 x_i  + \hat{\beta}_3 u_i + \hat{\varepsilon}_i ,\\
u_i &=&  \hat{\delta}_0 +  \hat{\delta}_1 z_i  + \hat{\delta}_2^{\T} x_i + \hat{v}_i.
\end{eqnarray*}
The omitted-variable bias formula states that 
$$
\tilde{\beta}_1 - \hat{\beta}_1 =  \hat{\beta}_3 \hat{\delta}_1.
$$ 
This formula is simple but may be difficult to interpret since $u$ is unobserved and its scale is unclear to researchers. 

Prove the following formula: 
\begin{equation}\label{eq::ols-sensitivity-jrssb}
|\tilde{\beta}_1 - \hat{\beta}_1|^2 = R^2_{Y.U\mid ZX} \times \frac{R^2_{Z.U\mid X}}{1 - R^2_{Z.U\mid X}}  
\times   \frac{ \rss(Y\sim 1_n+Z+X)  }{  \rss(Z\sim 1_n+X) }  . 
\end{equation}

Remark: \citet{cinelli2020making} suggested the partial $R^2$ parametrization for the omitted-variable bias formula. 
The formula \eqref{eq::ols-sensitivity-jrssb} has three factors: 
\begin{enumerate}[label=(F\arabic*), ref=F\arabic*]
\item
the first factor depends on the unknown {\it sensitivity parameters} $R^2_{Y.U\mid ZX} $, which measures the confounder-outcome relationship;
\item
the second factor depends on the unknown {\it sensitivity parameters} $R^2_{Z.U\mid X}$, which measures the treatment-confounder relationship;
\item
the third factor equals the ratio of the two residual sums of squares based on the observed data, which does not depend on the unmeasured confounder. 
\end{enumerate}
The beauty of \eqref{eq::ols-sensitivity-jrssb} is that the partial $R^2$ parameters $R^2_{Y.U\mid ZX} $ and $R^2_{Z.U\mid X}$ do not depend on the unknown scale of $u$. 
\citet{zhang2022interpretable} derived more general omitted-variables bias formulas based on partial $R^2$.

\chapter{Leverage Scores and Leave-One-Out Formulas}\label{chapter::leave-one-out}

This chapter will discuss two related topics: leverage scores and leave-one-out formulas. We have seen leverage scores in previous chapters, which are defined as the diagonal elements of the hat matrix  $H=X(X^{\T}X)^{-1}X^{\T} $ from OLS. This chapter will present more properties of the leverage scores. Intuitively, the leverage score of unit $i$ measures whether covariate $x_i$ is an outlier among other covariates. Consequently, units with larger leverage scores will have larger impact on the final OLS outputs, which can be quantified by various leave-one-out formulas proved in this chapter. 

Leave-one-out is a general idea in statistics and machine learning. It is a basic idea to avoid overfitting. It applies to all statistical models. The beauty of OLS is that we can derive leave-one-out formulas explicitly. Those formulas not only allow for fast computation but also provide insights into the properties of OLS.

From an exploratory data analysis perspective, we can use the idea of leave-one-out to assess the impact of individual observations, or equivalently, to assess the {\it stability} of the results with respect to deleting individual observations \citep{yu2020veridical}. 
In the linear model, plotting the leverage scores can be viewed as a shortcut to the leave-one-out. 

This chapter will provide more details for the above high-level statements.

\section{Leverage scores}

The hat matrix  $H=X(X^{\T}X)^{-1}X^{\T} $ is a key matrix for OLS. Because 
$$
H 
 = \begin{pmatrix}
x_1^{\T} \\
\vdots \\
x_{n}^{\T}
\end{pmatrix} 
(X^{\T}X)^{-1} 
\begin{pmatrix}
x_1& \cdots & x_n
\end{pmatrix},
$$
its $(i,j)$th element equals $h_{ij}=x_{i}^{\T}(X^{\T}X)^{-1}x_{j}  .$
In this chapter, we will pay special attention
to its diagonal elements 
\[
h_{ii}=x_{i}^{\T}(X^{\T}X)^{-1}x_{i}  \quad (i= 1, \ldots, n)
\]
 often called the {\it leverage scores}, which play important roles in many
discussions later. 

\subsection{The average leverage score equals $p/n$}

Because $H$ is a projection matrix of rank $p$, we have
\[
\sumn h_{ii}=\text{trace}(H)=\text{rank}(H)=p, 
\]
which implies that
\[
n^{-1}\sumn h_{ii}=p/n, 
\]
i.e., the average of the leverage scores equals $p/n$. Therefore, the maximum of the leverage scores must be larger than or equal to $p/n$:
$$
\max_{1\leq i \leq n} h_{ii} \geq  n^{-1}\sumn h_{ii}=p/n . 
$$
As $p / n$ becomes larger, we will observe more extreme leverage scores.

\subsection{The  leverage scores are all bounded between $0$ and $1$}

Because $H=H^{2}$ and $H=H^{\T}$, we have
\begin{eqnarray*}
h_{ii} &=& \sum_{j=1}^{n}h_{ij}h_{ji} \\
&=& \sum_{j=1}^{n}h_{ij}^{2} \\
&=& h_{ii}^{2}+\sum_{j\neq i}h_{ij}^{2} \\ 
&\geq & h_{ii}^2  , 
\end{eqnarray*}
which implies 
$$
h_{ii}\in[0,1] , 
$$ 
i.e., each leverage score is bounded between 0 and 1.\footnote{This also follows from Theorem \ref{theorem::rayleigh} in Appendix \ref{chapter::linear-algebra} since the eigenvalues of $H$ are $0$ and $1$.}

\subsection{The $i$th leverage score measures the impact of the $i$th observation in prediction}

Because $\hat{Y}=HY$, we have
\[
\hat{y}_{i}=\sum_{j=1}^{n}h_{ij}y_{j}=h_{ii}y_{i}+\sum_{j\neq i}h_{ij}y_{j},
\]
which implies that
\[
\frac{\text{\ensuremath{\partial\hat{y}_{i}}}}{\partial y_{i}}=h_{ii}.
\]
So $h_{ii}$ measures the contribution of $y_{i}$ in the predicted
value $\hat{y}_{i}.$ In general, we do not want the contribution
of $y_{i}$ in predicting itself to be too large, because this means
we do not borrow enough information from other observations, making
the prediction very noisy. This is also clear from the variance of the predicted value 
$\hat{y}_{i}=x_{i}^{\T}\hat{\beta}$ under the Gauss--Markov model:\footnote{We have already proved a more general result on the covariance matrix of $\hat{Y}$ in Theorem \ref{thm:GMcov}.}
\begin{eqnarray*}
\text{\var}(\hat{y}_{i}) 
&=& x_{i}^{\T}\cov(\hat{\beta})x_{i} \\
&=&\sigma^{2}x_{i}^{\T}(X^{\T}X)^{-1}x_{i} \\
&=&\sigma^{2}h_{ii}.
\end{eqnarray*}
So the variance of $\hat{y}_{i}$ increases with $h_{ii}$.

\subsection{The $i$th leverage score measures whether
$x_i$ is an outlier compared with other covariates}

The $h_{ii}$ measures whether
observation $i$ is an outlier based on its covariate value, that
is, whether $x_{i}$ is far from the center of the data. Partition the
design matrix as $X=\left(\begin{array}{cc}
1_{n} & X_{2}\end{array}\right)$ with $H_{1}=n^{-1}1_{n}1_{n}^{\T}.$ The covariates $X_{2}$ has sample mean $\bar{x}_{2}=n^{-1}\sumn x_{i2}$ and sample covariance 
\begin{eqnarray*}
S  
&=&  (n-1)^{-1}\sumn(x_{i2}-\bar{x}_{2})(x_{i2}-\bar{x}_{2})^{\T} \\
&=& (n-1)^{-1}X_{2}^{\T}(I_{n}-H_{1})X_{2}.
\end{eqnarray*}
The sample Mahalanobis distance between $x_{i2}$ and the center $\bar{x}_{2}$
is 
\[
D_{i}^{2}=(x_{i2}-\bar{x}_{2})^{\T}S^{-1}(x_{i2}-\bar{x}_{2}).
\]
Theorem \ref{thm::leverage-mdist} below shows that $h_{ii}$ is a monotone function
of $D_{i}^{2}$:

\begin{theorem}\label{thm::leverage-mdist}
Assume that we include the intercept in OLS.
We have 
\begin{equation}
 h_{ii}= \frac{1}{n} + \frac{D_{i}^{2}}{n-1}  ,\label{eq:leveragemdis}
\end{equation}
so $h_{ii} \geq 1/n$. 
\end{theorem}

\begin{myproof}{Theorem}{\ref{thm::leverage-mdist}}
The definition of $D_{i}^{2}$ implies that it is the $(i,i)$th element
of the following matrix:
\begin{align*}
 & \left(\begin{array}{c}
x_{12}-\bar{x}_{2}\\
\vdots\\
x_{n2}-\bar{x}_{2}
\end{array}\right)^{\T}S^{-1}\left(\begin{array}{ccc}
x_{12}-\bar{x}_{2} & \cdots & x_{n2}-\bar{x}_{2}\end{array}\right)\\
= & (I_{n}-H_{1})X_{2}\left\{ (n-1)^{-1}X_{2}^{\T}(I_{n}-H_{1})X_{2}\right\} ^{-1}X_{2}^{\T}(I_{n}-H_{1})\\
= &(n-1)\tilde{X}_{2}(\tilde{X}_{2}^{\T}\tilde{X}_{2})^{-1}\tilde{X}_{2}^{\T}\\
= &(n-1)\tilde{H}_{2}\\
= &(n-1)(H-H_{1}),
\end{align*}
recalling that $\tilde{X}_2 = (I_n - H_1) X_2$, $\tilde{H}_2 = \tilde{X}_2 (\tilde{X}_2^{\T} \tilde{X}_2)^{-1} \tilde{X}_2^{\T}$, and $H = H_1 + \tilde{H}_{2}$ by Lemma \ref{lemma::decompose-projection-matrices}. 
Therefore, 
\[
D_{i}^{2}=(n-1) (h_{ii}-1/n)
\]
which implies (\ref{eq:leveragemdis}).
\end{myproof}

\subsection{Other properties of the leverage scores}

Another advanced result on the leverage scores is due to 
\citet{huber1973robust}. He proved that in the linear model with non-Normal IID $\varepsilon_i$ with mean $0$ and variance $\sigma^2 < \infty$, all linear combinations of the OLS coefficient are asymptotically Normal if and only if the maximum leverage score converges to $0$. This is a very elegant asymptotic result on the OLS coefficient. I give more details in Appendix \ref{chapter::limiting-theorems} as an application of the Lindeberg--Feller CLT.

 These are the basic properties of the leverage scores. 
\citet{chatterjee1988sensitivity} provided an in-depth discussion of the properties of the leverage scores. 
We will see their roles frequently in later parts of this chapter.

\section{Leave-one-out formulas}

To measure the impact of the $i$th observation on the final OLS estimator,
a natural approach is to delete the $i$th row from the full data

\[
X=\left(\begin{array}{c}
x_{1}^{\T}\\
\vdots\\
x_{n}^{\T}
\end{array}\right),\quad Y=\left(\begin{array}{c}
y_{1}\\
\vdots\\
y_{n}
\end{array}\right),
\]
and check how much the OLS estimator changes. Let 
$$
X_{[-i]} = \left(\begin{array}{c}
x_{1}^{\T}\\
\vdots\\
x_{i-1}^{\T} \\
x_{i+1}^{\T} \\
\vdots\\
x_{n}^{\T}
\end{array}\right),\quad 
Y_{[-i]} = \left(\begin{array}{c}
y_{1}\\
\vdots\\
y_{i-1} \\
y_{i+1} \\
\vdots\\
y_{n}
\end{array}\right)
$$ 
denote the leave-$i$-out data, and define
\begin{equation}
\hat{\beta}_{[-i]}=(X_{[-i]}^{\T}X_{[-i]})^{-1}X_{[-i]}^{\T}Y_{[-i]}\label{eq:delete1ols}
\end{equation}
as the corresponding OLS estimator. We can fit $n$ OLS by deleting
the $i$th row $(i=1,\ldots,n)$. However, this is computationally
intensive especially when $n$ is large. Theorem \ref{thm::leave-one-out-beta} shows that we need only to fit OLS once and then compute all leave-one-out coefficients explicitly.

\begin{theorem}\label{thm::leave-one-out-beta}
Recall that $\hat{\beta}$ is the full data OLS, $\hat{\varepsilon}_{i}$
is the residual and $h_{ii}$ is the leverage score for the $i$th
observation. We have
\[
\hat{\beta}_{[-i]}=\hat{\beta}-(1-h_{ii})^{-1}(X^{\T}X)^{-1}x_{i}\hat{\varepsilon}_{i}
\]
if $X^{\T}X$ is invertible and $h_{ii} \neq 1$. 
\end{theorem}

The condition that $X^{\T}X$ is invertible ensures the full OLS has a unique solution. The additional condition $h_{ii} \neq 1$ ensures that the leave-$i$-out OLS has a unique solution; see Problem \ref{hw10::h-grammatrix}.

\begin{myproof}{Theorem}{\ref{thm::leave-one-out-beta}}
From (\ref{eq:delete1ols}), we need to invert 
\[
X_{[-i]}^{\T}X_{[-i]}=\sum_{i'\neq i}x_{i'}x_{i'}^{\T}=X^{\T}X-x_{i}x_{i}^{\T}
\]
 and calculate 
\[
X_{[-i]}^{\T}Y_{[-i]}=\sum_{i'\neq i}x_{i}y_{i}=X^{\T}Y-x_{i}y_{i},
\]
which are the original $X^{\T}X$ and $X^{\T}Y$ without the contribution of the $i$th observation. 
Using the following Sherman--Morrison formula in Problem \ref{hwmath1::inverse-block-matrix}:
\[
(A+uv^{\T})^{-1}=A^{-1}-(1+v^{\T}A^{-1}u)^{-1}A^{-1}uv^{\T}A^{-1}
\]
with $A=X^{\T}X,u=x_i$, and $v=-x_{i}$ we can invert $X_{[-i]}^{\T}X_{[-i]}$
as\footnote{See Problem \ref{hwmath::rank1psd} in Appendix \ref{chapter::linear-algebra} for a related linear algebra result.}
\begin{align*}
(X_{[-i]}^{\T}X_{[-i]})^{-1} & =(X^{\T}X)^{-1}+\left\{ 1-x_{i}^{\T}(X^{\T}X)^{-1}x_{i}\right\} ^{-1}(X^{\T}X)^{-1}x_{i}x_{i}^{\T}(X^{\T}X)^{-1}\\
 & =(X^{\T}X)^{-1}+\left(1-h_{ii}\right)^{-1}(X^{\T}X)^{-1}x_{i}x_{i}^{\T}(X^{\T}X)^{-1}.
\end{align*}
Therefore, 
\begin{align*}
\hat{\beta}_{[-i]} & =(X_{[-i]}^{\T}X_{[-i]})^{-1}X_{[-i]}^{\T}Y_{[-i]}\\
 & =\left\{ (X^{\T}X)^{-1}+\left(1-h_{ii}\right)^{-1}(X^{\T}X)^{-1}x_{i}x_{i}^{\T}(X^{\T}X)^{-1}\right\} \left(X^{\T}Y-x_{i}y_{i}\right)\\
 & =(X^{\T}X)^{-1}X^{\T}Y \\
 &\  -(X^{\T}X)^{-1}x_{i}y_{i}\\
 & \ +\left(1-h_{ii}\right)^{-1}(X^{\T}X)^{-1}x_{i}x_{i}^{\T}(X^{\T}X)^{-1}X^{\T}Y \\
 &\ -\left(1-h_{ii}\right)^{-1}(X^{\T}X)^{-1}x_{i}x_{i}^{\T}(X^{\T}X)^{-1}x_{i}y_{i}\\
 & =\hat{\beta}-(X^{\T}X)^{-1}x_{i}y_{i}+\left(1-h_{ii}\right)^{-1}(X^{\T}X)^{-1}x_{i}x_{i}^{\T}\hat{\beta}-h_{ii}\left(1-h_{ii}\right)^{-1}(X^{\T}X)^{-1}x_{i}y_{i}\\
 & =\hat{\beta}-\left(1-h_{ii}\right)^{-1}(X^{\T}X)^{-1}x_{i}y_{i}+\left(1-h_{ii}\right)^{-1}(X^{\T}X)^{-1}x_{i}\hat{y}_{i}\\
 & =\hat{\beta}-\left(1-h_{ii}\right)^{-1}(X^{\T}X)^{-1}x_{i}\hat{\varepsilon}_{i}.
\end{align*}
\end{myproof}

With the leave-$i$-out OLS estimator $\hat{\beta}_{[-i]}$, we can
define the predicted residual
\[
\hat{\varepsilon}_{[-i]}=y_{i}-x_{i}^{\T}\hat{\beta}_{[-i]},
\]
which is different from the original residual $\hat{\varepsilon}_{i}.$ The
predicted residual based on leave-one-out can better measure the performance
of the prediction because it mimics the real problem of predicting a 
future observation. In contrast, the original residual based on the full
data $\hat{\varepsilon}_i = y_i - x_i^{\T} \hat{\beta}$ gives an overly optimistic measure of the performance of the prediction. This is related
to the overfitting issue discussed in Chapter \ref{chapter::overfitting}. Under the Gauss--Markov
model, Theorem \ref{thm:GMcov} implies that the original residual
has mean zero and variance
\begin{equation}
\var(\hat{\varepsilon}_{i})=\sigma^{2}(1-h_{ii}),\label{eq:varresidualGMmodel}
\end{equation}
and we can show that the predicted residual has mean zero and variance
\begin{equation}
\var(\hat{\varepsilon}_{[-i]})=\var(y_{i}-x_{i}^{\T}\hat{\beta}_{[-i]})=\sigma^{2}+\sigma^{2}x_{i}^{\T}(X_{[-i]}^{\T}X_{[-i]})^{-1}x_{i}.\label{eq:varpresidual-1}
\end{equation}
Theorem \ref{theorem:looresidual} below further simplifies the predicted residual and
its variance.

\begin{theorem}
\label{theorem:looresidual}
Assume $h_{ii} \neq 1$. 
We have 
$$
\hat{\varepsilon}_{[-i]}=\hat{\varepsilon}_{i}/(1-h_{ii}),
$$
and under Assumption \ref{assume::gm-model}, we have 
\begin{equation}
\var(\hat{\varepsilon}_{[-i]})=\sigma^{2}/(1-h_{ii}).\label{eq:varpresidual-2}
\end{equation}
\end{theorem}

\begin{myproof}{Theorem}{\ref{theorem:looresidual}}
By definition and Theorem \ref{thm::leave-one-out-beta}, we have
\begin{align}
\hat{\varepsilon}_{[-i]} & =y_{i}-x_{i}^{\T}\hat{\beta}_{[-i]}\nonumber \\
 & =y_{i}-x_{i}^{\T}\left\{ \hat{\beta}-\left(1-h_{ii}\right)^{-1}(X^{\T}X)^{-1}x_{i}\hat{\varepsilon}_{i}\right\} \nonumber \\
 & =y_{i}-x_{i}^{\T}\hat{\beta}+\left(1-h_{ii}\right)^{-1}x_{i}^{\T}(X^{\T}X)^{-1}x_{i}\hat{\varepsilon}_{i}\nonumber \\
 & =\hat{\varepsilon}_{i}+h_{ii}\left(1-h_{ii}\right)^{-1}\hat{\varepsilon}_{i}\nonumber \\
 & =\hat{\varepsilon}_{i}/(1-h_{ii})\text{.}\label{eq:predresidualformula}
\end{align}
Combining (\ref{eq:varresidualGMmodel}) and (\ref{eq:predresidualformula}), we can derive its variance:
$$
\var(\hat{\varepsilon}_{[-i]})
= \var( \hat{\varepsilon}_{i} ) / (1-h_{ii})^2
= \sigma^{2}(1-h_{ii}) / (1-h_{ii})^2
=\sigma^{2}/(1-h_{ii}) . 
$$
\end{myproof}

Comparing formulas (\ref{eq:varpresidual-1}) and (\ref{eq:varpresidual-2}),
we obtain that 
\[
1+x_{i}^{\T}(X_{[-i]}^{\T}X_{[-i]})^{-1}x_{i}=(1-h_{ii})^{-1}=\left\{ 1 -  x_{i}^{\T}(X^{\T}X)^{-1}x_{i}\right\} ^{-1}.
\]
This  is not an obvious linear algebra identity, but it follows immediately from
the two ways of calculating the variance of the predicted residual.

\section{Applications of the leave-one-out formulas}

\subsection{Gauss updating formula}\label{sec::gauss-updating}

Consider an online setting in which the data points come sequentially
as illustrated by the figure below:

\begin{figure}[h]
\centering
\includegraphics[width = \textwidth]{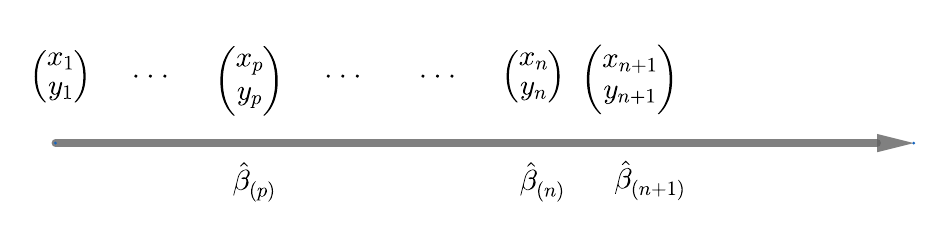}
\end{figure}

In this setting, we can update
the OLS estimator step by step: based on the first $n$ data points
$(x_{i},y_{i})_{i=1}^{n}$, we calculate the OLS estimator $\hat{\beta}_{(n)}$,
and with an additional data point $(x_{n+1},y_{n+1})$, we update
the OLS estimator as $\hat{\beta}_{(n+1)}$. These two OLS estimators
are closely related as shown in Theorem \ref{theorem:gaussupdate} below. 

\begin{theorem}
\label{theorem:gaussupdate}
Let $X_{(n)}$ be the design matrix and $Y_{(n)}$
be the outcome vector for the first $n$ observations. We have 
$$
\hat{\beta}_{(n+1)}=\hat{\beta}_{(n)}+\gamma_{(n+1)}\hat{\varepsilon}_{[n+1]},
$$
where $\gamma_{(n+1)}=(X_{(n+1)}^{\T}X_{(n+1)})^{-1}x_{n+1}$ and
$\hat{\varepsilon}_{[n+1]}=y_{n+1}-x_{n+1}^{\T}\hat{\beta}_{(n)}$
is the predicted residual of the $(n+1)$the outcome based on the
OLS of the first $n$ observations. 
\end{theorem}

\begin{myproof}{Theorem}{\ref{theorem:gaussupdate}}
This is the reverse form of the leave-one-out formula in Theorem \ref{thm::leave-one-out-beta}. We can view
the first $n+1$ data points as the full data, and $\hat{\beta}_{(n)}$
as the OLS estimator leaving the $(n+1)$the observation out.
Applying Theorem \ref{thm::leave-one-out-beta}, we have 
\begin{align*}
\hat{\beta}_{(n)} & =\hat{\beta}_{(n+1)}-(X_{(n+1)}^{\T}X_{(n+1)})^{-1}x_{n+1}\frac{\hat{\varepsilon}_{n+1}}{1-h_{n+1,n+1}}\\
 & =\hat{\beta}_{(n+1)}-\gamma_{(n+1)}\hat{\varepsilon}_{[n+1]},
\end{align*}
where $\hat{\varepsilon}_{n+1}$ is the $(n+1)$th residual based
on the full data OLS, and the $(n+1)$th predicted residual equals
$\hat{\varepsilon}_{[n+1]}=\hat{\varepsilon}_{n+1}/(1-h_{n+1,n+1})$
based on Theorem \ref{theorem:looresidual}.
\end{myproof}

Theorem \ref{theorem:gaussupdate} shows that to obtain $\hat{\beta}_{(n+1)}$
from $\hat{\beta}_{(n)}$, the adjustment depends on the predicted
residual $\hat{\varepsilon}_{[n+1]}$. If we have a perfect prediction
of the $(n+1)$th observation based on $\hat{\beta}_{(n)}$, then
we do not need to make any adjustment to obtain $\hat{\beta}_{(n+1)}$;
if the predicted residual is large, then we need to make a large
adjustment. 

Theorem \ref{theorem:gaussupdate} suggests an algorithm for sequentially
computing the OLS estimators. But it gives a formula that involves inverting $X_{(n+1)}^{\T}X_{(n+1)}$ at each step.
Using the Sherman--Morrison formula in Problem \ref{hwmath1::inverse-block-matrix}
for updating the inverse of $X_{(n+1)}^{\T}X_{(n+1)}$ based on the
inverse of $X_{(n)}^{\T}X_{(n)}$, we have an even simpler algorithm
below:
\begin{enumerate}[label=(G\arabic*), ref=G\arabic*]
\item\label{alg::gaussupdate1} Start with $V_{(n)}=(X_{(n)}^{\T}X_{(n)})^{-1}$ and $\hat{\beta}_{(n)}$.
\item Update 
\[
V_{(n+1)}=V_{(n)}-\left\{ 1+x_{n+1}^{\T}V_{(n)}x_{n+1}\right\} ^{-1}V_{(n)}x_{n+1}x_{n+1}^{\T}V_{(n)}.
\]
\item Calculate $\gamma_{(n+1)}=V_{(n+1)}x_{n+1}$ and $\hat{\varepsilon}_{[n+1]}=y_{n+1}-x_{n+1}^{\T}\hat{\beta}_{(n)}$.
\item\label{alg::gaussupdate4} Update $\hat{\beta}_{(n+1)}=\hat{\beta}_{(n)}+\gamma_{(n+1)}\hat{\varepsilon}_{[n+1]}$. 
\end{enumerate}

\subsection{Outlier detection based on residuals}

Under the Normal linear model $Y=X\beta+\varepsilon$ with $\varepsilon\sim\N(0,\sigma^{2}I_{n})$,
we know some basic probabilistic properties of the residual vector:
\[
E(\hat{\varepsilon})=0,\qquad\var(\hat{\varepsilon})=\sigma^{2}(I_{n}-H).
\]
At the same time, the residual vector is computable based on the data.
So it is sensible to check whether these properties of the residual
vector are plausible based on data, which in turn serves as modeling
checking for the Normal linear model. 

The first quantity is the standardized residual
\[
\text{standr}_{i}=\frac{\hat{\varepsilon}_{i}}{\sqrt{\hat{\sigma}^{2}(1-h_{ii})}}.
\]
We may hope that it has mean 0 and variance 1. However, because
of the dependence between $\hat{\varepsilon}_{i}$ and $\hat{\sigma}^{2}$, it is not easy to find the exact distribution
of $\text{standr}_{i}$. 

The second quantity is the studentized residual
based on the predicted residual:
\[
\text{studr}_{i}=\frac{\hat{\varepsilon}_{[-i]}}{\sqrt{\hat{\sigma}_{[-i]}^{2}/(1-h_{ii})}}=\frac{y_{i}-x_{i}^{\T}\hat{\beta}_{[-i]}}{\sqrt{\hat{\sigma}_{[-i]}^{2}/(1-h_{ii})}},
\]
where $\hat{\beta}_{[-i]}$ and $\hat{\sigma}_{[-i]}^{2}$ are the estimates of the coefficient and variance based
on the leave-$i$-out OLS. Because $(y_{i},\hat{\beta}_{[-i]},\hat{\sigma}_{[-i]}^{2})$
are mutually independent under the Normal linear model, we can show
that 
\begin{equation}
\text{studr}_{i}\sim t_{n-p-1}.\label{eq:studentized-t}
\end{equation}
I leave the rigorous proof of \eqref{eq:studentized-t} as Problem \ref{hw10::studentized-residual}. 
Because we know the  distribution of $\text{studr}_{i}$, we can compare it with the quantiles of the $t$ distribution. 

The third quantity is Cook's distance \citep{cook1977detection}:
\begin{align*}
\text{cook}_{i} & =(\hat{\beta}_{[-i]}-\hat{\beta})^{\T}X^{\T}X(\hat{\beta}_{[-i]}-\hat{\beta})/(p\hat{\sigma}^{2})\\
 & =(X\hat{\beta}_{[-i]}-X\hat{\beta})^{\T}(X\hat{\beta}_{[-i]}-X\hat{\beta})/(p\hat{\sigma}^{2}),
\end{align*}
 where the first form measures the change of the OLS estimator and
the second form measures the change in the predicted values based
on leaving-$i$-out. It has a slightly different motivation, but
eventually, it is related to the previous two residuals due to the leave-one-out formulas. 

\begin{theorem}\label{thm::cooksdist-standr}
Cook's distance is related to the standardized residual via:
\[
\textup{cook}_{i}=\textup{standr}_{i}^{2}\times\frac{h_{ii}}{p(1-h_{ii})}.
\]
\end{theorem}

I leave the proof of Theorem \ref{thm::cooksdist-standr}  as Problem \ref{hw10::cooks-standr}.

\begin{figure}[ht]
\centering
\includegraphics[width = \textwidth]{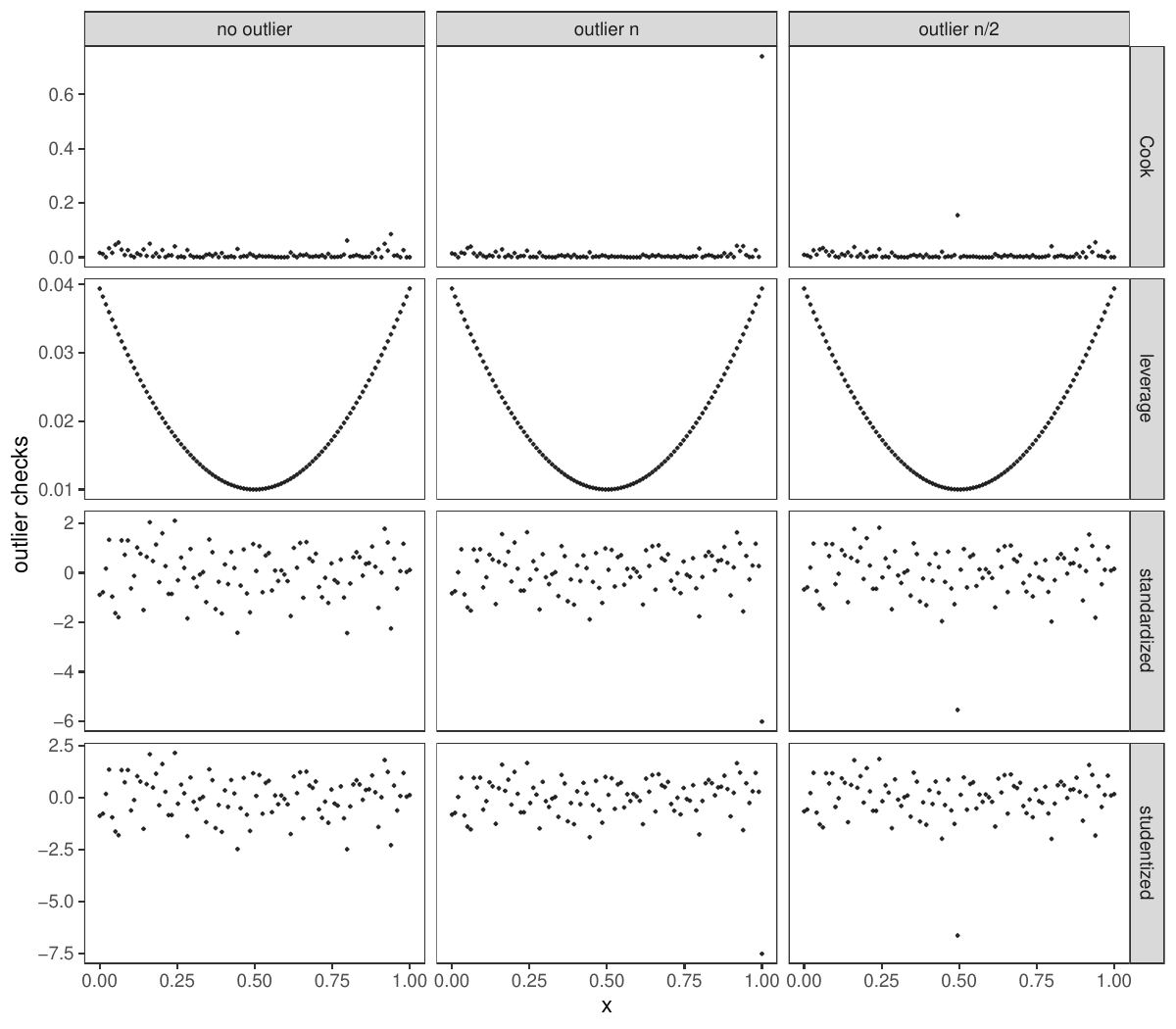}
\caption{Outlier detections}\label{fig::outlier-detections}
\end{figure}

I will end this subsubsection with two examples.  The first one is simulated. 
I generate data from a univariate Normal linear model without outliers. I then use \ri{hatvalues}, \ri{r.standard}, \ri{r.student} and \ri{cooks.distance} to an \ri{lm.object} to calculate the leverage scores, standardized residuals, studentized residuals, and Cook's distances. Their plots are in the first column of Figure \ref{fig::outlier-detections}. 
\begin{lstlisting}
n = 100
x = seq(0, 1, length = n)
y = 1 + 3*x + rnorm(n)
lmmod = lm(y ~ x)
hatvalues(lmmod)
rstandard(lmmod)
rstudent(lmmod)
cooks.distance(lmmod)
\end{lstlisting}
If I add $8$ to the outcome of the last observation, the plots change to the second column of Figure \ref{fig::outlier-detections}. If I add $8$ to the $50$th observation, the plots change to the last column of Figure \ref{fig::outlier-detections}. Both visually show the outliers. 
 In this example, the three residual plots give qualitatively the same pattern, so the choice among them does not matter much. In general cases, we may prefer $\text{studr}_{i}$ because it has a known distribution under the Normal linear model. 

\begin{figure}[ht]
\centering
\includegraphics[width = \textwidth]{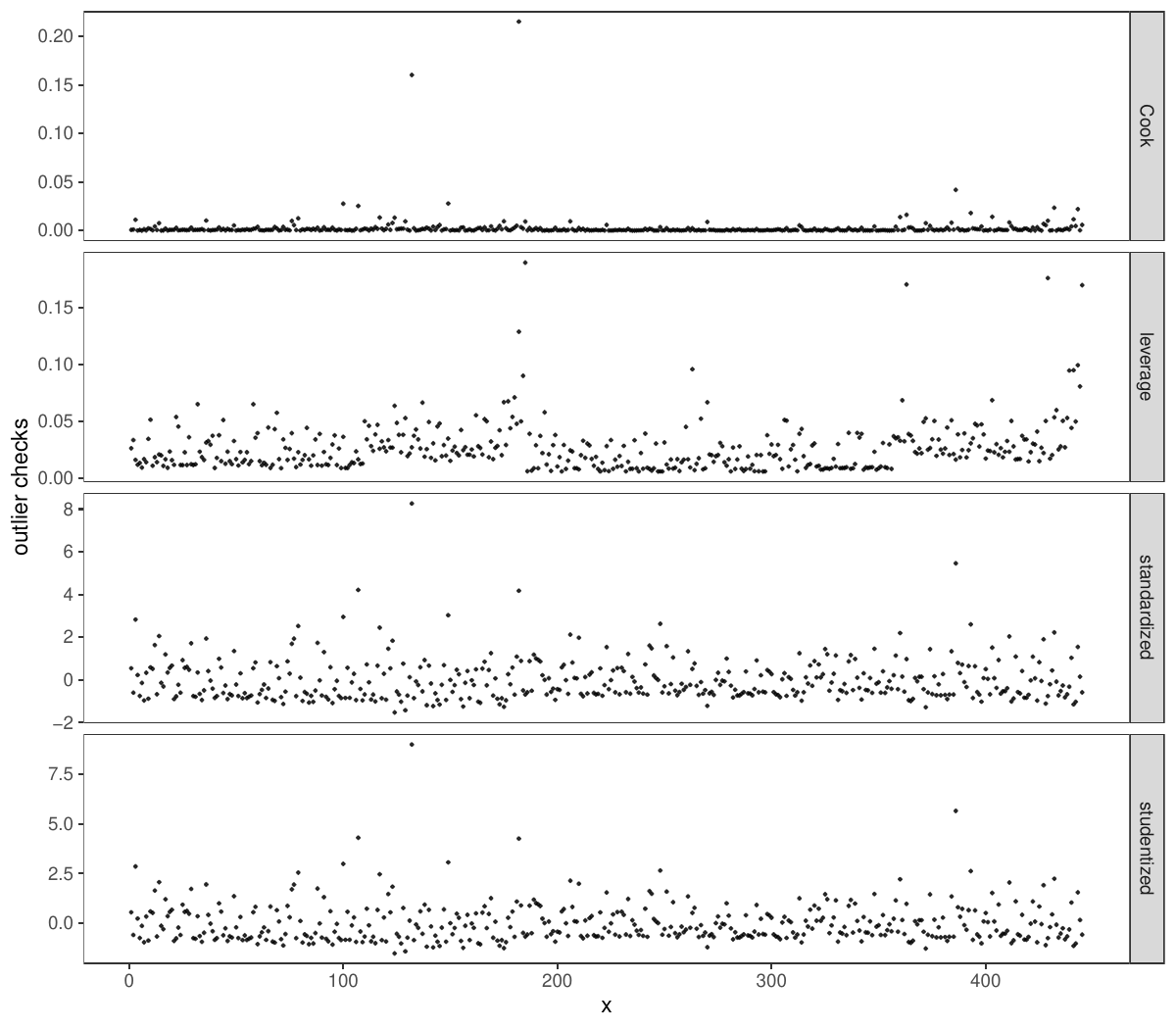}
\caption{Outlier detections in the LaLonde data}\label{fig::outlier-detections-lalonde}
\end{figure}

The second one is a further analysis of the Lalonde data. Based on the plots in Figure \ref{fig::outlier-detections-lalonde}, there are indeed some outliers in the data. It is worth investigating them more carefully. 

Although the outliers detection methods above are very classic, they are rarely implemented in modern data analyses. They are simple and useful diagnostics. I recommend using them at least as a part of the exploratory data analysis.

\subsection{Jackknife}

Jackknife is a general strategy for bias reduction and variance estimation proposed by \citet{quenouille1949approximate, quenouille1956notes} and popularized by  \citet{tukey1958bias}. Based on independent data $(Z_1,\ldots,Z_n)$, how to estimate the variance of a general estimator $\hat{\theta}(Z_1,\ldots,Z_n)$ of the parameter $\theta$? Define $\hat{\theta}_{[-i]}$ as the estimator without observation $i$, and the pseudo-value as 
$$
\tilde{\theta}_i = n \hat{\theta} - (n-1) \hat{\theta}_{[-i]} .
$$ 
The jackknife point estimator is $\hat{\theta}_\textsc{j} = n^{-1} \sumn \tilde{\theta}_i$, and the jackknife variance estimator is 
 $$
\hat{V}_\textsc{j}  = \frac{1}{n(n-1)}\sumn (\tilde{\theta}_i  - \hat{\theta}_\textsc{j}  )(\tilde{\theta}_i - \hat{\theta}_\textsc{j}  )^{\T}.
 $$
Where does the idea of jackknife come from? Honestly, I do not know. However, as a sanity check, we can test whether it makes sense in OLS. See details below.

We have already shown that the OLS coefficient is unbiased and derived several variance estimators for it. 
Here we focus on the jackknife in OLS using the leave-one-out formula for the coefficient. The pseudo-value is
\begin{eqnarray*}
\tilde{\beta}_i &=& n \hat{\beta} - (n-1) \hat{\beta}_{[-i]} \\
&=& n \hat{\beta} - (n-1) \left\{ \hat{\beta}-(1-h_{ii})^{-1}(X^{\T}X)^{-1}x_{i}\hat{\varepsilon}_{i} \right\} \\
&=& \hat{\beta}  + (n-1)  (1-h_{ii})^{-1}(X^{\T}X)^{-1}x_{i}\hat{\varepsilon}_{i} .
\end{eqnarray*}
The jackknife point estimator is 
$$
\hat{\beta}_\textsc{j} = \hat{\beta} + \frac{n-1}{n} \left( n^{-1} \sumn x_i x_i^{\T} \right)^{-1} 
\left(   n^{-1} \sumn x_{i}  \frac{  \hat{\varepsilon}_{i} }{ 1-h_{ii} }   \right) .
$$
 It is a little unfortunate that the jackknife point estimator is not identical to the OLS estimator, which is BLUE under the Gauss--Markov model.  
We can show that  $E(\hat{\beta}_\textsc{j} )  = \beta$ and it is a linear estimator. So the Gauss--Markov theorem ensures that $\cov( \hat{\beta}_\textsc{j} ) \succeq \cov( \hat{\beta})$. (Readers, make sure these two sentences make sense to you; see Problem \ref{hw::ols-jackknife}.)

 Nevertheless, the difference between $\hat{\beta}_\textsc{j}$ and $\hat{\beta}$  is quite small. I omit their difference in the following derivation. Assuming that $\hat{\beta}_\textsc{j}  \cong \hat{\beta} $, we can continue to calculate the approximate jackknife variance estimator:
\begin{eqnarray*}
\hat{V}_\textsc{j} &\cong& \frac{1}{n(n-1)} \sumn (\tilde{\beta}_i - \hat{\beta})(\tilde{\beta}_i  - \hat{\beta})^{\T} \\
&=& \frac{n-1}{n} (X^{\T}X)^{-1}  \sumn \left(  \frac{  \hat{\varepsilon}_{i}  }{ 1-h_{ii}} \right)^2 x_{i}x_{i}^{\T}  (X^{\T}X)^{-1},
\end{eqnarray*} 
which is almost identical to the HC3 form of the EHW covariance matrix introduced in Chapter \ref{sec::HCs}. \citet{miller1974unbalanced} first analyzed the jackknife in OLS but dismissed it immediately. \citet{hinkley1977jackknifing} modified the original jackknife and proposed a version that is identical to HC1, and \citet{wu1986jackknife} proposed some further modifications and proposed a version that is identical to HC2. 
\citet{weber1986jackknife} made connections between EHW and jackknife standard errors. 
However, \citet{long2000using}'s finite-sample simulation seems to favor the original jackknife or HC3.

\section{Homework problems}

\paragraph{Implementing the Gauss updating formula}
Implement the algorithm in \eqref{alg::gaussupdate1}--\eqref{alg::gaussupdate4}, and try it on simulated data.

\paragraph{The distribution of the studentized residual}\label{hw10::studentized-residual}

Prove (\ref{eq:studentized-t}).

\paragraph{Leave-one-out coefficient}\label{hw10::loo-coefficient}

Prove 
\[
\hat{\beta}=\sumn w_{i}\hat{\beta}_{[-i]},
\]
and find the weights $w_{i}$'s. Prove that they are positive and sum to
one. Does $\hat{\beta}=n^{-1}\sumn\hat{\beta}_{[-i]}$ hold in general?

\paragraph{Cook's distance and the standardized residual}\label{hw10::cooks-standr}

Prove Theorem \ref{thm::cooksdist-standr}.

\paragraph{The relationship between the standardized and studentized residual}\label{hw10::residuals-relationship}

Prove Theorem \ref{thm::standardized-studentized} below. 

\begin{theorem}\label{thm::standardized-studentized}
We have 
\begin{enumerate}
\item $(n-p-1)\hat{\sigma}_{[-i]}^{2}=(n-p)\hat{\sigma}^{2}-\hat{\varepsilon}_{i}^{2}/(1-h_{ii})$.

\item There is a monotone relationship between the standardized and studentized
residual:
\[
\text{studr}_{i}=\text{standr}_{i}\sqrt{\frac{n-p-1}{n-p-\text{standr}_{i}^{2}}}.
\]
\end{enumerate}
\end{theorem}

Remark: The right-hand side of the first result must be nonnegative so  $\sum_{k=1}^n   \hat{\varepsilon}_k^2 \geq \hat{\varepsilon}_{i}^{2}/(1-h_{ii})$, which 
implies the following inequality:
$$
h_{ii}  + \frac{  \hat{\varepsilon}_i^2 }{ \sum_{k=1}^n   \hat{\varepsilon}_k^2 } \leq 1.
$$
From this inequality, if $h_{ii} = 1$ then $\hat{\varepsilon}_i = 0$ which further implies that $h_{ij} = 0$ for all $j \neq i.$


\paragraph{More on the jackknife estimator in OLS}\label{hw::ols-jackknife}

Prove that under the Gauss--Markov model, 
$E(\hat{\beta}_\textsc{j} )  = \beta$ and  $\cov( \hat{\beta}_\textsc{j} ) \succeq \cov( \hat{\beta})$.

\paragraph{Subset and full-data OLS coefficients}
\label{hw10::subset-ols-coef}

Leaving one observation out, we have the OLS coefficient formula in Theorem \ref{thm::leave-one-out-beta}. Leave a subset of observations out, we have the OLS coefficient formula below. Partition the covariate matrix and outcome vector based on a subset $S$ of $\{1, \ldots, n\}$:
$$
X = \begin{pmatrix}
X_S\\
X_{\backslash S}
\end{pmatrix},\qquad 
Y = \begin{pmatrix}
Y_S\\
Y_{\backslash S}
\end{pmatrix} .
$$  
Without using the observations in set $S$, we have the OLS coefficient 
$$
\hat\beta_{\backslash S} = (X_{\backslash S}^{\T} X_{\backslash S})^{-1} X_{\backslash S}Y_{\backslash S}.
$$
The corresponding leave-$S$-out residual vector is
$$
\hat{\varepsilon}_{\backslash S} = Y_S - X_S \hat\beta_{\backslash S}. 
$$
Theorem \ref{thm::leave-subset-out-beta} below facilitates the computation of many $\hat\beta_{\backslash S} $'s and $\hat\beta_{\backslash S}$'s simultaneously, without running the OLS for each $S$.  It relies crucially on the subvector of the residuals 
$$
\hat{\varepsilon}_S = Y_S - X_S\hat\beta 
$$  
and the submatrix of $H$
$$
H_{SS} = X_{S}(X^{\T}X)^{-1}X_{S}^{\T}
$$
corresponding to the set $S$. 
Prove Theorem \ref{thm::leave-subset-out-beta}. 

\begin{theorem}\label{thm::leave-subset-out-beta}
Assume $X^{\T}X$ and $I-H_{SS}$ are both invertible, where $I$ is the identity matrix with the same dimension as  $H_{SS}$. 
Recall that $\hat{\beta}$ is the full data OLS. 
We have
\[
\hat\beta_{\backslash S}   =\hat{\beta}- (X^{\T}X)^{-1}  X_{S}^{\T} (I-H_{SS})^{-1} \hat{\varepsilon}_S 
\]
and
$$
\hat{\varepsilon}_{\backslash S}
= (I-H_{SS})^{-1} \hat{\varepsilon}_S . 
$$
\end{theorem}

\paragraph{Gauss updating formula with batches of data}
\label{hw10::gauss-updating-batch}

This problem extends the discussion in Chapter \ref{sec::gauss-updating}.
Assume the data come in batches, with $X_b \in \mathbb{R}^{n_b\times p}$ and $Y_b \in \mathbb{R}^{n_b}$ for $b=1,\ldots, B, B+1$. Derive the Gauss-type updating formula and design an algorithm to compute the OLS coefficients efficiently.

\paragraph{Bounding the leverage score}
\label{hw10::bounding-h-more}

With the intercept included in the OLS, Theorem \ref{thm::leverage-mdist} shows $ n^{-1} \leq h_{ii} \leq 1$ for all $i = 1, \ldots , n$.
Prove the following stronger result:
$$
n^{-1} \leq h_{ii} \leq   s_i^{-1}
$$
where $s_i$ is the number of rows that are identical to $x_i$ or $-x_i$.

\paragraph{More on the leverage score}
\label{hw10::h-grammatrix}

Prove Theorem \ref{thm::leave-one-out-det} below.   
 
\begin{theorem}
\label{thm::leave-one-out-det}
$\textup{det}(X_{[-i]}^{\T}X_{[-i]}) = (1-h_{ii}) \textup{det}(X^{\T} X). $
\end{theorem}

Remark: If $h_{ii} =1$, then $X_{[-i]}^{\T}X_{[-i]}$ is degenerate with determinant $0$. Therefore, if we delete an observation $i$ with leverage score $1$, the columns of the covariate matrix become linearly dependent.

\chapter{Population Ordinary Least Squares and Misspecified Linear Model}
 \label{chapter::populationOLS}
 
 Previous chapters assume fixed $X$ and random $Y$. We can also view each observation $(x_i, y_i)$ as IID draws from a population and discuss population OLS. The population OLS allows us to achieve the following goals:
 \begin{enumerate}[label=(G\arabic*), ref=G\arabic*]
 \item
 We can view the OLS from the level of random variables instead of data points. 
 \item
The population OLS  facilitates the discussion of the properties of misspecified linear models.
\item
The population OLS motivates a prediction procedure called {\it conformal prediction} without imposing any distributional assumptions. 
 \end{enumerate}

\section{Population OLS}

Assume that $(x_{i},y_{i})_{i=1}^{n}$ are IID with the same distribution
as $(x,y)$, where $x\in\mathbb{R}^{p}$ and $y\in\mathbb{R}$. 
Below I will use $(x,y)$ to denote a general observation, dropping the subscript $i$ for simplicity. 
Let
the joint distribution be $F(x,y)$, and $E(\cdot)$, $\var(\cdot)$,
and $\cov(\cdot)$ be the expectation, variance, and covariance under
this joint distribution. We want to use $x$ to predict $y$.
Theorem \ref{thm::equivalent-definition-of-conditional-mean} below states that the conditional expectation $E(y\mid x)$
is the best predictor in terms of the mean squared error.

\begin{theorem}\label{thm::equivalent-definition-of-conditional-mean}
For any function $m(x)$, we have the decomposition 
\begin{equation}
E\left[\left\{ y-m(x)\right\} ^{2}\right]=E\left[\left\{ E(y\mid x)-m(x)\right\} ^{2}\right]+ E\{ \var(y\mid x) \},\label{eq:decomposition-popols}
\end{equation}
provided the existence of the moments in \eqref{eq:decomposition-popols}. The decomposition \eqref{eq:decomposition-popols} implies
\[
E(y\mid x)=\arg\min_{m(\cdot)}E\left[\left\{ y-m(x)\right\} ^{2}\right]
\]
with the minimum value equaling $E\{ \var(y\mid x)\} ,$ the expectation of the conditional
variance of $y$ given $x$. 
\end{theorem}

Theorem \ref{thm::equivalent-definition-of-conditional-mean} is well known in probability theory. I relegate its proof as  Problem \ref{hw11::population-mean}. 
We have finite data points, but the function $E(y\mid x)$ lies in
an infinite dimensional space. Nonparametric estimation of $E(y\mid x)$
is generally a hard problem, especially with a multidimensional $x$.
As a starting point, we often use a linear function of $x$ to approximate
$E(y\mid x)$ and define the population OLS coefficient as
\[
\beta=\arg\min_{b\in\mathbb{R}^{p}} \mathcal{L}(b) ,
\]
where
\begin{align*}
\mathcal{L}(b) & =E\left\{ (y-x^{\T}b)^{2}\right\} \\
 & =E\left\{ y^{2}+b^{\T}xx^{\T}b-2yx^{\T}b\right\} \\
 & =E(y^{2})+b^{\T}E\left(xx^{\T}\right)b-2E\left(yx^{\T}\right)b 
\end{align*}
 is a quadratic function of $b$. From the first-order condition,
we have
\[
\frac{\partial\mathcal{L}(b)}{\partial b}\mid_{b=\beta}=2E\left(xx^{\T}\right)\beta-2E\left(xy\right)=0
\]
based on Proposition \ref{prop::vector-calculus} in Appendix \ref{chapter::linear-algebra}, 
so
\begin{equation}
\label{eq::pols-formula}
\beta= \{ E(xx^{\T}) \} ^{-1}E\left(xy\right) , 
\end{equation}
if $E(xx^{\T})$ is non-degenerate. 
From the second-order condition, 
\[
\frac{\partial^{2}\mathcal{L}(b)}{\partial b\partial b^{\T}}=2E\left(xx^{\T}\right) 
\]
is positive definite. 
So $\beta$ is the unique minimizer of $\mathcal{L}(b)$.

The above derivation shows that $x^{\T}\beta$ is the best linear predictor. Theorem \ref{thm:bestlinear-conditionalmean} below states  that $x^{\T}\beta$ is
the best linear approximation to the possibly nonlinear conditional mean $E(y\mid x).$
\begin{theorem}
\label{thm:bestlinear-conditionalmean}
If $E(xx^{\T})$ is non-degenerate, then
\begin{eqnarray*}
\beta &=& \arg\min_{b\in\mathbb{R}^{p}}E\left[\left\{ E(y\mid x)-x^{\T}b\right\} ^{2}\right] \\
&=& \{ E(xx^{\T} ) \} ^{-1}E\left\{ xE(y\mid x)\right\} .
\end{eqnarray*}
\end{theorem}

As a special case, when the covariate ``$x$'' in the above formulas contains $1$ and a scalar $x$,
the OLS coefficient has the following form.
\begin{corollary}
\label{corollary:scalar-pop-ols}
For scalar $x$ and $y$, we have 
\[
(\alpha,\beta)=\arg\min_{a,b}E(y-a-bx)^{2},
\]
where $\alpha=E(y)-E(x)\beta$ and 
\[
\beta=\frac{\cov(x,y)}{\var(x)} = \rho_{xy} \sqrt{   \frac{\var(y)}{\var(x)}  }
\]
\end{corollary}

Recall that 
$$
\rho_{xy}  = \frac{   \cov(x,y)   }{   \sqrt{   \var(x) \var(y)  }    }
$$
is the population Pearson correlation coefficient. So Corollary \ref{corollary:scalar-pop-ols} gives the population version of the Galtonian formula. 
I leave the proofs of Theorem \ref{thm:bestlinear-conditionalmean} and Corollary \ref{corollary:scalar-pop-ols} as Problems \ref{hw11::condition-mean-app} and \ref{hw11::univariate-ols}.

With $\beta$, we can define
\begin{equation}
\varepsilon=y-x^{\T}\beta\label{eq:populationresidualols}
\end{equation}
as the population residual. Because we usually do not use the upper-case letter $E$ for $\varepsilon$, this notation may cause confusion with previous discussion on OLS,  where $\varepsilon$ denotes the vector of the error terms. Here $\varepsilon$ is a scalar in \eqref{eq:populationresidualols}.   
By the definition of $\beta$, we can verify
\begin{equation}
E(x\varepsilon)=E\left\{ x(y-x^{\T}\beta)\right\} =E(xy)-E(xx^{\T})\beta=0.\label{eq:ols-uncorrelated}
\end{equation}
If we include $1$ as a component of $x$, then $E(\varepsilon)=0$,
which, coupled with (\ref{eq:ols-uncorrelated}), implies $\cov(x,\varepsilon)=0$.
So with an intercept in $\beta$, the mean of the population residual
must be zero, and it is uncorrelated with other covariates by construction. 

We can also rewrite (\ref{eq:populationresidualols}) as
\begin{equation}
y=x^{\T}\beta+\varepsilon,\label{eq:population-ols-decomposition}
\end{equation}
which holds by the definition of the population OLS coefficient and residual without any modeling
assumption. We call (\ref{eq:population-ols-decomposition}) with \eqref{eq::pols-formula} and \eqref{eq:populationresidualols} the {\it  population
OLS decomposition}.

\section{Population FWL Theorem and Cochran's formula}

Assume $(y_i, x_{i1}, x_{i2})_{i=1}^n$ are IID, where $y_i$ is a scalar, $x_{i1}$ has dimension $k$, and $x_{i2}$ has dimension $l$. 
With
$$
x = \begin{pmatrix}
x_1\\
x_2
\end{pmatrix}, 
\quad
\beta = \begin{pmatrix}
\beta_1 \\
\beta_2
\end{pmatrix},
$$
define the population OLS of $y$ on $x_1$ and $x_2$ as
\begin{equation}\label{eq::population-ols-long}
y = \beta^{\T} x =  \beta_1^{\T} x_1 +  \beta_2 ^{\T} x_2+ \varepsilon.
\end{equation}
To aid the interpretation of the population OLS coefficient, we have the following population FWL theorem.

\begin{theorem}
[Population FWL Theorem]
\label{thm::population-fwl}
Consider the population OLS decomposition \eqref{eq::population-ols-long}. 
Define $\tilde{x}_2$ as the residual of the component-wise\footnote{If $x_2$ is a vector, we run population OLS of each component of $x_2$ on $x_1$ to obtain the residual. Vectorize the residuals to obtain $\tilde{x}_2$.} population OLS of $x_2$ on $x_1$:
$$
\tilde{x}_2 = x_2 - E(x_2 x_1^{\T}) E(x_1 x_1^{\T})^{-1} x_1.
$$
Define $\tilde{y}$ as the residual of the population OLS of $y$ on $x_1$:
$$
\tilde{y} = y - E(y x_1^{\T}) E(x_1 x_1^{\T})^{-1} x_1.
$$

Then the coefficient $\beta_2$ has the following equivalent forms:
\begin{eqnarray}
\beta_2 &=&  [ E(xx^{\T})^{-1} E(xy)]_{\textup{the last $l$ components}} \label{eq::population-ols-definition1} \\
&=& E(\tilde{x}_2\tilde{x}_2^{\T})^{-1} E(\tilde{x}_2 y) \label{eq::population-ols-partial1} \\
&=& E(\tilde{x}_2\tilde{x}_2^{\T})^{-1} E(\tilde{x}_2 \tilde{y}) . \label{eq::population-ols-partial2}
\end{eqnarray} 
\end{theorem}

The form \eqref{eq::population-ols-definition1} is the definition of $\beta_2$ from the population OLS \eqref{eq::population-ols-long}. 
The form \eqref{eq::population-ols-partial1} states that $\beta_2$ equals the population OLS coefficient of $y$ on $\tilde{x}_2$. 
The form \eqref{eq::population-ols-partial2} states that $\beta_2$ equals the population OLS coefficient of $\tilde{y}$ on $\tilde{x}_2$. 
\citet{angrist2008mostly} provide a special case of Theorem \ref{thm::population-fwl} with $l=1$.

\begin{myproof}{Theorem}{\ref{thm::population-fwl}}
Introduce the notation:
$$
E\left( 
\begin{pmatrix}
y\\
x_1\\
x_2
\end{pmatrix}
\begin{pmatrix}
y & x_1^{\T} & x_2^{\T}
\end{pmatrix}
\right) 
= \begin{pmatrix}
E(y^2) & E(y x_1^{\T}) & E(y x_2^{\T}) \\
E(x_1 y) & E(x_1 x_1^{\T}) & E(x_1 x_2^{\T}) \\
E(x_2 y) & E(x_2 x_1^{\T}) & E(x_2 x_2^{\T}) \\
\end{pmatrix}
= \begin{pmatrix}
\Sigma_{00} & \Sigma_{01} & \Sigma_{02} \\
\Sigma_{10} & \Sigma_{11} & \Sigma_{12} \\
\Sigma_{20} & \Sigma_{21} & \Sigma_{22}
\end{pmatrix}.
$$
By definition of the population OLS of $y$ on $x_1$ and $x_2$, we have
$$
\begin{pmatrix}
\beta_1 \\
\beta_2
\end{pmatrix}
= \begin{pmatrix}
\Sigma_{11} & \Sigma_{12} \\
\Sigma_{21} & \Sigma_{22}
\end{pmatrix}^{-1} 
\begin{pmatrix}
\Sigma_{10} \\
\Sigma_{20}
\end{pmatrix}.
$$
Use the first form of the inverse of $2\times 2$ block matrix in Problem \ref{hwmath1::inverse-block-matrix} to obtain
\begin{eqnarray*}
\beta_2 
&=& \begin{pmatrix}
* & *\\
- \Sigma_{22\mid 1}^{-1} \Sigma_{21}  \Sigma_{11}^{-1} & \Sigma_{22\mid 1}^{-1}
\end{pmatrix}
\begin{pmatrix}
\Sigma_{10} \\
\Sigma_{20}
\end{pmatrix} \\
&=& - \Sigma_{22\mid 1}^{-1} \Sigma_{21}  \Sigma_{11}^{-1}  \Sigma_{10} 
+ \Sigma_{22\mid 1}^{-1}\Sigma_{20} \\
&=& \Sigma_{22\mid 1}^{-1} (\Sigma_{20}- \Sigma_{21}  \Sigma_{11}^{-1}  \Sigma_{10} ) , 
\end{eqnarray*}
where $\Sigma_{22\mid 1} = \Sigma_{22} -\Sigma_{21} \Sigma_{11}^{-1} \Sigma_{12} $ and $*$ signifies unimportant terms.

Now we consider the population OLS coefficient of $y$ on $\tilde{x}_2$:
\begin{eqnarray*}
&& E(\tilde{x}_2\tilde{x}_2^{\T})^{-1} E(\tilde{x}_2 y) \\ 
&=& [ E \{  (x_2 - \Sigma_{21} \Sigma_{11}^{-1} x_1) (x_2^{\T} - x_1 ^{\T} \Sigma_{11}^{-1} \Sigma_{12} ) \} ]^{-1}
E\{  (x_2 - \Sigma_{21} \Sigma_{11}^{-1} x_1)  y  \} \\
&=& (\Sigma_{22} -\Sigma_{21} \Sigma_{11}^{-1} \Sigma_{12} )^{-1} (\Sigma_{20} -\Sigma_{21} \Sigma_{11}^{-1} \Sigma_{10} ) \\
&=& \Sigma_{22\mid 1}^{-1}  (\Sigma_{20} -\Sigma_{21} \Sigma_{11}^{-1} \Sigma_{10} )\\
&=& \beta_2.
\end{eqnarray*}

Finally, we consider the population OLS coefficient of $\tilde{y}$ on $\tilde{x}_2$:
\begin{eqnarray*}
&& E(\tilde{x}_2\tilde{x}_2^{\T})^{-1} E(\tilde{x}_2 \tilde{y}) \\
&=& [ E \{  (x_2 - \Sigma_{21} \Sigma_{11}^{-1} x_1) (x_2^{\T} - x_1 ^{\T} \Sigma_{11}^{-1} \Sigma_{12} ) \} ]^{-1}
E\{  (x_2 - \Sigma_{21} \Sigma_{11}^{-1} x_1) (y - x_1 ^{\T} \Sigma_{11}^{-1} \Sigma_{10} ) \} \\
&=& (\Sigma_{22} -\Sigma_{21} \Sigma_{11}^{-1} \Sigma_{12} )^{-1} (\Sigma_{20} -\Sigma_{21} \Sigma_{11}^{-1} \Sigma_{10} ) \\
&=& \Sigma_{22\mid 1}^{-1}  (\Sigma_{20} -\Sigma_{21} \Sigma_{11}^{-1} \Sigma_{10} )\\
&=& \beta_2.
\end{eqnarray*}

\end{myproof}

We also have a population version of Cochran's formula. 
We have the following OLS decompositions of random variables
\begin{eqnarray}
y &=& \beta_1 ^{\T} x_{1} + \beta_2 ^{\T}x_{2} + \varepsilon , \label{eq::long}\\
y&=&\tilde{\beta}_2 ^{\T} x_{2} + \tilde{\varepsilon}, \label{eq::short}\\
x_{1} &=& \delta ^{\T} x_{2} + u . \label{eq::inter}
\end{eqnarray}
Equation \eqref{eq::long} defines the population long regression, and Equation \eqref{eq::short} defines the population short regression. In Equation \eqref{eq::inter}, $\delta$ is a $l \times k$ matrix because it is the OLS decomposition of a vector on a vector. We can view \eqref{eq::inter} as OLS decomposition of each component of $x_{i1}$ on $x_{i2}$. 
Theorem \ref{thm::population-cochran-formula} below states the population version of Cochran's formula.

\begin{theorem}[population Cochran's formula]
\label{thm::population-cochran-formula}
Based on \eqref{eq::long}--\eqref{eq::inter}, we have 
$$
\tilde{\beta}_2 = \beta_2 + \delta \beta_1.
$$
\end{theorem}

I relegate the proof of Theorem \ref{thm::population-cochran-formula} as Problem \ref{hw11::population-cochran}.

\section{Population $R^2$ and partial correlation coefficient}

Exclude $1$ from $x $ and assume $ x \in \mathbb{R}^{p-1}$ in this subsection. 
Assume that covariates and outcome are centered with mean zero and covariance matrix
\[
\cov\left(\begin{array}{c}
x\\
y
\end{array}\right)=\left(\begin{array}{cc}
\Sigma_{xx} & \Sigma_{xy}\\
\Sigma_{yx} & \sigma_{y}^{2}
\end{array}\right).
\]
There are multiple equivalent definitions of $R^2$. I start with the following definition
\begin{eqnarray}
R^{2}=\frac{\Sigma_{yx}\Sigma_{xx}^{-1}\Sigma_{xy}}{\sigma_{y}^{2}}, \label{eq::r2-population}
\end{eqnarray}
and will give several equivalent definitions below. 
Let $  \beta $ be the
population OLS coefficient of $y$ on $x$, and $\hat{y}=   x^{\T}\beta$ be the best
linear predictor. 

\begin{theorem}\label{thm::population-r2}
The $R^{2}$ defined in \eqref{eq::r2-population} has the following equivalent forms:
\begin{eqnarray}
R^{2} &=& \frac{\var(\hat{y})}{\var(y)} \label{eq::r2-form1} \\
&=& \max_{b   \in \mathbb{R}^{p-1}  }\rho ^{2}(y,x^{\T}b) \label{eq::r2-form2} \\
&=& \rho ^{2}(y,\hat{y}). \label{eq::r2-form3}
\end{eqnarray}
\end{theorem}

The form \eqref{eq::r2-form1} states that $R^{2}$ equals the ratio of the variance of the best linear predictor of $y$ and the variance of $y$ itself.
The form \eqref{eq::r2-form2} states that $R^{2}$ equals the maximum value of the squared Pearson correlation coefficient between $y$ and a linear combination of $x$, over all possible linear combinations of $x$. 
The form \eqref{eq::r2-form3} states that $R^{2}$ equals the squared Pearson correlation coefficient between $y$ and the best linear predictor of $y$.

\begin{myproof}{Theorem}{\ref{thm::population-r2}}
I first prove \eqref{eq::r2-form1}. 
Because of the centering of $x$, we have $\beta = \Sigma_{xx}^{-1}\Sigma_{xy}$ and 
\begin{align*}
\var(  \hat{y}  ) & =\beta^{\T} \Sigma_{xx} \beta\\
 & = \Sigma_{yx} \Sigma_{xx}^{-1}\Sigma_{xx}\Sigma_{xx}^{-1}\Sigma_{xy}\\
 &= \Sigma_{yx}\Sigma_{xx}^{-1}\Sigma_{xy}. 
\end{align*}
Therefore, $\var(  \hat{y}  ) / \var(y) = R^2$. 

I then prove \eqref{eq::r2-form2}. 
We have 
\[
\rho ^{2}(y,x^{\T}b)=\frac{\text{\cov}^{2}(y,x^{\T}b)}{\var(y)\var(x^{\T}b)}=\frac{b^{\T}\Sigma_{xy}\Sigma_{yx}b}{\sigma_{y}^{2}\times b^{\T}\Sigma_{xx}b}.
\]
Define $\gamma=\Sigma_{xx}^{1/2}b$ and $b=\Sigma_{xx}^{-1/2}\gamma$
such that $b$ and $\gamma$ have one-to-one mapping. So the maximum value of 
\[
\sigma_{y}^{2}\times\rho ^{2}(y,x^{\T}b)=\frac{\gamma^{\T}\Sigma_{xx}^{-1/2}\Sigma_{xy}\Sigma_{yx}\Sigma_{xx}^{-1/2}\gamma}{\gamma^{\T}\gamma} 
\]
equals the maximum eigenvalue of $\Sigma_{xx}^{-1/2}\Sigma_{xy}\Sigma_{yx}\Sigma_{xx}^{-1/2}$, attained when $\gamma$ equals the corresponding eigenvector; see Theorem \ref{theorem::rayleigh} in Appendix \ref{chapter::linear-algebra}. 
The matrix $\Sigma_{xx}^{-1/2}\Sigma_{xy}\Sigma_{yx}\Sigma_{xx}^{-1/2}$ is positive semi-definite and has rank one due to Proposition \ref{eq::matrix-product-inequality} in Appendix \ref{chapter::linear-algebra}, so it has exactly one non-zero eigenvalue which must equal its trace. So
\begin{align*}
\max_{b \in \mathbb{R}^{p-1} }\rho ^{2}(y,x^{\T}b) & =\sigma_{y}^{-2}\text{trace}(\Sigma_{xx}^{-1/2}\Sigma_{xy}\Sigma_{yx}\Sigma_{xx}^{-1/2})\\
 & =\sigma_{y}^{-2}\text{trace}(\Sigma_{xy}\Sigma_{yx}\Sigma_{xx}^{-1/2}\Sigma_{xx}^{-1/2})\\
 & =\sigma_{y}^{-2}\text{trace}(\Sigma_{yx}\Sigma_{xx}^{-1}\Sigma_{xy})\\
 & =\sigma_{y}^{-2}\Sigma_{yx}\Sigma_{xx}^{-1}\Sigma_{xy}\\
 & =R^{2}.
\end{align*}

I finally prove \eqref{eq::r2-form3}. We have
\begin{eqnarray*}
\rho^{2}(y,\hat{y}) &=&
\frac{ \cov(y, \hat{y})^2 }{ \var(y) \var(\hat{y}) } \\
&=& \frac{ \beta^{\T} \Sigma_{xy} \Sigma_{yx} \beta  }{  \sigma_y^2 \Sigma_{yx}\Sigma_{xx}^{-1}\Sigma_{xy} },
\end{eqnarray*}
by the results in the proofs of \eqref{eq::r2-form2} and \eqref{eq::r2-form2}. Use the formula $\beta = \Sigma_{xx}^{-1}\Sigma_{xy}$ to further simplify the above expression as
\begin{eqnarray*}
\rho^{2}(y,\hat{y}) &=&
\frac{ \Sigma_{yx}\Sigma_{xx}^{-1} \Sigma_{xy} \Sigma_{yx} \Sigma_{xx}^{-1}\Sigma_{xy}  }{  \sigma_y^2 \Sigma_{yx}\Sigma_{xx}^{-1}\Sigma_{xy} } \\
&=& \frac{ \Sigma_{yx}\Sigma_{xx}^{-1} \Sigma_{xy}   }{  \sigma_y^2  }  \\
&=& R^2.
\end{eqnarray*}
\end{myproof}

We can also define population partial correlation and $R^{2}.$ For
scalar $y$ and $x$ with another scalar or vector $w$, we can define
the population OLS decomposition based on $(1,w)$ as
\begin{equation}
\label{eq::population-ols-decomposition}
y=\hat{y}+\tilde{y},\qquad x=\hat{x}+\tilde{x},
\end{equation} 
where 
$$ 
 \tilde{y} =  \{ y- E(y) \} - \{ w - E(w) \}^{\T}\beta_{y},\qquad 
 \tilde{x} =  \{ x- E(x) \} - \{  w-E(w) \} ^{\T}\beta_{x} ,
 $$ 
 with $\beta_{y}$ and $\beta_{x}$ being the coefficients of $w$ in these population OLS. 
We then define the population partial correlation coefficient as
\[
\rho_{yx\mid w}=\rho_{\tilde{y}\tilde{x}}.
\]
If the marginal correlation and partial correlation have different signs, then we have Simpson's paradox at the population level.

With a scalar $w$, we have a more explicit formula below.
\begin{theorem}
\label{thm:populationpartialcorr}For scalar $(y,x,w)$, we have
\[
\rho_{yx\mid w}=\frac{\rho_{yx}-\rho_{xw}\rho_{yw}}{\sqrt{1-\rho_{xw}^{2}}\sqrt{1-\rho_{yw}^{2}}}.
\]
\end{theorem}

I relegate the proof of Theorem \ref{thm:populationpartialcorr} as  Problem \ref{hw11::population-partial-correlation}.

We can also extend the definition of $\rho_{yx\mid w}$ to the partial $R^2$ with a scalar $y$ and possibly vector $x$ and $w$. The population OLS decompositions \eqref{eq::population-ols-decomposition} still hold in the general case. Then we can define the partial $R^2$ between $y$ and $x$ given $w$ as the $R^2$ between $ \tilde{y}$ and $  \tilde{x}$: 
$$
R^2_{y.x|w} = R^2_{ \tilde{y}.  \tilde{x}  } . 
$$

\section{Inference for the population OLS}

\subsection{Inference with the Eicker--Huber--White standard errors}
Based on the IID data $(x_{i},y_{i})_{i=1}^{n}$, we can
obtain the moment estimator for the population OLS coefficient
\[
\hat{\beta}=\left(n^{-1}\sumn x_{i}x_{i}^{\T}\right)^{-1}\left(n^{-1}\sumn x_{i}y_{i}\right),
\]
and the residuals $ \hat{\varepsilon}_i = y_i - x_i^{\T} \hat{\beta} . $ Assume finite fourth moments of $(x,y)$. We can use the law of large numbers to show that 
\begin{eqnarray*}
n^{-1}\sumn x_{i}x_{i}^{\T} &\rightarrow & E(xx^{\T}), \\
n^{-1}\sumn x_{i}y_{i} &\rightarrow & E(xy) ,
\end{eqnarray*}
so $\hat{\beta}\rightarrow\beta$ in probability. We can use the central limit theorem (CLT) to show that $n^{-1/2}\sumn x_{i}\varepsilon_{i}\rightarrow\N(0,M)$ in distribution, where $M= E(\varepsilon^{2}xx^{\T})$,
so 
\begin{eqnarray}
\label{eq::ols-populationols-clt}
\sqrt{n}(\hat{\beta}-\beta)\rightarrow\N ( 0, V  )
\end{eqnarray}
in distribution, where $V =  B^{-1} M B^{-1}$ with $B =   E(xx^{\T}) $. 
The moment estimator for the asymptotic variance of $\hat{\beta}$ is again
 the Eicker--Huber--White (EHW) robust covariance
estimator:
\begin{eqnarray}
\label{eq::ehw-populationols}
\hat{V}_{\textsc{ehw}} =  
n^{-1}
\left(n^{-1}\sumn x_{i}x_{i}^{\T}\right)^{-1}
\left(n^{-1}\sumn  \hat{\varepsilon}_i^2  x_{i}x_{i}^{\T}\right) 
\left(n^{-1}\sumn x_{i}x_{i}^{\T}\right)^{-1} .
\end{eqnarray}
Following the almost the same proof of Theorem \ref{thm::ehw-fixeddesign-consistency}, we can show that $\hat{V}_{\textsc{ehw}} $ is consistent for the asymptotic covariance $V$. I summarize the formal results below, with the proof relegated as Problem \ref{hw11::population-ols-asymptotics}.

\begin{theorem}
\label{thm::population-ols}
Assume that $(x_i, y_i)_{i=1}^n \iidsim (x, y)$ with $E ( \| x \|^4 )  <\infty  $ and $E (y^4) <\infty$. We have \eqref{eq::ols-populationols-clt} and $n \hat{V}_{\textsc{ehw}}  \rightarrow  V$ in probability. 
\end{theorem}

So the EHW standard error is not only robust to the heteroskedasticity of the errors but also robust to the misspecification of the linear model \citep{huber::1967, white1980using, angrist2008mostly, buja2019models}. Of course, when the linear model is wrong, we need to modify the interpretation of $\beta$: it is the coefficient of $x$ in the best linear prediction of $y$ or the best linear approximation of the conditional mean function $E(y\mid x)$.

\section{To model or not to model?}

\subsection{Population OLS and the restricted mean model}

\begin{figure}[ht]
\centering
\includegraphics[width = \textwidth]{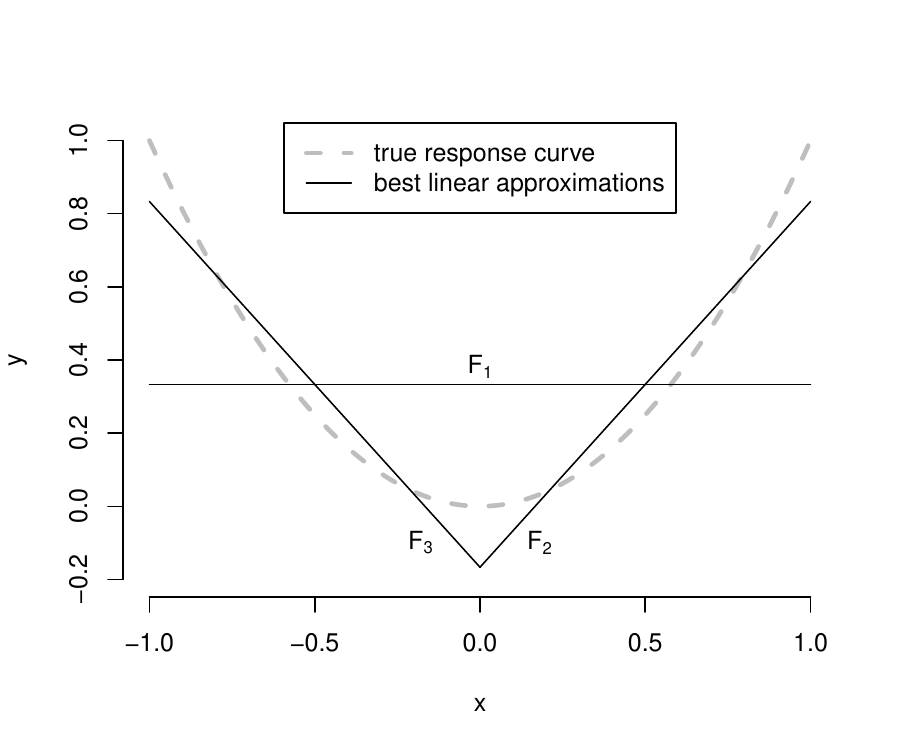}
\caption{Best linear approximations correspond to three different distributions of $x$.}\label{fig::bestlinearapproximation}
\end{figure}

I start with a simple example. In the following calculation, I will use the fact that the $k$th moment of a Uniform$(0,1)$ random variable equals $1/(k+1)$; see Problem \ref{hwmath2::uniform-moments}.

\begin{example}\label{eg::bestlinearapproximations}
Assume that $x\sim F(x)$, $\varepsilon\sim\N(0,1),x\ind\varepsilon$,
and $y=x^{2}+\varepsilon$.
\begin{enumerate}
\item
If $x\sim F_{1}(x)$ is Uniform$(-1,1)$, then the best linear
approximation is $1/3+0\cdot x$. We can see this result from
$$
\beta(F_{1})=\frac{\text{\cov}(x,y)}{\var(x)}=\frac{\cov(x,x^{2})}{\var(x)}=\frac{E(x^{3})}{\var(x)}=0,
$$
and $\alpha(F_{1})=E(y)=E(x^{2}) =1/3$. 

\item
If $x\sim F_{2}(x)$ is Uniform$(0,1)$, then the best linear approximation is $-1/6+x$. We can see this result from 
$$
\beta(F_{2})=\frac{\cov(x,y)}{\var(x)}=\frac{\cov(x,x^{2})}{\var(x)}=\frac{E(x^{3})-E(x)E(x^{2})}{E(x^{2})-\left\{ E(x)\right\} ^{2}}=\frac{1/4-1/2\times1/3}{1/3-(1/2)^{2}}=1,
$$
and $\alpha(F_{2})=E(y)-\beta E(x)=E(x^{2})-E(x)=1/3-1/2=-1/6$

\item 
If $x\sim F_{3}(x)$ is Uniform$(-1,0)$, then the best linear approximation is $-1/6-x.$ This result holds by symmetry. 
\end{enumerate}
Figure \ref{fig::bestlinearapproximation} shows the true conditional mean function $x^2$ and the best linear approximations. As highlighted in the notation above, the best linear approximation
depends on the distribution of $x$.
\end{example}

From Example \ref{eg::bestlinearapproximations}, we can see that the best linear approximation depends on the distribution of $X$. This complicates the interpretation of $\beta$ from the population OLS decomposition \citep{buja2019models}. Consequently, this can cause problems for {\it external validity} because $\beta$ will be different in a future environment with different distribution of $X$. \citet[][page 66]{sims2010but} pointed this out as a critique of \citet{angrist2008mostly}.

To ensure the stability of $\beta$ across different environments, we often invoke the following {\it restricted mean model}.

\begin{assumption}[restricted mean model]
\label{assume::model-restricted}
Assume \begin{eqnarray}
E(y\mid x)=x^{\T}\beta
\label{eq::restricted-mean-model}
\end{eqnarray}
or, equivalently, 
\[
y=x^{\T}\beta+\varepsilon,\qquad E(\varepsilon\mid x)=0.
\]
\end{assumption}

Assumption \ref{assume::model-restricted} restricts the conditional mean of $y$ given $x$ to be linear in $x$, justifying the name ``restricted mean model.'' Nevertheless, Assumption \ref{assume::model-restricted} imposes weak assumptions on the distribution of $y$ or $\varepsilon$. 
Under Assumption \ref{assume::model-restricted}, the population OLS coefficient equals the true parameter in the restricted mean model:
\begin{align*}
\left\{ E(xx^{\T})\right\} ^{-1}E(xy) & =\left\{ E(xx^{\T})\right\} ^{-1}E\left\{ xE(y\mid x)\right\} \\
 & =\left\{ E(xx^{\T})\right\} ^{-1}E(xx^{\T}\beta)\\
 & =\beta.
\end{align*}
Moreover, the population OLS coefficient does not depend on the distribution
of $x$. The above asymptotic inference applies to this model too.

\begin{figure}[th]
\centering
\includegraphics[width = \textwidth]{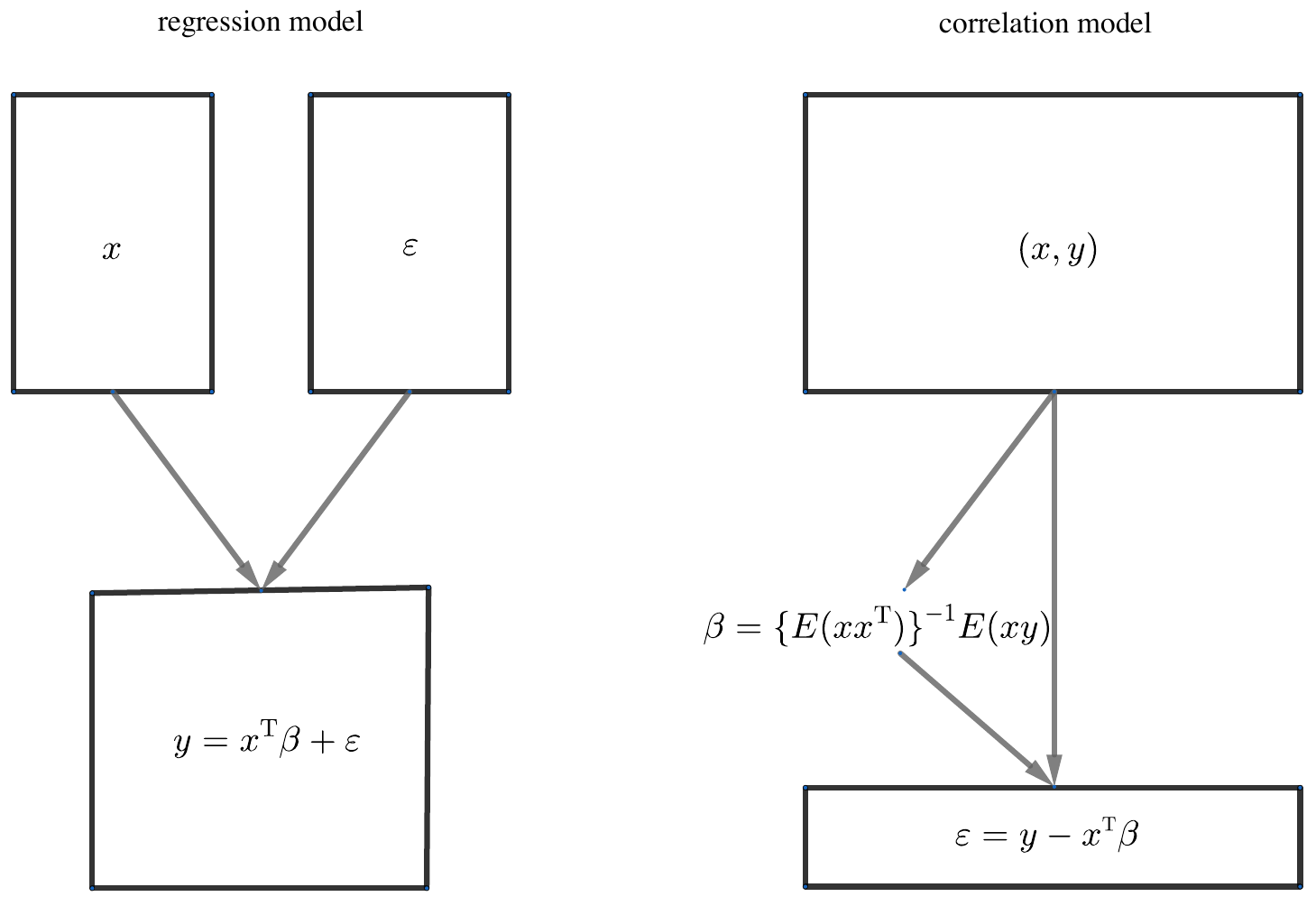}
\caption{Freedman's classification of OLS}\label{fig::freedman-classification}
\end{figure}

\citet{freedman1981bootstrapping} distinguished two types of OLS as shown in Figure \ref{fig::freedman-classification}: 
\begin{enumerate}[label=(M\arabic*), ref=M\arabic*]
\item
the {\it regression model}, as shown in left-hand side of Figure \ref{fig::freedman-classification};
\item
the {\it correlation model}, as shown in right-hand side of Figure \ref{fig::freedman-classification}. 
\end{enumerate}
In the regression model, we first generate $x$ and $\varepsilon$ under some restrictions, for example, $E(\varepsilon\mid x) = 0$, and then generate the outcome based on $y=x^{\T} \beta + \varepsilon$, a linear function of $x$ with error $\varepsilon$. 
In the correlation model, we start with a pair $(x,y)$, then decompose $y$ into the best linear predictor $x^{\T} \beta $ and the leftover residual $\varepsilon$. 
Compare the subtle difference between the $\varepsilon$ in the regression model and the correlation model. The regression model requires $E(\varepsilon\mid x) = 0$, whereas the correlation model ensures $E(\varepsilon x) = 0$. The regression model imposes a stronger assumption because $E(\varepsilon\mid x) = 0$ implies
$$
E(\varepsilon x) = E\{E(\varepsilon x \mid x)  \} =  E\{E(\varepsilon  \mid x)  x \}  = 0. 
$$

\subsection{More on residual plots}
\label{sec::residual-plot}

Most standard statistical theory for inference assumes a correctly specified linear model (e.g., Gauss--Markov model, Normal linear model, or restricted mean model). However, the corresponding inferential procedures are often criticized since it is challenging to ensure that the model is correctly specified. Alternatively, we can argue that without assuming the linear model, the OLS estimator is consistent for the coefficient in the best linear approximation of the conditional mean function $E(y\mid x)$, which is often a meaningful quantify even the linear model is misspecified. This can be misleading. Example \ref{eg::bestlinearapproximations} shows that the best linear approximation can be a bad approximation to a nonlinear conditional mean function, and it depends on the distribution of the covariates.

\begin{figure}
\centering
\includegraphics[width = \textwidth]{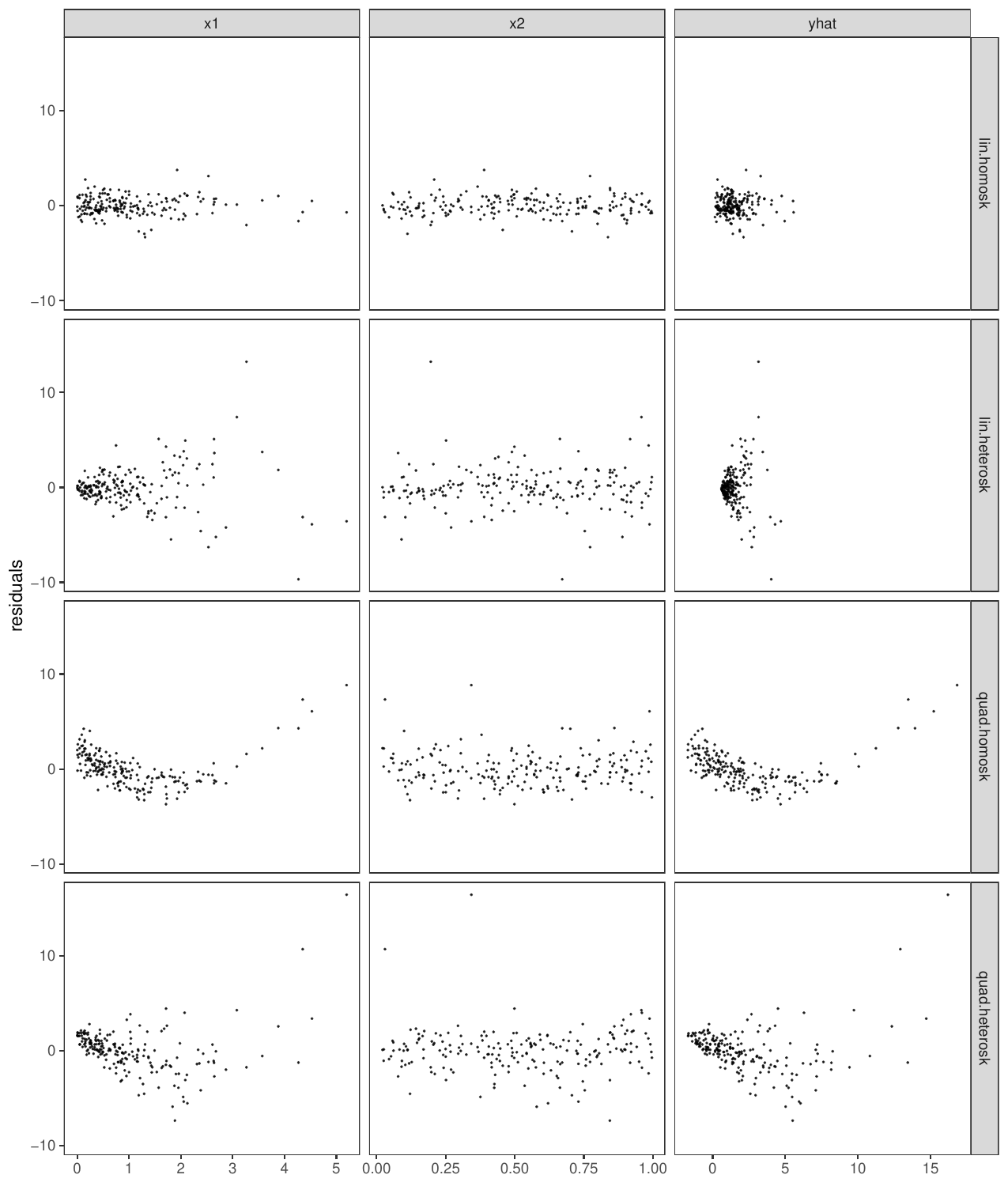}
\caption{Residual plots}\label{fig::residual-plots}
\end{figure}

A classic statistical approach is to check whether the residual $\hat\varepsilon_i$ has any nonlinear trend with respect to the covariates. With only a few covariates, we can plot the residual against each covariate; with many covariates, we can plot the residual against the fitted value $\hat{y}_i$. Figure \ref{fig::residual-plots} gives four examples. In these examples, the covariates are the same:
\begin{lstlisting}
n = 200
x1 = rexp(n)
x2 = runif(n)
\end{lstlisting}

The outcome models differ:
\begin{enumerate}
[label=(Y\arabic*), ref=Y\arabic*]
\item\label{modelY1}
linear homoskedastic: 
\ri{y = x1 + x2 + rnorm(n)}; 
\item\label{modelY2}
linear heteroskedastic: 
\ri{y = x1 + x2 + rnorm(n, 0, x1+x2)}; 
\item\label{modelY3}
quadratic homoskedastic: 
\ri{y = x1^2 + x2^2 + rnorm(n)}; 
\item\label{modelY4}
quadratic heteroskedastic: 
\ri{y = x1^2 + x2^2 + rnorm(n, 0, x1+x2)}.
\end{enumerate}

In the last two outcome models \eqref{modelY3} and \eqref{modelY4}, the residuals indeed show some nonlinear relationship with the covariates and the fitted value. This suggests that the linear function can be a poor approximation to the true conditional mean function.

\section{Conformal prediction based on exchangeability}\label{sec::conformal-prediction}

Chapter \ref{sec::prection-normallinear} discusses the prediction of a future outcome $y_{n+1}$ based on $x_{n+1}$ and $(X,Y)$. It requires the Normal linear model assumption. Chapter \ref{chapter::EHW} relaxes the Normality assumption on the error term in statistical inference but does not discuss prediction. Under the heteroskedastic linear model assumption in Chapter \ref{chapter::EHW}, we can predict the mean $E(y_{n+1}) = x_{n+1}^{\T} \beta$ by $x_{n+1}^{\T} \hat\beta$ with asymptotic standard error
$
( x_{n+1}^{\T}  \hat{V}_{\textsc{ehw}}  x_{n+1} )^{1/2},
$
where $ \hat{V}_{\textsc{ehw}}$ is the EHW covariance matrix for the OLS coefficient. However, it is fundamentally challenging to predict $y_{n+1}$ itself since the heteroskedastic linear model allows it to have a completely unknown variance $\sigma^2_{n+1}$.

Under the population OLS formulation, it seems even more challenging to predict the future outcome since we do not even assume that the linear model is correctly specified. In particular, $x_{n+1}^{\T} \hat\beta$ does not have the same mean as $y_{n+1}$ in general. Perhaps surprisingly, we can construct a prediction interval for $y_{n+1}$ based on $x_{n+1}$ and $(X,Y)$ using an idea called {\it conformal prediction} \citep{vovk2005algorithmic, lei2018distribution}. It leverages the {\it exchangeability}\footnote{Exchangeability is a technical term in probability and statistics. Random elements $z_1, \ldots, z_n$ are exchangeable if $(z_{\pi(1)}, \ldots, z_{\pi(n)} ) $ have the same distribution as $(z_1, \ldots, z_n)$, where $\pi(1), \ldots, \pi(n)$ is a permutation of the integers $1,\ldots, n$. In other words, a set of random elements are exchangeable if their joint distribution does not change under re-ordering. IID random elements are exchangeable.} of the data points 
$$
(x_1, y_1),\ldots,  (x_n, y_n), (x_{n+1}, y_{n+1}) .
$$ 
Pretending that we know the value $y_{n+1} = y^*$, we can fit OLS using $n+1$ data points and obtain residuals
$$
\hat{\varepsilon}_i( y^* ) = y_i - x_i^{\T} \hat{\beta}(  y^* ),\quad (i=1,\ldots, n+1)
$$
where we emphasize the dependence of the OLS coefficient and residuals on the unknown $y^*$. 
The absolute values of the residuals $|\hat{\varepsilon}_i(  y^*  )|$'s are also exchangeable, so the rank of $|\hat{\varepsilon}_{n+1}(  y^*  )|$, denoted by 
$$
\hat{R}_{n+1}(  y^*  )  = 1 +  \sum_{i=1}^{n}  1\{    |\hat{\varepsilon}_{i}(  y^*  )| \leq    |\hat{\varepsilon}_{n+1}(  y^*  )|   \}  ,
$$
must have a uniform distribution over $\{ 1, 2, \ldots,n,n+1 \}$, a known distribution not depending on anything else. It is a pivotal quantity satisfying
\begin{eqnarray}\label{eq::pivotal-conformal}
\pr\left\{   \hat{R}_{n+1}(  y^*   )  \leq \lceil  (1-\alpha)(n+1) \rceil \right\} \geq 1-\alpha.
\end{eqnarray}
Equivalently, this is a statement linking the unknown quantity  $y^*$ and observed data, so it gives a confidence set for $y^*$ at level $1-\alpha$. In practice, we can use a grid search to solve for the inequality \eqref{eq::pivotal-conformal} involving $y^*$.

Below we evaluate the leave-one-out prediction with the Boston housing data. 
\begin{lstlisting}
library("mlbench")
data(BostonHousing)
attach(BostonHousing)
n = dim(BostonHousing)[1]
p = dim(BostonHousing)[2] - 1
ymin = min(medv)
ymax = max(medv)
grid.y  = seq(ymin - 30, ymax + 30, 0.1)
BostonHousing = BostonHousing[order(medv), ]
detach(BostonHousing)

ols.fit.full = lm(medv ~ ., data = BostonHousing,
                  x = TRUE, y = TRUE, qr = TRUE)
beta     = ols.fit.full$coef
e.sigma  = summary(ols.fit.full)$sigma
X        = ols.fit.full$x
Y        = ols.fit.full$y
X.QR     = ols.fit.full$qr
X.Q      = qr.Q(X.QR)
X.R      = qr.R(X.QR)
Gram.inv = solve(t(X.R)%*%X.R)
hatmat   = X.Q%*%t(X.Q) 
resmat   = diag(n) - hatmat
leverage = diag(hatmat)
Resvec   = ols.fit.full$residuals

cvt  = qt(0.975, df = n-p-1)
cvr  = ceiling(0.95*(n+1))

loo.pred = matrix(0, n, 5)
loo.cov  = matrix(0, n, 2)
for(i in 1:n)
{
  beta.i = beta - Gram.inv%*%X[i, ]*Resvec[i]/(1-leverage[i])
  e.sigma.i = sqrt(e.sigma^2*(n - p) - 
                     (Resvec[i])^2/(1 - leverage[i]))/
              sqrt(n - p - 1)
  pred.i = sum(X[i, ]*beta.i) 
  lower.i = pred.i - cvt*e.sigma.i/sqrt(1 - leverage[i])
  upper.i = pred.i + cvt*e.sigma.i/sqrt(1 - leverage[i])
  loo.pred[i, 1:3] = c(pred.i, lower.i, upper.i)
  loo.cov[i, 1] = findInterval(Y[i], c(lower.i, upper.i))
  
  grid.r  = sapply(grid.y,
                   FUN = function(y){
                     Res = Resvec + resmat[, i]*(y - Y[i]) 
                     rank(abs(Res))[i]
                   })
  Cinterval = range(grid.y[grid.r<=cvr])
  loo.pred[i, 4:5] = Cinterval
  loo.cov[i, 2] = findInterval(Y[i], Cinterval)
  
}
\end{lstlisting}

In the above code, I use the QR decomposition to compute $X^{\T} X$ and $H$. Moreover, the calculations of \ri{lower.i}, \ri{upper.i}, and \ri{Res} use some tricks to avoid fitting $n$ OLS. I relegate the justification of them to Problem \ref{hw11::loo-conformal}. 

The variable \ri{loo.pred} has five columns corresponding to the point predictors, lower and upper intervals based on the Normal linear model and conformal prediction. 
\begin{lstlisting}
> colnames(loo.pred) = c("point", "G.l", "G.u", "c.l", "c.u")
> head(loo.pred)
         point        G.l       G.u   c.l  c.u
[1,]  6.633514  -2.941532 16.208559  -3.5 16.7
[2,]  8.806641  -1.349367 18.962649  -2.6 20.1
[3,] 12.044154   2.608290 21.480018   2.2 21.8
[4,] 11.025253   1.565152 20.485355   1.2 21.0
[5,] -5.181154 -14.819041  4.456733 -15.0  4.9
[6,]  8.324114  -1.382910 18.031138  -2.0 18.8
\end{lstlisting}

\begin{figure}
\centering
\includegraphics[width = 0.95\textwidth]{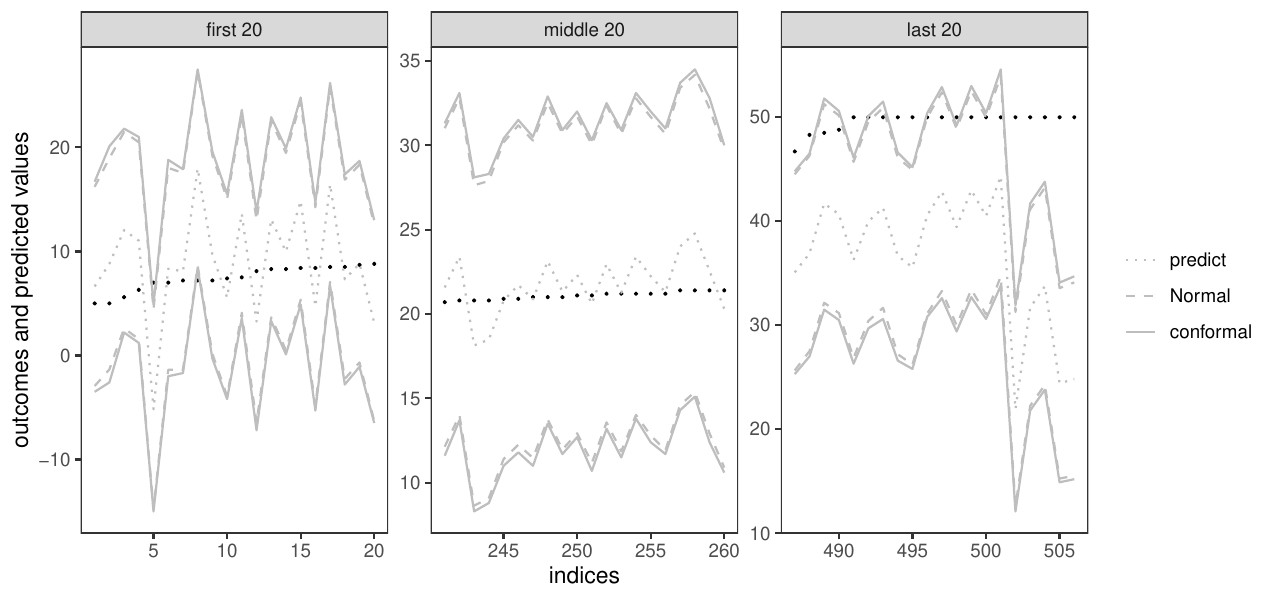}
\caption{Leave-one-out prediction intervals based on the Boston housing data.
The left, middle and right panels are for the 20 observations with the outcomes at the bottom, middle, and top, respectively.
Each panel shows the original outcomes, predicted outcomes, as well as the prediction intervals based on the Normal linear model and conformal prediction. 
}\label{fig::pi-boston-housing-conformal}
\end{figure}

Figure \ref{fig::pi-boston-housing-conformal} plots the observed outcomes and the prediction intervals for the 20 observations with the outcomes at the bottom, middle, and top. The Normal and conformal intervals are almost indistinguishable. For the observations with the highest outcome, the predictions are quite poor. Surprisingly, the overall coverage rates across observations are close to $95\%$ for both methods. 
\begin{lstlisting}
> apply(loo.cov==1, 2, mean)
[1] 0.9486166 0.9525692
\end{lstlisting}

\begin{figure}
\centering
\includegraphics[width = 0.95\textwidth]{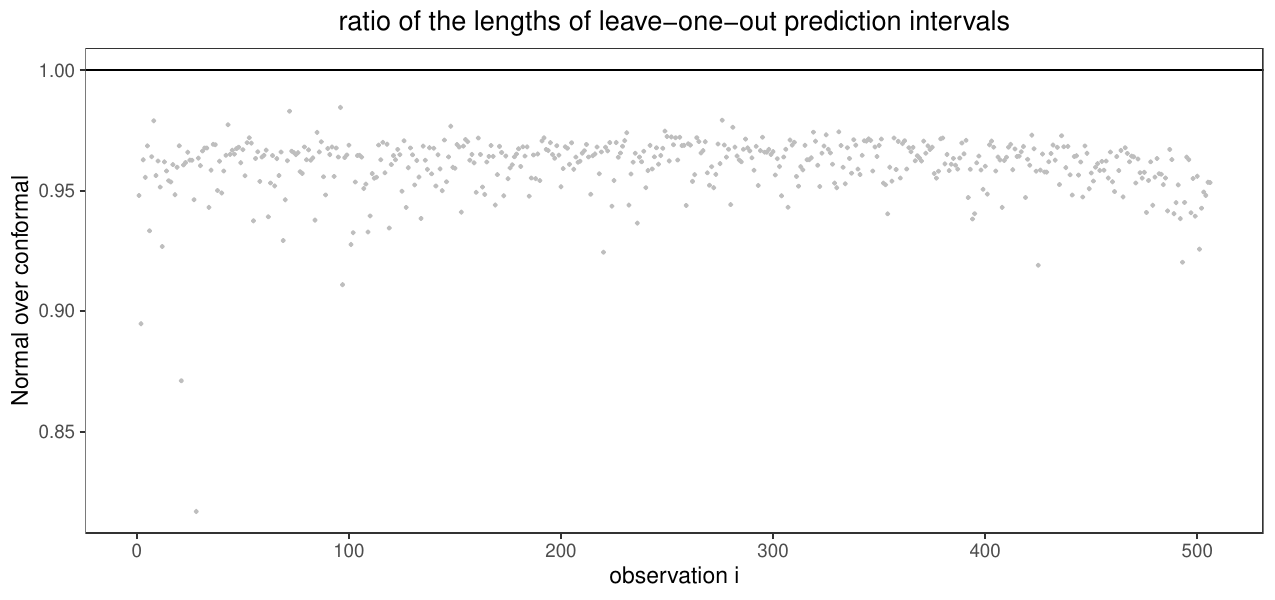}
\caption{Boston housing data}\label{fig::ratio-boston-housing-conformal}
\end{figure}

Figure \ref{fig::ratio-boston-housing-conformal} compares the lengths of the two prediction intervals. Although the conformal prediction intervals are slightly wider than the Normal prediction interval, the differences are rather small, with the ratio of the length above 0.96.

If you read the above argument for conformal prediction again, you will realize that the whole argument does not rely on using OLS as the predictor. You can replace OLS with an arbitrary predictor, without harming the theoretical guarantee of the conformal prediction interval. Conformal prediction is a powerful idea for using black-box predictors while maintaining confidence interval guarantees. See \citet{angelopoulos2023conformal} for its recent developments and applications.

Nevertheless, the confidence interval guarantees based on conformal prediction differ from that based on the Normal linear model. In particular, the conformal prediction interval covers $y_{n+1}$ with probability larger than or equal to $1-\alpha$, {\it averaged} over the randomness of the past data and future $x_{n+1}$, whereas the prediction interval based on the Normal linear model covers $y_{n+1}$ with probability larger than or equal to $1-\alpha$, {\it conditional} on all observed covariates. 
Therefore, the magic of conformal prediction comes from modifying the statistical model and the theoretical guarantees. 
Some practitioners may argue that the conditional coverage guarantee is more relevant than the average coverage guarantee. In those cases, the conformal prediction interval should be used with caution.

\section{Homework problems}

\paragraph{Conditional mean}\label{hw11::population-mean}

Prove Theorem \ref{thm::equivalent-definition-of-conditional-mean}. 

\paragraph{Best linear approximation}\label{hw11::condition-mean-app}

Prove Theorem \ref{thm:bestlinear-conditionalmean}.

Remark: It is similar to Problem \ref{hw7::sample-partial-correlation}.

\paragraph{Univariate population OLS}\label{hw11::univariate-ols}

Prove Corollary \ref{corollary:scalar-pop-ols}.

\paragraph{Asymptotics for the population OLS}\label{hw11::population-ols-asymptotics}

Prove Theorem \ref{thm::population-ols}.

\paragraph{Population Cochran's formula}\label{hw11::population-cochran}

Prove Theorem \ref{thm::population-cochran-formula}.

\paragraph{Canonical correlation analysis (CCA)}\label{hw11::cca}

Assume that $(x,y)$, where $x \in \mathbb{R}^p$ and $y \in \mathbb{R}^k$, has the joint non-degenerate covariance matrix:
$$
\begin{pmatrix}
\Sigma_{xx} & \Sigma_{xy} \\
\Sigma_{yx} & \Sigma_{yy}
\end{pmatrix}. 
$$
\begin{enumerate}
\item
Find the best linear combinations $(\alpha, \beta)$ that
give the maximum Pearson correlation coefficient:
\[
(\alpha, \beta) = 
\text{\ensuremath{\arg\max_{a \in \mathbb{R}^k ,  b \in \mathbb{R}^p}}}\ \rho(y^{\T}a,x^{\T}b).
\]
Note that you need to detail the steps in calculating $(\alpha, \beta)$ based on the covariance matrix above.

\item
Define the maximum value as $\textsc{cc}(x,y)$. Prove that $\textsc{cc}(x,y) \geq 0$, and $\textsc{cc}(x,y) = 0$ if $x\ind y$. 
\end{enumerate}

Remark: The maximum value $\textsc{cc}(x,y)$ is called the canonical correlation between $x$ and $y$. 
We can also define partial canonical correlation between $x$ and $y$ given $w$.

\paragraph{Population partial correlation coefficient}\label{hw11::population-partial-correlation}

Prove Theorem \ref{thm:populationpartialcorr}.

\paragraph{Independence and correlation}\label{hw11::independence-correlation}

With scalar random variables $x$ and $y$, it is well known that if $x\ind y$, then $\rho_{yx}=0$. However, with another random variable $w$, if $x\ind y\mid w$, then $\rho_{yx\mid w}=0$ may not hold. 

Give a counterexample in which  $x\ind y\mid w$ but $\rho_{yx\mid w}\neq 0$.

Remark: 
With scalar random variables $x$ and $y$, if $x\ind y$, then we have $\cov(y, x) = 0$,  which implies 
$$
\rho_{yx}=  \frac{ \cov(y, x)  }{ \sqrt{   \var(y)  \var(x)   }  } = 0.
$$
With another random variable $w$, if $x\ind y\mid w$, then we have $\cov(y, x\mid w) = 0$, which, however, does not imply $\rho_{yx\mid w}=0$ because $\rho_{yx\mid w}$ is not defined as 
$$
\frac{ \cov(y, x\mid w)  }{ \sqrt{   \var(y\mid w)  \var(x\mid w)   }  }. 
$$
\citet{shah2020hardness} had related discussion on this issue.

\paragraph{Best linear approximation of a cubic curve}\label{hw11::bestlinear-cubic}

Assume that $x\sim\N(0,1)$, $\varepsilon\sim\N(0,\sigma^{2})$, $x\ind\varepsilon$,
and $y=x^{3}+\varepsilon$. Find the best linear approximation of
$E(y\mid x)$ based on $(1, x)$.  Plot both $E(y\mid x)$ and its best linear approximation together and compare them.

\paragraph{Leave-one-out formula in conformal prediction}\label{hw11::loo-conformal}

Justify the calculations of \ri{lower.i}, \ri{upper.i}, and \ri{Res} in Section \ref{sec::conformal-prediction}.

\paragraph{Conformal prediction for multiple outcomes}
Assuming exchangeability of 
$$
(x_1,y_1),\ldots,  (x_n,y_n), (x_{n+1},y_{n+1}), \ldots , (x_{n+k},y_{n+k}).   
$$
Propose a method to construct joint conformal prediction regions for $(y_{n+1},\ldots, y_{n+k})$ based on $(X,Y)$ and $(x_{n+1},\ldots, x_{n+k})$.

\paragraph{Cox's theorem}\label{hw11::cox-theorem1960}

\citet{cox1960jrssb} considered the data-generating process
$$
x_1 \longrightarrow x_2  \longrightarrow y
$$
under the following linear models:
for $i=1,\ldots, n$, we have 
$$
x_{i2} = \alpha_0 + \alpha_1x_{i1} + \eta_i
$$
and
$$
y_i = \beta_0 + \beta_1 x_{i2} + \varepsilon_i
$$
where $\eta_i$ has mean 0 and variance $\sigma_\eta^2$, $\varepsilon_i$ has mean 0 and variance $\sigma_\varepsilon^2$, and the $\eta_i$s and  $\varepsilon_i$s are independent.
The linear model implies
$$
y_i = (\beta_0 + \beta_1\alpha_0) + (\beta_1 \alpha_1) x_{i1} +   (\varepsilon_i  +  \beta_1  \eta_i  )
$$
where $\varepsilon_i  +  \beta_1  \eta_i$s are independent with mean 0 and variance $\sigma_\varepsilon^2 + \beta_1^2 \sigma_\eta^2$. 

Therefore, we have two ways to estimate $\beta_1 \alpha_1$:
\begin{enumerate}[label=(W\arabic*), ref=W\arabic*]
\item
the first estimator is $\hat\gamma_1$,  the OLS estimator of the $y_i$'s on the $x_{i1}$'s with the intercept; 

\item
the second estimator is $\hat\alpha_1 \hat\beta_1$, the product of the OLS estimator of the $x_{i2}$'s on the $x_{i1}$'s with the intercept and that of the $y_i$'s on the $x_{i2}$'s with the intercept. 
\end{enumerate}

\citet{cox1960jrssb} reported Theorem \ref{thm::cox-theorem-1960} below. Prove Theorem \ref{thm::cox-theorem-1960}. 

\begin{theorem}
\label{thm::cox-theorem-1960}
Let $X_1 = (x_{11}, \ldots, x_{n1})$. 
We have 
$$
\var(\hat\alpha_1 \hat\beta_1 \mid X_1) \leq \var(\hat\gamma_1\mid X_1) ,
$$
and more precisely,
$$
\var(\hat\gamma_1\mid X_1) 
= \frac{   \sigma_\varepsilon^2     + \beta_1^2  \sigma_\eta^2   }{    \sum_{i=1}^n   (x_{i1} - \bar{x}_1)^2     }
$$
and
$$
\var(\hat\alpha_1 \hat\beta_1 \mid X_1) 
= \frac{   \sigma_\varepsilon^2   E(  \hat\rho_{12}^2 \mid X_1)   + \beta_1^2  \sigma_\eta^2   }{    \sum_{i=1}^n   (x_{i1} - \bar{x}_1)^2     }
$$
where $\hat\rho_{12} \in [-1,1]$ is the sample Pearson correlation coefficient between the $x_{i1}$'s and the $x_{i2}$'s. 
\end{theorem}

Remark: If we further assume that the error terms are Normal, then $\hat\alpha_1 \hat\beta_1$ is the maximum likelihood estimator for $\alpha_1 \beta_1$. Therefore, the asymptotic optimality theory for the maximum likelihood estimator justifies the superiority of $\hat\alpha_1 \hat\beta_1$ over $\hat\gamma_1$. Theorem \ref{thm::cox-theorem-1960} provides a stronger finite-sample result without assuming the Normality of the error terms.

\paragraph{Measurement error and Frisch's bounds}\label{hw11::measurement-error}

\begin{enumerate}
\item
Given scalar random variables $x$ and $y$, we can obtain the population OLS coefficient $(\alpha, \beta)$ of $y$ on $(1,x)$. However, $x$ and $y$ may be measured with errors, that is, we observe $x^* = x + u$ and $y^* = y + v$, where $u$ and $v$ are mean zero error terms satisfying $u\ind v$  and $(u ,v )\ind (x,y)$. We can obtain the population OLS coefficient $(\alpha^*, \beta^*)$ of $y^*$ on $(1,x^*)$ and the population OLS coefficient $(a^*, b^*)$ of $x^*$ on $(1,y^*)$. 

 Prove that if $\beta = 0$ then $ \beta^*  =b^* = 0 $; if $\beta \neq  0$ then
$$ 
| \beta^* | \leq |\beta | \leq 
1 / |  b^* |   .
$$

\item
Given scalar random variables $x, y$ and a random vector $w$, we can obtain the population OLS coefficient $(\alpha, \beta, \gamma)$ of $y$ on $(1,x,w)$. When $x$ and $y$ are measured with error as above with mean zero errors satisfying $u\ind v$ and  $(u ,v )\ind (x,y,w)$, we can obtain the population OLS coefficient $(\alpha^*, \beta^*, \gamma^*)$ of $y$ on $(1,x^*,w)$, and the population OLS coefficient $(a^*, b^*, c^*)$ of $x^*$ on $(1,y^*,w)$.

Prove that the same result holds as in the first part of the problem. 
\end{enumerate}

Remark: \citet{tamer2010partial} reviewed \citet{frisch1934statistical}'s upper and lower bounds for the univariate OLS coefficient based on the two OLS coefficients of the observables. The second part of the problem extends the result to the multivariate OLS with a covariate subject to measurement error. The lower bound is well documented in most books on measurement errors, but the upper bound is much less well known.

\paragraph{A three-way decomposition}\label{hw11::three-way-ols}

The main text of this chapter focuses on the two-way decomposition of the outcome:
$
y = x^{\T} \beta + \varepsilon ,
$
where $\beta$ is the population OLS coefficient and $\varepsilon$ is  the population OLS residual. However, $x^{\T} \beta$ is only the best linear approximation to the true conditional mean function $\mu(x) =  E(y \mid x)$. This suggests the following three-way decomposition of the outcome:
$$
y = x^{\T} \beta  + \{ \mu(x)  - x^{\T} \beta \} + \{ y -  \mu(x) \},
$$
which must hold without any assumptions. Introduce the notation for the linear term 
$$
\hat{y} = x^{\T} \beta,
$$ 
 the notation for the approximation error:
$$
\delta = \mu(x)  - x^{\T} \beta ,
$$
and the notation for the
 ``ideal residual'':
$$
e =  y -  \mu(x) .
$$
Then 
we can decompose the outcome as
$$
y = \hat{y}   +   \delta  + e
$$
and
 the population OLS residual as
$$
 \varepsilon  =  \{ \mu(x)  - x^{\T} \beta \} + \{ y -  \mu(x) \} =  \delta  + e. 
$$

\begin{enumerate}
\item
Prove  
$$
E(   \hat{y}    e \mid x ) = 0,\quad
E(   \delta   e \mid x ) = 0
$$
and
$$
E(   \hat{y}    e  ) = 0,\quad
E(   \delta   e   ) = 0,\quad
E(  \hat{y}   \delta  ) = 0.
$$
Prove
$$
E( \varepsilon^2 ) = E(  \delta ^2 ) + E(  e^{2}) . 
$$

\item
Introduce an intermediate quantity between the population OLS coefficient $\beta $ and the OLS coefficient $\hat{\beta} $:
$$
\tilde{\beta} = \left( n^{-1} \sumn x_i x_i^{\T}  \right)^{-1}  \left( n^{-1} \sumn x_i \mu(x_i)  \right).
$$
Equation \eqref{eq::ols-populationols-clt} states that 
$
\sqrt{n} (  \hat{\beta}  - \beta ) \rightarrow \N(0,  B^{-1} M B^{-1} )
$
in distribution, 
where $B = E(xx^{\T})$ and $M = E(\varepsilon^2 xx^{\T})$.

Prove that 
$$
\cov( \hat{\beta}  - \beta ) = \cov(  \hat{\beta}  - \tilde\beta ) + \cov(   \tilde\beta - \beta  ),
$$
and moreover, 
\begin{eqnarray*}
\sqrt{n} (  \hat{\beta}  - \tilde\beta ) & \rightarrow & \N(0,  B^{-1} M_1 B^{-1} ),\\ 
\sqrt{n} (    \tilde\beta - \beta )  & \rightarrow & \N(0,  B^{-1} M_2 B^{-1} )
\end{eqnarray*}
in distribution, where 
\begin{eqnarray*}
M_1 &=& E(e^{2} xx^{\T}),\\ 
M_2 &=& E( \delta^2 xx^{\T})
\end{eqnarray*}
Verify that $M = M_1 + M_2$. 
\end{enumerate}

Remark: To prove the result, you may find the law of total covariance formula in \eqref{theorem::law-total-cov} helpful. We can also write $M_1$ as $M_1 = E\{ \var(y\mid x) xx^{\T} \}$. So the meat matrix $M$ has two sources of uncertainty, one is from the conditional variance of $y$ given $x$, and the other is from the approximation error.

\part{Overfitting, Regularization, and Model Selection}\label{part::modelselection}
   
\chapter{Perils of Overfitting}\label{chapter::overfitting}

Previous chapters assume that the covariate matrix $X$ is given and the linear model, correctly specified or not, is also given. Although including useless covariates in the linear model results in less precise estimators, this problem is not severe when the total number of covariates is small compared with the sample size. In many modern applications, however, the number of covariates can be large compared with the sample size. Sometimes, it can be a nonignorable fraction of the sample size; sometimes, it can even be larger than the sample size. For instance, modern DNA sequencing technology often generates covariates of millions of dimensions, which is much larger than the usual sample size under study. In these applications, the theory in previous chapters is inadequate. This chapter introduces an important notion in statistics: overfitting.

\section{David Freedman's simulation}

\citet{freedman1983note} used a simple simulation to illustrate the problem with a large number of covariates. 
He simulated data from the following Normal linear model $Y=X\beta+\varepsilon$
with $\varepsilon\sim\N(0,\sigma^{2}I_{n})$ and $\beta=(\mu,0,\ldots,0)^{\T}$. He then computed the sample $R^{2}$. Since the covariates do not explain any variability of 
the outcome at all in the true model, we would expect $R^{2}$ to be extremely small over repeated sampling. However,
he showed, via both simulation and theory, that $R^{2}$ is surprisingly
large when $p / n$ is not close to 0.

\begin{figure} 
\centering
\includegraphics[width =  \textwidth]{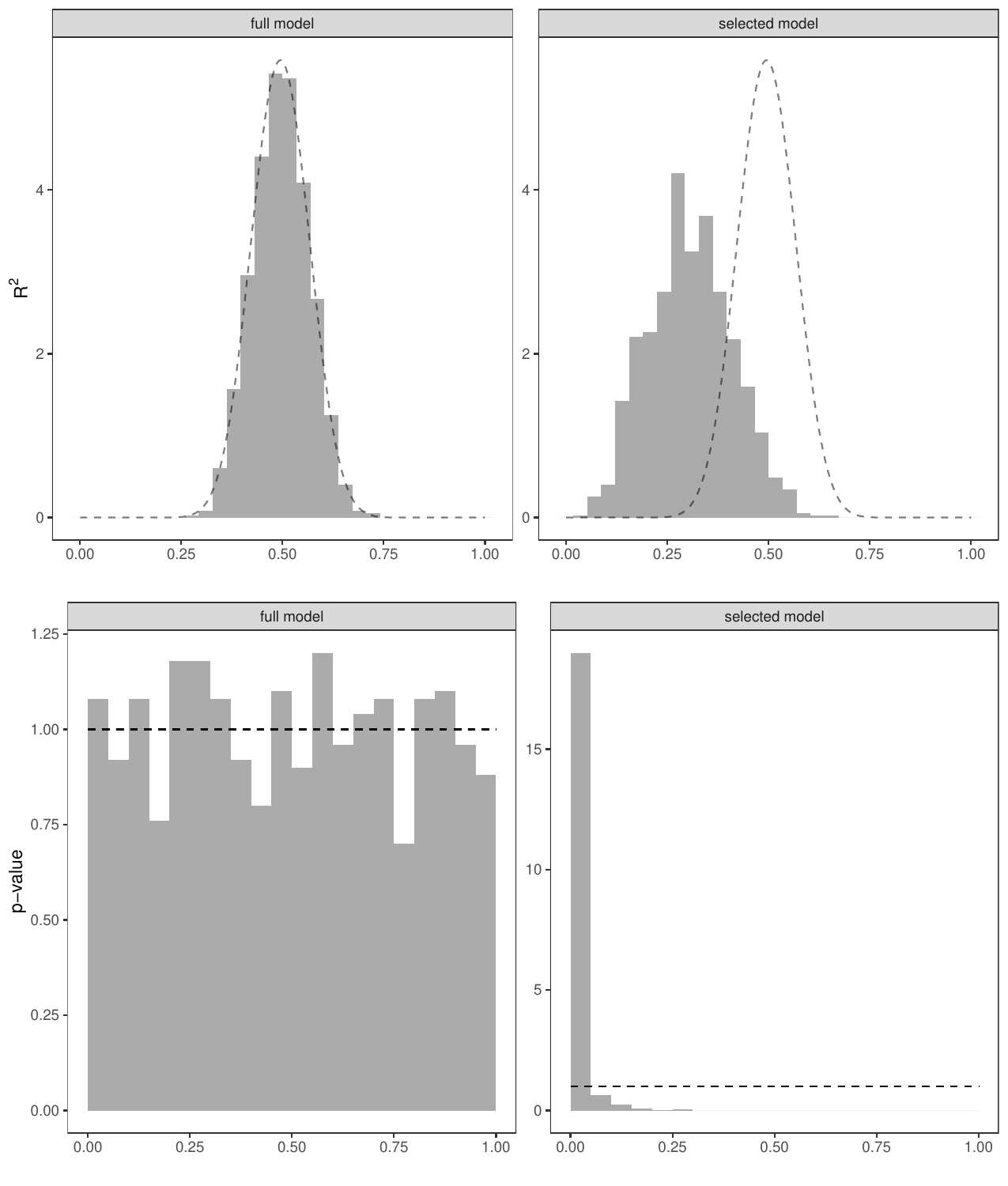}
\caption{Freedman's simulation. The first row shows the histograms of the $R^2$s, and the second row shows the histograms of the $p$-values in testing that all coefficients are 0. The first column corresponds to the full model without testing, and the second column corresponds to the selected model with testing at the significance level $0.25$.
}\label{fig::freedman-simulation}
\end{figure}

Figure \ref{fig::freedman-simulation} shows the results from Freedman's simulation setting with $n=100$ and $p=50$, over $1000$ replications. 
The $(1,1)$th subfigure shows the histogram of the $R^2$, which centers around $0.5$. 
This can be easily explained by the exact distribution of $R^{2}$ proved in Corollary \ref{coro::dist-r2}: 
\[
R^{2}\sim\textup{Beta}\left(\frac{p-1}{2},\frac{n-p}{2}\right),
\]
with the density shown in the $(1,1)$th and $(1,2)$th subfigure of Figure \ref{fig::freedman-simulation}. Based on the formulas in Proposition \ref{thm:beta-moments}, the beta distribution above has mean
\[
E(R^{2})=\frac{\frac{p-1}{2}}{\frac{p-1}{2}+\frac{n-p}{2}}=\frac{p-1}{n-1}
\]
 and variance
\begin{eqnarray*}
\var(R^{2}) 
&=& \frac{\frac{p-1}{2}\times\frac{n-p}{2}}{\left(\frac{p-1}{2}+\frac{n-p}{2}\right)^{2}\left(\frac{p-1}{2}+\frac{n-p}{2}+1\right)} \\
&=&\frac{2(p-1)(n-p)}{(n-1)^{2}(n+1)}.
\end{eqnarray*}
When $p/n\rightarrow0$, we have 
\[
E(R^{2})\rightarrow0,\quad\var(R^{2})\rightarrow0,
\]
so Markov's inequality implies that $R^{2}\rightarrow0$ in probability (see Proposition \ref{prop::markov-lln} for a related result).
However, when $p/n\rightarrow\gamma\in(0,1)$, we have 
\[
E(R^{2})\rightarrow\gamma,\quad\var(R^{2})\rightarrow0,
\]
so Markov's inequality implies that $R^{2}\rightarrow\gamma$ in probability.
This means that when $p$ has the same order as $n$, the sample
$R^{2}$ is close to the ratio $p/n$ even though there is no association
between the covariates and the outcome in the true data-generating
process. In Freedman's simulation, $\gamma=0.5$ so $R^{2}$ is close
to $0.5$.

The $(1,2)$th subfigure shows the histogram of the $R^2$ based on a model selection first step by dropping all covariates with $p$-values larger than 0.25. The $R^2$ in the $(1,2)$th subfigure are slightly smaller but still centered around $0.37$. The joint $F$ test based on the selected model does not generate uniform $p$-values in the $(2,2)$th subfigure, in contrast to the uniform $p$-values in the $(2,1)$th subfigure. With a model selection step, statistical inference becomes much more complicated. This is a topic called {\it selective inference}, which is beyond the scope of this book.

The above simulation and calculation give an important warning: we cannot
over-interpret the sample $R^{2}$, because it can be too optimistic
about model fitting. In many empirical research, $R^{2}$ is at most
$0.1$ with a large number of covariates, making us wonder whether those researchers are just chasing the noise rather than the signal. So we do not trust $R^{2}$ as a model-fitting measure with a
large number of covariates. In general, $R^2$ cannot avoid overfitting, and we must modify it for model selection.

\section{Variance inflation factor}
\label{section::vif-theorem}

Theorem \ref{theorem:varianceinflationfactor} below quantifies the potential problem of including too many covariates in OLS. 
It introduces the notion of variance inflation factor (VIF).

\begin{theorem}
\label{theorem:varianceinflationfactor}
Consider a fixed covariate matrix $X = (X_1, \ldots, X_p) \in \mathbb{R}^p$.
Let $\hat{\beta}_{j}$ be the coefficient of $X_{j}$ of the OLS fit
of $Y$ on $(1_{n},X_{j}: j \in S ) $, where $S$ is a subset of $\{1, \ldots, p \}$. Under the
model $y_{i}=f(x_{i})+\varepsilon_{i}$ with an unknown (and possibly nonlinear) $f(\cdot)$ and the $\varepsilon_{i}$'s
uncorrelated with mean zero and variance $\sigma^{2}$, the variance
of $\hat{\beta}_{j}$ equals  
\[
\var(\hat{\beta}_{j})=\frac{\sigma^{2}}{\sumn(x_{ij}-\bar{x}_{j})^{2}}\times\frac{1}{1-R_{j}^{2}},
\]
where $R_{j}^{2}$ is the sample $R^{2}$ from the OLS fit of $X_{j}$
on $1_{n}$ and all other covariates in $\{X_{j}: j \in S\}$. 
\end{theorem}

Theorem \ref{theorem:varianceinflationfactor} does not even assume that the true mean function is linear. It 
states that the variance
of $\hat{\beta}_{j}$ has two multiplicative components. If we run
a short regression of $Y$ on $1_{n}$ and $X_{j} = (x_{1j}, \ldots, x_{nj})^{\T}$, the coefficient equals
\[
\tilde{\beta}_{j}=\frac{\sumn(x_{ij}-\bar{x}_{j})y_{i}}{\sumn(x_{ij}-\bar{x}_{j})^{2}}
\]
where $\bar{x}_{j} = n^{-1} \sumn x_{ij}$. It
has variance
\begin{eqnarray*}
\var(\tilde{\beta}_{j}) &=& \var\left\{ \frac{\sumn(x_{ij}-\bar{x}_{j})y_{i}}{\sumn(x_{ij}-\bar{x}_{j})^{2}}\right\} \\
&=&\frac{\sumn(x_{ij}-\bar{x}_{j})^{2}\sigma^{2}}{\left\{ \sumn(x_{ij}-\bar{x}_{j})^{2}\right\} ^{2}} \\
&=&\frac{\sigma^{2}}{\sumn(x_{ij}-\bar{x}_{j})^{2}}.
\end{eqnarray*}
So the first component is the variance of the OLS coefficient in the
short regression. The second component $1/(1-R_{j}^{2})$ is called
the VIF. The VIF indeed inflates the variance
of $\tilde{\beta}_{j}$, and the more covariates are added into the
long regression, the larger the variance inflation factor is. 
In \ri{R}, the \ri{car} package provides the function \ri{vif} to compute the VIF for each covariate. 

The proof of Theorem \ref{theorem:varianceinflationfactor} below is based
on the FWL Theorem in Chapter \ref{chapter::FWL-theorem}.

\begin{myproof}{Theorem}{\ref{theorem:varianceinflationfactor}}
Let $\tilde{X}_{j} = (\tilde{x}_{1j}, \ldots, \tilde{x}_{nj})^{\T}$ be the residual vector from the OLS fit of $X_{j}$
on $1_{n}$ and all other covariates in $\{X_{j}: j \in S\}$, which have sample mean 0. The FWL Theorem implies that 
\[
\hat{\beta}_{j}=\frac{\sumn\tilde{x}_{ij}y_{i}}{\sumn\tilde{x}_{ij}^{2}} , 
\]
which has variance
\begin{equation}\label{eq::vif-variance1}
\var(\hat{\beta}_{j})=\frac{\sumn\tilde{x}_{ij}^{2}\sigma^{2}}{\left\{ \sumn\tilde{x}_{ij}^{2}\right\} ^{2}}=\frac{\sigma^{2}}{\sumn\tilde{x}_{ij}^{2}}.
\end{equation}
Because $\sumn\tilde{x}_{ij}^{2}$ is the residual sum of squares
from the OLS of $X_{j}$ on $1_{n}$ and all other covariates in $\{X_{j}: j \in S\}$, it is
related to $R_{j}^{2}$ via
\[
R_{j}^{2}=1-\frac{\sumn\tilde{x}_{ij}^{2}}{\sumn(x_{ij}-\bar{x}_{j})^{2}}
\]
or, equivalently,
\begin{equation}\label{eq::vif-variance2}
\sumn\tilde{x}_{ij}^{2}=(1-R_{j}^{2})\sumn(x_{ij}-\bar{x}_{j})^{2}.
\end{equation}
Combining \eqref{eq::vif-variance1} and \eqref{eq::vif-variance2} gives Theorem \ref{theorem:varianceinflationfactor}. 
\end{myproof}

\section{Bias-variance trade-off}

Theorem \ref{theorem:varianceinflationfactor} above characterizes the variance
of the OLS coefficient, but it does not characterize its bias. In
general,   a more complex model is closer to
the true mean function $f(x_i)$, and can then reduce the bias of approximating the mean function.
However, Theorem \ref{theorem:varianceinflationfactor} implies that a more complex model results in larger variances of the OLS
coefficients. So we face a bias-variance trade-off.

Consider a simple case where the true data-generating process is linear:
\begin{equation}
y_{i}=\beta_{0}+\beta_{1}x_{i1}+\cdots+\beta_{s}x_{is}+\varepsilon_{i} . \label{eq:correct-model}
\end{equation}
Ideally, we want
to use the model (\ref{eq:correct-model}) with exactly $s$ covariates. In practice, we may not know which covariates to include in the OLS. 
If we underfit the data using a short regression with $q<s$: 
\begin{equation}
y_{i}=\tilde{\beta}_0+\tilde{\beta}_{1}x_{i1}+\cdots+\tilde{\beta}_{q}x_{iq}+\tilde{\varepsilon}_{i},\qquad(i=1,\ldots,n)\label{eq:underfitted-model}
\end{equation}
then the OLS coefficients are biased. If we increase the complexity of
the model to overfit the data using a long regression with $p>s$: 
\begin{equation}
y_{i}=\hat{\beta}_0 +\hat{\beta}_{1}x_{i1}+\cdots+\hat{\beta}_{p}x_{ip}+\hat{\varepsilon}_{i},\qquad(i=1,\ldots,n)\label{eq:overfitted-model}
\end{equation}
then the OLS coefficients are unbiased. Theorem \ref{theorem:varianceinflationfactor},
however, shows that the OLS coefficients from the under-fitted model
(\ref{eq:underfitted-model}) have smaller variances than those from
the overfitted model (\ref{eq:overfitted-model}).

Example \ref{example::bias-variance-tradeoff} below illustrates the idea of overfitting under an ideal Normal linear model.

\begin{figure}[ht]
\centering 
\includegraphics[width = 0.8\textwidth]{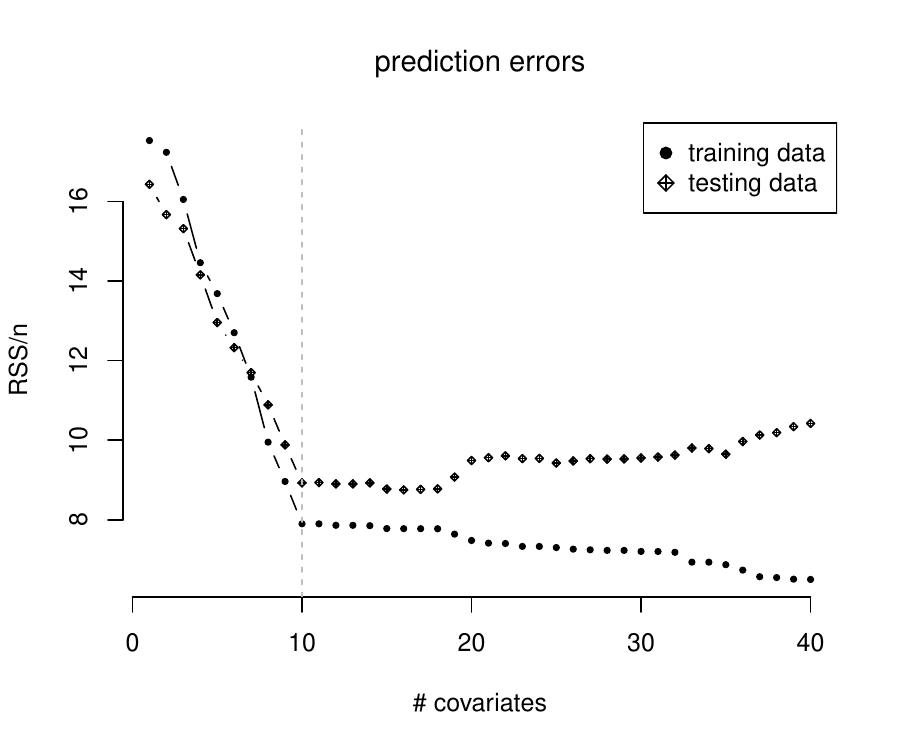}
\caption{Training and testing errors: linear mean function}\label{fig::bias-variance-tradeoff-linearmodel}
\end{figure}

\begin{example}\label{example::bias-variance-tradeoff}
In general, we have a sequence of models with increasing complexity.
For simplicity, we consider nested models containing $1_{n}$ and covariates
\[
\left\{ X_{1}\right\} \subset\left\{ X_{1},X_{2}\right\} \subset\cdots\subset\left\{ X_{1},\ldots,X_{p}\right\}
\]
in the following simulation setting. The true linear model is $y_i=x_i^{\T} \beta + \varepsilon_i$, $\varepsilon_i \stackrel{\textup{IID}}{\sim} \N(0,1)$ with $p=40$ but only the first 10 covariates have non-zero coefficients 1 and all other covariates have coefficients 0. 
We generate two datasets: both have sample size $n=200$, all covariates have IID $\N(0,1)$ entries, and the error terms are IID.
We use the first dataset to fit the OLS and thus call it the `` training dataset.'' We use the second dataset to assess the performance of the fitted OLS from the training dataset, and thus call it the ``testing dataset.''\footnote{Splitting a dataset into the {\it training dataset} and the {\it testing dataset} is a standard tool to assess the out-of-sample performance of proposed methods. It is important in statistics and machine learning.}
Figure \ref{fig::bias-variance-tradeoff-linearmodel} plots the residual sum of squares against the number of covariates in the training and testing datasets. By definition of OLS, the residual sum of squares decreases with the number of covariates in the training dataset, but it first decreases and then increases in the testing dataset with minimum value attained at $10$, the number of covariates in the true data generating process. 
\end{example}

Example \ref{eg::nonlinear-overfitting} below illustrates the idea of overfitting under a nonlinear model. Even though the true mean function is nonlinear, we still use OLS with polynomials of covariates to approximate the truth.\footnote{A celebrated theorem due to Weierstrass states that on a bounded interval, any continuous function can be approximated arbitrarily well by a polynomial function: 
\begin{theorem}[Weierstrass's theorem]
\label{thm::Weierstrass}
Suppose $f$ is a continuous function defined on the interval $[a, b]$. For every $\varepsilon  > 0$, there exists a polynomial $p$ such that for all $x \in [a, b]$, we have $|f(x) - p(x) | < \varepsilon $.
\end{theorem} 
}

\begin{figure}[ht]
\centering 
\includegraphics[width = 0.8 \textwidth]{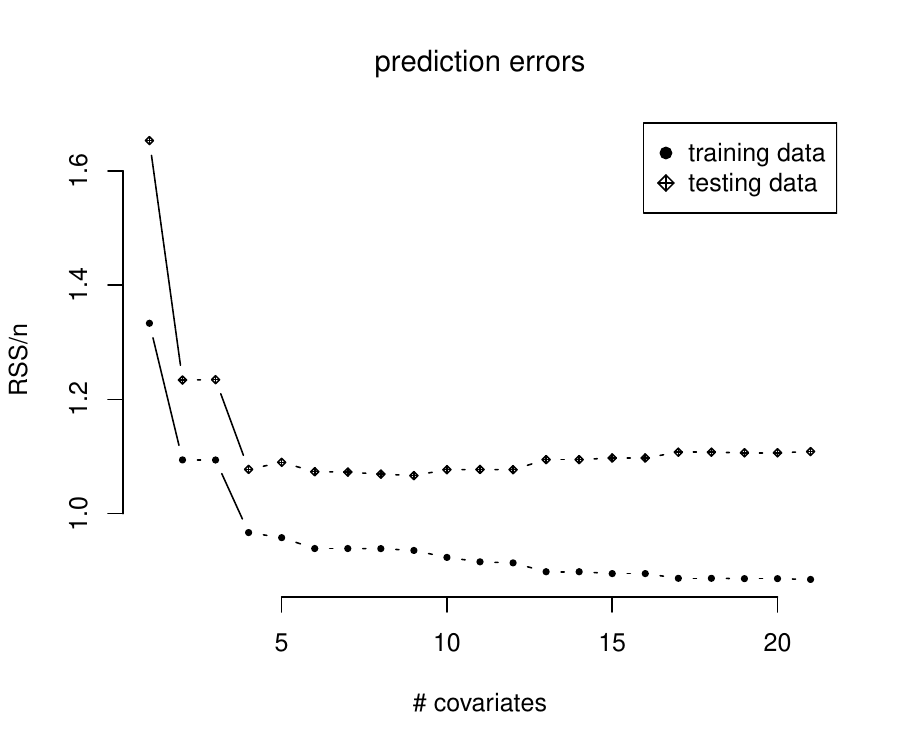}
\caption{Training and testing errors: nonlinear mean function}\label{fig::bias-variance-tradeoff-linearmodel-poly}
\end{figure}

\begin{example}\label{eg::nonlinear-overfitting}
The true nonlinear model is $y_i  = \sin(2\pi x_i) + \varepsilon_i$, $\varepsilon_i \stackrel{\textup{IID}}{\sim} \N(0,1)$ with the $x_i$'s equally spaced in $[0,1]$ and the error terms are IID. The training and testing datasets both have sample sizes $n=200$. Figure \ref{fig::bias-variance-tradeoff-linearmodel-poly} plots the residual sum of squares against the order of the polynomial in the OLS fit
$$
y_i = \sum_{j=0}^{p-1} \beta_j x_i^j + \varepsilon_i.
$$
By the definition of OLS, the residual sum of squares decreases with the order of polynomials in the training dataset, but it achieves the minimum near $p=5$ in the testing dataset. We can show that the residual sum of squares decreases to zero with $p=n$ in the training dataset; see Problem \ref{hw12::perfect-polynomial}. However, it is larger than that under $p=5$ in the testing dataset. 
\end{example}

\section{Model selection criteria}

With a large number of covariates $X_{1},\ldots,X_{\overline{p}}$, we want to
select a model that has the best performance for prediction. In total,
we have $2^{\overline{p}}$ possible models. Which one is the best? What is the
criterion for the best model? Practitioners often use the linear model for multiple purposes. A dominant criterion is the prediction performance of the linear model in a new dataset \citep{yu2020veridical}.
However, we do not have the new dataset yet in the statistical modeling stage. So we need to find criteria that are good proxies for the prediction performance.

\subsection{RSS, R-squared and adjusted R-squared}

The first obvious criterion is the residual sum of squares (\textsc{rss}), which, however, is not a good criterion because it favors the largest model. 
The sample $R^{2}$ has the same problem of favoring the largest model.
Most model selection criteria are in some sense modifications of \textsc{rss} or $R^2$.

The adjusted $R^{2}$ takes into account the complexity of the model:
\begin{align*}
\bar{R}^{2} 
 & =1-\frac{\hat{\sigma}^{2}}{\hat{\sigma}_{y}^{2}} \\ 
  & =1-\frac{\sumn(y_{i}-\hat{y}_{i})^{2}/(n-p)}{\sumn(y_{i}-\bar{y})^{2}/(n-1)}\\
& =1-\frac{n-1}{n-p}(1-R^{2}) . 
\end{align*}
So based on $\bar{R}^{2}$, the best model has the smallest $\hat{\sigma}^{2}$,
the estimator for the variance of the error term in the Gauss--Markov
model. Theorem \ref{theorem:Fstat-adjustedR2} below shows that $\bar{R}^{2}$ is closely related to the $F$ statistic in
testing two nested Normal linear models.

\begin{theorem}
\label{theorem:Fstat-adjustedR2} 
Consider the setting of Chapter \ref{sec::fwl-anova}.
Test two nested Normal linear
models:
\[
Y=X_{1}\beta_{1}+\varepsilon
\]
versus
\[
Y=X_{1}\beta_{1}+X_{2}\beta_{2}+\varepsilon,
\]
or, equivalently, test $\beta_{2}=0$. We can use the standard
$F$ statistic defined in Chapter \ref{sec::fwl-anova}, and we can also compare the adjusted $R^{2}$'s from
these two models: $\bar{R}_{1}^{2}$ and $\bar{R}_{2}^{2}$. 

They are related in the following sense: $F>1$ if and only if $\bar{R}_{1}^{2}<\bar{R}_{2}^{2}$. 
\end{theorem}

I leave the proof of Theorem \ref{theorem:Fstat-adjustedR2} as Problem \ref{hw12::f-r2}. 
From Theorem \ref{theorem:Fstat-adjustedR2}, $\bar{R}^{2}$ does not necessarily favor the largest model. However, $\bar{R}^{2}$ still
favors unnecessarily large models compared with the usual
hypothesis testing based on the Normal linear model because the mean of $F$ is approximately $1$, but the upper
quantile of $F$ is much larger than $1$ (for example, the 95\% quantile of $F_{1,n-p}$ is larger than 3.8, and the 95\% quantile of $F_{2,n-p}$ is larger than 2.9).

\subsection{Information criteria}
\label{chapter::aic-bic}

Taking into account the model complexity, we can find the model with the smallest \textsc{aic} or \textsc{bic}, defined as
\begin{eqnarray*}
\textsc{aic} &=& n\log\frac{\textsc{rss}}{n}+2p , \\
\textsc{bic} &=& n\log\frac{\textsc{rss}}{n}+p\log n,
\end{eqnarray*}
with full names
``Akaike's information criterion '' and 
``Bayes information criterion,'' respectively. The theoretical derivations of \textsc{aic} and \textsc{bic} are beyond the scope of this book. 

\textsc{aic} and \textsc{bic} are both monotone functions of the \textsc{rss} penalized by the number of parameters $p$ in the model. The penalty in \textsc{bic} is larger, so it favors smaller models than \textsc{aic}. \citet{shao1997asymptotic}'s results suggested that \textsc{bic} can consistently select the true model if the linear model is correctly specified, but \textsc{aic} can select the model that minimizes the prediction error if the linear model is misspecified. In most statistical practice, the linear model assumption cannot be justified, so we recommend using \textsc{aic}.

\subsection{Cross-validation (CV)}

We can use the leave-one-out cross-validation based on the predicted residual: 
$$
\textsc{press}=\sumn\hat{\varepsilon}_{[-i]}^{2}, 
$$
which is called the predicted residual error sum of squares (PRESS)
statistic. By Theorem \ref{theorem:looresidual}, it equals 
\begin{equation}
\label{eq::press-stat}
\textsc{press}
=\sumn\frac{\hat{\varepsilon}_i^{2}}{(1-h_{ii})^{2}},
\end{equation}
and therefore, it depends not only on the residuals but also on the leverage scores. 

Because the average value of $h_{ii}$ is $n^{-1}\sumn h_{ii}=p/n$,
we can approximate PRESS by the generalized cross-validation (GCV)
criterion:
\begin{eqnarray*}
\textsc{gcv} &=& \sumn\frac{\hat{\varepsilon}_i^{2}}{(1-p/n)^{2}}\\
&=&\textsc{rss}\times\left(1-\frac{p}{n}\right)^{-2}.
\end{eqnarray*}
When $p/n\approx0$, we have\footnote{The approximation is due to the Taylor expansion $\log(1+x) = x - x^2/2 + x^3/3 - \cdots \approx x$.}
\begin{eqnarray*}
\log\textsc{gcv} &=& \log\textsc{rss}-2\log\left(1-\frac{p}{n}\right)\\
&\approx &\log\textsc{rss}+\frac{2p}{n} \\
&=&\textsc{aic}/n+\log n,
\end{eqnarray*}
so GCV is approximately equivalent to AIC with small $p/n$. With large $p/n$, they may have large differences. 

GCV is not crucial for OLS, because it is easy to compute PRESS.
However, it is much more useful in other models where
we need to fit the data $n$ times to compute PRESS. 
For a general model without simple leave-one-out formulas, it is computationally intensive to obtain PRESS. The
$K$-fold cross-validation ($K$-CV) is computationally more attractive. The best model has the smallest $K$-CV, computed as follows:
\begin{enumerate}
\item randomly shuffle the observations;
\item split the data into $K$ folds;
\item for each fold $k$, use all other folds as the training data, and
compute the predicted errors on fold $k$ $(k=1,\ldots, K)$;
\item sum up the prediction errors across $K$ folds, denoted by $K$-CV.
\end{enumerate}

When $K  = 3$, we split the data into $3$ folds. Run OLS to obtain a fitted function with folds $2,3$ and use it to predict on fold 1, yielding prediction error $r_1$;  run OLS with folds $1,3$ and predict on fold $2$,  yielding prediction error $r_2$; run OLS with folds $1, 2$ and predict on fold $3$, yielding prediction error $r_3$. The total prediction error is $r = r_1  + r_2 + r_3$. We want to select a model that minimizes $r$. Usually, practitioners choose $K=5$ or $10$, but this can depend on the computational resource.

\section{Best subset and forward/backward selection}

Given a model selection criterion, we can select the best model. 

\begin{figure}[H]
\centering 
\includegraphics[width =  \textwidth]{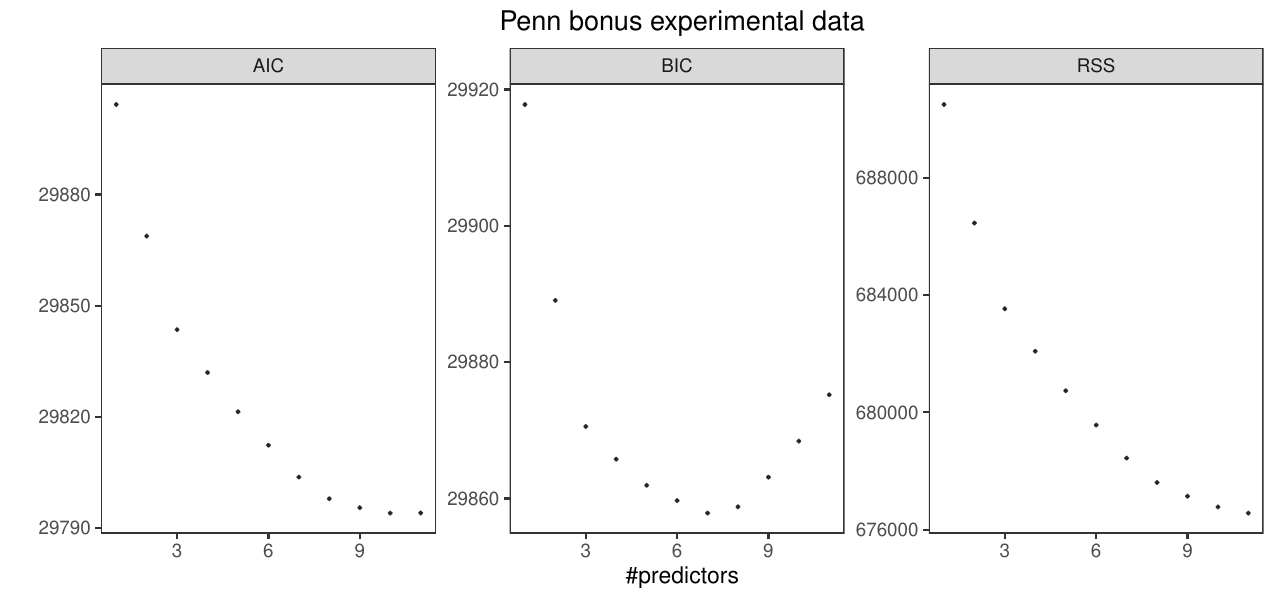}
\includegraphics[width = \textwidth]{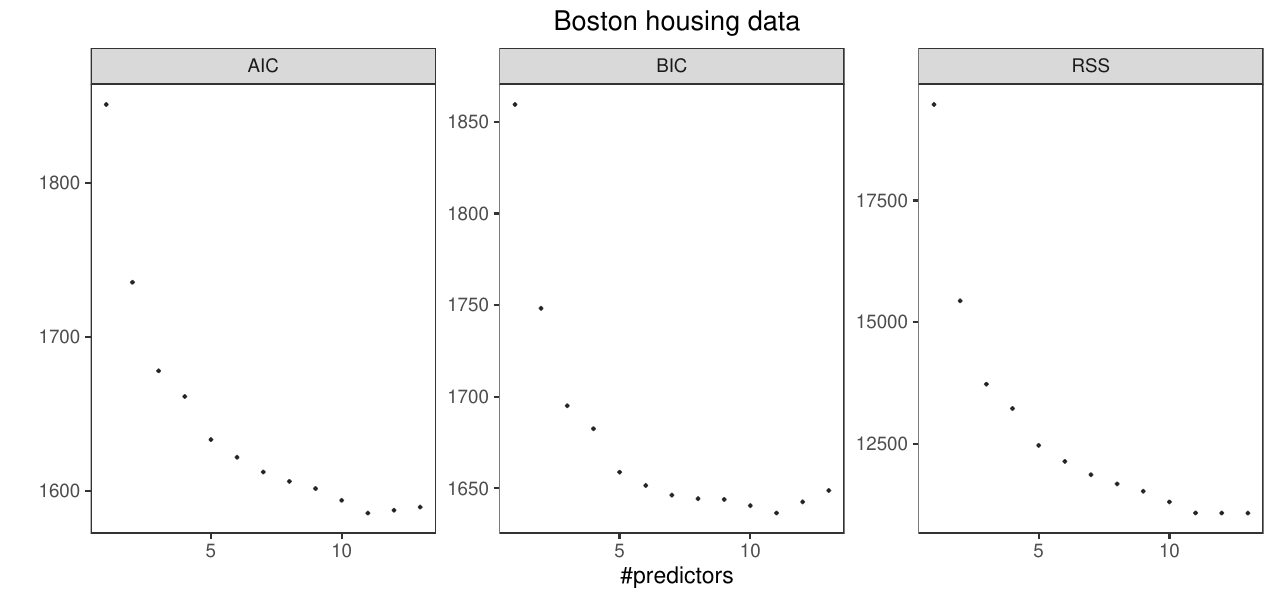} 
\caption{Best subset selection}\label{fig::best-subset-selection-penn-boston}
\end{figure}

For a small $\overline{p}$, we can enumerate all $2^{\overline{p}}$ models. The function \ri{regsubsets} in the \ri{R} package \ri{leaps} implements this.\footnote{Note that this function uses a definition of \textsc{bic} that differs from the above definition by a constant, but this does not change the model selection result.}
Figure \ref{fig::best-subset-selection-penn-boston} shows the results of the best subset selection in two applications. \textsc{rss} always favors the largest model. \textsc{bic} favors slightly smaller model than \textsc{aic}.

For large $\overline{p}$, we can use forward or backward regressions. 
Forward regression starts with a model with only the intercept. In step one, it finds the best covariate among the $\overline{p}$ candidates based on the prespecified criterion. In step two, it keeps this covariate in the model and finds the next best covariate among the remaining $\overline{p} - 1$ candidates. It proceeds by adding the next best covariate one by one.

The backward regression does the opposite. It starts with the full model with all $\overline{p}$ covariates. In step one, it drops the worst covariate among the $\overline{p}$ candidates based on the prespecified criterion. In step two, it drops the next worst covariate among the remaining $\overline{p} - 1$ candidates. It proceeds by dropping the next worst covariate one by one.

Both methods generate a sequence of models, and select the best one based on the prespecified criterion. Forward regression works in the case with $p\geq n$ but it stops at step $n-1$; backward regression works only in the case with $p<n.$ The functions \ri{step} or \ri{stepAIC} in the \ri{MASS} package implement them.

\section{Homework problems}

\paragraph{Inflation and deflation of the estimated variance}\label{hw12::inf-def-est-var}

This problem extends Theorem \ref{theorem:varianceinflationfactor} to the estimated variance. 

The covariate matrix $X$ has columns $1_n, X_1, \ldots, X_p$. 
Compare the coefficient of $X_1$ in the following long and short regressions:
$$
Y = \hat{\beta}_0 1_n  + \hat{\beta}_1 X_1 + \cdots + \hat{\beta}_p X_p  +\hat \varepsilon,
$$
and 
$$ 
Y =\tilde{\beta}_0 1_n  +  \tilde{\beta}_1 X_1 + \cdots + \tilde{\beta}_q  X_q   +\tilde \varepsilon,
$$
where $q < p$. Under the condition in Theorem \ref{theorem:varianceinflationfactor}, 
$$  \frac { \var( \hat{\beta}_1 ) }{  \var( \tilde{\beta}_1 ) }
 = \frac{  1- R^2_{X_1.X_2\cdots X_q}  }{  1- R^2_{X_1.X_2\cdots X_p}   } \geq 1 ,
$$
recalling that $R^2_{U.V}$ denotes the $R^2$ of $U$ on $V$. 
Now we compare the corresponding estimated variances $\hat \var( \hat{\beta}_1 )$ and $\tilde \var( \tilde{\beta}_1 )$ based on homoskedasticity. 

\begin{enumerate}
\item
Prove that
$$
\frac{\hat \var( \hat{\beta}_1 )}{  \tilde \var( \tilde{\beta}_1 ) }
=   \frac{   1 - R^2_{Y.X_1\cdots X_p}  }{  1 - R^2_{Y.X_1\cdots X_q}  } \times  \frac{  1- R^2_{X_1.X_2\cdots X_q}  }{  1- R^2_{X_1.X_2\cdots X_p}   }  \times \frac{n-q - 1}{n-p - 1}    . 
$$

\item
Using the definition of the partial $R^2$ in Problem \ref{hw09::partial-R2}, prove that
$$
\frac{\hat \var( \hat{\beta}_1 )}{  \tilde \var( \tilde{\beta}_1 ) }
=    \frac{   1 - R^2_{Y.X_{q+1} \cdots X_p  \mid  X_1\cdots X_q}  }{  1 - R^2_{ X_1.X_{q+1}\cdots X_p \mid  X_2\ldots X_q}  }  
\times  \frac{n-q - 1}{n-p - 1} .
$$
\end{enumerate}

Remark: The first result shows that the ratio of the estimated variances has three factors: 
\begin{enumerate}[label=(F\arabic*), ref=F\arabic*]
\item
the first factor corresponds to the $R^2$'s of the outcome on the covariates,
\item
the second factor equals the ratio of the true variances $\var( \hat{\beta}_1 )  /   \var( \tilde{\beta}_1 )$,
\item
the third factor corresponds to the degrees of freedom correction.
\end{enumerate}
The first factor deflates the estimated variance since the $R^2$ increases with more covariates included in the regression, and the second and the third factors inflate the estimated variance. Overall, whether adding more covariates inflate or deflate the estimated variance depends on the interplay of the three factors. The answer is not as definite as Theorem \ref{theorem:varianceinflationfactor}.

The variance inflation result in Theorem \ref{theorem:varianceinflationfactor} sometimes causes confusion. It only concerns the variance. When we view some covariates as random, then the bias term can also contribute to the variance of the OLS estimator. In this case, we should interpret Theorem \ref{theorem:varianceinflationfactor} with caution. See \citet{ding2019two} for a related discussion.

\paragraph{Inflation and deflation of the variance under heteroskedasticity}\label{hw12::inf-def-heteroskedasticity}

Revisit Section \ref{section::vif-theorem}. 
Relax the condition in Theorem \ref{theorem:varianceinflationfactor}  as $\var(\varepsilon_i) = \sigma_i^2$ to allow for heteroskedasticity with possibly different variances of the error terms.  Give a counterexample in which
$$
\var(\hat\beta_j)  > \var(\tilde\beta_j)
$$
for some $j$.

Remark: First derive the formulas of $\var(\hat\beta_j)  $ and $ \var(\tilde\beta_j)$ under heteroskedasticity, and then give a numerical example by specifying the $X$ matrix and the $\sigma_i^2$'s.

\paragraph{Equivalence of $F$ and $\bar{R}^{2}$}\label{hw12::f-r2}

Prove Theorem \ref{theorem:Fstat-adjustedR2}.

\paragraph{Using \textsc{press} to construct an unbiased estimator for $\sigma^2$}\label{hw12::press-unbiased-sigma2}

Prove that 
$$
\hat{\sigma}^{2}_{\textsc{press}} 
= 
\frac{  \textsc{press}  }{ \sumn (1-h_{ii})^{-1}   }
$$
is unbiased for $\sigma^2$ under the Gauss--Markov model in Assumption \ref{assume::gm-model}, recalling $\textsc{press} $ in \eqref{eq::press-stat} and the leverage score $h_{ii}$ of unit $i$.

Remark: Theorem \ref{thm:varianceestOLS} shows that $\hat{\sigma}^{2}=\textsc{rss}/(n-p)$ is unbiased for $\sigma^2$ under the Gauss--Markov model. 
\textsc{rss} is the ``in-sample'' residual sum of squares, whereas \textsc{press} is the ``leave-one-out'' residual sum of squares. 
The estimator $\hat{\sigma}^{2}$ is standard, whereas $\hat{\sigma}^{2}_{\textsc{press}} $ appeared in \citet{shen2023algebraic}.

\paragraph{Simulation with misspecified linear models}

Replicate the simulation in Example \ref{example::bias-variance-tradeoff} with correlated covariates and an outcome model with quadratic terms of covariates. 

\paragraph{Best subset selection in \ri{lalonde} data}

Produce the figure similar to the ones in Figure \ref{fig::best-subset-selection-penn-boston} based on the \ri{lalonde} data in the \ri{Matching} package. Report the selected model based on \textsc{aic}, \textsc{bic}, \textsc{press}, and \textsc{gcv}.

 \paragraph{Perfect polynomial}\label{hw12::perfect-polynomial}
 
Prove that given distinct $x_i \ (i=1, \ldots, n)$ within $[ 0,1 ]$ and any $y_i\ (i=1, \ldots, n)$, we can always find an $(n-1)$th order polynomial 
$$
p_n(x) = \sum_{j=0}^{n-1} b_j x^j
$$ 
such that 
$$
p_n(x_i) = y_i, \quad  (i=1, \ldots, n).
$$

Remark: Use the formula in \eqref{eq::det-Vandermonde} in Appendix \ref{chapter::linear-algebra}.

\chapter{Ridge Regression}
\label{chapter::ridge-regression}

The OLS estimator has many nice properties. For example, Chapter \ref{chapter::gauss-markov} shows that it is BLUE under the Gauss--Markov model, and Chapter \ref{chapter::normal-linear-model} shows that it follows a Normal distribution that allows for finite-sample exact inference under the Normal linear model.
However, OLS has problems with the columns of $X$ are highly correlated. This issue becomes salient when the number of covariates $p$ is large compared with the sample size $n$. In particular, when $p>n$, the OLS estimator is not unique because $X^{\T}X$ is not invertible. This chapter will first discuss these issues and then introduce ridge regression as a modification of OLS.

\section{Introduction to ridge regression}

The first motivation of ridge regression is straightforward from the linear algebra perspective.  From the formula
\[
\hat{\beta}=(X^{\T}X)^{-1}X^{\T}Y,
\]
if the columns of $X$ are highly correlated, then $X^{\T}X$ will
be nearly singular; more extremely, if the number of covariates is larger than the sample size, then $X^{\T}X$ has a rank smaller than
or equal to $n$ and thus is not invertible. So numerically, the OLS
estimator can be unstable due to inverting $X^{\T}X$. Because $X^{\T}X$
must be positive semi-definite, its smallest eigenvalue determines
whether it is invertible or not. \citet{hoerl1970ridge} proposed
the following ridge estimator as a modification of OLS:
\begin{equation}
\hat{\beta}^{\text{ridge}}(\lambda)=(X^{\T}X+\lambda I_{p})^{-1}X^{\T}Y, \qquad(\lambda>0)
\label{eq::ridge-hoerl1970form}
\end{equation}
which involves a positive tuning parameter $\lambda$. 
Because the smallest eigenvalue of $X^{\T}X+\lambda I_{p}$ is larger
than or equal to $\lambda>0$, the ridge estimator is always well
defined.

Now I turn to the second equivalent motivation. The 
OLS estimator minimizes the residual sum of squares 
$$
\textsc{rss}(b_0,b_1,\ldots,b_p) = 
\sumn(y_{i}-b_{0}-b_{1}x_{i1}-\cdots-b_{p}x_{ip})^{2}.
$$
From Theorem \ref{theorem:varianceinflationfactor} on the variance inflation factor, the variances of the OLS
estimators increase with additional covariates included in the regression,
leading to unnecessarily large estimators by chance. To avoid large OLS coefficients,
we can penalize the residual sum of squares criterion with the squared length of the coefficients\footnote{This is also called the Tikhonov regularization \citep{tikhonov1943stability}. See \citet{bickel2006regularization} for a review of the idea of regularization in statistics.}  
and use
\begin{equation}
\hat{\beta}^{\text{ridge}}(\lambda)=\arg\min_{b_{0},b_{1},\ldots,b_{p}}\left\{ \textsc{rss}(b_0,b_1,\ldots,b_p)+\lambda\sum_{j=1}^{p}b_{j}^{2}\right\} .\label{eq:ridge-definition-1}
\end{equation}
Again in (\ref{eq:ridge-definition-1}), $\lambda $ is a tuning parameter
that ranges from zero to infinity. We first discuss the ridge estimator
with a fixed $\lambda$ and then discuss how to choose it. When $\lambda=0$,
it reduces to OLS; when $\lambda=\infty$, all coefficients must be
zero except that $\hat{\beta}_{0}^{\text{ridge}}(\infty)=\bar{y}$.
With $\lambda \in (0,\infty)$, the ridge coefficients are generally
smaller than the OLS coefficients, and the penalty shrinks the OLS
coefficients toward zero. So the parameter $\lambda$ controls the
magnitudes of the coefficients or the ``complexity'' of the model.
In (\ref{eq:ridge-definition-1}), we only penalize the slope parameters
not the intercept.

As as equivalent form of \eqref{eq:ridge-definition-1}, we can also
define the ridge estimator as
\begin{align}
\hat{\beta}^{\text{ridge}}(t) & =\arg\min_{b_{0},b_{1},\ldots,b_{p}} \textsc{rss}(b_0,b_1,\ldots,b_p) \nonumber \\
 & \ \textup{such that }\sum_{j=1}^{p}b_{j}^{2}\leq t.\label{eq:ridge-definition-2}
\end{align}
Definitions (\ref{eq:ridge-definition-1}) and (\ref{eq:ridge-definition-2})
are equivalent because for a given $\lambda$, we can always find
a $t$ such that the solutions from (\ref{eq:ridge-definition-1})
and (\ref{eq:ridge-definition-2}) are identical. In fact, the corresponding $t$ and $\lambda$ satisfy $t = \|  \hat{\beta}^{\text{ridge}} (\lambda)  \|^2 .$

However, the ridge estimator has an obvious problem: it is not invariant
to linear transformations of $X$. In particular, it is not equivalent
under different scaling of the covariates. Intuitively, the $b_{j}$'s
depend on the scale of $X_{j}$'s, but the penalty term $\sum_{j=1}^{p}b_{j}^{2}$
puts equal weight on each coefficient. A convention in practice is
to standardize the covariates before applying the ridge estimator.\footnote{
I choose this standardization because it is the default choice in the function \ri{lm.ridge} in the \ri{R} package \ri{MASS}. 
In practical data analysis, the covariates may have concrete meanings. In those cases, you may not want to scale the covariates in the way as Condition \ref{condition::standardization}. 
However, the discussion below does not rely on the choice of scaling although it requires centering the covariates and outcome. 
} 

\begin{condition}
[standardization]\label{condition::standardization}
The covariates satisfy
\[
n^{-1}\sumn x_{ij}=0,\qquad n^{-1}\sumn x_{ij}^{2}=1,\qquad(j=1,\ldots,p)
\]
and the outcome satisfy $\bar{y} = 0$.
\end{condition}

With all covariates centered at zero, the ridge estimator for
the intercept, given any values of the slopes and the tuning parameter $\lambda$,
equals 
$
\hat{\beta}_{0}^{\text{ridge}}=\bar{y}.
$
So if we center the outcomes at mean zero, then we can drop the intercept
in the ridge estimators defined in (\ref{eq:ridge-definition-1})
and (\ref{eq:ridge-definition-2}).

For descriptive simplicity, I will assume Condition \ref{condition::standardization} and call it {\it standardization} from now on. Condition \ref{condition::standardization} allows us to drop the intercept. Using the matrix form, the ridge estimator minimizes
\[
 (Y-Xb)^{\T}(Y-Xb)+\lambda b^{\T}b,
\]
which is a quadratic function of $b$. From the first order condition, we have
$$
-2X^{\T}\{ Y-X\hat{\beta}^{\text{ridge}}(\lambda) \} +2\lambda\hat{\beta}^{\text{ridge}}(\lambda)=0.
$$
Solve the above linear equation of $\hat{\beta}^{\text{ridge}}(\lambda)$ to obtain
$$
\hat{\beta}^{\text{ridge}}(\lambda)
=(X^{\T}X+\lambda I_{p})^{-1}X^{\T}Y,
$$
which coincides with the definition in \eqref{eq::ridge-hoerl1970form}. We also have the second-order condition:
\[
2X^{\T}X+2\lambda I_{p}\succ0,
\]
is positive definite with any $\lambda>0$, 
which verifies that the ridge estimator is indeed the minimizer. 
The predicted vector is
\begin{eqnarray*}
\hat{Y}^{\text{ridge}}(\lambda) &=& X \hat{\beta}^{\text{ridge}}(\lambda) \\
&=& X (X^{\T}X+\lambda I_{p})^{-1}X^{\T}Y \\
&=& H(\lambda) Y,
\end{eqnarray*}
where 
$$
H(\lambda)  =  X (X^{\T}X+\lambda I_{p})^{-1}X^{\T} 
$$ 
is the hat matrix for ridge regression. When $\lambda = 0$, it reduces to the hat matrix for the OLS; when $\lambda >0$, it is not a projection matrix because $\{  H(\lambda) \} ^2 \neq H(\lambda).$

\section{Ridge regression via the SVD of the covariate matrix}

\begin{figure}[h]
\centering 
\includegraphics[width = \textwidth]{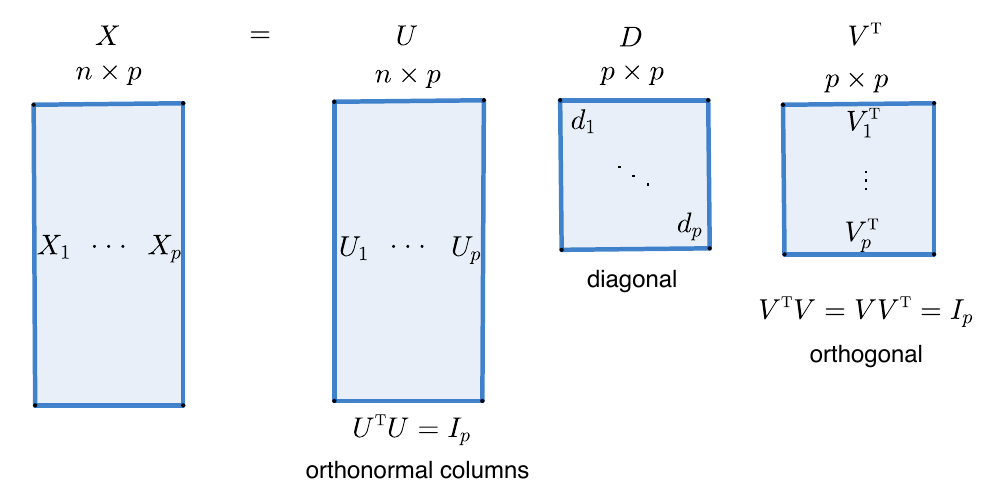}
\caption{SVD of $X$, when $n>p$}\label{fig::svd-X}
\end{figure}

I will focus on the case with $n \geq p$ and relegate the discussion of the case with $n \leq p$ to Problem \ref{hw13::compute-ridge-large-p}. 
To facilitate the presentation, I will use the singular value decomposition (SVD) of the covariate matrix:
$$
X = UDV^{\T}
$$
where $U \in \mathbb{R}^{n\times p}$ has orthonormal columns such that $U^{\T} U = I_p$, $V \in \mathbb{R}^{p\times p}$ is an orthogonal matrix with $V V^{\T} = V^{\T} V = I_p$, and $D \in \mathbb{R}^{p\times p}$ is a diagonal matrix consisting of the singular values. Figure \ref{fig::svd-X} illustrates the SVD. For readers who are not familiar with SVD, please review Appendix \ref{chapter::linear-algebra} before reading the remaining parts of this chapter.

The SVD of $X$ implies the eigen-decomposition of $X^{\T} X$:
$$
X^{\T} X = V D^2 V^{\T} 
$$
with eigenvectors $V_j$ being the column vectors of $V$ and eigenvalues $d_j^2$ being the squared singular values.
Lemma \ref{lemma::ridge-coefficient} below is crucial for simplifying the theory and computation.

\begin{lemma}
\label{lemma::ridge-coefficient}
The ridge coefficient equals
$$
 \hat{\beta}^{\textup{ridge}}(\lambda)  = 
V\textup{diag}\left(  \frac{d_j}{d_j^2 + \lambda }  \right)  U^{\T} Y,
$$
where the diagonal matrix is $p\times p$. 
\end{lemma}

\begin{myproof}{Lemma}{\ref{lemma::ridge-coefficient}}
The ridge coefficient equals
\begin{eqnarray*}
 \hat{\beta}^{\text{ridge}}(\lambda) &=& (X^{\T}X+\lambda I_{p})^{-1}X^{\T}Y \\
 &=& ( VDU^{\T} UD V^{\T}  + \lambda I_p)^{-1} VDU^{\T} Y \\
 &=& V( D^2 + \lambda I_p )^{-1} V^{\T} V D U^{\T} Y\\
 &=& V( D^2 + \lambda I_p )^{-1}   D U^{\T} Y\\
 &=& V\text{diag}\left(  \frac{d_j}{d_j^2 + \lambda }  \right) U^{\T} Y.
\end{eqnarray*}
\end{myproof}

\section{Statistical properties}

The Gauss--Markov theorem shows that the OLS estimator is BLUE under the Gauss--Markov
model: $Y=X\beta+\varepsilon$, where $\varepsilon$ has mean zero
and covariance $\sigma^{2}I_{n}$. Then in what sense, can ridge regression improve OLS? I will discuss the statistical properties of the
ridge estimator under the Gauss--Markov model.

Based on Lemma \ref{lemma::ridge-coefficient}, we can calculate the mean of the ridge estimator:
\begin{eqnarray*}
E\{  \hat{\beta}^{\text{ridge}}(\lambda)\} &=& V\text{diag}\left(  \frac{d_j}{d_j^2 + \lambda }  \right)  U^{\T} X\beta \\
&=&  V\text{diag}\left(  \frac{d_j}{d_j^2 + \lambda }  \right)  U^{\T} UDV^{\T}\beta \\
&=& V\text{diag}\left(  \frac{d_j^2}{d_j^2 + \lambda }  \right) V^{\T}\beta, 
\end{eqnarray*}
which does not equal $\beta$ in general. So the ridge estimator is biased. We can also calculate the covariance  matrix of the ridge estimator:
\begin{eqnarray*}
\cov\{  \hat{\beta}^{\text{ridge}}(\lambda)\} &=& \sigma^2 V\text{diag}\left(  \frac{d_j}{d_j^2 + \lambda }  \right)  U^{\T}  
U\text{diag}\left(  \frac{d_j}{d_j^2 + \lambda }  \right) V^{\T}  \\
&=&  \sigma^2 V\text{diag}\left(  \frac{d_j^2}{(d_j^2 + \lambda)^2 }  \right)   V^{\T}  . 
\end{eqnarray*}
The mean squared error (MSE)
is a measure capturing the bias-variance trade-off:
\[
\textsc{mse}(\lambda)=E\left[\left\{ \hat{\beta}^{\text{ridge}}(\lambda)-\beta\right\} ^{\T}\left\{ \hat{\beta}^{\text{ridge}}(\lambda)-\beta\right\} \right].
\]
Using Theorem \ref{thm::mean-quadratic-form} on the expectation of a quadratic form, we have
\[
\textsc{mse}(\lambda)=\underbrace{    [ E\{  \hat{\beta}^{\text{ridge}}(\lambda)\}  - \beta  ]^{\T} [ E\{  \hat{\beta}^{\text{ridge}}(\lambda)\} - \beta  ]   }_{C_{1}}
+\underbrace{\text{trace}[ \cov\{  \hat{\beta}^{\text{ridge}}(\lambda)\} ]  }_{C_{2}}.
\]
Theorem \ref{theorem:bias-variance-tradeoff-ridge} below simplifies $C_{1}$ and $C_{2}.$

\begin{theorem}
\label{theorem:bias-variance-tradeoff-ridge} Under  Assumption \ref{assume::gm-model}, the ridge estimator satisfies   
\begin{equation}
C_{1}=\lambda^{2}\sum_{j=1}^{p}\frac{\gamma_{j}^{2}}{( d_j^2 +\lambda)^{2}},\label{eq:term1-ridge-mse}
\end{equation}
where $\gamma=V^{\T}\beta=(\gamma_{1},\ldots,\gamma_{p})^{\T}$ has the $j$th coordinate $\gamma_j = V_j^{\T} \beta$,
and 
\begin{equation}
C_{2}=\sigma^{2}\sum_{j=1}^{p}\frac{ d_j^2 }{( d_j^2 +\lambda)^{2}}.\label{eq:term2-ridge-mse}
\end{equation}
\end{theorem}

\begin{myproof}{Theorem}{\ref{theorem:bias-variance-tradeoff-ridge}}
First, $\hat{\beta}^{\text{ridge}}(\lambda)$ is biased for estimating $\beta$:
$$
E\{\hat{\beta}^{\text{ridge}}(\lambda) \} - \beta = 
\left\{ V\text{diag}\left(  \frac{d_j^2}{d_j^2 + \lambda }  \right) V^{\T} - I_p \right\} \beta.
$$
Therefore, 
\begin{eqnarray*}
C_1 &=&     \beta^{\T}      \left\{   V\text{diag}\left(  \frac{d_j^2}{d_j^2 + \lambda }  \right) V^{\T} - I_p   \right\}^2 \beta \\
&=&    \beta^{\T}      V  \text{diag}\left(  \frac{\lambda^2}{ (d_j^2 + \lambda)^2 }  \right)   V^{\T}     \beta \\
&=& \gamma^{\T}  \text{diag}\left(  \frac{\lambda^2}{ (d_j^2 + \lambda)^2 }  \right)   \gamma \\
&=& \lambda^{2}\sum_{j=1}^{p}\frac{\gamma_{j}^{2}}{( d_j^2 +\lambda)^{2}} . 
\end{eqnarray*}


Second, we have 
\begin{eqnarray*}
C_2 &=&  \sigma^2 \text{trace} \left( V\text{diag}\left(  \frac{d_j^2}{(d_j^2 + \lambda)^2 }  \right)   V^{\T} \right) \\
&=& \sigma^2 \text{trace} \left( \text{diag}\left(  \frac{d_j^2}{(d_j^2 + \lambda)^2 }  \right ) \right) \\
&=& \sigma^{2}\sum_{j=1}^{p}\frac{ d_j^2 }{( d_j^2 +\lambda)^{2}}. 
\end{eqnarray*}

%
%
\end{myproof}

Theorem \ref{theorem:bias-variance-tradeoff-ridge} shows the bias-variance trade-off for the ridge estimator. The MSE is 
\begin{eqnarray*}
\textsc{mse}(\lambda) &=& C_{1}+C_{2} \\
&=& \lambda^{2}\sum_{j=1}^{p}\frac{\gamma_{j}^{2}}{( d_j^2 +\lambda)^{2}}+\sigma^{2}\sum_{j=1}^{p}\frac{ d_j^2 }{( d_j^2 +\lambda)^{2}}.
\end{eqnarray*}
When $\lambda=0$, the ridge estimator reduces to the OLS estimator:
the bias is zero and the variance $\sigma^{2}\sum_{j=1}^{p} d_j^{-2}$
dominates. When $\lambda=\infty$, the ridge estimator reduces to
zero: the bias $\sum_{j=1}^{p}\gamma_{j}^{2}$ dominates and the variance
is zero. As we increase $\lambda$ from zero, the bias increases and
the variance decreases. So we face a bias-variance trade-off.

\section{Selection of the tuning parameter}

\subsection{Based on parameter estimation}
For parameter estimation, we want to choose the $\lambda$ that minimizes the MSE.
So the optimal $\lambda$ must satisfy the following first-order condition:
$$
\frac{\partial\textsc{mse}(\lambda)}{\partial\lambda}  =2\sum_{j=1}^{p}\gamma_{j}^{2}\frac{\lambda}{ d_j^2 +\lambda}\frac{ d_j^2 +\lambda-\lambda}{( d_j^2 +\lambda)^{2}}-2\sigma^{2}\sum_{j=1}^{p}\frac{ d_j^2 }{( d_j^2 +\lambda)^{3}}=0
$$
which is equivalent to
\begin{equation}
\lambda  \sum_{j=1}^{p}\frac{\gamma_{j}^{2} d_j^2 }{( d_j^2 +\lambda)^{3}}=\sigma^{2}\sum_{j=1}^{p}\frac{ d_j^2 }{( d_j^2 +\lambda)^{3}}.\label{eq:optimal-ridge-tuning}
\end{equation}
However, (\ref{eq:optimal-ridge-tuning}) is not directly useful because
we do not know $\gamma$ and $\sigma^{2}$. Three methods below try
to solve (\ref{eq:optimal-ridge-tuning}) approximately. 

\citet{dempster1977simulation} used OLS to construct an unbiased estimator
$\hat{\sigma}^{2}$ and $\hat{\gamma}=V^{\T} \hat{\beta}$, and then
solve $\lambda$ from
\[
\lambda\sum_{j=1}^{p}\frac{\hat{\gamma}_{j}^{2} d_j^2 }{( d_j^2 +\lambda)^{3}}=\hat{\sigma}^{2}\sum_{j=1}^{p}\frac{ d_j^2 }{( d_j^2 +\lambda)^{3}},
\]
which is a nonlinear equation of $\lambda$. 

\citet{hoerl1975ridge} assumed that $X^{\T}X=I_{p}$. Then $ d_j^2 =1\ (j=1,\ldots,p)$
and $\gamma=\beta$, and solve $\lambda$ from
\[
\lambda\sum_{j=1}^{p}\frac{\hat{\beta}_{j}^{2}}{(1+\lambda)^{3}}=\hat{\sigma}^{2}\sum_{j=1}^{p}\frac{1}{(1+\lambda)^{3}},
\]
resulting in
\[
\lambda_{\textsc{hkb}}=p\hat{\sigma}^{2}/\|\hat{\beta}\|^{2}.
\]

\citet{jf1976simulation} used
\[
\lambda_{\textsc{lw}}=p\hat{\sigma}^{2}/\hat{\beta}^{\T}D^2 \hat{\beta}
\]
to weight the $\beta_j$'s based on the eigenvalues of $X^{\T} X$. 

But all these methods require estimating $(\beta,\sigma^{2})$. If
the initial OLS estimator is not reliable, then these estimates of
$\lambda$ are unlikely to be reliable. None of these methods work for the case with $p > n$.

\subsection{Based on prediction}
For prediction, we need slightly different criteria. 
Without estimating $(\beta,\sigma^{2})$, we can use the leave-one-out cross-validation. The leave-one-out formula for the ridge below is similar to that for OLS.

\begin{theorem}\label{thm::looformula-ridge}
Define $\hat{\beta}(\lambda) = (X^{\T} X + \lambda I_p)^{-1} X^{\T} Y$ as the ridge coefficient (dropping the superscript ``ridge'' for simplicity), $\hat{\varepsilon}(\lambda) = Y - X \hat{\beta}(\lambda)$ as the residual vector using the full data, and $ h_{ii} (\lambda) = x_i^{\T} (X^{\T} X + \lambda I_p)^{-1}  x_i$ as the $(i,i)$th diagonal element of $H  (\lambda)  =  X (X^{\T} X + \lambda I_p)^{-1}  X^{\T} $. Define  $\hat{\beta}_{[-i]} (\lambda)$ as the ridge coefficient without observation $i$, and $\hat{\varepsilon}_{[-i]} (\lambda) = y_i - x_i^{\T}  \hat{\beta}_{[-i]} (\lambda)$ as the predicted residual. The leave-one-out formulas for ridge regression are
$$
\hat{\beta}_{[-i]}(\lambda) =  \hat{\beta}(\lambda)  - \{ 1-h_{ii}  (\lambda)  \} ^{-1}  (X^{\T} X + \lambda I_p)^{-1} x_i \hat{\varepsilon}_i(\lambda)
$$
and
$$
\hat{\varepsilon}_{[-i]}(\lambda) = \hat{\varepsilon}_i (\lambda)/ \{  1-h_{ii}  (\lambda) \} .
$$
\end{theorem}

I leave the proof of Theorem \ref{thm::looformula-ridge} as Problem  \ref{hw13::loo-ridge}. 

By Theorem \ref{thm::looformula-ridge}, the PRESS statistic for ridge is 
$$
\textsc{press}  (\lambda)  = \sumn \left\{   \hat{\varepsilon}_{[-i]} (\lambda) \right \}^2 
= \sumn \frac{   \left\{  \hat{\varepsilon}_i (\lambda) \right\}^2 }{ \{ 1-h_{ii}  (\lambda)  \} ^2 }.
$$

\citet{golub1979generalized} proposed the GCV criterion to simplify the calculation of the PRESS statistic by replacing $h_{ii} (\lambda)$ with their average value $n^{-1} \text{trace}\{ H(\lambda) \}$:
\[
\textsc{gcv}(\lambda)=\frac{\sumn\left\{   \hat{\varepsilon}_i (\lambda)   \right\} ^{2}}{\left[1-n^{-1}\text{trace}\left\{ H(\lambda)\right\} \right]^{2}}.
\]
 
In the \ri{R} package \ri{MASS}, the function \ri{lm.ridge} implements the ridge regression, \ri{kHKB} and \ri{kLW} report two estimators for $\lambda$, and \ri{GCV} contains the GCV values for a sequence of $\lambda$.

\section{Computation of ridge regression}\label{sec::computation-ridge}

Lemma \ref{lemma::ridge-coefficient} gives the ridge coefficients. So
 the predicted vector equals
\begin{eqnarray*}
 \hat{Y}(\lambda) &=& X  \hat{\beta}^{\text{ridge}}(\lambda) \\
&=& UDV^{\T} V\text{diag}\left(  \frac{d_j}{d_j^2 + \lambda }  \right)  U^{\T} Y \\
&=& UD \text{diag}\left(  \frac{d_j}{d_j^2 + \lambda }  \right)  U^{\T} Y \\
&=& U \text{diag}\left(  \frac{d_j^2}{d_j^2 + \lambda }  \right)  U^{\T} Y,
\end{eqnarray*}
and the hat matrix equals
\begin{eqnarray*}
H  (\lambda)  = U \text{diag}\left(  \frac{d_j^2}{d_j^2 + \lambda }  \right)   U^{\T}.
\end{eqnarray*}
These formulas allow us to compute the ridge coefficient and predictor vector for many values of $\lambda$ without inverting each $X^{\T} X + \lambda I_p$. 
We have similar formulas for the case with $n<p$; see Problem \ref{hw13::compute-ridge-large-p}.

A subtle point is due to the standardization of the covariates of the outcome. In \ri{R}, the \ri{lm.ridge} function first computes the ridge coefficient based on the standardized covariates and outcome, and then transforms them back to the original scale. Let $\bar{x}_1,\ldots, \bar{x}_p, \bar{y}$ be the means of the covariates and outcome, and let $\text{sd}_j = \{ n^{-1} \sumn (x_{ij} - \bar{x}_j) ^2 \}^{1/2}$ $(j=1, \ldots, p)$ be the standard deviation of the covariates which are report as \ri{scales} in the output of \ri{lm.ridge}. From the ridge coefficients $\{ \hat{\beta}_1^\text{ridge}(\lambda) ,\ldots,  \hat{\beta}_p^\text{ridge}(\lambda) \}$ based on the standardized variables, we can obtain the predicted values based on the original variables as
$$
\hat{y}_i(\lambda ) - \bar{y} =  \hat{\beta}_1^\text{ridge}(\lambda)  (x_{i1} - \bar{x}_1)/\text{sd}_1 + \cdots +  \hat{\beta}_p^\text{ridge}(\lambda)  (x_{ip} - \bar{x}_p)/\text{sd}_p
$$
or, equivalently,
$$
\hat{y}_i(\lambda )  = \hat\alpha^\text{ridge}(\lambda) 
+  \hat{\beta}_1^\text{ridge}(\lambda) /  \text{sd}_1 \times  x_{i1} + \cdots +  \hat{\beta}_p^\text{ridge}(\lambda)  / \text{sd}_p \times x_{ip}   
$$
where 
$$
\hat\alpha^\text{ridge}(\lambda)  = 
\bar{y} - \hat{\beta}_1^\text{ridge}(\lambda) \bar{x}_1/ \text{sd}_1 -  \cdots - \hat{\beta}_p^\text{ridge}(\lambda) \bar{x}_p/ \text{sd}_p . 
$$

\section{Numerical examples}

We can use the following numerical example to illustrate the bias-variance trade-off in selecting $\lambda$ in the ridge.

\subsection{Uncorrelated covariates}
 I first simulate data from a Normal linear model with uncorrelated covariates. 
\begin{lstlisting}
library(MASS)
n = 200
p = 100
beta = rep(1/sqrt(p), p) 
sig  = 1/2
X    = matrix(rnorm(n*p), n, p)
X    = scale(X)
X    = X*sqrt(n/(n-1))
Y    = as.vector(X%*%beta + rnorm(n, 0, sig))
\end{lstlisting}

The following code calculates the theoretical bias, variance, and mean squared error, reported in the $(1,1)$th panel of Figure \ref{fig::bias-variance-tradeoff-ridge}.  
 
\begin{lstlisting}
eigenxx = eigen(t(X)%*%X)
xis     = eigenxx$values
gammas  = t(eigenxx$vectors)%*%beta

lambda.seq = seq(0, 70, 0.01)
bias2.seq  = lambda.seq
var.seq    = lambda.seq
mse.seq    = lambda.seq
for(i in 1:length(lambda.seq))
{
	   ll = lambda.seq[i]
	   bias2.seq[i]  = ll^2*sum(gammas^2/(xis + ll)^2)
	   var.seq[i]    = sig^2*sum(xis/(xis + ll)^2)
	   mse.seq[i]    = bias2.seq[i] + var.seq[i]
}
 
y.min = min(bias2.seq, var.seq, mse.seq)
y.max = max(bias2.seq, var.seq, mse.seq)
par(mfrow = c(2, 2))
plot(bias2.seq ~ lambda.seq, type = "l",
     ylim = c(y.min, y.max), 
     xlab = expression(lambda), main = "", 
     ylab = "bias-variance tradeoff", 
     lty = 2, bty = "n")
lines(var.seq ~ lambda.seq, lty = 3)
lines(mse.seq ~ lambda.seq, lwd = 3, lty = 1)
abline(v = lambda.seq[which.min(mse.seq)], 
       lty = 1, col = "grey")
legend("topright", c("bias", "variance", "mse"),
       lty = c(2, 3, 1), lwd = c(1, 1, 4), bty = "n")
\end{lstlisting}

The  $(1,1)$th panel also reported the $\lambda$'s based on different approaches. 

\begin{lstlisting}
ridge.fit = lm.ridge(Y ~ X, lambda = lambda.seq)
abline(v = lambda.seq[which.min(ridge.fit$GCV)], 
       lty = 2, col = "grey")
abline(v = ridge.fit$kHKB, lty = 3, col = "grey")
abline(v = ridge.fit$kLW, lty = 4, col = "grey")
legend("bottomright", 
       c("MSE", "GCV", "HKB", "LW"),
       lty = 1:4, col = "grey", bty = "n")
\end{lstlisting}

I also calculate the prediction error of the ridge estimator in the testing dataset, which follows the same data-generating process as the training dataset. The $(1,2)$th panel of Figure \ref{fig::bias-variance-tradeoff-ridge} shows its relationship with $\lambda$. Overall, GCV, HKB, and LW are similar, but the $\lambda$ selected by the MSE criterion is the worst for prediction. 

\begin{lstlisting}
X.new    = matrix(rnorm(n*p), n, p) 
X.new    = scale(X.new)
X.new    = X.new*matrix(sqrt(n/(n-1)), n, p)
Y.new    = as.vector(X.new%*%beta + rnorm(n, 0, sig))
predict.error = Y.new - X.new%*%ridge.fit$coef
predict.mse   = apply(predict.error^2, 2, mean)
plot(predict.mse ~ lambda.seq, type = "l",
     xlab = expression(lambda),
     ylab = "predicted MSE", bty = "n")
abline(v = lambda.seq[which.min(mse.seq)], 
       lty = 1, col = "grey")
abline(v = lambda.seq[which.min(ridge.fit$GCV)], 
       lty = 2, col = "grey")
abline(v = ridge.fit$kHKB, lty = 3, col = "grey")
abline(v = ridge.fit$kLW, lty = 4, col = "grey")
legend("bottomright", 
       c("MSE", "GCV", "HKB", "LW"),
       lty = 1:4, col = "grey", bty = "n")

mtext("independent covariates", side = 1,
      line = -58, outer = TRUE, font.main = 1, cex=1.5)
\end{lstlisting}

\subsection{Correlated covariates}

I then simulate data from a Normal linear model with correlated covariates. 
\begin{lstlisting}
n = 200
p = 100
beta = rep(1/sqrt(p), p) 
sig  = 1/2
## correlated Normals
X    = matrix(rnorm(n*p), n, p) + rnorm(n, 0, 0.5)
## standardize the covariates
X    = scale(X)
X    = X*matrix(sqrt(n/(n-1)), n, p)
Y    = as.vector(X%*%beta + rnorm(n, 0, sig))
\end{lstlisting}

\begin{figure} 
\centering 
\includegraphics[width = \textwidth]{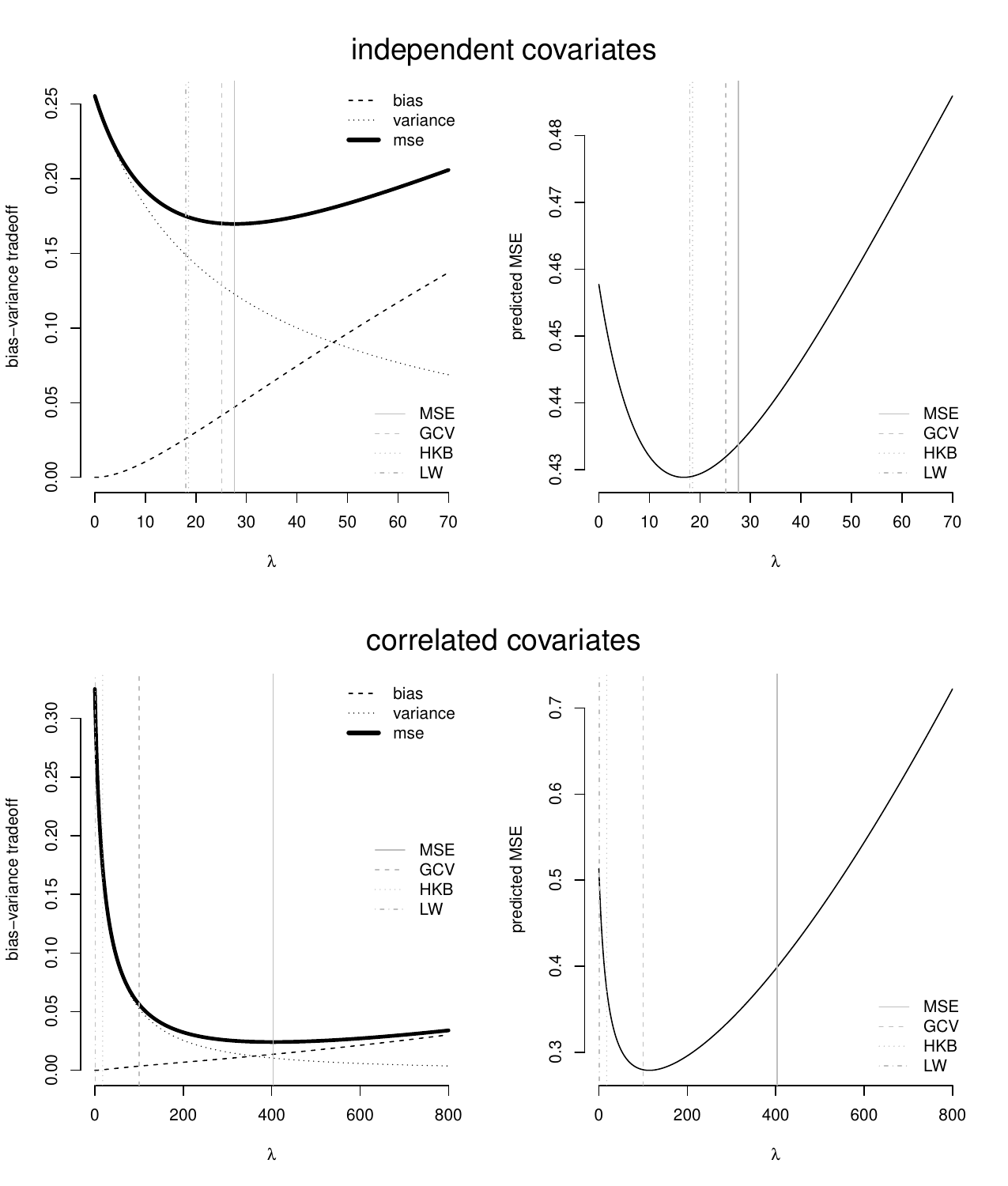}
\caption{Bias-variance trade-off in ridge regression}\label{fig::bias-variance-tradeoff-ridge}
\end{figure}

The second row of Figure \ref{fig::bias-variance-tradeoff-ridge} shows the bias-variance trade-off. Overall, GCV works the best for selecting $\lambda$ for prediction. 

\section{Further commments on OLS, ridge, and PCA}

The SVD of $X$ is closely related to the {\it principal component analysis} (PCA).
Assume that the columns of $X$ are centered, so $X^{\T} X = VD^2V^{\T}$ is proportional to the sample covariance matrix of $X$. Assume $d_1 \geq d_2 \geq \cdots$. PCA tries to find linear combinations of the covariate $x_i$ that contain maximal information.   For a vector $v \in \mathbb{R}^p$, the linear combination $v^{\T} x_i $ has sample variance proportional to
$$
Q(v) = 
v^{\T} X^{\T} X v.
$$ 
If we multiply $v$ by a constant $c$, the above sample variance will change by the factor $c^2$. So a meaningful criterion is to maximize $Q(v)$ such that $\|v\| = 1$. This is exactly the setting of Theorem \ref{thm::eigven-max-q}. The maximum value equals $d_1^2$ which is achieved by $V_1$, the first column of $V$. We call
$$
XV_1 = \begin{pmatrix}
x_1^{\T} V_1 \\
\vdots \\
x_n^{\T} V_1
\end{pmatrix}
$$
the first principal component of $X$. Similar to Theorem  \ref{thm::eigven-max-q}, we can further maximize $ Q(v)$ such that $\|v\| = 1$ and $v\perp V_1$, yielding the maximum value $d_2^2$ which is achieved by $V_2$. We call $XV_2$ the second principal component of $X$. By induction, we can define all the $p$ principal components, stacked in the following $n\times p$ matrix: 
$$
(XV_1, \ldots, X V_p) =  XV = UDV^{\T} V = UD.
$$
So $UD$ in the SVD decomposition contains the principal components of $X$. Since $D$ is a diagonal matrix that only changes the scales of the columns of $U$, we also call $U = (U_1, \ldots, U_p)$ the principal components of $X$. They are orthogonal since $U^{\T} U = I_p$.

Section \ref{sec::computation-ridge} shows that the ridge estimator yields the predicted value
\begin{eqnarray*}
\hat{Y}(\lambda)  &=&  U \text{diag}\left(  \frac{d_j^2}{d_j^2 + \lambda }  \right)  U^{\T} Y \\
&=& \sum_{j=1}^p   \frac{d_j^2}{d_j^2 + \lambda }  \langle U_j , Y\rangle     U_j 
\end{eqnarray*}
where $\langle U_j , Y\rangle = U_j^{\T} Y$ denotes the inner product of vectors $U_j $ and $ Y.$
As a special case with $\lambda = 0$, the OLS estimator yields the predicted value
\begin{eqnarray*}
\hat{Y}    &=& U  U^{\T} Y \\
&=& \sum_{j=1}^p    \langle U_j , Y\rangle     U_j ,
\end{eqnarray*}
which is identical to the predicted value based on OLS of $Y$ on the principal components $U$. Moreover, the principal components in $U$ are orthogonal and have unit length, so by Corollary \ref{cor:orthogonal}, the OLS fit of $Y$ on $U$ is equivalent to the component-wise OLS of $Y$ on $U_j$ with coefficient $ \langle U_j , Y\rangle$ $(j=1,\ldots, p)$. So the predicted value based OLS equals a linear combination of the principal components with coefficients  $ \langle U_j , Y\rangle$; the predicted value based on ridge also equals a linear combination of the principal components but the coefficients are shrunk by the factors $d_j^2 / (d_j^2 + \lambda)$.

When the columns of $X$ are not linearly independent, for example, $p>n$, we cannot run OLS of $Y$ on $X$ or OLS of $Y$ on $U$, but we can still run ridge regression. Motivated by the formulas above, another approach is to run OLS of $Y$ on the first $p^*$ principal components $ \tilde{U} = ( U_1, \ldots, U_{p^*})$ with $p^* < p$. This is called the {\it principal component regression} (PCR). The predicted value is
\begin{eqnarray*}
\hat{Y} (p^*) &=& (\tilde{U}^{\T} \tilde{U}  )^{-1} \tilde{U} ^{\T} Y \\
&=& \sum_{j=1}^{p^*}    \langle U_j , Y\rangle     U_j ,
\end{eqnarray*}
which truncates the summation in the formula of $\hat{Y} $ based on OLS. 
Compared with the predicted values of OLS and ridge regression, $\hat{Y} (p^*) $ effectively imposes zero weights on the principal components corresponding to small singular values. It depends on a tuning parameter $p^*$ similar to $\lambda$ in the ridge regression. Since $p^*$ must be a positive integer and $\lambda$ can be any positive real value, PCR is a discrete procedure while ridge regression is a continuous procedure.

\section{Homework problems}

\paragraph{Ridge coefficient as a posterior mode under a Normal prior}\label{hw13::ridge-bayes}

Assume fixed $X, \sigma^2$ and $\tau^2$. 
Prove that if
\begin{equation}
\label{eq::model-ridge-1}
Y \mid \beta \sim\N(X\beta,\sigma^{2} I_n )
\end{equation}
and
\begin{equation}
\label{eq::model-ridge-2}
\beta\sim\N(0,\tau^{2}I_{p}),
\end{equation}
then the mode of the posterior distribution of $\beta\mid Y$ equals $\hat{\beta}^{\text{ridge}}(\sigma^{2}/\tau^{2})$:
$$
\hat{\beta}^{\text{ridge}}(\sigma^{2}/\tau^{2}) = \arg \max_{\beta} f(  \beta\mid Y  )
$$
where $f(\beta\mid Y)$ is the posterior density of $\beta$ given $Y$.

Remark: In Bayesian statistics, \eqref{eq::model-ridge-1} is the Normal linear model, whereas \eqref{eq::model-ridge-2} is the {\it prior distribution} of the parameter $\beta$.

\paragraph{Derivative of the MSE}\label{hw13::derivative-of-mse}

Prove that 
$$
\frac{\partial\textsc{mse}(\lambda)}{\partial\lambda} \Big |_{\lambda = 0} <0 .
$$

Remark: This result ensures that the ridge estimator must have a smaller MSE than OLS in a neighborhood of $\lambda = 0$, which is coherent with the pattern in Figure \ref{fig::bias-variance-tradeoff-ridge}.

\paragraph{Ridge and OLS}\label{hw13::ridge-ols}

Prove that if  $X$ has linearly independent columns, then
\begin{eqnarray*}
\hat{\beta}^{\text{ridge}}(\lambda) 
&=& (X^{\T} X + \lambda I_p)^{-1} X^{\T} X \hat{\beta} \\
&=& V\text{diag}\left(    \frac{  d_j^2 }{  d_j^2 + \lambda   }  \right) V^{\T}  \hat{\beta}
\end{eqnarray*}
where $\hat{\beta}$ is the OLS coefficient.

\paragraph{Ridge as OLS with augmented data}\label{hw13::ridge-data-aug}

Prove that $\hat{\beta}^{\text{ridge}}(\lambda)$ equals the OLS coefficient of $\tilde{Y}$ on $\tilde{X}$ with
augmented data
\[
\tilde{Y}=\left(\begin{array}{c}
Y\\
0_{p}
\end{array}\right),\qquad
\tilde{X}=\left(\begin{array}{c}
X\\
\sqrt{\lambda}I_{p}
\end{array}\right) ,
\]
where $ \tilde{Y}$ is an $n+p$ dimensional vector and $ \tilde{X}$ is an $(n+p)\times p$ matrix.

Remark: The columns of $\tilde{X}$ must be linearly independent, so the inverse of $\tilde{X}^{\T} \tilde{X}$ always exists. 
This is a theoretical result of the ridge regression. It should not be used for computation especially when $p$ is large. 

\paragraph{Leave-one-out formulas for ridge}\label{hw13::loo-ridge}

Prove Theorem \ref{thm::looformula-ridge}.

Remark: You can use the result in Problem \ref{hw13::ridge-data-aug} and apply the leave-one-out formulas for OLS in Theorems \ref{thm::leave-one-out-beta} and \ref{theorem:looresidual}.

\paragraph{Generalized ridge regression}\label{hw13::general-ridge}

Covariates have different importance, so it is reasonable to use different weights in the
penalty term. Find the explicit formula for the ridge regression
with general quadratic penalty:
\[
\arg\min_{b\in\mathbb{R}^{p}}\left\{ (Y-Xb)^{\T}  (Y-Xb) +\lambda b^{\T}Qb\right\} 
\]
where $Q$ is a $p\times p$ positive definite matrix.

\paragraph{Degrees of freedom of ridge regression}

For a predictor $\hat{Y}$ for $Y$, define the degrees of freedom
of the predictor as $\sumn\cov(y_{i},\hat{y}_{i}) /  \sigma^2 .$ Calculate the
degrees of freedom of ridge regression in terms of the eigenvalues
of $X^{\T}X$.

\paragraph{Extending the simulation in Figure \ref{fig::bias-variance-tradeoff-ridge}}

Re-run the simulation that generates Figure \ref{fig::bias-variance-tradeoff-ridge}, and report the $\lambda$ selected by \citet{dempster1977simulation}'s method, PRESS, and $K$-fold CV. Extend the simulation to the case with $p>n.$

\paragraph{Unification of OLS, ridge, and PCR}\label{hw13::unification}

We can unify the predicted values of the OLS, ridge, and PCR as
$$
\hat{Y} = \sum_{j=1}^p  s_j  \langle U_j, Y \rangle U_j,
$$
where
$$
s_j  = 
\begin{cases}
1, & \text{OLS}, \\
\frac{  d_j^2 }{   d_j^2 + \lambda  },  & \text{ridge}, \\ 
1(j \leq p^*),  & \text{PCR} .
\end{cases}
$$
Based on the unified formula, show that under Assumption \ref{assume::gm-model}, we have 
$$
E (\hat{Y}  ) = \sum_{j=1}^p s_j d_j \gamma_j U_j
$$
with the $\gamma_j$'s defined in Theorem \ref{theorem:bias-variance-tradeoff-ridge}, and
$$
\cov (\hat{Y}  ) = \sigma^2  \sum_{j=1}^p  s_j^2 U_j U_j^{\T}. 
$$

\paragraph{An equivalent form of ridge coefficient}\label{hw13::ridge-equiv-computation}

Prove Theorem \ref{thm::ridge-two-forms} below which states two equivalent forms of the ridge coefficient.

\begin{theorem}\label{thm::ridge-two-forms}
For $\lambda > 0$, we have 
$$
\hat{\beta}^{\textup{ridge}}(\lambda) = 
(X^{\T}X+\lambda I_{p})^{-1}X^{\T}Y = X^{\T} (XX^{\T} + \lambda I_n)^{-1} Y.
$$
\end{theorem}

Remark: Theorem \ref{thm::ridge-two-forms} has several interesting implications. The first form of the ridge coefficient involves inverting a $p\times p$ matrix, and it is more useful when $p<n$. The second form of the ridge coefficient involves inverting an $n\times n$ matrix, so it is more useful when $p>n$. 
From the second form of the ridge coefficient, we can see that the ridge estimator lies in $\mathcal{C}(X^{\T})$, the row space of $X$. That is, the ridge estimator can be written as $X^{\T} \delta $, where $\delta = (XX^{\T} + \lambda I_n)^{-1} Y \in \mathbb{R}^{p}$. This always holds but is particularly interesting in the case with $p>n$ when the row space of $X$ is not the entire $\mathbb{R}^{p}$. 
Third, if $p>n$ and $XX^{\T}$ is invertible, then we can let $\lambda$ go to zero on the right-hand side, yielding
$$
\hat{\beta}^{\text{ridge}}(0) = X^{\T} (XX^{\T}  )^{-1} Y
$$
 which is the minimum norm OLS estimator; see Problem \ref{para::minimum-norm-rols}. Using the definition of the pseudoinverse in Appendix \ref{chapter::linear-algebra}, we can further show that
 $$
 \hat{\beta}^{\text{ridge}}(0) = X^{+} Y. 
 $$

\paragraph{Ridge estimator as the minimum norm OLS estimator with augmented features}\label{hw13::ridge-ols-augmentedX}
 
This problem is the dual problem of Problem \ref{hw13::ridge-data-aug}.

Prove that  $\hat{\beta}^{\text{ridge}}(\lambda)$ equals the first $p$ components of the minimum norm OLS estimator of $Y$ on $\tilde X$:
$$
\tilde X^{\T} (\tilde X \tilde X^{\T}  )^{-1} Y
$$
with the $n \times (p+n)$ matrix
$$
\tilde X = (X, \sqrt{\lambda} I_n) . 
$$

Remark: Use the results in Problem \ref{hw13::ridge-equiv-computation}.

\paragraph{Computation of ridge with $n<p$}\label{hw13::compute-ridge-large-p}
When $n<p$, $X$ has singular value decomposition $X = U D V^{\T}$, where $D \in \mathbb{R}^{n\times n}$ is a diagonal matrix containing the singular values, $U\in \mathbb{R}^{n\times n}$ is an orthogonal matrix with $UU^{\T} = U^{\T} U = I_n$, and $V \in \mathbb{R}^{p\times n}$ has orthonormal columns with $V^{\T} V = I_n$. 

Prove that the ridge coefficient, the predicted value, and the hat matrix have the same form as the case with $n>p$.  The only subtle difference is that the diagonal matrices have dimension $n\times n$. 

Remark: The above result also ensures that Theorem \ref{theorem:bias-variance-tradeoff-ridge} holds when $p > n$ if we modify the summation as ``from $j=1$ to $n$.''

\paragraph{Recommended reading}

To celebrate the 50th anniversary of  \citet{hoerl1970ridge}'s paper in {\it Technometrics}, the editor invited Roger W. Hoerl,  the son of Art Hoerl, to review the historical aspects of the original paper, and Trevor Hastie to review the essential role of the idea of ridge regression in data science. See \citet{hoerl2020ridge} and \citet{hastie2020ridge}.

\chapter{Lasso}\label{chapter::lasso}

Ridge regression works well for prediction, but it may be difficult
to interpret many small but non-zero ridge coefficients. \citet{tibshirani1996regression} proposed to use the lasso, the acronym for the Least Absolute Shrinkage
and Selection Operator, to achieve the ambitious goal of simultaneously
estimating parameters and selecting important variables in the linear
regression. By changing the penalty term in ridge regression,
the lasso automatically estimates some parameters as zero, dropping
them out of the model and thus selecting the remaining variables as
important predictors. 
This chapter introduces the lasso.

\section{Introduction to the lasso}

Recall the definition of the residual sum of squares 
$$
\textsc{rss}(b_0,b_1,\ldots,b_p) = 
\sumn(y_{i}-b_{0}-b_{1}x_{i1}-\cdots-b_{p}x_{ip})^{2}.
$$
\citet{tibshirani1996regression} defined the lasso as 
\begin{align}
\hat{\beta}^{\text{lasso}}(t) & =\arg\min_{b_{0},b_{1},\ldots,b_{p}}\textsc{rss}(b_0,b_1,\ldots,b_p)  \nonumber \\
 & \ \text{such that }\sum_{j=1}^{p}|b_{j}|\ensuremath{\leq t} .\label{eq:lasso-form2}
\end{align}
\citet{osborne2000lasso} studied its equivalent form
\begin{equation}
\hat{\beta}^{\text{lasso}}(\lambda)=\arg\min_{b_{0},b_{1},\ldots,b_{p}}\left\{\textsc{rss}(b_0,b_1,\ldots,b_p) +\lambda\sum_{j=1}^{p}|b_{j}|\right\} . \label{eq:lasso-form1}
\end{equation}
The two forms of lasso are equivalent in the sense that for a given $\lambda$ in \eqref{eq:lasso-form1}, there exists a $t$ such that the solution for \eqref{eq:lasso-form2} is identical to the solution for \eqref{eq:lasso-form1}. In particular, $t = \sum_{j=1}^p \hat{\beta}^{\text{lasso}}_j(\lambda) $. 
Technically, the minimizer of the lasso problem may not be unique especially when $p>n$, so the right-hand sides of the optimization problems \eqref{eq:lasso-form2} and \eqref{eq:lasso-form1} should be a set. Fortunately, even though the minimizer may not be unique, the resulting predictor is always unique. \citet{tibshirani2013lasso} clarifies this issue; see Problem \ref{hw14::uniqueness-lasso-pred}.

Both forms \eqref{eq:lasso-form2} and \eqref{eq:lasso-form1} are useful for understanding the lasso. 
We will use the form \eqref{eq:lasso-form2} for geometric intuition and use the form (\ref{eq:lasso-form1}) for computation. Similar to
ridge regression, the lasso is not invariant to the linear transformation
of $X$. We proceed after standardizing the covariates and outcome as Condition \ref{condition::standardization}. For the same reason as ridge regression, we can drop the intercept after standardization.

\section{Comparing the lasso and ridge: a geometric perspective}

The ridge and lasso are very similar: both minimize a penalized version of the residual sum of squares. They differ in the penalty term: ridge uses an $L_2$ penalty, i.e., the $L_2$ norm of the coefficient $ \| b\|^2 =  \sum_{j=1}^p b_j^2$, and lasso uses an $L_1$ penalty, i.e., the $L_1$ norm of the coefficient   $\| b\|_1 = \sum_{j=1}^p | b_j |$. Compared to the ridge, the lasso can give sparse solutions due to the non-smooth penalty term. That is, estimators of some coefficients are exactly zero.

\begin{figure}
\centering
\includegraphics[width = 0.7\textwidth]{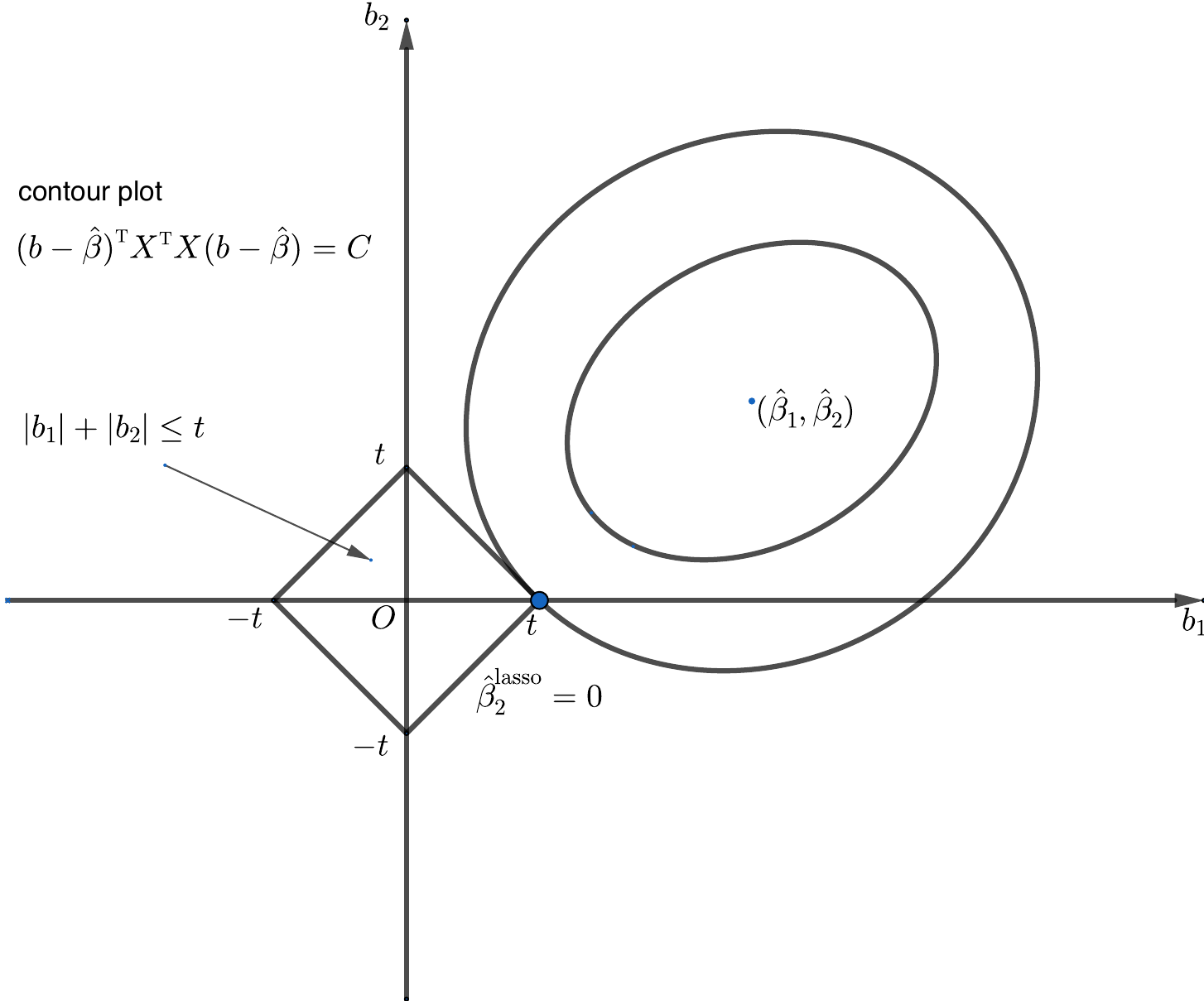}
\caption{Lasso with a sparse solution}\label{fig::lasso-sparse}
\end{figure}

\begin{figure}
\centering 
\includegraphics[width = 0.7\textwidth]{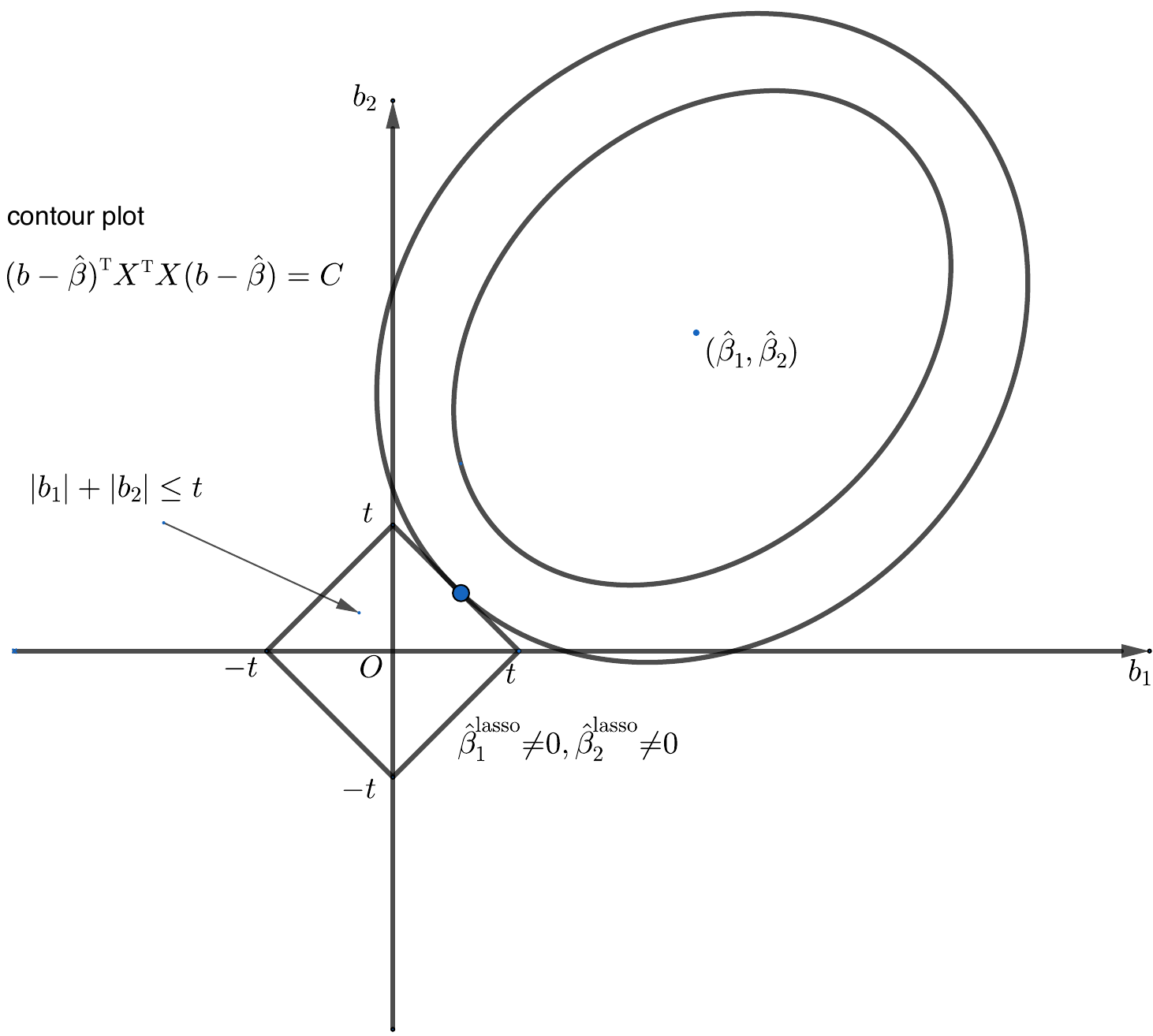}
\caption{Lasso with a non-sparse solution}\label{fig::lasso-non-sparse}

\end{figure}

Focus on the form \eqref{eq:lasso-form2}.  We can gain insights from the contour plot of the residual sum of squares as a function of $b$. With a well-defined OLS estimator $\hat{\beta}$, Theorem \ref{thm::geometryofols} ensures 
\begin{align*}
(Y-Xb)^{\T}(Y-Xb)  =(Y-X\hat{\beta})^{\T}(Y-X\hat{\beta})
  +(b-\hat{\beta})^{\T}X^{\T}X(b-\hat{\beta}),
\end{align*}
which equals a constant term plus a quadratic function centered at the OLS coefficient. Without any penalty, the minimizer is of course the OLS coefficient. With the $L_1$ penalty, the OLS coefficient may not be in the region defined by $\sum_{j=1}^{p}|b_{j}| \leq t$. If this happens, the intersection of the contour plot of $(Y-Xb)^{\T}(Y-Xb)$ and the border of the restriction region $\sum_{j=1}^{p}|b_{j}| \leq t$ can be at some axis. For example, Figure \ref{fig::lasso-sparse} shows a case with $p=2$, and the lasso estimator hits the x-axis, resulting in a zero coefficient for the second coordinate. However, this does not mean that lasso always generates sparse solutions because sometimes the intersection of the contour plot of $(Y-Xb)^{\T}(Y-Xb)$ and the border of the restriction region is at an edge of the region. For example, Figure \ref{fig::lasso-non-sparse} shows a case with a non-sparse lasso solution.

\begin{figure}
\centering 
\includegraphics[width = 0.7\textwidth]{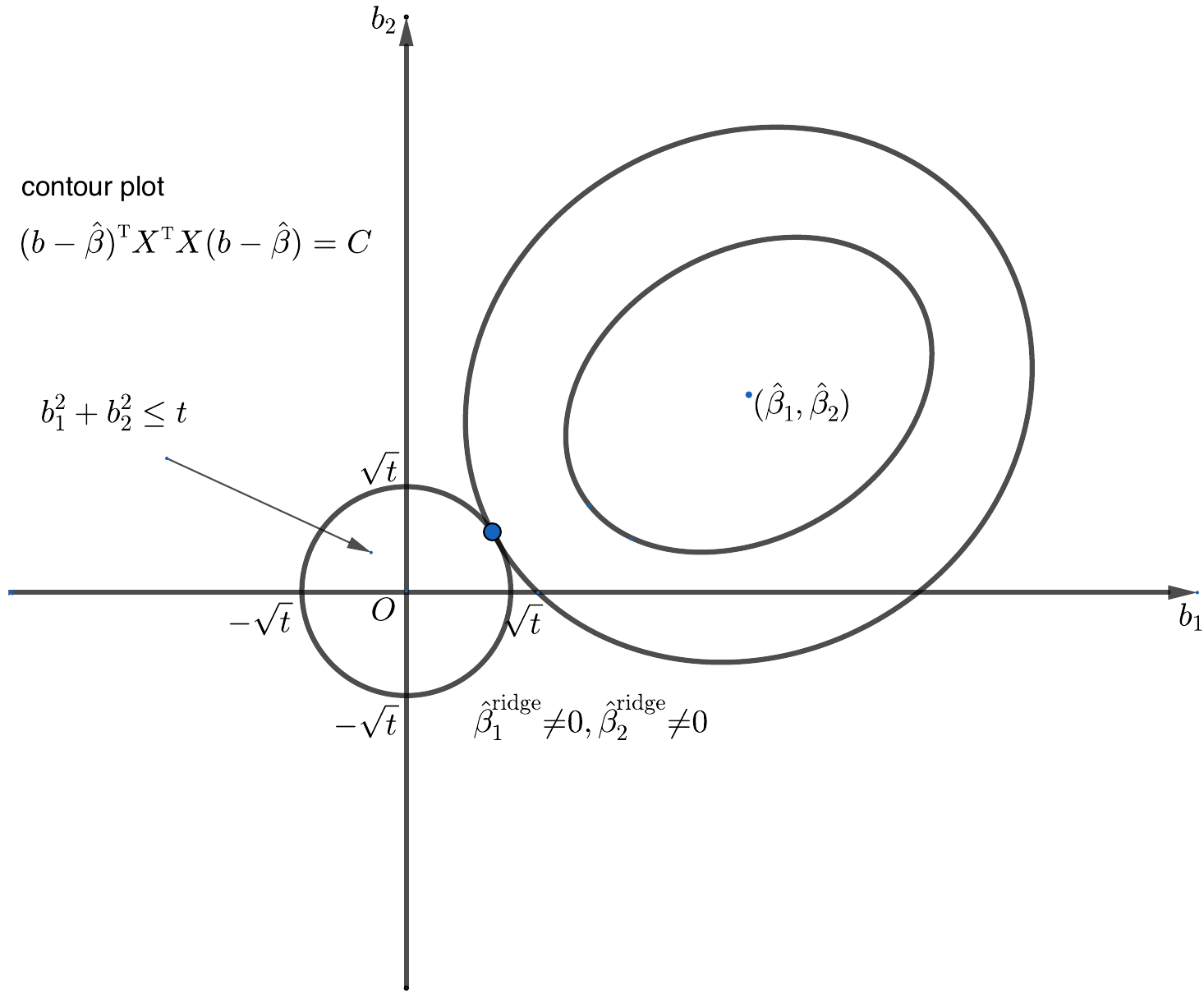}
\caption{Ridge regression}
\label{fig::ridge-non-sparse}

\end{figure}

In contrast, the restriction region of the ridge is a circle, so the ridge solution does not hit any axis unless the original OLS coefficient is zero. Figure \ref{fig::ridge-non-sparse} shows the general ridge estimator. 

\section{Computing the lasso coefficients via coordinate descent}

Many efficient algorithms can solve the lasso problem. The \ri{glmnet} package in \ri{R} uses the coordinate descent algorithm based on the form \eqref{eq:lasso-form1} \citep{friedman2007pathwise, friedman2010regularization}. I will first review a lemma which is the stepstone for the algorithm.

\subsection{The soft-thresholding lemma}

Let $\textup{sign}(x)$ denote the sign of a real number $x$, which equals $1, 0, -1$ if $x>0, x=0, x<0$, respectively. Let $(x)_{+}=\max(x,0)$
denote the positive part of a real number $x$.  

\begin{lemma}
\label{lemma:softthresholding-lemma}Given $b_{0}$ and $\lambda \geq 0$, we have 
\begin{align*}
\arg\min_{b\in \mathbb{R}}\frac{1}{2}(b-b_{0})^{2}+\lambda|b| & =\textup{sign}(b_{0})\left(|b_{0}|-\lambda\right)_{+}\\
 & =\begin{cases}
b_{0}-\lambda, & \text{if }b_{0}\geq\lambda,\\
0 & \text{if }-\lambda\leq b_{0}\leq\lambda,\\
b_{0}+\lambda & \text{if }b_{0}\leq-\lambda . 
\end{cases}
\end{align*}
\end{lemma}

\begin{figure}
\centering 
\includegraphics[width=\textwidth]{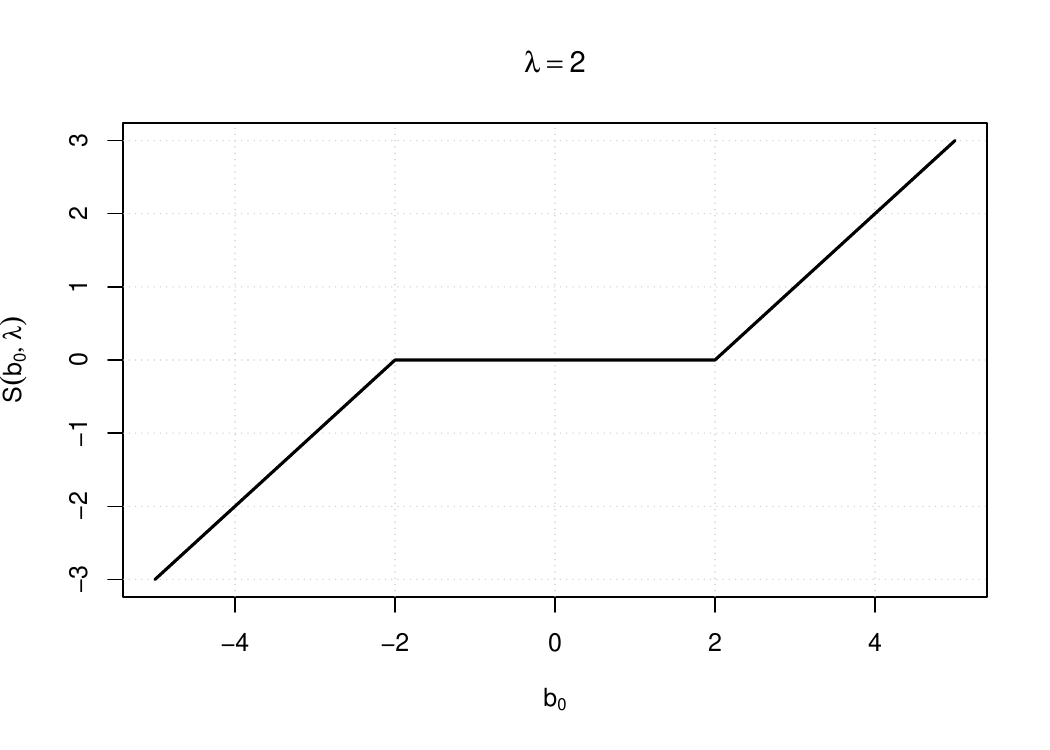}

\caption{Soft-thresholding}\label{fig::soft-threshholding-function}

\end{figure}

The solution in Lemma \ref{lemma:softthresholding-lemma} is a function
of $b_{0}$ and $\lambda$, and we will use the notation 
\[
S(b_{0},\lambda)=\textup{sign}(b_{0})\left(|b_{0}|-\lambda\right)_{+}
\]
in this chapter, where $S$ denotes the soft-thresholding operator. For a given $\lambda>0$, it is a function of $b_{0}$
illustrated by Figure \ref{fig::soft-threshholding-function}. The proof of Lemma \ref{lemma:softthresholding-lemma}
is to solve the optimization problem. It is tricky since we cannot naively solve the first-order condition due to the non-smoothness of $|b|$ at $0$. Nevertheless, it is only a one-dimensional optimization problem, and I relegate the proof to Problem \ref{hw14::soft-thresholding-lemma}.

\subsection{Coordinate descent for the lasso}

For a given $\lambda>0$, we can use the following algorithm: 
\begin{enumerate}
\item Standardize the data to satisfy Condition \ref{condition::standardization}. 
So we need to solve a lasso problem without the intercept. For simplicity
of derivation, we change the scale of the residual sum of squares
without essentially changing the problem:\footnote{It will change the scale of $\lambda$. However, $\lambda$ is a tuning parameter anyway, which will often be determined via cross-validation.}
\[
\min_{b_{1},\ldots,b_{p}}\frac{1}{2n}\sumn(y_{i} -b_{1}x_{i1}-\cdots-b_{p}x_{ip})^{2}+\lambda\sum_{j=1}^{p}|b_{j}|.
\]
Initialize $\hat{\beta}.$

\item Update $\hat{\beta}_{j}$ given all other coefficients. Define the
partial residual as $r_{ij}=y_{i}-\sum_{k\neq j}\hat{\beta}_{k} x_{ik}$.
Updating $\hat{\beta}_{j}$ is equivalent to minimizing
\[
\frac{1}{2n}\sumn(r_{ij}-b_{j}x_{ij})^{2}+\lambda|b_{j}|.
\]
Define 
\[
\hat{\beta}_{j,0}=\frac{\sumn x_{ij}r_{ij}}{\sumn x_{ij}^{2}}=n^{-1}\sumn x_{ij}r_{ij}
\]
as the OLS coefficient of the $r_{ij}$'s on the $x_{ij}$'s, so 
\begin{align*}
\frac{1}{2n}\sumn(r_{ij}-b_{j}x_{ij})^{2} & =\frac{1}{2n}\sumn(r_{ij}-\hat{\beta}_{j,0}x_{ij})^{2}+\frac{1}{2n}\sumn x_{ij}^{2}(b_{j}-\hat{\beta}_{j,0})^{2}\\
 & =\text{constant}+\frac{1}{2}(b_{j}-\hat{\beta}_{j,0})^{2}.
\end{align*}
Then updating $\hat{\beta}_{j}$ is equivalent to minimizing $\frac{1}{2}(b_{j}-\hat{\beta}_{j,0})^{2}+\lambda|b_{j}|$.
Lemma \ref{lemma:softthresholding-lemma} implies 
\[
\hat{\beta}_{j}=S(\hat{\beta}_{j,0},\lambda).
\]

\item Iterate until convergence. 
\end{enumerate}

Does the algorithm always converge? 
The theory of \citet{tseng2001convergence} ensures it converges, but this is beyond the scope of this book. 
We can start with a large $\lambda$ and all 0 lasso coefficients. We then
gradually decrease $\lambda$, and for each $\lambda$, we apply the
above algorithm. We finally select $\lambda$ via $K$-fold cross-validation. Since we gradually decrease $\lambda$, the initial values from the last step are very close to the minimizer and the algorithm converges fairly fast.

\section{Example: comparing OLS, ridge and lasso}\label{sec::lasso-example}

\begin{figure}[ht]
\centering
\includegraphics[width = \textwidth]{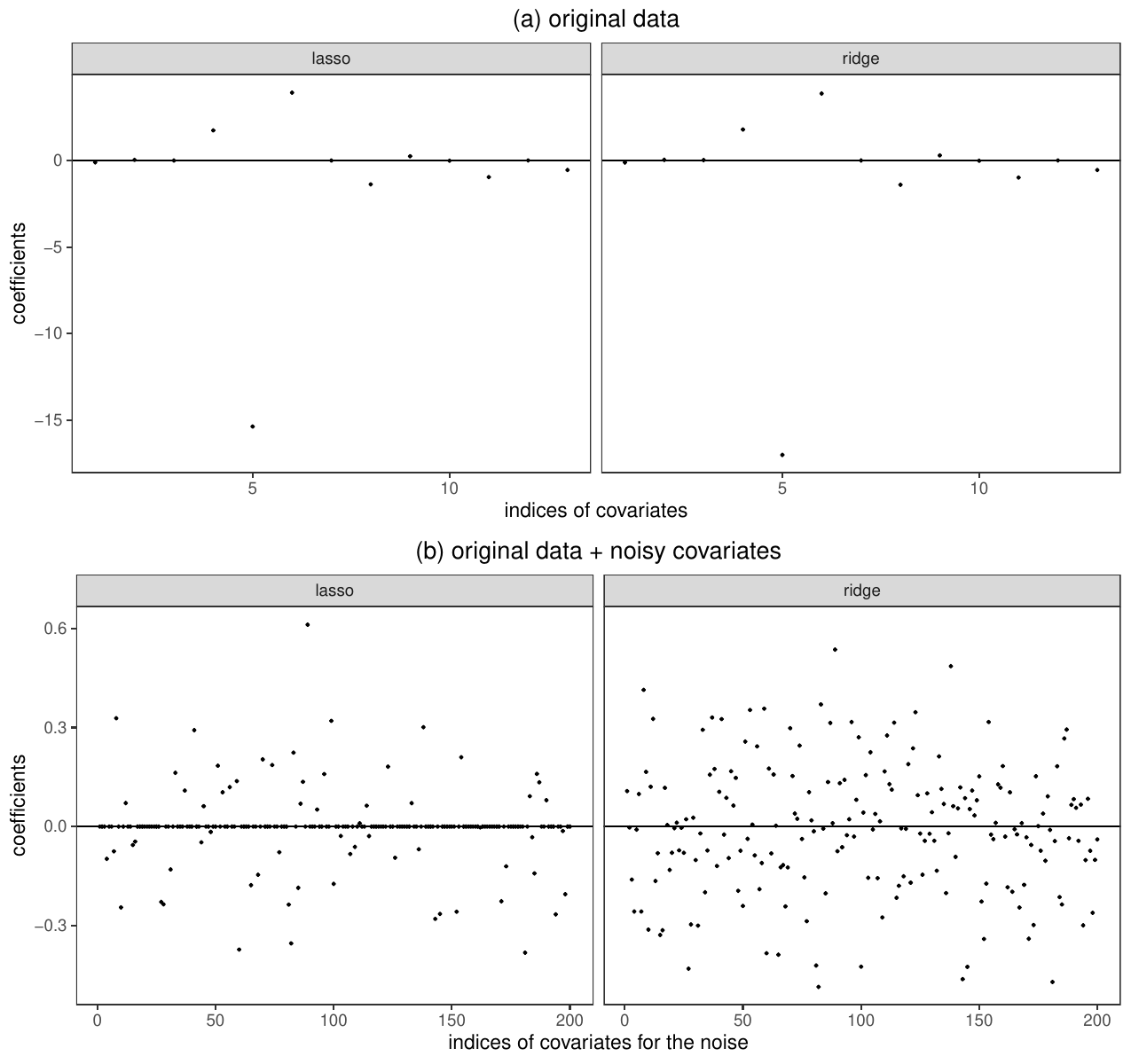}
\caption{Comparing ridge and lasso}
\label{fig::ridge-lasso-coefficients-boston}
\end{figure}

In the Boston housing data, the OLS, ridge, and lasso have similar performance in out-of-sample prediction. Lasso and ridge have similar coefficients. See Figure \ref{fig::ridge-lasso-coefficients-boston}(a). 

\begin{lstlisting}
> library("mlbench")
> library("glmnet")
> library("MASS")
> data(BostonHousing)
> 
> ## training and testing data
> set.seed(230)
> nsample = dim(BostonHousing)[1]
> trainindex = sample(1:nsample, floor(nsample*0.9))
> 
> xmatrix = model.matrix(medv ~ ., data = BostonHousing)[, -1]
> yvector = BostonHousing$medv 
> dat = data.frame(yvector, xmatrix)
> 
> ## linear regression
> bostonlm = lm(yvector ~ ., data = dat[trainindex, ])
> predicterror = dat$yvector[- trainindex] - 
+                     predict(bostonlm, dat[- trainindex, ])
> mse.ols = sum(predicterror^2)/length(predicterror)
> 
> ## ridge regression 
> lambdas= seq(0, 5, 0.01)
> lm0 = lm.ridge(yvector ~ ., data = dat[trainindex, ],
+                lambda = lambdas)
> coefridge = coef(lm0)[which.min(lm0$GCV), ]
> predicterrorridge = dat$yvector[- trainindex] -
+   cbind(1, xmatrix[- trainindex, ])%*%coefridge
> mse.ridge = sum(predicterrorridge^2)/length(predicterrorridge)
> 
> ## lasso 
> cvboston = cv.glmnet(x = xmatrix[trainindex, ], y = yvector[trainindex])
> coeflasso = coef(cvboston, s = "lambda.min")
> predicterrorlasso = dat$yvector[- trainindex] -
+                         cbind(1, xmatrix[- trainindex, ])%*%coeflasso
> mse.lasso = sum(predicterrorlasso^2)/length(predicterrorlasso)
> 
> c(mse.ols, mse.ridge, mse.lasso)
[1] 29.37365 29.07174 28.88161
\end{lstlisting}

But if we artificially add 200 columns of covariates of pure noise $\N(0,1)$, then the ridge and lasso perform much better. Lasso can automatically shrink many coefficients to zero. See   Figure \ref{fig::ridge-lasso-coefficients-boston}(b). 

\begin{lstlisting}
> ## adding more noisy covariates
> n.noise = 200
> xnoise  = matrix(rnorm(nsample*n.noise), nsample, n.noise)
> xmatrix = cbind(xmatrix, xnoise)
> dat = data.frame(yvector, xmatrix)
> 
> ## linear regression
> bostonlm = lm(yvector ~ ., data = dat[trainindex, ])
> predicterror = dat$yvector[- trainindex] - 
+   predict(bostonlm, dat[- trainindex, ])
> mse.ols = sum(predicterror^2)/length(predicterror)
> 
> ## ridge regression 
> lambdas= seq(100, 150, 0.01)
> lm0 = lm.ridge(yvector ~ ., data = dat[trainindex, ],
+                lambda = lambdas)
> coefridge = coef(lm0)[which.min(lm0$GCV), ]
> predicterrorridge = dat$yvector[- trainindex] -
+   cbind(1, xmatrix[- trainindex, ])%*%coefridge
> mse.ridge = sum(predicterrorridge^2)/length(predicterrorridge)
> 
> 
> ## lasso 
> cvboston = cv.glmnet(x = xmatrix[trainindex, ], y = yvector[trainindex])
> coeflasso = coef(cvboston, s = "lambda.min")
> 
> predicterrorlasso = dat$yvector[- trainindex] -
+   cbind(1, xmatrix[- trainindex, ])%*%coeflasso
> mse.lasso = sum(predicterrorlasso^2)/length(predicterrorlasso)
> 
> c(mse.ols, mse.ridge, mse.lasso)
[1] 41.80376 33.33372 32.64287
\end{lstlisting}

\section{Other shrinkage estimators}

\subsection{Bridge estimator}

A general class of shrinkage estimators is the bridge estimator \citep{frank1993statistical}: 
\[
\hat{\beta}(\lambda)=\arg\min_{b_{0},b_{1},\ldots,b_{p}}\left\{ \textsc{rss}(b_0,b_1,\ldots,b_p) +\lambda\sum_{j=1}^{p}|b_{j}|^{q}\right\} , 
\]
or, equivalently, 
\begin{align*}
\hat{\beta}(t) & =\arg\min_{b_{0},b_{1},\ldots,b_{p}} \textsc{rss}(b_0,b_1,\ldots,b_p)  \\
 & \ \text{such that }\sum_{j=1}^{p}|b_{j}|^{q}\ensuremath{\leq t.}
\end{align*}

\begin{figure}
\centering 
\subfloat[$0<q<1$]{\includegraphics[width=0.33\textwidth]{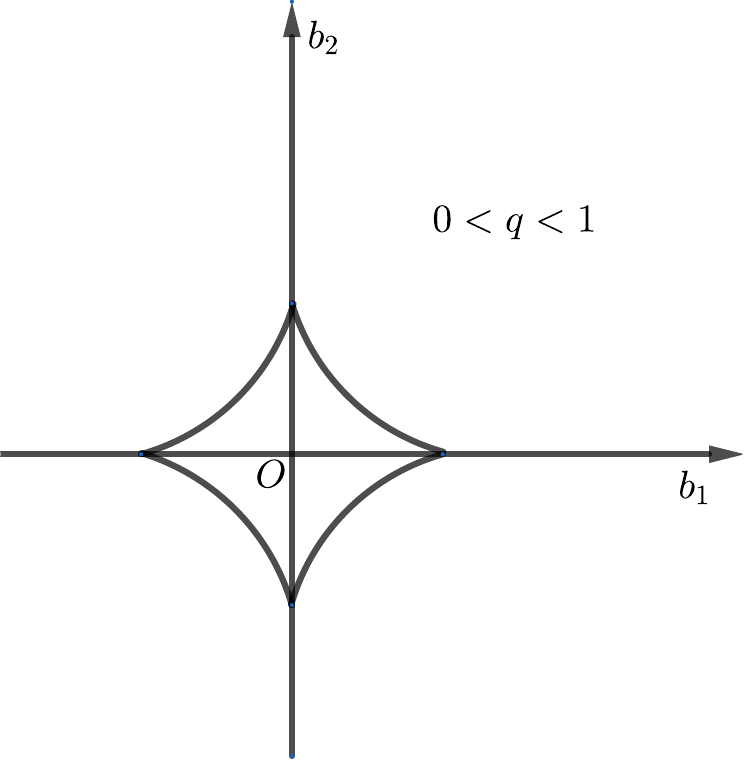}
}
\subfloat[$q=1$]{\includegraphics[width=0.33\textwidth]{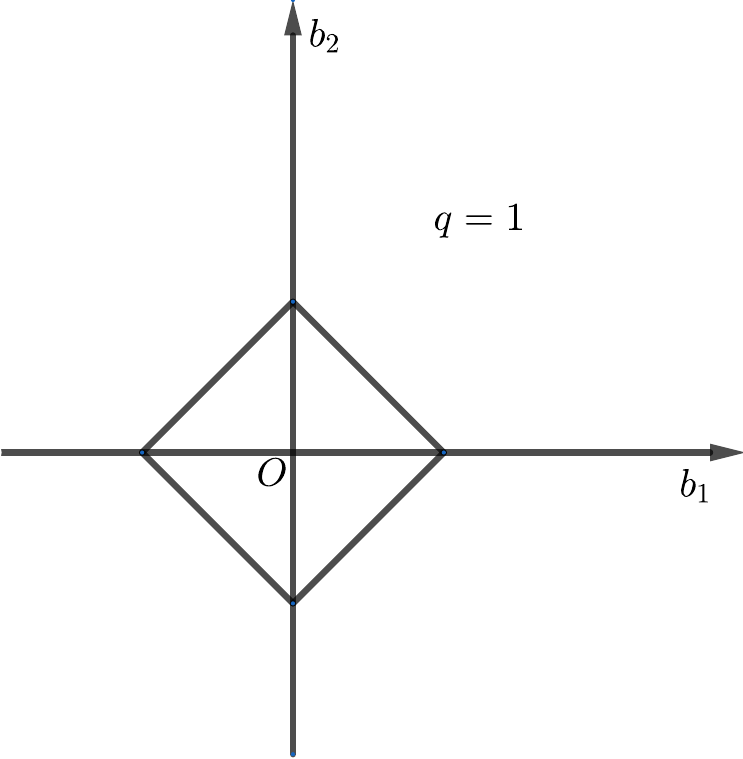}
}\subfloat[$q=2$]{\includegraphics[width=0.33\textwidth]{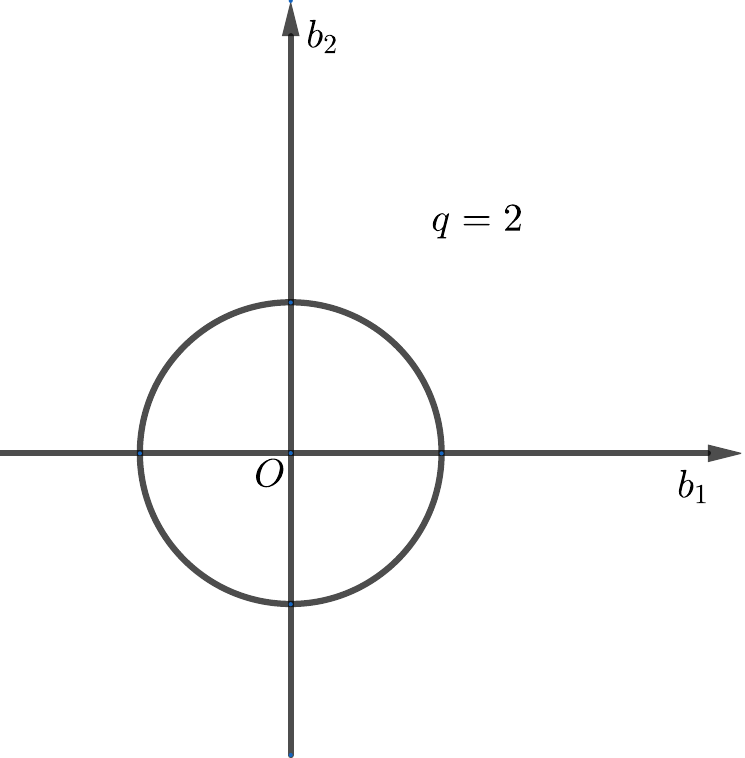}
}\caption{Shrinkage estimators}\label{fig::general-shrinkage-estimators}
\end{figure}

Figure \ref{fig::general-shrinkage-estimators} shows the constraints corresponding to different values of $q$.

\subsection{Elastic net}

\citet{zou2005regularization} proposed the elastic net, which combines the penalties of the lasso and ridge: 
\[
\hat{\beta}^{\text{enet}}(\lambda,\alpha)=\arg\min_{b_{0},b_{1},\ldots,b_{p}}\left[ \textsc{rss}(b_0,b_1,\ldots,b_p)  +\lambda\sum_{j=1}^{p}\left\{ \alpha b_{j}^{2}+(1-\alpha)|b_{j}|\right\} \right].
\]

\begin{figure}
\centering
\includegraphics[width = 0.6\textwidth]{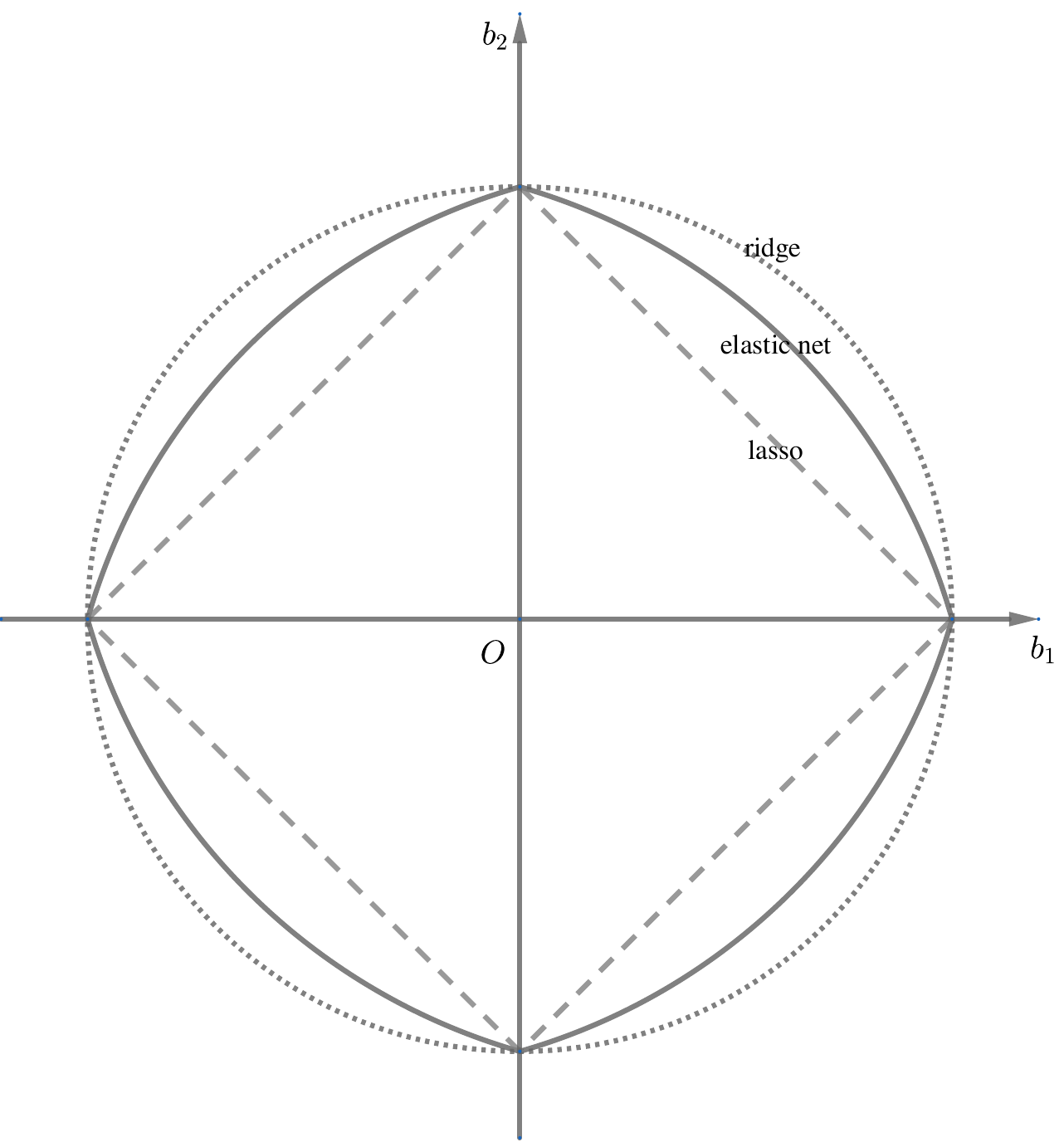}
\caption{Comparing the ridge, lasso, and elastic net}
\label{fig::elasticnet-penalty}
\end{figure}

Figure \ref{fig::elasticnet-penalty} compares the constraints corresponding to the ridge, lasso, and elastic net. Because the constraint of the elastic net is not smooth, it encourages sparse solutions in the same way as the lasso. Due to the ridge penalty, the elastic net can deal with the collinearity of the covariates better than the lasso. 

\citet{friedman2007pathwise} proposed to use the coordinate descent algorithm to solve for the elastic net estimator, and \citet{friedman2009glmnet} implemented it in an \ri{R} package \ri{glmnet}.

\section{Homework problems}

\paragraph{Uniqueness of the lasso prediction}\label{hw14::uniqueness-lasso-pred}

Consider the lasso problem: 
$$
\min_{b\in\mathbb{R}^{p}}   \|Y-Xb\|^{2}+\lambda\|b\|_{1} . 
$$
Prove that if $\hat{\beta}^{(1)}$ and $\hat{\beta}^{(2)}$ are two solutions, then $ \alpha  \hat{\beta}^{(1)} + (1-\alpha)  \hat{\beta}^{(2)}$ must also be a solution, and $X \hat{\beta}^{(1)} = X\hat{\beta}^{(2)}$ must hold, for any $0\leq  \alpha  \leq 1$.

Remark: 
The following facts may be useful for the problem.
The function $\| \cdot \|^2$ is strongly convex. That is, for any $v_1, v_2$ and $ 0< \alpha < 1$, we have
$$
\|   \alpha  v_1  +(1-\alpha ) v_2     \|^2 \leq \alpha \| v_1 \|^2 + (1- \alpha ) \| v_2 \|^2
$$
and the inequality holds when $v_1 \neq v_2$.  The function $\|\cdot \|$ is convex. That is, for any $v_1, v_2$ and $ 0< \alpha < 1$, we have
$$
\|   \alpha  v_1  +(1-\alpha ) v_2     \|_1 \leq \alpha \| v_1 \|_1 + (1- \alpha ) \| v_2 \|_1.
$$
\citet{tibshirani2013lasso} provided detailed discussion about the uniqueness of the lasso problem.

\paragraph{The soft-thresholding lemma}\label{hw14::soft-thresholding-lemma}

Prove Lemma \ref{lemma:softthresholding-lemma}. 

\paragraph{Penalized OLS with an orthogonal design matrix}\label{hw14::orthogonal-ols-pen}

Consider the special case with standardized and orthogonal design
matrix:
\[
X^{\T}1_{n}=0,\qquad X^{\T}X=I_{p}.
\]
For a fixed $\lambda\geq0$, find the explicit formulas of the $j$th coordinates of the following
estimators in terms of the corresponding $j$th coordinate of the OLS estimator $\hat{\beta}_j$ and
$\lambda$ $(j=1,\ldots, p)$:
\begin{align*}
\hat{\beta}^{\text{ridge}}(\lambda) & =\arg\min_{b\in\mathbb{R}^{p}}\left\{ \|Y-Xb\|^{2}+\lambda\|b\|^{2}\right\} ,\\
\hat{\beta}^{\text{lasso}}(\lambda) & =\arg\min_{b\in\mathbb{R}^{p}}\left\{ \|Y-Xb\|^{2}+\lambda\|b\|_{1}\right\} ,\\
\hat{\beta}^{\text{enet}}(\lambda) & =\arg\min_{b\in\mathbb{R}^{p}}\left\{ \|Y-Xb\|^{2}+\lambda  (  \alpha \|b\|^{2} + (1-\alpha) \|b\|_{1} )  \right\}  ,\\
\hat{\beta}^{\text{subset}}(\lambda) & =\arg\min_{b\in\mathbb{R}^{p}}\left\{ \|Y-Xb\|^{2}+\lambda\|b\|_{0}\right\} ,
\end{align*}
where 
\begin{eqnarray*}
 \|b\|^2 &=&  \sum_{j=1}^{p}b_{j}^{2},\\ 
 \|b\|_{1} &=& \sum_{j=1}^{p}|b_{j}|,\\ 
 \|b\|_{0} &=& \sum_{j=1}^{p}1(b_{j}\neq0).
\end{eqnarray*}

\paragraph{Standardization in the elastic net}\label{hw14::standardization-enet}

For fixed $\lambda$ and $\alpha$, prove that the intercept in 
$
\hat{\beta}^{\text{enet}}(\lambda,\alpha)
$
equals zero under the standardization in Condition \ref{condition::standardization}.

\paragraph{Coordinate descent for the elastic net}\label{hw14::enet-coordinate-descent}

Give the detailed coordinate descent algorithm for the elastic net.

\paragraph{Reducing elastic net to lasso}\label{hw14::enet-lasso}

Consider the following form of the elastic net:
$$
\arg\min_{ b \in\mathbb{R}^{p}}  \|  Y -Xb \|^2 + \lambda  \{  \alpha \| b\|^2 + (1-\alpha) \| b \|_1 \} .
$$
Prove that it reduces to the following lasso:
$$
\arg\min_{b\in\mathbb{R}^{p}}  \|  \tilde Y -  \tilde Xb \|^2 + \tilde  \lambda  \| b \|_1,
$$
where 
$$
\tilde Y = \begin{pmatrix}
Y \\
0_p
\end{pmatrix},\quad
\tilde X = \begin{pmatrix}
X \\
\sqrt{\lambda \alpha} I_p
\end{pmatrix},\quad
\tilde{\lambda} = \lambda  (1-\alpha).
$$

Remark: Use the result in Problem \ref{hw13::ridge-data-aug}.

\paragraph{Reducing lasso to iterative ridge}\label{hw14::lasso-iterative-ridge}

Based on the simple result 
$$
\min_{ac=b} (a^2+c^2)/2 = |b|,
$$
for scalars $a,b,c$, \citet{hoff2017lasso} rewrote the lasso problem 
$$
\min_{b\in  \mathbb{R}^p } \{   \| Y-Xb\|^2 + \lambda \|b\|_1 \}
$$
as
$$
\min_{u,v\in  \mathbb{R}^p } \{   \| Y-X(u\circ v)\|^2 + \lambda ( \|u\|^2 + \|v\|^2 )/2 \}
$$
where $\circ$ denotes the component-wise product of vectors. \citet[][Lemma 1]{hoff2017lasso} proved that a local minimum of the new problem must be a local minimum of the lasso problem. 

Prove that the new problem can be solved based on the following iterative ridge regressions:
\begin{enumerate}
\item
given $u$, we update $v$ based on the ridge regression of $Y$ on $X_u$ with tuning parameter $\lambda /2$, where $X_u = X \text{diag}(u_1,\ldots, u_p)$;
\item
given $v$, we update $u$ based on the ridge regression of $Y$ on $X_v$ with tuning parameter $\lambda/2$, where $X_v = X \text{diag}(v_1,\ldots, v_p)$.

\end{enumerate}

\paragraph{More noise in the Boston housing data}

The Boston housing data have $n=506$ observations. Add $p=n$ columns of covariates of random noise, and compare OLS, ridge, and lasso, as in Section \ref{sec::lasso-example}. Add $p=2n$ columns of covariates of random noise, and compare OLS, ridge, and lasso.

\paragraph{Recommended reading}

\citet{tibshirani2011regression}  gives a review of the lasso, as well as its history and recent developments. Two discussants, Professors Peter B\"uhlmann and Chris Holmes, make some excellent comments.

\part{Transformation and Weighting}
    
\chapter{Transformations in OLS}\label{chapter::transformation}
 
Transforming the outcome and covariates is fundamental in linear models. Whenever we specify a linear model $y_i = x_i^{\T} \beta + \varepsilon_i$, we implicitly have transformed the original $y$ and $x$, or at least we have chosen the scales of them. 
\citet{carroll1988transformation} is a textbook on transformations in OLS.  This chapter discusses some important special cases.

\section{Transformation of the outcome}

Although we can view
\[
y_{i}=x_{i}^{\T}\beta+\varepsilon_{i},\quad(i=1,\ldots,n)
\]
as a linear projection that works for any type of outcome $y_{i}\in\mathbb{R}$,
the linear model works the best for continuous outcomes and especially
for Normally distributed outcomes. Sometimes, the linear model can be a poor approximation of the original outcome but may perform well for certain transformations of the outcome. 

\subsection{Log transformation}

With positive, especially heavy-tailed outcomes, a standard transformation
is the log transformation. So we fit a linear model
\[
\log y_{i}=x_{i}^{\T}\beta+\varepsilon_{i},\quad(i=1,\ldots,n).
\]
The interpretation of the coefficients changes a little bit. Because
\[
\frac{\text{\ensuremath{\partial\log}}\hat{y}_{i}}{\partial x_{ij}}=\frac{\partial\hat{y}_{i}}{\hat{y}_{i}}\Big/\partial x_{ij}=\hat{\beta}_{j},
\]
we can interpret $\hat{\beta}_{j}$ in the following way: ceteris
paribus, if $x_{ij}$ increases by one unit, then the proportional
increase in the average outcome is $\hat{\beta}_{j}$. In economics, $\hat{\beta}_{j}$ is the {\it semi-elasticity} of $y$ on $x_j$ in the model with log transformation on the outcome.

Sometimes, we may apply the log transformation on both the outcome and a certain covariate:
$$
\log y_i = \beta_1 x_{i1}  +\cdots + \beta_j \log x_{ij} + \cdots + \beta_p x_{ip} + \varepsilon_i,\quad(i=1,\ldots,n).
$$
The $j$th fitted coefficient becomes
$$
\frac{\text{\ensuremath{\partial\log}}\hat{y}_{i}}{\partial \log x_{ij}} 
= \frac{\partial\hat{y}_{i}}{\hat{y}_{i}}\Big/ \frac{\partial x_{ij}}{x_{ij}}=\hat{\beta}_{j},
$$
so ceteris
paribus, if $x_{ij}$ increases by $1\%$, then   the average outcome increases by $\hat{\beta}_{j} \%$. In economics, $\hat{\beta}_{j}$ is the {\it $x_j$-elasticity} of $y$ in the model with log transformation on both the outcome and $x_{j}$.

The log transformation only works for positive variables. 
For a nonnegative outcome, we can modify the log transformation to $\log (y_{i} + 1).$

\subsection{Box--Cox transformation}

Power transformation is another important class.
The Box--Cox transformation unifies the log transformation and the
power transformation:
\[
g_{\lambda}(y)=\begin{cases}
\frac{y^{\lambda}-1}{\lambda}, & \lambda\neq0,\\
\log y, & \lambda=0.
\end{cases}
\]
L'H\^opital's rule implies that 
\[
\lim_{\lambda\rightarrow0}\frac{y^{\lambda}-1}{\lambda}=\lim_{\lambda\rightarrow0}\frac{\d y^{\lambda}/\d\lambda}{1}=\lim_{\lambda\rightarrow0}y^{\lambda}\log y=\log y,
\]
so as a function of $\lambda$, $g_{\lambda}(y)$ is continuous at
$\lambda = 0$. The log transformation is a limiting version of the power transformation. Can we choose $\lambda$ based on data? \citet{box1964analysis} proposed
a strategy based on the maximum likelihood under the Normal linear model:
\[
Y_{\lambda}=\left(\begin{array}{c}
y_{\lambda1}\\
\vdots\\
y_{\lambda n}
\end{array}\right)=\left(\begin{array}{c}
g_{\lambda}(y_{1})\\
\vdots\\
g_{\lambda}(y_{1})
\end{array}\right)\sim\N(X\beta,\sigma^{2}I_{n}).
\]
The density function of $Y_{\lambda}$ is 
\[
f(Y_{\lambda})=(2\pi\sigma^{2})^{-n/2}\exp\left\{ -\frac{1}{2\sigma^{2}}(Y_{\lambda}-X\beta)^{\T}(Y_{\lambda}-X\beta)\right\} .
\]
The Jacobian of the transformation from $Y$ to $Y_{\lambda}$ is
\[
\det\left(\frac{\partial Y_{\lambda}}{\partial Y}\right)=\det\left(\begin{array}{cccc}
y_{1}^{\lambda-1}\\
 & y_{2}^{\lambda-1}\\
 &  & \ddots\\
 &  &  & y_{n}^{\lambda-1}
\end{array}\right)=\prod_{i=1}^{n}y_{i}^{\lambda-1},
\]
so the density function of $Y$ is 
\[
f(Y)=(2\pi\sigma^{2})^{-n/2}\exp\left\{ -\frac{1}{2\sigma^{2}}(Y_{\lambda}-X\beta)^{\T}(Y_{\lambda}-X\beta)\right\} \prod_{i=1}^{n}y_{i}^{\lambda-1}.
\]
If we treat the density function of $Y$ as a function of $(\beta,\sigma^{2},\lambda)$,
then it is the likelihood function, defined as $L(\beta,\sigma^{2},\lambda).$
Given $(\sigma^{2},\lambda)$, maximizing the likelihood function
is equivalent to minimizing $(Y_{\lambda}-X\beta)^{\T}(Y_{\lambda}-X\beta)$,
i.e., we can run OLS of $Y_{\lambda}$ on $X$ to obtain 
\[
\hat{\beta}(\lambda)=(X^{\T}X)^{-1}X^{\T}Y_{\lambda}.
\]
Given $\lambda$, maximizing the likelihood function is equivalent
to first obtaining $\hat{\beta}(\lambda)$ and then obtaining $\hat{\sigma}^{2}(\lambda)=n^{-1}Y_{\lambda}^{\T} (I_{n}-H)Y_{\lambda}.$
The final step is to maximize the {\it profile likelihood} as a function
of $\lambda$:
\[
L(\hat{\beta}(\lambda),\hat{\sigma}^{2}(\lambda),\lambda)=\left\{ 2\pi\hat{\sigma}^{2}(\lambda)\right\} ^{-n/2}\exp\left\{ -\frac{n\hat{\sigma}^{2}(\lambda)}{2\hat{\sigma}^{2}(\lambda)}\right\} \prod_{i=1}^{n}y_{i}^{\lambda-1}.
\]
Dropping some constants, the log profile likelihood function of $\lambda$
is
\[
l_{\textsc{p}}(\lambda)=-\frac{n}{2}\log\hat{\sigma}^{2}(\lambda)+(\lambda-1)\sumn\log y_{i}.
\]

The \ri{boxcox} function in the  \ri{R} package \ri{MASS} plots $l_{\textsc{p}}(\lambda)$, finds it
maximizer $\hat{\lambda}$, and construct a 95\% confidence
interval $[\hat{\lambda}_{\textsc{l}},\hat{\lambda}_{\textsc{U}}]$
based on the following asymptotic pivotal quantity
\[
2\left\{ l_{\textsc{p}}(\hat{\lambda})-l_{\textsc{p}}(\lambda)\right\} \asim\chi_{1}^{2},
\]
which holds by Wilks' Theorem. In practice, we often use the $\lambda$
values within $[\hat{\lambda}_{\textsc{l}},\hat{\lambda}_{\textsc{U}}]$
that have scientific meanings.

I use two datasets to illustrate the Box--Cox transformation. 

\begin{figure}[ht]
\centering
\includegraphics[width = \textwidth]{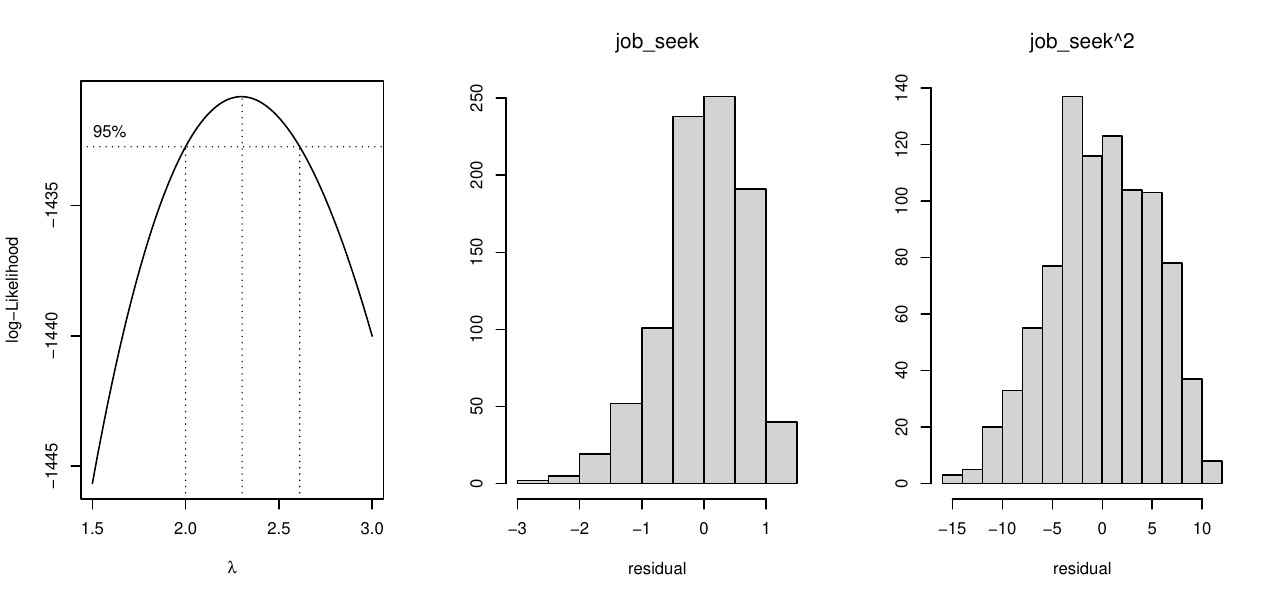}
\caption{Box--Cox transformation in the \ri{jobs} data} \label{fig::boxcox-jobs}
\end{figure}

\begin{example}
I use the \ri{jobs} data in the \ri{mediation} package \citep{tingley2014mediation} to illustrate the Box--Cox transformation. The outcome of interest is \ri{job_seek}, the level of job-search self-efficacy. 
The regressor \ri{treat} is a binary indicator for whether the participant was randomly selected for the JOBS II training program.
Descriptions of other covariates can be found in the \ri{R} package. 
In this example, $\lambda = 2$ seems a plausible value. See the \ri{R} code below and Figure \ref{fig::boxcox-jobs}. 
\end{example}

\begin{lstlisting}
library(MASS)
library(mediation)
par(mfrow = c(1, 3))
jobslm = lm(job_seek ~ treat + econ_hard + depress1 + sex + age + occp + marital + 
              nonwhite + educ + income, data = jobs)
boxcox(jobslm, lambda = seq(1.5, 3, 0.1), plotit = TRUE)
jobslm2 = lm(I(job_seek^2) ~ treat + econ_hard + depress1 + sex + age + occp + marital + 
               nonwhite + educ + income, data = jobs)
hist(jobslm$residuals, xlab = "residual", ylab = "", 
     main = "job_seek", font.main = 1)
hist(jobslm2$residuals, , xlab = "residual", ylab = "", 
     main = "job_seek^2", font.main = 1)
\end{lstlisting}

\begin{figure}[ht]
\centering
\includegraphics[width = \textwidth]{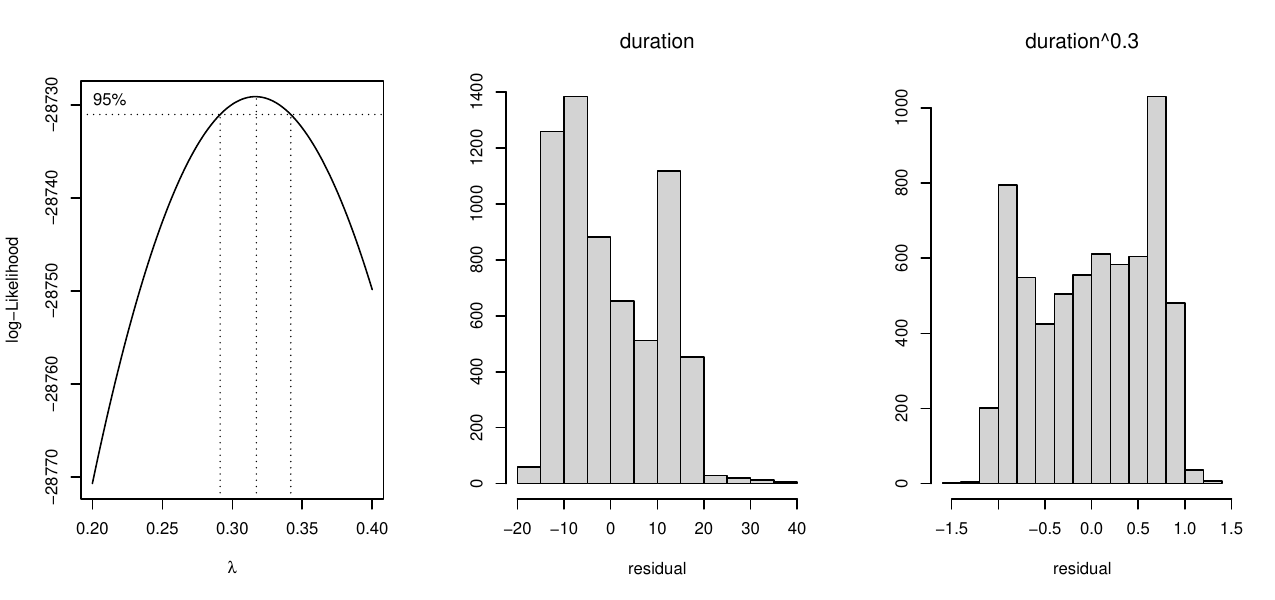}
\caption{Box--Cox transformation in the Penn bonus experiment data} \label{fig::boxcox-penn}
\end{figure}

\begin{example}
I use the Penn bonus experiment data to illustrate the Box--Cox transformation. The outcome of interest is \ri{duration}, the duration until employment. See \citet{koenker2002inference} for descriptions of other regressors. 
In this example, $\lambda = 0.3$ seems a plausible value. However, the residual plot does not seem Normal, making the Box--Cox transformation not very meaningful.  
See the \ri{R} code below and Figure \ref{fig::boxcox-penn}. 
\end{example}

\begin{lstlisting}
penndata = read.table("pennbonus.txt")
par(mfrow = c(1, 3))
pennlm = lm(duration ~ ., data = penndata)
boxcox(pennlm, lambda = seq(0.2, 0.4, 0.05), plotit = TRUE)

pennlm.3 = lm(I(duration^(0.3)) ~., data = penndata)

hist(pennlm$residuals, xlab = "residual", ylab = "", 
     main = "duration", font.main = 1)
hist(pennlm.3$residuals, xlab = "residual", ylab = "", 
     main = "duration^0.3", font.main = 1)
\end{lstlisting}

\section{Transformation of the covariates}

\subsection{Polynomial, basis expansion, and generalized additive model}\label{eq::basis-expansion}

Linear approximations may not be adequate, so we can consider a polynomial
specification. With one-dimensional $x$, we can use 
\[
y_{i}=\beta_{1}+\beta_{2}x_{i}+\beta_{3}x_{i}^{2}+\cdots+\beta_{p}x_{i}^{p-1}+\varepsilon_{i}.
\]
In economics, it is almost the default choice to include the quadratic term of working experience in the log wage equation. I give an example below using the data from \citet{angrist2006quantile}. The quadratic term of \ri{exper} is significant.

\begin{lstlisting}
> library(foreign)
> census00 = read.dta("census00.dta")
> head(census00)
  age educ    logwk     perwt exper exper2 black
1  48   12 6.670576 1.0850021    30    900     0
2  42   13 6.783905 0.9666383    23    529     0
3  49   13 6.762383 1.2132297    30    900     0
4  44   13 6.302851 0.4833191    25    625     0
5  45   16 6.043386 0.9666383    23    529     0
6  43   13 5.061138 1.0850021    24    576     0
> 
> census00ols1 = lm(logwk ~ educ + exper + black, 
+                   data = census00)
> census00ols2 = lm(logwk ~ educ + exper + I(exper^2) + black, 
+                   data = census00)
> round(summary(census00ols1)$coef, 4)
            Estimate Std. Error  t value Pr(>|t|)
(Intercept)   4.8918     0.0315 155.0540   0.0000
educ          0.1152     0.0012  99.1472   0.0000
exper         0.0002     0.0008   0.2294   0.8185
black        -0.2466     0.0085 -29.1674   0.0000
> round(summary(census00ols2)$coef, 4)
            Estimate Std. Error  t value Pr(>|t|)
(Intercept)   5.0777     0.0887  57.2254   0.0000
educ          0.1148     0.0012  97.6506   0.0000
exper        -0.0148     0.0067  -2.2013   0.0277
I(exper^2)    0.0003     0.0001   2.2425   0.0249
black        -0.2467     0.0085 -29.1732   0.0000
\end{lstlisting}

We can also include polynomial terms of more than one covariate, for example,
$$
(1,x_{1i},\ldots,x_{i1}^{d},x_{i2},\ldots,x_{i2}^{l})
$$
or 
$$
(1,x_{1i},\ldots,x_{i1}^{d},x_{i2},\ldots,x_{i2}^{l},x_{i1}x_{i2},\ldots,x_{i1}^{d}x_{i2}^{l}).
$$

\begin{figure}
\centering 
\includegraphics[width =  \textwidth]{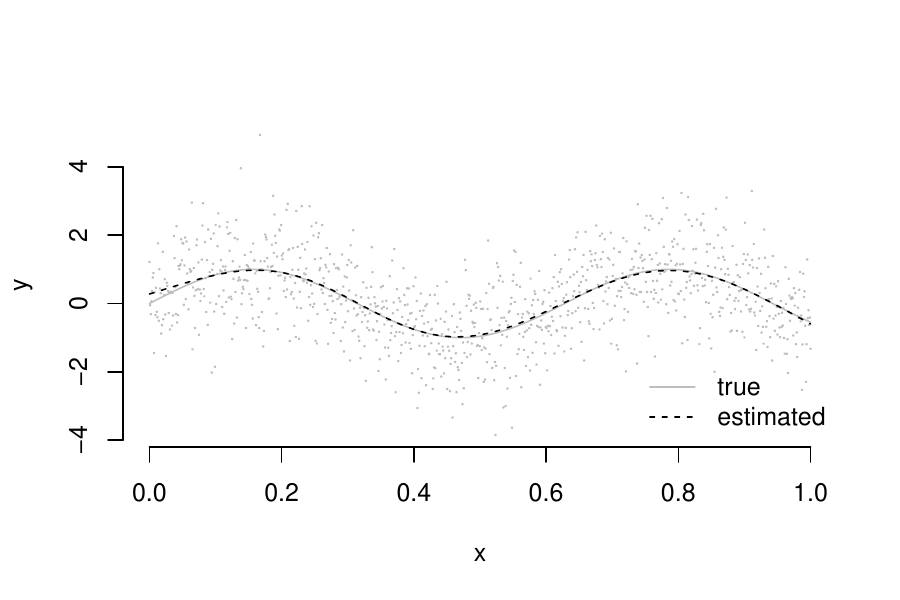}
\caption{Nonparametric regression using the basis expansion: simulated data}\label{fig::npreg-basis}
\end{figure}

We can also approximate the conditional mean function of the outcome by a linear combination of some basis functions:
\begin{eqnarray*}
y_{i} &=& f(x_{i})+\varepsilon_{i} \\
& \cong & \sum_{j=1}^{J}\beta_{j}S_{j}(x_{i})+\varepsilon_{i},
\end{eqnarray*}
where the $S_{j}(x_{i})$'s are basis functions. The \ri{gam} function in the \ri{mgcv} package uses this strategy including the automatic procedure of choosing the number of basis functions $J$. The following example has a sine function as the truth, and the basis expansion approximation yields reasonable performance with sample size $n=1000$. Figure \ref{fig::npreg-basis} plots both the true and estimated curves.

\begin{lstlisting}
library(mgcv)
n = 1000
dat = data.frame(x <- seq(0, 1, length.out = n),
                 true <- sin(x*10),
                 y <- true + rnorm(n))
np.fit = gam(y ~ s(x), data = dat)
plot(y ~ x, data = dat, bty = "n",
     pch = 19, cex = 0.1, col = "grey")
lines(true ~ x, col = "grey") 
lines(np.fit$fitted.values ~ x, lty = 2)
legend("bottomright", c("true", "estimated"), 
       lty = 1:2, col = c("grey", "black"), 
       bty = "n")
\end{lstlisting}

\begin{figure}
\centering 
\includegraphics[width =  \textwidth]{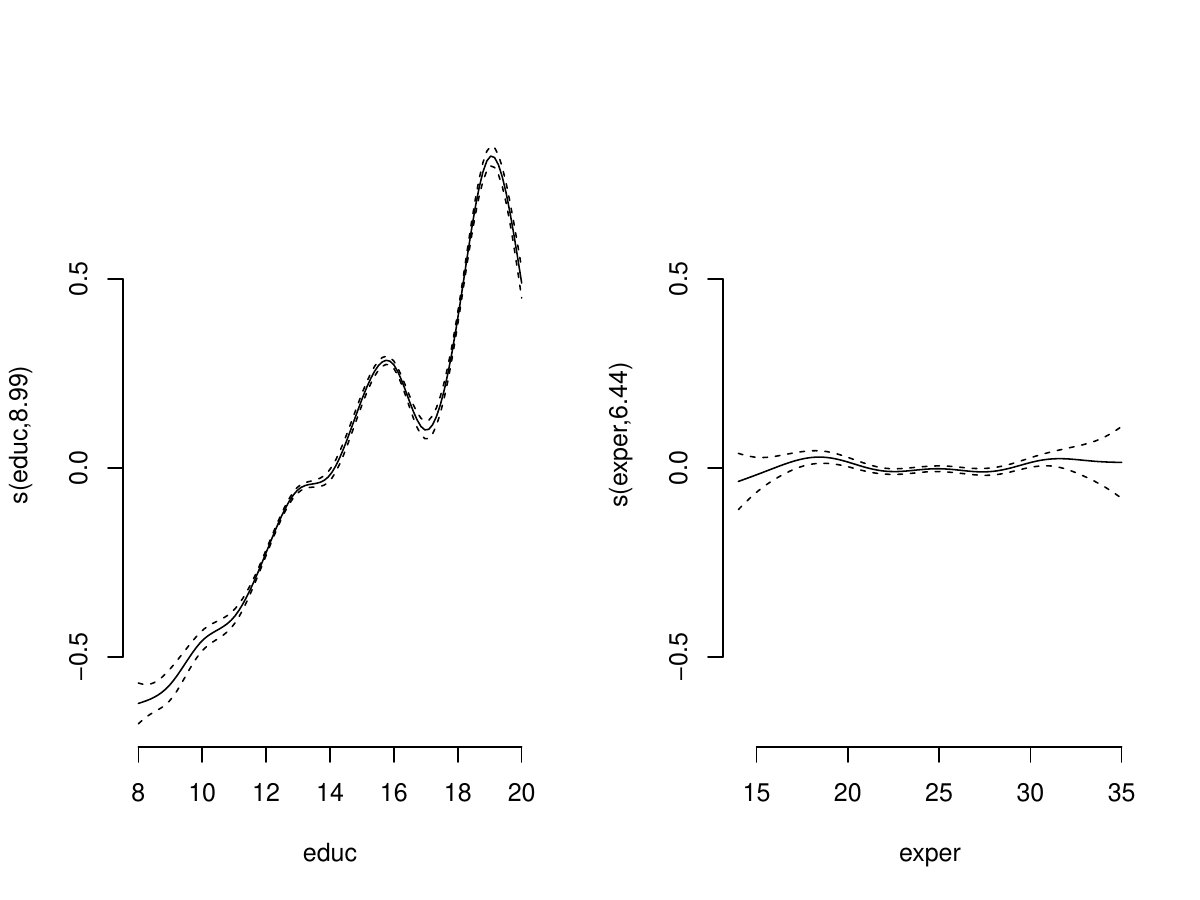}
\caption{Generalized additive model}\label{fig::gam-wage}
\end{figure}

The generalized additive model is an extension of the multivariate case: 
\begin{align*}
y_{i} & =f_{1}(x_{i1})+\cdots+f_{p}(x_{ip})+\varepsilon_{i}\\
 & \cong\sum_{j=1}^{J_{1}}\beta_{1j}S_{j}(x_{i1})+\cdots+\sum_{j=1}^{J_{p}}\beta_{pj}S_{j}(x_{ip})+\varepsilon_{i}.
\end{align*}
The \ri{gam} function in the \ri{mgcv} package implements this strategy. Again I use the dataset from \citet{angrist2006quantile} to illustrate the procedure with nonlinearity in \ri{educ} and \ri{exper} shown in Figure \ref{fig::gam-wage}.

\begin{lstlisting}
census00gam = gam(logwk ~ s(educ) + s(exper) + black, 
                  data = census00) 
summary(census00gam)
par(mfrow = c(1, 2))
plot(census00gam, bty = "n")
\end{lstlisting}

See \citet{wood2017generalized} for more details about the generalized additive model.

\subsection{Regression discontinuity and regression kink}

\begin{figure}
\centering 
\includegraphics[width =  \textwidth]{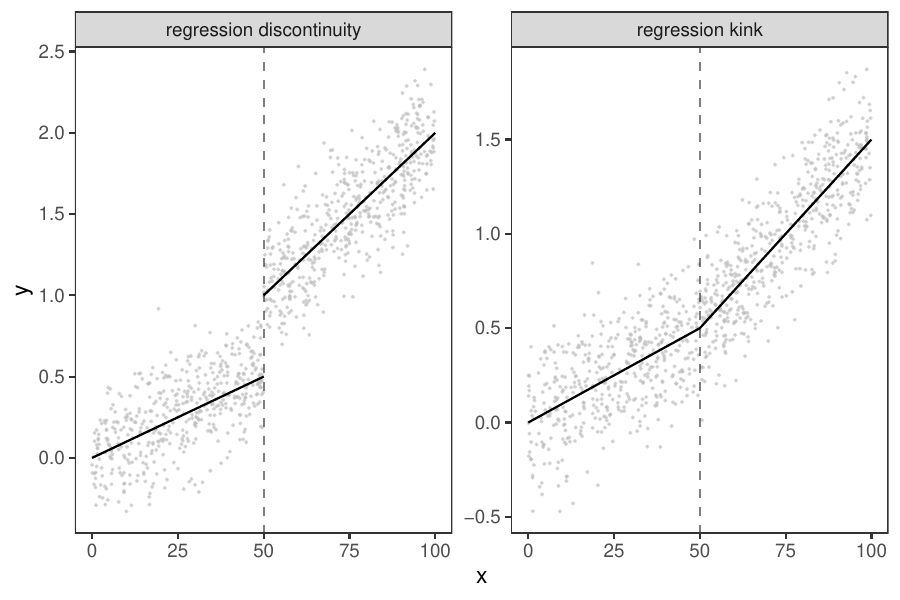}

\caption{Regression discontinuity and kink}\label{fig::reg-discon-kink}

\end{figure}

The left panel of Figure \ref{fig::reg-discon-kink} shows an example of regression discontinuity, where the linear functions before and after a cutoff point can differ with a possible jump. A simple way to capture the two regimes of linear regression is to fit the following model: 
\[
y_{i}=\beta_{1}+\beta_{2}x_{i}+\beta_{3}1\left(x_{i}>c\right)+\beta_{4}x_{i}1\left(x_{i}>c\right)+\varepsilon_{i}.
\]
So 
\[
y_{i}=\begin{cases}
\beta_{1}+\beta_{2}x_{i}+\varepsilon_{i} & x_{i}\leq c,\\
\left(\beta_{1}+\beta_{3}\right)+\left(\beta_{2}+\beta_{4}\right)x_{i}+\varepsilon_{i}, & x_{i}>c.
\end{cases}
\]
Testing the discontinuity at $c$ is equivalent to testing
\[
\left(\beta_{1}+\beta_{3}\right)+\left(\beta_{2}+\beta_{4}\right)c=\beta_{1}+\beta_{2}c ,
\]
which is equivalent to
\[
\beta_{3}+\beta_{4}c=0.
\]

If we center the covariates at $c$, then 
\[
y_{i}=\beta_{1}+\beta_{2}(x_{i}-c)+\beta_{3}1\left(x_{i}>c\right)+\beta_{4}(x_{i}-c)1\left(x_{i}>c\right)+\varepsilon_{i}
\]
and
\[
y_{i}=\begin{cases}
\beta_{1}+\beta_{2}(x_{i}-c)+\varepsilon_{i} & x_{i}\leq c,\\
\left(\beta_{1}+\beta_{3}\right)+\left(\beta_{2}+\beta_{4}\right)(x_{i}-c)+\varepsilon_{i}, & x_{i}>c.
\end{cases}
\]
So testing the discontinuity at $c$ is equivalent to testing $\beta_{3}=0$. 

The right panel of Figure \ref{fig::reg-discon-kink} shows an example of regression kink, where the linear functions before and after a cutoff point can differ but the whole regression line is continuous. A simple way to capture the two regimes of linear regression is to fit the following model: 
\[
y_{i}=\beta_{1}+\beta_{2}R_{c}(x_{i})+\beta_{3}(x_{i}-c)+\varepsilon_{i}
\]
using  
\[
R_{c}(x)=\max(0,x-c)=\begin{cases}
0, & x\leq c,\\
x-c, & x>c.
\end{cases}
\]
So 
\[
y_{i}=\begin{cases}
\beta_{1}+\beta_{3}(x_{i}-c)+\varepsilon_{i}, & x_{i}\leq c,\\
\beta_{1}+\left(\beta_{2}+\beta_{3}\right)(x_{i}-c)+\varepsilon_{i}, & x_{i}>c.
\end{cases}
\]
This ensures that the mean function is continuous at $c$ with both
left and right limits equaling $\beta_{1}$. Testing the kink is equivalent
to testing $\beta_{2}=0.$

These regressions have many applications in economics, but I omit the economic background. Readers can find more discussions in  \citet{angrist2008mostly} and \citet{card2015inference}.

\section{Homework problems}

\paragraph{Piecewise linear regression}

Generate data in the same way as the example in Figure \ref{fig::npreg-basis}, and fit a continuous piecewise linear function with cutoff points $0,0.2,0.4,0.6,0.8,1$.

\chapter{Interactions in OLS}\label{chapter::interaction}
 
 Interaction is an important notion in applied statistics. It measures the interplay of two or more variables acting simultaneously on an outcome. Epidemiologists find that cigarette smoking and alcohol consumption both increase the risks of many cancers. Then they want to measure how cigarette smoking and alcohol consumption jointly increase the risks. That is, does cigarette smoking increase the risks of cancers more in the presence of alcohol consumption than in the absence of it? Political scientists are interested in measuring the interplay of different get-out-to-vote interventions on voting behavior.

 This chapter will review many aspects of interaction in the context of linear regression. \citet{Cox:1984tx} and \citet{berrington2007interpretation} reviewed interaction from a statistical perspective. \citet{Vanderweele::2015} offers a textbook discussion on interaction with a focus on applications in epidemiology.

\section{Two binary covariates interact} 

Start with the simplest yet nontrivial example with two binary covariates $x_{i1}, x_{i2} \in \{0, 1\}$. We can fit the OLS:
\begin{equation}
\label{eq::2X2-ols}
y_i=\hat{\beta}_{0}+\hat{\beta}_{1} x_{i1}+\hat{\beta}_{2} x_{i2}+\hat{\beta}_{12} x_{i1} x_{i2}+\hat{\varepsilon}_i. 
\end{equation}
We can express the coefficients in terms of the means of the outcomes within four combinations of the covariates. The following proposition
is an algebraic result.

\begin{proposition}\label{proposition::2X2-ols}
From \eqref{eq::2X2-ols}, we have
\begin{eqnarray*} 
\hat{\beta}_{0} &=& \bar{y}_{00}, \\
\hat{\beta}_{1} &=& \bar{y}_{10} - \bar{y}_{00}, \\
\hat{\beta}_{2} &=& \bar{y}_{01}-\bar{y}_{00}, \\
\hat{\beta}_{12} &=& (\bar{y}_{11}-\bar{y}_{10})-(\bar{y}_{01}-\bar{y}_{00}),   
\end{eqnarray*} 
where $\bar{y}_{f_{1}f_{2}}$ is the average value of the $y_{i}$'s
with $x_{i1} = f_{1}$ and $x_{i2} = f_{2}$:
$$
\bar{y}_{f_{1}f_{2}} = \frac{ \sumn I(x_{i1} = f_{1}, x_{i2} = f_{2}) y_i }{\sumn I(x_{i1} = f_{1}, x_{i2} = f_{2})}.
$$
\end{proposition}

The proof of Proposition \ref{proposition::2X2-ols} is pure algebraic which is relegated to Problem \ref{hw15interaction::2X2ols}. 
The proposition generalizes to OLS with more than two binary covariates. See \citet{zhao2021regression} for more details.

The coefficient $\hat{\beta}_{12}$ equals the difference between $\bar{y}_{11}-\bar{y}_{10}$, the effect of $x_{i2}$ on $y_i$ holding $x_{i1}$ at level $1$, and $\bar{y}_{01}-\bar{y}_{00}$, the effect of $x_{i2}$ on $y_i$ holding $x_{i1}$ at level $0$. It also equals 
$$
\hat{\beta}_{12}  =  (\bar{y}_{11}-\bar{y}_{01})-(\bar{y}_{10}-\bar{y}_{00}),
$$
that is, the difference between $\bar{y}_{11}-\bar{y}_{01} $, the effect of $x_{i1}$ on $y_i$ holding $x_{i2}$ at level $1$, and $\bar{y}_{10}-\bar{y}_{00}$, the effect of $x_{i1}$ on $y_i$ holding $x_{i2}$ at level $0$. The formula shows the symmetry of $x_{i1} $ and $x_{i2}$ in defining interaction.

\section{A binary covariate interacts with a general covariate}

\subsection{Treatment effect heterogeneity}

In many studies, we are interested in the effect of a binary treatment $z_i$ on an outcome $y_i$, adjusting for some background covariates $x_i$. The covariates can play many roles in this problem. They may affect the treatment, enter the outcome model, and modify the effect of the treatment on the outcome. We can formulate the problem in terms of linear regression: 
\begin{eqnarray}
\label{eq::ols-interaction}
y_i =  \beta_0 + \beta_1 z_i + \beta_2^{\T} x_i + \beta_3^{\T} x_iz_i  + \varepsilon_i ,
\end{eqnarray}
where $E( \varepsilon_i \mid z_i, x_i ) = 0$. So
$$
E(y_i \mid z_i = 1, x_i) =  \beta_0 + \beta_1 + (\beta_2 + \beta_3)^{\T} x_i
$$
and
$$
E(y_i \mid z_i = 0, x_i) =  \beta_0   + \beta_2  ^{\T} x_i,
$$
which implies that
$$
E(y_i \mid z_i = 1, x_i)  - E(y_i \mid z_i = 0, x_i)  = \beta_1 + \beta_3^{\T} x_i.
$$
The difference between two conditional expectations, $E(y_i \mid z_i = 1, x_i)  - E(y_i \mid z_i = 0, x_i) $, is often called the conditional average treatment effect (CATE). Under model \eqref{eq::ols-interaction}, the CATE is a linear function of the covariates. As long as $\beta_3 \neq 0$, we have treatment effect heterogeneity, which is also called {\it effect modification}. A statistical test for $\beta_3 = 0$ is straightforward based on OLS and EHW standard error. 

Note that \eqref{eq::ols-interaction} includes the interaction of the treatment and all covariates. With prior knowledge, we may believe that the treatment effect varies with respect to a subset of covariates, or, equivalently, we may set some components of $\beta_3$ to be zero.

 \subsection{Johnson--Neyman technique}

\citet{johnson1936tests} proposed a technique to identify the region of covariates in which the conditional average treatment $\beta_1 + \beta_3^{\T} x$ is zero. 
For a given $x$, we can test the null hypothesis that $\beta_1 + \beta_3^{\T} x = 0$, which is a linear combination of the regression coefficients of \eqref{eq::ols-interaction}. If we fail to reject the null hypothesis, then this $x$ belongs to the region of zero CATE. 
See \citet{rogosa1981relationship} for more discussions.

\subsection{Blinder--Oaxaca decomposition}

The linear regression \eqref{eq::ols-interaction} also applies to descriptive statistics, when $z_i$ is a binary indicator for subgroups. For example, $z_i$ can be a binary indicator for age, racial, or gender groups, $y_i$ can be the log wage, and $x_i$ can be a vector of explanatory variables such as education, experience, industry, and occupation. 
Sometimes, it is more insightful to write \eqref{eq::ols-interaction} in terms of two possibly non-parallel linear regressions:
\begin{eqnarray}
\label{eq::ols-control}
y_i = \gamma_0 + \theta_0^{\T} x_i + \varepsilon_{i}, \quad 
E( \varepsilon_{i} \mid z_i =0, x_i ) = 0
\end{eqnarray}
for the group with $z_i = 0$, and
\begin{eqnarray}
\label{eq::ols-treatment}
y_i = \gamma_1 + \theta_1^{\T} x_i + \varepsilon_{i}, \quad
E( \varepsilon_{i} \mid z_i = 1, x_i ) = 0
\end{eqnarray}
for the group with $z_i = 1$. Regressions \eqref{eq::ols-control} and \eqref{eq::ols-treatment} are just a reparametrization of \eqref{eq::ols-interaction} with
$$
\gamma_0 = \beta_0,\quad
\theta_0 = \beta_2,\quad 
\gamma_1 =\beta_0 + \beta_1,\quad
\theta_1 = \beta_2 + \beta_3 . 
$$

Based on \eqref{eq::ols-control} and \eqref{eq::ols-treatment}, we can decompose the difference in the outcome means as
\begin{eqnarray*}
&& E(y_i \mid z_i = 1) - E(y_i \mid z_i = 0) \\
&=&  \{ \gamma_1 + \theta_1^{\T} E(x_i \mid z_i = 1)  \} - \{ \gamma_0 + \theta_0^{\T} E(x_i \mid z_i = 0 )  \} \\
&=& \theta_0^{\T}  \{ E(x_i \mid z_i = 1)  -   E(x_i \mid z_i = 0 ) \}  \\
&&  + (\theta_1 - \theta_0) ^{\T}   E(x_i \mid z_i = 0 )    +  \gamma_1 - \gamma_0  \\
&&  + (\theta_1 - \theta_0) ^{\T} \{ E(x_i \mid z_i = 1)  -   E(x_i \mid z_i = 0 ) \} \\
&=& \mathcal E  + \mathcal C  + \mathcal I . 
\end{eqnarray*}
The decomposition has three components:
\begin{enumerate}[label=(\textup{C}\arabic*), ref=\textup{C}\arabic*]
\item
The first component 
\begin{eqnarray*}
\mathcal E 
&=& 
\theta_0^{\T}  \{ E(x_i \mid z_i = 1)  -   E(x_i \mid z_i = 0 ) \}  \\
&=& \beta_2^{\T}  \{ E(x_i \mid z_i = 1)  -   E(x_i \mid z_i = 0 ) \} 
\end{eqnarray*}
measures the {\it endowment effect}, because it is due to the difference in the covariate means across groups;
\item 
The second component
\begin{eqnarray*}
\mathcal C 
&=& 
(\theta_1 - \theta_0) ^{\T}   E(x_i \mid z_i = 0 )    +  \gamma_1 - \gamma_0  \\
&=& \beta_3^{\T}   E(x_i \mid z_i = 0 ) + \beta_1
\end{eqnarray*}
measures the difference in coefficients;
\item 
The third component 
\begin{eqnarray*}
\mathcal I 
&=& 
(\theta_1 - \theta_0) ^{\T} \{ E(x_i \mid z_i = 1)  -   E(x_i \mid z_i = 0 ) \} \\
&=& \beta_3^{\T}  \{ E(x_i \mid z_i = 1)  -   E(x_i \mid z_i = 0 ) \}
\end{eqnarray*}
measures the interaction between the endowment and coefficients.
\end{enumerate}

The above decomposition is called the Blinder--Oaxaca decomposition. \citet{jann2008blinder} reviews other forms of the decomposition, extending the original forms in \citet{blinder1973wage} and \citet{oaxaca1973male}. 

Estimation and testing for $\mathcal E, \mathcal C$, and $\mathcal I$ are straightforward. Based on the OLS of \eqref{eq::ols-interaction} and the sample means $\bar{x}_1$ and $\bar{x}_0$ of the covariates, we have point estimators
\begin{eqnarray*}
\hat{\mathcal E} &=& \hat \beta_2^{\T} (\bar{x}_1 - \bar{x}_0),\\ 
\hat{\mathcal C} &=& \hat \beta_3^{\T}  \bar{x}_0 +  \hat \beta_1,\\ 
\hat{\mathcal I} &=& \hat \beta_3^{\T} (\bar{x}_1 - \bar{x}_0).
\end{eqnarray*}
Given the covariates, they are linear transformations of the OLS coefficients.

\subsection{Chow test}

\citet{chow1960tests} proposed to test whether the two regressions \eqref{eq::ols-control} and \eqref{eq::ols-treatment} are identical. Under the null hypothesis that $\gamma_0 = \gamma_1$ and $\theta_0 = \theta_1$, he proposed an $F$ test assuming homoskedasticity, which is called the {\it Chow test} in econometrics. In fact, this is just a special case of the standard $F$ test for the null hypothesis that $\beta_1=0$ and $\beta_3 = 0$ in \eqref{eq::ols-interaction}. Moreover, based on the OLS in \eqref{eq::ols-interaction}, we can also derive the robust test based on the EHW covariance estimator.

\citet{chow1960tests} discussed a subtle case in which one group has a small size rendering the OLS fit underdetermined. I relegate the details to Problem \ref{hw15interaction::chow-small}. Note that under the null hypothesis, $\mathcal C = \mathcal I  = 0,$ so the difference in the outcome means is purely due to the difference in the covariate means.

\section{Difficulties of interaction}

Practitioners also interpret the coefficient of the product term of two continuous variables as an interaction. However, this heuristics causes subtle issues. This section reviews some important issues.

\subsection{Removable interaction}

The first issue is about {\it removable interaction.} I use a simple numerical example to explain the idea. 

In the \ri{R} code below, the interaction term is not significant, which is coherent with the true data-generating process that the mean of $\log (y)$ is linear in $x_1$ and $x_2$ without interaction.

\begin{lstlisting}
> n  = 1000
> x1 = rnorm(n)
> x2 = rnorm(n)
> y  = exp(x1 + x2 + rnorm(n))
> ols.fit = lm(log(y) ~ x1*x2)
> summary(ols.fit)

Call:
lm(formula = log(y) ~ x1 * x2)

Residuals:
    Min      1Q  Median      3Q     Max 
-3.7373 -0.6822 -0.0111  0.7084  3.1039 

Coefficients:
             Estimate Std. Error t value Pr(>|t|)    
(Intercept)  0.003214   0.031286   0.103    0.918    
x1           1.056801   0.030649  34.480   <2e-16 ***
x2           1.009404   0.030778  32.797   <2e-16 ***
x1:x2       -0.017528   0.030526  -0.574    0.566    
\end{lstlisting}

In the \ri{R} code below, the interaction term is significant because the mean of $y$ is not linear in $x_1$ and $x_2$. 

\begin{lstlisting}
> ols.fit = lm(y ~ x1*x2)
> summary(ols.fit)

Call:
lm(formula = y ~ x1 * x2)

Residuals:
   Min     1Q Median     3Q    Max 
-35.95  -5.17  -0.97   2.34 513.35 

Coefficients:
            Estimate Std. Error t value Pr(>|t|)    
(Intercept)   5.2842     0.6686   7.903 7.17e-15 ***
x1            6.7565     0.6550  10.315  < 2e-16 ***
x2            4.9548     0.6577   7.533 1.11e-13 ***
x1:x2         7.3810     0.6524  11.314  < 2e-16 ***
\end{lstlisting}

From the example above, we can see that taking the log of the outcome removes the interaction. Therefore, the interaction in the OLS fit of $y$ on $(x_1,x_2, x_1x_2)$ is {\it removable}. The lesson here is that the significance of interaction in OLS depends on the scale of the outcome.

\subsection{Main effect in the presence of interaction}\label{sec::main-interaction-centering}

The second issue is that including the interaction term complicates the interpretation of the main effects. I also use a simple example to explain the idea.

In the OLS fit below, we observe significant main effects if we do not include the interaction term. 

\begin{lstlisting}
> ## data from "https://stats.idre.ucla.edu/stat/data/hsbdemo.dta"
> hsbdemo = read.table("hsbdemo.txt")
> ols.fit = lm(read ~ math + socst, data = hsbdemo)
> summary(ols.fit)

Call:
lm(formula = read ~ math + socst, data = hsbdemo)

Residuals:
     Min       1Q   Median       3Q      Max 
-18.8729  -4.8987  -0.6286   5.2380  23.6993 

Coefficients:
            Estimate Std. Error t value Pr(>|t|)    
(Intercept)  7.14654    3.04066   2.350   0.0197 *  
math         0.50384    0.06337   7.951 1.41e-13 ***
socst        0.35414    0.05530   6.404 1.08e-09 ***
\end{lstlisting}

Then we add the interaction term into the OLS, and suddenly we have
significant interaction but not significant main effects. 
\begin{lstlisting}
> ols.fit = lm(read ~ math*socst, data = hsbdemo)
> summary(ols.fit)

Call:
lm(formula = read ~ math * socst, data = hsbdemo)

Residuals:
     Min       1Q   Median       3Q      Max 
-18.6071  -4.9228  -0.7195   4.5912  21.8592 

Coefficients:
             Estimate Std. Error t value Pr(>|t|)   
(Intercept) 37.842715  14.545210   2.602  0.00998 **
math        -0.110512   0.291634  -0.379  0.70514   
socst       -0.220044   0.271754  -0.810  0.41908   
math:socst   0.011281   0.005229   2.157  0.03221 * 
\end{lstlisting}

However, if we center the covariates, the main effects are significant again. 
\begin{lstlisting}
> hsbdemo$math.c = hsbdemo$math - mean(hsbdemo$math)
> hsbdemo$socst.c = hsbdemo$socst - mean(hsbdemo$socst)
> ols.fit = lm(read ~ math.c*socst.c, data = hsbdemo)
> summary(ols.fit)

Call:
lm(formula = read ~ math.c * socst.c, data = hsbdemo)

Residuals:
     Min       1Q   Median       3Q      Max 
-18.6071  -4.9228  -0.7195   4.5912  21.8592 

Coefficients:
                Estimate Std. Error t value Pr(>|t|)    
(Intercept)    51.615327   0.568685  90.763  < 2e-16 ***
math.c          0.480654   0.063701   7.545 1.65e-12 ***
socst.c         0.373829   0.055546   6.730 1.82e-10 ***
math.c:socst.c  0.011281   0.005229   2.157   0.0322 *  
\end{lstlisting}

Then how should we interpret the main effects above? Clearly, the main effects depend on whether or not we include the interaction term, and depend on how we center the regressors. The following discussion proposes the notion of {\it average partial or marginal effect} to measure the main effect.

Based on the linear model with interaction
$$
E(y_i\mid x_{i1}, x_{i2}) = \beta_0 + \beta_1 x_{i1} + \beta_2 x_{i2} + \beta_{12} x_{i1} x_{i2},
$$
define the main effects as
$$
n^{-1} \sumn \frac{  \partial E(y_i\mid x_{i1}, x_{i2}) }{ \partial x_{i1} } = n^{-1} \sumn ( \beta_1 +  \beta_{12}   x_{i2}  )
= \beta_1 +  \beta_{12} \bar{x}_2
$$
and
$$
n^{-1} \sumn \frac{  \partial E(y_i\mid x_{i1}, x_{i2}) }{ \partial x_{i2} } = n^{-1} \sumn ( \beta_2 +  \beta_{12}   x_{i1}  )
= \beta_2 +  \beta_{12} \bar{x}_1 ,
$$
which are called the {\it average partial or marginal effects}. 
So when the covariates are centered at $\bar{x}_1 = \bar{x}_2 = 0$, we can interpret $\beta_1$ and $\beta_2$ as the main effects. 
In contrast, the interpretation of the interaction term does not depend on the centering of the covariates because
$$
\frac{  \partial^2 E(y_i\mid x_{i1}, x_{i2}) }{ \partial x_{i1} \partial x_{i2}}  = \beta_{12}.
$$

\subsection{Power}

The second issue is that usually, statistical tests for interaction do not have enough power. Proposition \ref{proposition::2X2-ols} provides a simple explanation. The variance of the interaction equals
$$
\var(\hat{\beta}_{12}) = \frac{\sigma_{11}^2}{n_{11}} + \frac{\sigma_{10}^2}{n_{10}}  +\frac{\sigma_{01}^2}{n_{01}}  +\frac{\sigma_{00}^2}{n_{00}} ,
$$
where $\sigma^2_{f_1f_2} = \var(y_i \mid x_{i1}   = f_1, x_{i2} = f_2)$. 
Therefore, its variance is driven by the smallest value of $n_{11}, n_{10}, n_{01}, n_{00}$. Even when the total sample size is large, one of the subgroup sample sizes can be small, resulting in a large variance of the estimator of the interaction.

\section{Homework problems}

\paragraph{Interaction and difference-in-differences}\label{hw15interaction::2X2ols}

Prove Proposition \ref{proposition::2X2-ols}. Moreover, simplify the HC0 and HC2 versions of the EHW standard errors of the coefficients in terms of $n_{f_1 f_2}$ and $\hat \sigma^2_{f_1f_2} ,$ where $n_{f_1 f_2}$ is the sample size and $\hat \sigma^2_{f_1f_2}$ is the sample variance of the outcomes for units with $x_{i1} = f_1$ and $x_{i2} = f_2$.

Remark: You can prove the proposition by inverting the $4\times 4$ matrix $X^{\T} X$. However, this method is a little too tedious. Moreover, this proof does not generalize to OLS with $K > 2$ binary covariates. So it is better to find alternative proofs.
For the EHW standard errors, you can use the results in Problems \ref{hw8::anova-ols-hc02} and \ref{hw08::invariance-ehw01234}.

\paragraph{Two OLS}\label{hw15interaction::2ols}

Consider the data $(x_i, z_i, y_i)_{i=1}^n$, where $x_i$ denotes the covariates, $z_i$ denotes the binary group indicator, and $y_i$ denotes the outcome. We can fit two separate OLS:
$$
\hat{y}_i =  \hat\gamma_1 + x_i^{\T} \hat \beta_1
$$
and
$$
\hat{y}_i =   \hat   \gamma_0 + x_i^{\T} \hat \beta_0
$$
with  data in group $1$ and group $0$, respectively. We can also fit a joint OLS using the pooled data:
$$
\hat{y}_i =   \hat \alpha_0 + \hat\alpha_z z_i + x_i^{\T} \hat\alpha_x  +  z_i x_i^{\T} \hat\alpha_{zx}.
$$

\begin{enumerate}
\item
Find $(\hat{\alpha}_0, \hat{\alpha}_z, \hat{\alpha}_x, \hat{\alpha}_{zx})$ in terms of $ (  \hat{\gamma}_1, \hat{\beta}_1, \hat{\gamma}_0, \hat{\beta}_0 )$.
\item
Prove that the fitted values $\hat y_i$'s are the same from the separate and the pooled OLS for all units $i=1, \ldots, n$. 
\item
Prove that the leverage scores $h_{ii}$'s are the same from the separate and the pooled OLS. 
\end{enumerate}

\paragraph{Chow test when one group size is too small}\label{hw15interaction::chow-small}

Assume \eqref{eq::ols-control} and \eqref{eq::ols-treatment} with homoskedastic Normal error terms. 
Let $n_1$ and $n_0$ denote the sample sizes of groups with $z_i = 1$ and $z_i = 0$. Consider the case with $n_0$ larger than the number of covariates but $n_1$ smaller than the number of covariates. So we can fit OLS and estimate the variance based on \eqref{eq::ols-control}, but we cannot do so based on \eqref{eq::ols-treatment}. The statistical test discussed in the main paper does not apply. \citet{chow1960tests} proposed the following test based on prediction.

Let $\hat{\gamma}_0$ and $\hat{\theta}_0$ be the coefficients, and $\hat{\sigma}^2_0$ be the variance estimate based on OLS with units $z_i = 0$. Under the null hypothesis that   $\gamma_0 = \gamma_1$ and $\theta_0 = \theta_1$,  predict the outcomes of the units $z_i=1$:
$$
\hat{y}_i = \hat{\gamma}_0 + \hat{\theta}_0^{\T} x_i 
$$
with the prediction error
$$
d_i = y_i - \hat{y}_i
$$
following a multivariate Normal distribution. Propose an $F$ test based on the $d_i$'s for units with $z_i = 1$.

Remark: It is more convenient to use the matrix form of OLS.

\paragraph{Invariance of the interaction}\label{hw15interaction::2X2-EHW}

In Section \ref{sec::main-interaction-centering}, the point estimate and standard error of the coefficient of the interaction term remain the same no matter whether we center the covariates or not. This result holds in general. This problem quantifies this phenomenon.

With scalars $x_{i1}, x_{i2}, y_i$ $(i=1,\ldots, n)$, we can fit the OLS
$$
y_i = \hat{\beta}_0  + \hat{\beta}_1 x_{i1}  +  \hat{\beta}_2 x_{i2} + \hat{\beta}_{12} x_{i1}  x_{i2}  + \hat{\varepsilon}_i.
$$
Under any location transformations of the covariates $  x_{i1}' =x_{i1}  -c_1 , x_{i2} ' = x_{i2} - c_2$, we can fit the OLS
$$
y_i = \tilde{\beta}_0  + \tilde{\beta}_1 x_{i1}'  +  \tilde{\beta}_2 x_{i2}' + \tilde{\beta}_{12} x_{i1} ' x_{i2}'  + \tilde{\varepsilon}_i.
$$

\begin{enumerate}
\item
Express $\hat{\beta}_0, \hat{\beta}_1, \hat{\beta}_2, \hat{\beta}_{12}$ in terms of $\tilde{\beta}_0 , \tilde{\beta}_1, \tilde{\beta}_2, \tilde{\beta}_{12}$. Verify that $\hat{\beta}_{12} = \tilde{\beta}_{12} .$

\item
Prove that the EHW standard errors for $\hat{\beta}_{12}$ and $\tilde{\beta}_{12} $ are identical.
\end{enumerate}

Remark: Use the results in Problems \ref{hw3::invariance-ols} and \ref{hw08::invariance-ehw01234}.

When $x_1$ and $x_2$ are binary covariates, there are three canonical choices of $(c_1, c_2)$:
\begin{enumerate}[label=(C\arabic*), ref=C\arabic*]
\item
$(c_1, c_2) = (0, 0)$: Then the interpretation of $\hat{\beta}_1$  is the effect of covariate $x_1$ holding the other covariate $x_2$ at 0, whereas the interpretation of $\hat{\beta}_2$ is the  effect of covariate $x_2$ holding the other covariate $x_1$ at 0.

\item
$(c_1, c_2) = (1/2, 1/2)$: Then the interpretation of $\tilde{\beta}_1$  is the average of the effects of covariate $x_1$ holding the other covariate $x_2$ at $0$ and $1$, whereas the interpretation of $\tilde{\beta}_2$  is the average of the effects of covariate $x_2$ holding the other covariate $x_1$ at $0$ and $1$.

\item
$(c_1, c_2) = (\bar{x}_1, \bar{x}_2)$: Then the interpretations of $\tilde{\beta}_1$ and $\tilde{\beta}_2$ are the average marginal effects. 
\end{enumerate} 
The coefficient of the interaction term is invariant to the  location transformations. 

See \citet{zhao2021regression} for more general discussions in the context of factorial experiments with multiple factors.

\chapter{Restricted OLS}
 \label{chapter::rols}

Assume that in the standard linear model $Y=X\beta  + \varepsilon$, the parameter has linear restrictions:
\begin{equation}
\label{eq::rOLS-restriction}
C \beta = r , 
\end{equation}
where $C $ is an $l\times p$ matrix and $r$ is a $l$ dimensional vector. Assume that $C $ has linearly independent row vectors; otherwise, some restrictions are redundant. 
We can use the  restricted OLS:  
$$
\hat\beta_\textup{r} = 
\arg \min_{b\in \mathbb{R}^p }  \|  Y - Xb \|^2
$$
under the restriction 
$$
C  b = r . 
$$

I first give some examples of linear models with restricted parameters, then derive the algebraic properties of the restricted OLS estimator $\hat\beta_\textup{r} $, and finally discuss statistical inference with restricted OLS. 

This chapter is mainly of theoretical interest. For readers who focus on practical data analysis, you can skip this chapter when you first read this book.

\section{Examples}

\begin{example}[Short regression]\label{eg::short-regression-rols}
Partition $X$ into $X_1$ and $X_2$ with $k$ and $l$ columns, respectively, with $p = k+l$. The short regression of $Y$ on $X_1$ yields OLS coefficient $\hat\beta_1$. So $(\hat\beta_1^{\T}, 0_l^{\T}) = \hat\beta_\textup{r} $ with 
$$
C  = (0_{l\times k}, I_{l\times l}),\quad r=0_l. 
$$
\end{example}

\begin{example}[Testing linear hypothesis]\label{eg::test-linear-hypothesis-rols}
Consider testing the linear hypothesis $C \beta = r$ in the linear model. We have discussed in Chapter \ref{chapter::normal-linear-model} the Wald test based on the OLS estimator and its estimated covariance matrix under the Normal linear model. An alternative strategy is to test the hypothesis based on comparing the residual sum of squares under the OLS and restricted OLS. Therefore, we need to compute both $\hat\beta$ and $\hat\beta_\textup{r} $. 
\end{example}

\begin{example}[One-way analysis of variance]\label{eg::one-way-anova-rols}
If $x_i  $ contains the intercept and $Q_1$ dummy variables of a discrete regressor of $Q_1$ levels, $(f_{i1}, \ldots, f_{iQ_1})^{\T}$, then we must impose a restriction on the parameter  in the linear model
$$
y_i = \alpha +\sum_{j=1}^{Q_1}   \beta_j f_{ij}   + \varepsilon_i.
$$
A canonical choice is $\beta_{Q_1} = 0$, which is equivalent to dropping the last dummy variable due to its redundancy. Another canonical choice is $\sum_{j=1}^{Q_1}   \beta_j  = 0$. This restriction keeps the symmetry of the regressors in the linear model and changes the interpretation of $\beta_j$ as the deviation from the ``effect'' of level $j$ with respect to the average ``effect.'' Both are special cases of restricted OLS.   
\end{example}

\begin{example}[Two-way analysis of variance]\label{eg::two-way-anova-rols}
With two factors of levels $Q_1$ and $Q_2$, respectively, the regressor $x_i$ contains the $Q_1$ dummy variables of the first factor, $(f_{i1}, \ldots, f_{iQ_1})^{\T}$, the $Q_2$ dummies of the second factor, $(g_{i1}, \ldots, g_{iQ_2})^{\T}$, and  the $Q_1Q_2$ dummy variables of the interaction terms, $(f_{i1}g_{i1}, \ldots, f_{iQ_1} g_{iQ_2})^{\T}$. We must impose restrictions on the parameters in the linear model
$$
y_i = \alpha + \sum_{j=1}^{Q_1}   \beta_j f_{ij} + \sum_{k=1}^{Q_2}  \gamma_k g_{ik} 
+ \sum_{j=1}^{Q_1}    \sum_{k=1}^{Q_2}  \delta_{jk} f_{ij} g_{ik}   + \varepsilon_i .
$$
Similar to the discussion in Example \ref{eg::one-way-anova-rols}, two canonical choices of restrictions are 
$$
\beta_{Q_1} = 0, \quad 
\gamma_{Q_2} = 0,\quad 
\delta_{Q_1, k} = \delta_{j, Q_2} = 0 ,\quad (j=1,\ldots, Q_1; k = 1, \ldots, Q_2) .
$$
and
$$
 \sum_{j=1}^{Q_1}   \beta_j = 0,\quad
 \sum_{k=1}^{Q_2}  \gamma_k = 0,\quad
  \sum_{j=1}^{Q_1}    \delta_{jk}  =  \sum_{k=1}^{Q_2}  \delta_{jk} =0,\quad (j=1,\ldots, Q_1; k = 1, \ldots, Q_2) .
$$

Problem \ref{hw7::lfe-panel-TWFE} already presented a more advanced example. 
\end{example}

\section{Algebraic properties}

I first give an explicit formula of the restricted OLS \citep{theil, rao}. For simplicity, Theorem \ref{thm::formula-rols} below assumes that $ X^{\T} X$ is invertible. This condition may not hold in general; see Examples \ref{eg::one-way-anova-rols} and \ref{eg::two-way-anova-rols}. \citet{greene} discussed the results without this assumption; see Problem \ref{para::rols-degenerate-greene} for more details.  

\begin{theorem}
\label{thm::formula-rols}
If $ X^{\T} X$ is invertible, then
$$
\hat{\beta}_\textup{r} = \hat{\beta} - (X^{\T} X)^{-1} C ^{\T} \{  C  (X^{\T} X)^{-1} C ^{\T} \}^{-1} (C  \hat{\beta} - r), 
$$
where $\hat{\beta}  $ is the unrestricted OLS coefficient. 
\end{theorem}

\begin{myproof}{Theorem}{\ref{thm::formula-rols}}
The Lagrangian for the restricted optimization problem is
$$
(Y - Xb)^{\T} (Y - Xb) - 2\lambda^{\T} (C b - r).
$$
So the first order condition is
$$
2X^{\T} (Y - Xb) - 2 C^{\T} \lambda = 0 
$$
which implies
\begin{equation}\label{eq::rls-linearsystem}
X^{\T}  X b = X^{\T} Y  -  C^{\T} \lambda .
\end{equation}
Solve the linear system in \eqref{eq::rls-linearsystem} to obtain
$$
b = (X^{\T}  X)^{-1}  (X^{\T} Y  -  C^{\T} \lambda).
$$
Using the linear restriction $C b = r$, we have
$$
C (X^{\T}  X)^{-1}  (X^{\T} Y  -  C^{\T} \lambda) = r
$$
which implies that
$$
\lambda =  
\{  C(X^{\T} X)^{-1} C^{\T} \}^{-1} (C  \hat{\beta} - r ).
$$
So the restricted OLS coefficient is
\begin{eqnarray*}
\hat{\beta}_\textup{r} 
&=& (X^{\T}  X)^{-1}  (X^{\T} Y  -  C^{\T} \lambda)  \\
&=&   \hat{\beta}  -  (X^{\T}  X)^{-1} C^{\T} \lambda  \\
&=&\hat{\beta} - (X^{\T} X)^{-1} C ^{\T} \{  C  (X^{\T} X)^{-1} C ^{\T} \}^{-1} (C  \hat{\beta} - r).
\end{eqnarray*}
Since the objective function is convex and the restrictions are linear, the solution from the first-order condition is indeed the minimizer. 
\end{myproof}

In the special case with $r=0$, Theorem \ref{thm::formula-rols} has a simpler form.

 \begin{corollary}
 \label{corollary::formula-rols-homo}
 Under the restriction \eqref{eq::rOLS-restriction} with $r=0$, we have 
 $$
\hat{\beta}_\textup{r}  =  M_\textup{r}  \hat{\beta} ,
 $$
 where 
  $$
 M_\textup{r} = I_p - (X^{\T} X)^{-1}  C ^{\T} \{  C  (X^{\T} X)^{-1} C ^{\T} \}^{-1}  C .
 $$ 
 Moreover, $ M_\textup{r} $ satisfies the following properties
\begin{eqnarray*}
 M_\textup{r}  (X^{\T} X)^{-1}  C ^{\T} &=& 0,
 \\ 
 C  M_\textup{r}  &=& 0,
 \\ 
 \{ I_p -  C^{\T} (C C^{\T})^{-1} C \}  M_\textup{r} &=&  M_\textup{r} .
\end{eqnarray*}
 \end{corollary}

The $M_\textup{r} $ matrix plays central roles below.

 The following result is also an immediate corollary of Theorem \ref{thm::formula-rols}. 
 
 \begin{corollary}
 \label{corollary::formula-rols}
 Under the restriction \eqref{eq::rOLS-restriction}, we have 
 $$
 \hat{\beta}_\textup{r} - \beta = M_\textup{r} ( \hat{\beta} - \beta  ) . 
 $$
 \end{corollary}

 I leave the proofs of Corollaries \ref{corollary::formula-rols-homo} and \ref{corollary::formula-rols} to Problem \ref{para::algebra-rols}.

 \section{Statistical inference}\label{sec::stat-inf-rols}

I first focus on the Gauss--Markov model with the restriction \eqref{eq::rOLS-restriction}.  As direct consequences of Corollary \ref{corollary::formula-rols}, we can show that the restricted OLS estimator is unbiased for $\beta$, and obtain its covariance matrix below.

\begin{corollary}
\label{corollary::moments-rols}
Assume the Gauss--Markov model under Assumption \ref{assume::gm-model} and the restriction \eqref{eq::rOLS-restriction}. We have 
\begin{eqnarray*}
E (\hat{\beta}_\textup{r}  )  & = & \beta    ,\\ 
\cov(  \hat{\beta}_\textup{r}  ) & = &
\sigma^2 M_\textup{r} (X^{\T}X)^{-1} M_\textup{r}^{\T}.
\end{eqnarray*}
\end{corollary}

Moreover, under the Normal linear model with the restriction \eqref{eq::rOLS-restriction}, we can derive the exact distribution of the restricted OLS estimator and propose an unbiased estimator for $\sigma^2$.

\begin{theorem}
\label{thm::distribution-normal-rols}
Assume the Normal linear model under Assumption \ref{assume::nlm} and the restriction \eqref{eq::rOLS-restriction}. We have 
$$
\hat{\beta}_\textup{r}  \sim \N (  \beta ,  \sigma^2 M_\textup{r} (X^{\T}X)^{-1} M_\textup{r}^{\T}). 
$$
An unbiased estimator for $\sigma^2$ is
$$
\hat{\sigma}^2_\textup{r} = \hat{\varepsilon}_\textup{r} ^{\T} \hat{\varepsilon}_\textup{r}  / (n-p+l),
$$
where $\hat{\varepsilon}_\textup{r}  =  Y - X \hat{\beta}_\textup{r} $. Moreover, 
$$
\hat{\beta}_\textup{r}   \ind \hat{\sigma}^2_\textup{r} .
$$
\end{theorem}

Corollary \ref{corollary::moments-rols} and Theorem \ref{thm::distribution-normal-rols} extend the results for the OLS estimator. I leave their proofs as Problem  \ref{para::moments-distribution-rols}. 
Based on the results in Theorem \ref{thm::distribution-normal-rols}, we can derive the $t$ and $F$ statistics for finite-sample inference of $\beta$ based on the estimator $\hat{\beta}_\textup{r} $ and the estimated covariance matrix
$$
\hat{\sigma}^2_\textup{r}  M_\textup{r} (X^{\T}X)^{-1} M_\textup{r}^{\T} . 
$$

 I then discuss statistical inference under the heteroskedastic linear model under Assumption \ref{assume::heteroskedasticity-lm} and the restriction \eqref{eq::rOLS-restriction}. Corollary \ref{corollary::formula-rols} implies that
 $$
 \cov(  \hat{\beta}_\textup{r}   )
=
 M_\textup{r} (X^{\T}X)^{-1}  X^{\T} \text{diag}\{\sigma_1^2, \ldots, \sigma_n^2 \}  X     (X^{\T}X)^{-1} M_\textup{r}^{\T}.
 $$
 Therefore, the EHW-type estimated covariance matrix is
 $$
 \hat{V}_{\textsc{ehw}, \textup{r}}
=
M_\textup{r} (X^{\T}X)^{-1}  
X^{\T} \text{diag}\{   \hat{\varepsilon}_{i, \textup{r}}^2 , \ldots, \hat{\varepsilon}_{n, \textup{r}}^2 \}  X    
 (X^{\T}X)^{-1} M_\textup{r}^{\T}.
 $$ 
 where the $\hat{\varepsilon}_{i, \textup{r}}$'s are the residuals from the restricted OLS.

\section{Final remarks}

This chapter follows \citet{theil} and \cite{rao}. \citet{bock1973statistical}, \citet{judge1974post}, and \citet{tarpey2000note} contained additional results on restricted OLS. 
\citet{zhao2023covariate} used restricted OLS to analyze factorial experiments with covariates.

\section{Homework problems}

\paragraph{Algebraic details of restricted OLS}
\label{para::algebra-rols}

Prove Corollaries \ref{corollary::formula-rols-homo} and \ref{corollary::formula-rols}.

\paragraph{Invariance of restricted OLS}\label{para::invariance-rls}

Consider an $N \times 1$ vector $ Y$  and two $N\times p$ matrices, $X$ and $X'$, that satisfy $X' = X\Gamma$ for some nonsingular $p\times p$ matrix $\Gamma$. 
The restricted OLS fits of 
$$
\begin{array}{ll}
Y =  X\hat{\beta}_\textup{r} + \hat{\epsilon}_\textup{r}  &\quad\text{subject to} \ \ C \hat{\beta}_\textup{r}  = r, 
 \\
Y =  \tilde X \tilde{\beta}_\textup{r}  +\tilde{\epsilon}_\textup{r}  &\quad \text{subject to} \ \ \tilde C  \tilde{\beta}_\textup{r}    = r ,
\end{array}
$$
with $\tilde X = X\Gamma$ and $\tilde C = C\Gamma$
yield $(\hat{\beta}_\textup{r} ,  \hat{\epsilon}_\textup{r},  \hat{V}_{\textsc{ehw}, \textup{r}})$ and $(\tilde{\beta}_\textup{r}  ,  \tilde{\epsilon}_\textup{r}  ,  \tilde{V}_{\textsc{ehw}, \textup{r}}  )$ as the coefficient vectors, residuals, and robust covariances.

Prove that  
$$
\hat{\beta}_\textup{r} = \Gamma \tilde{\beta}_\textup{r} ,\qquad
\hat{\epsilon}_\textup{r} = \tilde{\epsilon}_\textup{r}  ,\qquad
\hat{V}_{\textsc{ehw}, \textup{r}}= \Gamma \tilde{V}_{\textsc{ehw}, \textup{r}}   \Gamma^{\T}.
$$

\paragraph{Moments and distribution of restricted OLS}
\label{para::moments-distribution-rols}

Prove Corollary \ref{corollary::moments-rols} and Theorem \ref{thm::distribution-normal-rols}.

\paragraph{Gauss--Markov theorem for restricted OLS}
\label{para::gauss-markov-rols}

The Gauss--Markov theorem for $\hat{\beta}_\textup{r} $ holds, as an extension of Theorem \ref{thm:GMtheorem} for  $\hat{\beta} $.  Prove Theorem \ref{thm::gauss-markov-rols} below. 

\begin{theorem}[Gauss--Markov theorem for restricted OLS]
\label{thm::gauss-markov-rols}
Under the  Gauss--Markov model with the restrictions \eqref{eq::rOLS-restriction}, $\hat{\beta}_\textup{r} $ is the best linear unbiased estimator in the sense that $ \cov(\tilde{\beta}_\textup{r}) -  \cov( \hat{\beta}_\textup{r}  ) \succeq 0$ for any linear estimator $\tilde{\beta}_\textup{r} = \tilde{c} + \tilde{A}_\textup{r} Y$, with $\tilde{c} \in \mathbb{R}^p$ and $\tilde{A}_\textup{r} \in \mathbb{R}^{p\times n}$, that satisfies $E(\tilde{\beta}_\textup{r}) = \beta$ for all $\beta$ under constraint \eqref{eq::rOLS-restriction}. 
\end{theorem}

Remark: As a corollary of Theorem \ref{thm::gauss-markov-rols}, we have
$$
(X^{\T}X)^{-1} \succeq
M_\textup{r} (X^{\T}X)^{-1} M_\textup{r}^{\T} 
$$
because the restricted OLS estimator is BLUE whereas the unrestricted OLS is not, under the Gauss--Markov theorem with the restriction \eqref{eq::rOLS-restriction}.

\paragraph{Short regression as restricted OLS}
\label{para::short-regression-rols}

The short regression is a  special case of the formula of $\hat{\beta}_\textup{r} $. Show that 
$$
\hat{\beta}_\textup{r}  = 
\begin{pmatrix}
(X_1^{\T} X_1)^{-1} X_1^{\T} Y  \\
0_l 
\end{pmatrix}
$$ 
with
$$
C  = (0_{l\times k}, I_{l\times l}),\quad r=0_l. 
$$
In this special case, $p=k+l$. 

From the short regression, we can obtain the EHW estimated covariance matrix $\hat{V}_{\textsc{ehw}, 1}$. We can also obtain the EHW estimated covariance matrix $\hat{V}_{\textsc{ehw}, \textup{r}} $ from the restricted OLS. 

Prove that
$$
\hat{V}_{\textsc{ehw}, \textup{r}}  = \begin{pmatrix}
\hat{V}_{\textsc{ehw}, 1} & 0 \\
0 & 0 
\end{pmatrix} .
$$

\paragraph{Reducing restricted OLS to OLS}\label{para::rols-to-ols}

Consider the restricted OLS fit 
\begin{equation}
\label{eq::hw-rols}
Y = X  \hat{\beta}_\textup{r} + \hat{\varepsilon}_\textup{r}   \qquad \text{subject to} \ \  C \hat{\beta}_\textup{r} = 0,
\end{equation}
where $X \in \mathbb R^{n\times p}$ and $C \in \mathbb R^{l \times p}$. 

Let $C_\perp \in \mathbb R^{(p-l) \times p}$ be an orthogonal complement of $C$ in the sense that $(C_\perp^{\T}, C^{\T})$ is nonsingular with $ C_\perp C^{\T} = 0$. Define 
$$
X_{\perp} = X  C_\perp^{\T} (C_\perp C_\perp^{\T})^{-1} .
$$

Consider the corresponding unrestricted OLS fit
\begin{equation}
\label{eq::hw-ols-withoutres}
Y =X_{\perp} \hat\beta_\perp  +  \hat{\varepsilon}_\perp , 
\end{equation}

First, prove that the coefficient and residual vectors must satisfy
$$
\hat\beta_\perp = C_\perp  \hat{\beta}_\textup{r}  , \qquad 
\hat{\varepsilon}_\perp = \hat{\varepsilon}_\textup{r}  . 
$$

Second, prove  
$$
\hat{V}_{\textsc{ehw}, \perp} =  C_\perp \hat{V}_{\textsc{ehw}, \textup{r}}  C_\perp^{\T} , 
$$
where $\hat{V}_{\textsc{ehw}, \perp} $ is the EHW robust covariance matrix from \eqref{eq::hw-ols-withoutres} and  $\hat{V}_{\textsc{ehw}, \textup{r}} $ is the EHW robust covariance matrix from \eqref{eq::hw-rols}.

\paragraph{Minimum normal OLS estimator as restricted OLS}
\label{para::minimum-norm-rols}

An application of the formula of $\hat{\beta}_\textup{r} $ is the minimum norm estimator for under-determined linear equations. When $X$ has more columns than rows, $Y=X\beta$ can have infinitely many solutions, but we may only be interested in the solution with the minimum norm. Assume $p\geq n$ and the rows of $X$ are linearly independent. 

Prove that the solution to
$$
\min_{b\in \mathbb{R}^p} \| b\|^2 \text{ such that } Y = Xb
$$
is
$$
\hat{\beta}_\text{m} = X^{\T} (X X^{\T})^{-1} Y.
$$

\paragraph{Restricted OLS with degenerate design matrix}
\label{para::rols-degenerate-greene}

\citet{greene} pointed out that restricted OLS does not require that $X^{\T} X$ be invertible, although the proof of Theorem \ref{thm::formula-rols} does. Modify the proof to show that the restricted OLS and the Lagrange multiplier satisfy 
\begin{equation}\label{eq::r-ols-degenerate}
\begin{pmatrix}
\hat{\beta}_\textup{r}  \\
\lambda 
\end{pmatrix}
= 
W^{-1}
\begin{pmatrix}
X^{\T} Y \\
r
\end{pmatrix},
\end{equation}
as long as 
$$
W = 
\begin{pmatrix}
X^{\T} X &  C^{\T} \\
C & 0 
\end{pmatrix}
$$
is invertible.

Derive the statistical results in parallel with Section \ref{sec::stat-inf-rols}.

Remark:
If $X$ has full column rank $p$, then $W$ must be invertible. Even if $X$ does not have full column rank, $W$ can still be invertible. See Problem \ref{para::rols-degenerate-condition} below for more details.

\paragraph{Restricted OLS with degenerate design matrix: more algebra}
\label{para::rols-degenerate-condition}

This problem provides more algebraic details for Problem \ref{para::rols-degenerate-greene}. Prove Lemma \ref{lemma::invertible-W-rls} below.

\begin{lemma}
\label{lemma::invertible-W-rls}
Consider 
$$
W = 
\begin{pmatrix}
X^{\T} X &  C^{\T} \\
C & 0 
\end{pmatrix}
$$
where $X^{\T} X$ may not be invertible and $C$ has full row rank. 

The matrix $W$ is invertible if and only if
$
\begin{pmatrix}
X\\
C
\end{pmatrix}
$
has full column rank $p$. 
\end{lemma}

Remark: When $X$ has full column rank $p$, then 
$
\begin{pmatrix}
X\\
C
\end{pmatrix}
$
must have full column rank $p$, which ensures that $W$ is invertible by Lemma \ref{lemma::invertible-W-rls}. I made the comment in Problem \ref{para::rols-degenerate-greene}. 

The invertibility of $W$ plays an important role in other applications. See \citet{benzi2005numerical} and \citet{bai2013nonsingularity} for more general results.

\chapter{Weighted Least Squares}\label{chapter::WLS}

This chapter will discuss the weighted least squares (WLS),  a simple modification of OLS. Computationally, WLS minimizes the weighted average of the squared residuals of the individual observations, with the original OLS as a special case with equal weights. WLS is a useful tool in data analysis at least in the following two cases:
\begin{enumerate}[label=(C\arabic*), ref=C\arabic*]
\item
In the linear model with heteroskedastic errors, WLS can improve efficiency compared with OLS. 

\item
With survey data under non-uniform sampling probabilities, WLS can recover the targeted parameters by inverse probability weighting. 
\end{enumerate} 

Moreover, WLS is a powerful building block to understand other advanced statistics methods. This chapter will introduce the local linear regression, a powerful nonparametric regression method, based on WLS. Chapter \ref{chapter::binary-logit} will use WLS iteratively to compute the maximum likelihood estimate of the logistic regression.

\section{Generalized least squares}\label{section::GLS}

We can extend the Gauss--Markov model to allow for a general covariance structure of the error term. The following generalized Gauss--Markov model is due to \citet{aitkin1936least}.

\begin{assumption}
[Generalized Gauss--Markov model]
\label{assume::g-gaussmarkov}
We have 
$$
Y=X\beta+\varepsilon,
$$ 
with
$$
E(\varepsilon) = 0,\quad \cov(\varepsilon) = \sigma^2 \Sigma, 
$$
where $X$ is a fixed matrix with $p$ linearly independent columns.
The unknown parameters are  $\beta $ and $ \sigma^{2}$.
The $\Sigma$ is a known positive definite matrix.
\end{assumption}

Two leading cases of generalized least squares are
\begin{eqnarray}
\label{eq::diagonal-sigma}
\Sigma=\text{diag}\left\{ w_{1}^{-1},\ldots,w_{n}^{-1}\right\} ,
\end{eqnarray}
which corresponds to  a diagonal
covariance matrix, and 
\begin{eqnarray}
\label{eq::block-diagonal-sigma}
\Sigma = \text{diag}\left\{ \Sigma_1,\ldots, \Sigma_K\right\} , 
\end{eqnarray}
which corresponds to a block diagonal covariance matrix where $\Sigma_k$ is $n_k\times n_k$ and $\sum_{k=1}^K n_k = n$. 

Under Assumption \ref{assume::g-gaussmarkov}, we can still use the OLS estimator $\hat{\beta}=(X^{\T}X)^{-1}X^{\T}Y$. It is unbiased because
$$
E(\hat{\beta}) = (X^{\T}X)^{-1}X^{\T}E(Y) = (X^{\T}X)^{-1}X^{\T} X\beta = \beta 
$$
relies only on the assumption $E(\varepsilon) = 0$. 
It has covariance matrix
\begin{align}
\cov(\hat{\beta}) & =\cov\left\{ (X^{\T}X)^{-1}X^{\T}Y\right\} \nonumber \\
 & =(X^{\T}X)^{-1}X^{\T}\cov(Y)X(X^{\T}X)^{-1}\nonumber \\
 & =\sigma^{2}(X^{\T}X)^{-1}X^{\T}\Sigma X(X^{\T}X)^{-1}\label{eq:cov-ols-generalmodel}
\end{align}
because $\cov(Y) = \Sigma$. 
The OLS estimator is the BLUE under the Gauss--Markov model, but it is
not under the generalized Gauss--Markov model. Then what is the BLUE? We can transform the model under Assumption \ref{assume::g-gaussmarkov} into the Gauss--Markov model by standardizing the error
term to have mean $0$ and covariance $I_n$:  
\[
\Sigma^{-1/2}Y=\Sigma^{-1/2}X\beta+\Sigma^{-1/2}\varepsilon.
\]
Define $Y_* =\Sigma^{-1/2}Y, X_*=\Sigma^{-1/2}X$ and $ \varepsilon_* =\Sigma^{-1/2}\varepsilon$.
The model under Assumption \ref{assume::g-gaussmarkov} reduces to 
\[
Y_* = X_* \beta+ \varepsilon_*,
\]
with
\[ 
E(\varepsilon_*) = 0, \quad \cov(\varepsilon_*) = \sigma^{2}I_{n},
\]
which is the Gauss--Markov model for the transformed variables  $Y_*$ and $X_*$. 
The Gauss--Markov theorem ensures that the BLUE is
\begin{eqnarray*}
\hat{\beta}_{\Sigma}
&=& ({X}_*^{\T}{X}_*)^{-1}{X}_*^{\T}{Y}_* \\
&=& (X^{\T}\Sigma^{-1}X)^{-1}X^{\T}\Sigma^{-1}Y . 
\end{eqnarray*}
It is unbiased because
\begin{eqnarray*}
E(\hat{\beta}_{\Sigma}) 
&=& (X^{\T}\Sigma^{-1}X)^{-1}X^{\T}\Sigma^{-1}E(Y) \\
&=& (X^{\T}\Sigma^{-1}X)^{-1}X^{\T}\Sigma^{-1}X\beta \\
&=&\beta . 
\end{eqnarray*}
It has covariance matrix\footnote{The matrix $X^{\T}\Sigma^{-1} X$ is positive definite and thus invertible,
because
\begin{itemize}
\item 
$\Sigma^{-1} \succeq 0$ implies that  $\alpha^{\T}X^{\T}\Sigma X\alpha\geq 0$ for all $\alpha\in\mathbb{R}^{p}$; 

\item 
$\alpha^{\T}X^{\T}\Sigma X\alpha  =0$ if and only if $X\alpha=0$, which is equivalent to $\alpha=0$ since $X$ has linearly independent columns. 
\end{itemize}}
\begin{align}
\cov(\hat{\beta}_{\Sigma}) & =\cov \{ (X^{\T}\Sigma^{-1}X)^{-1}X^{\T}\Sigma^{-1}Y \} \nonumber \\
 & =(X^{\T}\Sigma^{-1}X)^{-1}X^{\T}\Sigma^{-1}\cov(Y)\Sigma^{-1}X(X^{\T}\Sigma^{-1}X)^{-1}\nonumber \\
 & =\sigma^{2}(X^{\T}\Sigma^{-1}X)^{-1}X^{\T}\Sigma^{-1}\Sigma\Sigma^{-1}X(X^{\T}\Sigma^{-1}X)^{-1}\nonumber \\
 & =\sigma^{2}(X^{\T}\Sigma^{-1}X)^{-1}.\label{eq:cov-wls-generalmodel}
\end{align}
In particular, $\cov(\hat{\beta}_{\Sigma})$ is
smaller than or equal to $\cov(\hat{\beta})$ in the matrix sense. 
So based on
(\ref{eq:cov-ols-generalmodel}) and (\ref{eq:cov-wls-generalmodel}),
we have the following pure linear algebra inequality:
\begin{corollary}\label{corollary::WLS-linear-algebra}
If $X$ has linearly independent columns and $\Sigma$ is invertible, then 
\[
(X^{\T}\Sigma^{-1}X)^{-1}\preceq(X^{\T}X)^{-1}X^{\T}\Sigma X(X^{\T}X)^{-1}.
\]
\end{corollary}

Problem \ref{hw16::linear-algebra-wls} gives a more general result.

\section{Weighted least squares}

This chapter focuses on the first covariance structure in \eqref{eq::diagonal-sigma} and Chapter \ref{chapter::gee} will discuss the second in \eqref{eq::block-diagonal-sigma}.  The $\Sigma$ in \eqref{eq::diagonal-sigma} results in the weighted least squares (WLS) estimator
\begin{eqnarray*}
\hat{\beta}_{w}  =\hat{\beta}_{\Sigma}
&=& (X^{\T}\Sigma^{-1}X)^{-1}X^{\T}\Sigma^{-1}Y \\
&=&\left(\sumn w_{i}x_{i}x_{i}^{\T}\right)^{-1}\sumn w_{i}x_{i}y_{i}.
\end{eqnarray*}
From the derivation above, we can also write the WLS estimator as
\begin{align*}
\hat{\beta}_{w}  
 & =\arg\min_{b\in \mathbb{R}^p}(Y-Xb)^{\T}\Sigma^{-1}(Y-Xb) \\
&  =\arg\min_{b\in \mathbb{R}^p}\sumn w_{i}(y_{i}-x_{i}^{\T}b)^{2}\\
  &=\arg\min_{b\in \mathbb{R}^p}( Y_*  - X_*  b)^{\T} ( Y_* -  X_* b)  \\
&  =\arg\min_{b\in \mathbb{R}^p}\sumn( y_{*i}  -x_{*i}^{\T} b)^{2},
\end{align*}
where
$$
Y_* = W^{1/2}Y, \quad   X_* = W^{1/2} X, \quad \textup{ with } W = \textup{diag}(w_1, \ldots, w_n),
$$
and
$$
y_{*i} =w_{i}^{1/2}y_{i},\quad x_{*i} =w_{i}^{1/2}x_{i} . 
$$
So the WLS is equivalent to the OLS with transformed variables, with the weights inversely proportional to the variances of the errors. 
By this equivalence, WLS inherits many properties of OLS. See the problems in Section \ref{sec::homework-problems} for more details.

Analogous to OLS, we can derive finite-sample exact inference based on the generalized Normal linear model:

\begin{assumption}[Generalized Normal Linear Model]
\label{assumption::G-nlm}
For $i=1, \ldots, n$, we have 
$$
y_{i}=x_{i}^{\T}\beta+\varepsilon_{i} ,
$$
where $x_i$ is fixed, $\varepsilon_{i} \sim \N(0, \sigma^2/w_i)$, and the $\varepsilon_{i}$'s are independent across units.  
\end{assumption}

The generalized Normal linear model in Assumption \ref{assumption::G-nlm} is equivalent to 
$$
y_{*i} = x_{*i}^{\T}\beta+ \varepsilon_{*i}
$$
with IID $\varepsilon_{*i} \sim \N(0, \sigma^2).$ That is, $(x_{*i}, y_{*i})$'s satisfy the classic Normal linear model.  
The \ri{lm} function with \ri{weights} reports the standard error, $t$-statistic, and $p$-value based on this model.

Assumption \ref{assumption::G-nlm} requires that the weights fully capture the heteroskedasticity, which is unrealistic in many problems. More generally, we can derive asymptotic inference based on the following heteroskedastic linear model:

\begin{assumption}
\label{assumption::G-hlm}
For $i=1, \ldots, n$, we have 
$$
y_{i}=x_{i}^{\T}\beta+\varepsilon_{i}, 
$$
where $x_i$ is fixed, and the $\varepsilon_{i}$'s are independent with mean zero and
variances $\sigma_{i}^{2}$.
\end{assumption}

Assumption \ref{assumption::G-hlm} is identical to the heteroskedastic linear model under Assumption \ref{assume::heteroskedasticity-lm}. I repeat it here for symmetric presentation. 
Under Assumption \ref{assumption::G-hlm}, it is possible that $w_{i}\neq1/\sigma_{i}^{2}$,
i.e., the variances used to construct the WLS estimator can be misspecified.
Even though there is no guarantee that $\hat{\beta}_{w}$ is BLUE,
it is still unbiased. From the decomposition
\begin{align*}
\hat{\beta}_{w} & =\left(\sumn w_{i}x_{i}x_{i}^{\T}\right)^{-1}\sumn w_{i}x_{i}y_{i}\\
 & =\left(\sumn w_{i}x_{i}x_{i}^{\T}\right)^{-1}\sumn w_{i}x_{i}(x_{i}^{\T}\beta+\varepsilon_{i})\\
 & =\beta+\left(n^{-1}\sumn w_{i}x_{i}x_{i}^{\T}\right)^{-1}\left(n^{-1}\sumn w_{i}x_{i}\varepsilon_{i}\right),
\end{align*}
we can apply the law of large numbers to show that $\hat{\beta}_{w}$
is consistent for $\beta$ and apply the CLT to
show that 
\[
\hat{\beta}_{w}\asim\N\left(\beta,V_w\right),
\]
where 
\[
V_w  = n^{-1}\left(n^{-1}\sumn w_{i}x_{i}x_{i}^{\T}\right)^{-1}\left(n^{-1}\sumn w_{i}^{2}\sigma_{i}^{2}x_{i}x_{i}^{\T}\right)\left(n^{-1}\sumn w_{i}x_{i}x_{i}^{\T}\right)^{-1}. 
\]
The EHW robust covariance generalizes to
\[
\hat{V}_{\ehw, w}=n^{-1}\left(n^{-1}\sumn w_{i}x_{i}x_{i}^{\T}\right)^{-1}\left(n^{-1}\sumn w_{i}^{2}\hat{\varepsilon}_{w,i}^{2}x_{i}x_{i}^{\T}\right)\left(n^{-1}\sumn w_{i}x_{i}x_{i}^{\T}\right)^{-1},
\]
where $\hat{\varepsilon}_{w,i}=y_{i}-x_{i}^{\T}\hat{\beta}_{w}$ is
the residual from the WLS. In the sandwich covariance above, $w_{i}$
appears in the ``bread'' matrix, but $w_{i}^{2}$ appears in the ``middle'' or ``meat'' matrix. This formula appeared in \citet{magee1998improving} and \citet{romano2017resurrecting}. The  function
\ri{hccm} in the 
\ri{R} package \ri{car} can compute various EHW covariance estimators based on WLS. To save space in the examples below, I report only the standard errors based on the generalized Normal linear model and leave the calculations of the EHW covariances to Problem \ref{hw16::ehw-wls-empirical}.

\section{WLS motivated by heteroskedasticity}


\subsection{Feasible generalized least squares}\label{section::fgls}

Assume that $\varepsilon$ has mean zero and covariance matrix $\text{diag}\{ \sigma_{1}^{2},\ldots,\sigma_{n}^{2} \} $.
If the $\sigma_{i}^{2}$'s are known, we can simply apply the WLS
above; if they are unknown, we need to estimate them first. This gives
the following feasible generalized least squares estimator (FGLS):
\begin{enumerate}[label=(S\arabic*), ref=S\arabic*]
\item Run OLS of $y_i$ on $x_i$ to obtain the residuals $\hat{\varepsilon}_i$. Then obtain the squared residuals $\hat{\varepsilon}_i^2.$

\item \label{enu:fgls-step2}
Run OLS of $\log(\hat{\varepsilon}^{2}_i)$ on
$x_i$ to obtain the fitted values and exponentiate them to obtain $(\hat{\sigma}_{i}^{2})_{i=1}^{n}$;
\item Run WLS of $y_i$ on $x_i$ with weights $ \hat{\sigma}_{i}^{-2} $
to obtain
\[
\hat{\beta}_{\textsc{fgls}}=\left(\sumn\hat{\sigma}_{i}^{-2}x_{i}x_{i}^{\T}\right)^{-1}\sumn\hat{\sigma}_{i}^{-2}x_{i}y_{i}.
\]
\end{enumerate}
In \eqref{enu:fgls-step2}, we can change the model based
on our understanding of heteroskedasticity. 
The above FGLS estimator is close to \citet[][Chapter 8]{wooldridge2012introductory}. \citet{romano2017resurrecting} propose to regress $\log ( \max(\delta^2, \hat \varepsilon_i^2) )$ on $ \log | x_{i1} |, \ldots, \log  |x_{ip}|$ to estimate the individual variances. Their modification has two features: first, they truncate the small residuals by a pre-specified positive number $\delta^2$; second, their regressors are the logs of the absolute values of the original covariates. 
\citet{rose1978nonparametric} and \citet{carroll1982adapting} proposed to estimate $\sigma_i^2$ by using the kernel regression estimator of the residual on covariates. 
\citet{robinson1987asymptotically} discussed the analog using the nearest neighbor method.

Here I use the Boston housing data to compare the OLS and FGLS.

\begin{lstlisting}
> library(mlbench)
> data(BostonHousing)
> ols.fit = lm(medv ~ ., data = BostonHousing)
> dat.res = BostonHousing
> ## log transformation of the squared residuals
> dat.res$medv = log((ols.fit$residuals)^2)
> t.res.ols = lm(medv ~ ., data = dat.res)
> w.fgls = exp(- t.res.ols$fitted.values)
> fgls.fit = lm(medv ~ ., weights = w.fgls, data = BostonHousing)
> round(summary(ols.fit)$coef, 3)
            Estimate Std. Error t value Pr(>|t|)
(Intercept)   36.459      5.103   7.144    0.000
crim          -0.108      0.033  -3.287    0.001
zn             0.046      0.014   3.382    0.001
indus          0.021      0.061   0.334    0.738
chas1          2.687      0.862   3.118    0.002
nox          -17.767      3.820  -4.651    0.000
rm             3.810      0.418   9.116    0.000
age            0.001      0.013   0.052    0.958
dis           -1.476      0.199  -7.398    0.000
rad            0.306      0.066   4.613    0.000
tax           -0.012      0.004  -3.280    0.001
ptratio       -0.953      0.131  -7.283    0.000
b              0.009      0.003   3.467    0.001
lstat         -0.525      0.051 -10.347    0.000
> round(summary(fgls.fit)$coef, 3)
            Estimate Std. Error t value Pr(>|t|)
(Intercept)    9.499      4.064   2.338    0.020
crim          -0.081      0.044  -1.825    0.069
zn             0.030      0.011   2.673    0.008
indus         -0.035      0.038  -0.922    0.357
chas1          1.462      1.119   1.306    0.192
nox           -7.161      2.784  -2.572    0.010
rm             5.675      0.364  15.588    0.000
age           -0.044      0.008  -5.501    0.000
dis           -0.927      0.139  -6.683    0.000
rad            0.170      0.051   3.312    0.001
tax           -0.010      0.002  -4.142    0.000
ptratio       -0.700      0.094  -7.447    0.000
b              0.014      0.002   6.545    0.000
lstat         -0.158      0.036  -4.380    0.000
\end{lstlisting}

Unfortunately, the coefficients, including the point estimates and standard errors, from OLS and FGLS are quite different for several covariates. This suggests that the linear model may be misspecified. Otherwise, both estimators are consistent for the same true coefficient, and they should not be so different even in the presence of randomness.

\citet{romano2017resurrecting} highlighted the efficiency gain from the FGLS compared with OLS in the presence of heteroskedasticity. 
\citet{diciccio2019improving} proposed some improved versions of the FGLS estimator even if the variance function is misspecified. However, it is unusual for practitioners to use FGLS even though it can be more efficient than OLS. There are several reasons. First, the EHW standard errors are convenient for correcting the standard error of OLS under heteroskedasticity. Second, the efficiency gain is usually small, and it is even possible that the FGLS is less efficient than OLS when the variance function is misspecified. Third, the linear model is very likely to be misspecified, and if so, OLS and FGLS estimate different parameters. The OLS has the interpretations as the best linear predictor and the best linear approximation of the conditional mean, but the FGLS has more complicated interpretations when the linear model is wrong. Based on these reasons, we need to carefully justify the choice of FGLS  over  OLS in practical data analyses.

\subsection{Aggregate data and ecological regression}
\label{sec::regression-aggregateddata}

In some case, $(y_{i},x_{i})$ come from aggregate data. For example,
$y_{i}$ can be the average test score and $x_{i}$ can be the average
parents' income of students within classroom $i$. If we believe that
the student-level test score and parents' income follow a homoskedastic linear model, then the model based on the classroom average must be heteroskedastic, with the variance inversely proportional to the classroom size. In this case, a natural choice of weight is $w_{i}=n_{i}$, the classroom size.

\begin{figure}
\centering
\includegraphics[width = \textwidth]{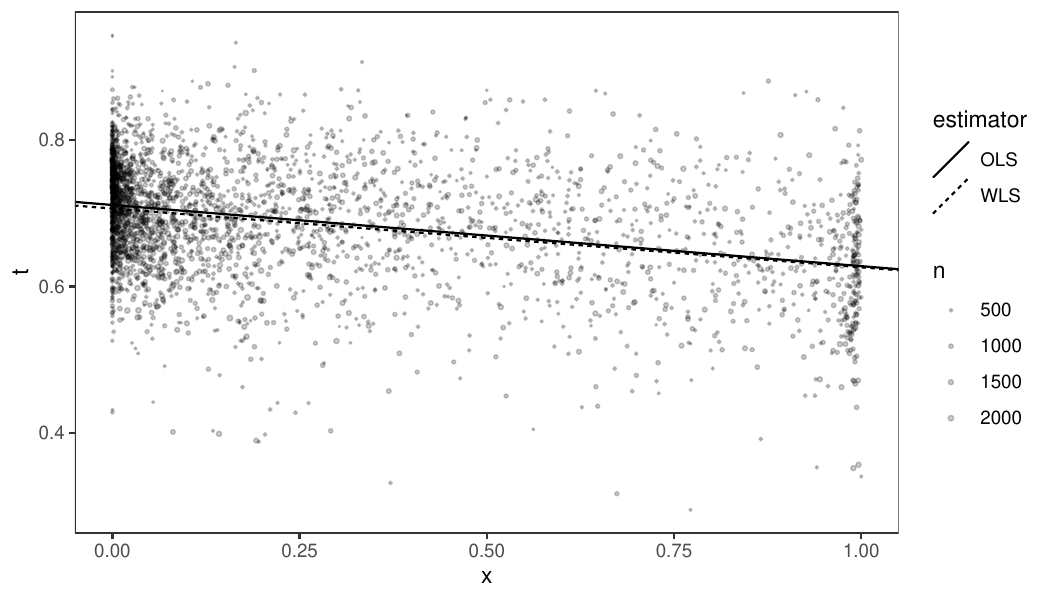}
\caption{Fulton data}\label{fig::lavoteall}
\end{figure}

Below I use the  \ri{lavoteall} dataset from the \ri{R} package \ri{ei}. It contains the  fraction of black registered voters \ri{x}, the fraction of voter turnout \ri{t}, and the total number of people \ri{n} in each Louisiana precinct. 
Figure \ref{fig::lavoteall} is the scatterplot. In this example, OLS and WLS give similar results although \ri{n} varies a lot across precincts.

\begin{lstlisting}
> lavoteall = read.csv("lavoteall.csv")
> ols.fit = lm(t ~ x, data = lavoteall)
> wls.fit = lm(t ~ x, weights = n, data = lavoteall)
> round(summary(ols.fit)$coef, 3)
            Estimate Std. Error t value Pr(>|t|)
(Intercept)    0.711      0.002 408.211        0
x             -0.083      0.004 -19.953        0
> round(summary(wls.fit)$coef, 3)
            Estimate Std. Error t value Pr(>|t|)
(Intercept)    0.706      0.002 421.662        0
x             -0.080      0.004 -19.938        0
\end{lstlisting}

In the above, we can interpret the coefficient of \ri{x} as the precinct-level relationship between the fraction of black registered voters and the fraction voting. Political scientists are interested in using aggregated data to infer individual voting behavior.
Hypothetically, the precinct $i$ has individual data $\{  x_{ij}, y_{ij}  : j = 1, \ldots, n_i\}$, where $x_{ij}$ and $y_{ij}$ are the binary racial and voting status of individual $(i, j)$ $(i=1,\ldots, n; j = 1, \ldots, n_i)$. However, we only observe the aggregated data $\{ \bar{x}_{i\cdot}, \bar{y}_{i\cdot}, n_i :  i = 1,\ldots, n\}$, where 
$$
\bar{x}_{i\cdot} = n_i^{-1} \sum_{j=1}^{n_i}  x_{ij},\quad
\bar{y}_{i\cdot} = n_i^{-1} \sum_{j=1}^{n_i}  y_{ij}
$$ 
are the fraction of black registered voters and the fraction voting, respectively. Can we infer the individual voting behavior based on the aggregated data? In general, this is almost impossible. Under some assumptions, we can make progress. Goodman's ecological regression below is one possibility.

Assume that for precinct $i = 1,\ldots, n$, we have
$$
y_{ij} \mid x_{ij} = 1 \iidsim \text{Bernoulli}(p_{i1}),\quad
y_{ij} \mid x_{ij} = 0 \iidsim \text{Bernoulli}(p_{i0}),\quad 
(j = 1, \ldots, n_i) . 
$$
This is the individual-level model, where the $p_{i1}$'s and $p_{i0}$'s measure the association between race and voting.
We further assume that they are random and independent of the $x_{ij}$'s, with means
\begin{equation}\label{eq::ecological-assumption-1}
E( p_{i1} ) = p_1,\quad
E( p_{i0} ) = p_0.
\end{equation}
Then we can decompose the aggregated outcome variable as 
\begin{eqnarray*}
\bar{y}_{i\cdot} &=&  n_i^{-1} \sum_{j=1}^{n_i}  y_{ij}  \\
&=&n_i^{-1} \sum_{j=1}^{n_i}   \{  x_{ij}  y_{ij}  + (1-x_{ij}) y_{ij}  \} \\
&=& n_i^{-1} \sum_{j=1}^{n_i}   \{  x_{ij}  p_{1}  + (1-x_{ij}) p_0  \}  + \varepsilon_i  \\
&=&  p_1 \bar{x}_{i\cdot}   + p_0 (1 -\bar{x}_{i\cdot} ) + \varepsilon_i,
\end{eqnarray*} 
where
$$
\varepsilon_i = n_i^{-1} \sum_{j=1}^{n_i}   \{  x_{ij}  (y_{ij} - p_{1} )   + (1-x_{ij}) ( y_{ij} -  p_0 ) \}.
$$
So we have a linear relationship between the aggregated outcome and covariate
$$
\bar{y}_{i\cdot}  = p_1 \bar{x}_{i\cdot}   + p_0 (1 -\bar{x}_{i\cdot} ) + \varepsilon_i,
$$
where 
$$
E(\varepsilon_i  \mid \bar{x}_{i\cdot}  ) = 0.
$$
\citet{goodman1953ecological} suggested to use the OLS of $\bar{y}_{i\cdot} $ on $\{ \bar{x}_{i\cdot}, (1 -\bar{x}_{i\cdot} )\}$ to estimate $(p_1, p_0)$, and \citet{goodman1959some} suggested to use the corresponding WLS with weight $n_i$ since the variance of $\varepsilon_i$ has the magnitude $n_i^{-1}$. Moreover, the variance of $\varepsilon_i$ has a rather complicated form of heteroskedasticity, so we should use the EHW standard error for inference. This is called {\it Goodman's regression} or {\it ecological regression}. The following \ri{R} code implements ecological regression based on the \ri{lavoteall} data.

\begin{lstlisting}
> ols.fit = lm(t ~ 0 + x + I(1-x), data = lavoteall)
> wls.fit = lm(t ~ 0 + x + I(1-x), weights = n, data = lavoteall)
> round(summary(ols.fit)$coef, 3)
         Estimate Std. Error t value Pr(>|t|)
x           0.628      0.003 188.292        0
I(1 - x)    0.711      0.002 408.211        0
> round(summary(wls.fit)$coef, 3)
         Estimate Std. Error t value Pr(>|t|)
x           0.626      0.003 194.493        0
I(1 - x)    0.706      0.002 421.662        0
\end{lstlisting}

The assumption in \eqref{eq::ecological-assumption-1} is crucial, which can be too strong when the precinct level $p_{i1}$'s and $p_{i0}$'s vary in systematic but unobserved ways. When the assumption is violated, it is possible that the ecological regression yields the opposite result compared to the individual regression. This is called the {\it ecological fallacy}. 

Another obvious problem of ecological regression is that the estimated coefficients may lie outside of the interval $[0,1]$. Problem \ref{hw16::ecological-regression2} gives an example. 

\citet{gelman2001models} gave an alternative set of assumptions justifying the ecological regression. \citet{king1997solution} proposed some extensions. \citet{robinson1950ecological}  warned that the ecological correlation might not inform individual correlation. \citet{freedman1991ecological} warned that the assumptions underlying the ecological regression might not be plausible in practice.

\section{WLS Motivated by Survey Weights}\label{sec::regression-surveydata}

WLS can be used in other settings unrelated to heteroskedasticity. A leading case is survey sampling. 
Most discussions in this book are based on IID samples, or, at
least, the sample represents the population of interest. Sometimes, researchers over sample some units and under sample some other
units from a population of interest. 

\begin{figure}
\centering
\includegraphics[width = \textwidth]{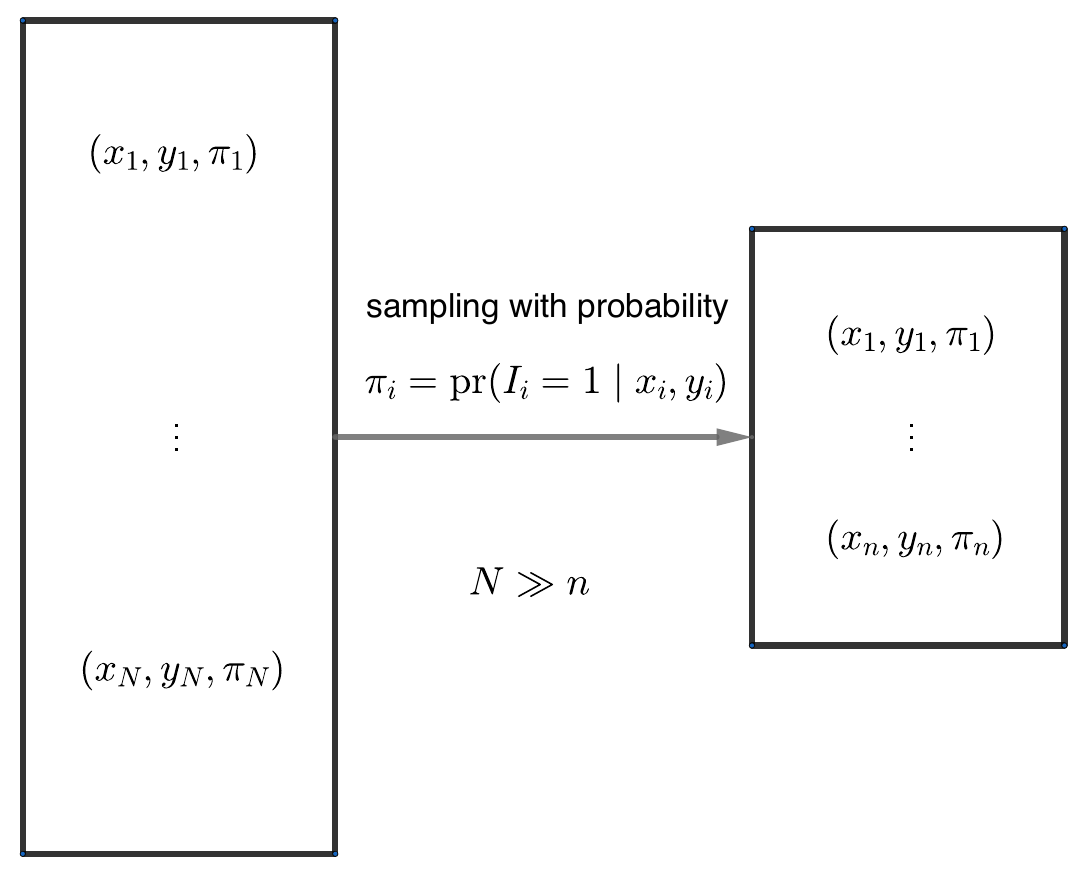}
\caption{Survey sampling}
\end{figure}

If we have the large population with size $N$, then the ideal OLS
estimator of the $y_i$'s on the $x_i$'s is
\[
\hat{\beta}_{\text{ideal}}=\left(\sum_{i=1}^{N}x_{i}x_{i}^{\T}\right)^{-1}\sum_{i=1}^{N}x_{i}y_{i}.
\]
However, we do not have all the data points in the large population,
but sample each data point independently with probability
\[
\pi_{i}=\pr(I_{i}=1\mid x_{i},y_{i}),
\]
where $I_{i}$ is a binary indicator for being included in the sample.
Conditioning on $X_N = (x_{i})_{i=1}^N$ and $Y_N = (y_{i})_{i=1}^{N}$, $\hat{\beta}_{\text{ideal}}$ is a fixed number, and an estimator is
the following WLS estimator\footnote{
The notation $\sum_{i=1}^{N}$ and $\sum_{i=1}^{n}$ can be confusing here. The summation $\sum_{i=1}^{N}$ is over all units in the population, whereas the summation $\sum_{i=1}^{n}$ is over all units in the observed data. 
}
\begin{eqnarray*}
\hat{\beta}_{1/\pi} 
&=& \left(\sum_{i=1}^{N}\frac{I_{i}}{\pi_{i}}x_{i}x_{i}^{\T}\right)^{-1}\sum_{i=1}^{N}\frac{I_{i}}{\pi_{i}}x_{i}y_{i}  \\
&=&\left(\sum_{i=1}^{n}\pi_{i}^{-1}x_{i}x_{i}^{\T}\right)^{-1}\sum_{i=1}^{n}\pi_{i}^{-1}x_{i}y_{i},
\end{eqnarray*}
with weights inversely proportional to the sampling probability. This
inverse probability weighting estimator is reasonable because 
\begin{align*}
E\left(\sum_{i=1}^{N}\frac{I_{i}}{\pi_{i}}x_{i}x_{i}^{\T}\mid X_N,Y_N\right) & =\sum_{i=1}^{N}x_{i}x_{i}^{\T},\\
E\left(\sum_{i=1}^{N}\frac{I_{i}}{\pi_{i}}x_{i}y_{i}\mid X_N,Y_N\right) & =\sum_{i=1}^{N}x_{i}y_{i}.
\end{align*}
The inverse probability weighting estimators are called
the Horvitz--Thompson estimators \citep{horvitz1952generalization}, which are the cornerstones of survey sampling.

Below I illustrate the use of sampling weight using the data from \citet{angrist2006quantile}. Below, \ri{perwt} represents sampling weight.

\begin{lstlisting}
> library(foreign)
> census00 = read.dta("census00.dta")
> ols.fit = lm(logwk ~ age + educ + exper + exper2 + black,
+              data = census00)
> wls.fit = lm(logwk ~ age + educ + exper + exper2 + black,
+              weights = perwt, data = census00)
> round(summary(ols.fit)$coef, 3)
            Estimate Std. Error t value Pr(>|t|)
(Intercept)    5.167      0.128  40.308    0.000
age           -0.015      0.007  -2.201    0.028
educ           0.130      0.007  19.712    0.000
exper2         0.000      0.000   2.243    0.025
black         -0.247      0.008 -29.173    0.000
> round(summary(wls.fit)$coef, 3)
            Estimate Std. Error t value Pr(>|t|)
(Intercept)    5.074      0.127  40.030    0.000
age           -0.008      0.007  -1.263    0.207
educ           0.123      0.007  18.807    0.000
exper2         0.000      0.000   1.294    0.196
black         -0.257      0.008 -32.013    0.000
\end{lstlisting}

\section{WLS as a Building Block for Local linear regression}

\begin{figure}
\includegraphics[width = \textwidth]{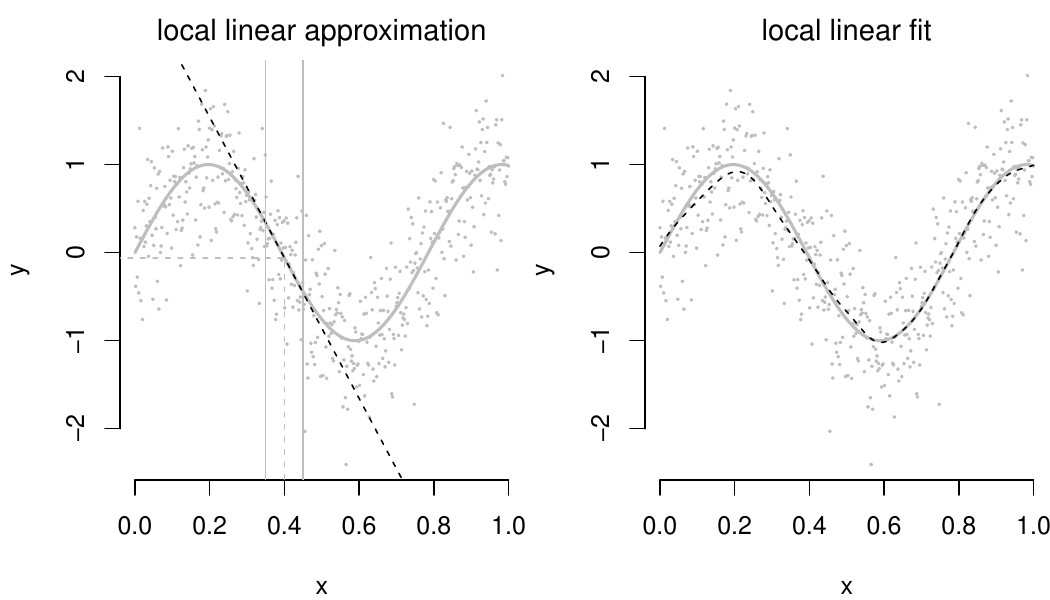}
\caption{Local linear regression}\label{fig::locallinearregression}
\end{figure}

Calculus tells us that locally, we can approximate any smooth function $f(x)$
by a linear function even though the original function can be highly
nonlinear:
$$
f(x) \approx  f(x_0) + f'(x_0) (x - x_0) ,
$$
when $x$ is near $x_0$. 
The left panel of Figure \ref{fig::locallinearregression} shows that in the neighborhood of $x_0=0.4$, even a sine function can be well approximated by a line. Based on data $(x_{i},y_{i})_{i=1}^{n}$, if we want to
predict the mean value of $y$ given $x=x_{0}$, then we can predict
based on a line with the local data points close to $x_{0}$.
It is also reasonable to down weight the points that are far from $x_{0}$,
which motivates the following WLS:
\[
(\hat{\alpha},\hat{\beta})=\arg\min_{a,b}\sumn w_{i}\left\{ y_{i}-a-b(x_{i}-x_{0})\right\} ^{2}
\]
with $w_{i}=K\left\{ (x_i -x_{0})/h\right\} $,  where $K(\cdot)$ is called
the kernel function and $h$ is called the bandwidth parameter. With
the fitted line $\hat{y}(x)=\hat{\alpha}+\hat{\beta}(x-x_{0})$,
the predicted value at $x=x_{0}$ is the intercept $\hat{\alpha}$. 

Technically, $K(\cdot)$ can be any density function, and two canonical
choices are the standard Normal density $K(t) = \frac{1}{ \sqrt{2\pi} } e^{-t^2/2} $ and the Epanechikov kernel
$K(t)=0.75(1-t^{2})1(|t|\leq1).$ The choice of the kernel does not
matter that much. 
The choice of the bandwidth matters much more.
With a larger bandwidth, we have a poorer linear approximation, leading to bias; with a smaller bandwidth, we have fewer data points, leading to larger variance. In practice, we face a bias-variance trade-off. In practice, we can either use cross-validation or other criterion to select $h.$

In general, we can approximate a smooth function by a polynomial:
$$
f(x) \approx  \sum_{k=0}^K  \frac{  f^{(k)}(x_0)  }{k!}  (x - x_0)^k,  
$$
when $x$ is near $x_0$. So
we can even fit a polynomial function locally, which is called the {\it  local polynomial regression} \citep{fan1996local}. 
In the \ri{R} package \ri{Kernsmooth}, the function \ri{locpoly} fits local polynomial regression, and the function \ri{dpill} selects $h$ based on \citet{ruppert1995effective}. The default specification of \ri{locpoly}  is the local linear regression. The right panel of Figure \ref{fig::locallinearregression} shows the local linear fit of the data.

\begin{lstlisting}
> library("KernSmooth")
KernSmooth 2.23 loaded
Copyright M. P. Wand 1997-2009
> n = 500
> x = seq(0, 1, length.out = n)
> fx= sin(8*x)
> y = fx + rnorm(n, 0, 0.5)
> library("KernSmooth")
> n = 500
> x = seq(0, 1, length.out = n)
> fx= sin(8*x)
> y = fx + rnorm(n, 0, 0.5)
> par(mfrow = c(1, 2), mar = c(4, 4, 2, 0.05))
> plot(y ~ x, pch = 19, cex = 0.2, col = "grey", bty = "n",
+      main = "local linear approximation", font.main = 1)
> lines(fx ~ x, lwd = 2, col = "grey")
> 
> x0 = 0.4
> y0 = sin(8*x0)
> segments(x0, -3, x0, y0, lty = 2, col = "grey")
> segments(-4, y0, x0, y0, lty = 2, col = "grey")
> 
> ylinear = sin(8*x0) + 8*cos(8*x0)*(x - x0)
> lines(ylinear ~ x, lty = 2)
> abline(v = x0 - 0.05, col = "grey")
> abline(v = x0 + 0.05, col = "grey")
> 
> 
> plot(y ~ x, pch = 19, cex = 0.2, col = "grey", bty = "n",
+      main = "local linear fit", font.main = 1)
> lines(fx ~ x, lwd = 2, col = "grey")
> h = dpill(x, y)
> locp.fit = locpoly(x, y, bandwidth = h)
> lines(locp.fit, lty = 2)
\end{lstlisting}

\section{Homework problems}
\label{sec::homework-problems}

\paragraph{A linear algebra fact related to WLS}\label{hw16::linear-algebra-wls} 
 
Theorem \ref{thm::efficiency-inequality} below extends Corollary \ref{corollary::WLS-linear-algebra}. Prove Theorem \ref{thm::efficiency-inequality}.

\begin{theorem}
\label{thm::efficiency-inequality}
We have 
$$
(X^{\T}  \Sigma^{-1}  X) ^{-1} 
\preceq
(X^{\T}  \Omega  X)^{-1}X^{\T}   \Omega  \Sigma   \Omega X(X^{\T}   \Omega  X)^{-1},
$$
as long as the inverse matrices exist. Moreover, the equality holds when $ \Omega = \Sigma^{-1}$. 
\end{theorem}

Remark: With $\Omega = I_n$, Theorem \ref{thm::efficiency-inequality} reduces to Corollary \ref{corollary::WLS-linear-algebra}. To prove Theorem \ref{thm::efficiency-inequality}, we can compare the covariance matrices of $\hat{\beta}_{\Sigma}$ and $\hat{\beta}_{\Omega}$ under the generalized Gauss--Markov model.

\paragraph{Generalized least squares with a block diagonal covariance}\label{hw16::gls-block-cov}

Partition $X$ and $Y$ into 
$$
X =  \begin{pmatrix}
X_1\\
\vdots \\
X_K
\end{pmatrix},\quad
Y = \begin{pmatrix}
Y_1\\
\vdots\\
Y_K
\end{pmatrix}
$$
corresponding to $\Sigma$ in \eqref{eq::block-diagonal-sigma} such that $X_k  \in \mathbb{R}^{ n_k\times p}$ and $Y_k\in  \mathbb{R}^{n_k}$. 

Prove that the generalized least squares estimator is
$$
\hat{\beta}_\Sigma = \left(\sum_{k=1}^K X_k ^{\T} \Sigma_k^{-1} X_k \right)^{-1} \left( \sum_{k=1}^K X_k ^{\T} \Sigma_k^{-1} Y_k \right).
$$

\paragraph{Univariate WLS}\label{hw16::univariate-wls}

Prove the following Galtonian formula for the univariate WLS:
$$
\min_{a,b}
\sumn w_i (y_i - a - b x_i)^2
$$
has the minimizer
\begin{eqnarray*}
\hat{\beta}_w &=& \frac{ \sumn w_i(x_i - \bar{x}_w) (y_i - \bar{y}_w) }{ \sumn w_i(x_i - \bar{x}_w)^2  },\\ 
\hat\alpha_w  &=& \bar{y}_w - \hat{\beta}_w  \bar{x}_w , 
\end{eqnarray*}
where $ \bar{x}_w = \sumn w_i x_i /  \sumn w_i   $ and $ \bar{y}_w = \sumn w_i y_i /  \sumn w_i $ are the weighted means of the covariate and outcome.

\paragraph{Difference-in-means with weights}\label{hw16::binaryx-wls}

This problem extends Problem \ref{hw16::univariate-wls}.

With a binary covariate $x_i$, show that the coefficient of $x_i$ in the WLS of $y_i$ on $(1,x_i)$ with weights $w_i$ $(i=1,\ldots, n)$ equals 
$
\bar{y}_{w,1} - \bar{y}_{w,0},
$
where
\begin{eqnarray*}
\bar{y}_{w,1}   &=&  \frac{ \sumn w_i x_i y_i  }{ \sumn w_i x_i } ,\\ 
\bar{y}_{w,0}  &=&  \frac{ \sumn w_i (1-x_i) y_i  }{ \sumn w_i (1-x_i) }
\end{eqnarray*}
are the weighted averages of the outcome under treatment and control, respectively.

\paragraph{Asymptotic Normality of WLS and robust covariance estimator}

Under the heteroskedastic linear model,  prove that $\hat{\beta}_{w}$ is consistent and asymptotically Normal,
and that $n\hat{V}_{w}$ is consistent for the asymptotic covariance
of $\sqrt{n}(\hat{\beta}_{w}-\beta).$ Specify the regularity conditions.

\paragraph{WLS in ANOVA}\label{hw16::wls-anova}

This problem extends Problems \ref{hw5:anova-f}  and \ref{hw8::anova-ols-hc02}. 

For units $i=1,\ldots, n$, assume $y_i$ denotes the outcome, $x_i$ denotes the $p$-vector with entries as the dummy variables for a discrete covariate with $p$ levels, $w_i > 0$ denotes a weight, and $\pi_i > 0$ denotes another weight that is a function of $x_i$ only (for example, $\pi_i = n_j/n$ if $x_i = e_j$). Run the following regressions:
\begin{enumerate}[label=(R\arabic*), ref=R\arabic*]
\item
WLS of $y_i$ on $x_i$ with weight $w_i$ for $i=1,\ldots, n$ to obtain the coefficient vector $\hat{\beta}$ and EHW covariance matrix $\hat{V}$;
\item
WLS of $y_i$ on $x_i$ with weight $w_i \pi_i$ for $i=1,\ldots, n$ to obtain the coefficient vector $\hat{\beta}'$ and EHW covariance matrix $\hat{V}'$.
\end{enumerate}

Prove that 
$
\hat{\beta} = \hat{\beta}'  
$ 
with the $j$th entry
$$
\hat\beta_j=\hat\beta_j' = \frac{  \sum_{i: x_i = e_j}  w_i y_i }{  \sum_{i: x_i = e_j}  w_i},
$$
and moreover, $\hat{V} = \hat{V}'$ are diagonal with the $(j,j)$th entry
$$
\hat{V}_{jj} = \hat{V}_{jj}' = \frac{\sum_{i: x_i = e_j}  w_i^2 (y_i - \hat\beta_j)^2 }{ ( \sum_{i: x_i = e_j}  w_i   )^2  }  .
$$

\paragraph{WLS with aggregate data}\label{para::WLSaggregatedata}

Due to privacy considerations, we may aggregate individual data $(x_i, y_i)_{i=1}^n$ into group means $(\bar{x}_j, \bar{y}_j)_{j=1}^m$ where 
$$
\bar{x}_j = \frac{1}{  |\mathcal{G}_j|  } \sum_{i \in \mathcal{G}_j} x_i,\quad
\bar{y}_j = \frac{1}{  |\mathcal{G}_j|  } \sum_{i \in \mathcal{G}_j} y_i
$$
with units $\{1, \ldots, n\}$ partitioned into disjoint sets $\mathcal{G}_1, \ldots, \mathcal{G}_m$. 

Assume that the individual data satisfy the Gauss--Markov model
$$
Y = X\beta  + \varepsilon
$$
with regression coefficient $\beta$ and homoskedastic variance $\sigma^2$. 
Based on the  individual data, we can obtain the OLS estimator $\hat\beta = (X^{\T} X)^{-1} X^{\T} Y $, which is BLUE.

Derive the BLUE for $\beta$ based on the aggregate data, denoted by $\tilde{\beta}$. Prove that $\cov(\tilde{\beta}) \geq \cov(\hat\beta)$.

Remark: \citet{prais1954grouping} provided a formal discussion of regression with aggregate data. \citet{lancaster1968grouping} discussed the possibility of $\cov(\tilde{\beta}) \leq \cov(\hat\beta)$ when the error terms in the individual model have different variances.

\paragraph{An infeasible generalized least squares estimator}

Can we skip Step \ref{enu:fgls-step2} in Section \ref{section::fgls} and directly apply the following
WLS estimator:
\[
\hat{\beta}_{\textsc{igls}}=\left(\sumn\hat{\varepsilon}_{i}^{-2}x_{i}x_{i}^{\T}\right)^{-1}\sumn\hat{\varepsilon}_{i}^{-2}x_{i}y_{i} , 
\]
with $\hat{\varepsilon}_i = y_i - x_i^{\T} \hat{\beta}$ is the residual from the OLS. 
If so, give a theoretical justification; if not, give a counterexample. Evaluate the finite-sample properties of $\hat{\beta}_{\textsc{igls}}$ using simulated data.

\paragraph{FWL theorem in WLS}\label{hw16::fwl-wls}

Theorem \ref{thm::fwl-wls} below extends Theorem \ref{thm:fwl}. Prove Theorem \ref{thm::fwl-wls}.

\begin{theorem}
\label{thm::fwl-wls}
Consider the WLS with an $n\times 1$ vector $Y$, an $n\times k$ matrix $X_1$, an $n\times l$ matrix $X_2$, and weights $w_{i}$'s. 
Consider the coefficient $\hat{\beta}_{w,2} $ in the long WLS fit
\[
Y=X_{1}\hat{\beta}_{w,1}+X_{2}\hat{\beta}_{w,2}+\hat{\varepsilon}_{w} . 
\]

It equals the coefficient of $\tilde{X}_{w,2}$ in the WLS fit of $Y$ on  $\tilde{X}_{w,2}$,
where $\tilde{X}_{w,2}$ are the residual vectors from the column-wise
WLS of $X_{2}$ on $X_{1}$. 
It also equals the coefficient of $\tilde{X}_{w,2}$ in the WLS fit of $\tilde{Y}_{w}$ on  $\tilde{X}_{w,2}$, where $\tilde{Y}_{w}$ is the residual vector from the WLS of $Y$ on $X_{1}$. 
\end{theorem}

\paragraph{Cochran's formula in WLS}\label{hw16::wls-sample-cochran-formula}

Theorem \ref{thm::cochran-wls} extends Theorem \ref{thm::cochran-formula}. Prove Theorem \ref{thm::cochran-wls}.

\begin{theorem}
\label{thm::cochran-wls}
Consider the WLS with an $n\times 1$ vector $Y$, an $n\times k$ matrix $X_1$, an $n\times l$ matrix $X_2$, and weights $w_{i}$'s. 
We can fit the following WLS:
\begin{eqnarray*}
Y &=& X_1 \hat{\beta}_{w,1} + X_2 \hat{\beta}_{w,2}+ \hat{\varepsilon}_{w},\\
Y &=& X_2 \tilde{\beta}_{w,2} + \tilde{\varepsilon}_w ,\\
X_1 &=& X_2 \hat{\delta}_w + \hat{U}_w,
\end{eqnarray*}
where $\hat{\varepsilon}_w, \tilde{\varepsilon}_w , \hat{U}_w$ are the residuals. 
The last WLS fit means the WLS fit of each column of $X_1$ on $X_2$. 

We have 
$$
\tilde{\beta}_{w,2} = \hat{\beta}_{w,2} +  \hat{\delta}_w \hat{\beta}_{w,1}.
$$
\end{theorem}

\paragraph{EHW robust covariance estimator in WLS}\label{hw16::ehw-wls}

We have proved in Section \ref{section::GLS} that the coefficients from WLS are identical to those from OLS with transformed variables.
Further, prove that the corresponding HC0 version of EHW covariance estimators are
also identical.

 \paragraph{Invariance of covariance estimators in WLS}\label{hw16::invariance-cov-wls}

Problem \ref{hw08::invariance-ehw01234} states the invariance of covariance estimators in OLS. Show that the same result holds for covariance estimators in WLS.

\paragraph{Ridge regression with weights}

Define the ridge regression with weights $w_{i}$'s, and derive the
the formula for the ridge coefficient. 

\paragraph{Coordinate descent algorithm in lasso with weights}

Define the lasso with weights $w_{i}$'s, and give the coordinate
descent algorithm for solving the weighted lasso problem.

\paragraph{General leave-one-out formula via WLS}
\label{hw16::loo-wls}

With data $(X,Y)$, we can define $\hat{\beta}_{[-i]}(w)$ as the WLS estimator of $Y$ on $X$ with weights $w_{i'} = 1(i' \neq i) + w 1(i'=i)$ for $i' = 1,\ldots, n$, where $ 0\leq w \leq 1$.  It reduces to the OLS estimator $\hat\beta$ when $w=1$ and the leave-one-out OLS estimator $\hat{\beta}_{[-i]}$ when $w=0$. 

Prove the general formula 
$$
\hat{\beta}_{[-i]}(w) = \hat{\beta}  - \frac{1-w}{1-(1-w)h_{ii}}  (X^{\T} X)^{-1} x_i \hat{\varepsilon}_i
$$
recalling that $h_{ii}$ is the leverage score and $\hat{\varepsilon}_i$ is the residual of observation $i$.

Remark: 
Based on the above formula, we can compute the derivative of $\hat{\beta}_{[-i]}(w) $ with respect to $w$:
$$
\frac{ \partial  \hat{\beta}_{[-i]}(w)  }{  \partial  w } = \frac{1}{\{ 1-(1-w)h_{ii} \}^2} (X^{\T} X)^{-1} x_i \hat{\varepsilon}_i,
$$
which reduces to 
$$
\frac{ \partial  \hat{\beta}_{[-i]}(0)  }{  \partial  w } = \frac{1}{( 1-h_{ii} )^2} (X^{\T} X)^{-1} x_i \hat{\varepsilon}_i
$$
at $w=0$ and 
$$
\frac{ \partial  \hat{\beta}_{[-i]}(1)  }{  \partial  w } =  (X^{\T} X)^{-1} x_i \hat{\varepsilon}_i
$$
at $w=1$. 
\citet{pregibon1981logistic} reviewed related formulas for OLS.
\citet{broderick2020automatic} discussed related formulas for general statistical models.

 \paragraph{Hat matrix and leverage score in WLS}
 \label{hw16::hat-matrix-leverage-scores}

 Based on the WLS estimator $\hat{\beta}_w = (X^{\T}  W X)^{-1} X^{\T} W Y $ with $W = \text{diag}(w_1, \ldots, w_n)$, we have the predicted vector 
 $$
 \hat{Y}_w = X \hat{\beta}_w  =  X (X^{\T}  W X)^{-1} X^{\T} W Y.
 $$
 This motivates the definition of the hat matrix 
 $$
H_w =   X (X^{\T}WX)^{-1} X^{\T}W 
$$
 such that $ \hat{Y}_w = H_w Y$. 
 
First, prove the following basic properties of the hat matrix:
\begin{eqnarray*}
WH_w &=& H_w^{\T} W,\\ 
X^{\T} W (I_n - H_w) &=& 0.
\end{eqnarray*}
(Note that $H_w$ is not a projection matrix in general.)

Second, prove an extended version of Theorem \ref{thm::leverage-mdist}: with $x_i = (1,x_{i2}^{\T} )^{\T} $, the $(i,i)$th diagonal element of $H_w$ satisfies
$$
h_{w,ii} = \frac{w_i}{  \sum_{i'=1}^n w_{i'} }  (1 + D_{w,i}^2)
$$ 
where 
$$
D_{w,i}^2 =   (x_{i2} - \bar{x}_{w,2})^{\T} S_w^{-1}  (x_{i2} - \bar{x}_{w,2})
$$
with $\bar{x}_{w,2} = \sumn  w_i x_{i2}  / \sumn  w_i $ being the weighted average of $x_{i2}$'s and  $S_w   =   \sumn w_i (x_{i2} - \bar{x}_{w,2})(x_{i2} - \bar{x}_{w,2})^{\T} / \sumn w_i  $ being the corresponding sample covariance matrix.

 Remark: 
\citet{li2009survey} presented the basic properties of $H_w$ for WLS in the context of survey data.

\paragraph{Leave-one-out formula for WLS}
\label{hw16::loo-for-wls}

Use the notation in Problem \ref{hw16::hat-matrix-leverage-scores}. 
Let $\hat\beta_w$ be the WLS estimator of $Y$ on $X$ with weights $w_i$'s. Let $\hat{\beta}_{w[-i]}$ be the WLS estimator without using the $i$th observation. 

Prove that 
$$
\hat{\beta}_{w[-i]} = \hat\beta_w - \frac{ w_i }{ 1-h_{w,ii} }  (X^{\T} W X)^{-1} x_i \hat{\varepsilon}_{w,i} .
$$

\paragraph{EHW standard errors in WLS}\label{hw16::ehw-wls-empirical}

Report the EHW standard errors in the examples in Sections \ref{section::fgls}, \ref{sec::regression-aggregateddata}, and \ref{sec::regression-surveydata}.

\paragraph{Another example of ecological inference}\label{hw16::ecological-regression2}

The \ri{fultongen} dataset in the \ri{ri} package contains aggregated data from 289 precincts in Fulton County, Georgia. The variable \ri{t} represents the fraction voting in 1994 and \ri{x} the fraction in 1992. The variable \ri{n} represents the total number of people. Run ecological regression similar to   Section \ref{sec::regression-aggregateddata}.


    \part{Generalized Linear Models}
  
\chapter{Logistic Regression for Binary Outcomes}
 \label{chapter::binary-logit}

Many applications have binary outcomes $y_i \in \{0, 1\}$. This chapter discusses statistical models of binary outcomes, focusing on the logistic regression, also called the logit regression for simplicity.

\section{Regression with binary outcomes}

\subsection{Linear probability model}

For simplicity, we can still use the linear model for a binary outcome. It is also called the {\it linear probability model}:
\[
y_{i}=x_{i}^{\T}\beta+\varepsilon_{i} \qquad \textup{ with } \qquad  E(\varepsilon_{i}\mid x_{i})=0
\]
because the conditional probability of $y_{i} = 1$ given $x_{i}$ is
a linear function of $x_{i}$: 
\[
\pr(y_{i}=1\mid x_{i})=E(y_{i}\mid x_{i})=x_{i}^{\T}\beta.
\]
An advantage of this linear model is  that the interpretation of the coefficient remains the same as linear models for general outcomes:
$$
\frac{ \partial  \pr(y_{i}=1\mid x_{i})  }{  \partial x_{ij} } = \beta_j,
$$
that is, $\beta_j$ measures the partial impact of $x_{ij}$ on the probability of $y_i$.

A minor technical issue is that the linear probability model implies heteroskedasticity because
\begin{eqnarray*}
\var(y_i\mid x_i) 
&=& \pr(y_{i}=1\mid x_{i}) (1- \pr(y_{i}=1\mid x_{i})) \\
&=& x_{i}^{\T}\beta (1-x_{i}^{\T}\beta).
\end{eqnarray*}
Therefore, we must use the EHW covariance based on OLS. We can also use the feasible generalized least squares (FGLS) in Chapter \ref{section::fgls} to improve efficiency over OLS.

A more severe problem with the linear probability model is its plausibility in general. We may not believe that a linear model is the correct model for a binary outcome because the probability $\pr(y_{i}=1\mid x_{i})$ on the left-hand side is bounded between 0 and 1, but the linear combination $x_{i}^{\T}\beta$ on the right-hand side can be unbounded for general covariates and coefficients. Nevertheless, the OLS decomposition $y_{i}=x_{i}^{\T}\beta+\varepsilon_{i}$
works for any $y_{i}\in\mathbb{R}$, so it is applicable for binary
$y_i$.

Sometimes, practitioners feel that the linear model is not natural for binary outcomes because
the predicted value can be outside the range of $[0,1]$. Therefore,
it is more reasonable to build a model that automatically accommodates
the binary feature of the outcome.

\subsection{General link functions}

A linear combination of general covariates may be outside the range
of $[0,1]$, but we can find a monotone transformation to force it
to lie within the interval $[0,1]$. This motivates us to consider the
following model:
\[
\pr(y_{i}=1\mid x_{i})=g(x_{i}^{\T}\beta),
\]
where $g(\cdot):\mathbb{R}\rightarrow[0,1]$ is a monotone function,
and its inverse is often called the link function. Mathematically,
the distribution function of any continuous random variable is a monotone function that maps from $\mathbb{R}$
to $[0,1]$. So we have infinitely many choices for $g(\cdot)$. Four
canonical choices ``logit'', ``probit'', ``cauchit'', and ``cloglog'' are below which are the standard options in \ri{R}:

\begin{tabular}{|c|c|}
name & functional form\tabularnewline
\hline 
\hline 
logit & $g(z)=\frac{e^{z}}{1+e^{z}}$\tabularnewline
\hline 
probit & $g(z)=\Phi(z)$\tabularnewline
\hline 
cauchit & $g(z)=\frac{1}{\pi}\arctan(z)+\frac{1}{2}$\tabularnewline
\hline 
cloglog & $g(z)=1-\exp(-e^{z})$\tabularnewline
\end{tabular}

The above $g(z)$'s correspond to different distribution functions. The $g(z)$ for the logit model\footnote{\citet{berkson1944application} was an early use of the logit model.}
is the distribution function of the standard logistic distribution
with density
\begin{equation}
g'(z)=\frac{e^{z}}{(1+e^{z})^{2}}=g(z)\left\{ 1-g(z)\right\} . \label{eq:densityoflogit}
\end{equation}
The $g(z)$ for the probit model\footnote{\citet{bliss1934method} was an early use of the probit model.} is the distribution function of a
standard Normal distribution. The $g(z)$ for the cauchit model is
the distribution function of the standard Cauchy distribution with
density
\[
g'(z)=\frac{1}{\pi(1+z^{2})} .
\]
The $g(z)$ for the cloglog model is the distribution function of the standard
log-Weilbull distribution with density
\[
g'(z)=\exp(z-e^{z}).
\]
I will give more motivations for the first three link functions in Section \ref{sec::binary-reg-latent} and for the fourth link function in Problem \ref{hw19::poisson-logistic}.

\begin{figure}
\centering
\includegraphics[width = \textwidth]{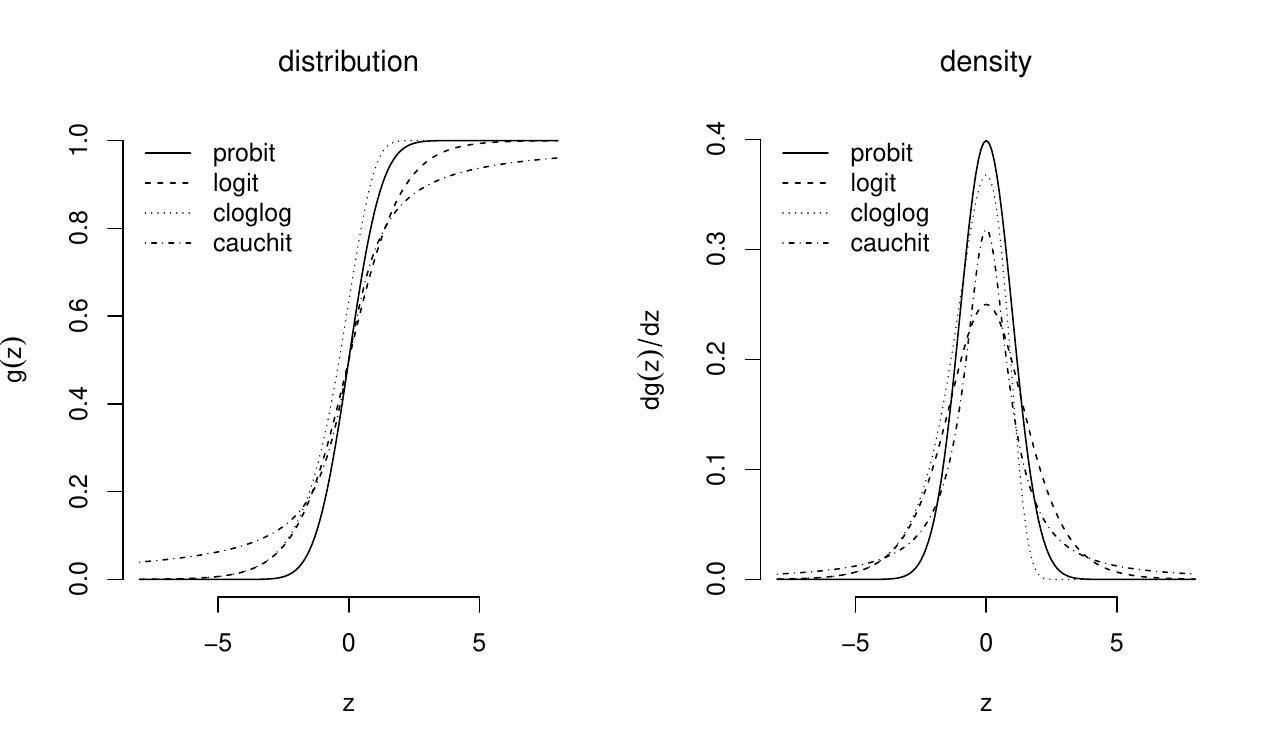}
\caption{Distributions and densities corresponding to the link functions}\label{fig::links-cdf=pdf}
\end{figure}

Figure \ref{fig::links-cdf=pdf} shows the distributions and densities of the corresponding link functions. The distribution functions are quite similar for all links, but the density for cloglog is asymmetric although all other three densities are symmetric.

This chapter will focus on the logit model, and extensions to other models are conceptually straightforward. We can also write the logit model as
\begin{equation}
\pr(y_{i}=1\mid x_{i})\equiv\pi(x_{i},\beta)=\frac{e^{x_{i}^{\T}\beta}}{1+e^{x_{i}^{\T}\beta}},\label{eq:logit-parametric-model}
\end{equation}
for the conditional probability of $y_{i}$ 
given $x_{i}$, 
or, equivalently,
\[
\text{logit}\left\{ \pr(y_{i}=1\mid x_{i})\right\} \equiv\log\frac{\pr(y_{i}=1\mid x_{i})}{1-\pr(y_{i}=1\mid x_{i})}=x_{i}^{\T}\beta,
\]
for the log of the odds of $y_{i}$ 
given $x_{i}$, with the logit function
$$
\text{logit}(\pi) = \log \frac{ \pi }{1 - \pi}. 
$$
Because $y_{i}$ is a binary random variable, its
probability completely determines its distribution. So we can also
write the logit model in the following form:

\begin{assumption}[binary logistic regression model]
\label{assume::logistic}
We have 
\[
y_{i}\mid x_{i}\sim\text{Bernoulli}\left(\frac{e^{x_{i}^{\T}\beta}}{1+e^{x_{i}^{\T}\beta}}\right) . 
\]
The observations are independent across units. The $\beta$ is the unknown parameter. 
\end{assumption}

Each coefficient $\beta_{j}$ measures the impact of $x_{ij}$ on the log odds of the outcome:
 $$
 \frac{ \partial  }{  \partial x_{ij} }  \text{logit} \{ \pr(y_{i}=1\mid   x_i ) \} = \beta_j .
 $$
Epidemiologists also call $\beta_{j}$ the
 conditional
log odds ratio\footnote{In probability theory, if $p$ is the probability, then $p/(1-p)$ is called the odds. It is a terminology from gambling.} because 
\begin{align*}
\beta_{j} & =\text{logit}\left\{ \pr(y_{i}=1\mid  \ldots,x_{ij}+1,\ldots )\right\} -\text{logit}\left\{ \pr(y_{i}=1\mid  \ldots,x_{ij},\ldots )\right\} \\
 & =\log\frac{\pr(y_{i}=1\mid  \ldots,x_{ij}+1 )}{1-\pr(y_{i}=1\mid  \ldots,x_{ij}+1 )} 
 - \log\frac{\pr(y_{i}=1\mid  \ldots,x_{ij},\ldots )}{1-\pr(y_{i}=1\mid  \ldots,x_{ij},\ldots )} \\
 &= \log \left\{
 \frac{\pr(y_{i}=1\mid \ldots,x_{ij}+1,\ldots )}{1-\pr(y_{i}=1\mid  \ldots,x_{ij}+1,\ldots )} 
 \Big/
\frac{\pr(y_{i}=1\mid  \ldots,x_{ij},\ldots )}{1-\pr(y_{i}=1\mid  \ldots,x_{ij},\ldots )} 
 \right\},
\end{align*}
that is, the change of the log odds of $y_{i}$ if we increase $x_{ij}$
by a unit holding other covariates unchanged. 
Qualitatively, if $\beta_j >0$, then larger values of $x_{ij}$ lead to larger probabilities of $y_i = 1$; if $\beta_j < 0$, then larger values of $x_{ij}$ lead to smaller probabilities of $y_i = 1$.

\section{Maximum likelihood estimator of the logistic model}

Because we have specified a fully parametric model for $y_{i}$ given
$x_{i}$, we can estimate $\beta$ using the maximum likelihood. With
independent observations,   the likelihood function for
general binary outcomes is\footnote{The notation can be confusing because $\beta$ denotes both the true parameter and the dummy variable for the likelihood function.
It is somewhat standard in statistics, when we do not want to introduce additional notation. In previous chapters of this book, I use $\beta$ to denote the true parameter in the linear model and $b$ to denote the parameter in the OLS objective function. 
}
\begin{eqnarray*}
L(\beta) &=& 
\prod_{i=1}^{n} f(y_i\mid x_i) \\
&=& 
\prod_{i=1}^{n}   \left\{  \pi(x_{i},\beta) \text{ if } y_i = 1 \text{ or } 1-\pi(x_{i},\beta) \text{ if } y_i = 0  \right\} \\
&=&
\prod_{i=1}^{n}\left\{ \pi(x_{i},\beta)\right\} ^{y_{i}}\left\{ 1-\pi(x_{i},\beta)\right\} ^{1-y_{i}}.
\end{eqnarray*}
Under the logit form (\ref{eq:logit-parametric-model}), the likelihood function simplifies to
\begin{align*}
L(\beta) & =\prod_{i=1}^{n}\left\{ \frac{\pi(x_{i},\beta)}{1-\pi(x_{i},\beta)}\right\} ^{y_{i}}\left\{ 1-\pi(x_{i},\beta)\right\} \\
 & =\prod_{i=1}^{n}\left(e^{x_{i}^{\T}\beta}\right)^{y_{i}}\frac{1}{1+e^{x_{i}^{\T}\beta}}\\
 & =\prod_{i=1}^{n}\frac{e^{y_{i}x_{i}^{\T}\beta}}{1+e^{x_{i}^{\T}\beta}}.
\end{align*}
The log-likelihood function is
\[
\log L(\beta)=\sumn\left\{ y_{i}x_{i}^{\T}\beta-\log(1+e^{x_{i}^{\T}\beta})\right\} ,
\]
the score function is
\begin{align*}
\frac{\partial\log L(\beta)}{\partial\beta} & =\sumn\left(x_{i}y_{i}-\frac{x_{i}e^{x_{i}^{\T}\beta}}{1+e^{x_{i}^{\T}\beta}}\right)\\
 & =\sumn x_{i}\left(y_{i}-\frac{e^{x_{i}^{\T}\beta}}{1+e^{x_{i}^{\T}\beta}}\right)\\
 &= \sumn x_{i}\left\{  y_{i}-  g(x_{i}^{\T} \beta)  \right\} \\
 & =\sumn x_{i}\left\{ y_{i}-\pi(x_{i},\beta)\right\} ,
\end{align*}
and the Hessian matrix
\begin{align*}
\frac{\partial^{2}\log L(\beta)}{\partial\beta\partial\beta^{\T}} & =\left(\frac{\partial^{2}\log L(\beta)}{\partial\beta_{j}\partial\beta_{j'}}\right)_{1\leq j,j'\leq p}\\
 & =-\sumn x_{i}\frac{\partial g(x_{i}^{\T}\beta)}{\partial\beta^{\T}}\\
 & \stackrel{(\ref{eq:densityoflogit})}{=}-\sumn x_{i}x_{i}^{\T}g(x_{i}^{\T}\beta)\left\{ 1-g(x_{i}^{\T}\beta)\right\} \\
 & =-\sumn\pi(x_{i},\beta)\left\{ 1-\pi(x_{i},\beta)\right\} x_{i}x_{i}^{\T}.
\end{align*}
For any $\alpha \in \mathbb{R}^p$, we have
$$
\alpha^{\T}  \frac{\partial^{2}\log L(\beta)}{\partial\beta\partial\beta^{\T}} \alpha
=-\sumn\pi(x_{i},\beta)\left\{ 1-\pi(x_{i},\beta)\right\} (\alpha^{\T} x_{i} )^2 \leq 0
$$
so the Hessian matrix is negative semi-definite.
If it is negative definite, then the likelihood function has a unique maximizer. 

The maximum likelihood estimate (MLE) must satisfy the following score or Normal equation:
\[
\sumn x_{i}\left\{ y_{i}-\pi(x_{i},\hat{\beta})\right\} =\sumn x_{i}\left(y_{i}-\frac{e^{x_{i}^{\T}\hat{\beta}}}{1+e^{x_{i}^{\T}\hat{\beta}}}\right)=0.
\]
If we view $\pi(x_{i},\hat{\beta})$ as the fitted probability
for $y_{i},$ then $y_{i}-\pi(x_{i},\hat{\beta})$ is the residual,
and the score equation is similar to that of OLS. Moreover, if $x_{i}$
contains $1$, then 
\[
\sumn\left\{ y_{i}-\pi(x_{i},\hat{\beta})\right\} =0 ,
\]
which implies
\[ 
n^{-1}\sumn y_{i}=n^{-1}\sumn\pi(x_{i},\hat{\beta}) . 
\]
That is, the average of the outcomes equals the average of their fitted values. 

However, the score equation is nonlinear, and in general, there is
no explicit formula for the MLE. We usually use Newton's method to
solve for the MLE based on the linearization of the score equation. Starting from the old value $\beta^{\text{old}}$, we can approximate the score equation by a linear equation:
\[
0=\frac{\partial\log L( \beta )}{\partial\beta}\cong\frac{\partial\log L(\beta^{\text{old}})}{\partial\beta}+\frac{\partial^{2}\log L(\beta^{\text{old}})}{\partial\beta\partial\beta^{\T}}(\beta-\beta^{\text{old}}),
\]
and then update
\[
\beta^{\text{new}}=\beta^{\text{old}}-\left\{ \frac{\partial^{2}\log L(\beta^{\text{old}})}{\partial\beta\partial\beta^{\T}}\right\} ^{-1}\frac{\partial\log L(\beta^{\text{old}})}{\partial\beta}.
\]
Using the matrix form, we can gain more insight from Newton's method.
Recall that 
$$
Y=\left(\begin{array}{c}
y_{1}\\
\vdots\\
y_{n}
\end{array}\right), \quad  X=\left(\begin{array}{c}
x_{1}^{\T}\\
\vdots\\
x_{n}^{\T}
\end{array}\right),
$$
and define 
$$
\Pi^{\text{old}}=\left(\begin{array}{c}
\pi(x_{1},\beta^{\text{old}})\\
\vdots\\
\pi(x_{n},\beta^{\text{old}})
\end{array}\right), \quad
 W^{\text{old}}=\text{diag}\left[\pi(x_{i},\beta^{\text{old}})\left\{ 1-\pi(x_{i},\beta^{\text{old}})\right\} \right]_{i=1}^{n}.
$$ 
Then 
\begin{align*}
\frac{\partial\log L(\beta^{\text{old}})}{\partial\beta} & =X^{\T}(Y  - \Pi^{\text{old}}),\\
\frac{\partial^{2}\log L(\beta^{\text{old}})}{\partial\beta\partial\beta^{\T}} & =-X^{\T}W^{\text{old}}X,
\end{align*}
and Newton's method simplifies to
\begin{align*}
\beta^{\text{new}} & =\beta^{\text{old}}+(X^{\T}W^{\text{old}}X)^{-1}X^{\T}(Y-\Pi^{\text{old}})\\
 & =(X^{\T}W^{\text{old}}X)^{-1}\left\{ X^{\T}W^{\text{old}}X\beta^{\text{old}}+X^{\T}(Y-\Pi^{\text{old}})\right\} \\
 & =(X^{\T}W^{\text{old}}X)^{-1}X^{\T}W^{\text{old}}Z^{\text{old}},
\end{align*}
where
\[
Z^{\text{old}}=X\beta^{\text{old}}+(W^{\text{old}})^{-1}(Y-\Pi^{\text{old}}).
\]
So we can obtain $\beta^{\text{new}}$ based on the WLS
fit of $Z^{\text{old}}$ on $X$ with weights $W^{\text{old}}$, the diagonal elements of which
are the conditional variances of the $y_{i}$'s given the $x_{i}$'s at $\beta^{\text{old}}$.
The \ri{glm} function in \ri{R} uses the Fisher scoring algorithm, which is
identical to Newton's method for the logit model.\footnote{The Fisher scoring algorithm uses a slightly different approximation:
$$
0=\frac{\partial\log L( \beta )}{\partial\beta}\cong\frac{\partial\log L(\beta^{\text{old}})}{\partial\beta}
+  E\left\{ \frac{\partial^{2}\log L(\beta^{\text{old}})}{\partial\beta\partial\beta^{\T}}  \mid X \right\} (\beta-\beta^{\text{old}}) ,
$$ 
with the expected Fisher information instead of the observed Fisher information. For other link functions, the Fisher scoring algorithm is different from Newton's method. 
} Sometimes, it is
also called the iteratively reweighted least squares algorithm.

\section{Statistics with the logit model}

\subsection{Inference}

Based on the general theory of MLE, $\hat{\beta}$ is consistent for
$\beta$ and is asymptotically Normal. Approximately, we can conduct
statistical inference based on
\[
\hat{\beta}\asim \N\left\{ \beta,\left(-\frac{\partial^{2}\log L(\hat{\beta})}{\partial\beta\partial\beta^{\T}}\right)^{-1}\right\} 
=\N\left\{ \beta, (X^{\T}\hat{W}X)^{-1} \right\},
\]
where 
\[
\hat{W}=\text{diag}\left[\pi(x_{i},\hat{\beta})\left\{ 1-\pi(x_{i},\hat{\beta} )\right\} \right]_{i=1}^{n}.
\]
Based on this, the \ri{glm} function reports the point estimate, standard error, $z$-value, and $p$-value for each coordinate of $\beta$. It is almost identical to the output of the \ri{lm} function, except that the interpretation of the coefficient becomes the conditional log odds ratio.

I use the data from \citet{hirano2000assessing} to illustrate logistic regression, where the main interest is the effect of the encouragement of receiving the flu shot via email on the binary indicator of flu-related hospitalization. We can fit a logistic regression using the \ri{glm} function in \ri{R} with \ri{family = binomial(link = logit)}. 

\begin{lstlisting}
> flu = read.table("fludata.txt", header = TRUE)
> flu = within(flu, rm(receive))
> assign.logit = glm(outcome ~ ., 
+                    family  = binomial(link = logit), 
+                    data    = flu)
> summary(assign.logit)

Call:
glm(formula = outcome ~ ., family = binomial(link = logit), data = flu)

Deviance Residuals: 
    Min       1Q   Median       3Q      Max  
-1.1957  -0.4566  -0.3821  -0.3048   2.6450  

Coefficients:
             Estimate Std. Error z value Pr(>|z|)    
(Intercept) -2.199815   0.408684  -5.383 7.34e-08 ***
assign      -0.197528   0.136235  -1.450  0.14709    
age         -0.007986   0.005569  -1.434  0.15154    
copd         0.337037   0.153939   2.189  0.02857 *  
dm           0.454342   0.143593   3.164  0.00156 ** 
heartd       0.676190   0.153384   4.408 1.04e-05 ***
race        -0.242949   0.143013  -1.699  0.08936 .  
renal        1.519505   0.365973   4.152 3.30e-05 ***
sex         -0.212095   0.144477  -1.468  0.14210    
liverd       0.098957   1.084644   0.091  0.92731    

(Dispersion parameter for binomial family taken to be 1)

    Null deviance: 1667.9  on 2860  degrees of freedom
Residual deviance: 1598.4  on 2851  degrees of freedom
AIC: 1618.4

Number of Fisher Scoring iterations: 5
\end{lstlisting}

Three subtle issues arise in the above code. First, \ri{flu = within(flu, rm(receive))} drops \ri{receive}, which is the indicator of whether a patient received the flu shot or not. The reason is that  \ri{assign} is randomly assigned but \ri{receive} is subject to selection bias, that is, patients receiving the flu shot can be quite different from patients not receiving the flu shot.

Second, the \ri{Null deviance} and \ri{Residual deviance} are defined as 
$
- 2 \log L(\tilde{\beta})
$
and
$
- 2 \log L(\hat{\beta}),
$
 respectively, where $\tilde{\beta}$ is the MLE assuming that all coefficients except the intercept are zero, and $\hat{\beta}$ is the MLE without any restrictions. They are not of independent interest, but their difference is: Wilks' theorem ensures that
 $$
 \{ - 2 \log L(\tilde{\beta}) \}  - \{- 2 \log L(\hat{\beta})\}
 =2 \log \frac{ L(\hat{\beta})  }{ L(\tilde{\beta}) } \asim \chi_{p-1}^2.
 $$
So we can test whether the coefficients of the covariates are all zero, which is analogous to the joint $F$ test in linear models. 

\begin{lstlisting}
> pchisq(assign.logit$null.deviance - assign.logit$deviance,
+        df = assign.logit$df.null - assign.logit$df.residual,
+        lower.tail = FALSE)
[1] 1.912952e-11
\end{lstlisting}

Third, the AIC is defined as $- 2 \log L(\hat{\beta}) + 2p$, where $p$ is the number of parameters in the logit model. This is also the general formula of AIC for other parametric models; recall its form under Normal linear model in Chapter \ref{chapter::aic-bic}.

\subsection{Prediction}

The logit model is often used for prediction or classification since the outcome is binary. With the MLE $\hat{\beta}$, we can predict the probability of being one as $\hat{\pi}_{n+1}  =  g(x_{n+1} ^{\T} \hat{\beta})$ for a unit with covariate value $x_{n+1} $, and we can easily dichotomize the fitted probability to predict the outcome itself by $\hat{y}_{n+1}  =  1(\hat{\pi}_{n+1}  \geq c)$, for example, with $c=0.5$.

We can even quantify the uncertainty in the fitted probability based on a linear approximation (i.e., the delta method in Proposition \ref{prop::delta-method}). Based on
\begin{eqnarray*}
\hat{\pi}_{n+1}   &=&  g(x_{n+1} ^{\T} \hat{\beta}) \\
&\cong&   g(x_{n+1} ^{\T} \beta) +  g'(x_{n+1} ^{\T}  \beta) x_{n+1} ^{\T} (\hat{\beta} - \beta) \\
&=&  g(x_{n+1} ^{\T} \beta) +  g(x_{n+1} ^{\T}  \beta )  \{1-g(x_{n+1} ^{\T}  \beta ) \} x_{n+1} ^{\T} (\hat{\beta} - \beta),
\end{eqnarray*}
we can approximate the asymptotic variance of $\hat{\pi}_{n+1}  $ by
$$
 [ g(x_{n+1} ^{\T}  \beta )  \{1-g(x_{n+1} ^{\T}  \beta ) \}  ]^2 x_{n+1} ^{\T}  (X^{\T}\hat{W}X)^{-1} x_{n+1} . 
$$

We can use the \ri{predict} function in \ri{R} to calculate the predicted values based on a \ri{glm} object in the same way as the linear model. If we specify \ri{type="response"}, then we obtain the fitted probabilities; if we specify \ri{se.fit = TRUE}, then we also obtain the standard errors of the fitted probabilities. In the following, I predict the probabilities of flu-related hospitalization if a patient receives the email encouragement or not, fixing other covariates at their empirical means.

\begin{lstlisting}
> emp.mean = apply(flu, 2, mean)
> data.ave = rbind(emp.mean, emp.mean)
> data.ave[1, 1] = 1
> data.ave[2, 1] = 0
> data.ave = data.frame(data.ave)
> data.ave
           assign    outcome      age      copd        dm    heartd      race
emp.mean        1 0.08528487 65.26949 0.2820692 0.2785739 0.5735757 0.6550157
emp.mean.1      0 0.08528487 65.26949 0.2820692 0.2785739 0.5735757 0.6550157
                renal       sex      liverd
emp.mean   0.01328207 0.6682978 0.003145753
emp.mean.1 0.01328207 0.6682978 0.003145753
> predict(assign.logit, newdata = data.ave,
+         type = "response", se.fit = TRUE)
$fit
  emp.mean emp.mean.1 
0.06981828 0.08378818 

$se.fit
   emp.mean  emp.mean.1 
0.006689665 0.007526307 

$residual.scale
[1] 1
\end{lstlisting}

\section{More on the interpretations of the coefficients}

Many practitioners find the coefficients in the logit model difficult to interpret. Another measure of the impact of the covariate on the outcome is the {\it average marginal effect or average partial effect}.\footnote{Recall its definition in Chapter \ref{sec::main-interaction-centering}.} For a continuous covariate $x_{ij}$, the average marginal effect is defined as 
\begin{eqnarray*}
\textsc{ame}_j 
&=& 
n^{-1}\sumn\frac{\partial\pr(y_{i}=1\mid x_{i})}{\partial x_{ij}} \\
&=& 
n^{-1}\sumn g'(x_{i}^{\T}\beta)\beta_{j},
\end{eqnarray*}
which reduces to the following form for the logit model
\[
\textsc{ame}_j = 
\beta_{j}\times n^{-1}\sumn\pi(x_{i},\beta)\left\{ 1-\pi(x_{i},\beta)\right\} .
\]
For a binary covariate $x_{ij}$, the average marginal effect is defined as 
$$
\textsc{ame}_j = 
n^{-1}\sumn  \{  \pr(y_{i}=1\mid  \ldots, x_{ij}=1, \ldots) - \pr(y_{i}=1\mid  \ldots, x_{ij}=0, \ldots )  \} . 
$$
The \ri{margins} function in the \ri{margins} package can compute the average marginal effects and the corresponding standard errors. In particular, the average marginal effect of \ri{assign} is not significant as shown below.

\begin{lstlisting}
> library("margins")
> ape = margins(assign.logit)
> summary(ape)
 factor     AME     SE       z      p   lower  upper
    age -0.0006 0.0004 -1.4322 0.1521 -0.0014 0.0002
 assign -0.0150 0.0103 -1.4480 0.1476 -0.0352 0.0053
   copd  0.0255 0.0117  2.1830 0.0290  0.0026 0.0485
     dm  0.0344 0.0109  3.1465 0.0017  0.0130 0.0559
 heartd  0.0512 0.0118  4.3441 0.0000  0.0281 0.0743
 liverd  0.0075 0.0822  0.0912 0.9273 -0.1536 0.1686
   race -0.0184 0.0109 -1.6958 0.0899 -0.0397 0.0029
  renal  0.1151 0.0278  4.1461 0.0000  0.0607 0.1696
    sex -0.0161 0.0110 -1.4660 0.1426 -0.0376 0.0054
\end{lstlisting}

The interaction term is much more complicated. Contradictory suggestions are given across fields. 
Consider
the following model
\[
\pr(y_{i}=1\mid x_{i1},x_{i2})=g(\beta_{0}+\beta_{1}x_{i1}+\beta_{2}x_{i2}+\beta_{12}x_{i1}x_{i2}).
\]
If the link is logit, then epidemiologists interpret $e^{\beta_{12}}$ as the interaction between $x_{i1}$ and $x_{i2}$ on the odds ratio scale. Consider a simple case with binary $x_{i1}$ and $x_{i2}$. Given $x_{i2} = 1$, the odds ratio of $x_{i1}$ on $y_i$ equals $e^{\beta_1 + \beta_{12}}$; given $x_{i2} = 0$, the odds ratio of $x_{i1}$ on $y_i$ equals $e^{\beta_1}$. Therefore, the ratio of the two odds ratio equals $e^{\beta_{12}}$.  When we measure effects on the odds ratio scale, the logistic model is a natural choice. The interaction term in the logistic model indeed measures the interaction of $x_{i1}$ and $x_{i2}$. 

\citet{ai2003interaction} gave a different suggestion. 
We have two ways to define the interaction effect: first,
\[
n^{-1}\sumn\frac{\partial\pr(y_{i}=1\mid x_{i1},x_{i2})}{\partial(x_{i1}x_{i2})}
=n^{-1}\sumn g'(  \beta_{0}+\beta_{1}x_{i1}+\beta_{2}x_{i2}+\beta_{12}x_{i1}x_{i2} )\beta_{12}; 
\]
second,
\begin{align*}
& n^{-1}\sumn\frac{\partial^{2}\pr(y_{i}=1\mid x_{i1},x_{i2})}{\partial x_{i1}\partial x_{i2}} \\
& =n^{-1}\sumn\frac{\partial}{\partial x_{i2}}\left\{ \frac{\partial\pr(y_{i}=1\mid x_{i1},x_{i2})}{\partial x_{i1}}\right\} \\
 & =n^{-1}\sumn\frac{\partial}{\partial x_{i2}}\left\{ g'(\beta_{0}+\beta_{1}x_{i1}+\beta_{2}x_{i2}+\beta_{12}x_{i1}x_{i2})(\beta_{1}+\beta_{12}x_{i2})\right\} \\
 & =n^{-1}\sumn\left\{ g''( \beta_{0}+\beta_{1}x_{i1}+\beta_{2}x_{i2}+\beta_{12}x_{i1}x_{i2} )(\beta_{2}+\beta_{12}x_{i1})(\beta_{1}+\beta_{12}x_{i2})+g'(z_i)\beta_{12}\right\} .
\end{align*}
Although the first one is more straightforward based on the definition of the average partial effect, the second one is more reasonable based on the natural definition of interaction based on the mixed derivative.  
Note that even if $\beta_{12} = 0$, the second definition of interaction 
does not necessarily equal 0 since
\begin{align*}
 n^{-1}\sumn\frac{\partial^{2}\pr(y_{i}=1\mid x_{i1},x_{i2})}{\partial x_{i1}\partial x_{i2}}  
 =n^{-1}\sumn   g''( \beta_{0}+\beta_{1}x_{i1}+\beta_{2}x_{i2} ) \beta_{1}  \beta_{2} . 
\end{align*}
 This is due to the nonlinearity of the link function. 
The second definition quantifies interaction based on the probability itself, whereas the parameter $\beta_{12}$ in the logistic model measures the interaction on the odds ratio scale. 

\section{Does the link function matter?}\label{chapter::link-matter}

First, I generate data from a simple one-dimensional logistic model.

\begin{lstlisting}
> n  = 100
> x  = rnorm(n, 0, 3)
> prob = 1/(1 + exp(-1 + x))
> y  = rbinom(n, 1, prob)
\end{lstlisting}

Then I fit the data with the linear probability model and binary models with four link functions. 
\begin{lstlisting}
> lpmfit    = lm(y ~ x)
> probitfit = glm(y ~ x, family = binomial(link = "probit"))
Warning message:
glm.fit: fitted probabilities numerically 0 or 1 occurred 
> logitfit  = glm(y ~ x, family = binomial(link = "logit"))
> cloglogfit= glm(y ~ x, family = binomial(link = "cloglog"))
Warning message:
glm.fit: fitted probabilities numerically 0 or 1 occurred 
> cauchitfit= glm(y ~ x, family = binomial(link = "cauchit"))
\end{lstlisting}

The coefficients are quite different because the coefficients measure the association between $x$ and $y$ on difference scales. These parameters are not directly comparable.  
Nevertheless, the signs of the coefficients are all negative. 

\begin{lstlisting}
> betacoef = c(lpmfit$coef[2], 
+              probitfit$coef[2], 
+              logitfit$coef[2], 
+              cloglogfit$coef[2], 
+              cauchitfit$coef[2])
> names(betacoef) = c("lpm", "probit", "logit", "cloglog", "cauchit")
> round(betacoef, 2)
    lpm  probit   logit cloglog cauchit 
  -0.10   -0.83   -1.47   -1.07   -2.09
\end{lstlisting}

However, if we care only about the prediction, then these five models give very similar results. 

\begin{lstlisting}
> table(y, lpmfit$fitted.values>0.5)
   
y   FALSE TRUE
  0    31    9
  1     5   55
> table(y, probitfit$fitted.values>0.5)
   
y   FALSE TRUE
  0    31    9
  1     5   55
> table(y, logitfit$fitted.values>0.5)
   
y   FALSE TRUE
  0    31    9
  1     5   55
> table(y, cloglogfit$fitted.values>0.5)
   
y   FALSE TRUE
  0    34    6
  1     7   53
> table(y, cauchitfit$fitted.values>0.5)
   
y   FALSE TRUE
  0    34    6
  1     7   53
  \end{lstlisting}

\begin{figure}
\centering
\includegraphics[width = \textwidth]{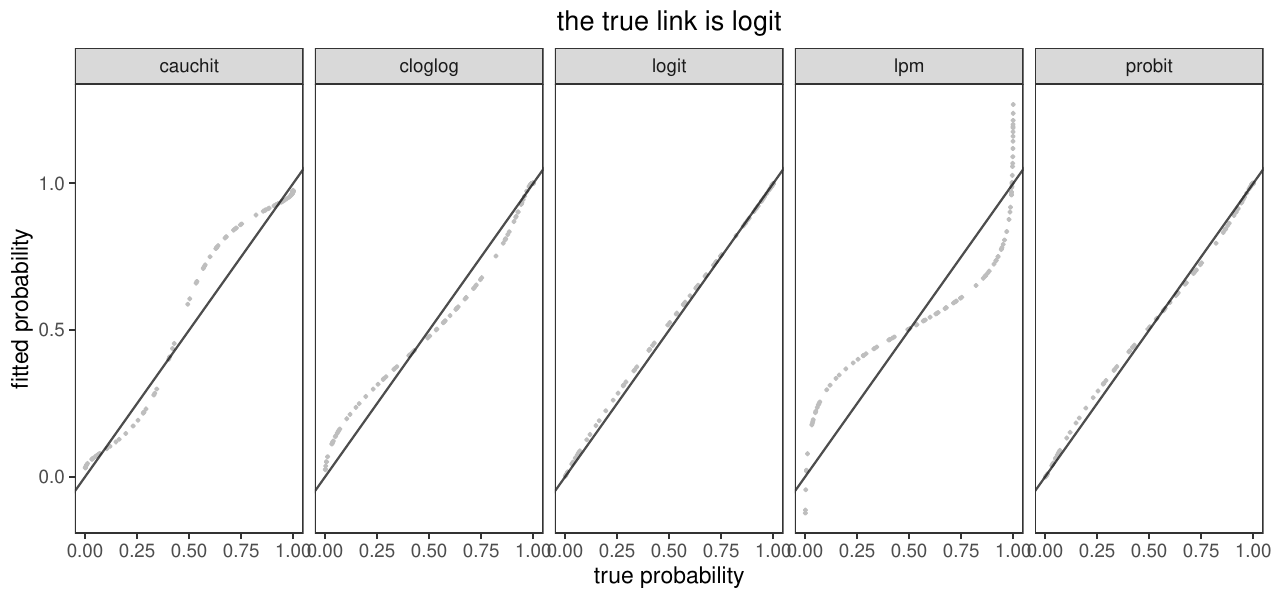}
\caption{Comparing the fitted probabilities from different link functions}\label{fig::compare-links}
\end{figure}

Figure \ref{fig::compare-links} shows the fitted probabilities versus the true probabilities $\pr(y_i =1\mid x_i)$. The patterns are quite similar although the linear probability model can give fitted probabilities outside $[0,1]$. When we use the cutoff point $0.5$ to predict the binary outcome, the problem of the linear probability model becomes rather minor.

An interesting fact is that the coefficients from the logit model approximately equal those from the probit model multiplied by 1.7, a constant that minimizes $\max_y | g_{\text{logit}}(by) - g_{\text{probit}}(y) |$. We can easily compute this constant numerically:
\begin{lstlisting}
> d.logit.probit = function(b){
+   x = seq(-20, 20, 0.00001)
+   max(abs(plogis(b*x) - pnorm(x)))
+ }
> optimize(d.logit.probit, c(-10, 10))
$minimum
[1] 1.701743

$objective
[1] 0.009457425

> 
> d.logit.cauchit = function(b){
+   x = seq(-20, 20, 0.00001)
+   max(abs(plogis(b*x) - pcauchy(x)))
+ }
> optimize(d.logit.cauchit, c(-10, 10))
$minimum
[1] 0.8590545

$objective
[1] 0.04945328

> 
> f.cloglog = function(z){
+   1 - exp(-exp(z))  
+ }
> d.logit.cloglog = function(b){
+   x = seq(-20, 20, 0.00001)
+   max(abs(plogis(b*x) - f.cloglog(x)))
+ }
> optimize(d.logit.cloglog, c(-10, 10))
$minimum
[1] 1.47175

$objective
[1] 0.1321229
\end{lstlisting} 
Based on the above calculation, the maximum difference is approximately 0.009. Therefore, the logit and probit link functions are extremely close up to the scaling factor 1.7.\footnote{See Problem \ref{hw17::logistic-moments} for a related heuristic argument. } However, $\min_b \max_y | g_{\text{logit}}(by) - g_{ * }(y) |$ is much larger for the link functions of cauchit and cloglog.

\section{Extensions of the logistic regression}

\subsection{Penalized logistic regression}

Similar to the high dimensional linear model, we can also extend the logit model to a penalized version. Since the objective function for the original logit model is the log-likelihood, we can minimize the following penalized log-likelihood function: 
\[
\arg\min_{\beta_0,\beta_1, \ldots, \beta_p}-\frac{1}{n}\sumn \ell_i(\beta )   +\lambda\sum_{j=1}^{p} \{  \alpha \beta_j^2 + (1-\alpha)  |\beta_{j}| \} ,
\]
where 
$$
\ell_i(\beta) = y_{i}(\beta_{0}+\beta_{1}x_{i1}+\cdots+\beta_{p}x_{ip})-\log(1+e^{\beta_{0}+\beta_{1}x_{i1}+\cdots+\beta_{p}x_{ip}})
$$ 
is the log-likelihood function based on the $i$the observation. When $\alpha = 1$, it gives the ridge analog of the logistic regression; when $\alpha = 0$, it gives the lasso analog; when $\alpha \in (0,1)$, it gives the elastic net analog. 
The \ri{R} package \ri{glmpath} uses the coordinate descent algorithm based on a quadratic
approximation of the log-likelihood function. We can select the tuning parameter $\lambda$ based on cross-validation. 

\subsection{Case-control study}\label{sec::case-control-study}

A nice property of the logit model is that it works not only for  the cohort study with data from conditional distribution $y_i\mid x_i$ but also for the case-control study with data from the conditional distribution $x_i\mid y_i$.\footnote{
\citet{breslow1996statistics} provided a scholarly review of the statistics of the case-control study in epidemiology.
} The former is a prospective study while the latter is a retrospective study. Below, I will explain the basic idea in \citet{prentice1979logistic}.

Assume that $(x_{i},y_{i},s_{i})_{i=1}^n $ are IID with 
\begin{eqnarray}
\label{eq::cc=1}
\pr(y_i = 1\mid x_i) =  \frac{e^{\beta_{0}+x_{i}^{\T}\beta}}{1+e^{\beta_{0}+x_{i}^{\T}\beta}} 
\end{eqnarray}
and
\begin{eqnarray}
\label{eq::cc=2}
\pr(s_{i}=1\mid x_{i},y_{i})=\pr(s_{i}=1\mid y_{i})=\begin{cases}
p_{1}, & \text{if }y_{i}=1,\\
p_{0}, & \text{if }y_{i}=0.
\end{cases}
\end{eqnarray}
But we only have data with $s_{i}=1$ with $p_{1}$ and $p_{0}$ often
unknown. Fortunately, conditioning on $s_{i}=1$, we have the following result. 

\begin{theorem}
\label{thm::case-control}
Under \eqref{eq::cc=1} and \eqref{eq::cc=2}, we have 
$$
\pr(y_{i}  =1\mid x_{i},s_{i}=1)
= \frac{e^{\delta+\beta_{0}+x_{i}^{\T}\beta}}{1+e^{\delta+\beta_{0}+x_{i}^{\T}\beta}}, 
$$
where $\delta=\log(p_{1}/p_{0})$. 
\end{theorem}

\begin{myproof}{Theorem}{\ref{thm::case-control}}
We have 
\begin{align*}
&\pr(y_{i}  =1\mid x_{i},s_{i}=1)\\
 & =\frac{\pr(y_{i}=1\mid x_{i})\pr(s_{i}=1\mid x_{i},y_{i}=1)}{\pr(y_{i}=1\mid x_{i})\pr(s_{i}=1\mid x_{i},y_{i}=1)+\pr(y_{i}=0\mid x_{i})\pr(s_{i}=1\mid x_{i},y_{i}=0)}
 \end{align*}
 by Bayes' formula. Under the logit model, we have
 \begin{align*}
\pr(y_{i}  =1\mid x_{i},s_{i}=1)
 & =\frac{\frac{e^{\beta_{0}+x_{i}^{\T}\beta}}{1+e^{\beta_{0}+x_{i}^{\T}\beta}}p_{1}}{\frac{e^{\beta_{0}+x_{i}^{\T}\beta}}{1+e^{\beta_{0}+x_{i}^{\T}\beta}}p_{1}+\frac{1}{1+e^{\beta_{0}+x_{i}^{\T}\beta}}p_{0}}\\
 & =\frac{e^{\beta_{0}+x_{i}^{\T}\beta}p_{1}}{e^{\beta_{0}+x_{i}^{\T}\beta}p_{1}+p_{0}}\\
 & =\frac{e^{\beta_{0}+x_{i}^{\T}\beta}p_{1}/p_{0}}{e^{\beta_{0}+x_{i}^{\T}\beta}p_{1}/p_{0}+1}\\
 & =\frac{e^{\delta+\beta_{0}+x_{i}^{\T}\beta}}{1+e^{\delta+\beta_{0}+x_{i}^{\T}\beta}} . 
\end{align*}
\end{myproof}

Theorem \ref{thm::case-control} ensures that conditioning on $s_{i}=1$, the model of $y_{i}$ given $x_{i}$ is still logit with the intercept changing from $\beta_{0}$ to $\beta_{0}+\log(p_{1}/p_{0})$. Although we cannot consistently estimate the intercept without knowing $(p_{1},p_{0})$, we can still estimate all the slopes consistently. \citet{kagan2001note} showed that the logistic link is the only one that enjoys this property.

I will end this subsubsection with a case study. 
\citet{samarani2019activating} hypothesized that variation in the inherited activating Killer-cell Immunoglobulin-like Receptor genes in humans is associated with their innate susceptibility/resistance to developing Crohn's disease. They used a case-control study from three cities (Manitoba,  Montreal, and Ottawa) in Canada to investigate the potential association. The following logistic regression uses all the data. 
Problem \ref{hw17::case-control-data-analysis} requires separate analyses based on \ri{center}.

\begin{lstlisting}
> dat  = read.csv("samarani.csv")
> pool.glm = glm(case_comb ~ ds1 + ds2 + ds3 + ds4_a + 
+                  ds4_b + ds5 + ds1_3 + center,
+                family = binomial(link = logit),
+                data = dat)
> summary(pool.glm)

Call:
glm(formula = case_comb ~ ds1 + ds2 + ds3 + ds4_a + ds4_b + ds5 + 
    ds1_3 + center, family = binomial(link = logit), data = dat)

Deviance Residuals: 
    Min       1Q   Median       3Q      Max  
-1.9982  -0.9274  -0.5291   1.0113   2.2289  

Coefficients:
               Estimate Std. Error z value Pr(>|z|)    
(Intercept)    -2.39681    0.21768 -11.011  < 2e-16 ***
ds1             0.55945    0.14437   3.875 0.000107 ***
ds2             0.42531    0.14758   2.882 0.003954 ** 
ds3             0.81377    0.14503   5.611 2.01e-08 ***
ds4_a           0.30270    0.30679   0.987 0.323802    
ds4_b           0.29199    0.17726   1.647 0.099511 .  
ds5             0.92049    0.14852   6.198 5.72e-10 ***
ds1_3           0.49982    0.14706   3.399 0.000677 ***
centerMontreal -0.05816    0.15889  -0.366 0.714316    
centerOttawa    0.14164    0.20251   0.699 0.484292    

(Dispersion parameter for binomial family taken to be 1)

    Null deviance: 1403.7  on 1020  degrees of freedom
Residual deviance: 1192.0  on 1011  degrees of freedom
AIC: 1212

Number of Fisher Scoring iterations: 3
\end{lstlisting}

\section{Other model formulations}

\subsection{Latent linear model}\label{sec::binary-reg-latent}

Let $y_{i}=1(y_{i}^{*}\geq0)$ where

\[
y_{i}^{*}=x_{i}^{\T}\beta + \varepsilon_{i}
\]
and $ - \varepsilon_{i}$ has distribution function $g(\cdot)$ and is
independent of $x_{i}$. From this latent linear model, we can verify
that 
\begin{align*}
\pr(y_{i}  =1\mid x_{i}) &=\pr(y_{i}^{*}\geq0\mid x_{i})\\
 & =\pr(x_{i}^{\T}\beta + \varepsilon_{i}\geq0\mid x_{i})\\
 & =\pr( - \varepsilon_{i}\leq x_{i}^{\T}\beta\mid x_{i})\\
 & =g(x_{i}^{\T}\beta).
\end{align*}
So the $g(\cdot)$ function can be interpreted as the distribution
function of the error term in the latent linear model.

This latent variable formulation provides another way to interpret the coefficients in the models for binary data. It is a powerful way to generate models for more complex data. We will see another example in the next chapter.

\subsection{Inverse model}\label{subsec:Inverse-model-gaussian-mixture}

Assume that 
\begin{eqnarray}
y_{i}\sim\text{Bernoulli}(q),
\label{eq::y-marginal}
\end{eqnarray}
and
\begin{eqnarray}
x_{i}\mid y_{i}=1\sim\N(\mu_{1},\Sigma),\quad x_{i}\mid y_{i}=0\sim\N(\mu_{0},\Sigma),
\label{eq::xgiveny}
\end{eqnarray} 
where $x_i$ does not contain $1$.
This is called the linear discriminant model. 
We can verify that $y_{i}\mid x_{i}$ follows a logit model as shown in Theorem \ref{thm::inverse-logit} below \citep{cornfield1961quantal, cornfield1962joint}.

\begin{theorem}
\label{thm::inverse-logit}
Under \eqref{eq::y-marginal} and \eqref{eq::xgiveny}, we have
$$
\textup{logit}\{  \pr(y_i =1\mid x_i) \} = \alpha + x_i^{\T} \beta, 
$$
where
\begin{eqnarray}
\alpha &=& \log\frac{q}{1-q}-\frac{1}{2}\left(\mu_{1}^{\T}\Sigma^{-1}\mu_{1}-\mu_{0}^{\T}\Sigma^{-1}\mu_{0}\right)  ,\\ 
\beta  &=& \Sigma^{-1}\left(\mu_{1}-\mu_{0}\right) . \label{eq::alpha-beta-lda}
\end{eqnarray}
\end{theorem}

\begin{myproof}{Theorem}{\ref{thm::inverse-logit}}
Using Bayes' formula, we have
\begin{align*}
\pr(y_{i}  =1\mid x_{i}) &=\frac{\pr(y_{i}=1,x_{i})}{\pr(x_{i})}\\
 & =\frac{\pr(y_{i}=1)\pr(x_{i}\mid y_{i}=1)}{\pr(y_{i}=1)\pr(x_{i}\mid y_{i}=1)+\pr(y_{i}=0)\pr(x_{i}\mid y_{i}=0)}\\
 &= \frac{ e^\Delta}{1+ e^\Delta},
\end{align*}
%
where
\begin{align*}
\Delta 
&= \log  \frac{\pr(y_{i}=1)\pr(x_{i}\mid y_{i}=1)}{ \pr(y_{i}=0)\pr(x_{i}\mid y_{i}=0) } \\
&=\log \frac{q\left\{ (2\pi)^p \text{det}(\Sigma)\right\} ^{-1/2}\exp\left\{ -(x_{i}-\mu_{1})^{\T}\Sigma^{-1}(x_{i}-\mu_{1})/2\right\}}
{  (1-q)\left\{ (2\pi)^p \text{det}(\Sigma)\right\} ^{-1/2}\exp\left\{ -(x_{i}-\mu_{0})^{\T}\Sigma^{-1}(x_{i}-\mu_{0})/2\right\} }\\
&=\log \frac{q\exp\left\{ -(x_{i}-\mu_{1})^{\T}\Sigma^{-1}(x_{i}-\mu_{1})/2\right\}}
{  (1-q) \exp\left\{ -(x_{i}-\mu_{0})^{\T}\Sigma^{-1}(x_{i}-\mu_{0})/2\right\}  }\\
& =\log\frac{q\exp\left\{ -\left(-2x_{i}^{\T}\Sigma^{-1}\mu_{1}+\mu_{1}^{\T}\Sigma^{-1}\mu_{1}\right)/2\right\} }{(1-q)\exp\left\{ -\left(-2x_{i}^{\T}\Sigma^{-1}\mu_{0}+\mu_{0}^{\T}\Sigma^{-1}\mu_{0}\right)/2\right\} }\\
 & = \log\frac{q}{1-q}-\frac{1}{2}\left(\mu_{1}^{\T}\Sigma^{-1}\mu_{1}-\mu_{0}^{\T}\Sigma^{-1}\mu_{0}\right)+x_{i}^{\T}\Sigma^{-1}\left(\mu_{1}-\mu_{0}\right) . 
\end{align*}
So $y_{i}\mid x_{i}$ follows a logistic model with
$\alpha$ and $\beta$ given in \eqref{eq::alpha-beta-lda}. 
\end{myproof}

We can easily obtain the moment estimators for the unknown parameters under \eqref{eq::y-marginal} and \eqref{eq::xgiveny}. Let $n_1 = \sumn y_i$ and $n_0 = n-n_1$. The moment estimator for $q$ is 
$
\hat{q} = n_1/n,
$
the sample mean of the $y_i$'s. The moment estimators for $\mu_1$ and $\mu_0$ are
\begin{eqnarray*}
\hat{\mu}_1 &=& n_1^{-1} \sumn  y_i x_i,\\ 
\hat{\mu}_0 &=& n_0^{-1} \sumn (1-y_i) x_i, 
\end{eqnarray*}
the sample means of the $x_i$'s for units with $y_i=1$ and $y_i=0$, respectively. The moment estimator for $\Sigma$ is 
$$
\hat{\Sigma} = \left[ 
\sumn y_i (x_i -\hat{\mu}_1  )(x_i-\hat{\mu}_1 )^{\T} +  \sumn (1-y_i) (x_i -\hat{\mu}_0  )(x_i-\hat{\mu}_0 )^{\T}
\right] /(n-2) ,
$$
the pooled covariance matrix, after centering the $x_i$'s by the $y$-specific means. 
Based on Theorem \ref{thm::inverse-logit}, we can obtain estimates $\hat{\alpha}$ and $\hat{\beta}$ by replacing the true parameters with their moment estimators. This gives us another way to fit the logistic model, which does not require iteration.

\citet{efron1975efficiency} compared the above moment estimator and the MLE derived from the logistic model. Since the linear discriminant model imposes stronger assumptions, the estimator based on Theorem \ref{thm::inverse-logit} is more efficient if the model is correct. In contrast, the MLE derived from the logistic model is more robust because it does not impose the Normality assumption on $x_i$.

\section{Homework problems}

\paragraph{Moments of the logistic distribution}
\label{hw17::logistic-moments}

Theorem \ref{thm::logistic-moments} states the first two moments of the logistic distribution. Prove Theorem \ref{thm::logistic-moments}.

\begin{theorem}\label{thm::logistic-moments}
Assume $\varepsilon$ is the logistic distribution with CDF
$$
\pr(\varepsilon \leq z) = \frac{e^z}{1 + e^z}.
$$
Then $\varepsilon$ has mean $E(\varepsilon) = 0$ and variance
$$
\var(\varepsilon) = \frac{\pi^2}{3}. 
$$
\end{theorem}

Remark: Theorem \ref{thm::logistic-moments} appeared in \citet{decani1986note}. 
The variance calculation involves non-trivial integrals. 
You may use Euler's formula $\sum_{k=1}^{\infty} \frac{1}{k^2} = \frac{\pi^2}{6}.$ For most readers, you can ignore this problem unless you want to review calculus.

An interesting implication of the variance is that with the same data, the logistic regression coefficients are about $\sqrt{\pi^2/3} \approx 1.8$ times the probit regression coefficients. Based on the latent linear model representations of the logistic and probit models,  the coefficients are only comparable when the corresponding $\varepsilon_i$ has the same variance. Of course, the logistic distribution and standard Normal distribution differ in many other ways. The calculation based on the square root of the variance ratio is only an approximation. Based on another heuristic argument, Chapter \ref{chapter::link-matter} derived $1.7$ as the approximated ratio.

\paragraph{Invariance of logistic regression}
\label{hw17::invariance-logistic}

This problem extends Problem \ref{hw3::invariance-ols}. 

Assume $\tilde{x}_i^{\T} = x_i^{\T} \Gamma$ with an invertible $\Gamma$. Run logistic regression of $y_i$'s on $x_i$'s to obtain the coefficient $\hat\beta$ and fitted probabilities $\hat\pi_i$'s. Run another logistic regression of $y_i$'s on $\tilde x_i$'s to obtain the coefficient $\tilde \beta$ and fitted probabilities $\tilde\pi_i$'s.

Prove that $\hat\beta = \Gamma \tilde \beta$ and $\hat\pi_i = \tilde\pi_i$ for all $i$'s.

\paragraph{Two logistic regressions}\label{hw17::2logistic}

This is an extension of Problem \ref{hw15interaction::2ols}. 

Given data $(x_i, z_i, y_i)_{i=1}^n$ where $x_i$ denotes the covariates, $z_i \in \{1,0 \}$ denotes the binary group indicator, and $y_i$ denotes the binary outcome. We can fit two separate logistic regressions:
$$
\text{logit}\{  \pr(y_i=1\mid z_i =1, x_i) \}=  \gamma_1 + x_i^{\T} \beta_1
$$
and
$$
\text{logit}\{  \pr(y_i=1\mid z_i =0, x_i) \}=  \gamma_0 + x_i^{\T} \beta_0
$$
with the treated and control data, respectively. We can also fit a joint logistic regression using the pooled data:
$$
\text{logit}\{  \pr(y_i=1\mid z_i , x_i) \} = \alpha_0 + \alpha_z z_i + x_i^{\T} \alpha_x  +  z_i x_i^{\T} \alpha_{zx}.
$$
Let the parameters with hats denote the MLEs, for example, $\hat{\gamma}_1$ is the MLE for $\gamma_1$. Find $(\hat{\alpha}_0, \hat{\alpha}_z, \hat{\alpha}_x, \hat{\alpha}_{zx})$ in terms of $ (  \hat{\gamma}_1, \hat{\beta}_1, \hat{\gamma}_0, \hat{\beta}_0 ).$

\paragraph{Likelihood for probit model}\label{hw17::probit-mle}

Write down the likelihood function for the probit model, and derive the steps for Newton's method and Fisher scoring for computing the MLE. How do we estimate the asymptotic covariance matrix of the MLE?

\paragraph{Logit and general exponential family}\label{hw17::logit-lda-exponential}
\citet{efron1975efficiency} pointed out an extension of Theorem \ref{thm::inverse-logit}. Prove Theorem \ref{thm::efron-logit} below and find the formulas of $ \alpha $ and $\beta$ in terms of $(\theta_1,\theta_0,\eta)$.

\begin{theorem}
\label{thm::efron-logit}
Under \eqref{eq::y-marginal} and 
$$
f(x_i\mid y_i = y) = g(\theta_y, \eta) h(x_i,\eta) \exp(x_i^{\T} \theta_y) ,\quad (y=0,1)
$$
with parameters $(\theta_1,\theta_0,\eta)$, 
we have
$$
\textup{logit}\{  \pr(y_i =1\mid x_i) \} = \alpha + x_i^{\T} \beta
$$
for some $\alpha$ and $\beta$. 
\end{theorem}

Remark: As a sanity check, you can compare this problem with Theorem \ref{thm::inverse-logit}.

\paragraph{Empirical comparison of logistic regression and linear discriminant analysis}

This problem is related to Chapter \ref{subsec:Inverse-model-gaussian-mixture}. 

Compare the performance of logistic regression and linear discriminant analysis in terms of prediction accuracy. You should simulate at least three cases: 
\begin{enumerate}[label=(C\arabic*), ref=C\arabic*]
\item
the model for linear discriminant analysis is correct;
\item
the model for linear discriminant analysis is incorrect but the model for logistic regression is correct;
\item
the model for logistic regression is incorrect. 
\end{enumerate}

\paragraph{Quadratic discriminant analysis}\label{hw17::QDA}

Theorem \ref{thm::inverse-logit-qda} below extends Theorem \ref{thm::inverse-logit}. Prove Theorem \ref{thm::inverse-logit-qda}.

\begin{theorem}
\label{thm::inverse-logit-qda}
Assume that 
$$
y_{i}\sim\textup{Bernoulli}(q),
$$
and
$$
x_{i}\mid y_{i}=1\sim\N(\mu_{1},\Sigma_1),\quad x_{i}\mid y_{i}=0\sim\N(\mu_{0},\Sigma_0),
$$ 
where $x_i \in \mathbb{R}^p$ does not contain $1$.

We have that
$$
 \textup{logit}\{  \pr(y_i =1\mid x_i) \} = \alpha + x_i^{\T} \beta + x_i^{\T} \Lambda x_i , 
$$
where
\begin{eqnarray*}
\alpha &=&  \log \frac{q}{1-q} - \frac{1}{2} \log \frac{  \text{det} (\Sigma_1) }{  \text{det} (\Sigma_0) }
- \frac{1}{2} ( \mu_{1}^{\T}\Sigma_1^{-1}\mu_{1}-\mu_{0}^{\T}\Sigma_0^{-1}\mu_{0}  ),\\
\beta &=& \Sigma^{-1}_1 \mu_{1}- \Sigma^{-1}_0\mu_{0} , \\
\Lambda &=& - \frac{1}{2} ( \Sigma^{-1}_1 - \Sigma^{-1}_0 ).
\end{eqnarray*} 
\end{theorem}

Remark: Theorem \ref{thm::inverse-logit-qda} extends the linear discriminant model in Section \ref{subsec:Inverse-model-gaussian-mixture} to the quadratic discriminant model by allowing for heteroskedasticity in the conditional Normality of $x$ given $y$ \citep{cornfield1962joint}. It implies the logistic model with the linear, quadratic, and interaction terms of the basic covariates.

\paragraph{Data analysis}\label{hw17::case-control-data-analysis}

Reanalyze the data in Chapter \ref{sec::case-control-study}, stratifying the analysis based on \ri{center}. Do the results vary significantly across centers?

\paragraph{$R^2$ in logistic regression}\label{hw17::r2-logistic}

Recall that $R^2$ in the linear model measures the linear dependence of the outcome on the covariates. 
However, the definition of $R^2$ is not obvious in the logistic model. The \ri{glm} function in \ri{R} does not return any $R^2$ for the logistic regression. 

Recall the following equivalent definitions of $R^2$ in the linear model
\begin{eqnarray*}
R^2 &=& \frac{  \sumn (\hat{y}_i - \bar{y})^2  }{   \sumn (y_i - \bar{y})^2  } \\
&=& 1 - \frac{  \sumn (y_i - \hat{y}_i )^2  }{   \sumn (y_i - \bar{y})^2  } \\
&=& \hat{\rho}^2_{y\hat{y}} = \frac{  \left(  \sumn (y_i - \bar{y}) (\hat{y}_i - \bar{y})  \right)^2 }
{ \sumn (y_i - \bar{y})^2  \sumn (\hat{y}_i - \bar{y})^2  } . 
\end{eqnarray*}
The fitted values are $\hat{\pi}_i = \pi(x_i, \hat{\beta})$ in the logistic model, which have mean $\bar{y}$ with the intercept included in the model. Analogously, we can define $R^2$ in the logistic model as
\begin{eqnarray*}
R^2_{\text{model}} &=&  \frac{  \textsc{ss}_\textsc{m}  }{   \textsc{ss}_\textsc{t}  } , \\  
R^2_{\text{residual}} &=& 1 - \frac{  \textsc{ss}_\textsc{r}   }{   \textsc{ss}_\textsc{t}  } , \\ 
R^2_{\text{correlation}} &=&  \hat{\rho}^2_{y\hat{\pi}} = \frac{  C_{y\hat{\pi}}^2  }
{  \textsc{ss}_\textsc{t}    \textsc{ss}_\textsc{m} }  , 
\end{eqnarray*}
where
\begin{eqnarray*}
\textsc{ss}_\textsc{t} &=&  \sumn (y_i - \bar{y})^2,\\ 
\textsc{ss}_\textsc{m} &=& \sumn (\hat{\pi}_i - \bar{y})^2,\\ 
\textsc{ss}_\textsc{r} &=&   \sumn (y_i - \hat{\pi}_i )^2,\\ 
C_{y\hat{\pi}} &=& \sumn (y_i - \bar{y}) (\hat{\pi}_i  - \bar{y}).
\end{eqnarray*}
These three definitions are not equivalent in general. In particular, $R^2_{\text{model}} $ differs from $R^2_{\text{residual}} $ since
$$
\textsc{ss}_\textsc{t}  = \textsc{ss}_\textsc{m}  + \textsc{ss}_\textsc{r}  + 2 C_{\hat{\varepsilon} \hat{\pi}}
$$
where 
$$
 C_{\hat{\varepsilon} \hat{\pi}} = \sumn (y_i - \hat{\pi}_i )  (\hat{\pi}_i - \bar{y}). 
$$

\begin{enumerate}
\item
Prove that $R^2_{\text{model}} \geq 0, R^2_{\text{correlation}}  \geq 0$ with equality holding if  $\hat{\pi}_i = \bar{y}$ for all $i$. Prove that $R^2_{\text{model}}  \leq 1, R^2_{\text{residual}}  \leq 1, R^2_{\text{correlation}}  \leq 1$ with equality holding if  $y_i =\hat{\pi}_i$ for all $i$. 

Note that $R^2_{\text{residual}}$ may be negative. Give an example. 

\item
Define 
$$
\bar{ \hat{\pi} }_1 = \frac{ \sumn y_i  \hat{\pi}_i  }{ \sumn y_i},\quad
\bar{ \hat{\pi} }_0 = \frac{ \sumn (1-y_i)  \hat{\pi}_i  }{ \sumn (1-y_i)}
$$
as the average of the fitted values for units with $y_i=1$ and $y_i=0$, respectively. Define
$$
D = \bar{ \hat{\pi} }_1  - \bar{ \hat{\pi} }_0.
$$
Prove that 
$$
D = (R^2_{\text{model}} +R^2_{\text{residual}}  )/2 = \sqrt{  R^2_{\text{model}}  R^2_{\text{correlation}}  }.
$$
Further prove that $D\geq 0$ with equality holding if   $\hat{\pi}_i = \bar{y}$ for all $i$, and $D\leq 1$ with equality holding if   $y_i =\hat{\pi}_i$ for all $i$. 

\item
\citet{mcfadden1974conditional} defined the following $R^2$:
$$
R^2_{\text{mcfadden}} = 1 - \frac{ \log L(\hat{\beta})  }{  \log L(\tilde{\beta}) }
$$
recalling that
$\tilde{\beta}$ is the MLE assuming that all coefficients except the intercept are zero, and $\hat{\beta}$ is the MLE without any restrictions.
This $R^2$ must lie within $[0,1]$. 

Verify that under the Normal linear model, the above formula does not reduce to the usual $R^2$ in OLS.

\item
\citet{cox1989analysis} defined the following $R^2$:
$$
R^2_{\text{CS}} = 1 - \left(  \frac{  L(\tilde{\beta}) }{  L(\hat{\beta}) }\right)^{2/n}.
$$

Verify that under the Normal linear model, the above formula reduces to the usual $R^2$. 
\end{enumerate}

Remark: \citet{tjur2009coefficients} gave an excellent discussion of $R^2_{\text{model}}   , R^2_{\text{residual}}   , R^2_{\text{correlation}} $ and $D$. \citet{nagelkerke1991note} pointed out that the upper bound of this $R^2_{\text{CS}}$ is $1 - (L(\tilde{\beta}))^{2/n} < 1$ and proposed to modify it as 
$$
R^2_{\text{nagelkerke}} = \frac{ 1 - \left(  \frac{  L(\tilde{\beta}) }{  L(\hat{\beta}) }\right)^{2/n}  }{   1 - \left(    L(\tilde{\beta}) \right)^{2/n}  }
$$
to ensure that its upper bound is $1$. 
However, this modification seems purely ad hoc.
Although $D$ is an appealing definition of $R^2$ for the logistic model, it does not generalize to other models. 
\citet{mckelvey1975statistical} suggested to define $R^2$ based on the latent linear representation of the logistic regression in Chapter \ref{sec::binary-reg-latent}:
$$
R^2 = \frac{ \hat{\beta}^{\T} S_{xx} \hat{\beta} }{  \hat{\beta}^{\T} S_{xx} \hat{\beta} + \pi^2/3},
$$
where $\hat{\beta}$ is the MLE of the coefficient, $S_{xx}$ is the sample covariance matrix of the $x_i$'s, and $\pi^2/3$ is the variance of the standard logistic distribution by Theorem \ref{thm::logistic-moments}. 
\citet{hu2006pseudo} studied the asymptotic properties of some of the $R^2$s above.

\chapter{Logistic Regressions for Categorical Outcomes} 
  \chaptermark{Multinomial and Proportional Odds Model}
\label{chapter::logit-categorical}

Categorical outcomes are common in empirical research.  There are two types of categorical outcomes:
\begin{enumerate}[label=(T\arabic*), ref=T\arabic*]
\item
Nominal.  For example, the outcome denotes the preference for fruits (apple, orange, and pear) or transportation services (Uber, Lyft, or BART if you live in the San Francisco Bay Area). 
\item
Ordinal.  For example, the outcome denotes the course evaluation at Berkeley (1, 2, \ldots, 7) or Amazon review (1 to 5 stars).  
\end{enumerate}
This chapter discusses statistical modeling strategies for categorical outcomes, including two classes of models corresponding to the nominal and ordinal outcomes, respectively.

\section{Multinomial distribution}

A categorical random variable $y$ taking values in $\left\{ 1,\ldots,K\right\} $
with probabilities $\pr(y=k)=\pi_{k}$  ($k=1,\ldots,K$) is
often called a multinomial distribution, denoted by
\begin{equation}
y\sim\text{Multinomial}\left\{ 1;(\pi_{1},\ldots,\pi_{K})\right\} ,\label{eq:multinomial-rv}
\end{equation}
where $\sum_{k=1}^{K}\pi_{k}=1.$
We can calculate the mean and covariance matrix of $y$:
\begin{proposition}
\label{proposition:positive-semi-definite-multinomial}If $y$ is the Multinomial
random variable in (\ref{eq:multinomial-rv}), then $(1(y=1),\ldots,1(y=K-1))$
has mean $(\pi_{1},\ldots,\pi_{K-1})$ and covariance matrix
\begin{equation}
\left(\begin{array}{cccc}
\pi_{1}(1-\pi_{1}) & -\pi_{1}\pi_{2} & \cdots & -\pi_{1}\pi_{K-1}\\
-\pi_{1}\pi_{2} & \pi_{2}(1-\pi_{2}) & \cdots & -\pi_{2}\pi_{K-1}\\
\vdots & \vdots & \ddots & \vdots\\
-\pi_{1}\pi_{K-1} & -\pi_{2}\pi_{K-1} & \cdots & \pi_{K-1}(1-\pi_{K-1})
\end{array}\right).\label{eq:cov-multinomial-rv}
\end{equation}
As a byproduct, the matrix in (\ref{eq:cov-multinomial-rv})
must be positive semi-definite.
\end{proposition}

\begin{myproof}{Proposition}{\ref{proposition:positive-semi-definite-multinomial}}
Without loss of generality, I will calculate the $(1,1)$th and the $(1,2)$th element of the matrix. First, $1(y=1)$ is Bernoulli with probability $\pi_1$, so the $(1,1)$th element equals var$( 1(y=1) ) = \pi_1 (1 - \pi_1)$. Similarly, the $(2,2)$th element equals var$( 1(y=2) ) = \pi_2 (1 - \pi_2)$.

Second, $1(y=1) +1(y=2)$ is Bernoulli with probability $\pi_1 + \pi_2$, so var$(1(y=1) +1(y=2)) = (\pi_1 + \pi_2)(1-\pi_1 -\pi_2)$. Therefore, the $(1,2)$-th element equals
\begin{eqnarray*}
&&\cov(1(y=1), 1(y=2) ) \\
&=& \left\{ \var( 1(y=1) +1(y=2) ) - \var( 1(y=1)) - \var(1(y=2)) \right\} / 2 \\
&=& \{  (\pi_1 + \pi_2)(1-\pi_1 -\pi_2) - \pi_1 (1 - \pi_1) - \pi_2 (1 - \pi_2) \} / 2 \\
&=& -\pi_1 \pi_2.
\end{eqnarray*}  
\end{myproof}

With independent samples of $(x_{i},y_{i})_{i=1}^{n}$, we want to
model $y_{i}$ based on covariates $x_{i}$\footnote{An alternative  strategy is to model $1(y_i=k)\mid x_i$ for each $k$. The advantage of this strategy is that it reduces to binary logistic models. The disadvantage of this strategy is that it does not model the whole distribution of $y_i$ and can lose efficiency in estimation.}:
\[
y_{i}\mid x_{i}\sim\textup{Multinomial}\left[1;\left\{ \pi_{1}(x_{i}),\ldots,\pi_{K}(x_{i})\right\} \right],
\]
where $\sum_{k=1}^{K}\pi_{k}(x_{i})=1$  for all $x_{i}.$
We can write the probability mass function of $\pr( y_{i}\mid x_{i} ) $ as
\begin{eqnarray*}
\pi_{y_i}(x_i) 
&=& 
  \prod_{k=1}^{K}  \left\{   \pi_{k}(x_{i}) \text{ if } y_{i}=k\right\}  \\
&=&  \prod_{k=1}^{K}\left\{ \pi_{k}(x_{i})\right\} ^{1(y_{i}=k) }.
\end{eqnarray*}
Here $\pi_{k}(x_{i} ) $ is a general function of $x_i$. The remaining parts of this chapter will discuss the canonical choices of $\pi_{k}(x_{i} ) $ for nominal and ordinal outcomes.

\section{Multinomial logistic model for nominal outcomes}

\subsection{Modeling}
Viewing category $K$ as the reference level, we can model the ratio of
the probabilities of categories $k$ and $K$ as
\[
\log\frac{\pi_{k}(x_{i})}{\pi_{K}(x_{i})}=x_{i}^{\T}\beta_{k}\qquad(k=1,\ldots,K-1)
\]
which implies that 
\[
\pi_{k}(x_{i})=\pi_{K}(x_{i})e^{x_{i}^{\T}\beta_{k}}\qquad(k=1,\ldots,K-1).
\]
Due to the normalization 
$$
1 = \sum_{k=1}^{K}\pi_{k}(x_{i})
= \sum_{k=1}^{K}\pi_{K}(x_{i})e^{x_{i}^{\T}\beta_{k}}
= \pi_{K}(x_{i})\sum_{k=1}^{K}e^{x_{i}^{\T}\beta_{k}},
$$
we have
$$
\pi_{K}(x_{i})
=
1\Big/\sum_{k=1}^{K}e^{x_{i}^{\T}\beta_{k}} .
$$
Therefore, 
$$
\pi_{k}(x_{i})
=
\frac{e^{x_{i}^{\T}\beta_{k}}}{\sum_{ l =1}^{K}e^{x_{i}^{\T}\beta_{ l }}} ,\qquad  (k=1,\ldots,K-1).
$$
A more compact form is the following definition of the multinomial logistic model.

\begin{definition}
[multinomial logistic regression model]\label{def::multinomial-logistic}
We have
\[
y_{i}\mid x_{i}\sim\textup{Multinomial}\left[1;\left\{ \pi_{1}(x_{i}),\ldots,\pi_{K}(x_{i})\right\} \right],
\]
with 
\begin{eqnarray}
\pi_{k}(x_{i})=\pi_{k}(x_{i},\beta)=\frac{e^{x_{i}^{\T}\beta_{k}}}{\sum_{ l =1}^{K}e^{x_{i}^{\T}\beta_{ l }}},\qquad(k=1,\ldots,K) . 
\label{eq::multinomial-logistic}
\end{eqnarray}
The observations are independent across units. The $\beta=(\beta_{1},\ldots,\beta_{K-1})$ denotes the unknown parameter,
with $\beta_{K}=0$ for the reference category. 
\end{definition}

From the ratio form of \eqref{eq::multinomial-logistic}, we can only identify $\beta_k - \beta_K$ for all $k= 1,\ldots, K$. So for convenience, we impose the restriction $\beta_{K}=0$ in Definition \ref{def::multinomial-logistic}.

Similar to the binary logistic regression model, we can interpret the coefficients as the conditional log odds ratio compared to the reference level:
\begin{align*}
\beta_{k,j} & =\log\frac{\pi_{k}( \ldots,x_{ij}+1,\ldots )}{\pi_{K}( \ldots,x_{ij}+1,\ldots )}-\log\frac{\pi_{k}( \ldots,x_{ij},\ldots )}{\pi_{K}( \ldots,x_{ij},\ldots )}\\
 & =\log\left\{ \frac{\pi_{k}( \ldots,x_{ij}+1,\ldots )}{\pi_{K}( \ldots,x_{ij}+1,\ldots )}\Big/\frac{\pi_{k}( \ldots,x_{ij},\ldots )}{\pi_{K}( \ldots,x_{ij},\ldots )}\right\} .
\end{align*}

\subsection{MLE}
The likelihood function for the multinomial logistic model is
\begin{align*}
L(\beta) & =\prod_{i=1}^{n}\prod_{k=1}^{K}\left\{ \pi_{k}(x_{i})\right\} ^{1(y_{i}=k)}\\
 & =\prod_{i=1}^{n}\prod_{k=1}^{K}\left\{ \frac{e^{x_{i}^{\T}\beta_{k}}}{\sum_{ l =1}^{K}e^{x_{i}^{\T}\beta_{ l }}}\right\} ^{1(y_{i}=k)}\\
 & =\prod_{i=1}^{n}
\left[  
 \left\{ \prod_{k=1}^{K}e^{1(y_{i}=k)x_{i}^{\T}\beta_{k}}\right\}  \Big/\sum_{k=1}^{K}e^{x_{i}^{\T}\beta_{k}}
 \right],
\end{align*}
and
the log-likelihood function is
\[
\log L(\beta)=\sumn
\left[ 
\sum_{k=1}^{K} 1(y_{i}=k)x_{i}^{\T}\beta_{k}-\log\left\{ \sum_{k=1}^{K}e^{x_{i}^{\T}\beta_{k}}\right\} \right].
\]
The score function is 
$$
\frac{\partial\log L(\beta)}{\partial\beta} = \begin{pmatrix}
\frac{\partial\log L(\beta)}{\partial\beta_{1}}\\
\vdots\\
\frac{\partial\log L(\beta)}{\partial\beta_{K-1}} 
\end{pmatrix}
\in \mathbb{R}^{p(K-1)}
$$
with 
\begin{align*}
\frac{\partial\log L(\beta)}{\partial\beta_{k}} & =\sumn\left\{ x_{i}1(y_{i}=k)-\frac{x_{i}e^{x_{i}^{\T}\beta_{k}}}{\sum_{ l =1}^{K}e^{x_{i}^{\T}\beta_{ l }}}\right\} \\
 & =\sumn x_{i}\left\{ 1(y_{i}=k)-\frac{e^{x_{i}^{\T}\beta_{k}}}{\sum_{ l =1}^{K}e^{x_{i}^{\T}\beta_{ l }}}\right\} \\
 & =\sumn x_{i}\left\{ 1(y_{i}=k)-\pi_{k}(x_{i},\beta)\right\}  \in \mathbb{R}^{p}  ,\quad(k=1,\ldots,K-1).
\end{align*}
The Hessian matrix is 
\begin{eqnarray}
\frac{\partial^{2}\log L(\beta)}{\partial\beta \partial\beta^{\T}}
=\begin{pmatrix}
\frac{\partial^{2}\log L(\beta)}{\partial\beta_{1}\partial\beta_{1}^{\T}}  & \frac{\partial^{2}\log L(\beta)}{\partial\beta_{1}\partial\beta_{2}^{\T}}  & \cdots & \frac{\partial^{2}\log L(\beta)}{\partial\beta_{1}\partial\beta_{K-1}^{\T}}  \\
\frac{\partial^{2}\log L(\beta)}{\partial\beta_{2}\partial\beta_{1}^{\T}}  & \frac{\partial^{2}\log L(\beta)}{\partial\beta_{2}\partial\beta_{2}^{\T}}  & \cdots & \frac{\partial^{2}\log L(\beta)}{\partial\beta_{2}\partial\beta_{K-1}^{\T}}  \\
\vdots &\vdots  &&\vdots \\
\frac{\partial^{2}\log L(\beta)}{\partial\beta_{K-1}\partial\beta_{1}^{\T}} & \frac{\partial^{2}\log L(\beta)}{\partial\beta_{K-1}\partial\beta_{2}^{\T}}  & \cdots & \frac{\partial^{2}\log L(\beta)}{\partial\beta_{K-1}\partial\beta_{K-1}^{\T}} 
\end{pmatrix}  
\in  \mathbb{R}^{p(K-1)\times p(K-1)}  \label{eq::hessian-logit-categorical}
\end{eqnarray}
with the $(k,k)$th block  
\begin{align*}
\frac{\partial^{2}\log L(\beta)}{\partial\beta_{k}\partial\beta_{k}^{\T}} & =-\sumn x_{i}\frac{\partial}{\partial\beta_{k}^{\T}}\left(\frac{e^{x_{i}^{\T}\beta_{k}}}{\sum_{ l =1}^{K}e^{x_{i}^{\T}\beta_{ l }}}\right)\\
 & =-\sumn x_{i}x_{i}^{\T}\frac{e^{x_{i}^{\T}\beta_{k}}\sum_{ l =1}^{K}e^{x_{i}^{\T}\beta_{ l }}-e^{x_{i}^{\T}\beta_{k}}e^{x_{i}^{\T}\beta_{k}}}{(\sum_{ l =1}^{K}e^{x_{i}^{\T}\beta_{ l }})^{2}}\\
 & =-\sumn\pi_{k}(x_{i},\beta)\left\{ 1-\pi_{k}(x_{i},\beta)\right\} x_{i}x_{i}^{\T} \in \mathbb{R}^{p\times p}   \quad(k=1,\ldots,K-1)
\end{align*}
and the $(k, l )$th block  
\begin{align*}
\frac{\partial^{2}\log L(\beta)}{\partial\beta_{k}\partial\beta_{ l }^{\T}} & =-\sumn x_{i}\frac{\partial}{\partial\beta_{ l }^{\T}}\left(\frac{e^{x_{i}^{\T}\beta_{k}}}{\sum_{ l =1}^{K}e^{x_{i}^{\T}\beta_{ l }}}\right)\\
 & =-\sumn x_{i}x_{i}^{\T}\frac{-e^{x_{i}^{\T}\beta_{k}}e^{x_{i}^{\T}\beta_{ l }}}{(\sum_{ l =1}^{K}e^{x_{i}^{\T}\beta_{ l }})^{2}}\\
 & =\sumn\pi_{k}(x_{i},\beta)\pi_{ l }(x_{i},\beta)x_{i}x_{i}^{\T}  \in \mathbb{R}^{p\times p}   \quad(k\neq  l :k, l =1,\ldots,K-1).
\end{align*}
We can verify that the Hessian matrix is negative semi-definite based
on Proposition \ref{proposition:positive-semi-definite-multinomial}, which
is left as Problem \ref{hw18::hessian-multinom}.

In \ri{R}, the function \ri{multinom} in the \ri{nnet} package uses Newton's method to fit the multinomial logistic model. We can make inference about the parameters based on the asymptotic Normality of the MLE. Based on a new observation with covariate $x_{n+1}$, we  can make prediction based on the fitted probabilities $\pi_k(x_{n+1}, \hat{\beta})$, and furthermore classify it into $K$ categories based on
$$
\hat{y}_{n+1} = \arg\max_{1\leq k \leq K}  \pi_k(x_{n+1}, \hat{\beta}) . 
$$

\section{A latent variable representation for the multinomial logistic regression}

We can view the multinomial logistic regression \eqref{eq::multinomial-logistic} as an extension of the binary logistic regression model. The binary logistic regression has a latent variable representation as shown in Section \ref{sec::binary-reg-latent}. The multinomial logistic regression model in Definition \ref{def::multinomial-logistic} also has a latent variable representation below.  

Assume
\[
\begin{cases}
U_{i1} & =  x_i^{\T} \beta_1 +\varepsilon_{i1},\\
\vdots\\
U_{iK} & =x_i^{\T} \beta_K +\varepsilon_{iK},
\end{cases}
\]
where $\varepsilon_{i1},\ldots,\varepsilon_{iK}$ are IID  standard Gumbel random variables.\footnote{See Appendix \ref{subsec::expo-gumbel} for a review of the Gumbel distribution.}
Using the language of economics, $(U_{i1} , \ldots, U_{iK})$ are the utilities associated with the choices $(1,\ldots, K)$. So unit $i$ chooses $k$ if $k$ has the highest utility:
$$
y_i = k  \quad 
\text{ if } 
U_{ik} > U_{il} \text{ for all }
l\neq k.
$$ 
We can show that this latent variable model implies \eqref{eq::multinomial-logistic}. This follows from the lemma below, which is due to \citet{mcfadden1974conditional}.\footnote{Daniel McFadden shared the 2000 Nobel Memorial Prize in Economic Sciences with James Heckman.
} When $K = 2$, it also gives another latent variable representation for the binary logistic regression, which is different from the one in Section \ref{sec::binary-reg-latent}.

\begin{lemma}
\label{lemma::mcfadden}
Assume
\[
\begin{cases}
U_{1} & =V_{1}+\varepsilon_{1},\\
\vdots\\
U_{K} & =V_{K}+\varepsilon_{K},
\end{cases}
\]
where $\varepsilon_{1},\ldots,\varepsilon_{K}$ are IID standard Gumbel. 
Define
\[
y=\arg\max_{1\leq l\leq K}U_{l}
\]
as the index corresponding to the maximum of the $U_{k}$'s. 
We have
\[
\pr(y=k)=\frac{e^{V_{k}}}{\sum_{l=1}^{K}e^{V_{l}}}.
\]
\end{lemma}

\begin{myproof}{Lemma}{\ref{lemma::mcfadden}}
Recall that the standard Gumbel random variable has 
CDF
$
F(z)=\exp(-e^{-z})
$
and density
$
f(z)=\exp(-e^{-z})e^{-z} . 
$

The event ``$y=k$'' is equivalent to the event ``$U_{k}>U_{l}$
for all $l\neq k$'', so
\begin{align*}
\pr(y  =k)=& \pr(U_{k}>U_{l},l=1,\ldots,k-1,k+1,\ldots,K)\\
= & \pr(V_{k}+\varepsilon_{k}>V_{l}+\varepsilon_{l},l=1,\ldots,k-1,k+1,\ldots,K)\\
= & \int_{-\infty}^{\infty}\pr(V_{k}+z>V_{l}+\varepsilon_{l},l=1,\ldots,k-1,k+1,\ldots,K)f(z) \diff z
\end{align*}
where the last line follows from conditioning on $\varepsilon_k$. By the independence of the $\varepsilon$'s, we have
\begin{align*}
\pr(y  =k)= & \int_{-\infty}^{\infty}\prod_{l\neq k}\pr(\varepsilon_{l}<V_{k}-V_{l}+z)f(z) \diff z\\
= & \int_{-\infty}^{\infty}\prod_{l\neq k}\exp(-e^{-V_{k}+V_{l}-z})\exp(-e^{-z})e^{-z}  \diff z\\
= & \int_{-\infty}^{\infty}\exp\left(-\sum_{l\neq k}e^{-V_{k}+V_{l}}e^{-z} \right)\exp(-e^{-z})e^{-z} \diff z.
\end{align*}
Changing of variables $t=e^{-z}$ with $\diff z=-1/t \diff t$, we obtain
\begin{eqnarray*}
\pr(y  =k) 
&=&\int_{0}^{\infty}\exp \left(-t\sum_{l\neq k}e^{-V_{k}+V_{l}} \right)\exp(-t)\diff t  \\
&=&\int_{0}^{\infty}\exp(-tC_{k})\diff t , 
\end{eqnarray*}
where
\[
C_{k}=1+\sum_{l\neq k}e^{-V_{k}+V_{l}}.
\]
The integral simplifies to $1/C_{k}$ due to the density of the exponential
distribution. Therefore,
\begin{eqnarray*}
\pr(y=k)  
&=&  \frac{1}{1+\sum_{l\neq k}e^{-V_{k}+V_{l}}}\\
&=&\frac{e^{V_{k}}}{e^{V_{k}}+\sum_{l\neq k}e^{V_{l}}} \\
&=&\frac{e^{V_{k}}}{\sum_{l=1}^{K}e^{V_{l}}}.
\end{eqnarray*}
\end{myproof}

Lemma \ref{lemma::mcfadden} is remarkable and elegant. 
I will use it again in Section \ref{sec::discretechioce}.

\section{Proportional odds model for ordinal outcomes}

For ordinal outcomes, we can still use the multinomial logistic model,
but by doing this, we discard the ordering information in the outcome.  Now I will introduce the proportional odds model for ordinal outcomes, which is more parsimonious and more interpretable than the multinomial logistic model.

Motivated by the latent linear representation of the binary logistic model in Chapter \ref{sec::binary-reg-latent},
we imagine that the ordinal outcome arises from discretizing a continuous
latent variable, as shown in the model below.

\begin{definition}
[proportional odds model for an ordinal outcome]\label{def::model-polr}
Assume that the latent variable $y_i^*$ satisfies the following linear model:
\begin{eqnarray}
\label{eq::latent-ordinal}
y_{i}^{*}  =x_{i}^{\T}\beta+\varepsilon_{i},
\end{eqnarray}
with $\varepsilon_{i} \ind x_i$,  where 
$$
\pr(\varepsilon_{i}\leq z\mid x_{i})=g(z)
$$
with $g(z) = e^z / (1+e^z)$ is the CDF of the standard logistic distribution. 
Discretize $y_{i}^{*} $ to obtain the observed $y_i$: 
\begin{eqnarray}
\label{eq::discretization-ordinal}
y_{i} =k,\quad\text{ if }\alpha_{k-1} <  y_{i}^{*}\leq\alpha_{k},\quad(k=1,\ldots,K)
\end{eqnarray}
where 
\[
-\infty=\alpha_{0}<\alpha_{1}<\cdots<\alpha_{K-1}<\alpha_{K}=\infty.
\]
The observations are independent across units. 
The unknown parameters are $(\beta,\alpha_{1},\ldots,\alpha_{K-1})$.
\end{definition}

\begin{figure}
\centering
\includegraphics[width = 0.95\textwidth]{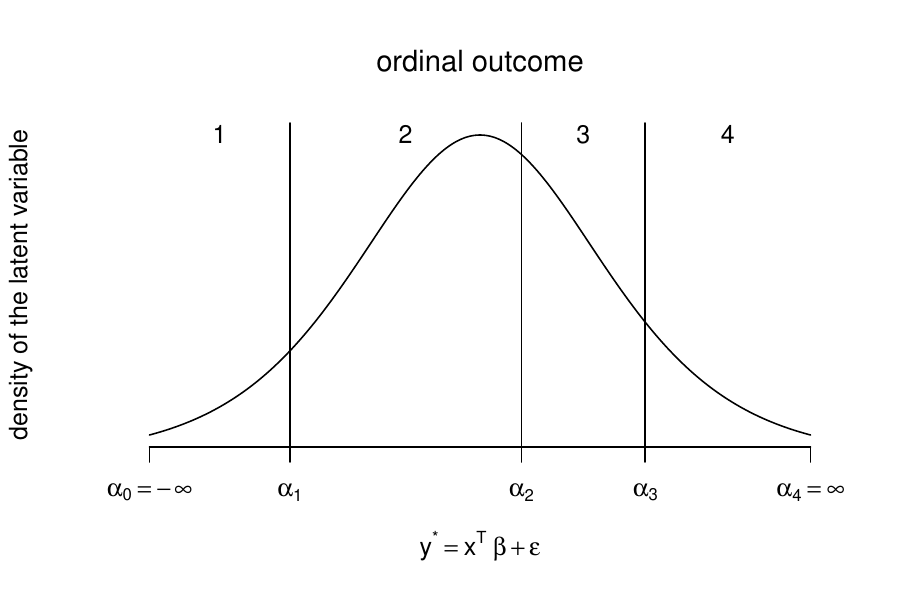}
\caption{Latent variable representation of the ordinal outcome}
\label{fig::latent-ordinal}
\end{figure}

In Definition \ref{def::model-polr}, the distribution of the latent error term $g(\cdot)$ is known and specified as the standard logistic distribution.  In general, $g(\cdot)$ can be the CDF of other random variables. For instance, when $g(\cdot)$ is the CDF of the standard Normal distribution, the model is called the ``ordered Probit model.'' I will focus on the proportional odds logistic model in the main text and defer the details for the ordered Probit model to Problem \ref{hw18::ordered-probit}.
Figure \ref{fig::latent-ordinal} illustrates the data generating process of Definition \ref{def::model-polr} with $K=4.$

Based on the latent linear model, we can compute 
\begin{align*}
\pr(y_{i}  \leq k\mid x_{i}) &=\pr(y_{i}^{*}\leq\alpha_{k}\mid x_{i})\\
 & =\pr(x_{i}^{\T}\beta+\varepsilon_{i}\leq\alpha_{k}\mid x_{i})\\
 & =\pr(\varepsilon_{i}\leq\alpha_{k}-x_{i}^{\T}\beta\mid x_{i})\\
 & =g(\alpha_{k}-x_{i}^{\T}\beta).
\end{align*}
With the proportional odds model, we have
\[
\pr(y_{i}\leq k\mid x_{i})=\frac{e^{\alpha_{k}-x_{i}^{\T}\beta}}{1+e^{\alpha_{k}-x_{i}^{\T}\beta}}
\]
or
\begin{equation}
\text{logit}\{  \pr(y_{i}\leq k\mid x_{i})  \} =\log\frac{\pr(y_{i}\leq k\mid x_{i})}{\pr(y_{i}>k\mid x_{i})}=\alpha_{k}-x_{i}^{\T}\beta . \label{eq:proportional-odds}
\end{equation}
The model has the ``proportional odds'' property because
$$
\frac{\pr(y_{i}\leq k\mid \ldots, x_{ij}+1,\ldots )}{\pr(y_{i}>k\mid \ldots, x_{ij}+1,\ldots)} 
\Big/ \frac{\pr(y_{i}\leq k\mid \ldots, x_{ij},\ldots)}{\pr(y_{i}>k\mid \ldots, x_{ij},\ldots)}
= e^{-\beta_j}
$$
which is a positive constant not depending on $k$.

The sign of $x_{i}^{\T}\beta$ is negative due to the latent variable
representation. Some textbooks and software packages use a positive
sign, but the function \ri{polr} in package \ri{MASS} of \ri{R} uses (\ref{eq:proportional-odds}).This book follows \ri{polr}.

The proportional odds model implies a quite complicated form
of the probability for each category:
$$
\pr(y_{i}=k\mid x_{i})=\frac{e^{\alpha_{k}-x_{i}^{\T}\beta}}{1+e^{\alpha_{k}-x_{i}^{\T}\beta}}-\frac{e^{\alpha_{k-1}-x_{i}^{\T}\beta}}{1+e^{\alpha_{k-1}-x_{i}^{\T}\beta}}, \quad  (k=1,\ldots,K).
$$
So the likelihood function is
\begin{eqnarray*}
L(\beta,\alpha_{1},\ldots,\alpha_{K-1}) 
&=& \prod_{i=1}^n \prod_{k=1}^K \{ \pr(y_{i}=k\mid x_{i})\}^{1(y_i = k)} \\
&=&  \prod_{i=1}^n \prod_{k=1}^K  \left( \frac{e^{\alpha_{k}-x_{i}^{\T}\beta}}{1+e^{\alpha_{k}-x_{i}^{\T}\beta}}-\frac{e^{\alpha_{k-1}-x_{i}^{\T}\beta}}{1+e^{\alpha_{k-1}-x_{i}^{\T}\beta}}\right)^{1(y_i = k)} .
\end{eqnarray*} 

The log-likelihood function is concave \citep{pratt1981concavity, burridge1981note}, and it is strictly concave in most cases. The function \ri{polr} in the \ri{R} package \ri{MASS} computes the MLE of the proportional odds model using the BFGS algorithm (which is similar to Newton's method but does not involve calculating the Hessian). It uses the explicit formulas of the gradient of the log-likelihood function, and computes the Hessian matrix numerically. I relegate the formulas of the gradient to Problem \ref{hw18::score-pom}. For more details of the Hessian matrix, see \citet{agresti2010analysis}, which is a textbook discussion on modeling ordinal data.

\section{A case study}\label{sec::multinomial-case-study}

I use an observational dataset from the Karolinska Institute in Stockholm, Sweden to illustrate the application of logistic regressions. \citet{rubin2008objective} used this dataset to investigate whether it is better for cardia cancer patients to be treated in a large-volume hospital compared with a small-volume hospital, where volume is defined by the number of patients with cardia cancer treated in recent years. I use the following variables: 
\begin{itemize}
\item
\ri{highdiag} indicating whether the patient was diagnosed at a high-volume hospital,
\item
\ri{hightreat} indicating whether the patient was treated at a high-volume hospital,
\item
\ri{age} representing the age of the patient,
\item
\ri{rural} indicating whether the patient was from a rural area,
\item
\ri{survival} representing the years of survival after diagnosis with three categories (``1'', ``2-4'', ``5+''). 
\end{itemize}

\begin{lstlisting}
karolinska = read.table("karolinska.txt", header = TRUE)
karolinska = karolinska[, c("highdiag", "hightreat",
                            "age", "rural", 
                            "male", "survival")]
\end{lstlisting}

\subsection{Binary logistic for the treatment}

We have two choices of the treatment: \ri{highdiag} and \ri{hightreat}. The logistic fit of \ri{highdiag} on covariates is shown below.

\begin{lstlisting}
> diagglm = glm(highdiag ~ age + rural + male, 
+               data = karolinska, 
+               family = binomial(link = "logit"))
> summary(diagglm)

Call:
glm(formula = highdiag ~ age + rural + male, family = binomial(link = "logit"), 
    data = karolinska)

Deviance Residuals: 
     Min        1Q    Median        3Q       Max  
-2.06147  -0.98645  -0.05759   1.01391   1.75696  

Coefficients:
            Estimate Std. Error z value Pr(>|z|)    
(Intercept)  3.46604    1.14545   3.026 0.002479 ** 
age         -0.03124    0.01481  -2.110 0.034854 *  
rural       -1.26322    0.34530  -3.658 0.000254 ***
male        -0.97524    0.41303  -2.361 0.018216 *  
\end{lstlisting}

The logistic fit of \ri{hightreat} is shown below.

\begin{lstlisting}
> treatglm = glm(hightreat ~ age + rural + male, 
+               data = karolinska, 
+               family = binomial(link = "logit"))
> summary(treatglm)

Call:
glm(formula = hightreat ~ age + rural + male, family = binomial(link = "logit"), 
    data = karolinska)

Deviance Residuals: 
    Min       1Q   Median       3Q      Max  
-2.2912  -0.9978   0.5387   0.8408   1.4810  

Coefficients:
            Estimate Std. Error z value Pr(>|z|)    
(Intercept)  6.44683    1.49544   4.311 1.63e-05 ***
age         -0.06297    0.01890  -3.332 0.000862 ***
rural       -1.28777    0.39572  -3.254 0.001137 ** 
male        -0.74856    0.45285  -1.653 0.098329 .  
\end{lstlisting}

Both treatments are associated with the covariates. \ri{hightreat} is more strongly associated with age.   \citet{rubin2008objective} argued that \ri{highdiag} is more random than  \ri{hightreat}, and may have weaker association with other hidden covariates. For each model below, I fit the data twice corresponding to two choices of treatment.  Overall, we should trust the results with \ri{highdiag} more based on \citet{rubin2008objective}'s argument.

\subsection{Binary logistic for the outcome}

I first fit binary logistic models for the dichotomized outcome indicating whether the patient survived longer than a year after diagnosis. 

\begin{lstlisting}
> karolinska$loneyear = (karolinska$survival != "1")
> loneyearglm = glm(loneyear ~ highdiag + age + rural + male,
+                  data = karolinska, 
+                  family = binomial(link = "logit"))
> summary(loneyearglm)

Call:
glm(formula = loneyear ~ highdiag + age + rural + male, family = binomial(link = "logit"), 
    data = karolinska)

Deviance Residuals: 
    Min       1Q   Median       3Q      Max  
-1.1755  -0.9936  -0.7739   1.3024   1.8557  

Coefficients:
            Estimate Std. Error z value Pr(>|z|)  
(Intercept) -1.22919    1.15545  -1.064   0.2874  
highdiag     0.13684    0.36586   0.374   0.7084  
age         -0.00389    0.01411  -0.276   0.7829  
rural        0.33360    0.35798   0.932   0.3514  
male         0.86706    0.44034   1.969   0.0489 *
\end{lstlisting}

The regressor \ri{highdiag} is not significant in the above regression.

\begin{lstlisting}
> loneyearglm = glm(loneyear ~ hightreat + age + rural + male, 
+                   data = karolinska, 
+                   family = binomial(link = "logit"))
> summary(loneyearglm)

Call:
glm(formula = loneyear ~ hightreat + age + rural + male, family = binomial(link = "logit"), 
    data = karolinska)

Deviance Residuals: 
    Min       1Q   Median       3Q      Max  
-1.3767  -0.9683  -0.6784   1.0813   2.0833  

Coefficients:
             Estimate Std. Error z value Pr(>|z|)   
(Intercept) -3.353977   1.317942  -2.545  0.01093 * 
hightreat    1.417458   0.455603   3.111  0.00186 **
age          0.008725   0.014840   0.588  0.55655   
rural        0.633278   0.368525   1.718  0.08572 . 
male         1.079973   0.452191   2.388  0.01693 * 
\end{lstlisting}

The regressor \ri{hightreat} is significant in the above regression.

\subsection{Multinomial logistic for the outcome}

I then fit multinomial logistic models for the outcome with three categories. 

\begin{lstlisting}
> library(nnet)
> yearmultinom = multinom(survival ~ highdiag + age + rural + male,
+                         data = karolinska)
# weights:  18 (10 variable)
initial  value 173.580742 
iter  10 value 134.331992
final  value 134.130815 
converged
> summary(yearmultinom)
Call:
multinom(formula = survival ~ highdiag + age + rural + male, 
    data = karolinska)

Coefficients:
    (Intercept)    highdiag          age     rural      male
2-4   -1.075818 -0.06973187 -0.004624030 0.1744256 0.5028786
5+    -4.180416  0.64036289 -0.001846453 0.7365111 2.1628717

Std. Errors:
    (Intercept)  highdiag        age     rural      male
2-4    1.286987 0.4113006 0.01596377 0.4014718 0.4716831
5+     2.003581 0.5816365 0.02148936 0.5741017 1.0741239

Residual Deviance: 268.2616 
AIC: 288.2616 
> predict(yearmultinom, type = "probs")[1:5, ]
          1       2-4         5+
1 0.5950631 0.2647047 0.14023222
2 0.5941802 0.2655369 0.14028293
3 0.8081376 0.1718963 0.01996613
4 0.5950631 0.2647047 0.14023222
5 0.6366929 0.2260086 0.13729849
\end{lstlisting}

The regressor \ri{highdiag} is not significant above.  The \ri{predict} function gives the fitted probabilities for all categories of the outcome. 

\begin{lstlisting}
> yearmultinom = multinom(survival ~ hightreat + age + rural + male,
+                         data = karolinska)
# weights:  18 (10 variable)
initial  value 173.580742 
iter  10 value 129.548642
final  value 129.283739 
converged
> summary(yearmultinom)
Call:
multinom(formula = survival ~ hightreat + age + rural + male, 
    data = karolinska)

Coefficients:
    (Intercept) hightreat         age     rural      male
2-4   -3.312433  1.326354 0.008527561 0.5186654 0.7514451
5+    -5.935172  1.627711 0.008978103 0.9063831 2.2780877

Std. Errors:
    (Intercept) hightreat        age     rural      male
2-4    1.463258 0.5141127 0.01660648 0.4085976 0.4806953
5+     2.190305 0.7320788 0.02244867 0.5645595 1.0739669

Residual Deviance: 258.5675 
AIC: 278.5675 
\end{lstlisting}

The regressor \ri{hightreat} is significant above.

\subsection{Proportional odds logistic for the outcome}

The multinomial logisitic model does not reflect the ordering information of the outcome.  I will fit the proportional odds models below. 

\begin{lstlisting}
> library(MASS)
> yearpo = polr(factor(survival) ~ highdiag + age + rural + male, 
+               Hess = TRUE, 
+               data = karolinska)
> summary(yearpo)
Call:
polr(formula = factor(survival) ~ highdiag + age + rural + male, 
    data = karolinska, Hess = TRUE)

Coefficients:
             Value Std. Error t value
highdiag  0.216755    0.35892  0.6039
age      -0.002881    0.01378 -0.2091
rural     0.371898    0.35313  1.0532
male      0.943955    0.43588  2.1656

Intercepts:
       Value   Std. Error t value
1|2-4   1.4079  1.1309     1.2450
2-4|5+  2.9284  1.1514     2.5434

Residual Deviance: 271.0778 
AIC: 283.0778 
> predict(yearpo, type = "probs")[1:5, ]
          1       2-4         5+
1 0.5862465 0.2800892 0.13366427
2 0.5855475 0.2804542 0.13399823
3 0.8087341 0.1421065 0.04915948
4 0.5862465 0.2800892 0.13366427
5 0.6205983 0.2615112 0.11789050
\end{lstlisting}

The regressor \ri{highdiag} is not significant above. The \ri{predict} function gives the fitted probabilities of three categories.

\begin{lstlisting}
> yearpo = polr(factor(survival) ~ hightreat + age + rural + male, 
+               Hess = TRUE, 
+               data = karolinska)
> summary(yearpo)
Call:
polr(formula = factor(survival) ~ hightreat + age + rural + male, 
    data = karolinska, Hess = TRUE)

Coefficients:
             Value Std. Error t value
hightreat 1.399538    0.44518  3.1438
age       0.008032    0.01438  0.5584
rural     0.638862    0.35450  1.8022
male      1.122698    0.44377  2.5299

Intercepts:
       Value  Std. Error t value
1|2-4  3.3273 1.2752     2.6092 
2-4|5+ 4.9258 1.3106     3.7583 

Residual Deviance: 260.2831 
AIC: 272.2831 
> predict(yearpo, type = "probs")[1:5, ]
          1       2-4         5+
1 0.7007473 0.2197669 0.07948581
2 0.7024290 0.2186709 0.07890008
3 0.7736340 0.1705068 0.05585924
4 0.7007473 0.2197669 0.07948581
5 0.5305401 0.3176927 0.15176720
\end{lstlisting}

The regressor \ri{hightreat} is significant above.

\section{Discrete choice models}\label{sec::discretechioce}

\subsection{Model}
The covariates in model \eqref{eq::multinomial-logistic} depend only on individuals. 
\citet{mcfadden1974conditional} extends it to allow for choice-specific covariates $z_{ik}$. His formulation is based on the latent utility representation: 
\[
\begin{cases}
U_{i1} & =   z_{i1}^{\T} \theta +\varepsilon_{i1},\\
\vdots\\
U_{iK} & =z_{iK}^{\T} \theta +\varepsilon_{iK},
\end{cases}
\]
where $\varepsilon_{i1},\ldots,\varepsilon_{iK}$ are IID standard Gumbel. Unit $i$ chooses $k$ if $k$ has the highest utility. Lemma \ref{lemma::mcfadden} implies that
\begin{eqnarray}
\pi_{k}(z_{i})=\pi_{k}(z_{i},\theta)=\frac{e^{  z_{ik}^{\T}\theta }  }{\sum_{ l =1}^{K}e^{ z_{i l }^{\T}\theta } } ,\qquad(k=1,\ldots,K) .
\label{eq::multinomial-logistic-mcfadden}
\end{eqnarray}

Model \eqref{eq::multinomial-logistic-mcfadden} seems rather similar to model \eqref{eq::multinomial-logistic}. However, there are many subtle differences. First, a component of $z_{ik}$ may vary only with choice $k$, for example, it can represent the price of choice $k$. Partition $z_{ik}$ into three types of covariates: $x_i$ that only vary cross individuals, $c_k$ that only vary across choices, and $w_{ik}$ that vary across both individuals and choices. Model \eqref{eq::multinomial-logistic-mcfadden} reduces to
$$
\pi_{k}(z_{i})  =\frac{e^{  x_{i}^{\T}\theta_x + c_k^{\T}\theta_c + w_{ik}^{\T} \theta_w }  }
{\sum_{ l =1}^{K}e^{ x_{i}^{\T}\theta_x + c_{ l }^{\T}\theta_c + w_{i l }^{\T} \theta_w }  }  
=\frac{e^{   c_k^{\T}\theta_c + w_{ik}^{\T} \theta_w }  }
{\sum_{ l =1}^{K}e^{   c_{ l }^{\T}\theta_c + w_{i l }^{\T} \theta_w } }  ,
$$
that is, the individual-specific covariates drop out. Therefore, $z_{ik}$ in model \eqref{eq::multinomial-logistic-mcfadden} does not contain covariates that vary only with individuals. In particular, $z_{ik}$ in model \eqref{eq::multinomial-logistic-mcfadden} does not contain the constant, but in contrast, the $x_i$ in model \eqref{eq::multinomial-logistic} ususally contains the intercept by default. 

Second, if we want to use individual-specific covariates in the model, they must have choice-specific coefficients. So a more general model unifying \eqref{eq::multinomial-logistic} and \eqref{eq::multinomial-logistic-mcfadden} is
\begin{eqnarray}
\pi_{k}(x_{i}, w_{ik}, \theta, \beta)=\frac{e^{  w_{ik}^{\T}\theta + x_i^{\T} \beta_k }  }{\sum_{ l =1}^{K}e^{ w_{i l }^{\T}\theta + x_i^{\T} \beta_{ l } }},\qquad(k=1,\ldots,K) .
\label{eq::multinomial-logistic-unify}
\end{eqnarray}
Equivalently, we can create pseudo covariates $z_{ik}$ as the original $w_{ik}$ together with interaction of $x_i$ and the dummy for choice $k$. For example, if $K = 3$ and $x_i$ contain the intercept and a scalar individual-specific covariate $x_i'$, then $(z_{i1}, z_{i2}, z_{i3})$ are
\begin{eqnarray*}
\begin{pmatrix}
z_{i1} \\
z_{i2} \\
z_{i3}
\end{pmatrix}
= 
\begin{pmatrix}
w_{i1} & 1 & 0 & x_i' & 0 \\
w_{i2} & 0 & 1 & 0 & x_i' \\
w_{i3} & 0 & 0 & 0 & 0
\end{pmatrix},
\end{eqnarray*}
where $K=3$ is the reference level. 
So with augmented covariates, the discrete choice model \eqref{eq::multinomial-logistic-mcfadden} is strictly more general than the multinomial logistic model \eqref{eq::multinomial-logistic}. In the special case with $K=2$, model \eqref{eq::multinomial-logistic-mcfadden} reduces to
$$
\pr(y_i = 1\mid   x_i ) = \frac{  e^{  x_i^{\T} \beta }   }{   1 + e^{  x_i^{\T} \beta }  }
$$
where $x_i = z_{i1} - z_{i2}$.

\subsection{MLE}

Based on the model specification \eqref{eq::multinomial-logistic-mcfadden}, the log likelihood function is
$$
\log L(\theta) 
= \sumn \sum_{k=1}^K 1(y_i = k)  \left(  z_{ik}^{\T}\theta -   \log \sum_{ l =1}^{K}e^{ z_{i l }^{\T}\theta }   \right).
$$
So the score function is
\begin{eqnarray*}
\frac{\partial }{ \partial \theta } \log L(\theta) 
&=& \sumn \sum_{k=1}^K 1(y_i = k)    \left(  z_{ik}  -   
\frac{   \sum_{ l =1}^{K}e^{ z_{i l }^{\T}\theta }  z_{i l } }{\sum_{ l =1}^{K}e^{ z_{i l }^{\T}\theta }  } \right) \\
&=& \sumn \sum_{k=1}^K 1(y_i = k)    \{  z_{ik}  -  E(z_{ik};\theta)    \},
\end{eqnarray*}
where $E(\cdot; \theta) $ is the average value of $\{ z_{i1}, \ldots, z_{iK} \}$ over the probability mass function
$$
p_k(\theta) = e^{z_{ik}^{\T}\theta} / \sum_{ l =1}^{K}e^{ z_{i l }^{\T}\theta }.
$$
The Hessian matrix is
\begin{eqnarray*}
&&\frac{\partial^2 }{ \partial \theta\partial \theta^{\T}  } \log L(\theta)  \\
&=& - \sumn \sum_{k=1}^K 1(y_i = k)  
\frac{   \sum_{ l =1}^{K}e^{ z_{i l }^{\T}\theta }  z_{i l } z_{i l }^{\T}\sum_{ l =1}^{K}e^{ z_{i l }^{\T}\theta }
- \sum_{ l =1}^{K}e^{ z_{i l }^{\T}\theta }  z_{i l } \sum_{ l =1}^{K}e^{ z_{i l }^{\T}\theta }  z_{i l }^{\T}
 }{  (\sum_{ l =1}^{K}e^{ z_{i l }^{\T}\theta })^2  }  \\
 &=& - \sumn \sum_{k=1}^K 1(y_i = k)   \cov(z_{ik}; \theta),
\end{eqnarray*}
where $\cov(\cdot; \theta) $ is the covariance matrix of $\{ z_{i1}, \ldots, z_{iK} \}$ over the probability mass function defined above. 
From these formulas, we can compute the MLE using Newton's method and obtain its asymptotic distribution based on the inverse of the Fisher information matrix. 

\subsection{Example}

The \ri{R} package \ri{mlogit} provides a function \ri{mlogit} to fit the general discrete logistic model \citep{croissant2020estimation}. 
Here I use an example from this package to illustrate the model fitting of \ri{mlogit}.

\begin{lstlisting}
> library("nnet")
> library("mlogit")
> data("Fishing")
> head(Fishing)
     mode price.beach price.pier price.boat price.charter
1 charter     157.930    157.930    157.930       182.930
2 charter      15.114     15.114     10.534        34.534
3    boat     161.874    161.874     24.334        59.334
4    pier      15.134     15.134     55.930        84.930
5    boat     106.930    106.930     41.514        71.014
6 charter     192.474    192.474     28.934        63.934
  catch.beach catch.pier catch.boat catch.charter   income
1      0.0678     0.0503     0.2601        0.5391 7083.332
2      0.1049     0.0451     0.1574        0.4671 1250.000
3      0.5333     0.4522     0.2413        1.0266 3750.000
4      0.0678     0.0789     0.1643        0.5391 2083.333
5      0.0678     0.0503     0.1082        0.3240 4583.332
6      0.5333     0.4522     0.1665        0.3975 4583.332
\end{lstlisting}

The dataset \ri{Fishing} is in the ``wide'' format, where \ri{mode} denotes the choice of four modes of fishing (beach, pier, boat and charter), \ri{price} and \ri{catch} denote the price and catching rates which are choice-specific, \ri{income} is individual-specific. 
We need to first transform the dataset into ``long'' format.

\begin{lstlisting}
> Fish  = dfidx(Fishing, 
+               varying = 2:9, 
+               shape = "wide", 
+               choice = "mode")
> head(Fish)
~~~~~~~
 first 10 observations out of 4728 
~~~~~~~
    mode   income   price  catch    idx
1  FALSE 7083.332 157.930 0.0678 1:each
2  FALSE 7083.332 157.930 0.2601 1:boat
3   TRUE 7083.332 182.930 0.5391 1:rter
4  FALSE 7083.332 157.930 0.0503 1:pier
5  FALSE 1250.000  15.114 0.1049 2:each
6  FALSE 1250.000  10.534 0.1574 2:boat
7   TRUE 1250.000  34.534 0.4671 2:rter
8  FALSE 1250.000  15.114 0.0451 2:pier
9  FALSE 3750.000 161.874 0.5333 3:each
10  TRUE 3750.000  24.334 0.2413 3:boat
\end{lstlisting}

Using only choice-specific covariates, we have the following fitted model:

\begin{lstlisting}
> summary(mlogit(mode ~ 0 + price + catch, data = Fish))

Call:
mlogit(formula = mode ~ 0 + price + catch, data = Fish, method = "nr")

Frequencies of alternatives:choice
  beach    boat charter    pier 
0.11337 0.35364 0.38240 0.15059 

nr method
6 iterations, 0h:0m:0s 
g'(-H)^-1g = 0.000179 
successive function values within tolerance limits 

Coefficients :
        Estimate Std. Error z-value  Pr(>|z|)    
price -0.0204765  0.0012231 -16.742 < 2.2e-16 ***
catch  0.9530982  0.0894134  10.659 < 2.2e-16 ***
\end{lstlisting}

If we do not enforce \ri{0 + price}, we allow for intercepts that vary across choices:

\begin{lstlisting}
> summary(mlogit(mode ~ price + catch, data = Fish))

Call:
mlogit(formula = mode ~ price + catch, data = Fish, method = "nr")

Frequencies of alternatives:choice
  beach    boat charter    pier 
0.11337 0.35364 0.38240 0.15059 

nr method
7 iterations, 0h:0m:0s 
g'(-H)^-1g = 6.22E-06 
successive function values within tolerance limits 

Coefficients :
                      Estimate Std. Error  z-value  Pr(>|z|)    
(Intercept):boat     0.8713749  0.1140428   7.6408 2.154e-14 ***
(Intercept):charter  1.4988884  0.1329328  11.2755 < 2.2e-16 ***
(Intercept):pier     0.3070552  0.1145738   2.6800 0.0073627 ** 
price               -0.0247896  0.0017044 -14.5444 < 2.2e-16 ***
catch                0.3771689  0.1099707   3.4297 0.0006042 ***
\end{lstlisting}

Using only individual-specific covariates, we have the following fitted model:

\begin{lstlisting}
> summary(mlogit(mode ~ 0 | income, data = Fish))

Call:
mlogit(formula = mode ~ 0 | income, data = Fish, method = "nr")

Frequencies of alternatives:choice
  beach    boat charter    pier 
0.11337 0.35364 0.38240 0.15059 

nr method
4 iterations, 0h:0m:0s 
g'(-H)^-1g = 8.32E-07 
gradient close to zero 

Coefficients :
                       Estimate  Std. Error z-value  Pr(>|z|)    
(Intercept):boat     7.3892e-01  1.9673e-01  3.7560 0.0001727 ***
(Intercept):charter  1.3413e+00  1.9452e-01  6.8955 5.367e-12 ***
(Intercept):pier     8.1415e-01  2.2863e-01  3.5610 0.0003695 ***
income:boat          9.1906e-05  4.0664e-05  2.2602 0.0238116 *  
income:charter      -3.1640e-05  4.1846e-05 -0.7561 0.4495908    
income:pier         -1.4340e-04  5.3288e-05 -2.6911 0.0071223 ** 
\end{lstlisting}

It is equivalent to fitting the multinomial logistic model using the original data. 

\begin{lstlisting}
> summary(multinom(mode ~ income, data = Fishing))
# weights:  12 (6 variable)
initial  value 1638.599935 
iter  10 value 1477.150646
final  value 1477.150569 
converged
Call:
multinom(formula = mode ~ income, data = Fishing)

Coefficients:
        (Intercept)        income
pier      0.8141506 -1.434028e-04
boat      0.7389178  9.190824e-05
charter   1.3412901 -3.163844e-05

Std. Errors:
         (Intercept)       income
pier    5.816490e-09 2.668383e-05
boat    3.209473e-09 2.057825e-05
charter 3.921689e-09 2.116425e-05

Residual Deviance: 2954.301 
AIC: 2966.301 
\end{lstlisting}

The most general model includes all covariates.

\begin{lstlisting}
> summary(mlogit(mode ~ price + catch | income, data = Fish))

Call:
mlogit(formula = mode ~ price + catch | income, data = Fish, 
    method = "nr")

Frequencies of alternatives:choice
  beach    boat charter    pier 
0.11337 0.35364 0.38240 0.15059 

nr method
7 iterations, 0h:0m:0s 
g'(-H)^-1g = 1.37E-05 
successive function values within tolerance limits 

Coefficients :
                       Estimate  Std. Error  z-value  Pr(>|z|)    
(Intercept):boat     5.2728e-01  2.2279e-01   2.3667 0.0179485 *  
(Intercept):charter  1.6944e+00  2.2405e-01   7.5624 3.952e-14 ***
(Intercept):pier     7.7796e-01  2.2049e-01   3.5283 0.0004183 ***
price               -2.5117e-02  1.7317e-03 -14.5042 < 2.2e-16 ***
catch                3.5778e-01  1.0977e-01   3.2593 0.0011170 ** 
income:boat          8.9440e-05  5.0067e-05   1.7864 0.0740345 .  
income:charter      -3.3292e-05  5.0341e-05  -0.6613 0.5084031    
income:pier         -1.2758e-04  5.0640e-05  -2.5193 0.0117582 *  
\end{lstlisting}

\subsection{More comments}

The assumption of Gumbel error terms is very strong. However, relaxing this assumption leads to much more complicated forms of the conditional probabilities of the outcome. The model \eqref{eq::multinomial-logistic-mcfadden} implies that 
$$
\frac{  \pi_k(z_i) }{ \pi_l(z_i) } = \exp\{  (z_{ik} - z_{il})^{\T} \theta   \},
$$
so the choice between $k$ and $l$ does not depend on the existence of other choices. This is called the
independence of irrelevant alternatives (IIA) assumption. This is often a plausible assumption. However, it may be violated. For example, with the apple and orange, someone chooses the apple; but with the apple, orange, and banana, they may choose the orange.

The model \eqref{eq::multinomial-logistic-mcfadden}  is the basic form of the discrete choice model. \citet{train2009discrete} is a monograph on this topic, which provides many extensions.

\section{Homework problems}

\paragraph{Inverse model for the  multinomial logit model}\label{hw18::inverse-multinom}

Theorem \ref{thm::inverse-multinom-logit} below extends Theorem \ref{thm::inverse-logit}. Prove Theorem \ref{thm::inverse-multinom-logit}. 

Assume   
\begin{eqnarray}
y_{i}\sim\text{Multinomial}(1; q_1, \ldots, q_K),
\label{eq::y-marginal-multinomial}
\end{eqnarray}
and
\begin{eqnarray}
x_{i}\mid y_{i}=k\sim\N(\mu_{k},\Sigma),
\label{eq::xgiveny-multinomial}
\end{eqnarray} 
where $x_i$ does not contain $1$.
We can verify that $y_{i}\mid x_{i}$ follows a multinomial logit model as shown in the theorem below.

\begin{theorem}
\label{thm::inverse-multinom-logit}
Under \eqref{eq::y-marginal-multinomial} and \eqref{eq::xgiveny-multinomial}, we have
$$
 \pr(y_i =k\mid x_i)   =   \frac{  e^{\alpha_k + x_i^{\T} \beta_k} }{   \sum_{l=1}^K e^{\alpha_l + x_i^{\T} \beta_l} } ,
$$
where
\begin{eqnarray*}
\label{eq::alpha-beta-lda}
\alpha_k = \log q_k -\frac{1}{2} \mu_k^{\T}\Sigma^{-1}\mu_k  ,\quad
\beta_k = \Sigma^{-1} \mu_k.
\end{eqnarray*}
\end{theorem}

\paragraph{Hessian matrix in the multinomial logit model}\label{hw18::hessian-multinom}

Prove that the Hessian matrix \eqref{eq::hessian-logit-categorical} of the log-likelihood function of the multinomial logit model is negative semi-definite.

Remark: Use Proposition \ref{proposition:positive-semi-definite-multinomial}. 

\paragraph{Iteratively reweighted least squares algorithm for the multinomial logit model}\label{hw18::irls-multinom}

Similar to the binary logistic model, Newton's method for computing the MLE for the multinomial logit model can be written as iteratively reweighted least squares. Give the details. 

\paragraph{Score function of the proportional odds model}\label{hw18::score-pom}

Derive the explicit formulas of the score function of the proportional odds model.

\paragraph{Ordered Probit regression}\label{hw18::ordered-probit}

If we choose $\varepsilon_i \mid x_i \sim \N(0,1)$ in \eqref{eq::latent-ordinal}, then the corresponding model is called the ordered Probit regression. Write down the likelihood function and derive the score function for this model. 

Remark: You can use the function \ri{polr} in \ri{R} to fit this model with the specification \ri{method = "probit"}. 

\paragraph{Case-control study and multinomial logistic model}\label{hw18::case-control-multinom}

Theorem \ref{thm::case-control-multinomial-logit} below extends Theorem \ref{thm::case-control}. Prove Theorem \ref{thm::case-control-multinomial-logit}.

\begin{theorem}
\label{thm::case-control-multinomial-logit}
Assume  
$$
\pr(y_i = k\mid x_i) = \frac{  e^{\alpha_k + x_i^{\T} \beta_k} }{ \sum_{ l =1}^K e^{\alpha_{ l  }+ x_i^{\T} \beta_{ l }} }
$$
and 
$$
\pr(s_i =1\mid y_i = k, x_i) = \pr(s_i =1\mid y_i = k ) = p_k 
$$
for $k=1,\ldots, K$. 
Then we have 
$$
\pr(y_i = k\mid x_i, s_i = 1) = \frac{  e^{\tilde \alpha_k  + x_i^{\T} \beta_k} }{ \sum_{ l =1}^K e^{\tilde \alpha_{ l  }   + x_i^{\T} \beta_{ l }} }
$$
with $\tilde \alpha_k = \alpha_k + \log p_k$ for $k=1,\ldots, K$. 
\end{theorem}

\chapter{Regression Models for Count Outcomes}
\label{chapter::count}
 
A random variable for counts can take values in $\left\{ 0,1,2,\ldots\right\} $. This type of variable is common in applied statistics. For example, it can represent 
\begin{enumerate}[label=(E\arabic*), ref=E\arabic*]
\item
how many times you visit the gym every week,
\item
how many lectures you have missed in the ``Linear Model'' course,
\item
how many traffic accidents happened in certain areas during certain periods. 
\end{enumerate}
This chapter focuses on statistical modeling of those outcomes given covariates. \citet{hilbe2014modeling} is a textbook focusing on count outcome regressions.

\section{Some random variables for counts\label{sec:Some-random-variables-counts}}

I first review four canonical choices of random variables for modeling count data.

\subsection{Poisson}

A random variable $y$ is Poisson$(\lambda)$, if its probability mass
function is
\[
\pr(y=k)=e^{-\lambda}\frac{\lambda^{k}}{k!}, \quad(k=0,1,2,\ldots)
\]
which sums to $1$ by the Taylor expansion formula $e^{\lambda} =  \sum_{k=0}^{\infty}  \frac{\lambda^{k}}{k!}$.
Propositions \ref{prop::poisson-1} and \ref{prop::poisson-2} below review the basic properties of Poisson$(\lambda)$ random variable. I relegate their proofs to Problem \ref{hw19::poisson12}.

\begin{proposition}\label{prop::poisson-1}
If $y\sim \textup{Poisson} (\lambda)$, then 
$$
E(y)=\var(y)=\lambda.
$$
\end{proposition}
%
%
\begin{proposition}\label{prop::poisson-2}
If $y_1, \ldots,  y_K$ are mutually independent with $y_k\sim \textup{Poisson}(\lambda_k)$ for $k=1,\ldots, K$,
then
$$
y_{1}+\cdots + y_{K}  
\sim  \textup{Poisson}(\lambda),
$$
and
$$
(y_{1}, \ldots, y_K) \mid y_{1}+ \cdots + y_{K} = n  
\sim  \textup{Multinomial}\left(n,  (\lambda_1/ \lambda, \ldots, \lambda_K / \lambda ) \right),
$$ 
where $\lambda = \lambda_{1}+ \cdots + \lambda_{K}$. 

Conversely, if $S\sim \textup{Poisson}(\lambda)$ with $\lambda = \lambda_{1}+ \cdots + \lambda_{K}$, and $(y_{1}, \ldots, y_K) \mid S=n \sim \textup{Multinomial}\left(n,   ( \lambda_1/ \lambda, \ldots, \lambda_K / \lambda ) \right)$, then $y_1, \ldots,  y_K$ are mutually independent with $y_k\sim \textup{Poisson}(\lambda_k)$ for $k=1,\ldots, K$. 
\end{proposition}

Where does the Poisson random variable come from? One way to generate Poisson is through independent   Bernoulli random variables. I will review \citet{le1960approximation}'s theorem below. Its proof is beyond the scope of this book.

\begin{theorem}\label{thm::lecam1960theorem}
Suppose $X_i$'s are independent Bernoulli random variables with probabilities $p_i$'s $(i=1,\ldots,n)$. Define $\lambda_n = \sumn p_i$ and $S_n = \sumn X_i$. Then
$$
\sum_{k=0}^{\infty}  \Big |   \pr(S_n = k) - e^{-\lambda_n} \frac{ \lambda_n^k }{ k!}  \Big | \leq 2\sumn p_i^2.
$$
\end{theorem}

As a special case, if $p_i = \lambda /  n$, then Theorem \ref{thm::lecam1960theorem} implies 
$$
\sum_{k=0}^{\infty}  \Big |   \pr(S_n = k) - e^{-\lambda} \frac{ \lambda^k }{ k!}  \Big | \leq 2\sumn (\lambda /  n)^2 = \lambda^2/n \rightarrow 0.
$$
So the sum of IID Bernoulli random variables is approximately Poisson,  if the probability has order $1/n.$ This is called the law of rare events, or Poisson limit theorem, or Le Cam's theorem. By Theorem \ref{thm::lecam1960theorem}, we can use Poisson as a model for the sum of many rare events.

\subsection{Negative-Binomial}

The Poisson distribution restricts that the mean must be the same
as the variance. It cannot capture the feature of overdispersed data
with the variance larger than the mean.\footnote{
This book focused on the issue of overdispersion. 
It is also possible to have underdispersed data with the variance smaller than the mean. See \citet{puig2023some} for the probabilistic mechanism that leads to underdispersion. 
} The Negative-Binomial is an extension
of the Poisson that allows for overdispersion. The definition below, due to \citet{fisher1943relation}, is different from its standard definition, but it is more natural as an extension of the Poisson.\footnote{With IID Bernoulli$(p)$ trials, the Negative-Binomial distribution,
denoted by $y\sim\text{NB}'(r,p)$, is the number of success before
the $r$th failure. 
Its probability mass function is
\[
\pr(y=k)=\left(\begin{array}{c}
k+r-1\\
k
\end{array}\right)(1-p)^{r}p^{k},\quad(k=0,1,2,\ldots)
\]
If $p=\mu/(\mu+\theta)$ and $r=\theta$ then these two definitions
coincide. 
This definition is more restrictive because $r$ must
be an integer. } Define $y$ as the Negative-Binomial random
variable, denoted by $\text{NB}(\mu,\theta)$ with $\mu>0$ and $\theta>0$,
if 
\begin{equation}
\begin{cases}
y\mid\lambda & \sim\text{Poisson}(\lambda),\\
\lambda & \sim\text{Gamma}(\theta,\theta/\mu).
\end{cases}\label{eq:definition-nb-rv}
\end{equation}
So the Negative-Binomial is the Poisson with a random Gamma intensity, that is, the Negative-Binomial is a scale mixture of the Poisson. If $\theta\rightarrow\infty$, then $\lambda$ is a point mass at $\mu$ and the Negative-Binomial reduces to Poisson$(\mu)$. Proposition \ref{proposition:The-Negative-Binomial-random} below gives the probability mass function and moments of the Negative-Binomial. I relegate its proof to Problem \ref{hw19::nb-properties}.

\begin{proposition}
\label{proposition:The-Negative-Binomial-random}The Negative-Binomial random
variable defined in (\ref{eq:definition-nb-rv}) has the probability
mass function
\[
\pr(y=k)=\frac{\Gamma(k+\theta)}{\Gamma(k+1)\Gamma(\theta)}\left(\frac{\theta}{\mu+\theta}\right)^{\theta}\left(\frac{\mu}{\mu+\theta}\right)^{k},\quad(k=0,1,2,\ldots),
\]
and moments
\begin{eqnarray*}
E(y)    &=& \mu,\\
\var(y) &=& \mu+\frac{\mu^{2}}{\theta} . 
\end{eqnarray*}
\end{proposition}

\begin{figure}[ht]
\centering
\includegraphics[width = 0.95\textwidth]{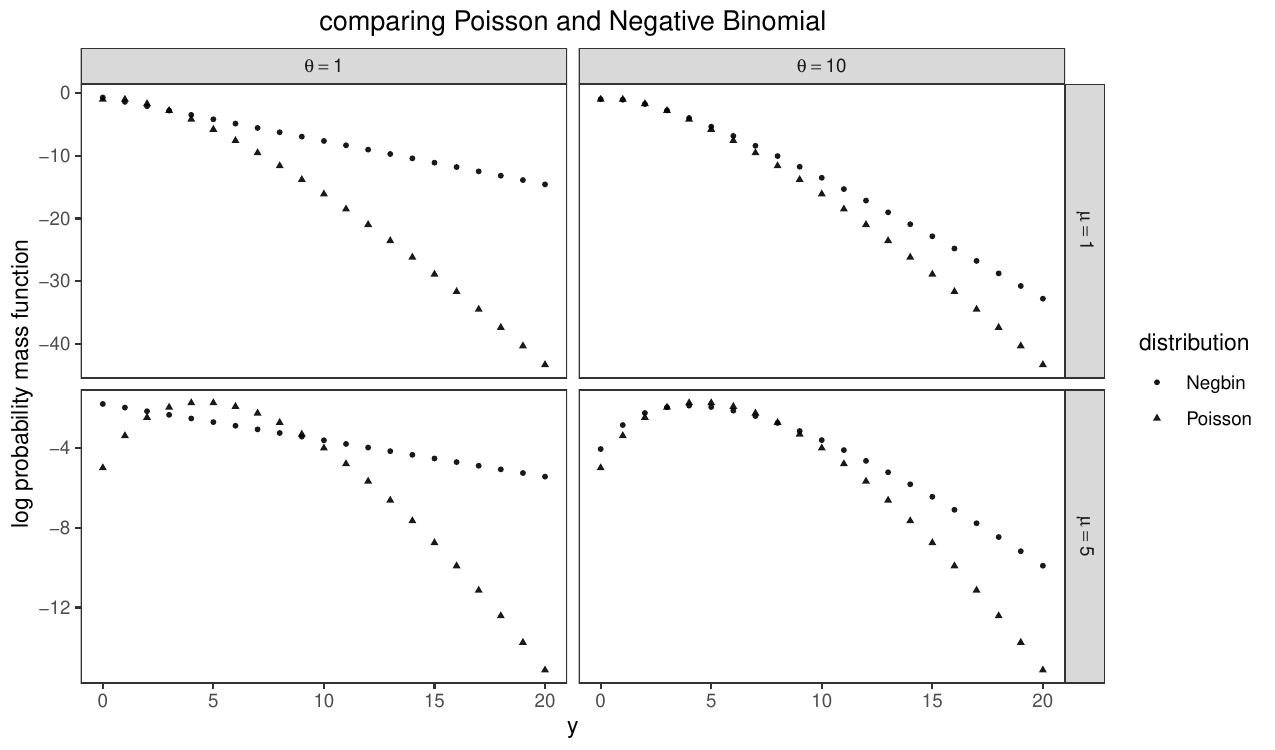}
\caption{Comparing the log probabilities of the Poisson and Negative-Binomial with the same mean}
\label{fig::poisson-negativebinomial}
\end{figure}

By Proposition \ref{proposition:The-Negative-Binomial-random}, the variance of Negative-Binomial is always larger than its mean, with a finite $\theta$. The dispersion parameter $\theta$ controls the variance of the Negative-Binomial. With the same mean, the Negative-Binomial has a larger variance than Poisson. Figure \ref{fig::poisson-negativebinomial} further compares the log probability mass functions of the Negative-Binomial and Poisson. It shows that the Negative-Binomial has a slightly higher probability at zero but much heavier tails than the Poisson.

\subsection{Zero-inflated count distributions}

Many count distributions have larger masses at zero compared to Poisson and Negative-Binomial. Therefore, it is also important to have more general distributions capturing this feature of empirical data. We can simply add an additional zero component to the Poisson or the Negative-Binomial.

A zero-inflated Poisson random variable $y$ is a mixture of two components:
a point mass at 0 and a Poisson$(\lambda)$ random variable, with
probabilities $p$ and $1-p$, respectively. So $y$ has the probability
mass function
\[
\pr(y=k)=\begin{cases}
p+(1-p)e^{-\lambda}, & \text{if }k=0,\\
(1-p)e^{-\lambda}\frac{\lambda^{k}}{k!}, & \text{if }k=1,2,\ldots.
\end{cases}
\]
It has the first two moments below:
\begin{proposition}
\label{proposition:moments-zeroinflated-poisson}
The zero-inflated Poisson random variable has the first two moments:
$$
E(y)=(1-p)\lambda,\quad \var(y)=(1-p)\lambda(1+p\lambda).
$$
\end{proposition}

A zero-inflated Negative-Binomial random variable $y$ is a mixture
of two components: a point mass at zero and a NB$(\mu,\theta)$ random
variable, with probabilities $p$ and $1-p$, respectively. So $y$
has probability mass function
\[
\pr(y=k)=\begin{cases}
p+(1-p)\left(\frac{\theta}{\mu+\theta}\right)^{\theta}, & \text{if }k=0,\\
(1-p)\frac{\Gamma(k+\theta)}{\Gamma(k+1)\Gamma(\theta)}\left(\frac{\theta}{\mu+\theta}\right)^{\theta}\left(\frac{\mu}{\mu+\theta}\right)^{k}, & \text{if }k=1,2,\ldots.
\end{cases}
\]
It has the first two moments below:
\begin{proposition}
\label{proposition:moments-zeroinflated-NB}
The zero-inflated Negative-Binomial random variable has the first two moments:
$$
E(y)=(1-p)\mu,\quad \var(y)=(1-p) \mu (1+\mu/\theta+p\mu).
$$
\end{proposition}

I leave the proofs of Propositions \ref{proposition:moments-zeroinflated-poisson} and \ref{proposition:moments-zeroinflated-NB} to Problem \ref{hw19::moment-0-poisson}.

\section{Regression models for counts}

To model a count outcome $y_{i}$ given $x_{i}$, we can still use
OLS. However, a problem with OLS is that the predicted value can be negative.
This can be easily fixed by running OLS of $\log(y_{i}+1)$ given
$x_{i}$. However, this still does not reflect the fact that $y_{i}$
is a count outcome. For example, these two OLS fits cannot easily make
a prediction for the probabilities $\pr(y_{i}\geq1\mid x_{i})$ or
$\pr(y_{i}>3\mid x_{i})$. A more direct approach is to model the
conditional distribution of $y_{i}$ given $ x_{i}$ using the distributions
reviewed in Section \ref{sec:Some-random-variables-counts}.

\subsection{Poisson regression}

I first discuss the Poisson regression model.

\begin{assumption}[Poisson regression model]
\label{assume::poisson-regression}
We have 
$$
y_{i}  \mid  x_{i}  \sim\textup{Poisson}(\lambda_{i})
$$
with 
$$
\lambda_{i}  = \lambda(x_{i},\beta)=e^{x_{i}^{\T}\beta}.
$$
The observations are independent across units. The $\beta$ is the unknown parameter. 
\end{assumption}

Under Assumption \ref{assume::poisson-regression}, the mean and variance of $y_{i}\mid x_{i}$ are
\[
E(y_{i}\mid x_{i})=\var(y_{i}\mid x_{i})=e^{x_{i}^{\T}\beta}.
\]
Because 
\[
\log E(y_{i}\mid x_{i})=x_{i}^{\T}\beta,
\]
this model is sometimes called the log-linear model, with the coefficient
$\beta_{j}$ interpreted as the conditional log mean ratio:
\[
\log\frac{E(y_{i}\mid  \ldots,x_{ij}+1,\ldots )}{E(y_{i}\mid  \ldots,x_{ij},\ldots )}=\beta_{j}.
\]

The likelihood function for independent Poisson random variables is\footnote{The notation $\propto$ means ``proportional to.'' Dropping those constants does not change the MLE.}
\[
L(\beta)=\prod_{i=1}^{n}e^{-\lambda_{i}}\frac{\lambda_{i}^{y_{i}}}{y_{i}!}\propto\prod_{i=1}^{n}e^{-\lambda_{i}}\lambda_{i}^{y_{i}},
\]
and omitting the constants, we can write the log-likelihood function
as
\begin{align*}
\log L(\beta) & =\sumn\left(-\lambda_{i}+y_{i}\log\lambda_{i}\right)=\sumn\left(-e^{x_{i}^{\T}\beta}+y_{i}x_{i}^{\T}\beta\right).
\end{align*}
The score function is
\begin{eqnarray*}
\frac{\partial\log L(\beta)}{\partial\beta} 
&=& \sumn\left(-x_{i}e^{x_{i}^{\T}\beta}+x_{i}y_{i}\right) \\
&=& \sumn x_{i}\left(y_{i}-e^{x_{i}^{\T}\beta}\right) \\
&=&\sumn x_{i}\left\{ y_{i}-\lambda(x_{i},\beta)\right\} ,
\end{eqnarray*}
and the Hessian matrix is
\begin{eqnarray*}
\frac{\partial^{2}\log L(\beta)}{\partial\beta\partial\beta^{\T}} 
& = &-\sumn x_{i}\frac{\partial}{\partial\beta^{\T}}\left(e^{x_{i}^{\T}\beta}\right) \\
&=&-\sumn e^{x_{i}^{\T}\beta}x_{i}x_{i}^{\T}, 
\end{eqnarray*}
which is negative semi-definite. When the Hessian is negative definite,
the MLE is unique. The MLE must satisfy that
\[
\sumn x_{i}\left(y_{i}-e^{x_{i}^{\T}\hat{\beta}}\right)=\sumn x_{i}\left\{ y_{i}-\lambda(x_{i},\hat{\beta})\right\} =0.
\]
We can solve this nonlinear equation using Newton's method:
\begin{align*}
\beta^{\text{new}} & =\beta^{\text{old}}-\left\{ \frac{\partial^{2}\log L(\beta^{\text{old}})}{\partial\beta\partial\beta^{\T}}\right\} ^{-1}\frac{\partial\log L(\beta^{\text{old}})}{\partial\beta}\\
 & =\beta^{\text{old}}-(X^{\T}W^{\text{old}}X)^{-1}X^{\T}(Y-\Lambda^{\text{old}}),
\end{align*}
where 
\[
X=\left(\begin{array}{c}
x_{1}^{\T}\\
\vdots\\
x_{n}^{\T}
\end{array}\right),\quad Y=\left(\begin{array}{c}
y_{1}\\
\vdots\\
y_{n}
\end{array}\right)
\]
and
\[
\Lambda^{\text{old}}=\left(\begin{array}{c}
\exp(x_{1}^{\T}\beta^{\text{old}})\\
\vdots\\
\exp(x_{n}^{\T}\beta^{\text{old}})
\end{array}\right),\quad W^{\text{old}}=\text{diag}\left\{ \exp(x_{i}^{\T}\beta^{\text{old}})\right\} _{i=1}^{n}.
\]
Similar to the derivation for the logit model, we can simplify Newton's
method to
\[
\beta^{\text{new}}=(X^{\T}W^{\text{old}}X)^{-1}X^{\T}W^{\text{old}}Z^{\text{old}},
\]
where 
\[
Z^{\text{old}}=X\beta^{\text{old}}+(W^{\text{old}})^{-1}(Y-\Lambda^{\text{old}}).
\]
So we have an iterative reweighted least squares algorithm. In \ri{R}, we can use the \ri{glm} function with ``\ri{family = poisson(link = "log")}'' to fit the Poisson regression, which uses Newton's method.

Statistical inference under Poisson regression relies on the Normal approximation to the MLE: 
\[
\hat{\beta}\asim\N\left\{ \beta,\left(-\frac{\partial^{2}\log L(\hat{\beta})}{\partial\beta\partial\beta^{\T}}\right)^{-1}\right\} =\N\left\{ \beta,(X^{\T}\hat{W}X)^{-1}\right\} ,
\]
where $\hat{W}=\text{diag} \{ \exp(x_{i}^{\T}\hat{\beta}) \} _{i=1}^{n}$.

After obtaining the MLE, we can predict the mean $E(y_i \mid x_i)$ by $ \hat{\lambda}_i =  e^{x_i^{\T} \hat{\beta}} $. Because Poisson regression is a fully parametrized model, we can also predict any other probability quantities involving $y_i \mid x_i$. For example, we can predict  $\pr(y_i = 0  \mid x_i)$ by $e^{- \hat{\lambda}_i }$, and $\pr(y_i \geq 3 \mid x_i) $ by $1 - e^{- \hat{\lambda}_i } (1 + \hat{\lambda}_i  + \hat{\lambda}_i ^2/2 )$. 
To account for the uncertainty, we can use the delta method\footnote{Review Appendix \ref{chapter::tools-limiting3} if you are unfamiliar with it.} to approximate the standard errors of the predictors.

\subsection{Negative-Binomial regression}

I then discuss the Negative-Binomial regression.

\begin{assumption}[Negative-Binomial regression model]
\label{assume::nb-regression}
We have 
$$
y_{i}  \mid x_{i} \sim\textup{NB}(\mu_{i},\theta)
$$
with 
$$
\mu_{i} =e^{x_{i}^{\T}\beta}.
$$
The observations are independent across units. The $(\beta, \theta)$ are the unknown parameters. 
\end{assumption}

Under Assumption \ref{assume::nb-regression}, the mean and variance of $y_{i}\mid x_{i}$ are
$$
E (y_{i}  \mid x_{i} ) = e^{x_{i}^{\T}\beta} ,
$$
and
$$
\var (y_{i}  \mid x_{i} ) = e^{x_{i}^{\T}\beta} (1+e^{x_{i}^{\T}\beta} / \theta) . 
$$
It is also a log-linear model. 

The log-likelihood function for Negative-Binomial regression is $\log L(\beta, \theta) = \sumn l_i(\beta, \theta)$ with
\begin{eqnarray*}
 l_i(\beta, \theta)  
&=&     \log \Gamma(y_i+\theta) - \log \Gamma(y_i + 1) - \log \Gamma(\theta)    \\
 && + \theta\log \left( \frac{\theta}{\mu_i +\theta} \right) + y_i \log \left( \frac{ \mu_i  }{ \mu_i  + \theta} \right) , 
\end{eqnarray*}
where $\mu_i  = e^{x_{i}^{\T}\beta}$ has partial derivative $\partial \mu_i / \partial \beta =  e^{x_{i}^{\T}\beta} x_i =  \mu_i x_i$. 
We can use Newton's algorithm or Fisher scoring algorithm to compute the MLE $(\hat\beta, \hat\theta)$ which requires deriving the first and second derivatives of $\log L(\beta, \theta)$ with respect to $(\beta, \theta)$. I will derive some important components and relegate other details to Problem \ref{hw19::derivatives-nb-reg}. First,
$$
\frac{ \partial \log L(\beta, \theta) }{\partial \beta} =
\sumn  (1+ \mu_i/\theta )^{-1} (y_i - \mu_i) x_i. 
$$
The corresponding first-order condition can be viewed as the estimating equation of Poisson regression with weights $  (1+ \mu_i/\theta )^{-1}$. Second,
$$
\frac{ \partial ^2 \log L(\beta, \theta) }{\partial \beta \partial \theta}
= \sumn  \frac{  \mu_i   }{  (\mu_i + \theta)^2  } (y_i - \mu_i) x_i.
$$
We can verify
$$
E\left\{ \frac{ \partial ^2 \log L(\beta, \theta) }{\partial \beta \partial \theta}  \mid X \right\} = 0
$$
since each term inside the summation has conditional expectation zero. This implies that the Fisher information matrix is diagonal, so $\hat\beta$ and $ \hat\theta$ are asymptotically independent. 

The \ri{glm.nb} in the \ri{MASS} package iterate between $\beta$ and $\theta$: given $\theta$, update $\beta$ based on Fisher scoring; given $\beta$, update $\theta$ based on Newton's algorithm. It reports standard errors based on the inverse of the Fisher information matrix.\footnote{
The command \texttt{rnbreg}   in \texttt{Stata} uses the BHHH algorithm by default, which may give slightly different numbers compared with \texttt{R}. The BHHH algorithm is similar to Newton's algorithm but avoids calculating the Hessian matrix. 
}

\subsection{Zero-inflated regressions}

I finally discuss the zero-inflated analogues of the Poisson and Negative-Binomial regressions. 

\begin{assumption}[zero-inflated Poisson regression model]
\label{assume::0-poisson-reg}
We have 
\[
y_{i}\mid x_{i}\sim\begin{cases}
0, & \text{with probability }p_{i},\\
\textup{Poisson}(\lambda_{i}), & \text{with probability }1-p_{i},
\end{cases}
\]
where 
$$
p_{i}  =\frac{e^{x_{i}^{\T}\gamma}}{1+e^{x_{i}^{\T}\gamma}},\quad 
\lambda_{i}  = e^{x_{i}^{\T}\beta}.
$$
The observations are independent across units. The $(\gamma, \beta)$ are the unknown parameters. 
\end{assumption}

\begin{assumption}[zero-inflated Negative-Binomial regression model]
\label{assume::0-nb-reg}
We have 
\[
y_{i}\mid x_{i}\sim\begin{cases}
0, & \text{with probability }p_{i},\\
\textup{NB}(\mu_{i},\theta), & \text{with probability }1-p_{i},
\end{cases}
\]
where 
$$
p_{i}  =\frac{e^{x_{i}^{\T}\gamma}}{1+e^{x_{i}^{\T}\gamma}},\quad 
\mu_{i}  = e^{x_{i}^{\T}\beta} .
$$
The observations are independent across units. The $(\gamma, \beta, \theta)$ are the unknown parameters. 
\end{assumption}

To avoid over-parametrization, we can also restrict some coefficients
to be zero. 
The  \ri{zeroinfl} function in the \ri{R} package \ri{pscl} can fit the zero-inflated Poisson and Negative-Binomial regressions.

\section{A case study}\label{sec::count-case-study}

I will use the dataset from \citet{royer2015incentives} to illustrate the regressions for count outcomes. From the regression formula below, I will estimate the effects of two treatments \ri{incentive_commit} and \ri{incentive} on the number of visits to the gym, controlling for two pretreatment covariates \ri{target} and \ri{member_gym_pre}. 

\begin{lstlisting}
library("ggplot2")
library("gridExtra")
library("foreign")
library("MASS")
gym1 = read.dta("gym_treatment_exp_weekly.dta")
f.reg = weekly_visit ~ incentive_commit + incentive + 
  target + member_gym_pre
\end{lstlisting}

\subsection{Linear, Poisson, and Negative-Binomial regressions}

\begin{figure}[ht]
\centering
\includegraphics[width = 0.95\textwidth]{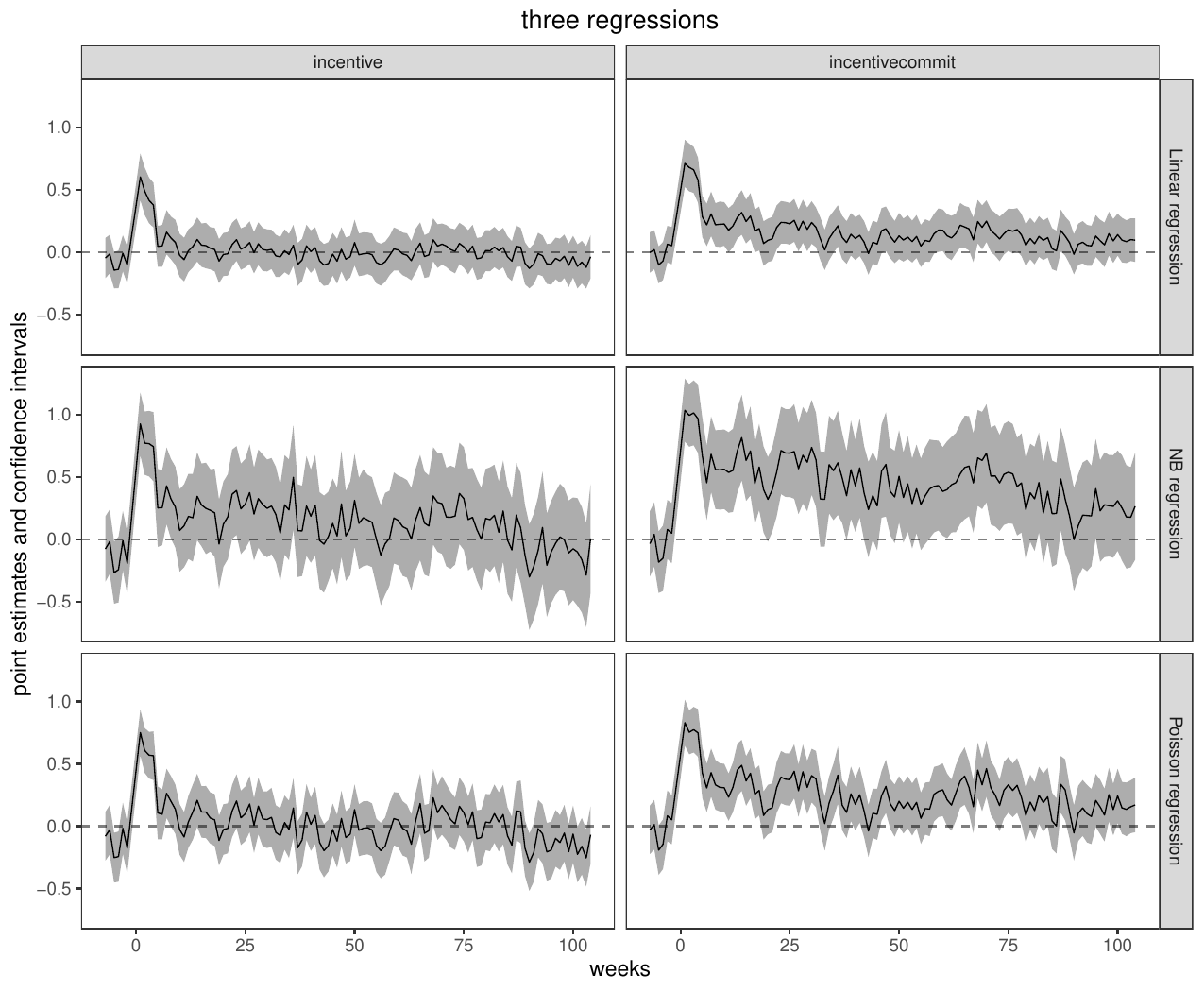}
\caption{Linear, Poisson, and Negative-Binomial regressions}\label{fig::three regressions}
\end{figure}

Each worker was observed over time. Therefore, I run regressions with the outcome data observed in each week. In the following, I compute the linear regression coefficients, standard errors, and AICs. 

\begin{lstlisting}
> weekids             = sort(unique(gym1$incentive_week))
> lweekids            = length(weekids)
> coefincentivecommit = 1:lweekids
> coefincentive       = 1:lweekids
> seincentivecommit   = 1:lweekids
> seincentive         = 1:lweekids
> AIClm               = 1:lweekids
> for(i in 1:lweekids)
+ {
+         gymweek = gym1[which(gym1$incentive_week == weekids[i]), ]   
+         regweek = lm(f.reg, data = gymweek)
+         regweekcoef = summary(regweek)$coef
+         
+         coefincentivecommit[i] = regweekcoef[2, 1]
+         coefincentive[i]       = regweekcoef[3, 1]
+         seincentivecommit[i]   = regweekcoef[2, 2]
+         seincentive[i]         = regweekcoef[3, 2]
+         
+         AIClm[i]               = AIC(regweek)
+ }
\end{lstlisting}

By changing the line with \ri{lm} to 
\begin{lstlisting}
regweek = glm(f.reg, family = poisson(link = "log"), data = gymweek)
\end{lstlisting}
and 
\begin{lstlisting}
regweek = glm.nb(f.reg, data = gymweek)
\end{lstlisting}
I obtain the corresponding results from Poisson and Negative-Binomial regressions, respectively. Figure \ref{fig::three regressions} compares the regression coefficients with the associated confidence intervals over time. Three regressions give very similar patterns: \ri{incentive_commit} has both short-term and long-term effects, but \ri{incentive} only has short-term effects.

\begin{figure}
\centering
\includegraphics[width = 0.8\textwidth]{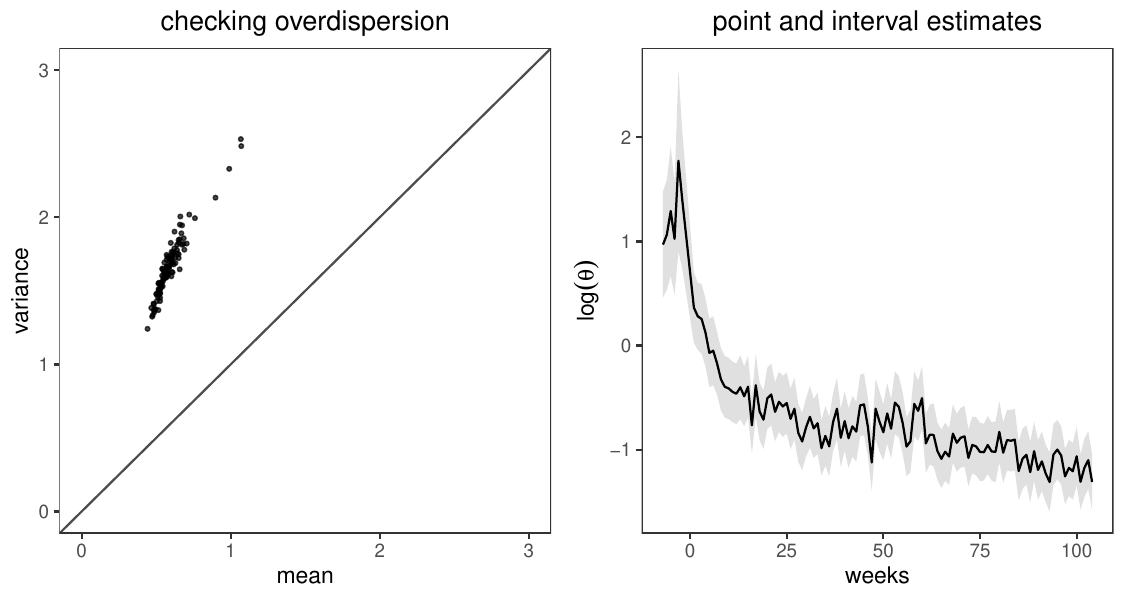}
\caption{Overdispersion of the data}\label{fig::overdispersion-gym}
\end{figure}

The left panel of Figure \ref{fig::overdispersion-gym} shows that variances are larger than the means for outcomes from all weeks, and the right panel of Figure \ref{fig::overdispersion-gym} shows the point estimates and confidence intervals of $\theta$ from Negative-Binomial regressions. Overall, overdispersion seems an important feature of the data.

\subsection{Zero-inflated regressions}

\begin{figure}[ht]
\centering
\includegraphics[width = 0.8\textwidth]{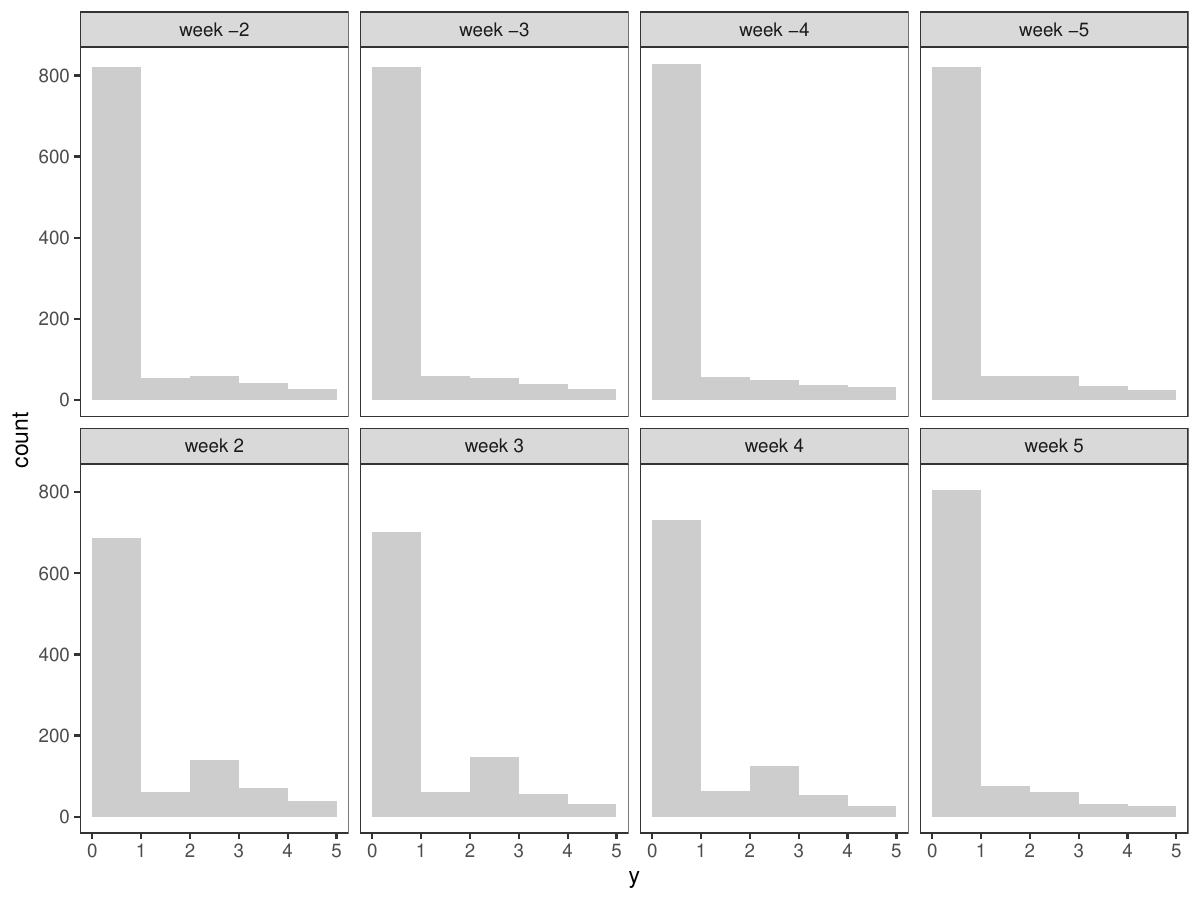}
\caption{Zero-inflation of the data}\label{fig::zeroinflation-gym}
\end{figure}

Figure \ref{fig::zeroinflation-gym} plots the histograms of the outcomes from four weeks before and four weeks after the experiment. Eight histograms all show severe zero inflation because most workers just did not go to the gym regardless of the treatments. Therefore, it seems crucial to accommodate the zeros in the models. 

\begin{figure}[ht]
\centering
\includegraphics[width = 0.95\textwidth]{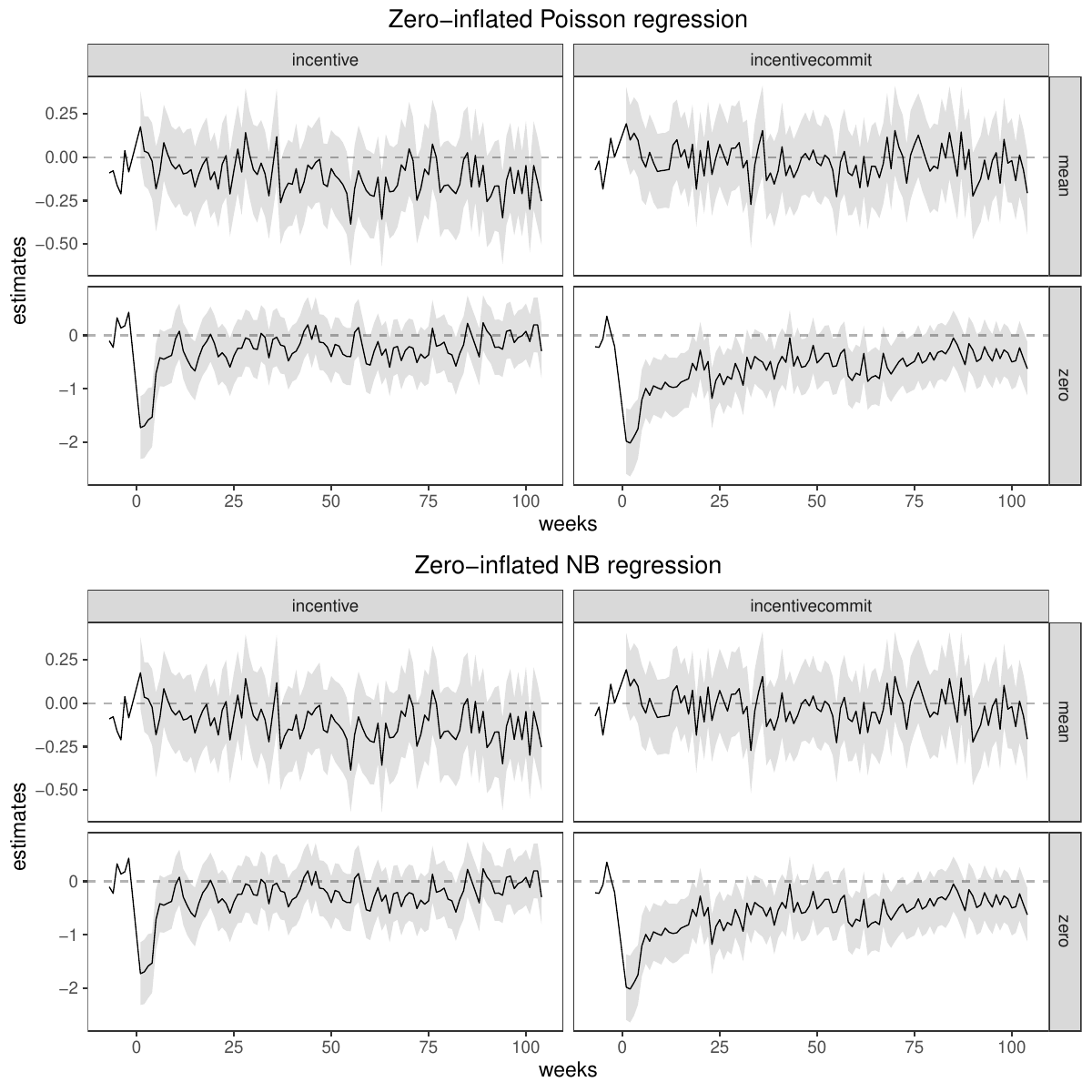}
\caption{Zero-inflated regressions}\label{fig::zeroinflated-regressions-gym}
\end{figure}

I now fit zero-inflated Poisson regressions. The model has parameters for the zero component and parameters for the Poisson components. 

\begin{lstlisting}
> library("pscl")
> coefincentivecommit0 = coefincentivecommit
> coefincentive0       = coefincentive
> seincentivecommit0   = seincentivecommit
> seincentive0         = seincentive
> AIC0poisson          = AICnb
> for(i in 1:lweekids)
+ {
+   gymweek = gym1[which(gym1$incentive_week == weekids[i]), ]   
+   regweek = zeroinfl(f.reg, dist = "poisson", data = gymweek)
+   regweekcoef = summary(regweek)$coef
+   
+   coefincentivecommit[i] = regweekcoef$count[2, 1]
+   coefincentive[i]       = regweekcoef$count[3, 1]
+   seincentivecommit[i]   = regweekcoef$count[2, 2]
+   seincentive[i]         = regweekcoef$count[3, 2]
+   
+   coefincentivecommit0[i] = regweekcoef$zero[2, 1]
+   coefincentive0[i]       = regweekcoef$zero[3, 1]
+   seincentivecommit0[i]   = regweekcoef$zero[2, 2]
+   seincentive0[i]         = regweekcoef$zero[3, 2]
+   
+   AIC0poisson[i]          = AIC(regweek)
+ }
\end{lstlisting}

Replacing the line with \ri{zeroinfl} by 
\begin{lstlisting}
  regweek = zeroinfl(f.reg, dist = "negbin", data = gymweek)
\end{lstlisting}
we can fit the corresponding zero-inflated Negative-Binomial regressions. Figure \ref{fig::zeroinflated-regressions-gym} plots the point estimates and the confidence intervals of the coefficients of the treatment. It shows that the treatments do not have effects on the Poisson or Negative-Binomial components, but have effects on the zero components. This suggests that the treatments affect the outcome mainly by changing the workers' behavior of whether to go to the gym.

Another interesting result is the large $\hat{\theta}$'s from the zero-inflated Negative-Binomial regression:
\begin{lstlisting}
> quantile(gymtheta, probs = c(0.01, 0.25, 0.5, 0.75, 0.99))
  1%  25%  50%  75%  99% 
12.3 13.1 13.7 14.4 15.7 
\end{lstlisting}

Once the zero-inflated feature has been modeled, it is not crucial to account for the overdispersion. It is reasonable because the maximum outcome is five, ruling out heavy-tailedness. This is further corroborated by the following comparison of the AICs from five regression models. 

\begin{lstlisting}
> diff.aic = AIC0nb - AIC0poisson
> quantile(diff.aic, probs = c(0.01, 0.25, 0.5, 0.75, 0.99))
      1%      25%      50%      75%      99% 
2.000000 2.000013 2.000024 2.000031 2.002898 
\end{lstlisting}

\begin{figure}[ht]
\centering
\includegraphics[width = \textwidth]{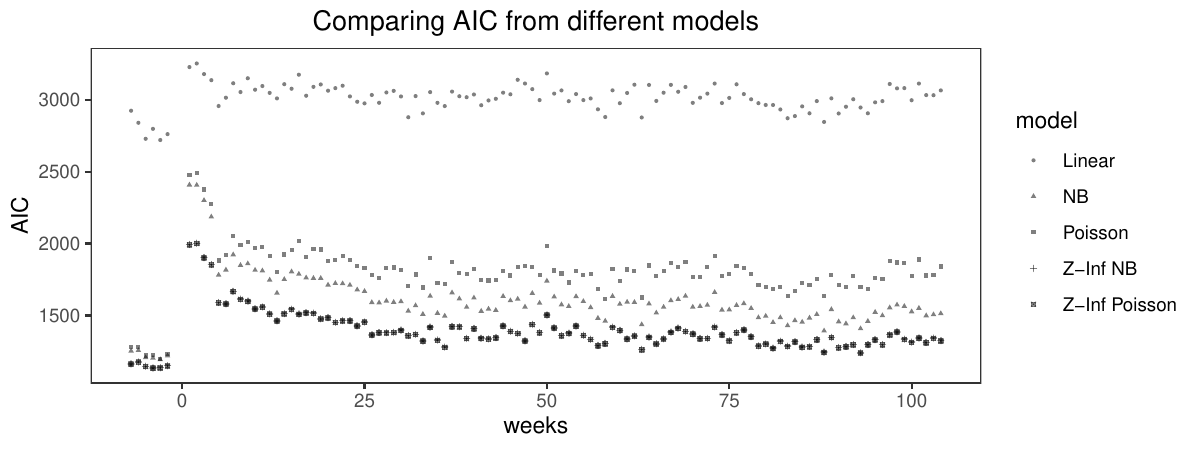}
\caption{Comparing AICs from five regression models}\label{fig::comparing-aic-gym}
\end{figure}

Figure \ref{fig::comparing-aic-gym} shows that zero-inflated Poisson regressions have the smallest AICs, beating the zero-inflated Negative-Binomial regressions, which are more flexible but have more parameters to estimate. 

\section{Homework problems}

 \paragraph{Basic properties of Poisson}\label{hw19::poisson12}
 
Prove Propositions \ref{prop::poisson-1} and \ref{prop::poisson-2}.

 \paragraph{Basic properties of Negative Binomial}\label{hw19::nb-properties}

Prove Proposition \ref{proposition:The-Negative-Binomial-random}.

\paragraph{Newton's method for Negative-Binomial regression}\label{hw19::derivatives-nb-reg}

Calculate the score function and Hessian matrix based on the log-likelihood function of the Negative-Binomial regression.
What is the joint asymptotic distribution of the MLE $(\hat\beta, \hat\theta)$?

\paragraph{Moments of Zero-inflated Poisson and Negative-Binomial}\label{hw19::moment-0-poisson}

Prove Propositions \ref{proposition:moments-zeroinflated-poisson} and \ref{proposition:moments-zeroinflated-NB}.

\paragraph{Overdispersion and zero-inflation}

Prove that for a zero-inflated Poisson, if $p \leq 1/2$ then $E(y) < \var(y)$ always holds. What is the condition for $E(y) < \var(y)$ when $p>1/2$?

\paragraph{Poisson latent variable and the binary regression model with the cloglog link}\label{hw19::poisson-logistic}

Assume that $y_{i}^* \mid x_{i}\sim\text{Poisson}(e^{x_{i}^{\T}\beta})$,
and define $y_{i}=1(y_{i}^*>0)$ as the indicator that $y_{i}^*$ is
not zero. 

Prove that $y_{i} \mid x_{i}$ follows a cloglog model, that
is,
\[
\pr(y_{i} =1\mid x_{i})=g(x_{i}^{\T}\beta),
\]
where $g(z)=1-\exp(-e^{z})$. 

Remark: The cloglog model for binary outcome arises naturally from a latent Poisson model. It was only briefly mentioned in Chapter \ref{chapter::binary-logit}.

\paragraph{Likelihood for the zero-inflated Poisson regression}

Write down the likelihood function for the Zero-inflated Poisson model,
and derive the steps for Newton's method. 

\paragraph{Likelihood for the Zero-inflated Negative-Binomial regression}

Write down the likelihood function for the Zero-inflated Negative-Binomial model, and
derive the steps for Newton's method.

\paragraph{Prediction in the Zero-inflated Negative-Binomial regression}

After obtaining the MLE $(\hat{\beta}, \hat{\gamma})$ and its asymptotic covariance matrix $\hat{V}$, predict the conditional mean $E(y_i \mid x_i)$, the conditional probability $\pr(y_i=0\mid x_i)$, and the conditional probability $\pr(y_i \geq 5\mid x_i)$. What are the associated asymptotic standard errors?

\paragraph{Data analysis}

\citet{zeileis2008regression} gives a tutorial on count outcome regressions using the dataset from \citet{deb1997demand}. Replicate and extend their analysis based on the discussion in this chapter.

\paragraph{Data analysis}

\citet{fisman2007corruption} was an application of Negative-Binomial regression, and \citet{albergaria2017narrow} replicated their study and argued that the zero-inflated Negative-Binomial regression was more appropriate. Replicate and extend their analysis based on the discussion in this chapter.

\chapter{Generalized Linear Models: A Unification}
   \chaptermark{GLM: a Unification}

This chapter unifies Chapters \ref{chapter::binary-logit}--\ref{chapter::count} under the formulation of the generalized linear model (GLM) by \citet{nelder1972generalized}.

\section{Generalized Linear Models}

So far we have discussed the following models for independent observations
$(y_{i},x_{i})_{i=1}^{n}$.

\begin{example}\label{eg::gaussianlinearmodel}
The Normal linear model for continuous outcomes assumes
\begin{equation}
y_{i}\mid x_{i}\sim\N(\mu_{i},\sigma^{2}),\quad\text{with }\mu_{i}=x_{i}^{\T}\beta.\label{eq:gaussianlinearmodel}
\end{equation}
\end{example}

\begin{example}\label{eg::binarylogisticmodel}
The logistic model for binary outcomes assumes
\begin{equation}
y_{i}\mid x_{i}\sim\textup{Bernoulli}(\mu_{i}),\quad\text{with }\mu_{i}=\frac{e^{x_{i}^{\T}\beta}}{1+e^{x_{i}^{\T}\beta}} . \label{eq:logisticlinearmodel}
\end{equation}
\end{example}

\begin{example}\label{eg::countpoissonmodel}
The Poisson model for count outcomes assumes 
\begin{equation}
y_{i}\mid x_{i}\sim\textup{Poisson}(\mu_{i}),\quad\text{with }\mu_{i}=e^{x_{i}^{\T}\beta} . \label{eq:poissonlinearmodel}
\end{equation}
\end{example}

\begin{example}\label{eg::countnbmodel}
The Negative-Binomial model for overdispersed count outcomes assumes
\begin{equation}
y_{i}\mid x_{i}\sim\textsc{NB}(\mu_{i},\delta),\quad\text{with }\mu_{i}=e^{x_{i}^{\T}\beta} . \label{eq: nblinearmodel}
\end{equation}
We use $\delta$ for the dispersion parameter to avoid
confusion because $\theta$ means something else below (Chapter \ref{chapter::count} uses $\theta$ for the dispersion parameter; \ri{R} also uses $\theta$ for the dispersion parameter). 
\end{example}

In the above models, $\mu_i$ denotes the conditional mean. This chapter 
will unify Examples \ref{eg::gaussianlinearmodel}--\ref{eg::countnbmodel} as GLMs. 

\subsection{Exponential family\label{subsec:Exponential-family}}

Consider a general conditional probability density or mass function:
\begin{equation}
f(y_{i}\mid x_{i};\theta_{i},\phi)=\exp\left\{ \frac{y_{i}\theta_{i}-b(\theta_{i})}{a(\phi)}+c(y_{i},\phi)\right\} ,\label{eq:natualexponentialfamily-dispersion}
\end{equation}
where $(\theta_{i},\phi)$ are unknown parameters, and $\left\{ a(\cdot),  b(\cdot),c(\cdot,\cdot)\right\} $
are known functions. The above conditional density (\ref{eq:natualexponentialfamily-dispersion})
is called the {\it natural exponential family with dispersion},\footnote{\citet{jorgensen1987exponential} used the name ``exponential dispersion model'' and provided a unified discussion.} where $\theta_{i}$
is the natural parameter depending on $x_{i}$, and $\phi$ is the
dispersion parameter. Sometimes, $a(\phi)=1$ and $c(y_{i},\phi)=c(y_{i})$,
simplifying the conditional density to a {\it natural exponential family}. Examples \ref{eg::gaussianlinearmodel}--\ref{eg::countnbmodel}
have a unified structure as (\ref{eq:natualexponentialfamily-dispersion}), as detailed below. 

\setcounter{example}{0}
\begin{example}[continued]\label{eg::gaussianlinearmodel-continue}
 Model (\ref{eq:gaussianlinearmodel}) has conditional probability
density function
\begin{align*}
f(y_{i}  \mid x_{i};\mu_{i},\sigma^{2}) &=(2\pi\sigma^{2})^{-1/2}\exp\left\{ -\frac{(y_{i}-\mu_{i})^{2}}{2\sigma^{2}}\right\} \\
 & =\exp\left\{ \frac{y_{i}\mu_{i}-\frac{1}{2}\mu_{i}^{2}}{\sigma^{2}}-\frac{1}{2}\log(2\pi\sigma^{2})-\frac{y_{i}^{2}}{2\sigma^{2}}\right\} ,
\end{align*}
with 
\[
\theta_{i}=\mu_{i},\quad b(\theta_{i})=\frac{1}{2}\theta_{i}^{2},
\]
and
$$
\phi=\sigma^{2},\quad a(\phi)=\sigma^{2} = \phi.
$$ 
\end{example}

\begin{example}[continued]\label{eg::binarylogisticmodel-continue}
 Model (\ref{eq:logisticlinearmodel}) has conditional probability
mass function
\begin{align*}
f(y_{i}  \mid x_{i};\mu_{i}) &=\mu_{i}^{y_{i}}(1-\mu_{i})^{1-y_{i}} \\
&=\left(\frac{\mu_{i}}{1-\mu_{i}}\right)^{y_{i}}(1-\mu_{i})\\
 & =\exp\left\{ y_{i}\log\frac{\mu_{i}}{1-\mu_{i}}-\log\frac{1}{1-\mu_{i}}\right\} ,
\end{align*}
with
\[
\theta_{i}=\log\frac{\mu_{i}}{1-\mu_{i}} \Longleftrightarrow \mu_i = \frac{e^{\theta_{i}}}{1 + e^{\theta_{i}}},
\quad b(\theta_{i})=\log\frac{1}{1-\mu_{i}}=\log(1+e^{\theta_{i}}),
\]
and
$$
 a(\phi)=1.
$$
\end{example}

\begin{example}[continued]\label{eg::countpoissonmodel-continue}
 Model (\ref{eq:poissonlinearmodel}) has conditional probability mass
function
\begin{align*}
f(y_{i}  \mid x_{i};\mu_{i}) &=e^{-\mu_{i}}\frac{\mu_{i}^{y_i}}{y_i!} \\
&  =\exp\left\{ y_{i}\log\mu_{i}-\mu_{i}-\log y_{i}!\right\} ,
\end{align*}
with
\[
\theta_{i}=\log\mu_{i},\quad b(\theta_{i})=\mu_{i}=e^{\theta_{i}},
\]
and
$$
a(\phi)=1.
$$
\end{example}

\begin{example}[continued]\label{eg::countnbmodel-continue}
 Model (\ref{eq: nblinearmodel}), for a {\it fixed} $\delta$, has conditional
probability mass function
\begin{align*}
&f(y_{i}  \mid x_{i};\mu_{i})  \\
& =\frac{\Gamma(y_{i}+\delta)}{\Gamma(\delta )\Gamma(y_{i}+1)}\left(\frac{\mu_{i}}{\mu_{i}+\delta}\right)^{y_{i}}\left(\frac{\delta}{\mu_{i}+\delta}\right)^{\delta}\\
 & =\exp\left\{ y_{i}\log\frac{\mu_{i}}{\mu_{i}+\delta}-\delta\log\frac{\mu_{i}+\delta}{\delta}  
  +\log\Gamma(y_{i}+\delta)-\log\Gamma(\delta)-\log\Gamma(y_{i}+1)\right\} ,
\end{align*}
with 
\[
\theta_{i}=\log\frac{\mu_{i}}{\mu_{i}+\delta}  \Longleftrightarrow  \frac{\delta}{  \mu_{i}+\delta  } = 1- e^{\theta_i}  ,
\quad b(\theta_{i})=\delta\log\frac{\mu_{i}+\delta}{\delta}= - \delta\log (1-e^{\theta_{i}}) ,
\]
and
$$
 a(\phi)=1.
$$
\end{example}

The logistic and Poisson models are simpler without the dispersion parameter.
The Normal linear model has a dispersion parameter for the variance.
The Negative-Binomial model is more complex: without fixing $\delta$
it does not belong to the exponential family with dispersion. 

The exponential family (\ref{eq:natualexponentialfamily-dispersion})
has elegant forms of the first two moments.

\begin{theorem}
\label{theorem:The-first-two-moments-NEF}The first two moments of (\ref{eq:natualexponentialfamily-dispersion})
are
$$
E(y_{i}\mid x_{i};\theta_{i},\phi)\equiv\mu_{i}=b'(\theta_{i})
$$
and
$$
\var(y_{i}\mid x_{i};\theta_{i},\phi)\equiv\sigma_{i}^{2}=b''(\theta_{i})a(\phi).
$$ 
\end{theorem}

\begin{myproof}{Theorem}{\ref{theorem:The-first-two-moments-NEF}}
We can use Bartlett's identities \citep{bartlett1953approximate} to prove the theorem; see Appendix \ref{sec::mle} for a review. 
The first two derivatives of the log conditional density are
\[
\frac{\partial\log f(y_{i}\mid x_{i};\theta_{i},\phi)}{\partial\theta_{i}}=\frac{y_{i}-b'(\theta_{i})}{a(\phi)},\]
and
\[
\quad\frac{\partial^{2}\log f(y_{i}\mid x_{i};\theta_{i},\phi)}{\partial\theta_{i}^{2}}=-\frac{b''(\theta_{i})}{a(\phi)}.
\]
Lemma \ref{lemma:bartlett-identity} implies that
\[
E\left\{ \frac{y_{i}-b'(\theta_{i})}{a(\phi)}\right\} =0,
\]
and
\[
 E\left[\left\{ \frac{y_{i}-b'(\theta_{i})}{a(\phi)}\right\} ^{2}\right]=\frac{b''(\theta_{i})}{a(\phi)},
\]
which further imply the first two moments of $y_{i}$ given $x_i$. 
\end{myproof}

\subsection{Generalized linear model}

\begin{figure}[ht]
\centering 
\includegraphics[width = \textwidth]{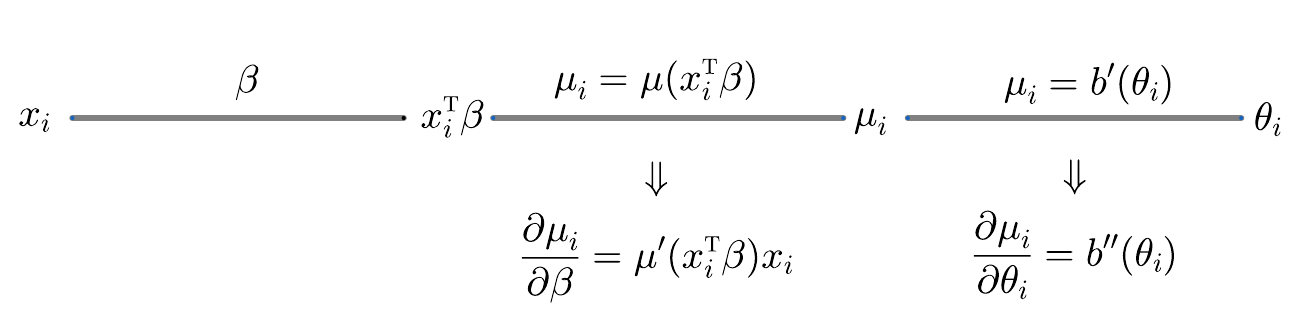}
\caption{Quantities in a GLM\label{fig:Quantities-in-a-glm}}
\end{figure}

Section \ref{subsec:Exponential-family} is general, allowing the
mean parameter $\mu_{i}$ to depend on $x_{i}$ in an arbitrary way.
This flexibility does not immediately generate a useful statistical
procedure. To borrow information across observations, we assume that
the relationship between $y_{i}$ and $x_{i}$ remain ``stable'' and
can be captured by a fixed parameter $\beta$. A simple starting point
is to use $x_{i}^{\T}\beta$ to approximate $\mu_{i}$, which, however,
works naturally only for outcomes taking values in a wide range of $(-\infty, \infty)$. For general outcome variables,
we can link its mean and the linear combination of covariates by
\[
\mu_{i}=\mu(x_{i}^{\T}\beta),
\]
where $\mu(\cdot)$ is a known function and $\beta$ is an unknown
parameter. The inverse of $\mu(\cdot)$ is called the link function. This is called a GLM, which
has the following components:
\begin{enumerate}[label=(C\arabic*), ref=C\arabic*] 
\item the conditional distribution (\ref{eq:natualexponentialfamily-dispersion})
as an exponential family with dispersion;
\item the conditional mean $\mu_{i}=b'(\theta_{i})$ and variance $\sigma_{i}^{2}=b''(\theta_{i})a(\phi)$
in Theorem \ref{theorem:The-first-two-moments-NEF};
\item the function linking the conditional mean and covariates $\mu_{i}=\mu(x_{i}^{\T}\beta)$. 
\end{enumerate}
Models (\ref{eq:gaussianlinearmodel})--(\ref{eq: nblinearmodel})
are the classical examples. 
Figure \ref{fig:Quantities-in-a-glm} illustrates the relationship among key quantities in a GLM.  In particular,
\begin{equation}
\theta_{i}=(b')^{-1}(\mu_{i})=(b')^{-1}\left\{ \mu(x_{i}^{\T}\beta)\right\} \label{eq:theta-beta-link}
\end{equation}
depends on $x_i$ and $\beta$, with $(b')^{-1}$ indicating the inverse function of $b'(\cdot)$.

\section{MLE for GLM}\label{sec::mle-glm-fisher}

The contribution of unit $i$ to the log-likelihood function is
\[
\ell_{i}=\log f(y_{i}\mid x_{i};\beta,\phi)=\frac{y_{i}\theta_{i}-b(\theta_{i})}{a(\phi)}+c(y_{i},\phi).
\]
The  contribution of unit $i$ to the score function is
\[
\frac{\partial\ell_{i}}{\partial\beta}=\frac{\partial\ell_{i}}{\partial\theta_{i}}\frac{\partial\theta_{i}}{\partial\mu_{i}}\frac{\partial\mu_{i}}{\partial\beta},
\]
where 
\begin{align*}
\frac{\partial\ell_{i}}{\partial\theta_{i}} & =\frac{y_{i}-b'(\theta_{i})}{a(\phi)},\\
\frac{\partial\theta_{i}}{\partial\mu_{i}} & =\frac{1}{b''(\theta_{i})}=\frac{a(\phi)}{\sigma_{i}^{2}}
\end{align*}
follow from Theorem \ref{theorem:The-first-two-moments-NEF}. 
So 
\[
\frac{\partial\ell_{i}}{\partial\beta}=\frac{y_{i}-b'(\theta_{i})}{\sigma_{i}^{2}}\frac{\partial\mu_{i}}{\partial\beta}=\frac{y_{i}-\mu_{i}}{\sigma_{i}^{2}}\frac{\partial\mu_{i}}{\partial\beta},
\]
leading to the following score equation for the MLE:
\begin{equation}
\sumn\frac{y_{i}-\mu_{i}}{\sigma_{i}^{2}}\frac{\partial\mu_{i}}{\partial\beta}=0,\label{eq:score-equation-mle-form1}
\end{equation}
or, more explicitly, 
\[
\sumn\frac{y_{i}-\mu (x_{i}^{\T}\beta) }{\sigma_{i}^{2}}\mu'(x_{i}^{\T}\beta)x_{i}=0 . 
\]

The general relationship (\ref{eq:theta-beta-link}) between $\theta_{i}$
and $\beta$ is quite complicated. A natural choice of $\mu(\cdot)$
is to cancel $(b')^{-1}$ in \eqref{eq:theta-beta-link} so that 
\[
\mu(\cdot)=b'(\cdot) ,
\]
which implies
\[
\theta_{i}=x_{i}^{\T}\beta.
\]
This link function $\mu(\cdot)$ is called the canonical link or the natural link, which leads
to further simplifications: By
\[ 
\mu_{i}  =b'(x_{i}^{\T}\beta),
\]
we have
\[
\frac{\partial\mu_{i}}{\partial\beta}=b''(x_{i}^{\T}\beta)x_{i}
= b''( \theta_i )x_{i}
=\frac{\sigma_{i}^{2}}{a(\phi)}x_{i}.
\] 
Therefore, 
\[
\frac{\partial\ell_{i}}{\partial\beta} = \frac{y_{i}-\mu_{i}}{\sigma_{i}^{2}}\frac{\sigma_{i}^{2}}{a(\phi)}x_{i} 
= a(\phi)^{-1} x_i (y_{i}-\mu_{i}) .
\]
The score equation simplifies to
\[ 
a(\phi)^{-1} \sumn  x_i (y_{i}-\mu_{i})    =0 ,
\]
which further simplifies to
\begin{eqnarray}
 \sumn x_{i}\left(y_{i}-\mu_{i}\right)=0.\label{eq:NormalEquation-NEF-GLM}
\end{eqnarray} 
We have shown that the MLEs of models (\ref{eq:gaussianlinearmodel})--(\ref{eq:poissonlinearmodel})
all satisfy (\ref{eq:NormalEquation-NEF-GLM}). However, the MLE of (\ref{eq: nblinearmodel})
does not because it does not use the natural link function resulting in $\mu(\cdot) \neq b'(\cdot) $: 
$$
\mu(*) = e^{*},\quad 
b'(*) = \delta \frac{e^*}{ 1 - e^*}. 
$$

Using Bartlett's second identity in Lemma \ref{lemma:bartlett-identity} in Appendix \ref{chapter::m-mle},
we can write the expected Fisher information conditional on covariates
as
\begin{align*}
\sumn E\left(\frac{\partial\ell_{i}}{\partial\beta}\frac{\partial\ell_{i}}{\partial\beta^{\T}}\mid x_{i}\right) & =\sumn E\left\{ \left(\frac{y_{i}-\mu_{i}}{\sigma_{i}^{2}}\right)^{2}\frac{\partial\mu_{i}}{\partial\beta}\frac{\partial\mu_{i}}{\partial\beta^{\T}}\mid x_{i}\right\} \\
 & =\sumn\frac{1}{\sigma_{i}^{2}}\frac{\partial\mu_{i}}{\partial\beta}\frac{\partial\mu_{i}}{\partial\beta^{\T}}\\
 & =\sumn\frac{1}{\sigma_{i}^{2}}\left\{ \mu'(x_{i}^{\T}\beta)\right\} ^{2}x_{i}x_{i}^{\T}\\
 & =X^{\T}WX,
\end{align*}
where 
\[
W=\text{diag}\left\{ \frac{1}{\sigma_{i}^{2}}\left\{ \mu'(x_{i}^{\T}\beta)\right\} ^{2}\right\} _{i=1}^{n}.
\]
With the canonical link, it further simplifies to
\begin{align*}
\sumn E\left(\frac{\partial\ell_{i}}{\partial\beta}\frac{\partial\ell_{i}}{\partial\beta^{\T}}\mid x_{i}\right) & =\sumn E\left\{ \left(\frac{y_{i}-\mu_{i}}{a(\phi)}\right)^{2}x_{i}x_{i}^{\T}\mid x_{i}\right\} \\
 & =\left\{ a(\phi)\right\} ^{-2}\sumn\sigma_{i}^{2}x_{i}x_{i}^{\T} . 
\end{align*}

We can obtain the estimated covariance matrix by replacing the unknown parameters with their estimates. 
Now we review the estimated covariance matrices of the classical GLMs with canonical links. 
\setcounter{example}{0}
\begin{example}[continued]\label{eg::gaussianlinearmodel-continue1}
In the Normal linear model,  
$
\hat{V} =  \hat{\sigma}^2 (X^{\T} X)^{-1}
$
with $ \sigma^2$ estimated further by the residual sum of squares. 
\end{example}

\begin{example}[continued]\label{eg::binarylogisticmodel-continue1}
In the binary logistic model,  
$
\hat{V} = (X^{\T} \hat{W} X)^{-1}
$
with  $\hat{W}= \textup{diag}\{ \hat{\pi}_i (1-\hat{\pi}_i) \}_{i=1}^n $, where $ \hat{\pi}_i  = e^{x_i^{\T} \hat{\beta}} / (1+e^{x_i^{\T} \hat{\beta}}).$
\end{example}

\begin{example}[continued]\label{eg::countpoissonmodel-continue1}
In the Poisson model,  
$
\hat{V} = (X^{\T} \hat{W} X)^{-1}
$
with  $\hat{W}= \textup{diag}\{ \hat{\lambda}_i   \}_{i=1}^n $, where $\hat{\lambda}_i  = e^{x_i^{\T} \hat{\beta}}.$
\end{example}

I relegate the derivation of the formula for the Negative-Binomial regression as Problem \ref{hw20::nb-sandwich}. It is a purely theoretical exercise since $\delta$ is usually unknown in practice.

\section{Other GLMs}

The \ri{glm} function in \ri{R} allows for the specification of the \ri{family} parameters, with the corresponding canonical link functions shown below: 
\begin{lstlisting}
binomial(link = "logit")
gaussian(link = "identity")
Gamma(link = "inverse")
inverse.gaussian(link = "1/mu^2")
poisson(link = "log")
quasi(link = "identity", variance = "constant")
quasibinomial(link = "logit")
quasipoisson(link = "log")
\end{lstlisting}
Examples \ref{eg::gaussianlinearmodel-continue1}--\ref{eg::countpoissonmodel-continue1} correspond to the second, the first, and the fifth choices above. Below I will briefly discuss the third choice for the Gamma regression and omit the discussion of other choices. See the help file of the \ri{glm} function and \citet{mccullagh1989generalized} for more details.

The Gamma$(\alpha, \beta)$ random variable is positive with mean $\alpha/\beta$ and variance $\alpha/\beta^2$. For convenience in modeling, we use a reparametrization Gamma$'(\mu, \nu)$, where  
$$
\begin{pmatrix}
\mu \\
\nu
\end{pmatrix}
= \begin{pmatrix}
\alpha/\beta \\
\alpha 
\end{pmatrix}
,
$$
which is equivalent to
$$
\begin{pmatrix}
\alpha \\
\beta
\end{pmatrix}
= \begin{pmatrix}
\nu\\
\nu / \mu 
\end{pmatrix}.
$$
So its mean equals $\mu$ and its variance equals $\mu^2 / \nu$ which is quadratic in $\mu$. A feature of the Gamma random variable is that its coefficient of variation equals $1 /  \sqrt{\nu}$, which does not depend on the mean. So  Gamma$'(\mu, \nu)$ is a parametrization based on the mean and the coefficient of variation \citep{mccullagh1989generalized}.\footnote{The coefficient of variation of a random variable $A$ equals  $\sqrt{\var(A)} / E(A)$. } 
Gamma$'(\mu, \nu)$ has density
$$
f(y)  =  \frac{\beta^\alpha }{ \Gamma (\alpha) } y^{\alpha - 1} e^{- \beta y} 
=  \frac{ (\nu / \mu )^\nu }{ \Gamma (\nu) } y^{\nu - 1} e^{- (\nu / \mu ) y} ,
$$ 
and we can verify that it belongs to the exponential family with dispersion. 
Gamma regression imposes Assumption \ref{assume::gamma-reg} below. 

\begin{assumption}[Gamma regression model]
\label{assume::gamma-reg}
We have 
$$
y_i \mid x_i \sim \textup{Gamma}' (\mu_i, \nu)
$$
with 
$$
\mu_i = e^{x_i^{\T} \beta}.
$$
The observations are independent across units. The $(\beta, \nu)$ are the unknown parameters. 
\end{assumption}
So the Gamma regression model is also a log-linear model. Assumption \ref{assume::gamma-reg} does not correspond to the canonical link. Instead, we should specify \ri{Gamma(link = "log")} to fit the log-linear Gamma regression model under Assumption \ref{assume::gamma-reg}. 

The log-likelihood function is
$$
\log L(\beta, \nu) = 
\sumn \left\{ 
- \frac{  \nu y_i }{  e^{x_i^{\T} \beta} }  + (\nu - 1)\log y_i + \nu \log \nu - \nu x_i^{\T} \beta - \log\Gamma(\nu)
\right\} .
$$
Then
\begin{eqnarray*}
\frac{ \partial  \log L(\beta, \nu) }{\partial \beta } 
&=& 
\sumn ( \nu y_i e^{ - x_i^{\T} \beta}  x_i  - \nu x_i )  \\
&=& 
\nu \sumn e^{ - x_i^{\T} \beta}  (y_i -  e^{x_i^{\T} \beta}) x_i  
\end{eqnarray*}
and
$$
\frac{ \partial^2  \log L(\beta, \nu) }{\partial \beta \partial \nu } =
 \sumn e^{ - x_i^{\T} \beta}  (y_i -  e^{x_i^{\T} \beta}) x_i  .
$$
So the MLE for $\beta$ solves the following estimating equation
$$
\sumn e^{ - x_i^{\T} \beta}  (y_i -  e^{x_i^{\T} \beta}) x_i   = 0.
$$
Moreover,  $ \partial^2  \log L(\beta, \nu)  / \partial \beta \partial \nu  $ has expectation zero
so the Fisher information matrix is diagonal. In fact, it is identically zero when evaluated at $\hat\beta$, because it is identical to the estimating equation for $\hat\beta$. 

I end this subsection with a comment on the estimating equation of $\beta$. It is similar to the Poisson score equation except for the additional weight $e^{ - x_i^{\T} \beta} $. For positive outcomes, it is also conventional to fit OLS of $\log y_i$ on $x_i$, resulting in the following estimating equation
$$
\sumn    (\log y_i - x_i^{\T} \beta) x_i   = 0.
$$
\citet{firth1988multiplicative} compared  Gamma and log-Normal regressions based on efficiency. However, these two models are not  comparable: Gamma regression assumes that the log of the conditional mean of $y_i$ given $x_i$ is linear in $x_i$, whereas log-Normal regression assumes that the conditional mean of $\log y_i$ given $x_i$ is linear in $x_i$. By Jensen's inequality, $\log E(y_i\mid x_i)   \geq  E( \log y_i\mid x_i)  .$

See Problem \ref{hw20::gamma-regression} for more discussions of Gamma regression model and its connection with the log-Normal regression model. Overall, OLS of $\log y_i$ on $x_i$ can be applied even when the Gamma regression model holds. The additional efficiency from Gamma regression seems a minor issue. In the applications I have seen, OLS of $\log y_i$ on $x_i$ is more common than Gamma regression.

\section{Homework problems}

\paragraph{MLE in GLMs with binary regressors}\label{hw20::mle-binary-z}

The MLEs for the parameters in Models (\ref{eq:gaussianlinearmodel})--(\ref{eq:poissonlinearmodel}) do not have explicit formulas in general. But in the special case with the covariate $x_{i}$ containing $1$ and a binary covariate $z_i \in \{0, 1\}$, their MLEs do have simple formulas.

Let $\alpha$ be the intercept and $\beta$ be the coefficient of $z$.  
Let $n_1 = \sumn z_i$ and $n_0=\sumn (1-z_i) $ denote the sample sizes for groups with $z_i=1$ and $z_i=0$, respectively. 
Define the sample means of the outcomes as 
$$
\bar{y}_1 = n_1^{-1} \sumn z_i y_i, \quad 
\bar{y}_0 = n_0^{-1}  \sumn (1-z_i) y_i ,
$$
and the sample variances of the outcome as 
$$
\hat{\sigma}_1^2 = (n_1-1)^{-1} \sumn z_i (y_i  - \bar{y}_1)^2 ,\quad 
\hat{\sigma}_0^2 = (n_0-1)^{-1} \sumn (1-z_i) (y_i  - \bar{y}_0)^2 .
$$
First,  find the MLEs $(\hat{\alpha}, \hat{\beta})$ in terms of sample means of the outcomes.
Then find the covariance estimators of $(\hat{\alpha}, \hat{\beta})$ in terms of the sample means and sample variances of the outcomes.

Remark: Use the formulas of $\hat{V}$ in Examples \ref{eg::gaussianlinearmodel-continue1}--\ref{eg::countpoissonmodel-continue1}.

\paragraph{Negative-Binomial covariance matrices}\label{hw20::nb-sandwich}

Assume that $\delta$ is known. 
Derive the estimated asymptotic covariance matrices of the MLE in the Negative-Binomial regression with $\mu_i = e^{x_i^{\T} \beta}$.

\paragraph{Gamma regression}

Verify that Gamma$'(\mu, \nu)$ belongs to the natural exponential family with dispersion. Derive the first- and second-order derivatives of the log-likelihood function and Newton's algorithm for computing the MLE. 
Derive the estimated asymptotic covariance matrices of the MLE.

\paragraph{Gamma regression model and OLS}
\label{hw20::gamma-regression}

Theorem \ref{thm::gamma-reg-ols} below gives the conditional mean and variance functions of $\log y_i$ given $x_i$, under the Gamma regression model. It demonstrates that under the Gamma regression model, $\log y_i$ given $x_i$ satisfies a homoskedastic linear model. Therefore, the OLS estimator of $\log y_i$ on $x_i$ is consistent for all components of $\beta$.  
Prove Theorem \ref{thm::gamma-reg-ols}. 

\begin{theorem}
\label{thm::gamma-reg-ols}
Under the Gamma regression model in Assumption \ref{assume::gamma-reg}, we have  
$$
E(\log y_i \mid x_i ) = \psi(\nu) - \log (\nu) + x_i^{\T} \beta
$$
and
$$
\var(\log y_i \mid x_i ) = \psi'(\nu)
$$
where $\psi(\nu) = \diff \log  \Gamma  (\nu) / \diff \nu$ is the digamma function and $\psi'(\nu)$ is the trigamma function. 
\end{theorem}

Remark: Use Proposition \ref{thm:gammamoments-log} to calculate the moments.

\chapter{Misspecified Generalized Linear Models: Restricted Mean Models and Sandwich Covariance Matrix}\label{chapter::sandwich}
   \chaptermark{GLM and Sandwich Covariance Matrix}

This chapter discusses the consequence of misspecified generalized linear models (GLMs), extending the Eicker--Huber--White (EHW) covariance estimator under OLS to its analogs under the GLMs. It serves as a stepping stone to the next chapter on the generalized estimating equation.

\section{Restricted mean model}
\label{sec::rmm-estimation-equation}

The logistic, Poisson, and Negative-Binomial models are extensions of
the Normal linear model. All of them are fully parametric models.
However, we have also discussed OLS as a restricted mean model
\[
E(y_{i}\mid x_{i})=x_{i}^{\T}\beta
\]
without imposing any additional assumptions (e.g., the variance) on
the conditional distribution. The restricted
mean model is a {\it semiparametric model}. Then a natural question is: what
are the analogs of the restricted mean model for the binary and count models? 

Binary outcome is too special because the conditional mean determines
the distribution. So if we assume that the conditional mean is $\mu_{i}=e^{x_{i}^{\T}\beta}/(1+e^{x_{i}^{\T}\beta})$,
then conditional distribution must be Bernoulli$(\mu_{i})$. Consequently, misspecification of the conditional mean function implies misspecification of the whole conditional distribution. 

For other outcomes, the conditional mean cannot determine the conditional
distribution. If we assume 
\[
E(y_{i}\mid x_{i})=\mu(x_{i}^{\T}\beta),
\]
we can verify that 
\[
E\left\{ \sumn\frac{y_{i}-\mu(x_{i}^{\T}\beta)}{\tilde{\sigma}^{2} (x_i, \beta)}\frac{\partial\mu(x_{i}^{\T}\beta)}{\partial\beta}\right\} =E\left[E\left\{ \sumn\frac{y_{i}-\mu(x_{i}^{\T}\beta)}{ \tilde{\sigma}^{2} (x_i, \beta) }\frac{\partial\mu(x_{i}^{\T}\beta)}{\partial\beta}\mid x_i \right\} \right]=0
\]
for any $\tilde{\sigma}^{2}$ that can be a function of $x_{i}$, the true parameter $\beta$, and maybe some other parameter $\phi$. So we can estimate $\beta$ by solving the estimating equation:
\begin{equation}
\sumn\frac{y_{i}-\mu(x_{i}^{\T}\beta)}{ \tilde{\sigma}^{2} (x_i, \beta)  }\frac{\partial\mu(x_{i}^{\T}\beta)}{\partial\beta}=0. \label{eq:estimating-equation-conditional-mean}
\end{equation}
If $\tilde{\sigma}^{2} (x_i, \beta) = \sigma^2(x_i) =\var(y_{i}\mid x_{i})$,
then the above estimating equation is the score equation derived from
the GLM of an exponential family. If not, \eqref{eq:estimating-equation-conditional-mean} is not a score function but it is still a valid estimating equation.
In the latter case, $ \tilde{\sigma}^{2} (x_i, \beta)  $ is a ``working''
variance. This has important implications for the practical data analysis:
\begin{enumerate}[label=(I\arabic*), ref=I\arabic*]
\item
We can interpret the MLE from a GLM more broadly: it is also
valid under a restricted mean model even if the conditional distribution
is misspecified.
\item
We can construct more general estimators
beyond the MLEs from GLMs.
\end{enumerate}
However, we must address the issue of variance
estimation since the inference based on the Fisher information matrix no longer
works in general. Appendix \ref{sec::mle} reviewed \citet{huber::1967}'s theory on the MLE with misspecified models, and this chapter applies his idea to misspecified GLMs.

\section{Sandwich covariance matrix}
\label{sec::rmm-sandwich-covariance}
 
 To simplify the notation, we assume $(x_i,y_i)_{i=1}^n$ are IID draws although we usually view the covariates as fixed. This additional assumption is innocuous as the final inferential procedures are identical.

 \begin{theorem}
 \label{theorem::glm-sandwich}
Assume $(x_i,y_i)_{i=1}^n$ are IID with $E(y_i \mid x_i) = \mu(x_{i}^{\T}\beta)$. We have
$$
\sqrt{n} (\hat{\beta}-\beta )\rightarrow\N(0,B^{-1}MB^{-1})
$$
with
\begin{align}
B & =E\left\{ \frac{1}{\tilde{\sigma}^{2}(x, \beta)}\frac{\partial\mu(x^{\T}\beta)}{\partial\beta}\frac{\partial\mu(x^{\T}\beta)}{\partial\beta^{\T}}\right\} , \label{eq::sandwich-var-b}\\
M & =E\left[\frac{\sigma^{2}(x)}{\left\{ \tilde{\sigma}^{2}(x,  \beta)\right\} ^{2}}\frac{\partial\mu(x^{\T}\beta)}{\partial\beta}\frac{\partial\mu(x^{\T}\beta)}{\partial\beta^{\T}}\right]. \label{eq::sandwich-var-m}
\end{align}
\end{theorem}

 \begin{myproof}{Theorem}{\ref{theorem::glm-sandwich}}
 Applying Theorem \ref{theorem:sandwich-theorem-cov-iid} in Appendix \ref{chapter::m-mle} to 
\[
w=(x,y),\quad m(w,\beta)=\frac{y-\mu(x^{\T} b)}{\tilde{\sigma}^{2}(x, \beta)}\frac{\partial\mu(x^{\T} \beta)}{\partial \beta },
\]
we can derive the asymptotic distribution of the $\hat{\beta}$. 

The bread matrix equals
\begin{align}
B & =-E\left\{ \frac{\partial m(w,\beta)}{\partial\beta^{\T}}\right\} \nonumber \\
 & =-E\left[\frac{\partial}{\partial\beta^{\T}}\left\{ \frac{y-\mu(x^{\T}\beta)}{\tilde{\sigma}^{2}(x,\beta)}\frac{\partial\mu(x^{\T}\beta)}{\partial\beta}\right\} \right]\nonumber \\
 & =-E\left[\frac{y-\mu(x^{\T}\beta)}{\tilde{\sigma}^{2}(x,\beta)}\frac{\partial^{2}\mu(x^{\T}\beta)}{\partial\beta\partial\beta^{\T}}+\frac{\partial\mu(x^{\T}\beta)}{\partial\beta}\frac{\partial}{\partial\beta^{\T}}\left\{ \frac{y-\mu(x^{\T}\beta)}{\tilde{\sigma}^{2}(x,\beta)}\right\} \right]\label{eq:first-term-zero}\\
 & =-E\left[\frac{\partial\mu(x^{\T}\beta)}{\partial\beta}\frac{\partial}{\partial\beta^{\T}}\left\{ \frac{y-\mu(x^{\T}\beta)}{\tilde{\sigma}^{2}(x,\beta)}\right\} \right]\nonumber \\
 & =E\left[\frac{\partial\mu(x^{\T}\beta)}{\partial\beta}\frac{\partial\mu(x^{\T}\beta)}{\partial\beta^{\T}}/\tilde{\sigma}^{2}(x,\beta)+\frac{\partial\mu(x^{\T}\beta)}{\partial\beta}\frac{\partial\tilde{\sigma}^{2}(x,\beta)}{\partial\beta^{\T}}\frac{y-\mu(x^{\T}\beta)}{\left\{ \tilde{\sigma}^{2}(x,\beta)\right\} ^{2}}\right]\label{eq:second-term-zero}\\
 & =E\left\{ \frac{\partial\mu(x^{\T}\beta)}{\partial\beta}\frac{\partial\mu(x^{\T}\beta)}{\partial\beta^{\T}}/\tilde{\sigma}^{2}(x,\beta)\right\} ,\nonumber 
\end{align}
where the first term of (\ref{eq:first-term-zero}) and the second
term of (\ref{eq:second-term-zero}) are both zero under the restricted
mean model. 

The meat or middle matrix equals
\begin{align*}
M & =E\left\{ m(w,\beta)m(w,\beta)^{\T}\right\} \\
 & =E\left[\left\{ \frac{y-\mu(x^{\T}\beta)}{\tilde{\sigma}^{2}(x)}\right\} ^{2}\frac{\partial\mu(x^{\T}\beta)}{\partial\beta}\frac{\partial\mu(x^{\T}\beta)}{\partial\beta^{\T}}\right]\\
 & =E\left[\frac{\sigma^{2}(x)}{\left\{ \tilde{\sigma}^{2}(x)\right\} ^{2}}\frac{\partial\mu(x^{\T}\beta)}{\partial\beta}\frac{\partial\mu(x^{\T}\beta)}{\partial\beta^{\T}}\right].
\end{align*}
 \end{myproof}

We can estimate the asymptotic variance by replacing $B$ and $M$
by their sample analogs.
With $\hat\beta$ and the residual $\hat{\varepsilon}_i = y_{i}-\mu(x_{i}^{\T}\hat{\beta} ) $, we can conduct statistical inference based on the following Normal approximation:
\[
\hat{\beta}\asim\N(\beta,\hat{V}),
\]
with $\hat{V} \equiv n^{-1}\hat{B}^{-1}\hat{M}\hat{B}^{-1}$, 
where 
\begin{align*}
\hat{B} & =n^{-1}\sumn\frac{1}{\tilde{\sigma}^{2}(x_{i}, \hat{\beta})}\frac{\partial\mu(x_{i}^{\T}\hat{\beta})}{\partial\beta}\frac{\partial\mu(x_{i}^{\T}\hat{\beta})}{\partial\beta^{\T}},\\
\hat{M} & =n^{-1}\sumn \frac{ \hat{\varepsilon}_i^2    }{\tilde{\sigma}^{4}(x_{i}, \hat{\beta})} \frac{\partial\mu(x_{i}^{\T}\hat{\beta})}{\partial\beta}\frac{\partial\mu(x_{i}^{\T}\hat{\beta})}{\partial\beta^{\T}}.
\end{align*}
As a special case, when the GLM is correctly specified with $\sigma^2(x) = \tilde{\sigma}^2(x, \beta)$, then $B=M$ and the asymptotic variance reduces to the inverse of the Fisher information matrix discussed in Section \ref{sec::mle-glm-fisher}.

\setcounter{example}{0}
\begin{example}[continued]\label{eg::gaussianlinearmodel-continue2}
In a working Normal linear model,   $\tilde{\sigma}^2(x_i, \beta)  = \tilde{\sigma}^2$ is constant and 
$  \partial\mu(x_{i}^{\T} \beta ) / \partial\beta  = x_i$. So
\begin{eqnarray*}
\hat{B} &=& n^{-1}\sumn\frac{1}{\tilde{\sigma}^{2} } x_i x_i^{\T},\\ 
\hat{M} &=& n^{-1}\sumn  \frac{   \hat{\varepsilon}_i^2 }{  (\tilde{\sigma}^{2})^2 }  x_i x_i^{\T}
\end{eqnarray*}
with residual $ \hat{\varepsilon}_i  = y_i - x_i^{\T} \hat{\beta}$, recovering the EHW variance estimator
$$
\hat{V} = \left( \sumn   x_i x_i^{\T} \right)^{-1}
\left(\sumn \hat{\varepsilon}_i^2 x_i x_i^{\T} \right)
 \left( \sumn x_i x_i^{\T} \right)^{-1} ,
$$
because the $\tilde{\sigma}^{2}$'s cancel out. 
\end{example}

\begin{example}[continued]\label{eg::binarylogisticmodel-continue2}
In a working binary logistic model, $\tilde{\sigma}^2(x_i, \beta)  = \pi(x_i, \beta) \{1-\pi(x_i, \beta) \} $ and $ \partial\mu(x_{i}^{\T} \beta ) / \partial\beta  = \pi(x_i, \beta) \{1-\pi(x_i, \beta)\} x_i$, where $\pi(x_i, \beta)  = \mu(x_i^{\T} \beta) = e^{x_i^{\T} \beta}  / (1 +e^{x_i^{\T} \beta} )$. So
\begin{eqnarray*}
\hat{B} &=& n^{-1}\sumn \hat{\pi}_i (1-\hat{\pi}_i )  x_i x_i^{\T},\\ 
\hat{M} &=& n^{-1}\sumn  \hat{\varepsilon}_i^2   x_i x_i^{\T}
\end{eqnarray*}
with fitted mean $\hat{\pi}_i = e^{x_i^{\T}  \hat{\beta}}  / (1 +e^{x_i^{\T}  \hat{\beta}} )  $ and residual $ \hat{\varepsilon}_i  = y_i - \hat{\pi}_i$, yielding a new covariance estimator
$$
\hat{V} = \left( \sumn \hat{\pi}_i (1-\hat{\pi}_i )   x_i x_i^{\T} \right)^{-1}
\left(\sumn \hat{\varepsilon}_i^2 x_i x_i^{\T} \right)
 \left( \sumn  \hat{\pi}_i (1-\hat{\pi}_i )   x_i x_i^{\T} \right)^{-1} . 
$$
\end{example}

\begin{example}[continued]\label{eg::countpoissonmodel-continue2}
In a working Poisson model, $\tilde{\sigma}^2(x_i, \beta)  = \lambda(x_i, \beta)   $ and $ \partial\mu(x_{i}^{\T} \beta ) / \partial\beta  = \lambda(x_i, \beta)  x_i$, where $\lambda(x_i, \beta)  = \mu(x_i^{\T} \beta) = e^{x_i^{\T} \beta}  $. So
\begin{eqnarray*}
\hat{B} &=& n^{-1}\sumn \hat{\lambda}_i  x_i x_i^{\T},\\ 
\hat{M} &=& n^{-1}\sumn  \hat{\varepsilon}_i^2   x_i x_i^{\T}
\end{eqnarray*}
with fitted mean $ \hat{\lambda}_i =  e^{x_i^{\T} \hat{\beta} }  $ and residual $ \hat{\varepsilon}_i  = y_i -  \hat{\lambda}_i $, yielding a new covariance estimator
$$
\hat{V} = \left( \sumn  \hat{\lambda}_i   x_i x_i^{\T} \right)^{-1}
\left(\sumn \hat{\varepsilon}_i^2 x_i x_i^{\T} \right)
 \left( \sumn \hat{\lambda}_i x_i x_i^{\T} \right)^{-1} . 
$$
\end{example}

Again, I relegate the derivation of the formulas for the Negative-Binomial regression as a homework problem. 
The \ri{R} package \ri{sandwich} implements the above covariance matrices \citep{zeileis2006object}.

\section{Applications of the sandwich standard errors}

\subsection{Linear regression}

In \ri{R}, several functions can compute the EHW standard error: the \ri{hccm} function in the \ri{car} package, and the \ri{vcovHC} and \ri{sandwich} functions in the \ri{sandwich} package. The first two are special functions for OLS, and the third one works for general models.

Below, I use these functions to compute various types of standard errors based on the Boston housing data. 

\begin{lstlisting}
> library("car")
> library("lmtest")
> library("sandwich")
> library("mlbench")
> 
> ## linear regression
> data("BostonHousing")
> lm.boston = lm(medv ~ ., data = BostonHousing)
> hccm0     = hccm(lm.boston, type = "hc0")
> sandwich0 = sandwich(lm.boston, adjust = FALSE)
> vcovHC0   = vcovHC(lm.boston, type = "HC0")
> 
> hccm1     = hccm(lm.boston, type = "hc1") 
> sandwich1 = sandwich(lm.boston, adjust = TRUE) 
> vcovHC1   = vcovHC(lm.boston, type = "HC1")
> 
> hccm3     = hccm(lm.boston, type = "hc3") 
> vcovHC3   = vcovHC(lm.boston, type = "HC3")
> 
> dat.reg = data.frame(hccm0     = diag(hccm0)^(0.5),
+                      sandwich0 = diag(sandwich0)^(0.5),
+                      vcovHC0   = diag(vcovHC0)^(0.5),
+                      
+                      hccm1     = diag(hccm1)^(0.5),
+                      sandwich1 = diag(sandwich1)^(0.5),
+                      vcovHC1   = diag(vcovHC1)^(0.5),
+                      
+                      hccm3     = diag(hccm3)^(0.5),
+                      vcovHC3   = diag(vcovHC3)^(0.5))
> round(dat.reg[-1, ], 2)
        hccm0 sandwich0 vcovHC0 hccm1 sandwich1 vcovHC1 hccm3 vcovHC3
crim     0.03      0.03    0.03  0.03      0.03    0.03  0.03    0.03
zn       0.01      0.01    0.01  0.01      0.01    0.01  0.01    0.01
indus    0.05      0.05    0.05  0.05      0.05    0.05  0.05    0.05
chas1    1.28      1.28    1.28  1.29      1.29    1.29  1.35    1.35
nox      3.73      3.73    3.73  3.79      3.79    3.79  3.92    3.92
rm       0.83      0.83    0.83  0.84      0.84    0.84  0.89    0.89
age      0.02      0.02    0.02  0.02      0.02    0.02  0.02    0.02
dis      0.21      0.21    0.21  0.21      0.21    0.21  0.22    0.22
rad      0.06      0.06    0.06  0.06      0.06    0.06  0.06    0.06
tax      0.00      0.00    0.00  0.00      0.00    0.00  0.00    0.00
ptratio  0.12      0.12    0.12  0.12      0.12    0.12  0.12    0.12
b        0.00      0.00    0.00  0.00      0.00    0.00  0.00    0.00
lstat    0.10      0.10    0.10  0.10      0.10    0.10  0.10    0.10
\end{lstlisting}

The \ri{sandwich} function can compute HC0 and HC1, corresponding to adjusting for the degrees of freedom or not; \ri{hccm} and \ri{vcovHC} can compute other HC standard errors.

\subsection{Logistic regression}

\subsubsection{An application}
Revisit the flu shot example in Chapter \ref{chapter::binary-logit}. In this example, two types of standard errors are rather similar. The simple logistic model does not seem to suffer from severe misspecification. 

\begin{lstlisting}
> flu = read.table("fludata.txt", header = TRUE)
> flu = within(flu, rm(receive))
> assign.logit = glm(outcome ~ ., 
+                    family  = binomial(link = logit), 
+                    data    = flu)
> summary(assign.logit)
Coefficients:
             Estimate Std. Error z value Pr(>|z|)    
(Intercept) -2.199815   0.408684  -5.383 7.34e-08 ***
assign      -0.197528   0.136235  -1.450  0.14709    
age         -0.007986   0.005569  -1.434  0.15154    
copd         0.337037   0.153939   2.189  0.02857 *  
dm           0.454342   0.143593   3.164  0.00156 ** 
heartd       0.676190   0.153384   4.408 1.04e-05 ***
race        -0.242949   0.143013  -1.699  0.08936 .  
renal        1.519505   0.365973   4.152 3.30e-05 ***
sex         -0.212095   0.144477  -1.468  0.14210    
liverd       0.098957   1.084644   0.091  0.92731    

> coeftest(assign.logit, vcov = sandwich)

z test of coefficients:

              Estimate Std. Error z value  Pr(>|z|)    
(Intercept) -2.1998145  0.4059386 -5.4191 5.991e-08 ***
assign      -0.1975283  0.1371785 -1.4399  0.149885    
age         -0.0079859  0.0057053 -1.3997  0.161590    
copd         0.3370371  0.1556781  2.1650  0.030391 *  
dm           0.4543416  0.1394709  3.2576  0.001124 ** 
heartd       0.6761895  0.1521105  4.4454 8.774e-06 ***
race        -0.2429488  0.1430957 -1.6978  0.089544 .  
renal        1.5195049  0.3659238  4.1525 3.288e-05 ***
sex         -0.2120954  0.1489435 -1.4240  0.154447    
liverd       0.0989572  1.1411133  0.0867  0.930894    
\end{lstlisting}

\subsubsection{A misspecified logistic regression}

\citet{freedman2006so} discussed the following misspecified logistic regression. The discrepancy between the two types of standard errors is a warning of the misspecification of the conditional mean function because it determines the whole conditional distribution. In this case, it is not meaningful to interpret the coefficients. 

\begin{lstlisting}
> n = 100
> x = runif(n, 0, 10)
> prob.x = 1/(1 + exp(3*x - 0.5*x^2))
> y = rbinom(n, 1, prob.x)
> freedman.logit = glm(y ~ x, family = binomial(link = logit))
> summary(freedman.logit)
Coefficients:
            Estimate Std. Error z value Pr(>|z|)    
(Intercept)  -6.6764     1.3254  -5.037 4.72e-07 ***
x             1.1083     0.2209   5.017 5.25e-07 ***

> coeftest(freedman.logit, vcov = sandwich)

z test of coefficients:

            Estimate Std. Error z value Pr(>|z|)   
(Intercept) -6.67641    2.46035 -2.7136 0.006656 **
x            1.10832    0.39672  2.7937 0.005211 **
\end{lstlisting}

\subsection{Poisson regression}

\subsubsection{A correctly specified Poisson regression}

I first generate data from a correctly specified Poisson regression. The two types of standard errors are very close.

\begin{lstlisting}
> n = 1000
> x = rnorm(n)
> lambda.x = exp(x/5)
> y = rpois(n, lambda.x)
> pois.pois = glm(y ~ x, family = poisson(link = log))
> summary(pois.pois)
Coefficients:
             Estimate Std. Error z value Pr(>|z|)    
(Intercept) -0.004386   0.032117  -0.137    0.891    
x            0.189069   0.031110   6.077 1.22e-09 ***

> coeftest(pois.pois, vcov = sandwich)

z test of coefficients:

              Estimate Std. Error z value  Pr(>|z|)    
(Intercept) -0.0043862  0.0311957 -0.1406    0.8882    
x            0.1890691  0.0299124  6.3208 2.603e-10 ***
\end{lstlisting}

\subsubsection{A Negative-Binomial regression model}

I then generate data from a Negative-Binomial regression model. The conditional mean function is still $E(y_i\mid x_i) = e^{x_i^{\T} \beta}$, so we can still use Poisson regression as a working model. The robust standard error doubles the classical standard error. 

\begin{lstlisting}
> library(MASS)
> theta = 0.2
> y = rnegbin(n, mu = lambda.x, theta = theta)
> nb.pois = glm(y ~ x, family = poisson(link = log))
> summary(nb.pois)
Coefficients:
            Estimate Std. Error z value Pr(>|z|)    
(Intercept) -0.07747    0.03315  -2.337   0.0194 *  
x            0.13847    0.03241   4.272 1.94e-05 ***

> coeftest(nb.pois, vcov = sandwich)

z test of coefficients:

             Estimate Std. Error z value Pr(>|z|)  
(Intercept) -0.077475   0.079431 -0.9754  0.32937  
x            0.138467   0.061460  2.2530  0.02426 *
\end{lstlisting}

Because the true model is the Negative-Binomial regression, we can use the correct model to fit the data. Theoretically, the MLE is the most efficient estimator. However, in this particular dataset, the robust standard error from Poisson regression is no larger than the one from Negative-Binomial regression. Moreover, the robust standard errors from the Poisson and Negative-Binomial regressions are very close. 

\begin{lstlisting}
> nb.nb = glm.nb(y ~ x)
> summary(nb.nb)
Coefficients:
            Estimate Std. Error z value Pr(>|z|)  
(Intercept) -0.08047    0.07336  -1.097   0.2727  
x            0.16487    0.07276   2.266   0.0234 *

> coeftest(nb.nb, vcov = sandwich)

z test of coefficients:

             Estimate Std. Error z value Pr(>|z|)   
(Intercept) -0.080467   0.079510  -1.012 0.311517   
x            0.164869   0.063902   2.580 0.009879 **
\end{lstlisting}

\subsubsection{Misspecification of the conditional mean}

When the conditional mean function is misspecified, the Poisson and Negative-Binomial regressions give different point estimates, and it is unclear how to compare the standard errors. 

\begin{lstlisting}
> lambda.x = x^2
> y = rpois(n, lambda.x)
> wr.pois = glm(y ~ x, family = poisson(link = log))
> summary(wr.pois)
Coefficients:
            Estimate Std. Error z value Pr(>|z|)    
(Intercept) -0.03760    0.03245  -1.159 0.246457    
x            0.11933    0.03182   3.751 0.000176 ***

> coeftest(wr.pois, vcov = sandwich)

z test of coefficients:

             Estimate Std. Error z value Pr(>|z|)
(Intercept) -0.037604   0.053033 -0.7091   0.4783
x            0.119331   0.101126  1.1800   0.2380

> 
> wr.nb = glm.nb(y ~ x)
There were 26 warnings (use warnings() to see them)
> summary(wr.nb)
Coefficients:
            Estimate Std. Error z value Pr(>|z|)    
(Intercept)  0.15984    0.05802   2.755  0.00587 ** 
x           -0.34622    0.05789  -5.981 2.22e-09 ***

> coeftest(wr.nb, vcov = sandwich)

z test of coefficients:

             Estimate Std. Error z value Pr(>|z|)   
(Intercept)  0.159837   0.061564  2.5963 0.009424 **
x           -0.346223   0.238124 -1.4540 0.145957   
\end{lstlisting}

Overall, for count outcome regression, it seems that Poisson regression suffices as long as we use the robust standard error. The Negative-Binomial is unlikely to offer more if only the conditional mean is of interest.

%
%

\subsection{How robust are the robust standard errors?}

Section \ref{sec::rmm-estimation-equation} discusses the restricted mean model as an extension of the GLM, allowing for misspecification of the GLM while still preserving the conditional mean. We can extend the discussion to other parametric models. 
\citet{huber::1967} started the literature on the statistical properties of the MLE in a misspecified model, and
\citet{white1982maximum} addressed detailed inferential problems. 
\citet{buja2019models2} reviewed this topic recently.

The discussion in Section \ref{sec::rmm-estimation-equation} is useful when the conditional mean is correctly specified. However, if we think the GLM is severely misspecified with a wrong conditional mean, then the robust sandwich standard errors are unlikely to be helpful, because the MLE converges to a wrong parameter in the first place \citep{freedman2006so}.

\section{Homework problems}

\paragraph{MLE in GLMs with binary regressors}\label{hw20::mle-binary-z-misspecified}

Continue with Problem \ref{hw20::mle-binary-z}. Find the covariance estimators of $(\hat{\alpha}, \hat{\beta})$ in terms of the sample means and sample variances of the outcomes, without assuming the models are correct.

Remark: Use the formulas of $\hat{V}$ in Examples \ref{eg::gaussianlinearmodel-continue2}--\ref{eg::countpoissonmodel-continue2}.

\paragraph{Negative-Binomial covariance matrices}\label{hw20::nb-sandwich-misspecified}

Continue with Problem \ref{hw20::nb-sandwich}. Derive the estimated asymptotic covariance matrices of the MLE without assuming the Negative-Binomial model is correct.

\paragraph{Robust standard errors in the Karolinska data}\label{hw20::robust-se-karolinska}

Report the robust standard errors in the case study of Section \ref{sec::multinomial-case-study} in Chapter \ref{chapter::logit-categorical}. 

Remark: 
For some models, the function \ri{coeftest(*, vcov = sandwich)} does not work. Do some diagnostics when it happens.

\paragraph{Robust standard errors in the gym data}

Report the robust standard errors in the case study of Section \ref{sec::count-case-study} in Chapter \ref{chapter::count}.

       
\chapter{Generalized Estimating Equation for Correlated Multivariate Data}\label{chapter::gee}
 
Previous chapters dealt with cross-sectional data, that is, we observe $n$ units at a particular time point, collecting various covariates and outcomes. In addition, we assume that these units are independent, and sometimes, we even assume they are IID. Many applications have correlated data. We have two canonical examples of correlated data:  
\begin{enumerate}[label=(E\arabic*), ref=E\arabic*]
\item
repeated measurements of the same set of units over  time, which are often called {\it longitudinal data} in biostatistics \citep{fitzmaurice2012applied} or {\it panel data} in econometrics \citep{wooldridge2010econometric};  
\item
clustered observations belonging to classrooms, villages, etc, which are common in cluster-randomized experiments in education \citep{schochet2013estimators} and public health \citep{turner2017reviewdesign, turner2017reviewanalysis}. 
\end{enumerate}
Many excellent textbooks cover modeling correlated data intensively. This chapter focuses on a simple yet powerful strategy, {\it generalized estimating equation (GEE)},  which is a natural extension of the GLM discussed in Chapter \ref{chapter::sandwich}. It was initially proposed in \citet{liang1986longitudinal}, the most cited paper published in {\it Biometrika} in the past one hundred years \citep{titterington2013biometrika}. 
For simplicity, we will use the term ``longitudinal data'' for general correlated data.

\section{Examples of correlated data}

\subsection{Longitudinal data}
\label{section::longitudinal-gym}

We have used the data from \citet{royer2015incentives} in Chapter \ref{chapter::count}. Each worker's number of gym visits was repeatedly measured over more than 100 weeks. It is a standard longitudinal dataset. In Chapter \ref{chapter::count}, we conducted analysis for each week separately, and in this chapter, we will accommodate the longitudinal structure of the data.

\subsection{Clustered data: a neuroscience experiment} 
\label{section::cluster-neuro}

\citet{moen2016analyzing} examined the effects of Pten knockdown and fatty acid delivery on soma size of neurons in the brain of a mouse. The useful variables for our analysis are the id of mouse \ri{mouseid}, the fatty acid level \ri{fa}, the Pten knockdown indicator \ri{pten}, the outcome \ri{somasize}, the number of neurons \ri{numpten} and \ri{numctrl} under Pten knockdown or not.

\begin{lstlisting}
> Pten = read.csv("PtenAnalysisData.csv")[, -(7:9)]
> head(Pten)
  mouseid fa pten somasize numctrl numpten
1       0  0    0   83.837      30      44
2       0  0    0   69.984      30      44
3       0  0    0   82.128      30      44
4       0  0    0   86.446      30      44
5       0  0    0   74.032      30      44
6       0  0    0   71.693      30      44
\end{lstlisting}

The three-way table below shows the treatment combinations for 14 mice, from which
we can see that the Pten knockdown indicator varies within mice, but the fatty acid level varies only between mice. 
\begin{lstlisting}
> table(Pten$mouseid, Pten$fa, Pten$pten)
, ,  = 0

    
      0  1  2
  0  30  0  0
  1  58  0  0
  2  18  0  0
  3   2  0  0
  4  56  0  0
  5   0 39  0
  6   0 33  0
  7   0 58  0
  8   0 60  0
  9   0  0 15
  10  0  0 27
  11  0  0  7
  12  0  0 34
  13  0  0 22

, ,  = 1

    
      0  1  2
  0  44  0  0
  1  68  0  0
  2  33  0  0
  3  11  0  0
  4  76  0  0
  5   0 55  0
  6   0 55  0
  7   0 75  0
  8   0 92  0
  9   0  0 34
  10  0  0 29
  11  0  0 20
  12  0  0 53
  13  0  0 38
  \end{lstlisting}

\subsection{Clustered data: a public health intervention} 
\label{section::cluster-publichealth}

Poor sanitation leads to morbidity and mortality in developing countries. In 2012, \citet{guiteras2015encouraging} conducted a cluster-randomized experiment in rural Bangladesh to evaluate the effectiveness of different policies on the use of hygienic latrines. To illustrate the theory and method of GEE, I use a subset of their original data and exclude the households not eligible for subsidies or with missing outcomes, resulting in 10125 households in total. The median, mean, and maximum of village size are 83, 119, and 500, respectively.

I choose the outcome $y_{it}$ as the binary indicator for whether the household $(i,t)$ had access to a hygienic latrine or not, measured in June 2013, and covariate $x_{it}$ as the baseline access rate to hygienic latrines in the community that household $(i,t)$ belonged to, measured in January 2012 before the experiment.

The useful variables below are \ri{z}, \ri{x}, \ri{y}, and \ri{vid}, which denote the binary treatment indicator, covariate $x_{it}$, the outcome, and the village id \ri{vid}, 
\begin{lstlisting}
> hygaccess = read.csv("hygaccess.csv")
> hygaccess = hygaccess[,c("r4_hyg_access", "treat_cat_1", 
+                          "bl_c_hyg_access", "vid", "eligible")]
> hygaccess = hygaccess[which(hygaccess$eligible=="Eligible"&
+                               hygaccess$r4_hyg_access!="Missing"),]
> hygaccess$y = ifelse(hygaccess$r4_hyg_access=="Yes", 1, 0)
> hygaccess$z = hygaccess$treat_cat_1
> hygaccess$x = hygaccess$bl_c_hyg_access
\end{lstlisting}

\section{Marginal model and the generalized estimating equation}

We will extend the restricted mean model to deal with longitudinal
data, where we observe outcome $y_{it}$ and covariate $x_{it}$ for
each unit $i=1,\ldots,n$ at time $t=1,\ldots,n_{i}.$ The $n_i$'s can vary across units. 

When $n_{i}=1$
for all units, we drop the time index and model the conditional mean
as 
\[
E(y_{i}\mid x_{i})=\mu(x_{i}^{\T}\beta),
\]
and use the following estimating equation  
to estimate the parameter $\beta$:
\begin{equation}
\sumn\frac{y_{i}-\mu(x_{i}^{\T}\beta)}{\tilde{\sigma}^{2}(x_{i}, \beta)}\frac{\partial\mu(x_{i}^{\T}\beta)}{\partial\beta}=0 . \label{eq:quasi-likelihood-gee-chapter}
\end{equation}
In \eqref{eq:quasi-likelihood-gee-chapter},  $\tilde{\sigma}^{2}(x_{i}, \beta)$ is a working variance function usually motivated by a GLM, which can be misspecified (recall Chapter \ref{chapter::sandwich}).

With an $n_{i}\times1$ vector outcome and  an $n_{i}\times p$ covariate matrix
\begin{eqnarray}
\label{eq::individual-y-x}
Y_{i} = \begin{pmatrix}
y_{i1} \\
\vdots \\
y_{in_{i}}
\end{pmatrix},\quad 
X_{i} = \begin{pmatrix}
x_{i1}^{\T} \\
\vdots \\
x_{in_{i}}^{\T}
\end{pmatrix},\quad
(i=1,\ldots, n)
\end{eqnarray}
we can extend the restricted mean model to
\begin{eqnarray}
E(Y_{i}\mid X_{i}) 
&\equiv& \left(\begin{array}{c}
E(y_{i1}\mid X_{i})\\
\vdots\\
E(y_{in_{i}}\mid X_{i})
\end{array}\right)  \label{eq::def1-gee} \\
&=& \left(\begin{array}{c}
E(y_{i1}\mid x_{i1})\\
\vdots\\
E(y_{in_{i}}\mid x_{in_{i}})
\end{array}\right)  \label{eq::gee-assumtion-1} \\
&=&\left(\begin{array}{c}
\mu(x_{i1}^{\T}\beta)\\
\vdots\\
\mu(x_{in_{i}}^{\T}\beta)
\end{array}\right)   \label{eq::gee-assumtion-2} \\
&\equiv&\mu(X_{i},\beta), \label{eq::def2-gee}
\end{eqnarray}
where \eqref{eq::def1-gee} and \eqref{eq::def2-gee} are definitions, and \eqref{eq::gee-assumtion-1} and \eqref{eq::gee-assumtion-2}
are the two key assumptions. Assumption \eqref{eq::gee-assumtion-1} requires that the conditional mean of $y_{it}$ depends only on $x_{it}$
but not on any other $x_{is}$ with $s\neq t$. Assumption \eqref{eq::gee-assumtion-2} requires
that the relationship between $x_{it}$ and $y_{it}$ is stable across units and time points with the function $\mu(\cdot)$ and the parameter
$\beta$ not varying with respect to $i$ or $t$. The model assumptions in \eqref{eq::gee-assumtion-1} and \eqref{eq::gee-assumtion-2} 
are really strong, and I defer the critiques to the end of this chapter.
Nevertheless, the marginal model attracts practitioners for 
\begin{enumerate}[label=(A\arabic*), ref=A\arabic*] 
\item\label{advantage-1}
its similarity to GLM and the restricted mean model, and
\item \label{advantage-2}
its simplicity of requiring only specification of the marginal conditional
means, not the whole joint distribution. 
\end{enumerate}
The advantage \eqref{advantage-1} facilitates
the interpretation of the coefficient, and the advantage \eqref{advantage-2} is crucial because of the lack of familiar multivariate distributions in statistics
except for the multivariate Normal. The generalized estimating equation
(GEE) for $\beta$ is the vector form of (\ref{eq:quasi-likelihood-gee-chapter}):
\begin{equation}
\sumn\underbrace{\frac{\partial\mu (X_{i},\beta)}{\partial\beta}}_{p\times n_{i}}
\underbrace{\tilde{V}^{-1}(X_{i},\beta)}_{n_{i}\times n_{i}}
\underbrace{\left\{ Y_{i}-\mu(X_{i},\beta)\right\} }_{n_{i}\times1}
= \underbrace{0}_{p\times 1},\label{eq:gee-liang-zeger-form}
\end{equation}
where (\ref{eq:gee-liang-zeger-form}) has a similar form as (\ref{eq:quasi-likelihood-gee-chapter})
with three terms organized to match the dimension so that matrix multiplications
are well-defined:
\begin{enumerate}[label=(T\arabic*), ref=T\arabic*] 
\item the last term 
\[
Y_{i}-\mu(X_{i},\beta)=\left(\begin{array}{c}
y_{i1} -\mu(x_{i1}^{\T}\beta)\\
\vdots\\
y_{in_{i}} -\mu(x_{in_{i}}^{\T}\beta)
\end{array}\right)
\]
represents the residual vector,
\item the second term is the inverse of $\tilde{V}(X_{i}, \beta)$, a working covariance
matrix of the conditional distribution of $Y_{i}$ given $X_{i}$
which may be misspecified:
\[
\tilde{V}(X_{i}, \beta)\neq V(X_{i})\equiv\cov(Y_{i}\mid X_{i}).
\]
It is relatively easy to specify the working variance $\tilde{\sigma}^2(x_{it}, \beta)$ for each marginal component, for example, based on the marginal GLM.  So the key is to specify the $n_i\times n_i$ dimensional correlation matrix $R_i$ to obtain 
$$
\tilde{V}(X_{i}, \beta) = \text{diag}\{ \tilde{\sigma} (x_{it}, \beta) \}_{i=1}^{n_i} R_i \text{diag}\{ \tilde{\sigma} (x_{it}, \beta) \}_{i=1}^{n_i}.
$$
We assume that the $R_i$'s are given now, and will discuss how to choose them in Chapter \ref{chapter::working-correlation} below. 
\item the first term is the partial derivative of an $n_i \times 1$ 
vector $\mu (X_{i},\beta)=(\mu(x_{i1}^{\T}\beta),\ldots,\mu(x_{in_{i}}^{\T}\beta))^{\T}$
with respect to a $p\times1$ vector $\beta=(\beta_{1},\ldots,\beta_{p})^{\T}$:
\begin{eqnarray*}
\frac{\partial\mu (X_{i},\beta)}{\partial\beta}
&=& \left(  \frac{\partial \mu(x_{i1}^{\T}\beta)}{\partial \beta},\ldots,  \frac{ \partial  \mu(x_{in_{i}}^{\T}\beta)}{\partial \beta} \right)  \\
&=&\left(\begin{array}{ccc}
\frac{\partial\mu(x_{i1}^{\T}\beta)}{\partial\beta_{1}} & \cdots & \frac{\partial\mu(x_{in_{i}}^{\T}\beta)}{\partial\beta_{1}}\\
\vdots &  & \vdots\\
\frac{\partial\mu(x_{i1}^{\T}\beta)}{\partial\beta_{p}} & \cdots & \frac{\partial\mu(x_{in_{i}}^{\T}\beta)}{\partial\beta_{p}}
\end{array}\right),
\end{eqnarray*}
which is a $p\times n_i$ matrix, denoted by $D_{i} (\beta).$
\end{enumerate}

\section{Statistical inference with GEE}

\subsection{Computation using the Gauss--Newton method\label{subsec:Computation-gee-gauss-newton}}

We can use Newton's method to solve the GEE (\ref{eq:gee-liang-zeger-form}).
However, calculating the derivative of the left-hand side of (\ref{eq:gee-liang-zeger-form})
involves calculating the second order derivative of $ \mu (X_{i},\beta)$ with respect to $\beta$.
A simpler alternative without calculating the second-order derivative
is the Gauss--Newton method based on the following approximation:
\begin{align*}
0 & =\sumn\frac{\partial\mu (X_{i},\beta)}{\partial\beta}\tilde{V}^{-1}(X_{i})\left\{ Y_{i}-\mu(X_{i},\beta)\right\} \\
 & \cong \sumn D_{i} (\beta^{\text{old}})\tilde{V}^{-1}(X_{i}, \beta^{\text{old}})\left[\left\{ Y_{i}-\mu(X_{i},\beta^{\text{old}})\right\} -D_{i}^{\T}(\beta^{\text{old}})(\beta-\beta^{\text{old}})\right]\\
 & =\sumn D_{i} (\beta^{\text{old}})\tilde{V}^{-1}(X_{i}, \beta^{\text{old}})\left\{ Y_{i}-\mu(X_{i},\beta^{\text{old}})\right\} \\
 & \quad -\sumn D_{i} (\beta^{\text{old}})\tilde{V}^{-1}(X_{i}, \beta^{\text{old}})D_{i}^{\T}(\beta^{\text{old}})(\beta-\beta^{\text{old}}) . 
\end{align*}
So given $\beta^{\text{old}}$, we update it as
\begin{eqnarray}
\beta^{\text{new}} &=& \beta^{\text{old}}
+\left\{ \sumn D_{i} (\beta^{\text{old}})\tilde{V}^{-1}(X_{i}, \beta^{\text{old}})D_{i}^{\T}(\beta^{\text{old}})\right\} ^{-1} \nonumber \\
&& \times 
\sumn D_{i} (\beta^{\text{old}})\tilde{V}^{-1}(X_{i}, \beta^{\text{old}})\left\{ Y_{i}-\mu(X_{i},\beta^{\text{old}})\right\} . 
\label{eq::gauss-newton-gee}
\end{eqnarray}

\subsection{Asymptotic inference\label{subsec:Asymptotic-inference-GEE}}

The asymptotic distribution of $\hat{\beta}$ follows from Theorem \ref{theorem:sandwich-theorem-cov-ind}. Similar to the proof of Theorem \ref{theorem::glm-sandwich}, we can verify that $\sqrt{n}(\hat{\beta}-\beta)\rightarrow\N(0,B^{-1}MB^{-1})$
in distribution where
\begin{align*}
B & =E\left\{ n^{-1}\sumn  D_{i} (\beta)\tilde{V}^{-1}(X_{i}, \beta)D_{i}^{\T}(\beta)\right\} ,\\
M & =E\left\{ n^{-1}\sumn D_{i} (\beta)\tilde{V}^{-1}(X_{i}, \beta)V(X_{i})\tilde{V}^{-1}(X_{i}, \beta)D_{i}^{\T}(\beta)\right\} .
\end{align*}
After obtaining $\hat{\beta}$ and the residual vector $\hat{\varepsilon}_{i}=Y_{i}-\mu(X_{i},\hat{\beta})$
for unit $i$ $(i=1,\ldots,n)$, we can conduct asymptotic inference
based on the Normal approximation 
\[
\hat{\beta}\asim\N(\beta,n^{-1}\hat{B}^{-1}\hat{M}\hat{B}^{-1}),
\]
where 
\begin{align*}
\hat{B} & =n^{-1}\sumn D_{i} (\hat{\beta})\tilde{V}^{-1}(X_{i}, \hat{\beta})D_{i}^{\T}(\hat{\beta}),\\
\hat{M} & =n^{-1}\sumn D_{i} (\hat{\beta})\tilde{V}^{-1}(X_{i}, \hat{\beta})\hat{\varepsilon}_{i}\hat{\varepsilon}_{i}^{\T}\tilde{V}^{-1}(X_{i}, \hat{\beta})D_{i}^{\T}(\hat{\beta}).
\end{align*}
This covariance estimator proposed by \citet{liang1986longitudinal}, is robust to the misspecification of the marginal variances and the correlation structure as long as the conditional mean of $Y_i$ given $X_i$ is correctly specified.

\subsection{Implementation: choice of the working covariance matrix}
\label{chapter::working-correlation}

We have not discussed the choice of the working correlation matrix
$R_i$. Different choices do not affect the consistency
but affect the efficiency of $\hat{\beta}$. A simple starting point
is the independent working correlation matrix $R_i = I_{n_i}$.
Under this correlation matrix, the GEE reduces to
\begin{eqnarray*}
\sumn \left(  \frac{ \partial  \mu(x_{i1}^{\T}\beta)}{\partial \beta},\ldots,  \frac{ \partial   \mu(x_{in_{i}}^{\T}\beta)}{\partial \beta} \right) 
\left(\begin{array}{ccc}
\tilde{\sigma}^{-2}(x_{i1}, \beta)\\
 & \ddots\\
 &  & \tilde{\sigma}^{-2}(x_{in_{i}}, \beta)
\end{array}\right) \\
\times 
\left(\begin{array}{c}
 y_{i1} -\mu(x_{i1}^{\T}\beta)\\
\vdots\\
 y_{in_{i}} -\mu(x_{in_{i}}^{\T}\beta)
\end{array}\right) =0,
\end{eqnarray*}
or, in a  more compact form,
\begin{eqnarray}
\label{eq::independent-working-cov-gee}
\sumn\sum_{t=1}^{n_{i}}\frac{y_{it}-\mu(x_{it}^{\T}\beta)}{\tilde{\sigma}^{2}(x_{it}, \beta)}\frac{\partial\mu(x_{it}^{\T}\beta)}{\partial\beta}=0.
\end{eqnarray}
The equation \eqref{eq::independent-working-cov-gee} is simply the estimating equation of a restricted mean model
treating all data points $(i,t)$ as independent observations. This
implies that the point estimate assuming the independent working correlation matrix is still consistent, although we must change the standard error
as in Section \ref{subsec:Asymptotic-inference-GEE}. 

With this simple starting point, we have a consistent yet inefficient
estimate of $\beta$, and then we can compute the residuals. The correlation
among the residuals contains information about the true covariance
matrix. With small and equal $n_{i}$'s, we can estimate the conditional
covariance without imposing any structure based on the residuals.
Using the estimated covariance matrix, we can update the GEE estimate
to improve efficiency. This leads to a two-step procedure. 

An important intermediate case is motivated by the exchangeability
of the data points within the same unit $i$, so the working covariance
matrix is $\tilde{V}(X_{i}, \beta)= \text{diag}\{\tilde{\sigma}(x_{it})\}_{i=1}^{n_i} R_i(\rho) \text{diag}\{\tilde{\sigma}(x_{it})\}_{i=1}^{n_i} $, where
\[
R_i(\rho) = \left(\begin{array}{cccc}
1 & \rho & \cdots & \rho\\
\rho & 1 & \cdots & \rho\\
\vdots & \vdots &  & \vdots\\
\rho & \rho & \cdots & 1
\end{array}\right) .
\]
We can estimate $\rho$ based on the residuals from the first step.

The above three choices of the working covariance matrix are called
``\ri{independent}'', ``\ri{unstructured}'', and ``\ri{exchangeable}'' in the ``\ri{corstr}''
parameter of the function \ri{gee} in the \ri{gee} package in \ri{R}. This function
also contains other choices proposed by \citet{liang1986longitudinal}.

A carefully chosen working covariance matrix can lead to efficiency gain compared with the simple independent covariance matrix. An efficient estimator requires a correctly specified working covariance matrix. This is often an infeasible goal, and what is more, the conditional covariance $\cov(Y_i\mid X_i)$ is a nuisance parameter if the conditional mean is the main parameter of interest. In practice, the independent working covariance suffices in many applications despite its potential efficiency loss. This is similar to the use of OLS in the presence of heteroskedasticity in linear models. Section \ref{sec::crse-econometrics} focuses on the independent working covariance, which is common in econometrics. Section \ref{sec::critiquesonGEEassumptions} gives further justifications for this simple strategy.

\section{A special case: cluster-robust standard error} 
\label{sec::crse-econometrics}

Importantly, \citet{liang1986longitudinal}'s standard error treats each cluster $i$ as an independent contributor to the uncertainty. In econometrics, this is called the {\it cluster-robust standard error}. 
I will discuss linear and logistic regressions in this section, and leave the technical details of Poisson regression to Problem \ref{hw21::poisson-crse}.

Stack the $Y_i$'s and $X_i$'s in \eqref{eq::individual-y-x} together to obtain 
$$
Y = \begin{pmatrix}
Y_1 \\
\vdots \\
Y_n
\end{pmatrix},\quad
X = \begin{pmatrix}
X_1 \\
\vdots \\
X_n
\end{pmatrix},
$$
which are the $N$ dimensional outcome vector and $N\times p$ covariate matrix, where $N = \sumn n_i.$

\subsection{OLS}\label{sec::crse-ols}

An important special case is the
marginal linear model with an independent working covariance matrix
and homoskedasticity, resulting in the following estimating equation:
\[
\sumn\sum_{t=1}^{n_{i}}x_{it}(y_{it}-x_{it}^{\T}\beta)=0.
\]
So the point estimator is just the pooled OLS using all data
points:
\begin{align*}
\hat{\beta} & =\left(\sumn\sum_{t=1}^{n_{i}}x_{it}x_{it}^{\T}\right)^{-1}\sumn\sum_{t=1}^{n_{i}}x_{it}y_{it}\\
 & =\left(\sumn X_{i}^{\T}X_{i}\right)^{-1}\sumn X_{i}^{\T}Y_{i} \\
 &= (X^{\T} X)^{-1} X^{\T} Y.
\end{align*}
The three forms of $\hat{\beta} $ above are identical: the first one is based on $N$ observations, the second one is based on $n$ independent units, and the last one is based on the matrix form with the pooled data. 
Although the point estimate is identical to the case with independent
data points, we must adjust for the standard error according to Section
\ref{subsec:Asymptotic-inference-GEE}. From
$$
D_i(\beta) = ( x_{i1}, \ldots, x_{in_i} ) = X_i^{\T},
$$
we can verify that
\[
\hat{\cov}(\hat{\beta})=\left(\sumn X_{i}^{\T}X_{i}\right)^{-1}\sumn X_{i}^{\T}\hat{\varepsilon}_{i}\hat{\varepsilon}_{i}^{\T}X_{i}\left(\sumn X_{i}^{\T}X_{i}\right)^{-1},
\]
where $\hat{\varepsilon}_{i}=Y_{i}-X_{i}\hat{\beta}=(\hat{\varepsilon}_{i1},\ldots,\hat{\varepsilon}_{in_{i}})^{\T}$
is the residual vector of unit $i$. This is called the (Liang--Zeger) cluster-robust
covariance matrix in econometrics. The square roots of the diagonal terms are called the cluster-robust standard errors. 
The cluster-robust
covariance matrix is often much larger than the
(Eicker--Huber--White) heteroskedasticity-robust covariance matrix assuming
independence of observations $(i,t)$:
\[
\hat{\cov}_{\textsc{ehw}}(\hat{\beta})=\left(\sumn\sum_{t=1}^{n_{i}}x_{it}x_{it}^{\T}\right)^{-1}\sumn\sum_{t=1}^{n_{i}}\hat{\varepsilon}_{it}^{2}x_{it}x_{it}^{\T}\left(\sumn\sum_{t=1}^{n_{i}}x_{it}x_{it}^{\T}\right)^{-1}.
\]
Note that 
$$
X^{\T} X = \sumn X_{i}^{\T}X_{i} = \sumn\sum_{t=1}^{n_{i}}x_{it}x_{it}^{\T},
$$ 
so the bread matrices in $\hat{\cov}(\hat{\beta})$ and $\hat{\cov}_{\textsc{ehw}}(\hat{\beta})$ are identical. The only difference is due to the meat matrices:
$$
\sumn X_{i}^{\T}\hat{\varepsilon}_{i}\hat{\varepsilon}_{i}^{\T}X_{i}
= \sumn  \left( \sum_{t=1}^{n_{i}}\hat{\varepsilon}_{it} x_{it}\right) \left( \sum_{t=1}^{n_{i}}\hat{\varepsilon}_{it} x_{it} \right)^{\T}
\neq \sumn\sum_{t=1}^{n_{i}}\hat{\varepsilon}_{it}^{2}x_{it}x_{it}^{\T}
$$
in general.

\subsection{Logistic regression}\label{sec::crse-logit}

For binary outcomes, we can use the marginal logistic model with an independent working covariance matrix,
resulting in the following estimating equation:
\[
\sumn\sum_{t=1}^{n_{i}}x_{it} \left\{ y_{it}-  \pi(x_{it}, \beta)\right\} =0 , 
\]
where $ \pi(x_{it}, \beta) =  e^{x_{it}^{\T}\beta} / (1+e^{x_{it}^{\T}\beta} )  .$
So the point estimator is  the pooled logistic regression using all data
points, but we must adjust for the standard error according to Section
\ref{subsec:Asymptotic-inference-GEE}. From
\begin{eqnarray*}
D_i(\beta) &=&  (
 \pi(x_{i1}, \beta) \{1- \pi(x_{i1}, \beta)\} x_{i1}, \ldots, 
 \pi(x_{in_i}, \beta)\{1- \pi(x_{in_i}, \beta)\} x_{in_i} 
) \\
&=&  X_i^{\T} \tilde{V}(X_i, \beta),
\end{eqnarray*}
with $ \tilde{V}(X_i, \beta) = \text{diag}\{  \pi(x_{it}, \beta) \{1- \pi(x_{it}, \beta)\} \}_{t=1}^{n_i}$,
we can verify that
\begin{eqnarray*}
\hat{B} &=& n^{-1} \sumn X_i^{\T}  \hat{V}_i X_i,\\ 
\hat{M} &=& n^{-1} \sumn X_i^{\T}  \hat{\varepsilon}_i   \hat{\varepsilon}_i^{\T} X_i,
\end{eqnarray*}
where $\hat{\varepsilon}_{i} = (\hat{\varepsilon}_{i1}, \ldots, \hat{\varepsilon}_{in_i})^{\T}$ with residual
$\hat{\varepsilon}_{it}=y_{it}-e^{x_{it}\hat{\beta}}/(1+e^{x_{it}\hat{\beta}}) $, and $\hat{V}_i = \text{diag}\{  \pi(x_{it}, \hat{\beta}) \{1- \pi(x_{it}, \hat{\beta})\} \}_{t=1}^{n_i}.$
So the cluster-robust covariance estimator for logistic regression is
\[
\hat{\cov}(\hat{\beta})=\left(\sumn X_{i}^{\T} \hat{V}_i X_{i}\right)^{-1}
\sumn X_{i}^{\T}\hat{\varepsilon}_{i}\hat{\varepsilon}_{i}^{\T}X_{i}
\left(\sumn X_{i}^{\T} \hat{V}_i X_{i}\right)^{-1}.
\]

I leave the cluster-robust covariance estimator for Poisson regression to Problem \ref{hw21::poisson-crse}.

\section{Application}

I will use the \ri{gee} package for all the analyses below.

\subsection{Clustered data: a neuroscience experiment}

I will analyze the data in Chapter \ref{section::cluster-neuro}. 
The original study was interested in the potential interaction between two treatments, so I always include the interaction term in the regression model.

From the simple specification below, \ri{pten} has a significant effect, but \ri{fa} and the interactions are not significant.
\begin{lstlisting}
 Pten.gee = gee(somasize ~ factor(fa)*pten, 
+                id = mouseid, 
+                family = gaussian, 
+                corstr = "independence",
+                data = Pten)
> summary(Pten.gee)$coef 
                 Estimate Naive S.E. Naive z Robust S.E. Robust z
(Intercept)        93.106      1.594 58.4216       3.059  30.4374
factor(fa)1         3.756      2.175  1.7268       3.174   1.1836
factor(fa)2         6.907      2.551  2.7078       5.407   1.2774
pten               11.039      2.082  5.3016       2.200   5.0166
factor(fa)1:pten    8.727      2.834  3.0795       5.023   1.7373
factor(fa)2:pten   -2.904      3.270 -0.8881       3.554  -0.8173
> 
> 
> Pten.gee = gee(somasize ~ factor(fa)*pten, 
+                id = mouseid, 
+                family = gaussian, 
+                corstr = "exchangeable",
+                data = Pten)
> summary(Pten.gee)$coef
                 Estimate Naive S.E. Naive z Robust S.E. Robust z
(Intercept)        90.900      3.532 25.7376       2.701  33.6535
factor(fa)1         4.921      5.115  0.9621       2.914   1.6889
factor(fa)2         6.408      5.066  1.2649       5.904   1.0853
pten               11.501      1.979  5.8120       2.190   5.2515
factor(fa)1:pten    8.807      2.688  3.2766       5.050   1.7439
factor(fa)2:pten   -1.525      3.113 -0.4898       2.703  -0.5641
\end{lstlisting}

Including two covariates, we have the following results. The covariates are predictive of the outcome, changing the significance level of the main effect of \ri{fa}.  The interaction terms between \ri{pten} and \ri{fa} are not significant. 
\begin{lstlisting}
> Pten.gee = gee(somasize ~ factor(fa)*pten + numctrl + numpten, 
+                id = mouseid, 
+                family = gaussian, 
+                corstr = "independence",
+                data = Pten)
> summary(Pten.gee)$coef 
                 Estimate Naive S.E. Naive z Robust S.E. Robust z
(Intercept)       81.9422      2.791 29.3602      4.0917   20.026
factor(fa)1        6.2267      2.237  2.7835      4.2429    1.468
factor(fa)2       14.8956      2.657  5.6053      4.1839    3.560
pten              12.3771      2.020  6.1272      2.2477    5.507
numctrl            0.8721      0.120  7.2672      0.3028    2.880
numpten           -0.4843      0.101 -4.7948      0.2381   -2.034
factor(fa)1:pten   7.7498      2.744  2.8240      5.1064    1.518
factor(fa)2:pten  -2.9629      3.166 -0.9359      3.3105   -0.895
>  
> 
> Pten.gee = gee(somasize ~ factor(fa)*pten + numctrl + numpten, 
+                id = mouseid, 
+                family = gaussian, 
+                corstr = "exchangeable",
+                data = Pten)
> summary(Pten.gee)$coef
                 Estimate Naive S.E. Naive z Robust S.E. Robust z
(Intercept)       85.3316     5.2872 16.1393      5.4095  15.7745
factor(fa)1        5.4952     4.2761  1.2851      4.0207   1.3667
factor(fa)2       12.2174     4.1669  2.9320      4.2363   2.8840
pten              11.8044     1.9718  5.9865      2.1946   5.3789
numctrl            0.9326     0.2867  3.2527      0.3479   2.6810
numpten           -0.5678     0.2504 -2.2674      0.2772  -2.0482
factor(fa)1:pten   8.5137     2.6777  3.1795      5.0612   1.6821
factor(fa)2:pten  -1.7755     3.0995 -0.5728      2.7547  -0.6445
\end{lstlisting}

From the regressions above, we observe that (1) two choices of the covariance matrix do not lead to fundamental differences; and (2) without using the cluster-robust standard error, the results can be misleading.

 \subsection{Clustered data: a public health intervention} 

I will analyze the data in Chapter \ref{section::cluster-publichealth}. 
We first fit simple GEE without using the covariate.
\begin{lstlisting}
> hygaccess.gee = gee(y ~ z, id = vid,
+                     family = binomial(link = logit),
+                     corstr = "independence", 
+                     data = hygaccess)
> summary(hygaccess.gee)$coef
                    Estimate Naive S.E. Naive z Robust S.E. Robust z
(Intercept)          -0.7568    0.04439 -17.049      0.1763  -4.2924
zLPP Only             0.1551    0.06657   2.330      0.2301   0.6741
zLPP+Subsidy          0.7562    0.05503  13.742      0.2027   3.7313
zLPP+Subsidy+Supply   0.7344    0.05444  13.490      0.2010   3.6546
zSupply Only          0.3568    0.07364   4.846      0.3091   1.1544
> 
> hygaccess.gee = gee(y ~ z, id = vid,
+                     family = binomial(link = logit),
+                     corstr = "exchangeable", 
+                     data = hygaccess)
> summary(hygaccess.gee)$coef
                    Estimate Naive S.E. Naive z Robust S.E. Robust z
(Intercept)          -0.7799     0.1314 -5.9371      0.1522  -5.1235
zLPP Only             0.1638     0.2042  0.8021      0.2290   0.7152
zLPP+Subsidy          0.7789     0.1500  5.1926      0.1790   4.3524
zLPP+Subsidy+Supply   0.7348     0.1506  4.8798      0.1760   4.1753
zSupply Only          0.2690     0.2207  1.2187      0.3011   0.8931
\end{lstlisting}

Without adjusting for the covariates, treatment levels \ri{zLPP+Subsidy}   and \ri{zLPP+Subsidy+Supply} are significant. The \ri{exchangeable} working covariance matrix does seem to improve the estimated precision.

We then fit GEE with a covariate. 

\begin{lstlisting}
> hygaccess.gee = gee(y ~ z + x, id = vid,
+                     family = binomial(link = logit),
+                     corstr = "independence", 
+                     data = hygaccess)
> summary(hygaccess.gee)$coef
                    Estimate Naive S.E. Naive z Robust S.E. Robust z
(Intercept)          -1.7526    0.06174 -28.386      0.1398  -12.538
zLPP Only             0.2277    0.06833   3.332      0.1393    1.635
zLPP+Subsidy          0.6850    0.05645  12.133      0.1191    5.749
zLPP+Subsidy+Supply   0.7389    0.05578  13.246      0.1361    5.430
zSupply Only          0.3614    0.07514   4.810      0.2426    1.490
x                     2.0488    0.08209  24.957      0.2158    9.492
> 
> hygaccess.gee = gee(y ~ z + x, id = vid,
+                     family = binomial(link = logit),
+                     corstr = "exchangeable", 
+                     data = hygaccess)
> summary(hygaccess.gee)$coef
                    Estimate Naive S.E. Naive z Robust S.E. Robust z
(Intercept)          -1.7976     0.1324 -13.575      0.1541  -11.667
zLPP Only             0.3038     0.1781   1.705      0.1946    1.561
zLPP+Subsidy          0.7227     0.1316   5.491      0.1271    5.688
zLPP+Subsidy+Supply   0.8547     0.1327   6.441      0.1247    6.855
zSupply Only          0.3236     0.1911   1.693      0.2398    1.350
x                     1.9497     0.1128  17.286      0.1947   10.016
\end{lstlisting}

Covariate adjustment improves efficiency and makes the choice of the working covariance matrix less important.

\subsection{Longitudinal data}

I will analyze the data in Chapter \ref{section::longitudinal-gym}. 
The regression formula \ri{f.reg} will remain the same although other parameters may vary. 

\begin{lstlisting}
> library("gee")
> library("foreign")
> gym1 = read.dta("gym_treatment_exp_weekly.dta")
> f.reg = weekly_visit ~ incentive_commit + incentive + target + member_gym_pre
\end{lstlisting}

Using all data, we find a significant effect of \ri{incentive_commit} but an insignificant effect of \ri{incentive}. 

\begin{lstlisting}
 normal.gee = gee(f.reg, id = id, 
+                family = gaussian, 
+                corstr = "independence", 
+                data = gym1)
> normal.gee = summary(normal.gee)$coef
> normal.gee
                 Estimate Naive S.E. Naive z Robust S.E. Robust z
(Intercept)      -0.69005   0.011136 -61.968     0.08672  -7.9572
incentive_commit  0.15666   0.008358  18.745     0.06376   2.4569
incentive         0.01022   0.008275   1.235     0.05910   0.1729
target            0.62666   0.007465  83.949     0.06773   9.2527
member_gym_pre    1.14919   0.007077 162.375     0.06252  18.3801
\end{lstlisting}

However, this pooled analysis can be misleading because we have seen from the analysis before that the treatments have no effects in the pre-experimental periods and smaller effects in the long term. A pooled analysis can dilute the short-term effects, missing the treatment effect heterogeneity across time. This can be fixed by the following subgroup analysis based on time.

\begin{lstlisting}
> normal.gee1 = gee(f.reg, id = id, 
+                  subset = (incentive_week<0),
+                  family = gaussian, 
+                  corstr = "independence", 
+                  data = gym1)
> normal.gee1 = summary(normal.gee1)$coef
> normal.gee1
                  Estimate Naive S.E.  Naive z Robust S.E.  Robust z
(Intercept)      -0.879374    0.04230 -20.7868     0.08739 -10.06224
incentive_commit -0.004241    0.03175  -0.1336     0.06243  -0.06794
incentive        -0.073884    0.03144  -2.3502     0.06223  -1.18728
target            0.742675    0.02836  26.1887     0.06701  11.08301
member_gym_pre    1.601569    0.02689  59.5664     0.06600  24.26763
>  
> 
> normal.gee2 = gee(f.reg, id = id, 
+                   subset = (incentive_week>0&incentive_week<15),
+                   family = gaussian, 
+                   corstr = "independence", 
+                   data = gym1)
> normal.gee2 = summary(normal.gee2)$coef  
> normal.gee2
                 Estimate Naive S.E. Naive z Robust S.E. Robust z
(Intercept)       -0.7925    0.03275 -24.194     0.08982   -8.823
incentive_commit   0.3662    0.02458  14.898     0.06895    5.311
incentive          0.1744    0.02434   7.166     0.06457    2.701
target             0.6735    0.02196  30.674     0.07159    9.408
member_gym_pre     1.4138    0.02082  67.914     0.06727   21.018
> 
> normal.gee3 = gee(f.reg, id = id, 
+                   subset = (incentive_week>=15),
+                   family = gaussian, 
+                   corstr = "independence", 
+                   data = gym1)
> normal.gee3 = summary(normal.gee3)$coef
> normal.gee3
                  Estimate Naive S.E. Naive z Robust S.E. Robust z
(Intercept)      -0.661500   0.012222  -54.13     0.09028  -7.3273
incentive_commit  0.134789   0.009173   14.69     0.06676   2.0189
incentive        -0.009716   0.009082   -1.07     0.06142  -0.1582
target            0.611635   0.008193   74.66     0.07042   8.6860
member_gym_pre    1.077874   0.007768  138.77     0.06494  16.5967
\end{lstlisting} 

\begin{figure}[ht]
\centering
\includegraphics[width=\textwidth]{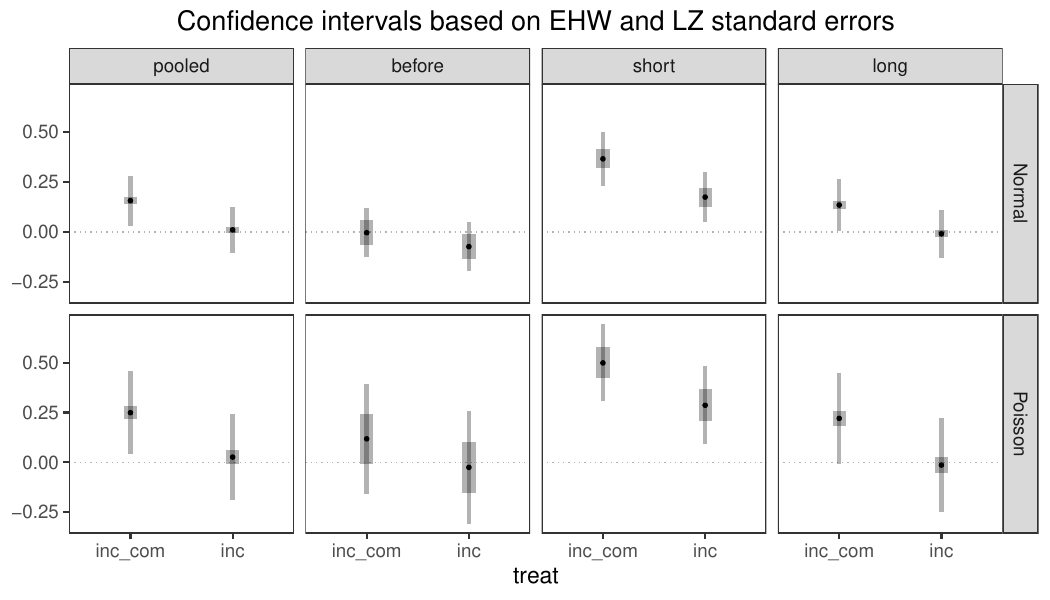}
\caption{GEE analysis of the gym data, where \ri{inc_com} and \ri{inc} correspond to the treatment levels \ri{incentive_commit} and \ri{incentive}, respectively. }\label{fig::gee-analysis-gym-data}
\end{figure}

Changing the \ri{family} parameter to \ri{poisson(link = log)}, we can fit a marginal log-linear model with independent Poisson covariance. Figure \ref{fig::gee-analysis-gym-data} shows the point estimates and confidence intervals based on the regressions above. The confidence intervals based on the cluster-robust standard errors are much wider than those based on the EHW standard errors. Without dealing with clustering, the confidence intervals are too narrow and give wrong inference.

\section{Critiques on the key assumptions}
\label{sec::critiquesonGEEassumptions}
Consider the simple case with $n_i=2$ for all $i$ below. 

\subsection{Assumption \eqref{eq::gee-assumtion-1}} 

I first discuss Assumption \eqref{eq::gee-assumtion-1}, It requires
$$
E(y_{it} \mid X_i) = E(y_{it} \mid x_{it}) , 
$$
which holds automatically if $x_{it} = x_i$ is time-invariant. With time-varying covariates, it effectively rules out the dynamics between $x$ and $y$. Assumption \eqref{eq::gee-assumtion-1} holds in the following data-generating process:
$$
\xymatrix{
x_{i1} \ar[r]\ar[d] &x_{i2} \ar[d] \\
y_{i1} & y_{i2}
}
$$
It does not hold if the lagged $x$ affects $y$ or the lagged $y$ affects $x$:
$$
\xymatrix{
x_{i1} \ar[r]\ar[d]\ar[dr] &x_{i2} \ar[d] \\
y_{i1} & y_{i2}
}
\qquad 
\text{ or }
\qquad 
\xymatrix{
x_{i1} \ar[r]\ar[d] &x_{i2} \ar[d] \\
y_{i1} \ar[ur] & y_{i2}
}
\qquad 
\text{ or }
\qquad 
\xymatrix{
x_{i1} \ar[r]\ar[d]\ar[dr] &x_{i2} \ar[d] \\
y_{i1} \ar[ur] & y_{i2}
}
$$
With more complex data generating processes, Assumption \eqref{eq::gee-assumtion-1} does not hold in general:
$$
\xymatrix{
x_{i1} \ar[r]\ar[d]\ar[dr] &x_{i2} \ar[d] \\
y_{i1} \ar[ur] \ar[r] & y_{i2}
}
$$

\citet{liang1986longitudinal} assumed fixed covariates, ruling out the dynamics of $x$.
\citet{sullivan1994cautionary} pointed out the importance of Assumption \eqref{eq::gee-assumtion-1} in GEE with random time-varying covariates. \citet{sullivan1994cautionary}  also showed that with an independent working covariance matrix, we can drop Assumption \eqref{eq::gee-assumtion-1} as long as the marginal conditional mean is correctly specified. That is, if $E(y_{it}\mid x_{it}) = \mu(x_{it}^{\T}\beta)$, then 
\begin{eqnarray*}
&&E\left\{ \sumn\sum_{t=1}^{n_{i}}\frac{y_{it}-\mu(x_{it}^{\T}\beta)}{\tilde{\sigma}^{2}(x_{it}, \beta)}\frac{\partial\mu(x_{it}^{\T}\beta)}{\partial\beta} \right\}  \\
 &=&\sumn\sum_{t=1}^{n_{i}}  E\left[ \frac{ E\{ y_{it}-\mu(x_{it}^{\T}\beta) \mid x_{it} \} }{\tilde{\sigma}^{2}(x_{it}, \beta)}\frac{\partial\mu(x_{it}^{\T}\beta)}{\partial\beta} \right] \\
 &=& 0.
\end{eqnarray*}
This gives another justification for the use of the independent working covariance matrix even though it can result in efficiency loss when Assumption \eqref{eq::gee-assumtion-1} holds.

\subsection{Assumption \eqref{eq::gee-assumtion-2}} 

I then discuss Assumption \eqref{eq::gee-assumtion-2}. It requires a ``stable'' relationship between $x$ and $y$ across clusters and time points:
$$
E(y_{it} \mid x_{it}) = \mu(x_{it}^{\T} \beta)
$$ 
where $\mu$ and $\beta$ do not depend on $i$ or $t$. 
For clustered data, we can justify this assumption by the exchangeability of the units within clusters. However, it is much harder to interpret or justify it for longitudinal data with complex outcome dynamics. 

We consider linear structural equations with a scalar time-invariant covariate. Without direct dependence of $y_{i2}$ on $y_{i1}$, the data generating process
\begin{eqnarray*}
y_{i1} &=& \alpha_1 + \beta  x_i  + \varepsilon_{i1},\\
y_{i2} &=& \alpha_2 + \beta x_i  + \varepsilon_{i2},
\end{eqnarray*}
corresponding to the graph
$$
\xymatrix{
&x_{i} \ar[dl]  \ar[dr] &  \\
y_{i1}   && y_{i2} 
}
$$
has conditional expectations $E(y_{it}\mid x_{i}) =  \alpha_t +  \beta  x_i   $ if 
\begin{equation}
\label{eq::exogenous-gee}
  E(\varepsilon_{it}\mid x_i) = 0.
\end{equation}
However, with direct dependence of $y_{i2}$ on $y_{i1}$, the data generating process
\begin{eqnarray*}
y_{i1} &=& \alpha_1 +  \beta x_i  + \varepsilon_{i1},\\
y_{i2} &=& \alpha_2 +  \gamma  y_{i1}   + \delta x_i   + \varepsilon_{i2},
\end{eqnarray*}
corresponding to the graph
$$
\xymatrix{
&x_{i} \ar[dl]  \ar[dr] &  \\
y_{i1}  \ar[rr] && y_{i2} 
}
$$
has conditional expectations $E(y_{i1}\mid x_{i}) = \alpha_1 +  \beta   x_i $ but 
$$
E(y_{i2}\mid x_{i}) = \alpha_2 +   \gamma  (\alpha_1 +  \beta   x_i)  + \delta x_i  = (\alpha_2 +   \gamma  \alpha_1) +    ( \delta + \beta\gamma )   x_i 
$$ 
if \eqref{eq::exogenous-gee} holds. The stability assumption requires 
$$
 \alpha_1 = \alpha_2 +   \gamma  \alpha_1,\quad  \beta = \beta\gamma + \delta , 
$$ 
which are strange restrictions on the model parameters. 

With time-varying covariates, this issue becomes even more subtle because Assumption \eqref{eq::gee-assumtion-1} is unlikely to hold in the first place.

\section{Final comments}

\subsection{Explanation versus prediction}

\citet{liang1986longitudinal}'s marginal model is more useful if the goal is to explain the relationship between $x$ and $y$, in particular, a component of $x_{it}$ represents the time-invariant treatment and $y_{it}$ represents the time-varying outcomes. 

If the goal is prediction, then the marginal model can be problematic. For instance, if we observe the covariate value for a future observation $x_{is}$, the marginal model gives predicted outcome $\mu(x_{is}^{\T} \hat\beta)$ with the associated standard error computed based on the delta method. We can see two obvious problems with this prediction. First, it does not depend on $s$. Consequently, predicting $s=10$ is the same as predicting $s=100$. However, the intuition is overwhelming that predicting the long-run outcome is much more difficult than predicting the short-run outcome, so we hope the standard error should be much larger for predicting the outcome at $s=100$. Second, the prediction does not depend on the lag outcomes because the marginal model ignores the dynamics of the outcome. With longitudinal observations, building a model with the lag outcomes may improve prediction.

\subsection{Small number of clusters}

In practice, it is possible that the number of ``clusters'' is small. If $n$ is small, then the Normal approximation of the GEE estimator can be inaccurate and the confidence interval based on the Liang--Zeger cluster-robust standard error can have poor coverage properties. This is a fundamentally challenging problem under the GEE formulation. 

Under the linear model,  \citet{zhang2025random} discussed an alternative formulation with  independent regressors $x_{it}$'s but correlated error terms $\varepsilon_{it}$'s. They showed that the non-clustered standard error may still work even when the clustered standard error cannot be applied (e.g., in the extreme case with a single cluster). This is beyond the scope of this book.

\section{Homework problems}

\paragraph{Sandwich asymptotic covariance matrix for GEE}\label{hw21::gee-sandwichvariance}

Verify the formulas of $B$ and $M$ in Section \ref{subsec:Asymptotic-inference-GEE}.

\paragraph{Cluster-robust standard error in OLS with a cluster-specific binary regressor}\label{hw21::crse-binary-x}

Consider a special case with $x_{it} = (1,x_i)^{\T}$ and $x_i \in \{0,1\}$ for $i=1,\ldots, n$, and view ``1'' as treatment and ``0'' as control. 

Prove  that the coefficient of $x_i$ in the pooled OLS fit of $y_{it}$ on $x_{it} $ equals $ \hat{\tau} =  \bar{y}_1 - \bar{y}_0$ where
\begin{eqnarray*}
\bar{y}_1 &=& \sumn \sum_{t=1}^{n_i} x_{i} y_{it} / N_1,\\ 
\bar{y}_0 &=& \sumn \sum_{t=1}^{n_i} (1-x_{i}) y_{it}/N_0,
\end{eqnarray*}
with $N_1 = \sumn  n_i x_{i}$ and $N_0 = \sumn  n_i (1-x_{i})$ denoting the total number of observations under treatment and control, respectively. Further, prove that the cluster-robust standard error of $ \hat{\tau}$ equals the square root of
$$
\frac{  \sumn x_i R_i^2 }{  N_1^2 }  
+  \frac{  \sumn (1-x_i) R_i^2 }{  N_0^2 } ,
$$
where 
$$
R_i = \left\{
\begin{array}{ccc}
 \sum_{t=1}^{n_i} (y_{it} - \bar{y}_1) , &&\text{if } x_i=1,\\
  \sum_{t=1}^{n_i} (y_{it} - \bar{y}_0) , &&\text{if } x_i=0. 
\end{array}
\right. 
$$

\paragraph{Cluster-robust standard error in GLM with a cluster-specific binary regressor}\label{hw21::crse-binary-x-GLM}

Inherit the setting from Problem \ref{hw21::crse-binary-x}. 

With a binary outcome $y_{it}$, prove that the coefficient of $x_i$ in the pooled logit regression of $y_{it}$ on $x_{it} $ equals $ \hat{\tau} =  \text{logit}\bar{y}_1 - \text{logit} \bar{y}_0$. Further, prove that the cluster-robust standard error of $ \hat{\tau}$ equals the square root of
$$
\frac{  \sumn x_i R_i^2 }{  N_1^2 \bar{y}_1(1-\bar{y}_1) }   
+ \frac{  \sumn (1-x_i) R_i^2 }{ N_0^2 \bar{y}_0 (1-\bar{y}_0) } . 
$$

With a count outcome $y_{it}$, prove that the coefficient of $x_i$ in the pooled Poisson regression of $y_{it}$ on $x_{it} $ equals $ \hat{\tau} =  \log \bar{y}_1 -  \log \bar{y}_0$. Further, prove that the cluster-robust standard error of $ \hat{\tau}$ equals the square root of
$$
\frac{  \sumn x_i R_i^2 }{ N_1^2 \bar{y}_1  }   +  \frac{ \sumn (1-x_i) R_i^2 }{  N_0^2 \bar{y}_0   } . 
$$

\paragraph{Cluster-robust standard error in ANOVA}\label{hw21::crse-anova}

This problem extends Problems \ref{hw5:anova-f}, \ref{hw8::anova-ols-hc02} and \ref{hw16::wls-anova}. 

Inherit the setting from Problem \ref{hw16::wls-anova}. 
If the units are clustered by a factor $c_i \in \{1, \ldots, M\}$ for $i=1,\ldots, n$, we can obtain the cluster-robust covariances $\hat{V}_\textsc{lz}$ and $\hat{V}_\textsc{lz}'$ from the two WLS fits. 

Prove that $\hat{V}_\textsc{lz} = \hat{V}_\textsc{lz}'.$

\paragraph{Cluster-robust standard error for Poisson regression}\label{hw21::poisson-crse}

Similar to Sections \ref{sec::crse-ols} and \ref{sec::crse-logit}, derive the cluster-robust covariance matrix for Poisson regression:
\[
\hat{\cov}(\hat{\beta})=\left(\sumn X_{i}^{\T}\hat{V}_{i}X_{i}\right)^{-1}\sumn X_{i}^{\T}\hat{\varepsilon}_{i}\hat{\varepsilon}_{i}^{\T}X_{i}\left(\sumn X_{i}^{\T}\hat{V}_{i}X_{i}\right)^{-1},
\]
where $\hat{\varepsilon}_{i}=Y_{i}- \mu(X_{i}, \hat{\beta})$ and $\hat{V}_{i}=\text{diag}\{e^{x_{it}^{\T}\hat{\beta}}\}_{t=1}^{n_{i}}.$

%

\paragraph{Data analysis}

Re-analyze the data from \citet{royer2015incentives} using the exchangeable working covariance matrix. Compare the corresponding results with Figure \ref{fig::gee-analysis-gym-data}.

        \part{Beyond Modeling the Conditional Mean}
            
\chapter{Quantile Regression}\label{chapter::quantile-regression}

Previous chapters focus on modeling the conditional mean of the outcome given covariates. Other features of the conditional distribution are also of interest. For instance, the conditional quantiles of the outcome given covariates can give insights into the conditional distribution beyond that from the conditional mean. This chapter focuses on modeling the conditional quantiles and introduces the idea of quantile regression.

\section{From the mean to the quantile}

For a random variable $y$, we can define its mean as 
\[
E(y)=\arg\min_{\mu  \in \mathbb{R} }E\left\{ (y-\mu)^{2}\right\} .
\]
With IID data $(y_{i})_{i=1}^{n}$, we can compute the sample mean
\[
\bar{y}=n^{-1}\sumn y_{i} = \arg\min_{\mu \in \mathbb{R} } n^{-1} \sumn (y_i-\mu)^{2} ,
\]
which satisfies the CLT: 
\[
\sqrt{n}(\bar{y}-E(y))\rightarrow\N(0, \var(y))
\]
in distribution if the variance $ \var(y)$ is finite.

\begin{figure}[ht]
\centering
\includegraphics[width=0.8\textwidth]{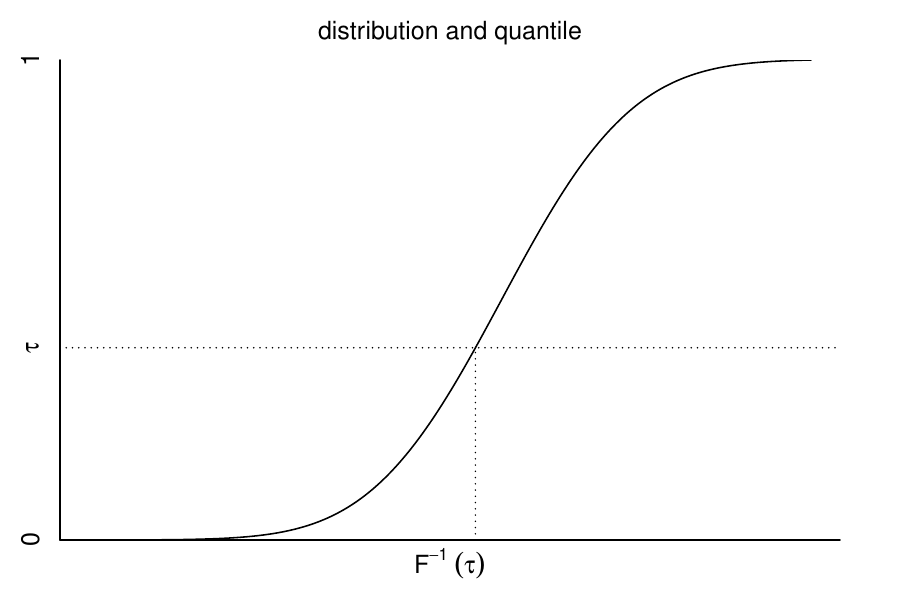}
\caption{CDF and quantile}\label{fig::cdf-quantile}
\end{figure}

\begin{figure}
\includegraphics[width=\textwidth]{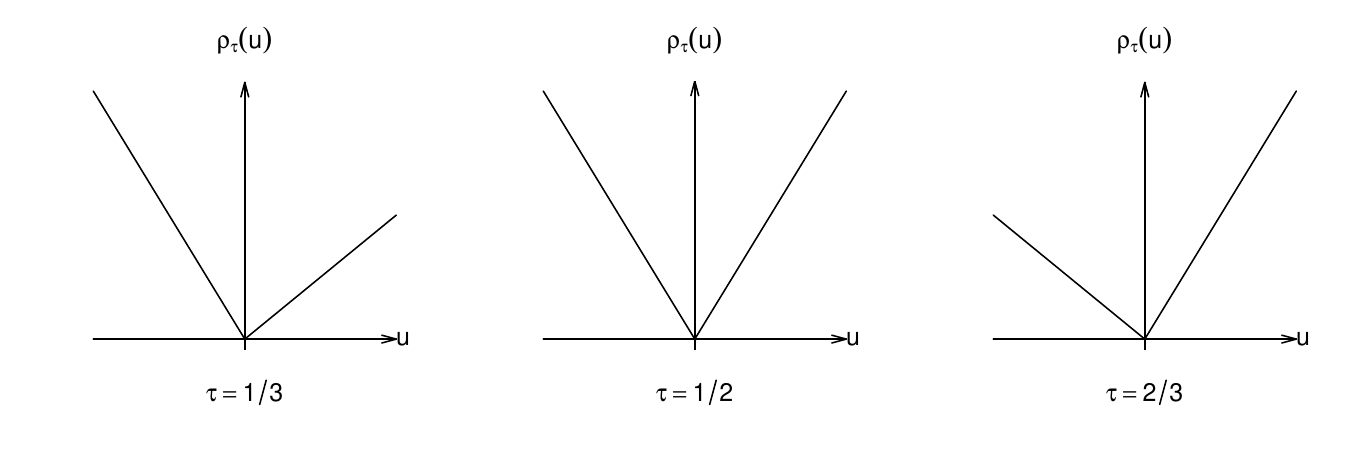}
\caption{Check function}\label{fig::check-function}
\end{figure}

However, the mean can miss important information about $y$. How about other features of the outcome $y$? Quantiles can
characterize the distribution of $y$. For a random variable $y$,
we can define its distribution function as $F(c)=\pr(y\leq c)$ and
its $\tau$th quantile as
\[
F^{-1}(\tau)=\inf\{ q:F(q)\geq\tau \} .
\]
This defines a quantile function $F^{-1}:[0,1]\rightarrow\mathbb{R}$.
If the distribution function is strictly monotone, then the quantile
function reduces to the inverse of the distribution function, and the $\tau$-th quantile solves $\tau=\pr(y\leq q)$ as an equation of $q$.
See Figure \ref{fig::cdf-quantile}. For simplicity, this chapter focuses on the case with a  monotone distribution
function. The definition above formulates the mean as the minimizer of an objective function. Similarly, we can define quantiles in an equivalent
way below.
\begin{proposition}
\label{proposition:definitionofquantile}With a monotone distribution function
and positive density at the $\tau$th quantile, we have
\[
F^{-1}(\tau)=\arg\min_{q \in \mathbb{R} }E\left\{ \rho_{\tau}(y-q)\right\} ,
\]
where 
\[
\rho_{\tau}(u)=u\left\{ \tau-1(u<0)\right\} =\begin{cases}
u\tau, & \text{if }u\geq0,\\
-u(1-\tau), & \text{if }u<0,
\end{cases}
\]
is the check function (the name comes from its shape; see Figure \ref{fig::check-function}). In particular,
the median of $y$ is 
\[
\textup{median}(y)=F^{-1}(0.5)=\arg\min_{q \in \mathbb{R}}E\{ |y-q|\} .
\]
\end{proposition}

\begin{myproof}{Proposition}{\ref{proposition:definitionofquantile}}
To simplify the proof, we further assume that $y$ has density function
$f(\cdot)$. We will use Leibniz's integral rule: 
\[
\frac{\d}{\d x}\left\{ \int_{a(x)}^{b(x)}f(x,t)\d t\right\} =f(x,b(x))b'(x)-f(x,a(x))a'(x)+\int_{a(x)}^{b(x)}\frac{\partial f(x,t)}{\partial x}\d t.
\]

We can write 
\[
E\left\{ \rho_{\tau}(y-q)\right\} =\int_{-\infty}^{q}(\tau-1)(c-q)f(c)\d c+\int_{q}^{\infty}\tau(c-q)f(c)\d c.
\]
To minimize it over $q$, we can solve the first-order condition
\[
\frac{\partial E\left\{ \rho_{\tau}(y-q)\right\} }{\partial q}=(1-\tau)\int_{-\infty}^{q}f(c)\d c-\tau\int_{q}^{\infty}f(c)\d c=0.
\]
So
\[
(1-\tau)\pr(y\leq q)-\tau\left\{ 1-\pr(y\leq q)\right\} =0
\]
which implies that
\[
\tau=\pr(y\leq q),
\]
so the $\tau$th quantile satisfies the first-order condition.
The second-order condition ensures it is the minimizer:
\[
\frac{\partial^{2}E\left\{ \rho_{\tau}(y-q)\right\} }{\partial q^{2}}\Big|_{q=F^{-1}(\tau)}=f\left\{ F^{-1}(\tau)\right\} >0
\]
by  Leibniz's integral rule again. 
\end{myproof}

The empirical distribution function is $\hat{F}(c)=n^{-1}\sumn1(y_{i}\leq c)$,
which is a step function, increasing but not strictly monotone. With
Proposition \ref{proposition:definitionofquantile}, we can easily define
the sample quantile as 
\[
\hat{F}^{-1}(\tau)=\arg\min_{q \in \mathbb{R}  }n^{-1}\sumn\rho_{\tau}(y_{i}-q),
\]
which may not be unique even though the population quantile is. We
can view $\hat{F}^{-1}(\tau)$ as a set containing all minimizers,
and with large samples the values in the set do not differ much. Similar
to the sample mean, the sample quantile also satisfies a CLT.

\begin{theorem}\label{thm::sample-quantiles-asymptotics}
Assume $(y_{i})_{i=1}^{n}\iidsim y$ with distribution function $F(\cdot)$
that is strictly increasing and density function $f(\cdot)$ that
is positive at the $\tau$th quantile. The sample quantile
is consistent for the true quantile and is asymptotically Normal:
\[
\sqrt{n}\left\{ \hat{F}^{-1}(\tau)-F^{-1}(\tau)\right\} \rightarrow\N\left(0,\frac{\tau(1-\tau)}{\left[f\left\{ F^{-1}(\tau)\right\} \right]^{2}}\right)
\]
in distribution. In particular, the sample median satisfies
\[
\sqrt{n}\left\{ \hat{F}^{-1}(0.5)-\textup{median}(y)\right\} \rightarrow\N\left(0,\frac{1}{4\left[f\left\{ \textup{median}(y)\right\} \right]^{2}}\right)
\]
in distribution. 
\end{theorem}

\begin{myproof}{Theorem}{\ref{thm::sample-quantiles-asymptotics}}
Based on the first-order condition in  Proposition \ref{proposition:definitionofquantile}, the population quantile solves
$$
E \{ m_\tau(y - q) \} =0,
$$
and the sample quantile solves
$$
n^{-1} \sumn m_\tau(y_i - q)  = 0,
$$
where the check function has a partial derivative with respect to
$u$ except for the point $0$: 
\begin{eqnarray*}
m_{\tau}(u) &=&  (\tau-1)1(u<0)+\tau1(u>0) \\
&=&\tau-1(u < 0).
\end{eqnarray*}
By Theorem \ref{theorem:sandwich-theorem-cov-iid} in Appendix \ref{chapter::m-mle},  we only need to find the ``bread'' and ``meat'' matrices, which are scalars now:
\begin{eqnarray*}
B &=& \frac{\partial   E \{ m_\tau(y - q) \}  }{ \partial q}\Big | _{q = F^{-1}(\tau)}\\
&=& \frac{\partial   E \{ \tau-1(y\leq q) \}  }{ \partial q}\Big | _{q =F^{-1}(\tau)}\\
&=& -\frac{\partial   F(q)   }{ \partial q}\Big | _{q =F^{-1}(\tau)}\\
&=& -f\{F^{-1}(\tau) \},
\end{eqnarray*}
and
\begin{eqnarray*}
M &=& E \left[ \{ m_\tau(y - q) \}^2 \right] \Big | _{q =F^{-1}(\tau)} \\
&=& E \left[ \left\{ \tau-1(y\leq q) \right\}^2 \right] \Big | _{q =F^{-1}(\tau)} \\
&=& E \left[ \tau^2 + 1(y\leq q) - 2\cdot  1(y\leq q) \tau \right] \Big | _{q =F^{-1}(\tau)} \\
&=& \tau^2 + \tau - 2\tau^2 \\
&=& \tau(1-\tau).
\end{eqnarray*}
Therefore, $\sqrt{n}\{ \hat{F}^{-1}(\tau)-F^{-1}(\tau)\}$ converges to Normal with mean zero and variance $M/B^2 = \tau(1-\tau)/ [f\{F^{-1}(\tau) \}]^2$. 
\end{myproof}

To conduct statistical inference for the quantile $F^{-1}(\tau)$, we need to estimate the density of $y$ at the $\tau$th quantile to obtain the estimated standard error of $\hat{F}^{-1}(\tau)$. However, kernel density estimation involves tuning parameters that may affect the performance of the final variance estimator. Alternatively, we can use the bootstrap to obtain the estimated standard error.\footnote{
As mentioned in the Preface, I do not cover the bootstrap carefully in this book. However, the bootstrap is very attractive for quantile estimation and quantile regression because the variance estimation is more complicated even with analytic formulas. 
} We will discuss the inference of quantiles in \ri{R} in Section \ref{section::numerical-examples-qr}.

\section{From the conditional mean to conditional quantile}

With an explanatory variable $x$ for outcome $y$, we can define
the conditional mean as
\[
E(y\mid x)=\arg\min_{m(\cdot)}E\left[\left\{ y-m(x)\right\} ^{2}\right].
\]
We can use a linear function $x^{\T}\beta$ to approximate the conditional
mean with the population OLS coefficient 
\[
\beta=\arg\min_{b \in \mathbb{R}^p  }E\left\{ (y-x^{\T}b)^{2}\right\} =\left\{ E(xx^{\T})\right\} ^{-1}E(xy),
\]
and the sample OLS coefficient
\[
\hat{\beta}=\left(n^{-1}\sumn x_{i}x_{i}^{\T}\right)^{-1}\left(n^{-1}\sumn x_{i}y_{i}\right).
\]
We have discussed the statistical properties of $\hat{\beta}$ in previous chapters.
Motivated by Proposition \ref{proposition:definitionofquantile}, we can
define the conditional quantile function as 
\[
F^{-1}(\tau\mid x)=\arg\min_{q(\cdot)}E\left[\rho_{\tau}\left\{ y-q(x)\right\} \right].
\]
We can use a linear function $x^{\T}\beta(\tau)$ to approximate the
conditional quantile function with
\[
\beta(\tau)=\arg\min_{b   \in \mathbb{R}^p  }E\left\{ \rho_{\tau}(y-x^{\T}b)\right\} 
\]
called the $\tau$th population regression quantile, and 
\begin{equation}
\hat{\beta}(\tau)=\arg\min_{b   \in \mathbb{R}^p }n^{-1}\sumn\rho_{\tau}(y_{i}-x_{i}^{\T}b)\label{eq:sample-regression-quantiles}
\end{equation}
called the $\tau$th sample regression quantile. 
As a special case, when $\tau = 0.5$, we have the regression median:
\[
\hat{\beta}(0.5)=\arg\min_{b  \in \mathbb{R}^p  } n^{-1}\sumn |y_{i}-x_{i}^{\T}b|,
\]
which is also called the least absolute deviations (LAD).

\citet{koenker1978regression} started the literature under a correctly specified conditional quantile model:
$$
F^{-1}(\tau\mid x) = x^{\T} \beta(\tau).
$$
The interpretation of the $j$-th coordinate of the coefficient, $\beta_j(\tau)$, is the partial influence of $x_{ij}$ on the $\tau$th conditional quantile of $y_i$ given $x_{i}$. 
\citet{angrist2006quantile} discussed quantile regression under misspecification, viewing it as the best linear approximation to the true conditional quantile function. This chapter will focus on the statistical properties of
the sample regression quantiles following \citet{angrist2006quantile}'s discussion of statistical inference allowing for the misspecification of the quantile regression model.

Before that, we first comment
on the population regression quantiles based on some generative models. Below assume that the $v_{i}$'s are IID with mean zero and distribution $g(c) = \pr(  v_i \leq c)$, and are independent of the covariates $x_i$'s.

\begin{example}
Under the linear model $y_{i}=x_{i}^{\T}\beta+\sigma v_{i}$, we can verify that 
$$
E(y_i\mid x_i) = x_{i}^{\T}\beta
$$
and
\[
F^{-1}(\tau\mid x_{i})=x_{i}^{\T}\beta+\sigma g^{-1}(\tau).
\]
Therefore, with the first regressor being $1$, we have
\[
\beta_{1}(\tau)=\beta_{1}+\sigma g^{-1}(\tau),\quad\beta_{j}(\tau)=\beta_{j},\quad(j=2,\ldots,p).
\]
In this case, both the true conditional mean and quantile functions are linear, and
the population regression quantiles are constant across $\tau$ except
for the intercept.
\end{example}

\begin{example}
Under a heteroskedastic linear model $y_{i}=x_{i}^{\T}\beta+(x_{i}^{\T}\gamma)v_{i}$
with   $x_{i}^{\T}\gamma>0$ for all $x_{i}$'s,
we can verify that 
$$
E(y_i\mid x_i) = x_{i}^{\T}\beta
$$
and 
\[
F^{-1}(\tau\mid x_{i})=x_{i}^{\T}\beta+x_{i}^{\T}\gamma g^{-1}(\tau).
\]
Therefore, 
\[
\beta(\tau)=\beta+\gamma g^{-1}(\tau).
\]
In this case, both the true conditional quantile functions are linear, and
all coordinates of the population regression quantiles vary with $\tau$. 
\end{example}

\begin{example}
Under the transformed linear model $\log y_{i}=x_{i}^{\T}\beta+\sigma v_{i}$, we can verify that 
$$
E(y_i\mid x_i) = \exp(x_{i}^{\T}\beta) M_v(\sigma),
$$
where $M_v(t) = E(e^{tv})$ is the moment generating function of $v$, 
and 
\[
F^{-1}(\tau\mid x_{i})=\exp\left\{ x_{i}^{\T}\beta+\sigma g^{-1}(\tau)\right\} .
\]
In this case, both the true conditional mean and quantile functions are log-linear in covariates. 
\end{example}

\section{Sample regression quantiles}

\subsection{Computation}

The sample regression quantiles (\ref{eq:sample-regression-quantiles}) do
not have explicit formulas in general, and we need to solve the optimization
problem numerically. Motivated by the piece-wise linear feature of
the check function, we decompose $y_{i}-x_{i}^{\T} b $ as the
difference between its positive part and negative part:
\[
y_{i}-x_{i}^{\T} b =u_{i}-v_{i},
\]
where 
\begin{eqnarray*}
u_{i} &=& \max(y_{i}-x_{i}^{\T} b ,0),\\ 
v_{i} &=& -\min(y_{i}-x_{i}^{\T} b,0).
\end{eqnarray*}
So the objective function simplifies to the summation of 
\[
\rho_{\tau}(y_{i}-x_{i}^{\T} b)=\tau u_{i}+(1-\tau)v_{i},
\]
which is simply a linear function of the $u_{i}$'s and $v_{i}$'s.
Of course, these $u_{i}$'s and $v_{i}$'s are not arbitrary because
they must satisfy the constraints by the data. Using the notation
\[
Y=\left(\begin{array}{c}
y_{1}\\
\vdots\\
y_{n}
\end{array}\right),\quad X=\left(\begin{array}{c}
x_{1}^{\T}\\
\vdots\\
x_{n}^{\T}
\end{array}\right),\quad u=\left(\begin{array}{c}
u_{1}\\
\vdots\\
u_{n}
\end{array}\right),\quad v=\left(\begin{array}{c}
v_{1}\\
\vdots\\
v_{n}
\end{array}\right),
\]
finding the $\tau$th regression quantile is equivalent to a linear
programming problem with linear objective function and linear constraints:
\begin{eqnarray*}
 & \min_{b \in \mathbb{R}^p ,u \in  \mathbb{R}^n,v \in \mathbb{R}^n} &  \tau1_{n}^{\T}u+(1-\tau)1_{n}^{\T}v  ,\\
 & \text{such that}& Y=Xb+u-v,\\
 &&  u_i \geq  0, v_i \geq  0 \quad  (i=1,\ldots, n).
\end{eqnarray*}
The function \ri{rq} in the \ri{R} package \ri{quantreg} can compute the regression quantiles with various methods. 

\subsection{Asymptotic inference}

Similar to the sample quantiles, the regression quantiles are also
consistent for the population regression quantiles and asymptotically
Normal. So we can conduct asymptotic inference based on  Theorem \ref{thm::regression-quantiles-asymptotics} below, which is due to \citet{angrist2006quantile}.

\begin{theorem}\label{thm::regression-quantiles-asymptotics}
Assume $(y_{i},x_{i})_{i=1}^{n}\iidsim(y,x)$. Under some regularity
conditions, we have
\[
\sqrt{n} \{ \hat{\beta}(\tau)-\beta(\tau) \} \rightarrow\N(0,B^{-1}MB^{-1})
\]
in distribution, where 
\[
B=E\left[f_{y\mid x}\left\{ x^{\T}\beta(\tau)\right\} xx^{\T}\right]
\]
and
\[
M=E\left[\left\{ \tau-1\left(y-x^{\T}\beta(\tau)\leq0\right)\right\} ^{2}xx^{\T}\right],
\]
with $f_{y\mid x}(\cdot)$ denoting the conditional density of $y$
given $x$.
\end{theorem}

\begin{myproof}{Theorem}{\ref{thm::regression-quantiles-asymptotics}}
The population regression quantile solves
\[
E\left\{ m_{\tau}(y-x^{\T}b)x\right\} =0,
\]
and the sample regression quantile solves
\[
n^{-1}\sumn m_{\tau}(y_{i}-x_{i}^{\T}b)x_{i}=0.
\]
By Theorem \ref{theorem:sandwich-theorem-cov-iid} in Appendix \ref{chapter::m-mle}, 
we only need to calculate the explicit forms of $B$
and $M$. Let $F_{y\mid x}(\cdot)$ and $f_{y\mid x}(\cdot)$ be the
conditional distribution and density functions. We have
\begin{align*}
E\left\{ m_{\tau}(y-x^{\T}b)x\right\}  & =E\left[\left\{ \tau-1(y-x^{\T}b\leq0)\right\} x\right]\\
 & =E\left[\left\{ \tau-F_{y\mid x}(x^{\T}b)\right\} x\right],
\end{align*}
so 
\[
\frac{\partial E\left\{ m_{\tau}(y-x^{\T}b)x\right\} }{\partial b^{\T}}
=-E\left\{ f_{y\mid x}(x^{\T}b)xx^{\T}\right\} .
\]
This implies the formula of $B$. The formula of $M$ follows from
\begin{align*}
M & =E\left\{ m_{\tau}^{2}(y-x^{\T}\beta(\tau))xx^{\T}\right\} \\
 & =E\left[\left\{ \tau-1(y-x^{\T}\beta(\tau)\leq0)\right\} ^{2}xx^{\T}\right].
\end{align*}
\end{myproof}

Based on Theorem \ref{thm::regression-quantiles-asymptotics}, we can estimate the asymptotic covariance matrix of $\hat{\beta}(\tau)$ by $n^{-1} \hat{B}^{-1} \hat{M}  \hat{B}^{-1}$, where
\[
\hat{M}=n^{-1}\sumn\left\{ \tau-1\left(y_{i}-x_{i}^{\T}\hat{\beta}(\tau)\leq0\right)\right\} ^{2}x_{i}x_{i}^{\T}
\]
and
\[
\hat{B}=\left(2nh\right)^{-1}\sumn1\left\{ |y_{i}-x_{i}^{\T}\hat{\beta}(\tau)|\leq h\right\} x_{i}x_{i}^{\T}
\]
for a carefully chosen $h$. \citet{powell1991estimation}'s theory suggests to use $h$ satisfying condition $h=O(n^{-1/3})$, but the theory is not so helpful since it only suggests the order of $h$. The \ri{quantreg} package in \ri{R} chooses a specific $h$ that satisfies this condition. 
In finite samples, the bootstrap often gives a better estimation of the asymptotic covariance matrix.

\section{Numerical examples}\label{section::numerical-examples-qr}

\subsection{Sample quantiles}

\begin{figure}[ht]
\centering
\includegraphics[width = \textwidth]{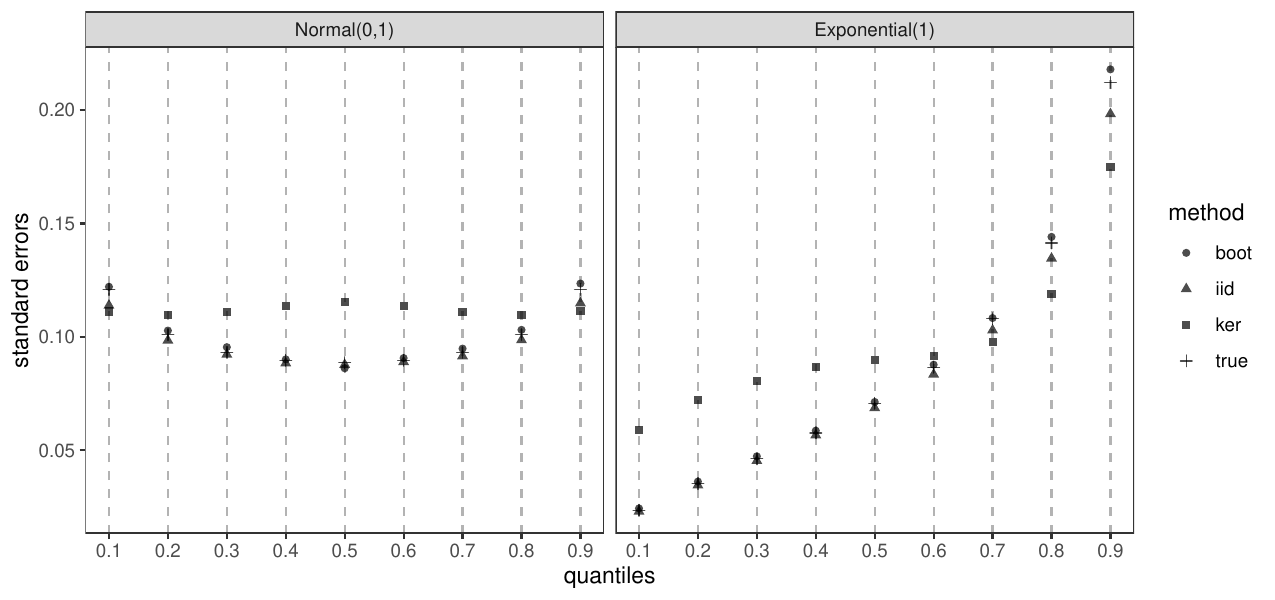}
\caption{Standard errors for sample quantiles}\label{fig::se-sample-quantiles}
\end{figure}

We can use the \ri{quantile} function to obtain the sample quantiles. However, it does not report standard errors. Instead, we can use the \ri{rq} function to compute sample quantiles by regressing the outcome on constant 1. These two functions may return different sample quantiles when they are not unique. The difference is often small with large sample sizes.

I use the following simulation to compare various methods for standard error estimation. The first data-generating process has a standard Normal outcome. 

\begin{lstlisting}
library(quantreg)
mc= 2000
n = 200
taus = (1:9)/10
get.se = function(x){x$coef[1,2]}
q.normal = replicate(mc,{
  y  = rnorm(n)
  qy = rq(y~1, tau = taus)
  se.iid = summary(qy, se = "iid")
  se.ker = summary(qy, se = "ker")
  se.boot= summary(qy, se = "boot")
  
  qy = qy$coef
  se.iid = sapply(se.iid, get.se)
  se.ker = sapply(se.ker, get.se)
  se.boot= sapply(se.boot, get.se)
  
  c(qy, se.iid, se.ker, se.boot)
})
\end{lstlisting}

In the above, \ri{se = "iid"}, \ri{se = "ker"}, and \ri{se = "boot"} correspond to the standard errors by \citet{koenker1978regression}, \citet{powell1991estimation}, and the bootstrap. 
I also run the same simulation but replace the Normal outcome with Exponential: \ri{y  = rexp(n)}. Figure \ref{fig::se-sample-quantiles} compares the estimated standard errors with the true asymptotic standard error in Theorem \ref{thm::sample-quantiles-asymptotics}. Bootstrap works the best, and the one involving kernel estimation of the density seems biased.

\subsection{OLS versus LAD}

I will use simulation to compare OLS and LAD. In \ri{rq}, the default value is \ri{tau=0.5}, fitting the LAD. 
I will consider four data-generating processes.

\begin{enumerate}[label=(D\arabic*), ref=D\arabic*]
\item 
The first data-generating process is a Normal linear model:

\begin{lstlisting}
x = rnorm(n)
simu.normal = replicate(mc, {
  y = 1 + x + rnorm(n)
  c(lm(y~x)$coef[2], rq(y~x)$coef[2])
})
\end{lstlisting}

\item 
The second data generating process replaces the error term to a Laplace distribution:\footnote{Note that the difference between two independent Exponentials has the same distribution as Laplace. See Proposition \ref{prop::expo-laplace} in Appendix \ref{chapter:appendix-rvs}, which justifies the \ri{R} code above.}

\begin{lstlisting}
simu.laplace = replicate(mc, {
  y = 1 + x + rexp(n) - rexp(n)
  c(lm(y~x)$coef[2], rq(y~x)$coef[2])
})
\end{lstlisting}

The OLS is the MLE under a Normal linear model according to Problem \ref{hw5:normal-mle}. The LAD is the MLE under a linear model with independent Laplace errors according to Problem \ref{hw5:laplace-mle}.

\item 
The third data-generating process replaces the error term with standard Exponential:

\begin{lstlisting}
simu.exp = replicate(mc, {
  y = 1 + x + rexp(n) 
  c(lm(y~x)$coef[2], rq(y~x)$coef[2])
})
\end{lstlisting}

\item 
The fourth data generating process has $y_i = 1+e_i x_i$ with $e_i$ IID Exponential, so 
$$
E(y_i \mid x_i) = 1+x_i,\quad \var(y_i\mid x_i) = x_i^2,
$$
which is a heteroskedastic linear model,
and
$$
F^{-1}(0.5\mid x_i) = 1 + \text{median}(e_i) x_i = 1+ ( \log 2) x_i,
$$
which is a linear quantile model. The coefficients are different in the conditional mean and quantile functions.

\begin{lstlisting}
x = abs(x)
simu.x = replicate(mc, {
  y = 1 + rexp(n)*x
  c(lm(y~x)$coef[2], rq(y~x)$coef[2])
})
\end{lstlisting}
\end{enumerate}

\begin{figure} 
\centering
\includegraphics[width = \textwidth]{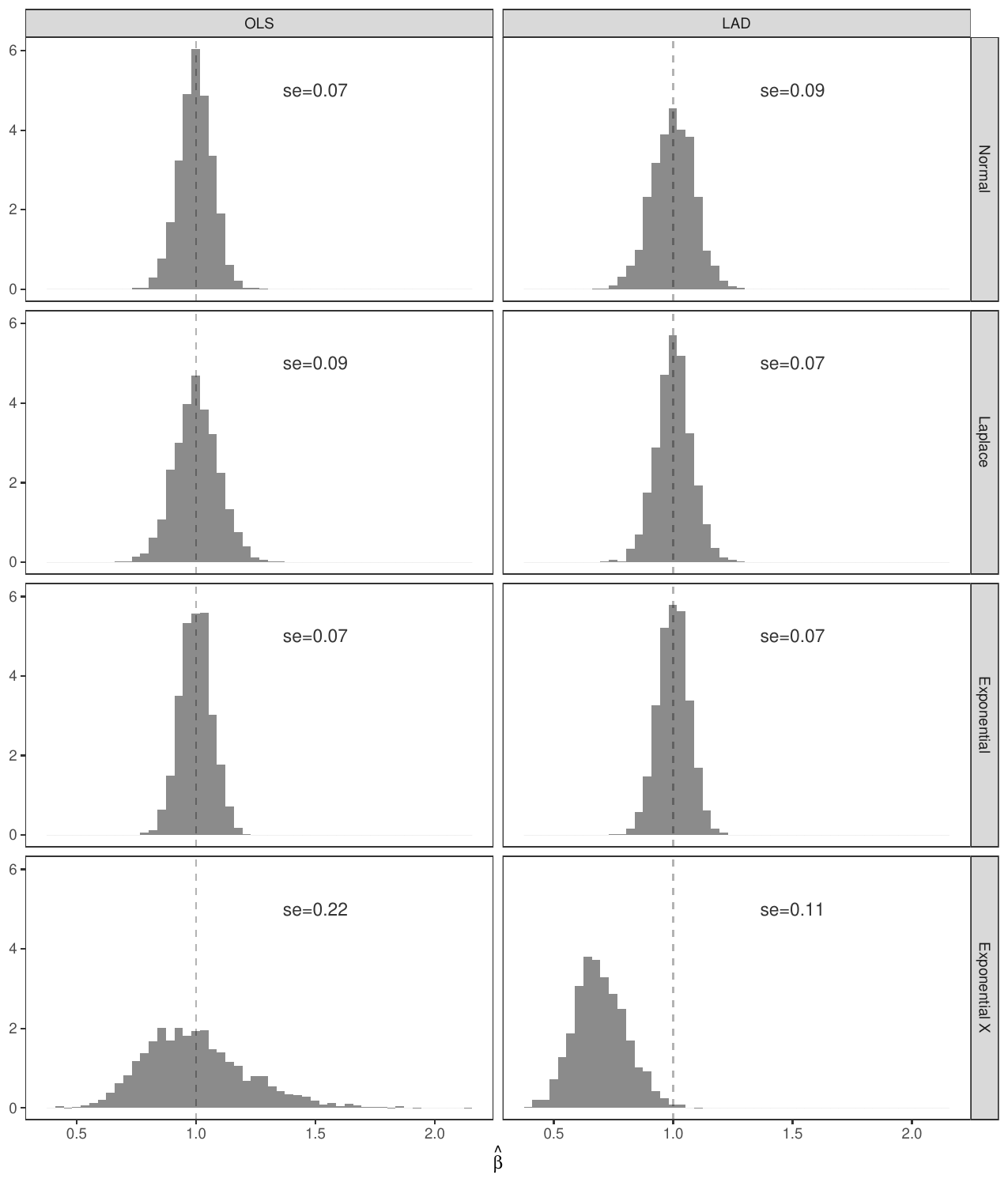}
\caption{Regression quantiles}\label{fig::simu-regression-quantiles}
\end{figure}

Figure \ref{fig::simu-regression-quantiles} compares OLS and LAD under the above four data-generating processes. With Normal errors, OLS is more efficient; with Laplace errors, LAD is more efficient. This confirms the theory of MLE. With Exponential errors, LAD is also more efficient than OLS. Under the fourth data-generating process, LAD is more efficient than OLS. In general, however, OLS and LAD target the conditional mean and conditional median, respectively. Since the parameters differ in general,   the comparison of the standard errors is not very meaningful. Both OLS and LAD give useful information about the data.

\section{Application}

\subsection{Parents' and children's heights}

I revisit Galton's data introduced in Chapter \ref{chapter::ols-1d}. The following \ri{R} code gives the coefficients for quantiles $0.1$ to $0.9$. 

\begin{lstlisting}
> library("HistData")
> taus     = (1:9)/10
> qr.galton = rq(childHeight ~ midparentHeight, 
+                tau = taus,
+                data = GaltonFamilies)
> coef.galton = qr.galton$coef
\end{lstlisting}

\begin{figure}[th]
\centering
\includegraphics[width = 0.8 \textwidth]{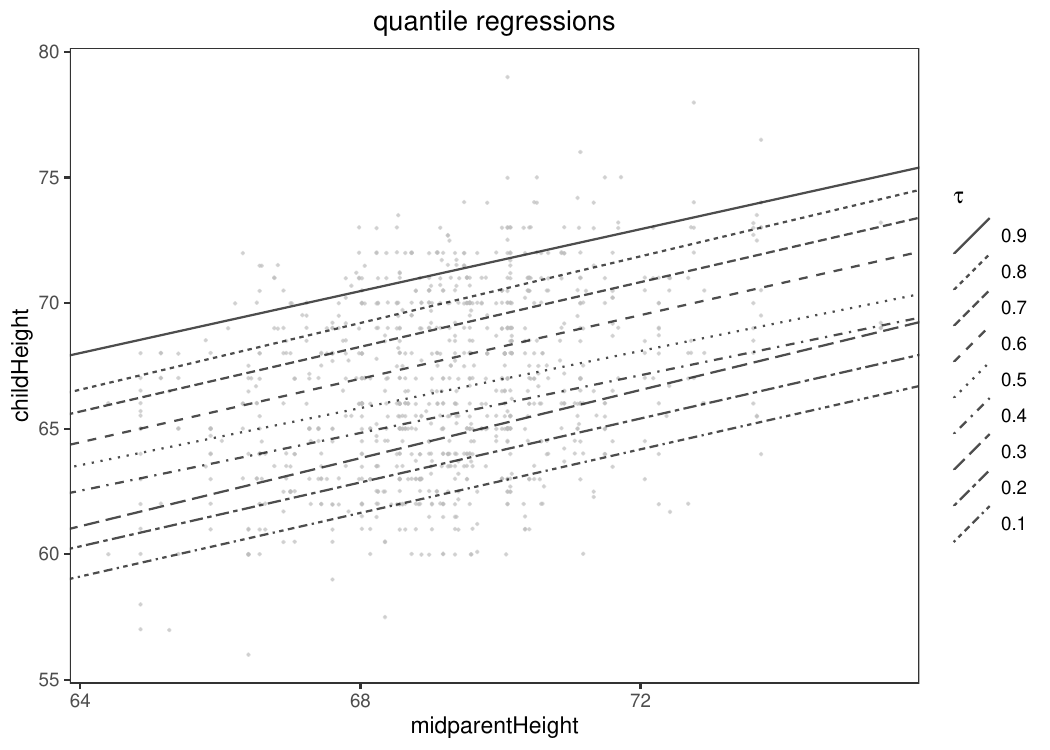}
\caption{Galton's data}\label{fig::galton-regression-quantiles}
\end{figure}

Figure \ref{fig::galton-regression-quantiles} shows the quantile regression lines, which are almost parallel with different intercepts. In Galton's data, $x$ and $y$ are very close to a bivariate Normal distribution. Theoretically, we can verify that with bivariate Normal $(x,y)$, the conditional quantile function $F^{-1}(\tau \mid x)$ is linear in $x$ with the same slope. See Problem \ref{hw22::qr-bivariate-Normal}.

\subsection{U.S. wage structure}

\citet{angrist2006quantile} used quantile regression to study the U.S. wage structure. They used census data in 1980, 1990, and 2000 to fit quantile regressions on log weekly wage on education and other variables. The following code gives the coefficients for quantile regressions with $\tau$ equaling $0.1$ to $0.9$. I repeat the regressions with data from three years. Due to the large sample size, I use the $m$-of-$n$ bootstrap with $m=500$.

\begin{lstlisting}
> library(foreign)
> census80 = read.dta("census80.dta")
> census90 = read.dta("census90.dta")
> census00 = read.dta("census00.dta")
> f.reg = logwk ~ educ + exper + exper2 + black
> 
> m.boot   = 500       
> rq80     = rq(f.reg, data = census80, tau = taus)
> rqlist80 = summary(rq80, se = "boot", 
+                    bsmethod= "xy", mofn = m.boot)
> rq90     = rq(f.reg, data = census90, tau = taus)
> rqlist90 = summary(rq90, se = "boot", 
+                    bsmethod= "xy", mofn = m.boot)
> rq00     = rq(f.reg, data = census00, tau = taus)
> rqlist00 = summary(rq00, se = "boot", 
+                    bsmethod= "xy", mofn = m.boot)
\end{lstlisting}

\begin{figure}[th]
\centering
\includegraphics[width = \textwidth]{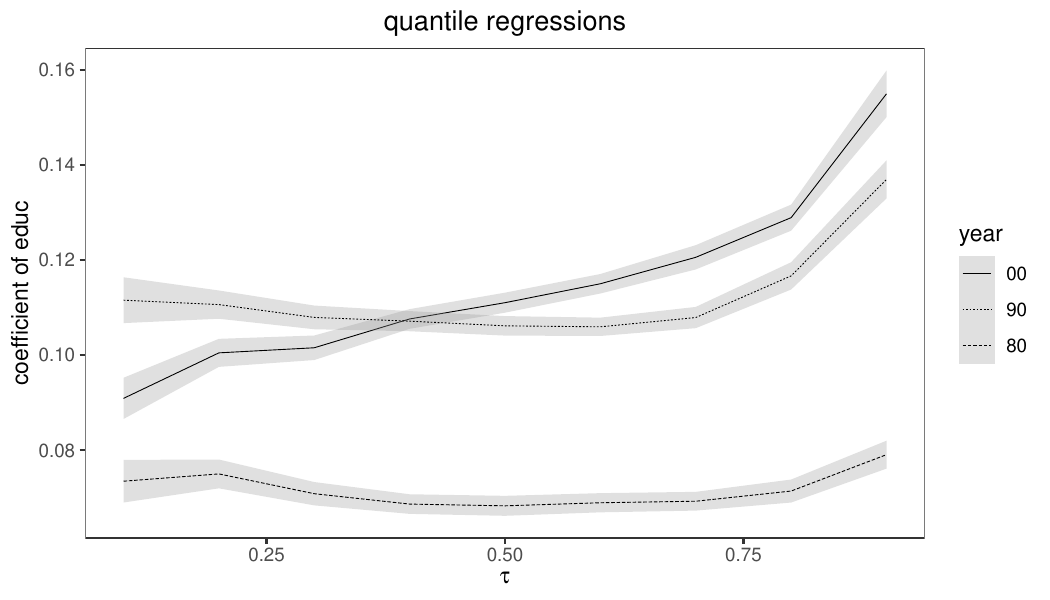}
\caption{\citet{angrist2006quantile}'s data}\label{fig::angrist-regression-quantiles}
\end{figure}

Figure \ref{fig::angrist-regression-quantiles} shows the coefficient of \ri{educ} across years and across quantiles. In the 1980 data, the coefficients are nearly constant across quantiles, showing no evidence of heterogeneity in the return of education. Compared with 1980, the return to education in 1990 increases across all quantiles, but it increases more at the upper quantiles. Compared with 1990, the return to education in 2000 decreases at the lower quantiles and increases at the upper quantiles, which shows more dramatic heterogeneity across quantiles.

The original data used by \citet{angrist2006quantile} contain weights due to sampling. Ideally, we should use the weights in the quantile regression. Like \ri{lm}, the \ri{rq} function also allows for specifying \ri{weights}. I relegate the details to Problem \ref{hw22::weighted-qr-application}.

\section{Extensions}

\subsection{Cluster-robust standard error for quantile regression}

With clustered data, we must use the cluster-robust standard error, which can be approximated by the clustered bootstrap with the \ri{rq} function. I use \citet{hagemann2017cluster}'s example below, where the students are clustered in classrooms.

\begin{lstlisting}
> star = read.csv("star.csv")
> star.rq  = rq(pscore ~ small + regaide + black + 
+                 girl + poor + tblack + texp + 
+                 tmasters + factor(fe),
+               data = star)
> res      = summary(star.rq, se = "boot")$coef[2:9, ]
> res.clus = summary(star.rq, se = "boot", 
+                    cluster = star$classid)$coef[2:9, ]
> round(res, 3)
           Value Std. Error t value Pr(>|t|)
small      6.500      1.122   5.795    0.000
regaide    0.294      1.071   0.274    0.784
black    -10.334      1.657  -6.237    0.000
girl       5.073      0.878   5.777    0.000
poor     -14.344      1.024 -14.011    0.000
tblack    -0.197      1.751  -0.113    0.910
texp       0.413      0.098   4.231    0.000
tmasters  -0.530      1.068  -0.497    0.619
> round(res.clus, 3)
           Value Std. Error t value Pr(>|t|)
small      6.500      1.662   3.912    0.000
regaide    0.294      1.627   0.181    0.857
black    -10.334      1.849  -5.588    0.000
girl       5.073      0.819   6.195    0.000
poor     -14.344      1.152 -12.455    0.000
tblack    -0.197      3.113  -0.063    0.949
texp       0.413      0.168   2.465    0.014
tmasters  -0.530      1.444  -0.367    0.713
\end{lstlisting}

\subsection{High-dimensional quantile regression}

With high dimensional covariates, we can use regularized quantile regression. For instance, the \ri{rq} function can implement the lasso version with \ri{method = "lasso"} and a prespecified \ri{lambda}. The \ri{R} package \ri{rqPen} implements various penalized quantile regressions.

\section{Homework problems}

\paragraph{Quantile regression with a binary regressor}\label{hw22::qr-binaryx}

For $i=1,\ldots, n$, the first $1/3$ observations have $x_i=1$ and the last $2/3$ observations have $x_i=0$;
$y_i\mid x_i =1$ follows an Exponential$(1)$, and $y_i\mid x_i = 0 $ follows an Exponential$(2)$. Find
$$
(\hat{\alpha}, \hat{\beta})  = \arg\min_{(a,b)} \sumn \rho_{1/2}(y_i - a - b x_i).
$$
and the joint asymptotic distribution.

\paragraph{Conditional quantile function in bivariate Normal}\label{hw22::qr-bivariate-Normal}

Prove that if $(x,y)$ follows a bivariate Normal, the conditional quantile function of $y$ given $x$ is linear in $x$ with the same slope across all $\tau$. 

\paragraph{Quantile range and variance}\label{hw22::quantile-range-variance}

A symmetric (around 0) random variable $y$ satisfies $y \sim -y$. 
Define the $1-\alpha$ quantile range of a symmetric random variable $y$ as the interval of its $\alpha/2$  and $1-\alpha/2$ quantiles.
Theorem \ref{thm::quantile-to-variance} shows that narrower quantile ranges imply smaller variance. Prove Theorem \ref{thm::quantile-to-variance}.

\begin{theorem}
\label{thm::quantile-to-variance}
Consider two symmetric random variables $y_1$ and $y_2$.
If the $1-\alpha$ quantile range of $y_1$ is wider than that of $y_2$ for all $\alpha$, then $\var(y_1) \geq \var(y_2)$.
\end{theorem}

Does the converse of Theorem \ref{thm::quantile-to-variance} hold? That is, does smaller variance imply narrow quantile ranges in general? If so, give a proof; if not, give a counterexample.

\paragraph{Interquartile range and estimation}\label{hw22::interquartile-range}

The interquartile range of a random variable $y$ equals the difference between its 75\% and 25\% quantiles. 
Based on IID data $(y_i)_{i=1}^n$, write a function to estimate the interquartile range and the corresponding standard error using the bootstrap. Use simulated data to evaluate the finite sample properties of the point estimate (e.g., bias and variance) and the 95\% confidence interval (e.g. coverage rate and length).

Find the asymptotic distribution of the estimator for the interquartile range.

\paragraph{Joint asymptotic distribution of the sample median and the mean}
\label{hw22::joint-mean-median}

Assume that $y_1,\ldots, y_n \sim y$ are IID. Find the joint asymptotic distribution of the sample mean $\bar{y}$ and median $\hat{m}$.

Remark: The mean $\mu$ and median $m$ satisfy the estimating equation with 
$$
w(y, \mu, m) = \begin{pmatrix}
y - \mu \\
0.5 - 1(y - m \leq 0)
\end{pmatrix}. 
$$

\paragraph{Weighted quantile regression and application}\label{hw22::weighted-qr-application}

Many real data contain weights due to sampling. For example, in \citet{angrist2006quantile}'s data, \ri{perwt} is the sampling weight. Define the weighted quantile regression problem theoretically and re-analyze \citet{angrist2006quantile}'s data with weights. Note that similar to \ri{lm} and \ri{glm}, the quantile regression function \ri{rq}  also has a parameter \ri{weights}.

\chapter{Modeling Time-to-Event Outcomes}
 \label{chapter::survival-analysis}

Time-to-event data are common in biomedical and social sciences. Statistical analysis of time-to-event data is called
{\it survival analysis} in biostatistics and {\it duration analysis} in econometrics.
The former name comes from biomedical applications, where the outcome
denotes the survival time or the time to the recurrence of the disease of interest. The
latter name comes from the economic applications, where the outcome
denotes the weeks unemployed or days until the next arrest after being released from incarceration. See \citet{kalbfleisch2011statistical} for biomedical applications and \citet{heckman1984econometric} for economic applications.

A key difficulty of analyzing time-to-event data is that the outcomes are often censored, that is, we can only observe that the outcomes are within certain intervals but not their precise values.  This chapter will introduce three powerful and popular statistical tools in analyzing time-to-event data in the presence of censoring:
\begin{enumerate}[label=(T\arabic*), ref=T\arabic*]
\item
the {\it Kaplan--Meier curve},  a nonparametric estimator of the {\it survival function} of the outcome, which is the analogue of the empirical cumulative distribution function without censoring;

\item
the {\it log-rank test}, a nonparametric two-sample test for equal {\it survival functions} across two treatment groups;

\item
the {\it Cox proportional hazards model}, which models the conditional {\it hazard function} of the outcome given covariates.
\end{enumerate}
These three tools frequently appear in medical journals.

\section{Examples}

\subsection{Survival analysis}

The Combined Pharmacotherapies and Behavioral Interventions study evaluated the efficacy of medication, behavioral therapy, and their combination for the treatment of alcohol dependence \citep{anton2006combined}. Between January 2001 and January 2004, $n=1224$ recently alcohol-abstinent volunteers were randomized to receive medical management with 16 weeks of naltrexone (100mg daily) or placebo, with or without a combined behavioral intervention. It was a two-by-two factorial experiment. The outcome of interest is the time to the first day of heavy drinking and other endpoints. I adopt the data from \citet{lin2016simultaneous}. 

\begin{lstlisting}
> COMBINE = read.table("combine_data.txt", header = TRUE)[, -1]
> head(COMBINE)
  AGE GENDER    T0_PDA NALTREXONE THERAPY   site relapse futime
1  31   male  3.333333          1       0 site_0       0    112
2  41 female 16.666667          1       1 site_0       1      8
3  44   male 73.333333          0       1 site_0       1     20
4  65   male 10.000000          1       0 site_0       0    112
5  39   male  0.000000          0       1 site_0       1      4
6  56   male 13.333333          0       0 site_0       1      1
\end{lstlisting}

\begin{figure}[th]
\centering
\includegraphics[width = \textwidth]{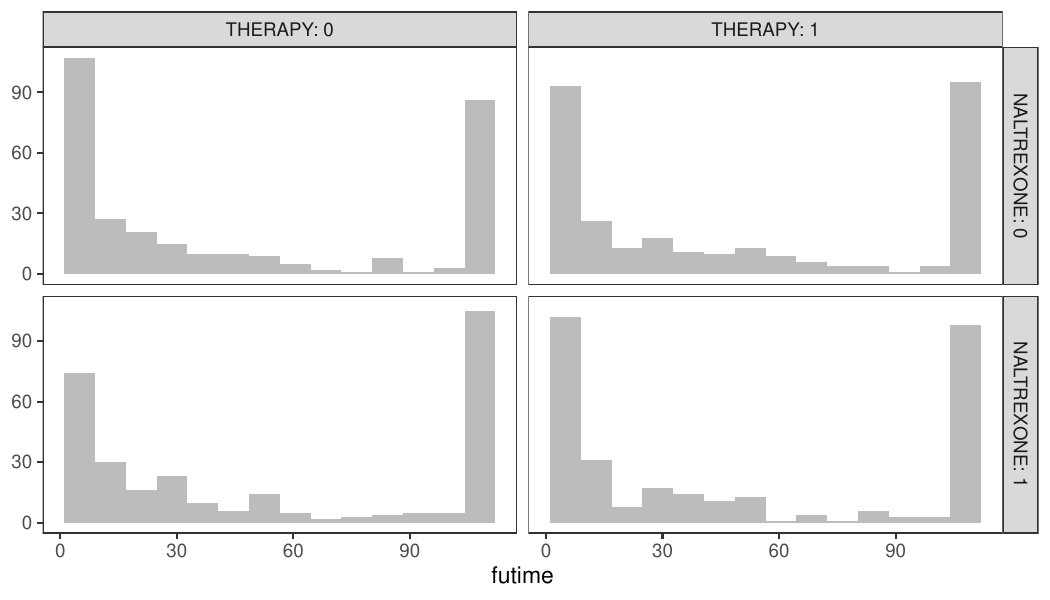}
\caption{Histograms of the time to event in the data from \citet{lin2016simultaneous}}\label{fig::histogram_combine_lindata}
\end{figure}

\ri{NALTREXONE} and \ri{THERAPY} are two treatment indicators. \ri{futime} is the follow-up time, which is censored if \ri{relapse} equals 0. For those censored observations, \ri{futime} equals 112, so it is administrative censoring. Figure \ref{fig::histogram_combine_lindata} shows the histograms of \ri{futime} in four treatment groups. A large number of patients have censored outcomes. 
Other variables are covariates.

\subsection{Duration analysis}

\citet{carpenter2002groups} asked the question: Why does the U.S. Food and Drug Administration approve some drugs more quickly than others? With data about 450 drugs reviewed from 1977 to 2000, he studied the dependence of review times on various covariates, including political influence, wealth of the richest organization representing the disease, media coverage, etc. I use the version of data analyzed by
\citet{keele2010proportionally}. 
The outcome \ri{acttime} is censored indicated by \ri{censor}.
The original paper contains more detailed explanations of the variables.

\begin{lstlisting}
> fda <- read.dta("fda.dta")
> names(fda)
 [1] "acttime"  "censor"   "hcomm"    "hfloor"   "scomm"   
 [6] "sfloor"   "prespart" "demhsmaj" "demsnmaj" "orderent"
[11] "stafcder" "prevgenx" "lethal"   "deathrt1" "hosp01"  
[16] "hospdisc" "hhosleng" "acutediz" "orphdum"  "mandiz01"
[21] "femdiz01" "peddiz01" "natreg"   "natregsq" "wpnoavg3"
[26] "vandavg3" "condavg3" "_st"      "_d"       "_t"      
[31] "_t0"      "caseid"  
\end{lstlisting}

An obvious feature of time-to-event data is that the outcome is non-negative. This can be easily dealt with by the log transformation. However, the outcomes are often censored, resulting in inadequate tail information. With right censoring, modeling the mean involves extrapolation in the right tail.  Therefore,  for censored time-to-event data, we need different statistical modeling tools beyond the conditional mean.

\section{Time-to-event data}

Let $T\geq0$ denote the outcome of interest. We can characterize
a non-negative continuous $T$ using its 
\begin{itemize}
\item
density $f(t)$, 
\item
cumulative distribution function $F(t)$, 
\item
survival function $S(t)=1-F(t)=\pr(T>t)$,   
\item
hazard function
\[
\lambda(t)=\lim_{\Delta t\downarrow0}\pr(t\leq T<t+\Delta t\mid T\geq t)/\Delta t.
\]
\end{itemize}
The hazard function plays a central role in survival analysis.  Within a small time interval $[t,t+\Delta t]$, we have the approximation
\[
\pr(t\leq T<t+\Delta t\mid T\geq t)\cong\lambda(t)\Delta t,
\]
so the hazard function denotes the death rate within a small interval conditioning on surviving up to time $t$.
Both the survival and hazard functions are commonly used to describe a positive random
variable. 
First, the survival function has a simple relationship with the expectation.

\begin{proposition}\label{prop::survival-mean}
For a non-negative random variable $T$, we have 
$$
E(T) = \int_{0}^{\infty} S(t) \d t .
$$
\end{proposition}

Proposition \ref{prop::survival-mean} holds for both continuous and discrete non-negative random variables.
It states that the expectation of a nonnegative random variable equals the area under the survival function. It does not require the existence of the density function of $T$. I leave its proof as Problem \ref{hw23::survival-expectation}.

Second, the survival and hazard functions can determine each other in the following way.

\begin{proposition}\label{prop::hazard-pdf-survival}
For a non-negative continuous random variable $T$,  the survival function determines the hazard function by
$$
\lambda(t)  =\frac{f(t)}{S(t)} = -\frac{\d}{\d t}\log S(t),
$$
and vice versa, the hazard function determines the survival function by
$$
S(t)  =\exp\left\{ -\int_{0}^{t}\lambda(s)\d s\right\} .
$$
\end{proposition}

\begin{myproof}{Proposition}{\ref{prop::hazard-pdf-survival}}
First, I show that the survival function determines the hazard function. 
By definition,
\begin{eqnarray*}
\lambda(t) &=&\lim_{\Delta t\downarrow0}\frac{\pr(t\leq T<t+\Delta t)}{\Delta t}\frac{1}{\pr(T\geq t)}  \\
&=&\lim_{\Delta t\downarrow0}\frac{F(t+\Delta t)-F(t)}{\Delta t}\frac{1}{S(t)}\\
&=&\frac{f(t)}{S(t)}.
\end{eqnarray*}
We can further write the above equation as
\begin{eqnarray*}
\lambda(t)  &=& \frac{f(t)}{S(t)} \\
&=& -\frac{\d S(t)/\d t}{S(t)} \\
&=&  -\frac{\d}{\d t}\log S(t).
\end{eqnarray*}

Second, I show that the hazard function determines the survival function.  From the first part, we have 
\[
\d\log S(t)=-\lambda(t)\d t
\]
which, by the Newton--Leibniz formula, implies
\[
\log S(t) - \log S(0)=-\int_{0}^{t}\lambda(s)\d s . 
\]
Because $\log S(0)=0$, we have $\log S(t) = -\int_{0}^{t}\lambda(s)\d s$, giving the final result. 
\end{myproof}

\section{Some examples of random variables for time to event}

I first review four canonical models for time to event. 

\begin{example}
[Exponential]
The $\textup{Exponential}(\lambda)$ random variable $T$ has
\begin{itemize}
\item
density $f(t) = \lambda e^{-\lambda t}$, 
\item
survival function $S(t) = e^{-\lambda t}$,  
\item
hazard function
$
\lambda(t)= \lambda,
$
which is constant. 
\end{itemize}
An important feature of the Exponential random variable is its memoryless property as shown in Proposition \ref{prop::memoryless}. 
\end{example}

\begin{figure}
\centering
\includegraphics[width = 0.49\textwidth]{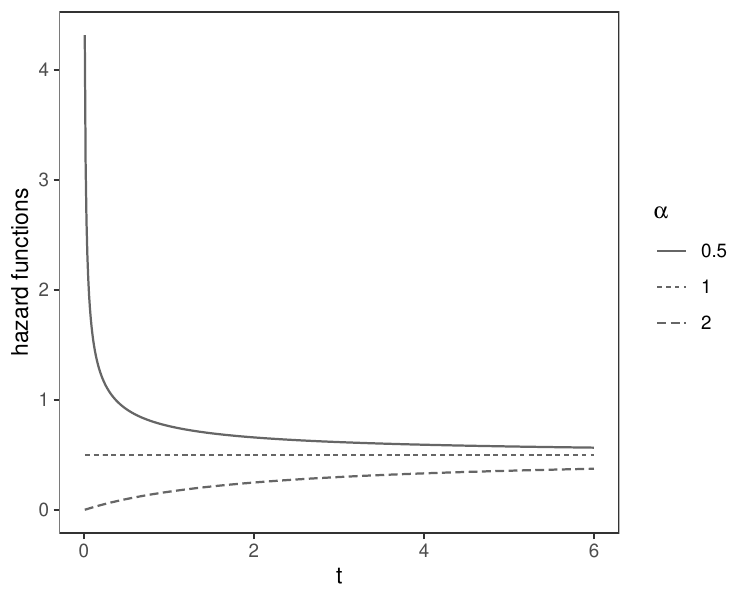}
\includegraphics[width = 0.49\textwidth]{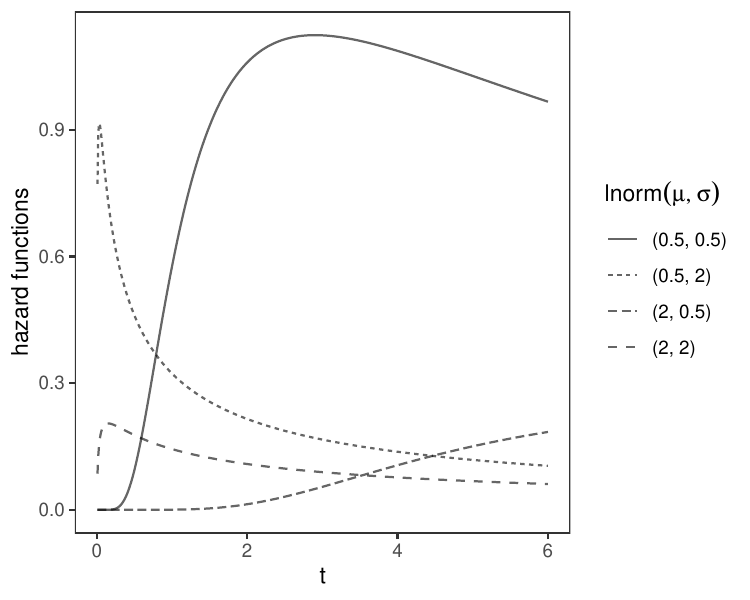}
\caption{Left: Gamma$(\alpha, \beta = 2)$ hazard functions; Right: Log-Normal$(\mu,\sigma^2)$ hazard functions}\label{fig::gamma-lnorm-hazard-functions}
\end{figure}

\begin{example}
[Gamma]
The $\textup{Gamma}(\alpha, \beta)$ random variable $T$ has density 
$$
f(t) =   \beta^\alpha t^{\alpha-1} e^{-\beta t}/ \Gamma(\alpha). 
$$
When $\alpha = 1$, it reduces to Exponential$(\beta)$ with a constant hazard function. In general, the survival function and hazard function do not have simple forms, but we can use \ri{dgamma} and \ri{pgamma} to compute them numerically. The left panel of Figure \ref{fig::gamma-lnorm-hazard-functions} plots the hazard functions of Gamma$(\alpha, \beta)$. When $\alpha <1$, the hazard function is decreasing; when $\alpha >1$, the hazard function is increasing. 
\end{example}

\begin{example}
[Log-Normal]
The Log-Normal random variable $T \sim \textup{Log-Normal}(\mu,\sigma^2)$ equals exponential of $\N(\mu,\sigma^2)$. The right panel of Figure \ref{fig::gamma-lnorm-hazard-functions} plots the hazard functions with four different parameter combinations. 
\end{example}

\begin{example}
[Weibull]\label{eg::weibull}
The Weibull distribution has many different parametrizations. Here I follow the \ri{R} function \ri{dweibull}, which has a \ri{shape} parameter $a>0$ and \ri{scale} parameter $b>0$. The $\textup{Weibull}(a,b)$ random variable $T$ can be generated by
\begin{eqnarray}
\label{eq::weibull-representation}
T = b Z^{1/a}
\end{eqnarray}
which is equivalent to
\[
\log T = \log b + a^{-1} \log Z,
\]
where $Z\sim  \textup{Exponential}(1)$. We can verify that  $T$ has 
\begin{itemize}
\item 
density function
$$
f(t)=  \frac{a}{b} \left(  \frac{t}{b}   \right)^{a-1} \exp\left\{  - \left(  \frac{t}{b}   \right)^a \right\},
$$
\item
survival function
$$
S(t)  = \exp\left\{  - \left(  \frac{t}{b}   \right)^a \right\},
$$
\item 
hazard function
$$
\lambda(t) = \frac{a}{b} \left(  \frac{t}{b}   \right)^{a-1}.
$$
\end{itemize}
So when $a=1$, the Weibull distribution reduces to the Exponential distribution with constant hazard function. When $a>1$, the hazard function increases; when $a<1$, the hazard function decreases. 
\end{example}

\begin{figure}
\centering
\includegraphics[width=0.9\textwidth]{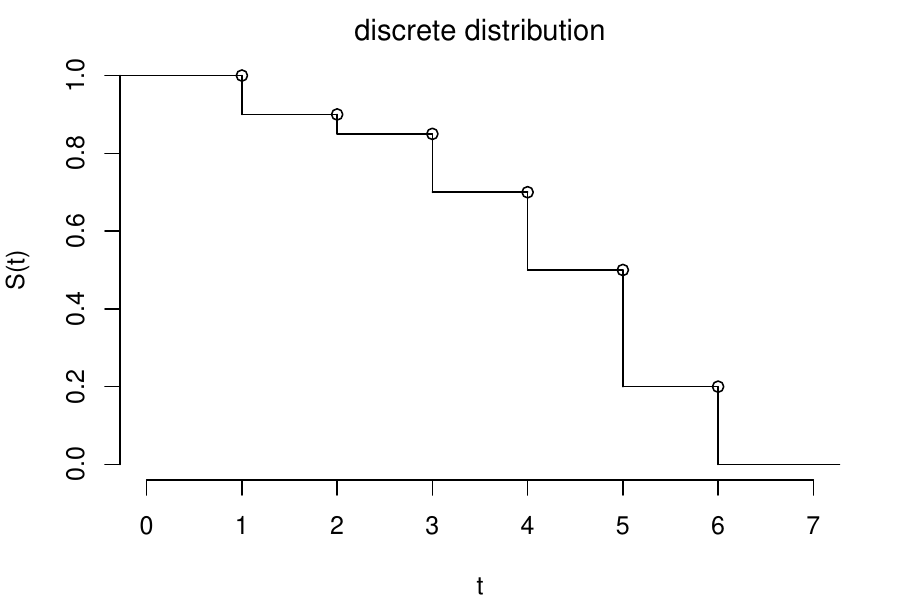}
\caption{Discrete survival function with masses $(0.1, 0.05, 0.15, 0.2, 0.3, 0.2)$ at $(1,2,3,4,5,6)$}
\label{fig::discrete-survival-fn}
\end{figure}

\begin{example}
[discrete  random variable for time to event]
We can characterize a positive discrete  random variable $T\in\left\{ t_{1},t_{2},\ldots\right\} $
by its 
\begin{itemize}
\item 
probability mass function $f(t_{k})=\pr(T=t_{k})$, 
\item
distribution
function $F(t)=\sum_{k:t_{k}\leq t}f(t_{k})$, 
\item
survival function $S(t)=\sum_{k:t_{k}>t}f(t_{k})$,
\item 
discrete hazard function
\[
\lambda_{k}=\pr(T=t_{k}\mid T\geq t_{k})=\frac{f(t_{k})}{S(t_{k}-)},
\]
where $S(t_{k}-)$ denotes the left limit of the function $S(t)$
at $t_{k}$. 
\end{itemize}
Figure \ref{fig::discrete-survival-fn} shows an example of a survival function for a discrete random variable, which shows that $S(t)$ is a step function and right-continuous with left limits. 
\end{example}

The discrete hazard and survival functions have the following
connection, which will be useful for Chapter \ref{chapter::km-curve} below.

\begin{proposition}
\label{proposition:discrete-hazard}For a positive discrete random variable
$T$, its survival function is a step function determined by
\[
S(t)=\pr(T>t)=\prod_{k:t_{k}\leq t}(1-\lambda_{k}).
\]
\end{proposition}

Note that $S(t)$ is a step function decreasing at each $t_k$,  because $\lambda_k$ is probability and thus bounded between zero and one.

\begin{myproof}{Proposition}{\ref{proposition:discrete-hazard}}
By definition, 
\begin{eqnarray*}
1-\lambda_{k} &=& 1 - \pr(T=t_{k}\mid T\geq t_{k}) \\
&=&\pr(T>t_{k}\mid T\geq t_{k})
\end{eqnarray*}
is the probability of surviving longer than $t_{k}$ conditional on
surviving at least as long as $t_{k}$. We can verify Proposition
\ref{proposition:discrete-hazard} within each interval of the $t_{k}$'s.
For example, if $t<t_{1}$, then 
$$
S(t)=\pr(T>t)=1;
$$ 
if $t_{1}\leq t<t_{2}$,
then 
\begin{eqnarray*}
S(t) &=&\pr(T>t_{1}) \\
&=& \pr(T>t_{1}, T\geq t_{1}) \\
&=&\pr(T>t_{1}\mid T\geq t_{1}) \pr(T\geq t_{1}) \\
&=&1-\lambda_{1};
\end{eqnarray*}
if $t_{2}\leq t<t_{3}$, then 
\begin{eqnarray*}
S(t) &=&\pr(T>t_{2}) \\
&=&\pr(T>t_{2}, T \geq t_{2}) \\
&=&\pr(T>t_{2}\mid T\geq t_{2})\pr(T\geq t_2)\\
&=&(1-\lambda_{2})(1-\lambda_{1}).
\end{eqnarray*}
We can also verify other values of $S(t)$ by induction. 
\end{myproof}

\section{Kaplan--Meier survival curve}\label{chapter::km-curve}

Without censoring, estimating the cumulative distribution function or the survival function is rather straightforward. With IID data $(T_1, \ldots, T_n)$, we can estimate the cumulative distribution function by $\hat{F}(t) =  n^{-1} \sumn 1(T_i \leq t) $ and the survival function by $\hat{S}(t) = 1 - \hat{F}(t)$.

\begin{figure}
\centering
\includegraphics[width = \textwidth]{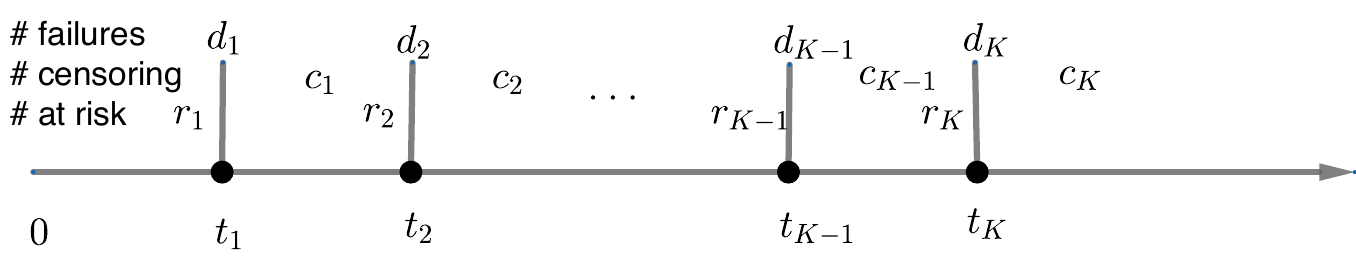}
\caption{Data structure for the Kaplan--Meier curve}\label{fig::kmcurve-data}
\end{figure}

Figure \ref{fig::kmcurve-data} shows the common data structure with censoring in survival analysis:
\begin{enumerate}[label=(S\arabic*), ref=S\arabic*]
\item
$t_1,\ldots, t_K$ are the death times, and $d_1,\ldots, d_K$ are the corresponding number of deaths;

\item
$r_1, \ldots, r_K$ are the number of patients at risk, that is, $r_1$ patients are not dead or censored right before time $t_1$, and so on;

\item
$c_1,\ldots, c_K$ are the number of censored patients within interval $[t_1, t_2 ) ,\ldots, [d_K, \infty)$.
\end{enumerate}

\citet{kaplan1958nonparametric} proposed the following simple estimator for the survival function.

\begin{definition}
[Kaplan--Meier curve]\label{def::km-curve}
First estimate the discrete hazard function at the failure times $\left\{ t_{1},\ldots,t_{K}\right\} $
as $\hat{\lambda}_{k}=d_{k}/r_{k}\ (k=1,\ldots,K)$ and then estimate the survival
function as
\[
\hat{S}(t)=\prod_{k:t_{k}\leq t}(1-\hat{\lambda}_{k}).
\]
\end{definition}

The $\hat{S}(t)$ in Definition \ref{def::km-curve}  is also called the product-limit estimator
of the survival function due to its mathematical form.

To quantify the uncertainty in the Kaplan--Meier curve,  we can use the famous Greenwood's formula \citep{greenwood1926report}. 
\begin{definition}
[Greenwood's formula]\label{def::greenwood-formula}
We can approximate the variance of $\log\hat{S}(t) $ by 
$$
\sum_{k:t_{k}\leq t}\frac{d_{k}}{r_{k}(r_{k}-d_{k})}.
$$
\end{definition}

Below I will give some intuition for Greenwood's formula. 
At each failure time $t_{k}$, we view $d_{k}$ as the result of $r_{k}$
Bernoulli trials with probability $\lambda_{k}$. So $\hat{\lambda}_{k}=d_{k}/r_{k}$
has variance $\lambda_{k}(1-\lambda_{k})/r_{k}$ which can be estimated
by 
\[
\hat{\var}(\hat{\lambda}_{k})=\hat{\lambda}_{k}(1-\hat{\lambda}_{k})/r_{k}.
\]
We can estimate the variance of the survival function using the delta method (see Proposition \ref{prop::delta-method} in Appendix \ref{chapter::limiting-theorems}). We can approximate the variance of
\begin{eqnarray*}
\log\hat{S}(t) 
&=& \sum_{k:t_{k}\leq t}\log(1-\hat{\lambda}_{k}) \\ 
&\cong & \sum_{k:t_{k}\leq t}\log(1-\lambda_{k})-\sum_{k:t_{k}\leq t}(1-\lambda_{k})^{-1}(\hat{\lambda}_{k}-\lambda_{k})
\end{eqnarray*}
by
\begin{eqnarray*}
\hat{\var}\left\{ \log\hat{S}(t)\right\}  &=&\sum_{k:t_{k}\leq t}(1-\lambda_{k})^{-2}\hat{\var}(\hat{\lambda}_{k}) \\
&=&\sum_{k:t_{k}\leq t}(1-\hat{\lambda}_{k})^{-2}\hat{\lambda}_{k}(1-\hat{\lambda}_{k})/r_{k}\\
&=&\sum_{k:t_{k}\leq t}\frac{d_{k}}{r_{k}(r_{k}-d_{k})} .
\end{eqnarray*}
A hidden assumption above is the independence of the $\hat{\lambda}_k$'s. This assumption cannot be justified due to the dependence of the events. However, a deeper theory of counting processes shows that  Greenwood's formula is valid even without the independence \citep{fleming2011counting}.

Based on Greenwood's formula, we can construct a confidence interval for $\log S(t)$:
\[
\log\hat{S}(t)\pm z_{\alpha}\sqrt{\hat{\var}\left\{ \log\hat{S}(t)\right\} },
\]
which implies a confidence interval for $S(t)$. However, this interval
can be outside of range $[0,1]$, because the log transformation $\log S(t)$ is in the range of $(-\infty, 0)$ but the Normal approximation is in the range $(-\infty, \infty).$ 
A better transformation is log-log:
\[
v(t)=\log\left\{ -\log S(t)\right\} ,\quad\hat{v}(t)=\log\left\{ -\log\hat{S}(t)\right\} .
\]
Using Taylor expansion, we can approximate the variance of 
\[
\hat{v}(t)\cong\log\left\{ -\log S(t)\right\} -\frac{1}{\log S(t)}\left\{  \log \hat{S}(t)- \log S(t)\right\} 
\]
by 
\[
\frac{\hat{\var}\left\{ \log\hat{S}(t)\right\} }{\left\{ \log\hat{S}(t)\right\} ^{2}}.
\]
Based on this formula and Greenwood's formula above, we can construct
a confidence interval for $v(t)$:
\[
\log\left\{ -\log\hat{S}(t)\right\} \pm z_{\alpha}\sqrt{\hat{\var}\left\{ \log\hat{S}(t)\right\} }/\log\hat{S}(t),
\]
which implies another confidence interval for $S(t)$. 
In the \ri{R} package \ri{survival}, the function \ri{survfit} can fit the Kaplan--Meier curve, where the specifications \ri{conf.type = "log"} and \ri{conf.type = "log-log"} return confidence intervals based on the log and log-log transformations, respectively.

\begin{figure}[th]
\centering
\includegraphics[width = \textwidth]{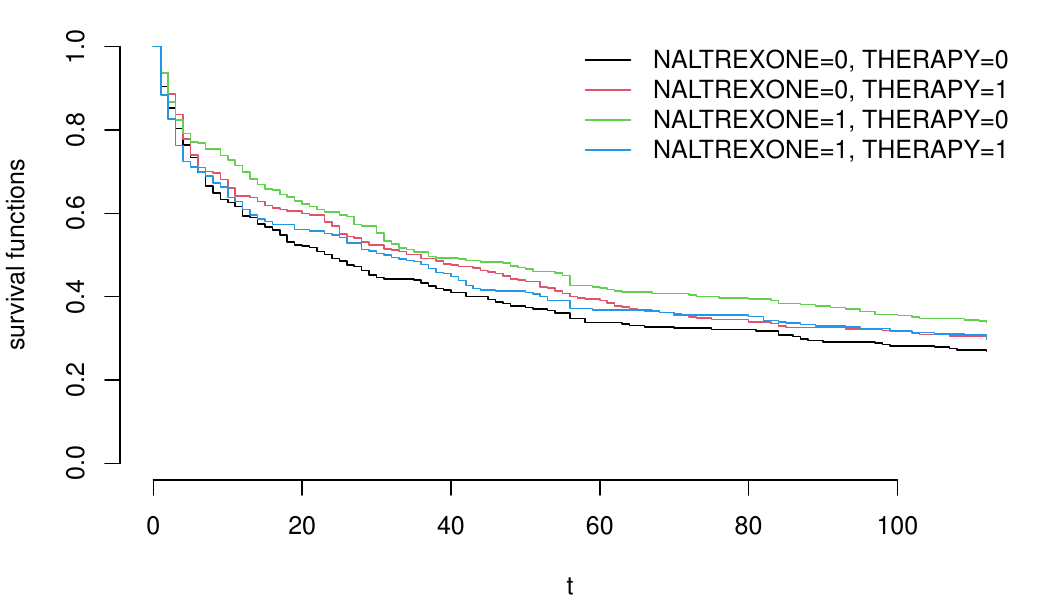}
\caption{\citet{lin2016simultaneous}'s data}\label{fig::kmcurve2x2-lin}
\end{figure}

Figure \ref{fig::kmcurve2x2-lin} plots four curves based on the combination of \ri{NALTREXONE} and  \ri{THERAPY} using the data of \citet{lin2016simultaneous}. I do not show the confidence intervals due to the large overlap. 

\begin{lstlisting}
> km4groups = survfit(Surv(futime, relapse) ~ NALTREXONE + THERAPY,
+                     data = COMBINE)
> plot(km4groups, bty = "n", col = 1:4,
+      xlab = "t", ylab = "survival functions")
> legend("topright",
+        c("NALTREXONE=0, THERAPY=0", 
+          "NALTREXONE=0, THERAPY=1", 
+          "NALTREXONE=1, THERAPY=0", 
+          "NALTREXONE=1, THERAPY=1"),
+        col = 1:4, lty = 1, bty = "n")
\end{lstlisting}

The above discussion on the Kaplan--Meier curve is rather heuristic. More fundamentally, what is the underlying censoring mechanism that ensures the possibility that the distribution of the survival time can be recovered by the observed data? It turns out that we have implicitly assumed that the survival time and the censoring time are independent. Problem \ref{hw23::identification-km} gives a theoretical statement.

\section{Cox model for time-to-event outcome}

Another important problem is to model the dependence of the survival time $T$ on covariates $x$. The major challenge is that the survival time is often censored. Let $C_i$ be the censoring time of unit $i$, and we can only observe the minimum value of the survival time and the censoring time. So the observed data are
$
(x_{i}, y_{i},\delta_{i})_{i=1}^{n}, 
$
where 
\[
y_{i}=\min(T_{i},C_{i}),\quad\delta_{i}=1(T_{i}\leq C_{i})
\]
are the event time and the censoring indicator, respectively. A key assumption is that the censoring mechanism is noninformative:
\begin{assumption}
[noninformative censoring]\label{assume::non-informative-censoring}
$
T_{i}\ind C_{i}\mid x_{i}.
$
\end{assumption}
Assumption \ref{assume::non-informative-censoring} is crucial for survival analysis. If it does not hold, we cannot obtain consistent estimators for the parameters in the conditional distribution $T_i \mid x_i$. See Problem \ref{hw23::identification-km} for the discussion of a fundamental problem of {\it identifiability}.

We can start with parametric models. 

\begin{example}\label{ex::log-normal-regression}
Assume $T_i \mid x_i \sim \textup{Log-Normal}(x_i^{\T}\beta, \sigma^2)$. Equivalently,
$$
\log T_i = x_i^{\T}\beta + \varepsilon_i , 
$$
where the $\varepsilon_i$'s are IID $\N(0,\sigma^2)$ independent of the $x_i$'s. This is a Normal linear model on $\log T_i$. 
\end{example}

 \begin{example}\label{ex::weibull-regression}
 Assume that $T_i \mid x_i \sim \textup{Weibull}(a, b= e^{x_i^{\T}\beta})$. Based on the definition of the Weibull distribution in \eqref{eq::weibull-representation}, we have
 $$
 \log T_i = x_i^{\T} \beta+ \varepsilon_i
 $$
 where the $\varepsilon_i$'s are IID $a^{-1} \log \textup{Exponential}(1)$, independent of the $x_i$'s. 
 \end{example}

The \ri{R} package \ri{survival} contains the function \ri{survreg} to fit parametric survival models including the choices of \ri{dist = "lognormal"}, \ri{dist = "weibull"}, etc. However, these parametric models are not commonly used unless the sample sizes are small and the models can be justified by the domain knowledge. The parametric forms can be too strong, and due to right censoring, the inference can be driven by extrapolation to the right tail.

\subsection{Cox model and its interpretation}

\citet{cox1972regression} proposed to model the conditional hazard function
\begin{eqnarray*}
\lambda(t\mid x) 
&=& \lim_{\Delta t\downarrow0}\pr(t\leq T<t+\Delta t\mid T\geq t,x)/\Delta t \\
&=&\frac{f(t\mid x)}{S(t\mid x)} .
\end{eqnarray*}
His celebrated proportional hazards model has the following form.

\begin{assumption}
[Cox proportional hazards model]\label{assume::cox-ph-model}
Assume the conditional hazard ratio function has the form
\begin{eqnarray}\label{eq::proportional-hazards-model}
\lambda(t\mid x)=\lambda_{0}(t)\exp(x^{\T}\beta) .
\end{eqnarray}
The observations are independent across units. The $\beta$ is an unknown parameter and $\lambda_{0}(\cdot)$ is an unknown function. 
\end{assumption}

Assumption \ref{assume::cox-ph-model} is equivalent to
$$
\log \lambda(t\mid x) = \log \lambda_{0}(t) + x^{\T}\beta .
$$
Unlike other regression models, $x$ does not contain the intercept in \eqref{eq::proportional-hazards-model}. If the first component of $x$ is $1$, then we can write 
\begin{eqnarray*}
\lambda(t\mid x) 
&=&\lambda_{0}(t)  \exp(x_1\beta_1 + \cdots + x_p \beta_p) \\
&=& \lambda_{0}(t) e^{\beta_1}  \exp(x_2\beta_2 + \cdots + x_p \beta_p)
\end{eqnarray*}
and redefine $\lambda_{0}(t) e^{\beta_1}$ as another unknown function. With an intercept, we cannot identify $\lambda_{0}(t) $ and $\beta_1$ separately.  So we drop the intercept to ensure identifiability. 

From the log-linear form of the conditional hazard function, we have
$$
\log \frac{\lambda(t\mid x')}{ \lambda(t\mid x)} = (x' - x)^{\T} \beta .
$$
Therefore, each coordinate of $\beta$ measures the log conditional hazard ratio holding other covariates constant. 
Because of this, \eqref{eq::proportional-hazards-model} is called the proportional hazards model. 
A positive $\beta_j$ suggests a ``positive'' effect on the hazard function and thus a ``negative'' effect on the survival time itself. 
Consider a special case with a binary $x_i$ to gain insights into the Cox model.

\begin{figure}
\centering
\includegraphics[width = \textwidth]{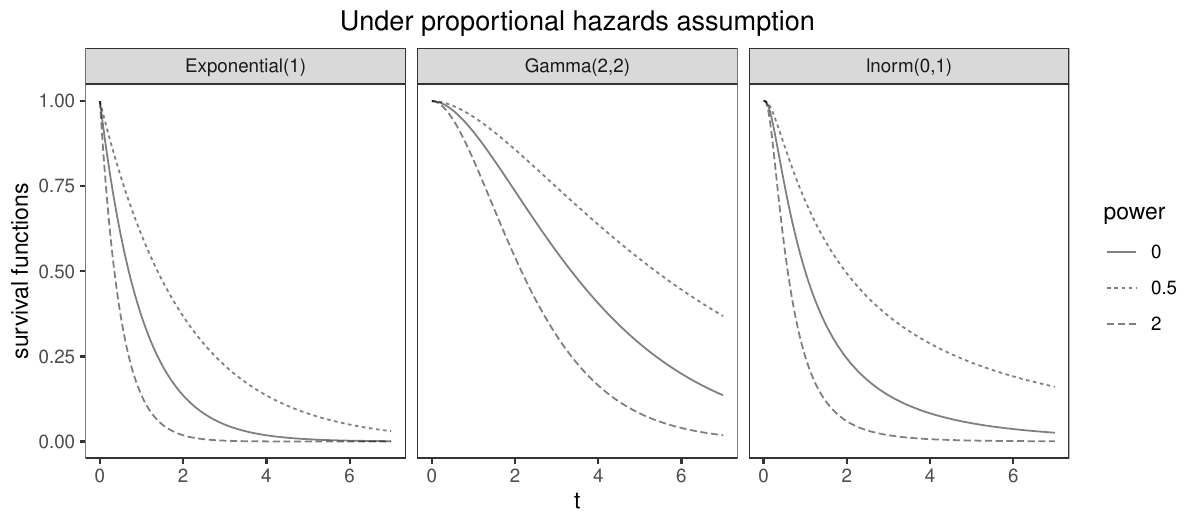}
\caption{Proportional hazards assumption with different baseline survival functions, where the power equals $\gamma = \exp(\beta)$.
}\label{fig::ph-assumptions}
\end{figure}

\begin{example}
[Cox proportional hazards model with a binary regressor]\label{eg::cox-binary-x}
With a binary $x$, the Cox proportional hazards model $\lambda(t\mid x) = \lambda_0(t)e^{x\beta}$ implies that $\lambda(t\mid 1) = \gamma  \lambda(t\mid 0)$ with $\gamma=\exp(\beta)$. Therefore, the survival functions satisfy
\begin{eqnarray*}
S(t\mid 1) 
&=& \exp\left\{  - \int_0^t  \lambda(u\mid 1) \d u  \right\} \\
&=&  \exp\left\{  - \gamma  \int_0^t  \lambda(u\mid 0) \d u  \right\}  \\
&=& \left\{  S(t\mid 0) \right\}^\gamma,
\end{eqnarray*}
which is a power transformation. 
Qualitatively, we have the following two cases:
\begin{enumerate}[label=(\textup{PH}\arabic*), ref=\textup{PH}\arabic*]
\item
$\beta <0$: so the hazard ratio $\gamma = \exp(\beta) < 1$ and $S(t\mid 1) \geq  S(t\mid 0)$ for all $t$, which implies a longer survival time under treatment;
\item 
$\beta >0$: so the hazard ratio $\gamma = \exp(\beta) > 1$ and $S(t\mid 1) \leq  S(t\mid 0)$ for all $t$, which implies a shorter survival time under treatment.
\end{enumerate}
Figure \ref{fig::ph-assumptions} shows some survival functions satisfying the proportional hazards assumption, none of which cross each other within the interval $t\in (0,\infty).$ When the two survival functions cross, the proportional hazards assumption does not hold. 
\end{example}

Theoretically, we can allow the covariates to be time-dependent, that is, $x_i(t)$ can depend on $t$ and thus is a stochastic process. However, the interpretation of the coefficient becomes challenging \citep{fisher1999time}. This chapter focuses on the simple case with time-invariant covariates.

\subsection{Partial likelihood}

The likelihood function is rather complicated, which depends on an unknown function $\lambda_0(t)$. 
Assuming no ties among the failure time $t_k$'s,\footnote{
With ties, there are subtle issues in defining the partial likelihood. One simple strategy is to add some small random noise to the failure times. Most statistical software packages implement various sophisticated methods to deal with ties.
} \citet{cox1972regression} proposed to use the partial likelihood function
to estimate $\beta$:
\[
L(b)=\prod_{k=1}^{K}\frac{\exp(x_{k}^{\T} b)}{\sum_{l\in R(t_{k})}\exp(x_{l}^{\T} b)},
\]
where the product is over $K$ time points with failures, $x_k$ is the covariate value of the failure at time $t_k$, and $R(t_k)$ contains the indices of the units at risk at time $t_k$, i.e., the units not censored or failed right before the time $t_k$.

\citet{freedman2008survival} gives a heuristic explanation of the partial likelihood based on the following results which extends Proposition \ref{prop::minimum-expo} in Appendix \ref{chapter:appendix-rvs} on the Exponential distribution. 

\begin{theorem}
\label{thm::freedman-explain-cox}
If $T_1,\ldots, T_n$ are independent with hazard function $\lambda_i(t)$ $(i=1,\ldots, n)$, then their minimum value $\underline{T} = \min_{1\leq i \leq n} T_i$ has hazard function $\sumn \lambda_i(t)$.
Moreover, if $\lambda_i(t) = c_i \lambda(t)$, then 
$$
\pr (  T_i = \underline{T}  ) = \frac{c_i}{ \sum_{i'=1}^n  c_{i'} }.
$$
\end{theorem} 

\begin{myproof}{Theorem}{\ref{thm::freedman-explain-cox}}
The survival function of $\underline{T} $ is
\begin{eqnarray*}
\pr (  \underline{T} > t  ) 
&=& \pr(  T_1>t,\ldots, T_n>t ) \\
&=& \prod_{i=1}^n S_i(t) ,
\end{eqnarray*}
so Proposition \ref{prop::hazard-pdf-survival} implies that its hazard function is
\begin{eqnarray*}
- \frac{\d}{\d t} \log \pr ( \underline{T} > t )  
&=& \sumn \left\{ - \frac{\d}{\d t} \log   S_i(t) \right\} \\
&=& \sumn \lambda_i(t). 
\end{eqnarray*}
So the first conclusion follows.

As a byproduct of the above proof, the density of $\underline{T} $ is $\sumn \lambda_i(t) \prod_{i=1}^n S_i(t)$ based in Proposition \ref{prop::hazard-pdf-survival}. It must have integral one; with $\lambda_i(t) = c_i \lambda(t)$, this implies 
\begin{equation}
\left( \sumn c_i  \right) \int_0^{\infty} \lambda(t) \prod_{i=1}^n S_i(t) \d t = 1.
\label{eq::integral1-minimum-Ts}
\end{equation}
Therefore, we have 
\begin{eqnarray*}
\pr ( T_i = \underline{T}   ) 
&=& \pr\{ T_i \leq T_{i'} \text{ for all } i' \neq i  \} \\
&=& \int_0^{\infty}   \prod_{i' \neq i} S_{i'}(t)  f_i(t) \d t\\
&=&  \int_0^{\infty}   \prod_{i'=1}^n S_{i'}(t)  \lambda_i(t) \d t \\
&=& c_i  \int_0^{\infty}  \lambda(t)   \prod_{i'=1}^n S_{i'}(t)  \d t \\
&=& c_i  \Big /  \sumn c_{i'},
\end{eqnarray*}
where the last equality holds due to \eqref{eq::integral1-minimum-Ts}. 
\end{myproof}

Theorem \ref{thm::freedman-explain-cox} explains each of the $K$ components in the partial likelihood. At time $t_k$, the units in $R(t_k)$ are all at risk, and unit $k$ fails, assuming no ties. The probability that unit $k$ has the smallest failure time among units in $R(t_k)$ is 
$$
 \frac{   \exp(x_k^{\T} \beta)  }{ \sum_{l\in R(t_k)}    \exp(x_l^{\T} \beta) } 
$$
according to Theorem \ref{thm::freedman-explain-cox}. The product in the partial likelihood is based on the independence of the events at the $K$ failure times, which is more difficult to justify. A rigorous justification relies on the deeper theory of counting processes \citep{fleming2011counting} or semiparametric statistics \citep{tsiatis2007semiparametric}.

The log-likelihood function is
\[
\log L( b )=\sum_{k=1}^{K}\left\{ x_{k}^{\T} b -\log\sum_{l\in R(t_{k})}\exp(x_{l}^{\T} b)\right\} ,
\]
and the score function is
\[
\frac{\partial\log L(b)}{\partial b}=\sum_{k=1}^{K}\left\{ x_{k}-\frac{\sum_{l\in R(t_{k})}\exp(x_{l}^{\T} b)x_{l}}{\sum_{l\in R(t_{k})}\exp(x_{l}^{\T} b)}\right\} .
\]
Define 
\[
\pi_{b}(l\mid R_{k})=  \frac{  \exp(x_{l}^{\T} b) }{  \sum_{l\in R(t_{k})}\exp(x_{l}^{\T} b) } ,\quad(l\in R(t_{k}))
\]
which sum to one, so they induce a probability measure leading to
expectation $E_{b}(\cdot\mid R_{k})$ and covariance $\cov_{b}(\cdot\mid R_{k})$.
With this notation, the score function simplifies to
\[
\frac{\partial\log L(b)}{\partial b}=\sum_{k=1}^{K}\left\{ x_{k}-E_{b}(x\mid R_{k})\right\} ,
\]
where 
$
E_{b}(x\mid R_{k}) = \sum_{l\in R(t_{k})}  \pi_{l}(b\mid R_{k}) x_l;
$
the Hessian matrix simplifies to
$$
\frac{\partial^{2}\log L(b)}{\partial b\partial b^{\T}}  =-\sum_{k=1}^{K}\cov_{b}(x\mid R_{k})\preceq 0,
$$
where
\begin{eqnarray*}
&&\cov_{b}(x\mid R_{k}) \\
&=& \frac{ 
\sum_{l\in R(t_{k})}\exp(x_{l}^{\T} b)x_{l}x_{l}^{\T}\sum_{l\in R(t_{k})}\exp(x_{l}^{\T} b) 
-\sum_{l\in R(t_{k})}\exp(x_{l}^{\T} b)x_{l}\sum_{l\in R(t_{k})}\exp(x_{l}^{\T} b)x_{l}^{\T} 
}
{  \left\{ \sum_{l\in R(t_{k})}\exp(x_{l}^{\T}b)\right\} ^{2} } \\
&=&
\sum_{l\in R(t_{k})}\pi_{b}(l\mid R_{k})x_{l}x_{l}^{\T}-\sum_{l\in R(t_{k})}\pi_{b}(l\mid R_{k})x_{l}\sum_{l\in R(t_{k})}\pi_{b}(l\mid R_{k})x_{l}^{\T} . 
\end{eqnarray*}

The \ri{coxph} function in the \ri{R} package \ri{survival} uses Newton's method to compute the maximizer $\hat{\beta}$ of the partial likelihood function, and uses the inverse of the observed Fisher information, $\left(  - \frac{\partial^{2}\log L( \hat\beta )}{\partial b\partial b^{\T}} \right)^{-1}$, to approximate the asymptotic covariance matrix of $\hat{\beta}$.

\citet{lin1989robust} proposed a sandwich covariance estimator to allow for the misspecification of the Cox model.  
It is similar to the form of the sandwich covariance estimator discussed in Chapter \ref{chapter::sandwich} and Appendix \ref{chapter::m-mle}. I omit the details. 
The \ri{coxph} function with \ri{robust = TRUE} reports the corresponding robust standard errors.

\subsection{Examples}
Using \citet{lin2016simultaneous}'s data, we have the following results. 

\begin{lstlisting}
> cox.fit <- coxph(Surv(futime, relapse) ~ NALTREXONE*THERAPY + 
+                    AGE + GENDER + T0_PDA + site,
+                  data=COMBINE)
> summary(cox.fit)
Call:
coxph(formula = Surv(futime, relapse) ~ NALTREXONE * THERAPY + 
    AGE + GENDER + T0_PDA + site, data = COMBINE)

  n= 1226, number of events= 856 

                        coef exp(coef)  se(coef)      z Pr(>|z|)    
NALTREXONE         -0.249719  0.779020  0.097690 -2.556  0.01058 *  
THERAPY            -0.167050  0.846158  0.096102 -1.738  0.08217 .  
AGE                -0.015540  0.984580  0.003559 -4.366 1.27e-05 ***
GENDERmale         -0.140621  0.868818  0.075368 -1.866  0.06207 .  
T0_PDA              0.002550  1.002553  0.001368  1.863  0.06242 .  
sitesite_1         -0.091853  0.912239  0.167261 -0.549  0.58290    
sitesite_10        -0.227185  0.796774  0.175427 -1.295  0.19531    
sitesite_2          0.121236  1.128892  0.160052  0.757  0.44876    
sitesite_3         -0.084483  0.918987  0.161121 -0.524  0.60004    
sitesite_4         -0.471612  0.623996  0.175203 -2.692  0.00711 ** 
sitesite_5         -0.128286  0.879602  0.161782 -0.793  0.42780    
sitesite_6         -0.240563  0.786185  0.161958 -1.485  0.13745    
sitesite_7          0.372004  1.450639  0.157616  2.360  0.01827 *  
sitesite_8          0.067700  1.070045  0.160876  0.421  0.67388    
sitesite_9          0.267373  1.306528  0.154911  1.726  0.08435 .  
NALTREXONE:THERAPY  0.337539  1.401495  0.137441  2.456  0.01405 * 
\end{lstlisting}

\ri{NALTREXONE} has a significant negative log hazard ratio, but \ri{THERAPY} has a nonsignificant negative log hazard ratio. More interestingly, their interaction \ri{NALTREXONE:THERAPY} has a significant positive log hazard ratio. This suggests that combining \ri{NALTREXONE} and \ri{THERAPY} is worse than using \ri{NALTREXONE} alone to delay the first time of heavy drinking and other endpoints. This is also coherent with the survival curves in Figure \ref{fig::kmcurve2x2-lin}, in which the best Kaplan--Meier curve corresponds to \ri{NALTREXONE=1, THERAPY=0}.

We can also obtain the robust standard errors. In this example, the robust standard errors do not differ much from the original standard errors. 
\begin{lstlisting}
> cox.fit <- coxph(Surv(futime, relapse) ~ NALTREXONE*THERAPY + 
+                    AGE + GENDER + T0_PDA + site,
+                  robust = TRUE, 
+                  data=COMBINE)
> summary(cox.fit)
Call:
coxph(formula = Surv(futime, relapse) ~ NALTREXONE * THERAPY + 
    AGE + GENDER + T0_PDA + site, data = COMBINE, robust = TRUE)

  n= 1226, number of events= 856 

                        coef exp(coef)  se(coef) robust se      z Pr(>|z|)    
NALTREXONE         -0.249719  0.779020  0.097690  0.097171 -2.570  0.01017 *  
THERAPY            -0.167050  0.846158  0.096102  0.096716 -1.727  0.08413 .  
AGE                -0.015540  0.984580  0.003559  0.003560 -4.365 1.27e-05 ***
GENDERmale         -0.140621  0.868818  0.075368  0.077656 -1.811  0.07017 .  
T0_PDA              0.002550  1.002553  0.001368  0.001358  1.877  0.06052 .  
sitesite_1         -0.091853  0.912239  0.167261  0.172910 -0.531  0.59527    
sitesite_10        -0.227185  0.796774  0.175427  0.176713 -1.286  0.19858    
sitesite_2          0.121236  1.128892  0.160052  0.169120  0.717  0.47346    
sitesite_3         -0.084483  0.918987  0.161121  0.161148 -0.524  0.60010    
sitesite_4         -0.471612  0.623996  0.175203  0.174456 -2.703  0.00686 ** 
sitesite_5         -0.128286  0.879602  0.161782  0.161614 -0.794  0.42732    
sitesite_6         -0.240563  0.786185  0.161958  0.155850 -1.544  0.12270    
sitesite_7          0.372004  1.450639  0.157616  0.165148  2.253  0.02429 *  
sitesite_8          0.067700  1.070045  0.160876  0.171414  0.395  0.69288    
sitesite_9          0.267373  1.306528  0.154911  0.155482  1.720  0.08550 .  
NALTREXONE:THERAPY  0.337539  1.401495  0.137441  0.138325  2.440  0.01468 * 
\end{lstlisting}

Using \citet{keele2010proportionally}'s data, we have the following results:
\begin{lstlisting}
> cox.fit <- coxph(Surv(acttime, censor) ~ 
+                    hcomm + hfloor + scomm + sfloor + 
+                    prespart + demhsmaj + demsnmaj + 
+                    prevgenx + lethal + 
+                    deathrt1 + acutediz + hosp01  + 
+                    hospdisc  + hhosleng + 
+                    mandiz01 + femdiz01 + peddiz01 + orphdum + 
+                    natreg + I(natreg^2) + vandavg3 + wpnoavg3 + 
+                    condavg3 + orderent + stafcder, 
+                data=fda)
> summary(cox.fit)
Call:
coxph(formula = Surv(acttime, censor) ~ hcomm + hfloor + scomm + 
    sfloor + prespart + demhsmaj + demsnmaj + prevgenx + lethal + 
    deathrt1 + acutediz + hosp01 + hospdisc + hhosleng + mandiz01 + 
    femdiz01 + peddiz01 + orphdum + natreg + I(natreg^2) + vandavg3 + 
    wpnoavg3 + condavg3 + orderent + stafcder, data = fda)

  n= 408, number of events= 262 

                  coef  exp(coef)   se(coef)      z Pr(>|z|)    
hcomm        3.642e-01  1.439e+00  2.951e+00  0.123 0.901775    
hfloor       7.944e+00  2.819e+03  8.173e+00  0.972 0.331071    
scomm        4.716e-01  1.603e+00  1.898e+00  0.248 0.803771    
sfloor       2.604e+00  1.352e+01  2.370e+00  1.099 0.271877    
prespart     8.038e-01  2.234e+00  3.042e-01  2.643 0.008226 ** 
demhsmaj     1.363e+00  3.909e+00  1.917e+00  0.711 0.476890    
demsnmaj     1.217e+00  3.377e+00  5.606e-01  2.171 0.029940 *  
prevgenx    -9.915e-04  9.990e-01  7.779e-04 -1.275 0.202459    
lethal       7.872e-02  1.082e+00  2.378e-01  0.331 0.740605    
deathrt1     6.537e-01  1.923e+00  2.435e-01  2.685 0.007253 ** 
acutediz     1.994e-01  1.221e+00  2.262e-01  0.882 0.377896    
hosp01       4.280e-02  1.044e+00  2.495e-01  0.172 0.863768    
hospdisc    -1.238e-06  1.000e+00  5.278e-07 -2.345 0.019002 *  
hhosleng    -1.273e-02  9.874e-01  1.988e-02 -0.640 0.521891    
mandiz01    -1.177e-01  8.889e-01  3.800e-01 -0.310 0.756711    
femdiz01     9.032e-01  2.468e+00  3.497e-01  2.583 0.009799 ** 
peddiz01    -3.401e-02  9.666e-01  5.112e-01 -0.067 0.946968    
orphdum      5.540e-01  1.740e+00  2.109e-01  2.626 0.008630 ** 
natreg      -2.221e-02  9.780e-01  8.282e-03 -2.682 0.007318 ** 
I(natreg^2)  1.029e-04  1.000e+00  4.567e-05  2.253 0.024276 *  
vandavg3    -2.014e-02  9.801e-01  1.536e-02 -1.311 0.189802    
wpnoavg3     5.220e-03  1.005e+00  1.426e-03  3.660 0.000252 ***
condavg3     9.628e-03  1.010e+00  2.271e-02  0.424 0.671637    
orderent    -1.810e-02  9.821e-01  8.147e-03 -2.222 0.026296 *  
stafcder     8.013e-04  1.001e+00  7.986e-04  1.003 0.315719    
\end{lstlisting}

We can also obtain the robust standard errors. In this example, the robust standard errors do not differ much from the original standard errors. 
\begin{lstlisting}
> cox.fit <- coxph(Surv(acttime, censor) ~ 
+                    hcomm + hfloor + scomm + sfloor + 
+                    prespart + demhsmaj + demsnmaj + 
+                    prevgenx + lethal + 
+                    deathrt1 + acutediz + hosp01  + 
+                    hospdisc  + hhosleng + 
+                    mandiz01 + femdiz01 + peddiz01 + orphdum + 
+                    natreg + I(natreg^2) + vandavg3 + wpnoavg3 + 
+                    condavg3 + orderent + stafcder, 
+                  robust = TRUE, 
+                  data=fda)
> summary(cox.fit)
Call:
coxph(formula = Surv(acttime, censor) ~ hcomm + hfloor + scomm + 
    sfloor + prespart + demhsmaj + demsnmaj + prevgenx + lethal + 
    deathrt1 + acutediz + hosp01 + hospdisc + hhosleng + mandiz01 + 
    femdiz01 + peddiz01 + orphdum + natreg + I(natreg^2) + vandavg3 + 
    wpnoavg3 + condavg3 + orderent + stafcder, data = fda, robust = TRUE)

  n= 408, number of events= 262 

                  coef  exp(coef)   se(coef)  robust se      z Pr(>|z|)    
hcomm        3.642e-01  1.439e+00  2.951e+00  2.844e+00  0.128 0.898105    
hfloor       7.944e+00  2.819e+03  8.173e+00  7.300e+00  1.088 0.276494    
scomm        4.716e-01  1.603e+00  1.898e+00  1.674e+00  0.282 0.778110    
sfloor       2.604e+00  1.352e+01  2.370e+00  2.672e+00  0.975 0.329803    
prespart     8.038e-01  2.234e+00  3.042e-01  3.041e-01  2.643 0.008206 ** 
demhsmaj     1.363e+00  3.909e+00  1.917e+00  1.671e+00  0.816 0.414540    
demsnmaj     1.217e+00  3.377e+00  5.606e-01  4.665e-01  2.609 0.009082 ** 
prevgenx    -9.915e-04  9.990e-01  7.779e-04  8.246e-04 -1.202 0.229195    
lethal       7.872e-02  1.082e+00  2.378e-01  2.250e-01  0.350 0.726466    
deathrt1     6.537e-01  1.923e+00  2.435e-01  2.147e-01  3.045 0.002325 ** 
acutediz     1.994e-01  1.221e+00  2.262e-01  1.992e-01  1.001 0.316750    
hosp01       4.280e-02  1.044e+00  2.495e-01  2.930e-01  0.146 0.883839    
hospdisc    -1.238e-06  1.000e+00  5.278e-07  4.775e-07 -2.592 0.009532 ** 
hhosleng    -1.273e-02  9.874e-01  1.988e-02  2.436e-02 -0.523 0.601242    
mandiz01    -1.177e-01  8.889e-01  3.800e-01  4.496e-01 -0.262 0.793436    
femdiz01     9.032e-01  2.468e+00  3.497e-01  3.502e-01  2.579 0.009899 ** 
peddiz01    -3.401e-02  9.666e-01  5.112e-01  5.759e-01 -0.059 0.952917    
orphdum      5.540e-01  1.740e+00  2.109e-01  1.923e-01  2.881 0.003961 ** 
natreg      -2.221e-02  9.780e-01  8.282e-03  8.510e-03 -2.610 0.009051 ** 
I(natreg^2)  1.029e-04  1.000e+00  4.567e-05  4.646e-05  2.214 0.026810 *  
vandavg3    -2.014e-02  9.801e-01  1.536e-02  1.528e-02 -1.318 0.187530    
wpnoavg3     5.220e-03  1.005e+00  1.426e-03  1.557e-03  3.352 0.000801 ***
condavg3     9.628e-03  1.010e+00  2.271e-02  2.181e-02  0.441 0.658859    
orderent    -1.810e-02  9.821e-01  8.147e-03  8.358e-03 -2.166 0.030321 *  
stafcder     8.013e-04  1.001e+00  7.986e-04  7.275e-04  1.101 0.270728 
\end{lstlisting}

\subsection{Log-rank test as a score test from Cox model}

A standard problem in clinical trials is to compare the survival times under treatment and control. Assume no ties in the failure times, and let $x$ denote the binary indicator for treatment. 
Under the proportional hazards assumption, the
control group has hazard $\lambda_{0}(t)$, and the treatment group has
hazard $\lambda_{1}(t) = \lambda_{0}(t)e^{\beta}.$ We are interested in testing the null hypothesis 
$$
H_0: \beta=0 .
$$
The null hypothesis $H_0$ is equivalent to
$$
H_0:  \lambda_{1}(t) = \lambda_{0}(t),
$$
which is further equivalent to
$$
H_0:  S_{1}(t) = S_{0}(t).
$$

Under $H_0$, the score function reduces to
\begin{eqnarray*}
\frac{\partial\log L(0)}{\partial b}  
&=&\sum_{k=1}^{K}\left\{ x_{k}-E_{b=0}(x\mid R_{k})\right\} \\
&=&\sum_{k=1}^{K}\left(x_{k}-\frac{r_{k1}}{r_{k}}\right),
\end{eqnarray*}
because 
\[
E_{b=0}(x\mid R_{k})=\frac{\sum_{l\in R(t_{k})}x_{l}}{\sum_{l\in R(t_{k})}1}=\frac{r_{k1}}{r_{k}}
\]
equaling the ratio of the number of treated units at risk $r_{k1}$ over the
number of units at risk $r_k$, at time $t_k$. The Fisher information at the null is
\begin{eqnarray*}
-\frac{\partial^{2}\log L( 0 )}{\partial b \partial b^{\T}} 
&=& \sum_{k=1}^{K}\cov_{b=0}(x\mid R_{k}) \\
&=& \sum_{k=1}^{K}\frac{r_{k1}}{r_{k}}\left(1-\frac{r_{k1}}{r_{k}}\right).
\end{eqnarray*}

The score test for classical parametric models relies on  Bartlett's identities that under the null hypothesis,  $\frac{\partial\log L( 0 )}{\partial b}$ has mean 0 and variance $-\frac{\partial^{2}\log L( 0 )}{\partial b\partial b^{\T}}  $. If further the CLT holds, then 
$$
\frac{ \frac{\partial\log L( 0 )}{\partial b}  }{  \sqrt{ -\frac{\partial^{2}\log L( 0 )}{\partial b \partial b^{\T}}   } }
\asim\N(0,1),
$$
which, under the Cox model, reduces to
\[
\text{LR} = \frac{\sum_{k=1}^{K}\left(x_{k}-\frac{r_{k1}}{r_{k}}\right)}{\sqrt{\sum_{k=1}^{K}\frac{r_{k1}}{r_{k}}\left(1-\frac{r_{k1}}{r_{k}}\right)}}\asim\N(0,1).
\]
So we reject the null at level $\alpha$, if $|\text{LR} | $ is larger than the upper $1-\alpha/2$ quantile of standard Normal. This
is almost identical to the log-rank test without ties. Allowing for ties, \citet{mantel1966evaluation} proposed a more general form of the log-rank test.\footnote{\citet{peto1972asymptotically} popularized the name ``log-rank test.''}

The \ri{survdiff} function in the \ri{survival} package implements various tests including the log-rank test as a special case. Below, I use the \ri{gehan} dataset in the \ri{MASS} package to illustrate the log rank test. The data were from a matched-pair experiment of 42 leukaemia patients \citep{gehan1965generalized}. Treated units received the drug 6-mercaptopurine, and the rest are controls. For illustration purposes, I ignore the pair indicators.

\begin{lstlisting}
> library(MASS)
> head(gehan)
  pair time cens   treat
1    1    1    1 control
2    1   10    1    6-MP
3    2   22    1 control
4    2    7    1    6-MP
5    3    3    1 control
6    3   32    0    6-MP
> survdiff(Surv(time, cens) ~ treat,
+          data = gehan)
Call:
survdiff(formula = Surv(time, cens) ~ treat, data = gehan)

               N Observed Expected (O-E)^2/E (O-E)^2/V
treat=6-MP    21        9     19.3      5.46      16.8
treat=control 21       21     10.7      9.77      16.8

 Chisq= 16.8  on 1 degrees of freedom, p= 4e-05 
 \end{lstlisting}

\begin{figure}[th]
\centering
\includegraphics[width = \textwidth]{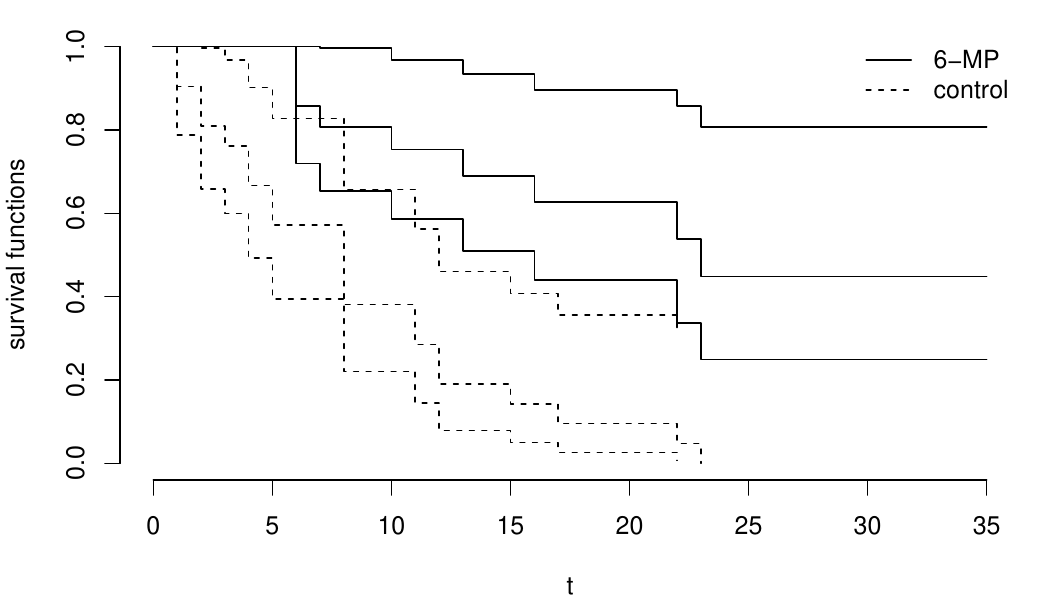}
\caption{Kaplan--Meier curves with 95\% confidence intervals based on \citet{gehan1965generalized}'s data}\label{fig:: gehan_kmcurve-data}
\end{figure}

The treatment was quite effective, yielding an extremely small $p$-value even with moderate sample size. It is also clear from the Kaplan--Meier curves in Figure \ref{fig:: gehan_kmcurve-data} and the results from fitting the Cox proportional hazards model. 

\begin{lstlisting}
> cox.gehan = coxph(Surv(time, cens) ~ treat,
+                     data = gehan)
> summary(cox.gehan)
Call:
coxph(formula = Surv(time, cens) ~ treat, data = gehan)

  n= 42, number of events= 30 

               coef exp(coef) se(coef)     z Pr(>|z|)    
treatcontrol 1.5721    4.8169   0.4124 3.812 0.000138 ***

             exp(coef) exp(-coef) lower .95 upper .95
treatcontrol     4.817     0.2076     2.147     10.81

Concordance= 0.69  (se = 0.041 )
Likelihood ratio test= 16.35  on 1 df,   p=5e-05
Wald test            = 14.53  on 1 df,   p=1e-04
Score (logrank) test = 17.25  on 1 df,   p=3e-05
\end{lstlisting}

The log-rank test is a standard tool in survival analysis. However, what it delivers is just a special case of the Cox proportional hazards model. The $p$-value from the log-rank test is close to the $p$-value from the score test of the Cox proportional hazards model with only a binary treatment indicator. The latter can also adjust for other pretreatment covariates.

\section{Extensions}

\subsection{Stratified Cox proportional hazards model}

Many randomized trials are stratified. The Combined Pharmacotherapies and Behavioral Interventions study reviewed at the beginning of this chapter is an example with \ri{site} indicating the strata. The previous analysis includes the dummy variables of  \ri{site} in the Cox model. An alternative more flexible model is to allow for different baseline hazard functions across strata. 

\begin{definition}
[Stratified Cox proportional hazards model]\label{def::stratified-cox-model}
Assume
$$
\lambda_s(t \mid x) = \lambda_s(t) \exp(\beta^{\T} x)
$$
for strata $s = 1,\ldots, S$. The observations are independent across units. The $\beta$ is an unknown parameter and $\{ \lambda_1(\cdot), \ldots, \lambda_S(\cdot) \}$ are unknown functions. 
\end{definition}

Under Definition \ref{def::stratified-cox-model}, within each stratum $s$, the proportional hazards assumption holds; across strata, the proportional hazard assumption may not hold. Within stratum $s$, we can obtain the partial likelihood $L_s(\beta)$; by independence of the data across strata, we can obtain the joint partial likelihood 
$$
\prod_{s=1}^S L_s(\beta) .
$$
Based on the standard procedure, we can obtain the MLE and conduct inference based on the large-sample theory. The \ri{coxph} function can naturally allow for stratification with the \ri{ + strata()} in the regression formula.

Revisit the data from \citet{lin2016simultaneous}. We change \ri{ + site} to \ri{+ strata(site)} in the regression formula to allows for different baseline hazard functions. In this example, the coefficients of other covariates do not change much. 

\begin{lstlisting}
> cox.fit <- coxph(Surv(futime, relapse) ~ NALTREXONE*THERAPY + 
+                    AGE + GENDER + T0_PDA + strata(site), 
+                  robust = TRUE, 
+                  data=COMBINE)
> summary(cox.fit)
Call:
coxph(formula = Surv(futime, relapse) ~ NALTREXONE * THERAPY + 
    AGE + GENDER + T0_PDA + strata(site), data = COMBINE, robust = TRUE)

  n= 1226, number of events= 856 

                        coef exp(coef)  se(coef) robust se      z Pr(>|z|)
NALTREXONE         -0.252239  0.777059  0.097788  0.096437 -2.616  0.00891
THERAPY            -0.173456  0.840754  0.096258  0.095958 -1.808  0.07066
AGE                -0.015104  0.985010  0.003554  0.003512 -4.301  1.7e-05
GENDERmale         -0.139837  0.869500  0.075388  0.076580 -1.826  0.06785
T0_PDA              0.002747  1.002751  0.001369  0.001350  2.035  0.04182
NALTREXONE:THERAPY  0.335671  1.398879  0.137676  0.136890  2.452  0.01420
                      
NALTREXONE         ** 
THERAPY            .  
AGE                ***
GENDERmale         .  
T0_PDA             *  
NALTREXONE:THERAPY *  

                   exp(coef) exp(-coef) lower .95 upper .95
NALTREXONE            0.7771     1.2869    0.6432    0.9387
THERAPY               0.8408     1.1894    0.6966    1.0147
AGE                   0.9850     1.0152    0.9783    0.9918
GENDERmale            0.8695     1.1501    0.7483    1.0103
T0_PDA                1.0028     0.9973    1.0001    1.0054
NALTREXONE:THERAPY    1.3989     0.7149    1.0697    1.8294

Concordance= 0.561  (se = 0.011 )
Likelihood ratio test= 35.24  on 6 df,   p=4e-06
Wald test            = 33.85  on 6 df,   p=7e-06
Score (logrank) test = 34.94  on 6 df,   p=4e-06,   Robust = 34.15  p=6e-06
\end{lstlisting}

\subsection{Clustered Cox model}

 With clustered data, we must adjust for the standard errors. The \ri{coxph} function reports the cluster-robust standard errors with the specification of \ri{cluster}. A canonical example of clustered data is from the matched-pair design if we view the pairs as clusters. The example below uses the data from \citet{huster1989modelling}, in which two eyes of a patient were either assigned to treatment or control. 
 
First, I ignore the clustering structure in the data.  
 \begin{lstlisting}
> library("timereg")
> data(diabetes)
> pair.cox = coxph(Surv(time, status) ~ treat + agedx,
+                  robust = TRUE, 
+                  data = diabetes)
> summary(pair.cox)
Call:
coxph(formula = Surv(time, status) ~ treat + agedx, data = diabetes, 
    robust = TRUE)

  n= 394, number of events= 155 

           coef exp(coef)  se(coef) robust se      z Pr(>|z|)    
treat -0.782157  0.457418  0.168972  0.170054 -4.599 4.24e-06 ***
agedx  0.004029  1.004037  0.005473  0.005595  0.720    0.471    

      exp(coef) exp(-coef) lower .95 upper .95
treat    0.4574      2.186    0.3278    0.6384
agedx    1.0040      0.996    0.9931    1.0151

Concordance= 0.591  (se = 0.025 )
Likelihood ratio test= 22.91  on 2 df,   p=1e-05
Wald test            = 21.16  on 2 df,   p=3e-05
Score (logrank) test = 22.78  on 2 df,   p=1e-05,   Robust = 21.84  p=2e-05
 \end{lstlisting}

Second, I specify \texttt{cluster = id} to obtain the robust standard errors clustered at the individual level.

 \begin{lstlisting}
> pair.cox = coxph(Surv(time, status) ~ treat + agedx,
+                  robust = TRUE, 
+                  cluster = id,
+                  data = diabetes)
> summary(pair.cox)
Call:
coxph(formula = Surv(time, status) ~ treat + agedx, data = diabetes, 
    robust = TRUE, cluster = id)

  n= 394, number of events= 155 

           coef exp(coef)  se(coef) robust se      z Pr(>|z|)    
treat -0.782157  0.457418  0.168972  0.148338 -5.273 1.34e-07 ***
agedx  0.004029  1.004037  0.005473  0.006256  0.644     0.52    

      exp(coef) exp(-coef) lower .95 upper .95
treat    0.4574      2.186    0.3420    0.6118
agedx    1.0040      0.996    0.9918    1.0164

Concordance= 0.591  (se = 0.023 )
Likelihood ratio test= 22.91  on 2 df,   p=1e-05
Wald test            = 27.84  on 2 df,   p=9e-07
Score (logrank) test = 22.78  on 2 df,   p=1e-05,   Robust = 26.37  p=2e-06

  (Note: the likelihood ratio and score tests assume independence of
     observations within a cluster, the Wald and robust score tests do not).  
 \end{lstlisting}

\subsection{Penalized Cox model}

The \ri{glmnet} function in the \ri{glmnet} package in \ri{R} implements the penalized version of the Cox model. You need to specify \ri{fammily = "cox"}  in the function.

\section{Critiques on survival analysis}

The Kaplan--Meier curve and the Cox proportional hazards model are standard tools for analyzing medical data with censored survival times. They are among the most commonly used methods in medical journals. \citet{kaplan1958nonparametric} and \citet{cox1972regression} are two of the most cited papers in statistics. 

\citet{freedman2008survival} criticized these two standard tools. Both rely on the critical assumption of noninformative censoring that censoring and survival time are independent or conditionally independent given covariates. When censoring is due to administrative constraints, this is a reasonable assumption. The data from \citet{lin2016simultaneous} is a convincing example of noninformative censoring.  However, many other studies have more complex censoring mechanisms, for example, one may drop out of the study, and another may be killed by an irrelevant cause. The Cox model relies on an additional assumption of proportional hazards. This particular functional form facilitates the interpretation of the coefficients as log conditional hazard ratios if the model is correctly specified. However, its interpretation becomes obscure when the model is mis-specified.  Two survival curves based on \citet{lin2016simultaneous}'s data cross each other, which makes the proportional hazards assumption dubious.

\citet{hernan2010hazards} offered a more fundamental critique on hazard-based survival analysis. For example, in a randomized treatment-control experiment, the hazard ratio at time $t$ is the ratio of the instantaneous probability of death conditioning on the event that the patients have survived up to time $t$:
$$
\frac{   \lim_{\Delta t\downarrow0}\pr(t\leq T<t+\Delta t\mid x=1, T\geq t)/\Delta t  }
{  \lim_{\Delta t\downarrow0}\pr(t\leq T<t+\Delta t\mid x=0, T\geq t)/\Delta t  } . 
$$
This ratio is difficult to interpret because patients who have survived up to time $t$ can be quite different in treatment and control groups, especially when the treatment is effective. Even though patients are randomly assigned at the baseline, the survivors up to time $t$ are not. \citet{hernan2010hazards} suggested focusing on the comparison of the survival functions.

\section{Homework problems}

\paragraph{From survival function to expectation}\label{hw23::survival-expectation}

Prove Proposition \ref{prop::survival-mean}.

Remark: Proposition \ref{prop::survival-mean} is similar to the following theorem in probability theory.

\begin{theorem}
\label{thm::discrete-expection-survival}
If the random variable $T$ takes only positive integer values, then
$$
E(T) = \sum_{t=1}^{\infty} \pr(T \geq t).
$$
\end{theorem}

\paragraph{Identifiability of the survival time under independent censoring}\label{hw23::identification-km}

Assume the survival time $T$ and censoring time $C$ are continuous and independent random variables. But we can only observe $y = \min(T, C)$ and $\delta = 1(T \leq C)$. 

Prove that the hazard function of $T$ can be identified by the following formula: 
$$
\lambda_T(t) =    \frac{   \pr(y = t, \delta = 1)   }{  \pr(y \geq t)  } . 
$$

Remark: The term ``identifiability'' is technical. It means that the parameter of interest can be recovered by the distribution of the observed variables. On the left-hand side of the formula above, $\lambda_T(t)$ is the hazard function of $T$, which is not fully observed. On the right-hand side of the formula above, $\pr(y = t, \delta = 1) $ and $\pr(y \geq t) $ only involve the distribution of the observed variables $(y, \delta)$. Therefore, we can say that even in the presence of censoring, the hazard function of $T$ can be identified by the observed data.

\paragraph{Log-Normal regression model}

Does the log-Normal regression model in Example \ref{ex::log-normal-regression} satisfy the proportional hazards assumption?
Based on IID data $(y_{i},x_{i},\delta_{i})_{i=1}^{n}$, what is the likelihood function under Assumption \ref{assume::non-informative-censoring}? 
Compare it with the partial likelihood function.

\paragraph{Weibull random variable}
\label{hw23::weibull-rv}

Using \eqref{eq::weibull-representation} to prove the formulas of density, survival, and hazard functions. Calculate its mean and variance. 

Remark: Use the Gamma function to express the moments.

\paragraph{Weibull regression model}
\label{hw23::weibull-regression}

Find the distribution of $\varepsilon_i$ in the Weibull regression model in Example \ref{ex::weibull-regression}.  Show $\log E(T \mid x )$ is linear in $x$, and $E(\log T \mid x)$ is linear in $x$. Does it satisfy the proportional hazards assumption? Based on
$(y_{i},x_{i},\delta_{i})_{i=1}^{n}$, what is the likelihood function under Assumption \ref{assume::non-informative-censoring}? 
Compare it with the partial likelihood function.

\paragraph{Invariance of the proportional hazards model}\label{hw23::invariance-ph-model}

Assume that $T \mid x$ follows a proportional hazards model. 

Prove that any non-negative and strictly increasing transformation $g(T)\mid x$ also follows a proportional
hazards model.

\appendix  

\part{Appendices}
\renewcommand{\theparagraph}{\Alph{chapter}.\arabic{paragraph}}

\chapter{Linear Algebra}\label{chapter::linear-algebra}

Linear algebra is crucial for understanding the theory of the linear model. This Appendix reviews the basics of linear algebra that are closely related to this book.

All vectors are column vectors in this book. This is coherent with \ri{R}.

\section{Basics of vectors and matrices}
 \label{sec::basics-vectors-matrices}

\paragraph*{Euclidean space}

The $n$-dimensional Euclidean space $\mathbb{R}^{n}$ is a set of
all $n$-dimensional vectors equipped with an inner product: 
$$
\langle x,y \rangle =x^{\T}y=\sumn x_{i}y_{i}, 
$$
where $x=(x_{1},\ldots,x_{n})^{\T}$ and $y=(y_{1},\ldots,y_{n})^{\T}$
are two $n$-dimensional vectors. The length of a vector $x$
is defined as 
$$
\|x\|=\sqrt{\langle x,x\rangle }=\sqrt{x^{\T}x}.
$$ 
The Cauchy--Schwarz inequality states that the inner product of $x$ and $y$ is bounded from above by the product of their length. 

\begin{proposition}
\label{prop::cauchy-schwarz}
For two $n$-dimensional vectors $x$ and $y$, we have 
\[
|\langle x,y \rangle |\leq\|x\|\cdot\|y\|,
\]
or, more transparently,
$$
\left(  \sumn x_i y_i  \right)^2  \leq  \left(  \sumn x_i^2  \right) \left(  \sumn y_i^2  \right)  . 
$$
The equality holds if and only if $ a y_i = bx_i$ for some $a$ and $b$, for all $i=1,\ldots, n.$
\end{proposition}

We can use the Cauchy--Schwarz inequality to prove the triangle inequality that the length of the summation of $x$ and $y$ is bounded from above by the summation of their length. 

\begin{proposition}
\label{proo::triangle}
For two $n$-dimensional vectors $x$ and $y$, we have 
$$
\| x + y \| \leq \| x\| + \| y\| .  
$$
\end{proposition}

We say that $x$ and $y$ are {\it orthogonal}, denoted by $x\perp y$, if $\langle x,y \rangle=0$. We call a set of vectors $v_1, \ldots, v_m \in \mathbb{R}^{n}$ {\it orthonormal} if they all have unit length and are mutually orthogonal.

Geometrically, we can define the cosine of the angle between two vectors $x, y \in \mathbb{R}^n$ as
$$
\cos \angle (x, y) = \frac{ \langle x, y \rangle    }{   \| x \| \| y \|  }
= \frac{  \sumn x_i y_i  }{    \sqrt{  \sumn x_i^2 \sumn y_i^2  }  }.
$$
For unit vectors, it reduces to the inner product. 
When both $x$ and $y$ are orthogonal to $1_n$, that is, $\bar{x} = n^{-1} \sumn x_i =0$ and $\bar{y} = n^{-1} \sumn y_i =0$, the formula of the cosine of the angle is identical to the sample Pearson correlation coefficient
$$
\hat{\rho}_{xy} =  \frac{  \sumn (x_i - \bar{x})  (y_i - \bar{y} )  }{  \sqrt{   \sumn (x_i - \bar{x} ) ^2 \sumn (y_i - \bar{y})^2  }  } .
$$
Sometimes, we simply say that the cosine of the angle of two vectors measures their correlation even when they are not orthogonal to $1_n$.

\paragraph*{Column space of a matrix}

Given an $n\times m$ matrix $A$, we can view it in terms of all elements 
$$
A = (a_{ij})
= \begin{pmatrix}
a_{11} & \cdots & a_{1m} \\
\vdots &  & \vdots \\
a_{n1} & \cdots & a_{nm}
\end{pmatrix},
$$ 
or row vectors
$$
A = \begin{pmatrix}
a_1^{\T} \\
\vdots \\
a_n^{\T}
\end{pmatrix},
$$
where $a_i \in \mathbb{R}^m$ $(i=1, \ldots, n)$,  or column vectors
$$
A=(A_{1},\ldots,A_{m}),
$$
where $A_j \in \mathbb{R}^n$ $j=1,\ldots, m$. 
In statistics, the rows correspond to the units, so the $i$th row vector is the vector observations for unit $i$. Moreover, viewing $A$ in terms of its column vectors can give more insights.

\begin{definition}
[column space]\label{def::column-space}
For an $n\times m$ matrix $A= (A_{1},\ldots,A_{m})$, define the  column space of $A$ as 
\[
\mathcal{C}(A)=\left\{ \alpha_{1}A_{1}+\cdots+\alpha_{m}A_{m}: \alpha_{1},\ldots,\alpha_{m}\in\mathbb{R}\right\} . 
\]
\end{definition}

The column space of $A$ is the set of all linear combinations of the column vectors  $A_{1},\ldots,A_{m}$.   The column space is important because  we can write $A\alpha$, with $\alpha = (\alpha_1, \ldots, \alpha_m)^{\T}$, as
$$
A\alpha = (A_{1},\ldots,A_{m}) \begin{pmatrix}
\alpha_1 \\
\vdots \\
\alpha_m
\end{pmatrix}
 = \alpha_{1}A_{1}+\cdots+\alpha_{m}A_{m}, 
$$
which is in $\mathcal{C}(A).$

We define the row space of $A$ as the column space of $A^{\T}$.

\paragraph{Matrix product}

Given an $n\times m$ matrix $A = (a_{ij})$ and an $m\times r$ matrix $B = (b_{ij})$, we can define their product as $C = AB$ where the $n\times r$ matrix $C = (c_{ij})$ has the $(i, j)$th element
$$
c_{ij} = \sum_{k=1}^m a_{ik} b_{kj} .
$$
In terms of the row vectors of $A$ or column vectors of $B$, we have
$$
c_{ij} = a_i^{\T} B_j ,
$$
that is, $c_{ij}$ equals the inner product of the $i$th row vector of $A$ and the $j$th column vector of $B$. Moreover, the matrix product satisfies 
\begin{equation}\label{eq::matrix-product-1}
AB = A (B_1, \ldots, B_r) = (AB_1, \ldots, AB_r)
\end{equation}
so the column vectors of $AB$ belongs to the column space of $A$; it also satisfies 
\begin{equation}\label{eq::matrix-product-2}
AB = \begin{pmatrix}
a_1^{\T} \\
\vdots \\
a_n^{\T}
\end{pmatrix} B
= \begin{pmatrix}
a_1^{\T} B \\
\vdots \\
a_n^{\T} B
\end{pmatrix}
\end{equation}
so the row vectors of $AB$ belong to the column space of $B^{\T}$, or equivalently, the row space of $B$.

\paragraph{Linearly independent vectors and rank}

We call a set of vectors $A_1, \ldots, A_m \in \mathbb{R}^{n}$ {\it linearly independent} if 
$$
x_1 A_1 + \cdots + x_m A_m = 0 
$$
must imply $x_1= \cdots = x_m = 0$. 
We call $A_{j_1}, \ldots, A_{j_k}$ maximally linearly independent if adding another vector makes them linearly dependent. Define $k$ as the rank of $\{A_1, \ldots, A_m\}$ and also define $k$ as the rank of the matrix $A = (A_1, \ldots, A_m)$.

A set of vectors may have different subsets of vectors that are maximally linearly independent. But the rank $k$ is unique. We can also define the rank of a matrix in terms of its row vectors. A remarkable theorem in linear algebra is that it does not matter whether we define the rank of a matrix in terms of its column vectors or row vectors.

From the matrix product formulas \eqref{eq::matrix-product-1} and \eqref{eq::matrix-product-2}, we have the following result.

\begin{proposition}\label{eq::matrix-product-inequality}
$
\textup{rank}(AB) \leq  \min\{  \textup{rank}(A),  \textup{rank}(B) \} . 
$
\end{proposition}

The rank decomposition of a matrix decomposes $A$ into the product of two matrices of full ranks. 

\begin{proposition}\label{eq::rank-decomposition}
If an $n\times m$ matrix $A$ has rank $k$, then $A = BC$ for some $n\times k$ matrix $B$ and $k\times m$ matrix $C$. 
\end{proposition}

\begin{myproof}{Proposition}{\ref{eq::rank-decomposition}}
Let $A_{j_1}, \ldots, A_{j_k}$ be the maximally linearly independent column vectors of $A$. Stack them into an $n\times k$ matrix $B = (A_{j_1}, \ldots, A_{j_k}) $. They can linearly represent all column vectors of $A$:
\begin{eqnarray*}
A &=& (c_{11} A_{j_1} + \cdots + c_{k1} A_{j_k}, \ldots, c_{1m} A_{j_1} + \cdots + c_{km} A_{j_k} ) \\
&=&  (BC_1, \ldots, B C_m) \\ 
&=& BC,
\end{eqnarray*}
where $C =  ( C_1, \ldots,  C_m)$ is an $k\times m$ matrix with column vectors
$$
C_1 = \begin{pmatrix}
c_{11}  \\
\vdots\\
 c_{k1}
\end{pmatrix},\cdots 
C_m = \begin{pmatrix}
 c_{1m} \\
 \vdots \\
 c_{km}
\end{pmatrix}.
$$
\end{myproof}

Proposition \ref{eq::matrix-product-inequality} ensures that the $B$ and $C$ in Proposition \ref{eq::rank-decomposition} must satisfy rank$(B) \geq k$ and rank$(C) \geq k$, so they must both have rank $k$. The decomposition in Proposition \ref{eq::rank-decomposition} is not unique since the choice of the maximally linearly independent column vectors of $A$ is not unique.

\paragraph*{Some special matrices}

An $n\times n$ matrix $A$ is symmetric if $A^{\T}=A$.

An $n\times n$ diagonal matrix $A$ has zero off-diagonal elements,
denoted by $A=\text{diag}\{a_{11},\ldots,a_{nn}\}.$
Diagonal matrices are symmetric. 

An $n\times n$
matrix is orthogonal if $A^{\T}A=AA^{\T}=I_{n}$. The column vectors of an orthogonal matrix are orthonormal; so are its row vectors. If $A$ is orthogonal, then 
$$
\| A x \|  = \|x\|
$$
for any vector $x\in \mathbb{R}^n$. That is, multiplying a vector by an orthogonal matrix does not change the length of the vector. Geometrically, an orthogonal matrix corresponds to rotation.

An $n\times n$ matrix $A$ is upper triangular if $a_{ij} = 0$ for $i > j$ and lower triangular if $a_{ij} = 0$ for $i < j$. 

\paragraph*{Determinant}

The original definition of the determinant of a matrix $A = (a_{ij}) \in \mathbb{R}^{n\times n}$, due to Leibniz, is quite complicated, which relies on the notation of permutation. A permutation $\sigma$  on $\{1, \ldots, n\} $ is a one-to-one mapping from $\{1, \ldots, n\} $ to $\{1, \ldots, n\} $. Let $\textup{sgn}(\sigma)$ denote the sign of the permutation $\sigma$, which equals 1 if $\sigma$ can be obtained via an even number of transpositions and 0 if $\sigma$ can be obtained via an odd number of transpositions. Define
\[
\textup{det}(A) = \sum_{\sigma } \textup{sgn}(\sigma) \prod_{i=1}^{n} a_{i,\sigma(i)},
\]
where the summation is over all possible permutations, and
$a_{i,\sigma(i)}$ is the $(i,\sigma(i))$-th element of $A$.



The determinant of a $2\times 2$ matrix has a simple form:
\begin{equation}\label{eq::det-2X2}
\textup{det}\begin{pmatrix}
a & b \\
c& d
\end{pmatrix}
= ad - bc.
\end{equation}
The determinant of the Vandermonde matrix has the following formula:
\begin{equation}
\label{eq::det-Vandermonde}
\textup{det}\begin{pmatrix}
1& x_1 & x_1^2 & \cdots & x_1^{n-1} \\
1& x_2 & x_2^2 & \cdots & x_2^{n-1} \\
\vdots & \vdots & \vdots & & \vdots \\
1&x_n & x_n^2 & \cdots & x_n^{n-1}
\end{pmatrix} 
= \prod_{1\leq i, j \leq n} (x_j - x_i).
\end{equation}

This book will not use the above definition of the determinant. 
The properties of the determinant are more useful. I will review two.

\begin{proposition}
For two square matrices $A$ and $B$, we have
$$
\textup{det}(AB) = \textup{det}(A) \textup{det}(B)  = \textup{det}(BA). 
$$
\end{proposition}

\begin{proposition}
For two square matrices $A \in \mathbb{R}^{m\times m}$ and $B \in \mathbb{R}^{n \times n}$, we have
$$
\textup{det}
\begin{pmatrix}
A & 0 \\
C& B
\end{pmatrix}
= 
\textup{det}
\begin{pmatrix}
A & D \\
0& B
\end{pmatrix}
=
\textup{det}(A) \textup{det}(B) . 
$$
\end{proposition}

\paragraph*{Inverse of a matrix}

Let $I_{n}$ be the $n\times n$ identity matrix. An $n\times n$
matrix $A$ is invertible or nonsingular if there exists an $n\times n$ matrix
$B$ such that $AB=BA=I_{n}.$ We call $B$ the inverse of $A$, denoted
by $A^{-1}.$  If $A$ is an orthogonal matrix, then $A^{\T} =A^{-1}.$

A square matrix is invertible if and only if det$(A) \neq 0.$

The inverse of a $2\times 2$ matrix is
\begin{equation}
\label{eq::2X2-inverse}
\begin{pmatrix}
a & b \\
c& d
\end{pmatrix}^{-1}
= \frac{1}{ad - bc} \begin{pmatrix}
d & -b \\
-c& a
\end{pmatrix}.
\end{equation}
The inverse of a $3\times 3$ lower triangular matrix is
\begin{equation}
\label{eq::3X3-lower-inverse}
\left(\begin{array}{lll}
a & 0 & 0 \\
b & c & 0 \\
d & e & f
\end{array}\right)^{-1}=\frac{1}{a c f}\left(\begin{array}{ccc}
c f & 0 & 0 \\
-b f & a f & 0 \\
b e-c d & -a e & a c
\end{array}\right). 
\end{equation}

A useful identity is
$$
(AB)^{-1}  = B^{-1} A^{-1}, 
$$
if both $A$ and $B$ are invertible.

\paragraph*{Eigenvalues and eigenvectors}

For an $n\times n$ matrix $A$, if there exists a pair of $n$-dimensional, non-zero
vector $x$ and a scalar $\lambda$ such that 
$$
Ax=\lambda x,
$$ 
then we call $\lambda$ an eigenvalue and $x$ the associated eigenvector of $A$. From
the definition, eigenvalue and eigenvector always come in pairs. The
following eigen-decomposition theorem is fundamental for real symmetric
matrices.

\begin{theorem}\label{thm::eigendecomposition}
If $A$ is an $n\times n$ symmetric matrix, then there exists
an orthogonal matrix $P$ such that 
\begin{equation}
\label{eq::eigendecompose-1}
P^{\T}AP=\textup{diag}\{\lambda_{1},\ldots,\lambda_{n}\},
\end{equation}
where the $\lambda$'s are the $n$ eigenvalues of $A$, and the column vectors of $P=(\gamma_{1},\cdots,\gamma_{n})$ are the corresponding eigenvectors. 
\end{theorem}

If we multiply \eqref{eq::eigendecompose-1} by $P$ from the left, then we can write the eigendecomposition as 
$$
AP=P\text{diag}\{\lambda_{1},\ldots,\lambda_{n}\}
$$
or, equivalently,
$$
A(\gamma_{1},\cdots,\gamma_{n}) = (\lambda_1 \gamma_{1},\cdots, \lambda_{n}\gamma_{n}),
$$
then $(\lambda_i,\gamma_i)$ must be a pair of eigenvalue and eigenvector. Moreover, the eigendecomposition in Theorem \ref{thm::eigendecomposition} is unique up to the permutation of the columns of $P$ and the corresponding $\lambda_i$'s. 

\begin{corollary}
If a real symmetric matrix $A$ has eigen-decomposition $P^{\T}AP=\textup{diag}\{\lambda_{1},\ldots,\lambda_{n}\}$, then 
$$
A=P\textup{diag}\{\lambda_{1},\ldots,\lambda_{n}\}P^{\T},
$$
and therefore,
$$
A^{k}=A A\cdots A=P\textup{diag}\{\lambda_{1}^{k},\ldots,\lambda_{n}^{k}\}P^{\T}.
$$
If the eigenvalues of $A$ are nonzero, then 
$$A^{-1}=P\textup{diag}\{1/\lambda_{1},\ldots,1/\lambda_{n}\}P^{\T}.$$
\end{corollary}

The eigen-decomposition is also useful for defining the square root
of an $n\times n$ symmetric matrix. In particular, if the eigenvalues
of $A$ are nonnegative, then we can define
\[
A^{1/2}=P\text{diag}\{\sqrt{\lambda_{1}},\ldots,\sqrt{\lambda_{n}}\}P^{\T} . 
\]
By definition, $A^{1/2}$ is a symmetric matrix satisfying $A^{1/2}A^{1/2}=A.$
There are other definitions of the square root of a symmetric matrix,
but we adopt this form in this book.

From \eqref{eq::eigendecompose-1}, we can write $A$ as 
\begin{align*}
A & =P\text{diag}\{\lambda_{1},\ldots,\lambda_{n}\}P^{\T}\\
 & =(\gamma_{1},\cdots,\gamma_{n})\text{diag}\{\lambda_{1},\ldots,\lambda_{n}\}\left(\begin{array}{c}
\gamma_{1}^{\T}\\
\vdots\\
\gamma_{n}^{\T}
\end{array}\right)\\
 & =\sumn\lambda_{i}\gamma_{i}\gamma_{i}^{\T}. 
\end{align*}

For an $n\times n$ symmetric matrix $A$, its rank equals the number of non-zero eigenvalues
and its determinant equals the product of all eigenvalues. The matrix
$A$ is of full rank if all its eigenvalues are non-zero, which implies
that its rank equals $n$ and its determinant is non-zero.

\paragraph*{Quadratic form}

For an $n\times n$ symmetric matrix $A=(a_{ij})$ and an $n$-dimensional
vector $x$, we can define the quadratic form as

\[
x^{\T}Ax=\langle x,Ax \rangle =\sumn\sum_{j=1}^{n}a_{ij}x_{i}x_{j}.
\]

We always consider a symmetric matrix in the quadratic form without
loss of generality. Otherwise, we can  symmetrize $A$ as $ \tilde{A} =  (A+A^{\T})/2$
without changing the value of the quadratic form because
\[
x^{\T}Ax=x^{\T}  \tilde{A}  x.
\]

We call $A$ positive semi-definite, denoted by $A\succeq 0,$ if $x^{\T}Ax\ge0$
for all $x$; we call $A$ positive definite, denoted by $A\succ 0,$
if $x^{\T}Ax>0$ for all nonzero $x.$

We can also define the partial order between matrices. We call $A\succeq B$
if and only if $A-B\succeq0$, and we call $A\succ B$ if and only
if $A-B\succ0$. This is important in statistics because we often
compare the efficiency of estimators based on their variances or covariance matrices. Given two unbiased estimators $\hat\theta_1$ and $\hat\theta_2$ for a scalar parameter $\theta$, we say that $\hat\theta_1$ is at least as efficient as $\hat\theta_2$ if var$(\hat\theta_2) \geq $ var$(\hat\theta_1)$. In the vector case, we say that $\hat\theta_1$ is at least as efficient as $\hat\theta_2$  if cov$(\hat\theta_2) \succeq $ cov$(\hat\theta_1)$, which is equivalent to var$(\ell^{\T} \hat\theta_2) \geq $ var$(\ell^{\T}\hat\theta_1)$ for any linear combination of the estimators.

The eigenvalues of a symmetric matrix determine whether it is positive semi-definite or positive definite.

\begin{theorem}
For a symmetric matrix $A$, it is positive semi-definite
if and only if all its eigenvalues are nonnegative, and it is positive
definite if and only if all its eigenvalues are positive.
\end{theorem}

An important result is the relationship between the eigenvalues and the extreme values of the quadratic form. Assume that the eigenvalues are rearranged in decreasing order such that $\lambda_1 \geq \cdots \geq \lambda_n$. 
For a unit vector $x$ with length $\| x \| = 1$, we have that
$$
x^{\T} A x = x^{\T} \sumn\lambda_{i}\gamma_{i}\gamma_{i}^{\T} x 
= \sumn\lambda_{i}  \alpha_{i}^2  
$$
where 
$$
\alpha 
= \begin{pmatrix}
\alpha_1 \\
\vdots \\
\alpha_n
\end{pmatrix}
= \begin{pmatrix}
\gamma_1^{\T} x \\
\vdots \\
\gamma_{n}^{\T} x
\end{pmatrix}
= P^{\T} x
$$
has length $ \|\alpha \|^2  = \| x\|^2 = 1$.
Then the maximum value of $x^{\T} A x $ is $\lambda_1$, which is achieved at $\alpha_1 = 1$ and $\alpha_2 = \cdots = \alpha_n = 0$ (for example, if $x = \gamma_1$, then $\alpha_1 = 1$ and $\alpha_2 = \cdots = \alpha_n = 0$). For a unit vector $x$ that is orthogonal to $\gamma_1$, we have that
$$
x^{\T} A x  = \sum_{i=2}^n \lambda_{i}  \alpha_{i}^2 
$$
where $\alpha = P^{\T} x$ has unit length with $\alpha_1 = 0$. The maximum value of $x^{\T} A x $ is $\lambda_2$, which is achieved at $\alpha_2 = 1$ and $\alpha_1 = \alpha_3 = \cdots = \alpha_n = 0$, for example, $x = \gamma_2$. By induction, we have the following theorem.

\begin{theorem}
\label{thm::eigven-max-q}
Suppose that an $n\times n$ symmetric matrix has eigen-decomposition $\sumn\lambda_{i}\gamma_{i}\gamma_{i}^{\T}$ where $\lambda_1 \geq \cdots \geq \lambda_n$. 
\begin{enumerate}
\item
The optimization problem
$$
\max_{x \in \mathbb{R}^n} x^{\T} A x \text{ such that } \|x\| = 1
$$
has maximum $\lambda_1$, which can be achieved by $\gamma_1$.
\item
The optimization problem
$$
\max_{x \in \mathbb{R}^n} x^{\T} A x \text{ such that } \|x\| = 1,  x \perp \gamma_1 
$$
has maximum $\lambda_2$, which can be achieved  by $\gamma_2$.
\item
The optimization problem
$$
\max_{x \in \mathbb{R}^n} x^{\T} A x \text{ such that } \|x\| = 1,  x \perp \gamma_1 ,\ldots, x\perp \gamma_k
$$
has maximum $\lambda_{k+1}$, which can be achieved by $\gamma_{k+1}$ $(k=1, \ldots, n-1)$. 
\end{enumerate}
\end{theorem}

Theorem \ref{thm::eigven-max-q} implies the following theorem on the  Rayleigh quotient
$$
r(x)=x^{\T}Ax/x^{\T}x \qquad  (x\in \mathbb{R}^n).
$$

\begin{theorem}
(Rayleigh quotient and eigenvalues) \label{theorem::rayleigh}
The maximum and minimum eigenvalues of  an $n\times n$ symmetric matrix $A$ equals
\[
\lambda_{\max}(A)=\max_{x\neq0}r(x),\qquad\lambda_{\min}(A)=\min_{x\neq0}r(x)
\]
with the maximizer and minimizer being the eigenvectors corresponding to the maximum and minimum eigenvalues, respectively. 
\end{theorem}

An immediate consequence of Theorem \ref{theorem::rayleigh} is that the diagonal elements of $A$ are bounded by the smallest and largest eigenvalues of $A$. This follows by taking $x = (0, \ldots, 1, \ldots, 0)^{\T}$, where only the $i$th element equals $1$. 

\paragraph*{Trace}

The trace of an $n\times n$ matrix $A=(a_{ij})$ is the sum of all
its diagonal elements, denoted by 
$$
\text{trace}(A)=\sumn a_{ii}.
$$

The trace operator has two important properties that can sometimes
help to simplify calculations.

\begin{proposition}\label{prop::trace-property1}
$\textup{trace}(AB)=\textup{trace}(BA)$ as long as $AB$ and $BA$
are both square matrices.
\end{proposition}

We can verify
Proposition \ref{prop::trace-property1} by definition. It states that $AB$ and $BA$ have the same trace although $AB$ differs from $BA$ in general. In fact, it is particularly useful if the dimension of $BA$
is much lower than the dimension of $AB$. For example, if both $A=(a_{1},\ldots,a_{n})^{\T}$
and $B=(b_{1},\ldots,b_{n})$ are vectors, then $\text{trace}(AB)=\text{trace}(BA)=\langle B^{\T},A \rangle =\sumn a_{i}b_{i}.$ 

\begin{proposition}\label{proposition::trace-eigenvalues}
The trace of an $n\times n$ symmetric matrix $A$ equals the sum
of its eigenvalues: $\textup{trace}(A)=\sumn\lambda_{i}$. 
\end{proposition}

\begin{myproof}{Proposition}{\ref{proposition::trace-eigenvalues}}
It follows from the eigen-decomposition and Proposition \ref{prop::trace-property1}. Let $\Lambda=\text{diag}\{\lambda_{1},\ldots,\lambda_{n}\}$. Then we have
$$
\text{trace}(A)  =\text{trace}(P\Lambda P^{\T})
  =\text{trace}(\Lambda P^{\T}P)
  =\text{trace}(\Lambda)
  =\sumn\lambda_{i}.
$$
\end{myproof}

\paragraph*{Projection matrix}

An $n\times n$ matrix $H$ is a projection matrix, if it is symmetric and $H^{2}=H.$ The eigenvalues of $H$ must be either $1$ or $0$. To see this, we assume that $H x = \lambda x$ for some nonzero vector $x$, and use two ways to calculate $H^2x$:
\begin{eqnarray*}
H^2x &=& Hx = \lambda x,\\ 
H^2x &=& H(Hx) = H(\lambda x ) = \lambda Hx =  \lambda^2 x.
\end{eqnarray*}
So $ (\lambda - \lambda^2) x = 0$ which implies that $\lambda - \lambda^2=0$, i.e., $\lambda = 0$ or $1$. 
%
So the trace of
$H$ equals its rank:
\[
\text{trace}(H)=\text{rank}(H).
\]

Why is this a reasonable definition of a ``projection matrix''? Or, why must a projection matrix
satisfy $H^{\T}=H$ and $H^{2}=H$? First, it is reasonable
to require that $Hx_{1}=x_{1}$ for any $x_{1} \in \mathcal{C}(H)$,
the column space of $H.$ Since $x_{1}=H\alpha$ for some $\alpha$,
we indeed have $Hx_{1}=H(H\alpha)=H^{2}\alpha=H\alpha=x_{1}$ because
of the property $H^{2}=H$. Second, it is reasonable to require that $x_{1}\perp x_{2}$
for any vector $x_{1}=H\alpha \in\mathcal{C}(H)$ and $x_{2}$ such that $Hx_{2}=0$.
So we need $\alpha^{\T}H^{\T}x_{2}=0$ which is true if $H=H^{\T}.$
Therefore, the two conditions are natural for the definition of a
projection matrix.

More interestingly, a project matrix has a more explicit form as stated below.

\begin{theorem}\label{thm::projection-matrix-form}
If an $n\times p$ matrix $X$ has $p$ linearly independent columns, then $H = X  (X^{\T} X)^{-1} X^{\T}$ is a projection matrix. Conversely, if an $n\times n$ matrix $H$ is a projection matrix with rank $p$, then $H = X  (X^{\T} X)^{-1} X^{\T}$ for some $n\times p$ matrix $X$ with linearly independent columns.
\end{theorem}

It is relatively easy to verify the first part of Theorem \ref{thm::projection-matrix-form}; see Chapter \ref{chapter::ols-vector}. The second part of Theorem \ref{thm::projection-matrix-form} follows from the eigen-decomposition of $H$, with the first $p$ eigen-vectors being the column vectors of $X$.

\paragraph*{Cholesky decomposition}

An $n\times n$ positive semi-definite matrix $A$ can be decomposed
as $A=LL^{\T}$ where $L$ is an $n\times n$ lower triangular matrix
with non-negative diagonal elements. 
If $A$ is positive definite,  the decomposition is unique. In general, it is not.
Take an arbitrary orthogonal
matrix $Q$, we have $A=LQQ^{\T}L^{\T}=CC^{\T}$ where $C=LQ$. So
we can decompose a positive semi-definite matrix $A$ as $A=CC^{\T}$, but
this decomposition is not unique.

\paragraph*{Singular value decomposition (SVD)}

Any $n\times m$ matrix $A$ can be decomposed as
$$
A = UDV^{\T}
$$
where $U$ is $n\times n$ orthogonal matrix, $V$ is $m\times m$ orthogonal matrix, and $D$ is $n\times m$ matrix with all zeros for the non-diagonal elements. 
For a tall matrix with $n \geq m$, the diagonal matrix $D$ has many zeros, so we can also write
$$
A = UDV^{\T}
$$
where $U$ is $n\times m$ matrix with orthonormal columns ($U^{\T} U = I_m$), $V$ is $m\times m$ orthogonal matrix, and $D$ is $m\times m$ diagonal matrix. Similarly, for a wide matrix with $n\leq m$, we can write
$$
A = UDV^{\T}
$$
where $U$ is $n\times n$ orthogonal matrix, $V$ is $m\times n$ matrix with orthonormal columns ($V^{\T} V = I_n$), and $D$ is $n\times n$ diagonal matrix. 

If $D$ has only $r \leq \min(m,n)$ nonzero elements, then we can further simplify the decomposition as
$$
A = U D  V^{\T}
$$
where $U$ is $n\times r$ matrix with orthonormal columns ($U^{\T} U = I_r$), $V$ is $m \times r$ matrix with orthonormal columns ($V^{\T} V = I_r$), and $D$ is $r\times r$ diagonal matrix. With more explicit forms of 
$$
U = (U_{1}, \ldots, U_{r}), \quad D = \text{diag}(d_1, \ldots, d_r),\quad V =  (V_{1}, \ldots, V_{r}),
$$
we can write $A$ as
$$
A = (U_{1}, \ldots, U_{r}) \begin{pmatrix}
d_1 & &  \\
 &  \ddots & \\
 & & d_r
\end{pmatrix}
\begin{pmatrix}
V_1^{\T} \\
\vdots \\
V_r^{\T}
\end{pmatrix}
= \sum_{k=1}^r  d_k U_k V_k^{\T}. 
$$

The SVD implies that
$$
AA^{\T} = UDD^{\T} U^{\T},\quad
A^{\T} A = V D^{\T}  D V^{\T} ,
$$
which are the eigen decompositions of $AA^{\T}$ and $A^{\T} A $. This ensures that $AA^{\T} $ and $A^{\T} A $ have the same non-zero eigenvalues.

An application of the SVD is to define the pseudoinverse of any matrix. Define $D^{+}$ as the pseudoinverse of $D$ with the non-zero elements inverted but the zero elements intact at zero. Define
$$
A^{+} = V D^{+} U^{\T} =  \sum_{k=1}^r d_k^{-1} V_k U_k^{\T} 
$$
as the pseudoinverse of $A$. The definition holds even if $A$ is not a square matrix. We can verify that 
$$
AA^{+} A = A, \quad 
A^{+}AA^{+} = A^{+}.
$$
If $A$ is a square nondegenerate matrix, then $A^{+} = A^{-1}$ equals the standard definition of the inverse. 
In the special case with a symmetric $A$, its SVD is identical to its eigen decomposition $A = P\text{diag}(\lambda_1, \ldots, \lambda_n) P^{\T}  $. If $A = P\text{diag}(\lambda_1, \ldots, \lambda_k, 0, \ldots, 0) P^{\T} $ is not invertible, its pseudoinverse equals
$$
A^{+} = P\text{diag}(\lambda_1^{-1}, \ldots, \lambda_k^{-1}, 0, \ldots, 0) P^{\T} 
$$
if rank$(A) = k < n$ and $\lambda_1,\lambda_1, \ldots, \lambda_k$ are the nonzero eigen-values.

Another application of the SVD is the  {\it polar decomposition} for any square matrix $A$. Since $A = UDV^{\T} = UDU^{\T} UV^{\T}$ with orthogonal $U$ and $V$, we have
\begin{equation}
\label{eq::polar-decompose}
A = (AA^{\T})^{1/2}\Gamma,
\end{equation}
where $ (AA^{\T})^{1/2}\ = UDU^{\T} $ and $\Gamma =  UV^{\T}$ is an orthogonal matrix.

\section{Vector calculus}

If $f(x)$ is a function from $\mathbb{R}^p$ to $\mathbb{R}$, then we use the notation 
$$
\frac{\partial f(x)}{\partial  x} \equiv  \begin{pmatrix}
\frac{\partial f(x)}{\partial x_1} \\
\vdots \\
 \frac{\partial f(x)}{\partial  x_p}  
\end{pmatrix}
$$
for the component-wise partial derivative, which must have the same dimension as $x$. It is often called the {\it gradient} of $f.$
For example, for a linear function $f(x) = x^{\T} a = a^{\T} x$ with $a,x\in \mathbb{R}^p$, we have
\begin{eqnarray}\label{eq::diff-linear}
\frac{\partial a^{\T} x  }{\partial  x} = \begin{pmatrix}
\frac{\partial a^{\T} x}{\partial x_1} \\
\vdots \\
 \frac{\partial a^{\T} x}{\partial  x_p}  
\end{pmatrix}
= \begin{pmatrix}
 \frac{\partial \sum_{j=1}^p  a_j x_j }{\partial x_1} \\
\vdots \\
 \frac{\partial \sum_{j=1}^p  a_j x_j }{\partial  x_p}  
\end{pmatrix}
= \begin{pmatrix}
a_1 \\
\vdots \\
a_p
\end{pmatrix}
= a;
\end{eqnarray}
for a quadratic function $f(x) = x^{\T} A x $ with a symmetric $A\in \mathbb{R}^{p\times p}$  and $x \in \mathbb{R}^p$, we have 
$$
\frac{ \partial  x^{\T} A x }{ \partial  x} = \begin{pmatrix}
\frac{\partial x^{\T} A x}{\partial x_1} \\
\vdots \\
 \frac{\partial x^{\T} A x}{\partial  x_p}  
\end{pmatrix}
= \begin{pmatrix}
\frac{\partial   \sum_{i=1}^p \sum_{j=1}^p a_{ij} x_i x_j }{\partial x_1} \\
\vdots \\
 \frac{\partial  \sum_{i=1}^p \sum_{j=1}^p a_{ij} x_i x_j   }{\partial  x_p}  
\end{pmatrix} 
= \begin{pmatrix}
2 a_{11} x_1 + \cdots + 2a_{1p} x_p \\
\vdots \\
2 a_{p1} x_1 + \cdots + 2a_{pp} x_p
\end{pmatrix} 
=2 Ax .
$$
These are two important rules of vector calculus used in this book, summarized below.

\begin{proposition}\label{prop::vector-calculus}
We have
\begin{eqnarray*}
\frac{\partial a^{\T} x  }{\partial  x}  &=& a, \\
\frac{ \partial  x^{\T} A x }{ \partial  x} &=& 2 Ax . 
\end{eqnarray*}
\end{proposition}

We can also extend the definition to vector functions. If $f(x) = (f_1(x), \ldots, f_q(x))^{\T}$ is a function from $\mathbb{R}^p$ to $\mathbb{R}^q$, then we use the notation
\begin{eqnarray}
\label{eq::diff-vector-veector}
\frac{\partial f(x)}{\partial  x}  \equiv  \left( \frac{\partial f_1(x)}{\partial  x}  ,\cdots, \frac{\partial f_q(x)}{\partial  x}  \right)
=\begin{pmatrix}
 \frac{\partial f_1(x)}{\partial  x_1} & \cdots &  \frac{\partial f_q(x)}{\partial  x_1} \\
 \vdots & & \vdots \\
  \frac{\partial f_1(x)}{\partial  x_p} & \cdots &  \frac{\partial f_q(x)}{\partial  x_p} 
\end{pmatrix},
\end{eqnarray} 
which is a $p\times q$ matrix with rows corresponding to the elements of $x$ and the columns corresponding to the elements of $f(x)$. 
We can easily extend the first result of Proposition \ref{prop::vector-calculus}.

\begin{proposition}\label{prop::vector-calculus-matrix}
For $B\in \mathbb{R}^{p \times q}$ and $x\in \mathbb{R}^p$, we have 
$$
\frac{\partial B^{\T} x  }{\partial  x}  = B. 
$$
\end{proposition}

 \begin{myproof}{Proposition}{\ref{prop::vector-calculus-matrix}}
 Partition $B= (B_1, \ldots, B_q)$ in terms of its column vectors. 
The $j$th element of $B^{\T}x$ is $B_j^{\T}x$ so the $j$-th column of $ \partial B^{\T}x / \partial x$ is $B_j$ based on Proposition \ref{prop::vector-calculus}. This verifies that $ \partial B^{\T}x / \partial x$ equals $B$. 
 \end{myproof}

Some authors define $\partial f(x) / \partial  x$ as the transpose of \eqref{eq::diff-vector-veector}. I adopt this form for its natural  connection with 
\eqref{eq::diff-linear} when $q=1$. 
Sometimes, it is indeed more convenient to work with the transpose of $\partial f(x) / \partial  x$. Then I will use the notation
$$
\frac{\partial f(x)}{\partial  x^{\T}}  = \left( \frac{\partial f(x)}{\partial  x}   \right)^{\T} 
=  \left( \frac{\partial f(x)}{\partial  x_1}  ,\cdots, \frac{\partial f(x)}{\partial  x_p}  \right), 
$$
which puts the transpose notation on $x$.

The above formulas become more powerful in conjunction with the chain rule. For example, for any differentiable function $h(z)$ mapping from $\mathbb{R} $ to $ \mathbb{R}$ with derivative $h'(z)$, we have
\begin{eqnarray*}
\frac{ \partial h (a^{\T} x)  }{\partial  x}  &=& h'(a^{\T} x)  a,\\ 
\frac{ \partial  h( x^{\T} A x) }{ \partial  x} &=& 2h'( x^{\T} A x) Ax.
\end{eqnarray*}
For any differentiable function $h(z)$ mapping from $ \mathbb{R}^q $ to $ \mathbb{R}$ with gradient 
$\partial h(z) / \partial z$, we have
\begin{eqnarray*}
\frac{\partial h(B^{\T} x)  }{\partial  x} 
&=& \frac{\partial h(B_1^{\T} x, \ldots, B_q^{\T}x)  }{\partial  x} \\
&=& \sum_{j=1}^q \frac{\partial h(B_1^{\T} x, \ldots, B_q^{\T}x)  }{\partial  z_j} B_j \\
&=&  B \frac{ \partial h(B^{\T} x) }{  \partial z } . 
\end{eqnarray*}

Moreover, we can also define the Hessian matrix of a function $f(x)$ mapping from $\mathbb{R}^p$ to $\mathbb{R}$:
$$
\frac{  \partial^2 f(x) }{ \partial x \partial x^{\T} } = \left( \frac{  \partial^2 f(x) }{ \partial x_i \partial x_j} \right)_{1\leq i, j \leq p}
= \frac{ \partial }{ \partial x^{\T} } \left(  \frac{  \partial f(x) }{ \partial x }  \right). 
$$

\section{Homework problems}

\paragraph{Triangle inequality of the inner product}\label{hwmath1::triangle-inner-product}

With three unit vectors $u, v, w  \in \mathbb{R}^n$, prove that
$$
\sqrt{ 1 -  \langle  u, w  \rangle } \leq \sqrt{  1 -   \langle  u, v  \rangle } + \sqrt{  1 -   \langle   v, w  \rangle }.
$$

Remark: The result is a direct consequence of the standard triangle inequality but it has an interesting implication. If $ \langle  u, v  \rangle  \geq 1-\epsilon$ and  $  \langle   v, w  \rangle \geq 1 - \epsilon$, then $\langle  u, w  \rangle \geq 1-4\epsilon$. This implied inequality is mostly interesting when $\epsilon $ is small. It states that when $u$ and $v$ are highly correlated and $v$ and $w$ are highly correlated, then $u$ and $w$ must also be highly correlated. Note that we can find counterexamples for the following relationship:
$$
\langle  u, v  \rangle > 0, \quad  \langle   v, w  \rangle  > 0 \quad   \text{ but }\quad   \langle  u, w  \rangle = 0. 
$$

\paragraph{Van der Corput inequality}
\label{hwmath1::vandercorputineq}

Assume that $v, u_1, \ldots, u_m \in \mathbb{R}^n$ have unit length. Prove  that 
$$
\left(  \sum_{i=1}^m   \langle v, u_i \rangle    \right)^2 \leq \sum_{i=1}^m \sum_{j=1}^m \langle u_i, u_j \rangle. 
$$

Remark: This result is not too difficult to prove, but it says something fundamentally interesting. If $v$ is correlated with many vectors $u_1, \ldots, u_m$, then at least some vectors in  $u_1, \ldots, u_m$ must be correlated.

\paragraph{Inverse of a block matrix}\label{hwmath1::inverse-block-matrix}

Prove that 
\begin{align*}
&\left(\begin{array}{cc}
A & B\\
C & D
\end{array}\right)^{-1} \\
& =\left(\begin{array}{cc}
A^{-1}+A^{-1}B(D-CA^{-1}B)^{-1}CA^{-1} & -A^{-1}B(D-CA^{-1}B)^{-1}\\
-(D-CA^{-1}B)^{-1}CA^{-1} & (D-CA^{-1}B)^{-1}
\end{array}\right)\\
 & =\left(\begin{array}{cc}
(A-BD^{-1}C)^{-1} & -(A-BD^{-1}C)^{-1}BD^{-1}\\
-D^{-1}C (A-BD^{-1}C)^{-1} & D^{-1}+D^{-1}C(A-BD^{-1}C)^{-1}BD^{-1}
\end{array}\right),
\end{align*}
provided that all the inverses of the matrices exist. The two forms of the inverse imply the Woodbury formula:
\[
(A-BD^{-1}C)^{-1}=A^{-1}+A^{-1}B(D-CA^{-1}B)^{-1}CA^{-1},
\]
which further implies the Sherman--Morrison formula:
\[
(A+uv^{\T})^{-1}=A^{-1}-(1+v^{\T}A^{-1}u)^{-1}A^{-1}uv^{\T}A^{-1},
\]
where $A$ is an invertible square matrix,  and $u$ and $v$ are two column vectors.

\paragraph{Matrix determinant lemma}\label{hwmath1::matrix-determinat-matrix}

Prove that  given the identity matrix $I_n$ and two $n$-vectors $u$ and $v$, we have
$$
\text{det}(I_n + uv^{\T}) = 1 + v^{\T} u. 
$$
Further prove that if $I_n$ is replaced by an $n\times n$ invertible matrix $A$, we have
$$
\text{det}(A + uv^{\T}) = (1 +  v^{\T} A^{-1} u)  \cdot  \text{det}(A). 
$$

\paragraph{Symmetric rank one update of the identity matrix}\label{hwmath1::symmetric-rank-one-update}

Given a real number $c$ and a vector $x\in \mathbb{R}^n$, consider the matrix $ I_n + c xx^{\T}$, which is a symmetric $n\times n$ matrix. Assume $1 + c \| x\|^2  > 0$.

\begin{enumerate}
\item
Find the eigenvalues and eigenvectors of $I_n + c xx^{\T}$.
\item
Use the eigendecomposition to prove that 
$$
\text{det}( I_n + c xx^{\T} ) 
= 1 + c \| x\|^2.
$$

Remark: You can compare this result with Problem \ref{hwmath1::matrix-determinat-matrix}. 

\item
Use the eigendecomposition to prove that
$$
(I_n + c xx^{\T})^{-1} 
=  I_n  - \frac{ c }{ 1 + c \| x\|^2  } xx^{\T}.
$$

Remark: You can compare this result with the Sherman--Morrison formula in Problem \ref{hwmath1::inverse-block-matrix}.

\item
Use the eigendecomposition to prove that
$$
(I_n + c xx^{\T})^{1/2} 
=  I_n  + \frac{ c  }{  1+ \sqrt{ 1 + c\| x\|^2}  } xx^{\T}.
$$

\item
Use the eigendecomposition to prove that
$$
(I_n + c xx^{\T})^{ - 1/2} 
=  I_n  - \frac{ c }{ \sqrt{1 + c \| x\|^2} (1+\sqrt{1 + c \| x\|^2})  } xx^{\T}. 
$$

\end{enumerate}

\paragraph{Rank one update and positive definiteness}\label{hwmath::rank1psd}

Assume $A$ is an $n\times n$ positive definite matrix. Assume $b$ is an $n$-dimensional vector. 

Prove that $A - bb^{\T}$ is positive definite if and only if $b^{\T} A^{-1} b <1$, and  $A - bb^{\T}$ is positive semi-definite if and only if $b^{\T} A^{-1} b \leq 1$.

Remark: \citet[][Appendix]{farebrother1976further} gave this result. It is not directly used in this book but is related to the leave-one-out formula in Theorem \ref{thm::leave-one-out-beta}.

\paragraph{Positive definiteness of the difference of inverses}\label{hwmath::psd-diff-inverse}

With scalars $a \geq  b > 0$, we know that $a^{-1} \leq b^{-1}$. A similar result hold for matrices.

Assume $A$ and $B$ are positive definite matrices. First, prove that if $A - I$ is positive definite, then $ I-  A^{-1}$ is positive definite; if $A - I$ is positive semi-definite, then $ I-  A^{-1}$ is positive semi-definite.
Second, prove that if $A - B$ is positive definite, then $ B^{-1} -  A^{-1}$ is positive definite; if $A - B$ is positive semi-definite, then $ B^{-1} -  A^{-1}$ is positive semi-definite.

\paragraph{Decomposition of a positive semi-definite matrix}\label{hwmath::decompose-psd}

Prove that if $A$ is positive semi-definite, then there exists a matrix $C$ such that $A = CC^{\T}$.

\paragraph{Trace of the product of two matrices}
\label{hwmath1::trace-product}
Prove that if $A$ and $B$ are two $n\times n$ positive semi-definite matrices, then trace$(AB) \geq 0.$

Remark: 
Use the eigen-decomposition of $A = \sum_{i=1}^n \lambda_i \gamma_i \gamma_i^{\T}$ to prove the result. 

In fact, a stronger result holds. If two $n\times n$ symmetric matrices $A$ and $B$ have eigenvalues
$$
\lambda_1 \geq \cdots \geq \lambda_n, \quad
\mu_1 \geq \cdots \geq \mu_n
$$
respectively, then 
$$
\sumn \lambda_i \mu_{n+1-i} \leq 
\textup{trace}(AB) \leq 
\sumn \lambda_i \mu_i. 
$$
The result is
due to \citet{von1937some} and \citet{ruhe1970perturbation}. See also \citet[][Lemma 4.12]{chen2019model}.

\paragraph{Trace of the product of two matrices and positive semi-definiteness}
\label{hwmath1::trace-product-psd}

This problem gives the other direction of Problem \ref{hwmath1::trace-product}.

Assume $A$ is a symmetric matrix. Prove that $A$ is positive semi-definite if and only if trace$(AB) \geq 0$ for all positive semi-definite matrices $B$.

Remark: One direction of this statement is in Problem \ref{hwmath1::trace-product}. We only need to prove the other direction. \citet{theobald1974generalizations} used it to analyze ridge regression.

\paragraph{Vector calculus}\label{hwmath1::vector-calculus-asymmetric}
What is the formula for $ \partial  x^{\T} A x / \partial  x$ if $A$ is not symmetric in Proposition 
\ref{prop::vector-calculus}?

\chapter{Random Variables}\label{chapter:appendix-rvs}

This Appendix reviews the basics of random variables.   
Let ``IID'' denote ``independent and identically distributed'', ``$\iidsim$'' denote a sequence of random variables that are IID with some common distribution, and ``$\ind$'' denote independence between random variables.

Define Euler's Gamma function as 
$$
\Gamma(z)=\int_{0}^{\infty}x^{z-1}e^{-x}\d x ,\qquad (z>0),
$$
which is a natural extension of the factorial since $\Gamma(n) = (n-1)!$. 
Further, define
the digamma function  as  $\psi(z) = \diff \log  \Gamma (z) / \diff z$ and the trigamma function $\psi'(z)$ as the derivative of $\psi(z)$.  
In \ri{R}, we can use 
\begin{rc}
gamma(z)
lgamma(z)
digamma(z)
trigamma(z)
 \end{rc}
 to compute $\Gamma(z)$, $\log \Gamma(z)$, $\psi(z)$, and $\psi'(z)$.

\section{Some important univariate random variables}

\subsection{Normal, chi-squared, t and F}

The standard Normal random variable $Z\sim\N(0,1)$ has density 
\[
f(z)=(2\pi)^{-1/2}\exp\left(-z^{2}/2\right).
\]
A Normal random variable $X$ has mean $\mu$ and variance $\sigma^{2}$, denoted by $\N(\mu,\sigma^2)$, 
if $X=\mu+\sigma Z$. We can show that $X$ has density 
\[
f(x)=(2\pi)^{1/2}\exp\left\{ -(x-\mu)^{2}/(2\sigma^{2})\right\} .
\]

A chi-squared random variable with degrees of freedom $n,$ denoted
by $Q_{n}\sim\chi_{n}^{2},$ can be represented as 
$$
Q_{n}=\sumn Z_{i}^{2} , 
$$
where $Z_{i} \iidsim \N(0,1)$. We can show that $Q_{n}$ has density 
\begin{equation}
f_{n}(q)=q^{n/2 - 1}\exp(-q/2)\Big/\left\{ 2^{n/2}\Gamma(n/2)\right\} ,\qquad(q>0).\label{eq:chisqPDF}
\end{equation}
We can verify that the above density \eqref{eq:chisqPDF} is well-defined even if we change
the integer $n$ to be an arbitrary positive real number $\nu$, and
call the corresponding random variable $Q_{\nu}$ a chi-squared random
variable with degrees of freedom $\nu$, denoted by $Q_\nu\sim \chi^2_\nu$.

A $t$ random variable with degrees of freedom $\nu$ can be represented
as 
$$
t_{\nu}=\frac{ Z}{ \sqrt{Q_{\nu}  / \nu }    } 
$$ 
where $Z\sim\N(0,1),Q_{\nu}\sim\chi_{\nu}^{2}$, 
and $Z\ind Q_{\nu}.$ 

An $F$ random variable with degrees of freedom $(r,s)$ can be represented
as
\[
F=\frac{Q_{r}/r}{Q_{s}/s}
\]
where $Q_{r}\sim\chi_{r}^{2},Q_{s}\sim\chi_{s}^{2}$, and $Q_{r}\ind Q_{s}$. 

\subsection{Beta--Gamma duality}

The $\text{Gamma}(\alpha,\beta)$ random variable with parameters
$\alpha,\beta>0$ has density
\begin{equation}
f(x)=\frac{\beta^{\alpha}}{\Gamma(\alpha)}x^{\alpha-1}e^{-\beta x},\quad(x>0).\label{eq:gammaPDF}
\end{equation}
The $\text{Beta}(\alpha,\beta)$ random variable with parameters $\alpha,\beta>0$
has density
\[
f(x)=\frac{\Gamma(\alpha+\beta)}{\Gamma(\alpha)\Gamma(\beta)}x^{\alpha-1}(1-x)^{\beta-1},\quad(0<x<1).
\]
These two random variables are closely related as shown in
Theorem \ref{thm:beta-gamma-duality} below.

\begin{theorem}
[Beta--Gamma duality]
\label{thm:beta-gamma-duality}If $X\sim\textup{Gamma}(\alpha,\theta),Y\sim\textup{Gamma}(\beta,\theta)$
and $X\ind Y$, then
\begin{enumerate}
\item $X+Y\sim\textup{Gamma}(\alpha+\beta,\theta),$
\item $X/(X+Y)\sim\textup{Beta}(\alpha,\beta),$
\item $X+Y\ind X/(X+Y).$
\end{enumerate}
\end{theorem}

Another simple but useful fact is that $\chi^{2}$ is a special Gamma
random variable. Comparing the densities in (\ref{eq:chisqPDF}) and
(\ref{eq:gammaPDF}), we obtain the following result.
\begin{proposition}\label{prop::chi2-gamma}
$\chi_{n}^{2}\sim\textup{Gamma}(n/2,1/2).$
\end{proposition}
We can also calculate the moments of the Gamma and Beta distributions.
\begin{proposition}
\label{thm:gammamoments}If $X\sim\textup{Gamma}(\alpha,\beta)$,
then
\begin{eqnarray*}
E(X) &=&\frac{\alpha}{\beta},\\ 
\var(X) &=&\frac{\alpha}{\beta^{2}}.
\end{eqnarray*}
\end{proposition}

\begin{proposition}
\label{thm:gammamoments-log}If $X\sim\textup{Gamma}(\alpha,\beta)$,
then
\begin{eqnarray*}
E(\log X)  &=&  \psi(\alpha) - \log \beta ,\\ 
\var(\log X) &=& \psi'(\alpha). 
\end{eqnarray*}
\end{proposition}

\begin{proposition}
\label{thm:beta-moments}If $X\sim\textup{Beta}(\alpha,\beta),$ then
\begin{eqnarray*}
E(X)  &=&  \frac{\alpha}{\alpha+\beta},\\
\var(X) &=& \frac{\alpha\beta}{(\alpha+\beta)^2(\alpha+\beta+1)}.
\end{eqnarray*}
\end{proposition}

\begin{proposition}
\label{thm:beta-moments-log}If $X\sim\textup{Beta}(\alpha,\beta),$ then
\begin{eqnarray*}
E(\log X)  &=& \psi(\alpha) - \psi(\alpha + \beta) ,\\ 
\var(\log X) &=& \psi ' (\alpha) - \psi ' (\alpha + \beta) . 
\end{eqnarray*}
\end{proposition}

I leave the proofs of the above propositions to Problem \ref{hwmath2::beta-gamma-moments}.

\subsection{Exponential, Laplace, and Gumbel distributions}
\label{subsec::expo-gumbel}

An Exponential$(\lambda)$ random variable $X \geq 0$ has density $f(x) = \lambda e^{-\lambda x}$, mean $1/\lambda$, median $\log 2/\lambda $ and variance $1/\lambda^2$.  The standard Exponential random variable $X_0$ has $\lambda = 1$, and $X_0 / \lambda$ generates Exponential$(\lambda)$.

An important feature of Exponential$(\lambda)$ is the memoryless property.

\begin{proposition}
[memoryless property of Exponential]
\label{prop::memoryless}
If $X\sim \textup{Exponential}(\lambda)$, then 
$$
\pr(X \geq x+c \mid X\geq c) =\pr(X \geq x).
$$
\end{proposition}

Proposition \ref{prop::memoryless} states that if $X$ represents the survival time, then the probability of surviving another $x$ time is always the same no matter how long the existing survival time is.  I leave the proof of Proposition \ref{prop::memoryless} to Problem \ref{hwmath2::memoryless-exponential}.

The minimum of independent exponential random variables also follows an exponential distribution. 

\begin{proposition}
\label{prop::minimum-expo}
Assume that $X_{i}\sim\textup{Exponential}(\lambda_{i})$ are independent $(i=1,\ldots,n)$. Then
$$
\underline{X} = \min(X_{1},\ldots,X_{n} ) \sim  \textup{Exponential}(  \lambda_{1}+\cdots+\lambda_{n}  )
$$
and
\[
\pr (X_{i}=\underline{X}  ) =\frac{\lambda_{i}}{\lambda_{1}+\cdots+\lambda_{n}}.
\]
\end{proposition}

I leave the proof of Proposition \ref{prop::minimum-expo} to Problem \ref{hwmath2::min-exponential}. Theorem \ref{thm::freedman-explain-cox} states a more general result without assuming the Exponential distribution.

The difference between two IID exponential random variables follows the Laplace distribution. 

\begin{proposition}
\label{prop::expo-laplace}
If $y_1$ and $y_2$ are two IID Exponential$(\lambda)$, then $y = y_1 - y_2$ has  density 
$$
 \frac{\lambda}{2}  \exp( - \lambda | c| ),\quad -\infty < c<\infty
$$ 
which is the density of a Laplace distribution with mean $0$ and variance $2/\lambda^2$. 
\end{proposition}

I leave the proof of Proposition \ref{prop::expo-laplace} to Problem \ref{hwmath2::laplace-diff-exponential}.

If $X_0 $ is the standard exponential random variable, then we define the Gumbel$(\mu, \beta)$ random variable as 
$$
Y = \mu - \beta \log X_0.
$$
The standard Gumbel distribution has $\mu =0$ and $\beta  =1$, with cumulative distribution function (CDF)
\[
F(y)=\exp(-e^{-y}),\quad y\in\mathbb{R}
\]
and density
\[
f(y)=\exp(-e^{-y})e^{-y},\quad y\in\mathbb{R}.
\]

By definition and Proposition \ref{prop::minimum-expo}, we can verify that the maximum of IID Gumbels is also Gumbel. 

\begin{proposition}
\label{prop::max-gumbel}
If $Y_1, \ldots, Y_n$ are IID $\textup{Gumbel}(\mu,\beta)$, then
$$
\max_{1\leq i \leq n}  Y_i \sim \textup{Gumbel}(\mu + \beta \log n, \beta ).
$$
If $Y_1, \ldots, Y_n$ are independent $\textup{Gumbel}(\mu_i, 1)$, then
$$
\max_{1\leq i \leq n}  Y_i \sim \textup{Gumbel}\left(\log \sumn e^{\mu_i}, 1 \right). 
$$
\end{proposition}

I leave the proof to Problem \ref{hwmath2::max-gumbel}.

\section{Multivariate distributions}

A random vector $(X_{1},\ldots,X_{n})^{\T}$ is a vector consisting of $n$ random
variables. If all components are continuous, we can define its joint
density $f_{X_{1}\cdots X_{n}}(x_{1},\ldots,x_{n}).$ 

For a random vector $\binom{X}{Y}$ with $X$ and $Y$ possibly being
vectors, if it has joint density $f_{XY}(x,y)$, then we can obtain
the marginal distribution of $X$   
$$
f_{X}(x)=\int f_{XY}(x,y)\d y
$$
and define the conditional density  
\[
f_{Y|X}(y\mid x)=\frac{f_{XY}(x,y)}{f_{X}(x)}\qquad \text{if }  f_{X}(x)\neq 0. 
\]
Based on the conditional density, we can define the conditional expectation
of any function of $Y$  as 
\[
E\left\{ g(Y)\mid X=x\right\} =\int g(y)f_{Y|X}(y\mid x)\d y
\]
and the conditional variance as 
\[
\var\left\{ g(Y)\mid X=x\right\} =E\left[\left\{ g(Y)\right\} ^{2}\mid X=x\right]-\left[E\left\{ g(Y)\mid X=x\right\} \right]^{2}.
\]
 In the above definitions, the conditional mean and variance are both deterministic
functions of $x$. We can replace $x$ by the random variable $X$ to define 
$E\left\{ g(Y)\mid X\right\} $
and $\var\left\{ g(Y)\mid X\right\} $, which are functions of the
random variable $X$ and are thus random variables. 

Below are two important laws of conditional expectation and variance. 

\begin{theorem}
[Law of total expectation] We have 
\[
E( Y ) =E\left\{  E ( Y \mid X ) \right\}  . 
\]
\end{theorem}

\begin{theorem}
[Law of total variance or analysis of variance] 
\label{theorem::law-total-var}
We have 
\[
\var (Y) =E\left\{  \var( Y \mid X) \right\}  +\var\left\{  E(Y \mid X) \right\}  . 
\]
\end{theorem}

\paragraph*{Independence}

Random variables $(X_{1},\ldots,X_{n})$ are mutually independent
if 
$$
f_{X_{1}\cdots X_{n}}(x_{1},\ldots,x_{n})=f_{X_{1}}(x_{1})\cdots f_{X_{n}}(x_{n}).
$$
In the definition of independence, each of $(X_{1},\ldots,X_{n})$ can
be vectors. We have the following rules under independence.

\begin{proposition}
 If $X\ind Y$, then $h(X)\ind g(Y)$ for any functions
$h(\cdot)$ and $g(\cdot)$.
\end{proposition}

\begin{proposition}
If $X\ind Y$, then
\begin{align*}
f_{XY}(x,y) & =f_{X}(x)f_{Y}(y),\\
f_{Y\mid X}(y\mid x) & =f_{Y}(y),\\
E\left\{ g(Y)\mid X\right\}  & =E\left\{ g(Y)\right\} ,\\
E\left\{ g(Y)h(X)\right\}  & =E\left\{ g(Y)\right\} E\left\{ h(X)\right\} .
\end{align*}
\end{proposition}

\paragraph*{Expectations of random vectors or random matrices}

For a random matrix $W=(w_{ij})$, we define $E(W)=(E(w_{ij}))$.
For constant matrices $A$ and $C$, we can verify that
\begin{align*}
E(AW+C) & =AE(W)+C,\\
E(AWC) & =AE(W)C.
\end{align*}

\paragraph*{Covariance between two random vectors}

If $W\in\mathbb{R}^{r}$ and $Y\in\mathbb{R}^{s},$ then their covariance 
$$
\cov(W,Y)=E\left[\left\{ W-E(W)\right\} \left\{ Y-E(Y)\right\} ^{\T}\right]
$$
is an $r\times s$ matrix. 
As a special case, 
$$
\cov(Y)=\cov(Y,Y)=E\left[\left\{ Y-E(Y)\right\} \left\{ Y-E(Y)\right\} ^{\T}\right]
= E(YY^{\T}) - E(Y) E(Y)^{\T}. 
$$
For a scalar random variable, $\cov(Y)=\var(Y).$

\begin{proposition}\label{proposition::covariance-lineartrans}
For $A\in \mathbb{R}^{r\times n},Y\in\mathbb{R}^{n}$ and
$C\in\mathbb{R}^{r}$, we have 
$$
\cov(AY+C)=A\cov(Y)A^{\T} .
$$
\end{proposition}

Using Proposition \ref{proposition::covariance-lineartrans},
we can verify that for any $n$-dimensional random vector, $\cov(Y)\succeq0$
because for all $x\in\mathbb{R}^{n}$, we have 
\[
x^{\T}\cov(Y)x=\cov(x^{\T}Y)=\var(x^{\T}Y)\geq0.
\]

\begin{proposition}
For two random vectors $W$ and $Y$, we have
\[
\cov(AW+C,BY+D)=A\cov(W,Y)B^{\T}
\]
and 
\[
\cov(AW+BY)=A\cov(W)A^{\T}+B\cov(Y)B^{\T}+A\cov(W,Y)B^{\T}+B\cov(Y,W)A^{\T}.
\]
\end{proposition}

Similar to Theorem \ref{theorem::law-total-var}, we have the following decomposition of the covariance. 

\begin{theorem}
[Law of total covariance] 
\label{theorem::law-total-cov}
We have 
\[
\cov\left(   Y, W    \right)  =E\left\{  \cov\left( Y ,W\mid X\right) \right\} +\cov\left\{ E(Y\mid X) , E(W\mid X) \right\}. 
\]
\end{theorem}

\section{Multivariate Normal and its properties}

I use a generative definition of the multivariate Normal random vector.
First, $Z$ is a standard Normal random vector if $Z=(Z_{1},\ldots,Z_{n})^{\T}$
has components $Z_{i}\iidsim \N(0,1)$. Given a mean vector $\mu$ and a
positive semi-definite covariance matrix $\Sigma$, define a Normal
random vector $Y\sim\text{N(\ensuremath{\mu,\Sigma})}$ with mean
$\mu$ and covariance $\Sigma$ if $Y$ can be represented as  
\begin{eqnarray}
Y=\mu+AZ, \label{eq::mvn-definitions}
\end{eqnarray}
where $A$ satisfies $\Sigma=AA^{\T}$. We can verify that $\cov(Y)=\Sigma$,
so indeed $\Sigma$ is its covariance matrix. If $\Sigma\succ0$,
then we can verify that $Y$ has density
\begin{equation}\label{equation::normal-density}
f_{Y}(y)=(2\pi)^{-n/2}\left\{ \det(\Sigma)\right\} ^{-1/2}\exp\left\{ -(y-\mu)^{\T}\Sigma^{-1}(y-\mu)/2\right\} .
\end{equation}

We can easily verify the following result by calculating the density.

\begin{proposition}\label{prop::rotation-mvn}
If $Z\sim\textup{N}(0,I_{n})$ and $\Gamma$ is an orthogonal matrix, then
$\Gamma Z\sim \N(0,I_{n}).$
\end{proposition}

I do not define multivariate Normal based on the density \eqref{equation::normal-density} because it is only well defined with a positive definite $\Sigma$. I do not define multivariate Normal based on the characteristic function because it is more advanced than the level of this book.
Definition \eqref{eq::mvn-definitions} does not require $\Sigma$ to be positive definite and is more elementary. 
However, it has a subtle issue of uniqueness. Although
the decomposition $\Sigma=AA^{\T}$ is not unique, the resulting distribution
$Y=\mu+AZ$ is. We can verify this using the Polar decomposition in \eqref{eq::polar-decompose}. 
Because $A=\Sigma^{1/2}\Gamma$ where $\Gamma$ is an orthogonal matrix,
we have $Y=\mu+\Sigma^{1/2}\Gamma Z=\mu+\Sigma^{1/2}\tilde{Z}$ where
$\tilde{Z}=\Gamma Z$ is a standard Normal random vector by Proposition \ref{prop::rotation-mvn}.
Importantly, although the definition \eqref{eq::mvn-definitions} can be general, we usually use the following representation
$$
Y=\mu+ \Sigma^{1/2} Z.
$$

\begin{theorem}
\label{thm:Normal-ind-uncorr}Assume that
\[
\left(\begin{array}{c}
Y_{1}\\
Y_{2}
\end{array}\right)\sim\textup{N}\left(\left(\begin{array}{c}
\mu_{1}\\
\mu_{2}
\end{array}\right),\left(\begin{array}{cc}
\Sigma_{11} & \Sigma_{12}\\
\Sigma_{21} & \Sigma_{22}
\end{array}\right)\right).
\]
Then $Y_{1}\ind Y_{2}$ if and only if $\Sigma_{12}=0.$
\end{theorem}
I leave the proof of Theorem \ref{thm:Normal-ind-uncorr} as Problem \ref{hwmath2::independence-uncorr-normal}.

\begin{proposition}\label{prop::linear-normal}
If $Y\sim\textup{N(\ensuremath{\mu,}\ensuremath{\Sigma})},$ then $BY+C\sim\textup{N\ensuremath{(B\mu+C,B\Sigma B^{\T}})}$,
that is, any linear transformation of a Normal random vector is also
a Normal random vector.
\end{proposition}

I leave the proof of Proposition \ref{prop::linear-normal} as Problem \ref{hwmath2::linear-normal}.

An obvious corollary of Proposition \ref{prop::linear-normal} is that if $X_1 \sim \N(\mu_1, \sigma_1^2)$ and $X_2 \sim \N(\mu_2, \sigma_2^2)$ are independent, then $X_1+X_2 \sim \N(  \mu_1 + \mu_2,  \sigma_1^2 + \sigma_2^2 )$. So the summation of two independent Normals is also Normal. Remarkably, the reverse of the result is also true. 

\begin{theorem}[Levy--Cramer]
\label{thm::levy-cramer}
 If $X_1 \ind X_2$ and $X_1 + X_2$ is Normal, then both $X_1$ and $X_2$ must be Normal.
\end{theorem}

The statement of Theorem \ref{thm::levy-cramer} is extremely simple. But its proof is non-trivial and beyond the scope of this book. See \citet{benhamou2018three} for a proof.

\begin{theorem}
Assume 
\[
\left(\begin{array}{c}
Y_{1}\\
Y_{2}
\end{array}\right)\sim\textup{N}\left(\left(\begin{array}{c}
\mu_{1}\\
\mu_{2}
\end{array}\right),\left(\begin{array}{cc}
\Sigma_{11} & \Sigma_{12}\\
\Sigma_{21} & \Sigma_{22}
\end{array}\right)\right).
\]
\begin{enumerate}
\item The marginal distributions are Normal:
\begin{align*}
Y_{1} & \sim\textup{N}\left(\mu_{1},\Sigma_{11}\right),\\
Y_{2} & \sim\textup{N}\left(\mu_{2},\Sigma_{22}\right).
\end{align*}
\item If $\Sigma_{22}\succ0,$ then the conditional distribution is Normal:
\[
Y_{1}\mid Y_{2}=y_{2}\sim\textup{N}\left(\mu_{1}+\Sigma_{12}\Sigma_{22}^{-1}(y_{2}-\mu_{2}),\Sigma_{11}-\Sigma_{12}\Sigma_{22}^{-1}\Sigma_{21}\right);
\]
$Y_{2}$ is independent of the residual 
\[
Y_{1}-\Sigma_{12}\Sigma_{22}^{-1}(Y_{2}-\mu_{2})\sim\textup{N}\left(\mu_{1},\Sigma_{11}-\Sigma_{12}\Sigma_{22}^{-1}\Sigma_{21}\right).
\]
\end{enumerate}
\end{theorem}

I review some other results of the multivariate Normal below. 
 
\begin{proposition}
Assume $Y\sim\textup{N}(\mu,\sigma^{2}I_{n})$. If $AB^{\T}=0$, then
$AY\ind BY$.
\end{proposition}
\begin{proposition}
Assume 
\[
\left(\begin{array}{c}
Y_{1}\\
Y_{2}
\end{array}\right)\sim\textup{N}\left(\left(\begin{array}{c}
\mu_{1}\\
\mu_{2}
\end{array}\right),\left(\begin{array}{cc}
\sigma_{1}^{2} & \rho\sigma_{1}\sigma_{2}\\
\rho\sigma_{1}\sigma_{2} & \sigma_{2}^{2}
\end{array}\right)\right),
\]
where $\rho$ is the correlation coefficient defined as
\[
\rho=\frac{\cov(Y_{1},Y_{2})}{\sqrt{\var(Y_{1})\var(Y_{2})}}.
\]
Then the conditional distribution is
\[
Y_{1}\mid Y_{2}=y_{2}\sim\textup{N}\left(\mu_{1}+\rho\frac{\sigma_{1}}{\sigma_{2}}(y_{2}-\mu_{2}),\sigma_{1}^{2}(1-\rho^{2})\right).
\]
\end{proposition}

\section{Quadratic forms of random vectors}

Given a random vector $Y$ and a symmetric matrix $A$, we can define
the quadratic form $Y^{\T}AY$, which is a random variable playing
an important role in statistics. The first theorem is about its mean.

\begin{theorem}\label{thm::mean-quadratic-form}
If $Y$ has mean $\mu$ and covariance $\Sigma$, then 
\[
E(Y^{\T}AY)=\textup{trace}(A\Sigma)+\mu^{\T}A\mu.
\]
\end{theorem}

The proof below uses the following three basic facts.
\begin{enumerate}[label=(\textup{F}\arabic*), ref=\textup{F}\arabic*]
\item\label{item::fact1} $E(YY^{\T})=\cov(Y)+E(Y)E(Y^{\T})=\Sigma+\mu\mu^{\T}.$
\item\label{item::fact2} For an $n\times n$ symmetric random matrix $W=(w_{ij})$, we have $E\left\{ \text{trace}(W)\right\} =\text{trace}\left\{ E(W)\right\} $ because $E\left(\sumn w_{ii}\right)=\sumn E(w_{ii}) .$
\item\label{item::fact3}
If $BC$ and $CB$ are both well defined, then $\text{trace}(BC)=\text{trace}(CB)$. 
\end{enumerate}

\begin{myproof}{Theorem}{\ref{thm::mean-quadratic-form}}
The conclusion follows from
\begin{eqnarray*}
E(Y^{\T}AY) & = &E\{ \text{trace}(Y^{\T}AY) \} \quad (\text{because } Y^{\T}AY \text{ is a scalar}) \\
 & = &E\{ \text{trace}(AYY^{\T}) \} \quad 
 (\text{by }\eqref{item::fact3}) \\
 & =& \text{trace}\{ E(AYY^{\T}) \} \quad
 (\text{by }\eqref{item::fact2})\\
 & = &\text{trace}\{ AE(YY^{\T}) \} \\
 & =& \text{trace}\{ A(\Sigma+\mu\mu^{\T}) \} \quad
 (\text{by }\eqref{item::fact1})\\
 & = &\text{trace}(A\Sigma) + \text{trace}(A\mu\mu^{\T})\\
 & =&\text{trace}(A\Sigma)+\text{trace}(\mu^{\T}A\mu) \quad 
 (\text{by }\eqref{item::fact3})\\
 & =&\text{trace}(A\Sigma)+\mu^{\T}A\mu. \quad (\text{because } \mu^{\T}A\mu \text{ is a scalar}) 
\end{eqnarray*}
\end{myproof}

The variance of the quadratic form is much more complicated for a general
random vector. For the multivariate Normal random vector, we have the
following formula.
\begin{theorem}
\label{thm:varianceofquadraticforms}
If $Y\sim\textup{N}(\mu,\Sigma)$,
then
\[
\var(Y^{\T}AY)=2\textup{trace}(A\Sigma A\Sigma)+4\mu^{\T}A\Sigma A\mu.
\]
\end{theorem}

I relegate the proof as Problem \ref{hw00math2::variance-quadratic}.

From its definition, $\chi_n^2$ is the summation of the squares of $n$ IID standard Normal random variables. It is closely related to
quadratic forms of multivariate Normals. 

\begin{theorem}\label{thm::normal-chisq}
We have the following results on the $\chi^2$ random variables. 
\begin{enumerate}
\item If $Y\sim\textup{N}(\mu,\Sigma)$ is an $n$-dimensional random vector
with $\Sigma\succ 0$, then 
\[
(Y-\mu)^{\T}\Sigma^{-1}(Y-\mu)\sim\chi_{n}^{2}.
\]
If rank$(\Sigma) = k \leq n$, then 
$$
(Y-\mu)^{\T}\Sigma^{+}(Y-\mu)\sim\chi_{k}^{2}.
$$

\item If $Y\sim\textup{N}(0,I_{n})$ and $H$ is a projection matrix of rank
$K$, then 
\[
Y^{\T}HY\sim\chi_{K}^{2}.
\]
\item If $Y\sim\textup{N}(0,H)$ where $H$ is a projection matrix of rank
$K$, then
\[
Y^{\T}Y\sim\chi_{K}^{2}.
\]
\end{enumerate}
\end{theorem}

\begin{myproof}{Theorem}{\ref{thm::normal-chisq}}
\begin{enumerate}
\item 
I only prove the general result with rank$(\Sigma) = k \leq n$.
By definition, $Y=\mu+\Sigma^{1/2}Z$ where $Z$ is a standard Normal
random vector, then
\begin{eqnarray*}
(Y-\mu)^{\T}\Sigma^{+}(Y-\mu)
&=&  Z^{\T}\Sigma^{1/2}\Sigma^{+}\Sigma^{1/2}Z \\
&=&  \sum_{i=1}^k Z_i^2  \sim  \chi_k^{2}.
\end{eqnarray*}
\item Using the eigendecomposition of the projection matrix
\[
H=P\text{diag}\left\{ 1,\ldots,1,0,\ldots,0\right\} P^{\T}
\]
with $K$ $1$'s in the diagonal matrix, we have
\begin{align*}
Y^{\T}HY & =Y^{\T}P\text{diag}\left\{ 1,\ldots,1,0,\ldots,0\right\} P^{\T}Y\\
 & =Z^{\T}\text{diag}\left\{ 1,\ldots,1,0,\ldots,0\right\} Z,
\end{align*}
where $Z=(Z_{1},\ldots,Z_{n})^{\T}=P^{\T}Y\sim\text{N}(0,P^{\T}P)=\text{N}(0,I_{n})$
is a standard Normal random vector. So
\[
Y^{\T}HY=\sum_{i=1}^{K}Z_{i}^{2}\sim\chi_{K}^{2}.
\]
\item Writing $Y=H^{1/2}Z$ where $Z$ is a standard Normal random vector,
we have 
\[
Y^{\T}Y=Z^{\T}H^{1/2}H^{1/2}Z=Z^{\T}HZ\sim\chi_{K}^{2}
\]
using the second result. 
\end{enumerate}
\end{myproof}

\section{Homework problems}

\paragraph{Uniform moments}\label{hwmath2::uniform-moments}

Let $X$ be Uniform$(0,1)$. Find $E(X^k)$ for $k=1, 2, \ldots$.

\paragraph{Beta-Gamma duality}\label{hwmath2::beta-gamma-dual}

Prove Theorem \ref{thm:beta-gamma-duality}.

Remark: Calculate the joint density of $(X+Y, X/(X+Y))$. 

\paragraph{Gamma and Beta moments}\label{hwmath2::beta-gamma-moments}

Prove Propositions \ref{thm:gammamoments}--\ref{thm:beta-moments-log}.

\paragraph{Memoryless property of Exponential}\label{hwmath2::memoryless-exponential}

Prove Proposition \ref{prop::memoryless}.

\paragraph{Minimum of independent Exponentials}\label{hwmath2::min-exponential}

Prove Proposition \ref{prop::minimum-expo}.

\paragraph{Laplace as the difference between two IID Exponentials}\label{hwmath2::laplace-diff-exponential}

Prove Proposition \ref{prop::expo-laplace}.

\paragraph{Maximums of Gumbels}\label{hwmath2::max-gumbel}

Prove Proposition \ref{prop::max-gumbel}.

\paragraph{Independence and uncorrelatedness in the multivariate Normal}\label{hwmath2::independence-uncorr-normal}

Prove Theorem \ref{thm:Normal-ind-uncorr}.

\paragraph{Linear transformation of Normal}\label{hwmath2::linear-normal}

Prove Proposition \ref{prop::linear-normal}.

\paragraph{Transformation of bivariate Normal}

Prove that if $(Y_{1},Y_{2})^{\T}$ follows
a bivariate Normal distribution
\[
\left(\begin{array}{c}
Y_{1}\\
Y_{2}
\end{array}\right)\sim\textup{N}\left(\left(\begin{array}{c}
0\\
0
\end{array}\right),\left(\begin{array}{cc}
1 & \rho\\
\rho & 1
\end{array}\right)\right),
\]
then 
$$
Y_{1}+Y_{2}\ind Y_{1}-Y_{2}.
$$

Remark: This result holds for arbitrary $\rho$.

\paragraph{Normal conditional distributions}\label{hwmath2::normal-conditionals}

Suppose that $(X_1, X_2)$ has the joint distribution
$$
f_{X_1X_2}\left(x_{1}, x_{2}\right) \propto C_0 \exp \left\{  -\frac{1}{2}\left( A x_{1}^{2} x_{2}^{2}+x_{1}^{2}+x_{2}^{2} -2 B x_{1} x_{2}-2 C_{1} x_{1}-2 C_{2} x_{2} \right) \right\},
$$
where $C_0$ is the normalizing constant depending on $(A,B,C_1, C_2)$. To ensure that this is a well-defined density, we need $A \geq 0$, and if $A=0$ then $|B| < 1$. 

Prove that the conditional distributions are
\begin{eqnarray*}
X_{1} \mid X_2 =  x_{2}  &\sim& \N\left(\frac{B x_{2}+C_{1}}{A x_{2}^{2}+1}, \frac{1}{A x_{2}^{2}+1}\right) , \\
X_{2} \mid X_1 =  x_{1}  &\sim& \N\left(\frac{B x_{1}+C_{2}}{A x_{1}^{2}+1}, \frac{1}{A x_{1}^{2}+1}\right).
\end{eqnarray*}

Remark: 
For a bivariate Normal distribution, the two conditional distributions are both Normal. The converse of the statement is not true. That is, even if the two conditional distributions are both Normal, the joint distribution may not be bivariate Normal. 
\citet{gelman1991note} reported this result.

\paragraph{Inverse of covariance matrix and conditional independence in multivariate Normal}\label{hwmath2::inverse-cov-conind-normal}

Assume $X = (X_1, \ldots, X_p)^{\T} \sim \N(\mu, \Sigma)$. Denote the inverse of its covariance matrix by $\Sigma^{-1}  = (\sigma^{jk})_{1\leq j, k\leq p}$. 

Prove that for any pair of $j \neq k$, we have 
$$
\sigma^{jk} = 0 \Longleftrightarrow  X_j \ind X_k \mid X_{\backslash (j,k)} , 
$$
where $X_{\backslash (j,k)}$ contains all the variables except $X_j$ and $X_k$.

Remark: This basic property of multivariate Normal motivates the Gaussian Graphical Model, which uses an undirected graph to illustrate the conditional independence relationship among random variables $X_1, \ldots, X_p$ \citep{dempster1972covariance}. In particular, $\sigma^{jk} = 0 $ if and only if the edge between $X_j$ and $X_k$ is missing.

\paragraph{Independence of linear and quadratic functions of the multivariate Normal}\label{hw00math2::independence-linear-quadratic}

Assume $Y\sim\text{N}(\mu,\sigma^{2}I_{n})$. For an $n$ dimensional
vector $a$ and two $n\times n$ symmetric matrices $A$ and $B$, prove that
\begin{enumerate}
\item if $ Aa =0$, then $a^{\T}Y\ind Y^{\T}AY$;
\item if $AB=BA=0$, then $Y^{\T}AY\ind Y^{\T}BY$.
\end{enumerate}

Remark: To simplify the proof, you can use the pseudoinverse of $A$ which satisfies $AA^{+}A=A$.
In fact, a strong result holds. \citet{ogasawara1951independence} proved the following theorem; see also \citet[][Theorem 5]{styan1970notes}.

\begin{theorem}
Assume $Y\sim\text{N}(\mu,\Sigma )$. Define quadratic forms $Y^{\T}AY$ and $ Y^{\T}BY$ for two symmetric matrices $A$ and $B$. The $Y^{\T}AY$ and $ Y^{\T}BY$ are independent if and only if 
$$
\Sigma A \Sigma B \Sigma = 0,\quad
\Sigma A \Sigma B \mu = \Sigma B \Sigma A \mu = 0,\quad
\mu^{\T} A \Sigma B \mu = 0.
$$
\end{theorem}

\paragraph{Independence of the sample mean and variance of IID Normals}\label{hw00math2::independence-mean-variance}

Theorem \ref{thm::iid-normal-mean-variance} below is a fundamental result on IID Normals. Prove Theorem \ref{thm::iid-normal-mean-variance}. 

\begin{theorem}
\label{thm::iid-normal-mean-variance}
If $X_{1},\ldots,X_{n} \iidsim \N(\mu,\sigma^{2}),$ then $\bar{X}\ind S^{2}$,
where $\bar{X}=n^{-1}\sumn X_{i}$ and $S^{2}=(n-1)^{-1}\sumn(X_{i}-\bar{X})^{2}$.
\end{theorem}

Remark: A remarkable result due to \citet{geary1936distribution} ensures the reverse of Theorem \ref{thm::iid-normal-mean-variance}. That is, if $X_{1},\ldots,X_{n} $ are IID and $\bar{X}\ind S^{2}$, then $X_{1},\ldots,X_{n} $ must be Normals. See \citet{lukacs1942characterization} and \citet{benhamou2018three} for proofs.

\paragraph{Variance of the quadratic form of the multivariate Normal}\label{hw00math2::variance-quadratic}

First prove Theorem \ref{thm:varianceofquadraticforms} with a symmetric $A$. 
Then prove Theorem \ref{thm::cov-quadraticform} below. 

\begin{theorem}
\label{thm::cov-quadraticform}
Assume $A_1$ and $A_2$ are symmetric matrices. If $Y\sim\textup{N}(\mu,\Sigma)$, then
$$
\cov(  Y^{\T}A_1Y, Y^{\T}A_2Y)
= 2\textup{trace}(A_1\Sigma A_2\Sigma)+4\mu^{\T}A_1\Sigma A_2\mu . 
$$
\end{theorem}

Remark: 
Theorem \ref{thm:varianceofquadraticforms} is a special case of Theorem \ref{thm::cov-quadraticform}. You can prove Theorem \ref{thm::cov-quadraticform} and then Theorem \ref{thm:varianceofquadraticforms} follows immediately. You can also first prove Theorem \ref{thm:varianceofquadraticforms} and then use it to prove Theorem \ref{thm::cov-quadraticform}. You can write $Y = \mu + \Sigma^{1/2} Z$ and reduce the problem to calculating the moments of standard Normals.

\chapter{Limiting Theorems and Basic Asymptotics}
\label{chapter::limiting-theorems}

This chapter reviews the basics of limiting theorems and asymptotic analyses that are useful for this book. See \citet{newey1994large} and \citet{van2000asymptotic} for in-depth discussions. 

\section{Convergence in probability}

\begin{definition}
Random vectors $Z_{n}\in\mathbb{R}^{K}$ converge to $Z$ in probability,
denoted by $Z_{n}\rightarrow Z$ in probability, if 
\[
\pr\left\{ \|Z_{n}-Z\|>c\right\} \rightarrow0,\quad n\rightarrow\infty 
\]
for all $c>0$. 
\end{definition}

This definition incorporates the classic definition of convergence
of non-random vectors:
\begin{proposition}
If non-random vectors $Z_{n}\rightarrow Z$, the convergence also
holds in probability.
\end{proposition}
Convergence in probability for random vectors is equivalent to element-wise
convergence because of Proposition \ref{prop:converge-prob-nonrandom} below:

\begin{proposition}\label{prop:converge-prob-nonrandom}
If $Z_{n}\rightarrow Z$ and $W_{n}\rightarrow W$ in probability,
then $(Z_{n},W_{n})\rightarrow(Z,W)$ in probability. 
\end{proposition}

Proposition \ref{prop:converge-prob-nonrandom} above does not require any conditions on the joint distribution of $(Z_n, W_n)$.

For an IID sequence of random vectors, we have the following weak law of
large numbers:
\begin{proposition}[Khintchine's weak law of large numbers]
If $Z_{1},\ldots,Z_{n}$ are \textup{IID} with mean $\mu\in\mathbb{R}^{K},$
then $  n^{-1}\sumn Z_{i}\rightarrow\mu$ in probability. 
\end{proposition}

A more elementary tool is Markov's inequality:
\begin{equation}
\pr\left\{ \|Z_{n}-Z\|>c\right\} \leq E\left\{ \|Z_{n}-Z\|\right\} /c\label{eq:1stmoment}
\end{equation}
or 
\begin{equation}
\pr\left\{ \|Z_{n}-Z\|>c\right\} \leq E\left\{ \|Z_{n}-Z\|^{2}\right\} /c^{2}.\label{eq:2ndmoment}
\end{equation}

Inequality (\ref{eq:1stmoment}) is useful if $E\left\{ \|Z_{n}-Z\|\right\} $
converges to zero, and inequality (\ref{eq:2ndmoment}) is useful
if $E\left\{ \|Z_{n}-Z\|^{2}\right\} $ converges to zero. The latter
gives a standard tool for establishing convergence in probability
by showing that the covariance matrix converges to zero.

\begin{proposition}\label{prop::markov-lln}
If random vectors $Z_{n}\in\mathbb{R}^{K}$ have mean zero and covariance
$\cov(Z_{n})=a_{n}C_{n}$ where $a_{n}\rightarrow0$ and $C_{n}\rightarrow C<\infty$,
then $Z_{n}\rightarrow0$ in probability. 
\end{proposition}

\begin{myproof}{Proposition}{\ref{prop::markov-lln}}
Using (\ref{eq:2ndmoment}), we have
\begin{align*}
\pr\left\{ \|Z_{n}\|>c\right\}  & \leq c^{-2}E\left\{ \|Z_{n}\|^{2}\right\} \\
 & =c^{-2}E\left(Z_{n}^{\T}Z_{n}\right)\\
 & =c^{-2}\text{trace}\left\{ E\left(Z_{n}Z_{n}^{\T}\right)\right\} \\
 & =c^{-2}\text{trace}\left\{ \cov(Z_{n})\right\} \\
 & =c^{-2}a_{n}\text{trace}(C_{n})\rightarrow0,
\end{align*}
which implies that $Z_{n}\rightarrow0$ in probability. 
\end{myproof}

For example, we usually use Proposition \ref{prop::markov-lln} to show the weak law of large numbers for the sample mean of independent random variables $\bar{Z}_n = n^{-1} \sumn Z_i$. If we can show that
\begin{eqnarray}
\cov( \bar{Z}_n   ) = n^{-2} \sumn \cov(Z_i) \rightarrow 0,
\label{eq::covariance-of-average}
\end{eqnarray}
then we can conclude that $\bar{Z}_n - n^{-1} \sumn E(Z_i) \rightarrow 0$ in probability. The condition in \eqref{eq::covariance-of-average} holds if $n^{-1} \sumn \cov(Z_i) $ converges to a constant matrix.

Note that convergence in probability does not imply convergence of moments in general. Proposition \ref{prop::DCT} below gives a sufficient condition.

\begin{proposition}[dominant convergence theorem]\label{prop::DCT}
If $Z_n \rightarrow Z$ in probability and $\| Z_n \| \leq \| Y \|$ with $E\| Y\| < \infty$, then $E(Z_n) \rightarrow E(Z)$. 
\end{proposition}

\section{Convergence in distribution}

\begin{definition}
Random vectors $Z_{n}\in\mathbb{R}^{K}$ converge to $Z$ in distribution,
if for all every continuous point $z$ of the function $t\rightarrow\pr(Z\leq t)$, we have 
\[
\pr(Z_{n}\leq z)\rightarrow\pr(Z\leq z),\quad n\rightarrow\infty.
\]
\end{definition}

When the limit is a constant, we have an equivalence of convergences
in probability and distribution:

\begin{proposition}
If $c$ is a non-random vector, then $Z_{n}\rightarrow c$ in probability
is equivalent to $Z_{n}\rightarrow c$ in distribution.
\end{proposition}

For IID sequences of random vectors, we have the Lindeberg--L\'{e}vy central limit theorem
(CLT):
\begin{proposition}[Lindeberg--L\'{e}vy CLT]
If random vectors $Z_{1},\ldots,Z_{n}$ are \textup{IID} with mean $\mu$
and covariance $\Sigma$, then $n^{1/2}(\bar{Z}_{n}-\mu)=n^{-1/2}\sumn(Z_{i}-\mu)\rightarrow\N(0,\Sigma)$
in distribution. 
\end{proposition}

The more general Lindeberg--Feller CLT holds for independent sequences
of random vectors:
\begin{proposition}\label{prop::lf-clt}
For each $n$, let $Z_{n1},\ldots,Z_{n,k_{n}}$ be independent random
vectors with finite variances such that
\begin{enumerate}[label=(LF\arabic*), ref=LF\arabic*]
\item\label{enum::LF1} $\sum_{i=1}^{k_{n}}E\left[\|Z_{ni}\|^{2}1\left\{ \|Z_{ni}\|>c\right\} \right]\rightarrow0$
for every $c>0$;
\item\label{enum::LF2} $\sum_{i=1}^{k_{n}}\cov(Z_{ni})\rightarrow\Sigma.$ 
\end{enumerate}
Then $\sum_{i=1}^{k_{n}}\left\{ Z_{ni}-E(Z_{ni})\right\} \rightarrow\N(0,\Sigma)$
in distribution.
\end{proposition}

Condition \eqref{enum::LF2} often holds by proper standardization, so the key is to verify Condition \eqref{enum::LF1}. 
Condition \eqref{enum::LF1} is general but it looks cumbersome. 
In many cases, we impose a stronger moment condition that is easier to verify:
\begin{enumerate}[label=(LF\arabic*'), ref=LF\arabic*']
\item\label{enum::LF'1} 
$\sum_{i=1}^{k_{n}}E \|Z_{ni}\|^{2+\delta} \rightarrow 0$ for some $\delta >0$. 
\end{enumerate}

We can prove that \eqref{enum::LF'1} implies that \eqref{enum::LF1}:
\begin{eqnarray*}
\sum_{i=1}^{k_{n}}E\left[\|Z_{ni}\|^{2}1\left\{ \|Z_{ni}\|>c\right\} \right]
&=& \sum_{i=1}^{k_{n}}E\left[\|Z_{ni}\|^{2+\delta}  \|Z_{ni}\|^{-\delta} 1\left\{ \|Z_{ni}\|^{\delta}>c^{\delta} \right\} \right] \\
&\leq & \sum_{i=1}^{k_{n}}E \|Z_{ni}\|^{2+\delta} c^{- \delta} \\ &\rightarrow & 0.
\end{eqnarray*}

Condition \eqref{enum::LF'1} is called the Lyapunov condition.

A beautiful application of the Lindeberg--Feller CLT is the proof of \citet{huber1973robust}'s theorem on OLS mentioned in Chapter \ref{chapter::leave-one-out}. I first review the theorem and then give a proof.

\begin{theorem}
\label{thm::huber-ols}
Assume $Y= X \beta + \varepsilon $ where the covariates are fixed and  error terms $\varepsilon = (\varepsilon_1, \ldots, \varepsilon_n)^{\T}$ are IID non-Normal with mean zero and finite variance $\sigma^2$. Recall the OLS estimator $\hat{\beta} = (X^{\T} X)^{-1} X^{\T} Y$. Any linear combination of $\hat{\beta}$ is asymptotically Normal  if and only if
$$
\max_{1\leq i \leq n} h_{ii} \rightarrow 0 ,
$$
where $h_{ii}$ is the $i$th diagonal element of the hat matrix $H  =  X (X^{\T} X)^{-1} X^{\T}$. 
\end{theorem}

In the main text, $h_{ii}$ is called the {\it leverage score} of unit $i$. The maximum leverage score 
$$
\kappa = \max_{1\leq i \leq n} h_{ii}
$$ 
plays an important role in analyzing the properties of OLS.

Theorem \ref{thm::huber-ols} assumes that the errors are not Normal because the asymptotic Normality under Normal errors is a trivial result (See Chapter \ref{chapter::normal-linear-model}). 
It is slightly different from the asymptotic analysis in Chapter \ref{chapter::EHW}. Theorem \ref{thm::huber-ols} only concerns any linear combination of the OLS estimator alone, but the results in  Chapter \ref{chapter::EHW} allow for the joint inference of several linear combinations of the OLS estimator. An implicit assumption of  Chapter \ref{chapter::EHW} is that the dimension $p$ of the covariate matrix is fixed, but Theorem \ref{thm::huber-ols} allows for a diverging $p$. The leverage score condition implicitly restricts the dimension and moments of the covariates. Another interesting feature of Theorem \ref{thm::huber-ols} is that the statement is coordinate-free, that is, it holds up to a non-singular transformation of the covariates (See also Problems \ref{hw3::invariance-ols} and \ref{hw03::invariance-of-H}). The proof of sufficiency follows \citet{huber1973robust} closely, and the proof of necessity was suggested by Professor Peter Bickel.

\begin{myproof}{Theorem}{\ref{thm::huber-ols}}
I first simplify the notation without essential loss of generality. 
By the invariance of the OLS estimator and the hat matrix in Problems \ref{hw3::invariance-ols} and \ref{hw03::invariance-of-H}, we can also assume $X^{\T} X = I_p$. So
$$
\hat{\beta} - \beta = (X^{\T} X)^{-1} X^{\T} \varepsilon = X^{\T} \varepsilon
$$
and the hat matrix 
$$
H = X (X^{\T} X)^{-1} X^{\T} = X  X^{\T}
$$
has diagonal elements $h_{ii} = x_i^{\T} x_i = \|x_i \|^2$ and non-diagonal elements $h_{ij} =x_i^{\T} x_j  $. We can also assume $\sigma^2 = 1$. 

Consider a fixed vector $a \in \mathbb{R}^p$ and assume $\|a\|^2 = 1$. 
We have
$$
a^{\T}  \hat{\beta} - a^{\T}  \beta = a^{\T}  X^{\T} \varepsilon \equiv s^{\T} \varepsilon, 
$$
where 
$$
s = X a  \Longleftrightarrow   s_i = x_i^{\T} a \quad (i=1,\ldots, n)
$$ 
satisfies 
$$
\| s \|^2 = a^{\T}  X^{\T}  X a = \|a\|^2 = 1
$$ 
and 
$$
s_i^2 = (x_i^{\T} a)^2 \leq \| x_i\|^2 \| a \|^2 = \| x_i\|^2 = h_{ii}
$$
by the Cauchy--Schwarz inequality.

I first prove the sufficiency. 
The key term $a^{\T}  \hat{\beta} - a^{\T}  \beta$ is a linear combination of the IID errors, and it has mean $0$ and variance 
$
\var( s^{\T} \varepsilon ) = \| s \|^2 =1.
$
We only need to verify Condition (LF\ref{enum::LF1}) to establish the CLT. It holds because for any fixed $c>0$, we have 
\begin{eqnarray} 
 \sumn E\left[   s_i^2 \varepsilon_i^2 1\left\{    |s_i  \varepsilon_i |  > c \right\}  \right] 
&\leq &  \sumn  s_i^2 \max_{1\leq i \leq  n } E\left[   \varepsilon_i^2 1\left\{    |s_i  \varepsilon_i |  > c \right\}  \right]   \label{eq::huber-max} \\
&=&   \max_{1\leq i \leq  n } E\left[   \varepsilon_i^2 1\left\{    |s_i  \varepsilon_i |  > c \right\}  \right]  \label{eq::huber-sum1} \\
&\leq &   E\left[   \varepsilon_i^2 1\left\{  \kappa^{1/2}    |  \varepsilon_i |  > c \right\}  \right]  \label{eq::huber-upper} \\
&\rightarrow & 0, \label{eq::huber-convergence}
\end{eqnarray}  
where \eqref{eq::huber-max} follows from the property of $\max$, \eqref{eq::huber-sum1} follows from the fact $\| s \|^2 =1$, \eqref{eq::huber-upper} follows from the fact that $ |s_i| \leq | h_{ii} |  \leq \kappa^{1/2}$, and \eqref{eq::huber-convergence} follows from $\kappa \rightarrow 0$ and the dominant convergence theorem in Proposition \ref{prop::DCT}.

I then prove the necessity. 
Pick one $i^*$ from $ \arg\max_{1\leq i\leq n}  h_{ii} $. 
Consider a special linear combination of the OLS estimator: $\hat{y}_{i^*} = x_{i^*}^{\T} \hat{\beta}$, which is the fitted value of the $i^*$th observation and has the form
\begin{eqnarray*}
\hat{y}_{i^*}   
&=&  x_{i^*}^{\T} \hat \beta  \\
&=&  x_{i^*}^{\T} (\beta + X^{\T} \varepsilon ) \\
&=& x_{i^*}^{\T}  \beta +  \sum_{j=1}^n x_{i^*}^{\T} x_j \varepsilon_j \\
&=& x_{i^*}^{\T}  \beta +  h_{i^*i^*} \varepsilon_{i^*} + \sum_{j\neq i^*} h_{i^*j} \varepsilon_j . 
\end{eqnarray*}
If $\hat{y}_{i^*} $ is asymptotically Normal, then both $ h_{i^*i^*} \varepsilon_{i^*} $ and $\sum_{j\neq i^*} h_{i^*j} \varepsilon_j$ must have Normal limiting distributions by Theorem \ref{thm::levy-cramer}. Therefore, $ h_{i^*i^*} $ must converge to zero because $\varepsilon_{i^*} $ has a non-Normal distribution. So $\max_{1\leq i \leq n} h_{ii}$ must converge to zero. 
\end{myproof}

\section{Tools for proving convergence in probability and distribution}\label{chapter::tools-limiting3}

The first tool is the continuous mapping theorem:
\begin{proposition}[continuous mapping theorem]
Let $f:\mathbb{R}^{K}\rightarrow\mathbb{R}^{L}$ be continuous except
on a set $O$ with $\pr(Z\in O)=0$. Then $Z_{n}\rightarrow Z$ implies
$f(Z_{n})\rightarrow f(Z)$ in probability (and in distribution).
\end{proposition}

The second tool is Slutsky's theorem:
\begin{proposition}[Slutsky's theorem]
Let $Z_{n}$ and $W_{n}$ be random vectors. If $Z_{n}\rightarrow Z$
in distribution, and $W_{n}\rightarrow c$ in probability (or in distribution)
for a constant $c$, then
\begin{enumerate}
\item $Z_{n}+W_{n}\rightarrow Z+c$ in distribution;
\item $W_{n}Z_{n}\rightarrow cZ$ in distribution;
\item $W_{n}^{-1} Z_{n}\rightarrow c^{-1}Z$ in distribution if $c\neq0$.
\end{enumerate}
\end{proposition}

The third tool is the delta method. I will present a special case below for asymptotically Normal random vectors. Heuristically, it states that if $T_n$ is asymptotically Normal, then a well-behaved function of $T_n$ is also asymptotically Normal. This is true because any function is a locally linear function by the first-order Taylor expansion. 

\begin{proposition}[delta method]\label{prop::delta-method}
Let $f(z)$ be a function from $\mathbb{R}^p$ to $\mathbb{R}^q$, and $\partial f(z) / \partial z \in \mathbb{R}^{p\times q}$ be the partial derivative matrix. 
If $\sqrt{n} (Z_n - \theta) \rightarrow \N(0, \Sigma)$ in distribution, then 
$$
\sqrt{n} \{ f(Z_n) - f(\theta) \} \rightarrow \N\left( 
0, \frac{ \partial f(\theta) }{  \partial z^{\T} } \Sigma \frac{ \partial f(\theta) }{  \partial z}
\right)
$$
in distribution. 
\end{proposition}

\begin{myproof}{Proposition}{\ref{prop::delta-method}}
I will give an informal proof. Using Taylor expansion, we have
$$
\sqrt{n} \{ f(Z_n)   - f(\theta) \}\cong  \frac{ \partial f(\theta) }{  \partial z^{\T} } \sqrt{n} (Z_n - \theta),
$$
which is a linear transformation of $\sqrt{n} (Z_n - \theta)$. Because $\sqrt{n} (Z_n - \theta) \rightarrow \N(0, \Sigma)$ in distribution, we have
\begin{eqnarray*}
\sqrt{n} \{ f(Z_n)   - f(\theta) \} 
&\rightarrow &  \frac{ \partial f(\theta) }{  \partial z^{\T} }\N(0, \Sigma) \\
&=& \N\left( 
0, \frac{ \partial f(\theta) }{  \partial z^{\T} } \Sigma \frac{ \partial f(\theta) }{  \partial z}
\right)
\end{eqnarray*}
in distribution. 
\end{myproof}

Proposition \ref{prop::delta-method} above is more useful when $\partial f(\theta) /   \partial z \neq 0$. Otherwise, we need to invoke higher-order Taylor expansion to obtain a more accurate asymptotic approximation.

\chapter{M-Estimation and MLE}
\label{chapter::m-mle}

A wide range of statistics estimation problems can be formulated as an estimating equation:
$$
\bar{m}(W,b) = n^{-1}\sumn m(w_{i},b) = 0, 
$$
where $m(\cdot, \cdot)$ is a vector function with the same dimension as $b$, and $W = \left\{ w_{i}\right\} _{i=1}^{n}$ are the observed data. The notation $\bar{m}(W,b)$ emphasizes that it is a function of the observed data and $b$. 
Let $\hat{\beta} $ denote the solution of $\bar{m}(W,b) = 0$ in the sense that $\bar{m}(W, \hat{\beta}) = 0$. 
Let $\beta$ denote the solution of $E\{ \bar{m}(W,b) \} = 0$ in the sense that $E\{ \bar{m}(W,  \beta) \} = 0$. 
Then $\hat{\beta} $ is an estimator of $\beta$. Under mild regularity conditions,  $\hat{\beta} $ is consistent for $\beta$ and asymptotically Normal.\footnote{There are counterexamples in which $\hat{\beta} $ is inconsistent; see \citet{freedman1982inconsistent}. The examples in this book are all regular.}
This is the classical theory of M-estimation. I will review it below. See \citet{stefanski2002calculus} for a reader-friendly introduction that contains many interesting and important examples. The proofs below are not rigorous. 
See \citet{newey1994large} for rigorous proofs.

\section{M-estimation}

I start with the simple case with IID data.

\begin{theorem}
\label{theorem:sandwich-theorem-cov-iid}Assume that $W = \left\{ w_{i}\right\} _{i=1}^{n}$ are IID with the same distribution as $w$.
The true parameter $\beta\in\mathbb{R}^{p}$ is the unique solution of  
\[
E\left\{ m(w,b)\right\} =0,
\]
and the estimator $\hat{\beta}\in\mathbb{R}^{p}$ is the solution of
\[
\bar{m}(W,b)  =0 . 
\]
Under regularity conditions, 
\[
\sqrt{n}(\hat{\beta}-\beta)\rightarrow\N(0,B^{-1}MB^{-\T})
\]
in distribution, where 
$$
B   = -  \frac{\partial E\left\{ m(w,\beta) \right\} }{\partial b^{\T}}
$$
and
$$
M = E\{  m(w,\beta) m(w,\beta)^{\T}\}  .
$$
\end{theorem}

\begin{myproof}{Theorem}{\ref{theorem:sandwich-theorem-cov-iid}}
I give a ``physics'' proof.  When I use approximations, I mean the
error terms are of higher orders under some regularity conditions. The
consistency follows from swapping the order of ``solving equation''
and ``taking the limit based on the law of large numbers'':
\begin{align*}
\lim_{n\rightarrow\infty}\hat{\beta} & =\lim_{n\rightarrow\infty}\left\{ \text{solve } \bar{m}(W,b)=0\right\} \\
 & =\text{solve } \left\{ \lim_{n\rightarrow\infty} \bar{m}(W,b) =0 \right\} \\
 & =\text{solve } \left[ E\left\{ m(w,b)\right\} =0 \right] \\
 &=\beta.
\end{align*}

The asymptotic Normality follows from three steps. First, from the Taylor expansion
$$
0  = \bar{m}(W,\hat{\beta}) \cong \bar{m}(W,\beta)
+ \frac{\partial \bar{m}(W,\beta)}{\partial b^{\T}} (\hat{\beta}-\beta)
$$
we obtain 
$$
 \sqrt{n}\left(\hat{\beta}-\beta\right) \cong  \left\{  -   \frac{\partial \bar{m}(W,\beta)}{\partial b^{\T}} \right\} ^{-1}\left\{ \frac{1}{\text{\ensuremath{\sqrt{n}}}}\sumn m(w_{i},\beta)\right\} .
$$
Second, the law of large numbers ensures that 
\[
- \frac{\partial \bar{m}(W,\beta)}{\partial b^{\T}}  
\rightarrow - \frac{\partial E\left\{ m(w,\beta) \right\} }{\partial b^{\T}} =B
\]
in probability, and the CLT ensures that
$$
n^{-1/2}\sumn m(w_{i},\beta)\rightarrow\ \N(0,M)
$$
in distribution. 
Finally, Slutsky's theorem implies the result. 
\end{myproof}

The above result also holds with independent but non-IID data.

\begin{theorem}
\label{theorem:sandwich-theorem-cov-ind}Assume that $\left\{ w_{i}\right\} _{i=1}^{n}$ are independent observations.
The true parameter $\beta\in\mathbb{R}^{p}$ is the unique solution to  
\[
E\left\{\bar{m}(W,b)\right\} =0,
\]
 and the estimator $\hat{\beta}\in\mathbb{R}^{p}$ is the solution to
\[
\bar{m}(W,b)  =0,
\]
Under some regularity conditions, 
\[
\sqrt{n}\left(\hat{\beta}-\beta\right)\rightarrow\N(0,B^{-1}MB^{-\T})
\]
in distribution, where 
$$
B  = - \lim_{n\rightarrow \infty } n^{-1} \sumn  
 \frac{\partial E\left\{ m(w_i,\beta) \right\} }{\partial b^{\T}} ,\quad 
M =   \lim_{n\rightarrow \infty } n^{-1} \sumn \cov\{ m(w_{i},\beta) \}  .
$$
\end{theorem}

For both cases above, we can further construct the following sandwich covariance estimator:
$$
\left( \sumn  
 \frac{\partial E \{ m(w_i, \hat\beta) \} }{\partial b^{\T}}  \right)^{-1} 
 \left( \sumn  m(w_{i},  \hat\beta) m(w_{i},  \hat\beta)^{\T}  \right)
  \left( \sumn  
 \frac{\partial E\{ m(w_i,  \hat\beta) \} }{\partial b}  \right)^{-1}  ,
$$
which is a plug-in estimator of the asymptotic covariance of $\hat{\beta}$. 
For non-IID data, the above covariance estimator can be conservative unless $ E\{ m(w_i, \beta) \}  = 0$ for all $i=1, \ldots, n$.

\begin{example}
\label{eg::sample-variance}
Assume that $x_1,\ldots, x_n \iidsim x$ with mean $\mu$. The sample mean $\bar{x}$ solves the estimating equation
$$
n^{-1} \sumn (x_i -  \mu) = 0.
$$
Apply Theorem \ref{theorem:sandwich-theorem-cov-iid} to obtain $B = -1$ and $M = \sigma^2$, which imply
$$
\sqrt{n} (\bar{x}  -\mu   ) \rightarrow  \N( 0, \sigma^2)
$$
in distribution, the standard CLT for the sample mean. Moreover, the sandwich covariance estimator is
$$
\hat{V}  = 
n^{-2} \sumn ( x_i -  \bar{x}  )^2 ,
$$
which equals the sample variance of $x$, multiplied by $(n-1)/n^2 \approx 1/n$. This is a standard result. 

If we only assume that $x_1,\ldots, x_n$ are independent with the same mean $\mu$ but possibly different variances $\sigma_i^2\ (i=1,\ldots, n)$, the sample mean $\bar{x}$ is still an unbiased estimator for $\mu$, which solves the same estimating equation above. Moreover, the sandwich covariance estimator $\hat{V} $ is still a consistent estimator for the true variance of $\bar{x}$. This is less standard.

If we assume that $x_i$'s are independent with  mean $\mu_i$ and variance $\sigma_i^2$, we can still use $\bar{x}$ to estimate $\mu = n^{-1} \sumn \mu_i $. The estimating equation remains the same as above. The sandwich covariance estimator $\hat{V}$ becomes conservative since $E(x_i - \mu) \neq 0$ in general. 

Problem \ref{hw-math4::mean} gives more details.
\end{example}

\begin{example}
\label{eg::sample-meanandvariance}
Assume that $x_1,\ldots, x_n \iidsim x$ with mean $\mu$ and variance $\sigma^2$. The sample mean and variance $(\bar{x}, \hat{\sigma}^2)$ jointly solves the estimating equation with
$$
m(x, \mu, \sigma^2) = \begin{pmatrix}
x - \mu \\
(x-\mu)^2 - \sigma^2
\end{pmatrix}
$$
ignoring the difference between $n$ and $n-1$ in the definition of the sample variance. Apply Theorem \ref{theorem:sandwich-theorem-cov-iid} to obtain 
$$
B = \begin{pmatrix}
-1 & 0 \\ 
0 & -1
\end{pmatrix},\quad
M = \begin{pmatrix}
\sigma^2 & \mu_3 \\
\mu_3 & \mu_4 - \sigma^4
\end{pmatrix}
$$
where $\mu_k = E\{ (x-\mu)^k\}$, which imply 
$$
\sqrt{n} \begin{pmatrix}
\bar{x} - \mu \\
\hat{\sigma}^2 - \sigma^2
\end{pmatrix} \rightarrow 
\N(0, M)
$$
in distribution. 
\end{example}

\begin{example}
Assume that $(x_i, y_i)_{i=1}^n$ are IID draws from $(x,y)$ with mean $(\mu_x, \mu_y)$. Use $\bar{x} / \bar{y}$ to estimate $\gamma = \mu_x / \mu_y$. It satisfies the estimating equation with 
$$
m(x,y,\gamma) = x - \gamma y .
$$
Apply Theorem \ref{theorem:sandwich-theorem-cov-iid} to obtain $B = -\mu_y$ and $M = \var(x-\gamma y)$, which imply 
$$
\sqrt{n} \left(   \frac{ \bar{x} }{ \bar{y} }  - \frac{ \mu_x }{   \mu_y }  \right)
\rightarrow
\N\left(0,   \frac{  \var(x-\gamma y) }{  \mu_y^2  }   \right)
$$
in distribution if $\mu_y \neq 0$. 
\end{example}

\section{Maximum likelihood estimator}\label{sec::mle}

As an important application of Theorem \ref{theorem:sandwich-theorem-cov-iid}, we can derive the asymptotic properties of the maximum likelihood estimator (MLE) $\hat{\theta}$ under \textup{IID} sampling from a parametric model 
$$
 y_1,\ldots, y_n \iidsim  f(y\mid \theta).
 $$
The following Bartlett's identities are fundamental for understanding the MLE.

\begin{lemma}[Bartlett's identities]
\label{lemma:bartlett-identity}
Given a probability density or mass
function $f(y\mid\theta)$ indexed by the parameter $\theta$, if we can
change the order of expectation and differentiation, then
$$
E\left(\frac{\partial\log f(y\mid\theta)}{\partial\theta}\right)=0
$$
and
$$
E\left\{  \frac{\partial \log f(y\mid \theta)}{ \partial \theta}  \frac{\partial \log f(y\mid \theta)}{ \partial \theta^{\T}}   \right\}
=
E\left\{ -  \frac{\partial^2 \log f(y\mid \theta)}{ \partial \theta  \partial \theta^{\T} }   \right\} .
$$
\end{lemma} 
 
\begin{myproof}{Lemma}{\ref{lemma:bartlett-identity}}
Define $\ell (y\mid \theta) = \log f(y\mid \theta)$ as the log likelihood function, so $e^{\ell (y\mid \theta)}$ is the density satisfying 
\begin{equation}
\label{eq::bartlett-1}
\int e^{\ell (y\mid \theta)} \d y  = \int f(y\mid \theta)\d y = 1
\end{equation}
by the definition of a probability density function (we can replace the integral by summation for a probability mass function). Differentiate \eqref{eq::bartlett-1} to obtain
$$ 
\frac{\partial }{\partial \theta} \int e^{\ell (y\mid \theta)} \d y = 0 .
$$
If we can swap $\frac{\partial }{\partial \theta}$ and $\int$, then
$$
  \int  \frac{\partial }{\partial \theta} e^{\ell (y\mid \theta)} \d y = 0 .
$$
Therefore,   
\begin{equation}
\label{eq::bartlett-2}
 \int  e^{\ell (y\mid \theta)} \frac{\partial }{\partial \theta} \ell (y\mid \theta)  \d y = 0 ,
\end{equation}
which implies Bartlett's first identity.

Differentiate \eqref{eq::bartlett-2} to obtain
$$
\frac{\partial }{\partial \theta^{\T}}  \int  e^{\ell (y\mid \theta)} \frac{\partial }{\partial \theta} \ell (y\mid \theta)  \d y = 0 .
$$
If we can swap $\frac{\partial }{\partial \theta^{\T}}$ and $\int$, then
$$
\int  \left[ e^{\ell (y\mid \theta)}  \frac{\partial }{\partial \theta} \ell (y\mid \theta)   \frac{\partial }{\partial \theta} \ell (y\mid \theta^{\T})  
+e^{\ell (y\mid \theta)} \frac{\partial^2 }{\partial \theta \partial \theta^{\T}} \ell (y\mid \theta) \right] 
   \d y = 0 ,
$$
which implies Bartlett's second identity. 
\end{myproof}

The MLE satisfies the following estimating equation:
\begin{eqnarray}
E\left\{  \frac{\partial \log f(y\mid \theta)}{ \partial \theta}   \right\} = 0,
\label{eq::bartlett-vector1}
\end{eqnarray}
which is Bartlett's first identity. 
Under regularity conditions, $\sqrt{n} (\hat{\theta} - \theta)$ converges in distribution to Normal with mean zero and covariance $B^{-1} M B^{-1}$, where
\begin{eqnarray*}
B 
&=& - \frac{\partial }{\partial \theta^{\T}} E\left\{  \frac{\partial \log f(y\mid \theta)}{ \partial \theta}   \right\} \\
&=&
E\left\{ -  \frac{\partial^2 \log f(y\mid \theta)}{ \partial \theta  \partial \theta^{\T} }   \right\} 
\end{eqnarray*}
is called the Fisher information matrix, denoted by $I(\theta)$,
and
$$
M = E\left\{  \frac{\partial \log f(y\mid \theta)}{ \partial \theta}  \frac{\partial \log f(y\mid \theta)}{ \partial \theta^{\T}}   \right\}
$$
is sometimes also called the Fisher information matrix, denoted by  $J(\theta)$.

If the model is correct,
Bartlett's second identity ensures that 
\begin{eqnarray}
I(\theta)=J(\theta),
\label{eq::bartlett-vector2}
\end{eqnarray}
and therefore  $\sqrt{n} (\hat{\theta} - \theta)$ converges in distribution to Normal with mean zero and covariance $I(\theta)^{-1} = J(\theta)^{-1}$. So a covariance matrix estimator for the MLE is $I_n(\hat{\theta})^{-1}$ or $J_n(\hat{\theta}) ^{-1}$, where 
$$
I_n(\hat{\theta}) =  - \sumn   \frac{\partial^2 \log f(y_i\mid \hat{\theta} )}{ \partial \theta \partial \theta^{\T} }  
$$   
and
$$
J_n(\hat{\theta}) = \sumn \frac{\partial \log f(y_i \mid \hat{\theta})}{ \partial \theta}  \frac{\partial \log f(y_i \mid \hat{\theta})}{ \partial \theta^{\T}}. 
$$
\citet{fisher1925theory} pioneered the asymptotic theory of the MLE under correctly specified models. 

If the model is incorrect, $I(\theta)$ can be different from $J(\theta)$ but the sandwich covariance $B^{-1} M B^{-1}$ still holds. So a covariance matrix estimator for the MLE under misspecification is 
$$
I_n(\hat{\theta})^{-1} J_n(\hat{\theta}) I_n(\hat{\theta})^{-1}.
$$
\citet{huber::1967} studied the asymptotic theory of the MLE under  misspecified models. He focused on the case with IID observations and derived the sandwich covariance formula.\footnote{Other forms of the sandwich covariance formula appeared also in earlier work of \citet{godambe1960optimum} and \citet{cox1961tests} for slightly different purposes.
}

Perhaps a more important question is what is the parameter if the model is misspecified.  The population analog of the MLE is the minimizer of
$$
- E \{  \log f(y\mid \theta) \} ,
$$
where the expectation is over true but unknown distribution $y \sim g(y)$.  We can rewrite the population objective function as
$$
 -\int g(y) \log f(y\mid \theta)  \diff y
= \int g(y) \log  \frac{ g(y) }{ f(y\mid \theta)  }  \diff y
-\int g(y) \log  g(y)  \diff y .
$$
The first term is called the Kullback--Leibler divergence or relative entropy of $g(y)$ and $f(y\mid \theta)$, whereas the second term is called the entropy of $g(y)$. The first term depends on $\theta$ whereas the second term does not. Therefore, the targeted parameter of the MLE is the minimizer of the Kullback--Leibler divergence. By Gibbs' inequality, the  Kullback--Leibler divergence is non-negative in general and is $0$ if $g(y) = f(y\mid \theta)$. Therefore, if the model is correct, then the true $\theta$ indeed minimizes the Kullback--Leibler divergence with minimum value $0$.

The following simple model provides insights into the variance estimators of the MLE.

\begin{example}
Assume that $y_1,\ldots, y_n \iidsim  \N(\mu ,1 )$. The log-likelihood contributed by unit $i$ is 
$$
\log f(y_i \mid \mu) = -\frac{1}{2} \log (2\pi) - \frac{1}{2}  (y_i -  \mu)^2,
$$
so
$$
\frac{ \partial \log f(y_i \mid \mu) }{ \partial \mu } =  y_i -  \mu ,\quad
\frac{ \partial^2 \log f(y_i \mid \mu) }{ \partial \mu^2 } =  - 1. 
$$
The MLE is $\hat{\mu} = \bar{y}$. 
If the model is correctly specified, we can use 
$$
I_n(\hat{\mu})^{-1} = n^{-1} \text{ or }
J_n(\hat{\mu})^{-1} = 1 / \sumn  (y_i -  \hat\mu)^2
$$
to estimate the variance of $\hat{\mu}$. If the model is misspecified, we can use
$$
I_n(\hat{\mu})^{-1} J_n(\hat{\mu})I_n(\hat{\mu})^{-1} =  \sumn  (y_i -  \hat\mu)^2/n^2
$$
to estimate the variance of $\hat{\mu}$. 

The sandwich variance estimator seems the best overall. The Normal model can be totally wrong but it is still meaningful to estimate the mean parameter $\mu = E(y)$. The MLE is just the sample moment estimator, which has variance $\textup{var}(y)/n$. Since the sample variance $s^2 = \sumn  (y_i -  \hat\mu)^2/(n-1)$ is unbiased for $\textup{var}(y)$, a natural unbiased estimator for $\textup{var}(\hat{\mu})$ is $s^2/n$, which is close to the sandwich variance estimator. 
\end{example}

The above discussion extends to the case with independent but non-IID data. The covariance estimators still apply by replacing each $f$ by $f_i$ within the summation. Note that the sandwich covariance estimator is conservative in general. \citet{white1982maximum} pioneered the asymptotic analysis of the MLE with misspecified models in econometrics but made a mistake for the $M$ term. \citet{chow1984maximum} corrected his error, and \citet{abadie2014inference} developed a more general theory. \citet{agresti2002measures} discussed measures of model fit with possibly misspecified models.

A leading application is the MLE under a misspecified Normal linear model. The Eicker--Huber--White (EHW) robust covariance arises naturally in this case.

\begin{example}
\label{eg::huber-EHW}
The Normal linear model has individual log-likelihood: 
$$
l_i  = -\frac{1}{2} \log(2\pi\sigma^2) - \frac{1}{2\sigma^2} (y_i - x_i^{\T} \beta)^2 ,\quad (i=1, \ldots, n)
$$
with the simplification $l_i = \log  f(y_i \mid x_i, \beta, \sigma^2).$
So the first-order derivatives are
$$
\frac{\partial l_i   }{ \partial \beta } = \frac{1}{\sigma^2} x_i (y_i - x_i^{\T} \beta), \quad 
\quad 
\frac{\partial l_i  }{ \partial \sigma^2} = -\frac{1}{2 \sigma^2} +  \frac{1}{2(\sigma^2)^2}  (y_i - x_i^{\T} \beta)^2 ;
$$
the second derivative is
$$
\frac{\partial^2 l_i   }{ \partial \beta^2 } =
-  \frac{1}{\sigma^2} x_i x_i^{\T},\quad
\frac{\partial^2 l_i   }{ \partial (\sigma^2)^2 } =
\frac{1}{2(\sigma^2)^2} - \frac{1}{(\sigma^2)^3} (y_i - x_i^{\T} \beta)^2
$$
and
$$
\frac{\partial^2 l_i  }{ \partial \beta \partial \sigma^2} =
-\frac{1}{(\sigma^2)^2}  x_i (y_i - x_i^{\T} \beta).
$$
The MLE of $\beta$ is the OLS estimator $\hat{\beta}$ and the MLE of $\sigma^2$ is $\tilde{\sigma}^2 = \sumn \hat{\varepsilon}_i^2 /n$, where $ \hat{\varepsilon}_i = y_i - x_i^{\T} \hat{\beta}$  is the residual.  

We have
$$
I_n(\hat\beta, \tilde\sigma^2 ) = \textup{diag}\left( \frac{1}{  \tilde{\sigma}^2}  \sumn x_i x_i^{\T} , \frac{n}{  2 (\tilde{\sigma}^2)^2 } \right), 
$$
and
$$
J_n(\hat\beta, \tilde\sigma^2 ) = \begin{pmatrix}
\frac{1}{  (\tilde{\sigma}^2)^2 }  \sumn  \hat{\varepsilon}_i^2 x_i x_i^{\T}  & *\\
*&*
\end{pmatrix},
$$
where the $*$ terms do not matter for the later calculations. If the Normal linear model is correctly specified, we can use the $(1,1)$th block of $I_n(\hat\beta, \tilde\sigma^2 )^{-1}$ as the covariance estimator for $\hat{\beta}$, which equals
$$
 \tilde{\sigma}^2  \left( \sumn x_i x_i^{\T} \right)^{-1}.
$$
If the Normal linear model is  misspecified, we can use the $(1,1)$th block of $I_n(\hat\beta, \tilde\sigma^2 )^{-1} J_n(\hat\beta, \tilde\sigma^2 ) I_n(\hat\beta, \tilde\sigma^2 )^{-1}$ as the covariance estimator for $\hat{\beta}$, which equals
$$
\left( \sumn x_i x_i^{\T} \right)^{-1}
\left(   \sumn  \hat{\varepsilon}_i^2 x_i x_i^{\T}  \right) 
\left( \sumn x_i x_i^{\T} \right)^{-1},
$$
the EHW robust covariance estimator introduced in Chapters \ref{chapter::EHW} and \ref{chapter::populationOLS}. 
\end{example}

%
%
%
%

\section{Homework problems}

\paragraph{Estimating the mean}\label{hw-math4::mean}

Example \ref{eg::sample-variance} concerns the asymptotic properties. This problem supplements it with more finite-sample results.

Slightly modify the sandwich covariance estimator to 
$$
\tilde{V} = \frac{1}{n(n-1)} \sumn (x_i - \bar{x})^2.
$$
Prove that $E(\tilde{V} ) = \var(\bar{x})$, if $x_1,\ldots, x_n$ are independent with the same mean $\mu$, and $E(\tilde{V} ) \geq  \var(\bar{x})$, if the $x_i $'s are independent but have different means $\mu_i$'s and variances $\sigma_i^2$'s.

\paragraph{Sample Pearson correlation coefficient}\label{hw-math4::pearson-cor-asym}

Assume that $(x_i, y_i)_{i=1}^n$ are IID draws from $(x,y)$ with mean $(\mu_x, \mu_y)$ and fourth moments. Derive the asymptotic distribution of the sample Pearson correlation coefficient. Express the asymptotic variance in terms of
$$
\mu_{kl} = E\{  (x-\mu_x)^k (y-\mu_y)^l \},
$$
for example, 
$$
\var(x) = \mu_{20}, \quad  \var(y) = \mu_{02},   \quad \cov(x,y) = \mu_{11},  \quad \rho = \mu_{11} / \sqrt{ \mu_{20}   \mu_{02} } .
$$

Remark: Use the fact that $\beta = (\mu_x, \mu_y, \mu_{20}, \mu_{02}, \rho)^{\T}$ satisfies the estimating equation with
$$
m(x,y,\beta) = \begin{pmatrix}
x - \mu_x \\
y - \mu_y \\
(x - \mu_x)^2 - \mu_{20} \\
(y - \mu_y)^2 - \mu_{02} \\
(x - \mu_x)(y - \mu_y) - \rho \sqrt{ \mu_{20}  \mu_{02} }
\end{pmatrix}
$$
and the sample moments and Pearson correlation coefficient are the corresponding estimators. 
You may also find the formula \eqref{eq::3X3-lower-inverse} useful.


\paragraph{A misspecified Exponential model}\label{hw00math3::sandwich-exponential}
Assume that $y_1,\ldots, y_n \iidsim $ Exponential distribution with  mean $\mu  $. Find the MLE of $\mu$ and its asymptotic variance estimators under correctly specified and incorrectly specified models.

\bibliographystyle{apalike}
\bibliography{causal}

\printindex

\end{document}